\pdfoutput=1
\documentclass{article}
\usepackage{epubtk}
\usepackage{booktabs}
\usepackage{tabularx}
\usepackage{amsthm}
\usepackage{latexsym}
\usepackage{multibox}
\usepackage{amssymb}
\usepackage{amsfonts}
\usepackage{array}
\usepackage{amsbsy}
\usepackage[centertags,intlimits]{amsmath}
\usepackage{graphicx}
\usepackage{eucal}
\usepackage{lscape}
\usepackage{makeidx}

\makeindex
\usepackage{ifpdf}

\newcommand{\al}{\alpha}
\newcommand{\be}{\beta}
\newcommand{\ga}{\gamma}
\newcommand{\de}{\delta}
\newcommand{\si}{\sigma}

\newcommand{\f}{\frac}
\newcommand{\mc}{\mathcal}
\newcommand{\la}{\lambda}

\newcommand{\om}{\omega}
\newcommand{\pa}{\partial}

\newcommand{\ti}{\tilde}

\newcommand{\mf}{\mathfrak}

\newcommand{\ph}{\phantom}

\newcommand{\hb}{\hat{\be}}

\newcommand{\hvp}{\hat{\varphi}}
\newcommand{\ha}{\hat{\alpha}}

\newcommand{\hm}{\hat{m}}
\newcommand{\mbb}{\mathbb}
\newcommand{\bpm}{\begin{pmatrix}}
\newcommand{\epm}{\end{pmatrix}}
\newcommand{\g}{\mathrm{g}}

\newcommand{\invisible}[2]{{\vrule height #1 truemm depth #2 truemm width 0mm}}
\newcommand{\bpi}{\boldsymbol{\pi}}

\newcommand{\CC}{\mathbb C}
\newcommand{\RR}{\mathbb R}
\newcommand{\ZZ}{\mathbb Z}
\newcommand{\QQ}{\mathbb Q}
\newcommand{\NN}{\mathbb N}
\newcommand{\HH}{\mathbb H}

\newcommand{\GO}{{\mathfrak{g}}_{0}}
\newcommand{\HO}{{\mathfrak{h}}_{0}}

\newcommand{\AO}{{\mathfrak{a}}_{0}}
\newcommand{\TO}{{\mathfrak{t}}_{0}}
\newcommand{\cG}{{\mathfrak{g}}}
\newcommand{\cH}{{\mathfrak{h}}}
\newcommand{\cK}{{\mathfrak{k}}}
\newcommand{\cA}{{\mathfrak{a}}}
\newcommand{\cN}{{\mathfrak{n}}}

\newcommand{\cM}{{\mathfrak{m}}}
\newcommand{\MC}{{\mathfrak{m}}^{\CC}}
\newcommand{\KO}{{\mathfrak{k}}_{0}}
\newcommand{\cP}{{\mathfrak{p}}}
\newcommand{\cS}{{\mathfrak{s}}}
\newcommand{\cT}{{\mathfrak{t}}}
\newcommand{\TC}{{\mathfrak{t}}^{\CC}}
\newcommand{\PO}{{\mathfrak{p}}_{0}}
\newcommand{\GC}{{\mathfrak{g}}^{\CC}}
\newcommand{\Gs}{{\mathfrak{g}}_{\sigma}}
\newcommand{\Uh}{{\mathfrak{u}}_{\theta}}
\newcommand{\Ut}{{\mathfrak{u}}_{\tau}}
\newcommand{\GR}{{\mathfrak{g}}^{\RR}}
\newcommand{\HC}{{\mathfrak{h}}^{\CC}}
\newcommand{\HR}{{\mathfrak{h}}^{\RR}}
\newcommand{\fHr}{{\mathfrak{h}}_{\RR}}

\newcommand{\agl}{{\bf g}_{\lambda}}
\newcommand{\aga}{{\bf g}_{\alpha}}
\newcommand{\agb}{{\bf g}_{\beta}}
\newcommand{\bag}{{\bf g}}
\newcommand{\ag}[1]{{\bf g}_{#1}}

\newcommand{\op}{\oplus}

\newcommand{\aD}{\alpha\in\Delta}

\newcommand{\Lb}[2]{\left[#1,\,#2\right ]}

\newcommand{\p}{\prime}
\newcommand{\pp}{\prime\prime}
\newcommand{\ppp}{\prime\prime\prime}
\newcommand{\mcg}{\mc{G}}
\newcommand{\mck}{\mc{K}(\mc{G})}
\newcommand{\mcgk}{\mc{G}/\mc{K}(\mc{G})}
\newcommand{\mfVV}{\pa_{\mu}\mathrm{V}(x)\mathrm{V}(x)^{-1}}
\newcommand{\tV}{\mathrm{V}}
\newcommand{\mcBe}{\mc{B}_{\mc{M}_{\be}}}

\newcommand{\mgb}{\bar{\mf{g}}}
\newcommand{\mh}{\mf{h}}
\newcommand{\mA}{\mc{A}}
\newcommand{\mK}{\mc{K}}
\newcommand{\mG}{\mc{G}}
\newcommand{\mN}{\mc{N}}

\def\ch{{\cal H}}
\def\cn{{\cal N}}
\def\ca{{\cal A}}
\def\ce{{\cal E}}
\def\cf{{\cal F}}

\newenvironment{theorem}
{\vspace{1 em} \noindent{\bf Theorem:}}
{\vspace{1 em}}

\newenvironment{maintheorem}
{\vspace{1 em} \noindent{\bf Main theorem:}}
{\vspace{1 em}}

\newenvironment{newproof}
{\vspace{1 em} \noindent{\bf Proof:}}
{\vspace{1 em}}

\newenvironment{proposition}
{\vspace{1 em} \noindent{\bf Proposition:}}
{\vspace{1 em}}

\newcounter{lemma}
\newenvironment{lemma}
{\refstepcounter{lemma}\vspace{1 em} \noindent{\bf Lemma~\thelemma:}}
{\vspace{1 em}}

\DeclareMathOperator{\diag}{diag}
\DeclareMathOperator{\Exp}{Exp}
\DeclareMathOperator{\spn}{span}
\DeclareMathOperator{\rank}{rank}
\DeclareMathOperator{\ad}{ad}
\DeclareMathOperator{\Ad}{Ad}
\DeclareMathOperator{\Real}{Re}
\DeclareMathOperator{\Tr}{Tr}
\DeclareMathOperator{\mult}{mult}
\DeclareMathOperator{\htx}{ht}

\begin{document}

\title{Spacelike Singularities and Hidden Symmetries of Gravity}

\author{\epubtkAuthorData{Marc Henneaux}
        {Physique Th\'eorique et Math\'ematique \\
         Universit{\'e}\ Libre de Bruxelles \& International Solvay Institutes \\
         Boulevard du Triomphe, ULB -- C.P. 231 \\
         B-1050 Bruxelles, Belgium}
        {henneaux@ulb.ac.be}
        {}
        \and
        \epubtkAuthorData{Daniel Persson}
        {Physique Th\'eorique et Math\'ematique \\
         Universit{\'e}\ Libre de Bruxelles \& International Solvay Institutes \\
         Boulevard du Triomphe, ULB -- C.P. 231 \\
         B-1050 Bruxelles, Belgium}
        {dpersson@ulb.ac.be}
        {}
        \and
        \epubtkAuthorData{Philippe Spindel}
        {Service de M\'ecanique et Gravitation \\
         Universit{\'e}\ de Mons-Hainaut, Acad\'emie Wallonie-Bruxelles \\
         Avenue du Champ de Mars 6,
         B-7000 Mons, Belgium}
        {philippe.spindel@umh.ac.be}
        {}}

\date{}
\maketitle

\begin{abstract}
  We review the intimate connection between (super-)gravity close to a
  spacelike singularity (the ``BKL-limit'') and the theory of Lorentzian
  Kac--Moody algebras. We show that in this limit the gravitational
  theory can be reformulated in terms of billiard motion in a region of
  hyperbolic space, revealing that the dynamics is completely determined
  by a (possibly infinite) sequence of reflections, which are elements
  of a Lorentzian Coxeter group. Such Coxeter groups are the Weyl groups
  of infinite-dimensional Kac--Moody algebras, suggesting that these
  algebras yield symmetries of gravitational theories. Our presentation
  is aimed to be a self-contained and comprehensive treatment of the
  subject, with all the relevant mathematical background material
  introduced and explained in detail. We also review attempts at making
  the infinite-dimensional symmetries manifest, through the construction
  of a geodesic sigma model based on a Lorentzian Kac--Moody algebra. An
  explicit example is provided for the case of the hyperbolic algebra
  $E_{10}$, which is conjectured to be an underlying symmetry of
  M-theory. Illustrations of this conjecture are also discussed in the
  context of cosmological solutions to eleven-dimensional supergravity.
\end{abstract}

\epubtkKeywords{hidden symmetries, duality, Kac-Moody algebras}

\newpage

\tableofcontents

\listoftables

\newpage

\section{Introduction}
\label{section:Introduction}
\setcounter{equation}{0}

It has been realized long ago that spacetime singularities are
generic in classical general relativity~\cite{HawkingEllis}.
However, their exact nature is still far from being well
understood. Although it is expected that spacetime singularities
will ultimately be resolved in a complete quantum theory of gravity, understanding
their classical structure is likely to shed interesting light and
insight into the nature of the mechanisms at play in the singularity
resolution. Furthermore, analyzing general relativity close to such
singularities also provides important information on
the dynamics of gravity within the regime where it breaks
down. Indeed, careful investigations of the field equations in this
extreme regime has revealed
interesting and unexpected symmetry properties of
gravity.

In the late 1960's, Belinskii, Khalatnikov and Lifshitz
(``BKL'')~\cite{BKL} gave a general description of spacelike
singularities in the context of the four-dimensional vacuum
Einstein theory. They provided convincing evidence that the
generic solution of the dynamical Einstein equations, in the
vicinity of a spacelike singularity, exhibits the following
remarkable properties:

\begin{itemize}
\item The spatial points dynamically decouple, i.e., the partial
  differential equations governing the dynamics of the spatial metric
  asymptotically reduce, as one goes to the singularity, to ordinary
  differential equations with respect to time (one set of ordinary
  differential equations per spatial point).
\item The solution exhibits strong chaotic properties of the type
  investigated independently by Misner~\cite{Misner} and called
  ``mixmaster behavior''. This chaotic behavior is best seen in the
  hyperbolic billiard reformulation of the dynamics due to
  Chitre~\cite{Chitre} and Misner~\cite{Misnerb} (for pure gravity in
  four spacetime dimensions).
\end{itemize}


\subsection{Cosmological billiards and hidden symmetries of gravity}

This important work has opened the way to many further fruitful
investigations in theoretical cosmology. Recently, a new -- and
somewhat unanticipated -- development has occurred in the field,
with the realisation that for the gravitational theories that have
been studied most (pure gravity and supergravities in various
spacetime dimensions) the dynamics of the gravitational field
exhibits strong connections with Lorentzian Kac--Moody
algebras, as discovered by Damour and
Henneaux~\cite{ArithmeticalChaos}, suggesting that these might be
``hidden'' symmetries of the theory.

These connections appear for the cases at hand because in the
BKL-limit, not only can the equations of motion be reformulated as
dynamical equations for billiard motion in a region of hyperbolic
space, but also this region possesses unique features: It is the
fundamental Weyl chamber of some Kac--Moody algebra. The dynamical
motion in the BKL-limit is then a succession of reflections in the
walls bounding the fundamental Weyl chamber and defines ``words'' in
the Weyl group of the Kac--Moody algebra.

Which billiard region of hyperbolic space \index{hyperbolic space}
actually emerges -- and
hence which Kac--Moody algebra is relevant -- depends on the
theory at hand, i.e., on the spacetime dimension, the menu of matter
fields, and the dilaton couplings. The most celebrated case is
eleven-dimensional supergravity, for which the billiard region is
the fundamental region of $E_{10} \equiv E_8^{++}$, one of the
four hyperbolic Kac--Moody algebras of highest rank 10. The root
lattice of $E_{10}$ is furthermore one of the few even,
Lorentzian, self-dual lattices -- actually the only one in 10
dimensions -- a fact that could play a key role in our ultimate
understanding of M-theory.

Other gravitational theories lead to other billiards characterized
by different algebras. These algebras are closely connected to
the hidden duality groups that emerge upon dimensional reduction
to three dimensions~\cite{InvarianceUnderCompactification,
  HenneauxJulia}.

That one can associate a regular billiard and an infinite discrete
reflection group (Coxeter group) to spacelike singularities of a
given gravitational theory in the BKL-limit is a robust fact (even
though the BKL-limit itself is yet to be fully understood), which,
in our opinion, will survive future developments. The mathematics
necessary to appreciate the billiard structure and its connection to
the duality groups in three dimensions involve hyperbolic Coxeter
groups, Kac--Moody algebras and real forms of Lie algebras.

The appearance of infinite Coxeter groups related to Lorentzian
Kac--Moody algebras has triggered fascinating conjectures on the
existence of huge symmetry structures underlying
gravity~\cite{DHN2}. Similar conjectures based on different
considerations had been made earlier in the pioneering
works~\cite{Julia:1980gr, E11andMtheory}. The status of these
conjectures, however, is still somewhat unclear since, in particular,
it is not known how exactly the symmetry would act.

The main purpose of this article is to explain the emergence of
infinite discrete reflection groups in gravity in a self-contained
manner, including giving the detailed mathematical background
needed to follow the discussion. We shall avoid, however,
duplicating already existing reviews on BKL billiards.

Contrary to the main core of the review, devoted to an explanation of
the billiard Weyl groups, which is indeed rather complete, we shall
also discuss some paths that have been taken towards revealing the
conjectured infinite-dimensional Kac--Moody symmetry. Our goal here
will only be to give a flavor of some of the work that has been done
along these lines, emphasizing its dynamical relevance. Because we
feel that it would be premature to fully review this second subject,
which is still in its infancy, we shall neither try to be
exhaustive nor give detailed treatments.


\subsection{Outline of the paper}

Our article is organized as follows. In Section~\ref{section:BKL}, we
outline the key features of the BKL phenomenon, valid in any
number of dimensions, and describe the billiard formulation which
clearly displays these features. Since the derivation of these
aspects have been already reviewed in~\cite{DHNReview}, we give
here only the results without proof. Next, for completeness, we
briefly discuss the status of the BKL conjecture -- assumed to be
valid throughout our review.

In Sections~\ref{section:Coxeter} and~\ref{section:KacMoody}, we
develop the mathematical tools necessary for apprehending those
aspects of Coxeter groups and Kac--Moody algebras that are needed in
the BKL analysis. First, in Section~\ref{section:Coxeter}, we provide
a primer on Coxeter groups (which are the mathematical structures that
make direct contact with the BKL billiards). We then move on to
Kac--Moody algebras in Section~\ref{section:KacMoody}, and we discuss,
in particular, some prominent features of \emph{hyperbolic} Kac--Moody
algebras.

In Section~\ref{section:KMBilliardsI} we then make use of these
mathematical concepts to relate the BKL billiards to Lorentzian
Kac--Moody algebras. We show that there is a simple connection between
the relevant Kac--Moody algebra and the U-duality algebras that appear
upon toroidal dimensional reduction to three dimensions, when these
U-duality algebras are split real forms. The Kac--Moody algebra is then
just the standard overextension of the U-duality algebra in question.

To understand the non-split case requires an understanding of real
forms of finite-dimensional semi-simple Lie algebras. This
mathematical material is developed in
Section~\ref{section:FiniteRealLieAlgebras}. Here, again, we have
tried to be both rather complete and explicit through the use of many
examples. We have followed a pedagogical approach privileging
illustrative examples over complete proofs (these can be
found in any case in the references given in the text). We
explain the complementary Vogan and Tits--Satake approaches,
where maximal compact and maximal noncompact Cartan subalgebras play
the central roles, respectively. The concepts of restricted root
systems and of the Iwasawa decomposition, central for understanding the
emergence of the billiard, have been given particular attention. For
completeness we provide tables listing all real forms of finite Lie
algebras, both in terms of Vogan diagrams and in terms of Tits--Satake
diagrams. In Section~\ref{section:KMBilliardsII} we use these
mathematical developments to relate the Kac--Moody billiards in the
non-split case to the U-duality algebras appearing in three
dimensions.

Up to (and including) Section~\ref{section:KMBilliardsII}, the
developments present well-established results. With
Section~\ref{section:LevelDecomposition} we initiate a journey into
more speculative territory. The presence of hyperbolic Weyl groups
suggests that the corresponding infinite-dimensional Kac--Moody
algebras might, in fact, be true underlying symmetries of the
theory. How this conjectured symmetry should actually act on the
physical fields is still unclear, however. We explore one approach in
which the symmetry is realized nonlinearly on a $(1+0)$-dimensional
sigma model based on $\mc{E}_{10}/\mc{K}(\mc{E}_{10})$, which is the
case relevant to eleven-dimensional supergravity. To this end, in
Section~\ref{section:LevelDecomposition} we introduce the concept of a
level decomposition of some of the relevant Kac--Moody algebras in
terms of finite regular subalgebras. This is necessary for studying
the sigma model approach to the conjectured infinite-dimensional
symmetries, a task undertaken in Section~\ref{section:sigmamodels}. We
show that the sigma model for $\mc{E}_{10}/\mc{K}(\mc{E}_{10})$
spectacularly reproduces important features of eleven-dimensional
supergravity. However, we also point out important limitations of the
approach, which probably does not constitute the final word on the
subject.

In Section~\ref{section:cosmologicalsolutions} we show that the
interpretation of eleven-dimensional supergravity in terms of a
manifestly $\mc{E}_{10}$-invariant sigma model sheds interesting and
useful light on certain cosmological solutions of the theory. These
solutions were derived previously but without the Kac--Moody algebraic
understanding. The sigma model approach also suggests a new method of
uncovering novel solutions. Finally, in
Section~\ref{section:conclusions} we present a concluding discussion
and some suggestions for future research.

\newpage


\section{The BKL Phenomenon}
\label{section:BKL}
\setcounter{equation}{0}

In this section, we explain the main ideas of the billiard 
\index{cosmological billiard|bb} description of the BKL behavior. Our
approach is based on the billiard review~\cite{DHNReview}, from which
we adopt notations and conventions. We shall here only outline the
logic and provide the final results. No attempt will be made to
reproduce the (sometimes heuristic) arguments underlying the
derivation.


\subsection{The general action}

We are interested in general theories describing Einstein gravity
coupled to bosonic ``matter'' fields. The only known bosonic matter
fields that consistently couple to gravity are $p$-form fields, so
our collection of fields will contain, besides the metric,
$p$-form fields, including scalar fields ($p=0$). The action reads 
\begin{equation}
  S\left[ g_{\mu \nu}, \phi, A^{(p)}\right] =
  \int d^D x \, \sqrt{-{}^{(D)}\g}
  \left[R - \partial_\mu \phi \, \partial^\mu \phi -
  \frac{1}{2} \sum_p \frac{e^{\lambda^{(p)} \phi}}{(p+1)!}
  F^{(p)}_{\mu_1 \cdots \mu_{p+1}}
  F^{(p) \, \mu_1 \cdots \mu_{p+1}} \right] + \mbox{``more''}, 
  \label{keyaction}
\end{equation}
where we have chosen units such that $16 \pi G = 1$. The
spacetime dimension is left unspecified. The Einstein metric
$\g_{\mu \nu}$ has Lorentzian signature $(-, +, \cdots, +)$ and is
used to lower or raise the indices. Its determinant is
$\!^{(D)}\g$, where the index $D$ is used to avoid any confusion
with the determinant of the spatial metric introduced below. We
assume that among the scalars, there is only one
dilaton\epubtkFootnote{This is done mostly for notational convenience.
  If there were other dilatons among the 0-forms, these should be
  separated off from the $p$-forms because they play a distinct
  role. They would appear as additional scale factors and would
  increase the dimensions of the relevant hyperbolic billiard (they
  define additional spacelike directions in the space of scale
  factors).}, denoted $\phi$, whose kinetic term is normalized with weight 1 
with respect to the Ricci scalar. The real parameter
$\lambda^{(p)}$ measures the strength of the coupling to the
dilaton. The other scalar fields, sometimes called axions, are
denoted $A^{(0)}$ and have dilaton coupling $\lambda^{(0)}
\not=0$. The integer $p \geq 0$ labels the various $p$-forms
$A^{(p)}$ present in the theory, with field strengths $F^{(p)}=dA^{(p)}$, 
\begin{equation}
  F^{(p)}_{\mu_1 \cdots \mu_{p+1}} =
  \partial_{\mu_1} A^{(p)}_{\mu_2 \cdots \mu_{p+1}} \pm
  p\ \mbox{permutations}.
  \label{fieldstrength}
\end{equation}
We assume the form degree $p$ to be strictly smaller than $D-1$, since a
$(D-1)$-form in $D$ dimensions carries no local degree of freedom.
Furthermore, if $p = D-2$ the $p$-form is dual to a scalar and we
impose also $\lambda^{(D-2)} \not=0$.

The field strength, Equation~(\ref{fieldstrength}), could be modified by
additional coupling terms of Yang--Mills or Chapline--Manton
type~\cite{pvnetal, Chapline:1982ww} (e.g., $F_C = dC^{(2)} - C^{(0)}
dB^{(2)}$ for two 2-forms $C^{(2)}$ and $B^{(2)}$ and a 0-form
$C^{(0)}$, as it occurs in ten-dimensional type IIB supergravity), but
we include these additional contributions to the action in
``more''. Similarly, ``more'' might contain Chern--Simons terms, as in
the action for eleven-dimensional supergravity~\cite{CJS}.

We shall at this stage consider arbitrary dilaton couplings and
menus of $p$-forms. The billiard derivation given below remains
valid no matter what these are; all theories described by the
general action Equation~(\ref{keyaction}) lead to the billiard picture.
However, it is only for particular $p$-form menus, spacetime
dimensions and dilaton couplings that the billiard region is
regular and associated with a Kac--Moody algebra. This will be
discussed in Section~\ref{section:KMBilliardsI}. Note that the action,
Equation~(\ref{keyaction}), contains as particular cases the bosonic
sectors of all known supergravity theories.


\subsection{Hamiltonian description}

We assume that there is a spacelike singularity at a finite
distance in proper time. We adopt a spacetime slicing adapted to
the singularity, which ``occurs'' on a slice of constant time. We
build the slicing from the singularity by taking pseudo-Gaussian
coordinates defined by $N = \sqrt{\g}$ and $N^i = 0$, where $N$ is
the lapse and $N^i$ is the shift~\cite{DHNReview}. Here, $\g
\equiv \det (\g_{ij})$. Thus, in some spacetime patch, the
metric reads\epubtkFootnote{Note that we have for convenience chosen
  to work with a coordinate coframe $dx^{i}$, with the imposed
  constraint $N= \sqrt{\g}$. In general, one may of course use an
  arbitrary spatial coframe, say $\theta^{i}(x)$, for which the
  associated gauge choice reads $N=w(x)\sqrt{\g}$, with $w(x)$ being a
  density of weight $-1$. Such a frame will be used in
  Section~\ref{section:SpatialHomogeneity}. This general kind of
  spatial coframe was also used extensively in the recent
  work~\cite{Damour:2007nb}.}
\begin{equation}
  ds^2 = - \g (dx^0)^2 + \g_{ij}(x^0, x^i) \, dx^i \, dx^j ,
\end{equation}
where the local volume $\g$ collapses at each spatial
point as $x^0 \rightarrow +\infty$, in such a way that the proper
time $dT = - \sqrt{\g}\, dx^0$ remains finite (and tends
conventionally to $0^+$). Here we have assumed the singularity to
occur in the past, as in the original BKL analysis, but a similar
discussion holds for future spacelike singularities.


\subsubsection{Action in canonical form}

In the Hamiltonian description of the dynamics, the canonical
variables are the spatial metric components $\g_{ij}$, the dilaton
$\phi$, the spatial $p$-form components $A^{(p)}_{m_1 \cdots m_p}$
and their respective conjugate momenta $\pi^{ij}$, $\pi_\phi$ and
$\pi_{(p)}^{m_1 \cdots m_p}$. The Hamiltonian action in the
pseudo-Gaussian gauge is given by 
\begin{equation}
  S \left[ \g_{ij}, \pi^{ij}\!, \phi, \pi_\phi,
  A^{(p)}_{m_1 \cdots m_p}, \pi_{(p)}^{m_1 \cdots m_p}\right] \!=\!
  \int \! dx^0 \left[\int d^d x \left( \! \pi^{ij} \dot{\g_{ij}} +
  \pi_\phi \dot{\phi} + \sum_p \pi_{(p)}^{m_1 \cdots m_p}
  \dot{A}^{(p)}_{m_1 \cdots m_p} \! \right) - H \right]\!, \quad
  \label{GaussAction}
\end{equation}
where the Hamiltonian is 
\begin{equation}
  \begin{array}{rcl}
    H &=& \displaystyle
    \int d^dx \, \ch,
    \\ [1.0 em]
    \ch &=& K' + V', 
    \\ [0.2 em]
    K' &=& \displaystyle
    \pi^{ij}\pi_{ij} - \frac{1}{d-1} (\pi^i_i)^2 +
    \frac{1}{4} (\pi_\phi)^2 +
    \sum_p \frac{(p!) e^{- \lambda^{(p)} \phi}}{2} \,
    \pi_{(p)}^{m_1 \cdots m_p} \pi_{(p) \, m_1 \cdots m_p},
    \\ [0.5 em]
    V' &=& \displaystyle
    - R \g + \g^{ij} \g \, \partial_i \phi \, \partial_j \phi +
    \sum_p \frac{e^{ \lambda^{(p)} \phi}}{2 \, (p+1)!} \, \g \,
    F^{(p)}_{m_1 \cdots m_{p+1}} F^{(p) \, m_1 \cdots m_{p+1}}. 
  \end{array}
\end{equation}
In addition to
imposing the coordinate conditions $N = \sqrt{\g}$ and $N^i = 0$,
we have also set the temporal components of the $p$-forms equal to
zero (``temporal gauge'').

The dynamical equations of motion are obtained by varying the
above action w.r.t.\ the canonical variables. Moreover, there are
constraints on the dynamical variables, which are 
\begin{equation}
  \begin{array}{rcl}
    \ch &=& 0 \qquad \mbox{(``Hamiltonian constraint'')}, \index{Hamiltonian constraint}
    \\ [0.2 em]
    \ch_i &=& 0 \qquad \mbox{(``momentum constraint'')}, 
    \\ [0.2 em]
    \varphi_{(p)}^{m_1 \cdots m_{p-1}} &=& 0 \qquad
    (\mbox{``Gauss law'' for each }p\mbox{-form, }p>0).
    \label{Gauss}
  \end{array}
\end{equation}
Here we have set 
\begin{equation}
  \begin{array}{rcl}
    \ch_i &=& \displaystyle
    -2 \pi_{i \; \; \vert j}^j + \pi_\phi \partial_i \phi +
    \sum_p \pi_{(p)}^{m_1 \cdots m_p} F^{(p)}_{i m_1 \cdots m_{p}},
    \\ [0.5 em]
    \varphi_{(p)}^{m_1 \cdots m_{p-1}} &=& \displaystyle
    - p \, \pi^{m_1 \cdots m_{p-1} m_p}_{(p)}{}_{\vert m_p},
  \end{array}
\end{equation}
where the subscript $| m_p$ denotes the spatially covariant
derivative. These constraints are preserved by the dynamical evolution
and need to be imposed only at one ``initial'' time, say at $x^0 = 0$.


\subsubsection{Iwasawa change of variables}

In order to study the dynamical behavior of the fields as $x^0
\rightarrow \infty$ ($\g \rightarrow 0$) and to exhibit the billiard
\index{cosmological billiard}
picture, it is particularly convenient to perform the Iwasawa
decomposition \index{Iwasawa decomposition|bb}of the spatial
metric. Let $\g(x^{0},x^{i})$ be the matrix with entries $\g_{ij}(x^{0},x^{i})$. We set 
\begin{equation}
  \g = \cn^{T} \ca^2 \, \cn, 
  \label{IwasawaII}
\end{equation}
where $\cn$ is an upper triangular matrix with $1$'s on the diagonal
($\cn_{ii} =1$, $\cn_{ij} = 0$ for $i>j$) and $\ca$ is a
diagonal matrix with positive elements, which we parametrize as
\begin{equation}
  \ca = \exp(-\beta),
  \qquad \beta = \diag (\beta^1, \beta^2, \cdots, \beta^d).
\end{equation}
Both $\cn$ and $\ca$ depend on the spacetime
coordinates. The spatial metric $d\sigma^2$ becomes
\begin{equation}
  d\sigma^2 = \g_{ij} \, dx^i \, dx^j =
  \sum_{k=1}^d e^{( -2\beta^k)} (\omega^k)^2
\end{equation}
with
\begin{equation}
  \omega^k = \sum_i \cn_{k \, i} \, dx^i.
\end{equation}
The variables $\beta^i$ of the Iwasawa decomposition give the
(logarithmic) scale factors in the new, orthogonal, basis. The
variables $\cn_{ij}$ characterize the change of basis that
diagonalizes the metric and hence they parametrize the off-diagonal
components of the original $\g_{ij}$.

We extend the transformation Equation~(\ref{IwasawaII}) in configuration
space to a canonical transformation in phase space through the
formula 
\begin{equation}
  \pi^{ij} d\g_{ij} = \pi^i d \beta_i + \sum_{i<j} P_{ij} \, d\cn_{ij} .
\end{equation}

Since the scale factors and the off-diagonal variables play very
distinct roles in the asymptotic behavior, we split off the
Hamiltonian as a sum of a kinetic term for the scale factors
(including the dilaton), 
\begin{equation}
  K = \frac{1}{4} \left[ \sum_{i=1}^d \pi_i^2 - \frac{1}{d-1}
  \left( \sum_{i=1}^d \pi_i \right)^2 + \pi_\phi^2 \right],
\end{equation}
plus the rest, denoted by $V$, which will
act as a potential for the scale factors. The Hamiltonian then becomes
\begin{equation}
  \begin{array}{rcl}
    \ch &=& \displaystyle
    K + V,
    \\ [0.5 em]
    V &=& \displaystyle
    V_S + V_G + \sum_p V_{p} + V_\phi,
    \\ [1.5 em]
    V_S &=& \displaystyle
    \frac{1}{2} \sum_{i<j} e^{-2(\beta^j - \beta^i)}
    \Bigl( \sum_m P_{im}\cn_{jm} \Bigr)^2,
    \\ [1.5 em]
    V_G &=& \displaystyle
    - R \g,
    \\ [0.5 em]
    V_{(p)} &=& \displaystyle
    V_{(p)}^\mathrm{el} + V_{(p)}^\mathrm{magn},
    \\ [1.0 em]
    V_{(p)}^\mathrm{el} &=& \displaystyle
    \frac{p! e^{- \lambda^{(p)} \phi}} {2} \,
    \pi_{(p)}^{m_1 \cdots m_p} \pi_{(p) \, m_1 \cdots m_p},
    \\ [1.0 em]
    V_{(p)}^\mathrm{magn} &=&  \displaystyle
    \frac{e^{\lambda^{(p)} \phi}}{2 \, (p+1)!}
    \, \g \, F^{(p)}_{m_1 \cdots m_{p+1}} F^{(p) \, m_1 \cdots m_{p+1}},
    \\ [1.5 em]
    V_\phi &=& \displaystyle
    \g^{ij} \g \, \partial_i \phi \, \partial_j \phi.
  \end{array}
\end{equation}
The kinetic term $K$ is quadratic in the momenta conjugate to the
scale factors and defines the inverse of a metric in the space of
the scale factors. \index{scale-factor space|bb} Explicitly, this metric reads 
\begin{equation}
  \sum_i(d\beta^i)^2 - \Bigl( \sum d \beta^i \Bigr)^2 + (d \phi)^2.
  \label{metricscalefactors}
\end{equation}
Since the metric coefficients do not depend on the scale factors, that
metric in the space of scale factors is flat, and, moreover, it is of
Lorentzian signature. A conformal transformation where all scale
factors are scaled by the same number ($\beta^i \rightarrow \beta^i +
\epsilon$) defines a timelike direction. It will be convenient in the
following to collectively denote all the scale factors (the $\beta^i$'s
and the dilaton $\phi$) as $\beta^\mu$, i.e., $(\beta^\mu) = (\beta^i,
\phi)$.

The analysis is further simplified if we take for new $p$-form
variables the components of the $p$-forms in the Iwasawa basis of
the $\omega^k$'s, 
\begin{equation}
  {\cal A}^{(p)}_{i_1 \cdots i_p} = \!\!\!\!
  \sum_{m_1, \cdots,m_{p}} \!\!\!\!
  (\cn^{-1})_{m_1 i_1} \cdots
  (\cn^{-1})_{m_{p} i_{p}} A_{(p) m_1 \cdots m_{p}},
\end{equation}
and again extend this configuration space transformation to a point
canonical transformation in phase space,
\begin{equation}
  \left( \cn_{ij}, P_{ij}, A^{(p)}_{m_1 \cdots m_p},
  \pi_{(p)}^{m_1 \cdots m_p} \right)
  \quad \rightarrow \quad
  \left( \cn_{ij}, P'_{ij}, {\cal A}^{(p)}_{m_1 \cdots m_p},
  \ce_{(p)}^{i_1 \cdots i_p} \right),
\end{equation}
using the formula $\sum p \, dq = \sum p'\, dq'$, which reads 
\begin{equation}
  \sum_{i<j} P_{ij} \dot{\cn}_{ij} + \sum_p \pi_{(p)}^{m_1 \cdots m_p}
  \dot{A}^{(p)}_{m_1 \cdots m_p} =
  \sum_{i<j} P'_{ij} \dot{\cn}_{ij} + \sum_p \ce_{(p)}^{i_1 \cdots i_p}
  \dot{{\cal A}}^{(p)}_{m_1 \cdots m_p}.
\end{equation}
Note that the scale factor variables are unaffected, while the momenta
$P_{ij}$ conjugate to $\cn_{ij}$ get redefined by terms involving
$\ce$, $\cn$ and $\ca$ since the components ${\cal A}^{(p)}_{m_1
 \cdots m_p}$ of the $p$-forms in the Iwasawa basis involve the
$\cn$'s. On the other hand, the new $p$-form momenta, i.e., the
components of the electric field $\pi_{(p)}^{m_1 \cdots m_p}$ in the
basis $\{\omega^k \}$ are simply given by 
\begin{equation}
  \ce^{i_1 \cdots i_p}_{(p)} = \!\!\!\!
  \sum_{m_1,\cdots,m_p} \cn_{i_1 m_1} \!\!\!\!
  \cn_{i_2 m_2} \cdots \cn_{i_p m_p} \pi_{(p)}^{m_1 \cdots m_p}.
\end{equation}
In terms of the new variables, the electromagnetic potentials
become 
\begin{equation}
  \begin{array}{rcl}
    V_{(p)}^\mathrm{el} &=& \displaystyle
    \frac{p!}{2} \!\!
    \sum_{i_1, i_2,\cdots, i_p} \!\!\!\!\!\!
    e^{-2 e_{i_1 \cdots i_p}(\beta)} (\ce^{i_1 \cdots i_p}_{(p)})^2,
    \\ [1.5 em]
    V_{(p)}^\mathrm{magn} &=& \displaystyle
    \frac{1}{2 \, (p+1)!}
    \sum_{i_1, i_2, \cdots,i_{p+1}} \!\!\!\!\!\!
    e^{-2 m_{i_{1} \cdots i_{p+1}}(\beta)} (\cf_{(p) \, i_1\cdots i_{p+1}})^2.
  \end{array}
\end{equation}
Here, $e_{i_1 \cdots i_p}(\beta)$ are the electric linear forms 
\begin{equation}
  e_{i_1 \cdots i_p}(\beta) =
  \beta^{i_1} + \cdots + \beta^{i_p} + \frac{\lambda^{(p)}}{2} \phi
\end{equation}
(the indices $i_j$ are all distinct because $\ce^{i_1 \cdots i_p}_{(p)}$ is
completely antisymmetric) while $\cf_{(p) \, i_1 \cdots i_{p+1}}$
are the components of the magnetic field $F_{(p) m_1 \cdots
m_{p+1}}$ in the basis $\{\omega^k\}$, 
\begin{equation}
  \cf_{(p) \, i_1 \cdots i_{p+1}} = \!\!\!\!
  \sum_{m_1, \cdots,m_{p+1}} \!\!\!\! (\cn^{-1})_{m_1 i_1} \cdots
  (\cn^{-1})_{m_{p+1} i_{p+1}} F_{(p) m_1 \cdots m_{p+1}},
\end{equation}
and $m_{i_{1} \cdots i_{p+1}}(\beta)$ are the magnetic linear forms
\begin{equation}
  m_{i_{1} \cdots i_{p+1}}(\beta) = \!\!\!\!\!\!
  \sum_{j \notin \{i_1,i_2,\cdots i_{p+1} \}} \!\!\!\!\!\!
  \beta^j -\frac{\lambda^{(p)}}{2} \phi.
\end{equation}
One sometimes rewrites $m_{i_{1} \cdots i_{p+1}}(\beta)$ as
$b_{i_{p+2} \cdots i_d}(\beta)$, where $\{i_{p+2}, i_{p+3},
\cdots, i_d \}$ is the set complementary to $\{i_1,i_2, \cdots
i_{p+1} \}$, e.g., 
\begin{equation}
  b_{1 \, 2 \, \cdots \, d-p-1}(\beta) = \beta^1 + \cdots +
  \beta^{d-p-1} -\frac{\lambda^{(p)}}{2} \phi = m_{d-p \, \cdots \, d}.
\end{equation}
The exterior derivative $\cf$ of
$\ca$ in the non-holonomic frame $\{\omega^k\}$ involves of course
the structure coefficients $C^i{}_{jk}$ in that frame, i.e.,
\begin{equation}
  \cf_{(p) \, i_1 \cdots i_{p+1}} =
  \partial_{[i_1} {\cal A}_{i_2 \cdots i_{p+1}]} +
  \mbox{``}C{\cal A}\mbox{''-terms},
\end{equation}
where 
\begin{equation}
  \partial_{i_1} \equiv
  \sum_{m_1} (\cn^{-1})_{m_1 i_1} (\partial/\partial x^{m_1})
\end{equation}
is here the frame derivative. Similarly, the potential $V_\phi$ reads 
\begin{equation}
  V_\phi = \sum_i e^{-2 \bar{m}_i(\beta)} (\cf_i)^2, 
\end{equation}
where $\cf_i$ is 
\begin{equation}
  \cf_i=(\cn^{-1})_{ji}\partial_j \phi
\end{equation}
and 
\begin{equation}
  \bar{m}_i(\beta) = \sum_{j\not=i} \beta^j. 
\end{equation}


\subsection{Decoupling of spatial points close to a spacelike singularity}

So far we have only redefined the variables without making any
approximation. We now start the discussion of the BKL-limit, \index{BKL-limit|bb} which
investigates the leading behavior of the fields as $x^0 \rightarrow
\infty$ ($\g \rightarrow 0$). Although the more recent
``derivations'' of the BKL-limit treat both elements at
once~\cite{DH1, DH2, ArithmeticalChaos, DHNReview}, it appears useful
-- especially for rigorous justifications -- to separate two aspects
of the BKL conjecture\epubtkFootnote{The Hamiltonian heuristic
  derivation of~\cite{DHNReview} shares many features in common with
  the work of~\cite{Kirillov1993, IKM94, IvKiMe94, KiMe}, extended to
  some higher-dimensional models in~\cite{IvMe, Ivashchuk:1999rm}. The
  central feature of~\cite{DHNReview} is the Iwasawa decomposition \index{Iwasawa decomposition}
  which enables one to clearly see the role of off-diagonal
  variables.}.

The first aspect is that the spatial points decouple in the limit $x^0
\rightarrow \infty$, in the sense that one can replace the Hamiltonian
by an effective ``ultralocal'' Hamiltonian $H^\mathrm{UL}$ involving no
spatial gradients and hence leading at each point to a set of
dynamical equations that are ordinary differential equations with
respect to time. The ultralocal effective Hamiltonian has a form
similar to that of the Hamiltonian governing certain spatially
homogeneous cosmological models, as we shall explain in this section.

The second aspect of the BKL-limit is to take the sharp wall limit \index{billiard wall} of
the ultralocal Hamiltonian. This leads directly to the billiard \index{cosmological billiard}
description, as will be discussed in
Section~\ref{section:DynamicsBilliardHyp}.


\subsubsection{Spatially homogeneous models}
\label{section:SpatialHomogeneity}

In spatially homogeneous models, the fields depend only on time in
invariant frames, e.g., for the metric
\begin{equation}
  ds^2 = \g_{ij}(x^0) \psi^i \psi^j,
  \label{coframe}
\end{equation}
where the invariant forms fulfill $$d\psi^i = -\frac{1}{2}
f^i{}_{jk} \psi^j \wedge \psi^k .$$ Here, the $f^i{}_{jk}$ are the
structure constants of the spatial homogeneity group.
Similarly, for a $1$-form and a $2$-form, 
\begin{equation}
  A^{(1)} = A_i(x^0) \psi^i,
  \qquad
  A^{(2)} = \frac{1}{2} A_{ij}(x^0) \psi^i \wedge \psi^j,
  \qquad \mbox{etc.}
\end{equation}
The Hamiltonian constraint yielding the field equations in the spatially
homogeneous context\epubtkFootnote{This Hamiltonian exists if
  $f^i{}_{ik}= 0$, as we shall assume from now on.} is obtained by
substituting the form of the fields in the general Hamiltonian
constraint and contains, of course, no explicit spatial gradients since the
fields are homogeneous. Note, however, that the structure constants
$f^i{}_{ik}$ contain implicit spatial gradients. The Hamiltonian can
now be decomposed as before and reads
\begin{equation}
  \begin{array}{rcl}
    \ch^\mathrm{UL} &=& \displaystyle
    K + V^\mathrm{UL},
    \\ [0.5 em]
    V^\mathrm{UL} &=& \displaystyle
    V_S + V_G^\mathrm{UL} +
    \sum_p \left( V_{(p)}^\mathrm{el} + V_{(p)}^\mathrm{UL,magn} \right), 
  \end{array}
  \label{calH1}
\end{equation}
where $K$, $V_S$ and
$V_{(p)}^\mathrm{el}$, which do not involve spatial gradients, are
unchanged and where $V_\phi$ disappears since $\pa_i \phi = 0$. The
potential $V_G$ is given by~\cite{Demaret} 
\begin{equation}
  V_G \equiv -\g R =
  \frac{1}{4} \!\! \sum_{i\not= j,i \not= k, j \not= k} \!\!\!\!
  e^{-2\alpha_{ijk}(\beta)} (C^i{}_{jk})^2 +
  \frac{1}{2}\sum_j e^{-2 \bar{m}_j(\beta)}
  \left(C^i{}_{jk} \, C^k{}_{ji} + \mbox{``more''} \right), 
  \label{formulaforR}
\end{equation}
where the linear forms $\alpha_{ijk}(\beta)$ (with $i, j, k$ distinct)
read 
\begin{equation}
  \alpha_{ijk}(\beta) = 2 \beta^i + \!\!\!\!\!\!
  \sum_{m \, : \, m \not= i, m \not= j, m \not= k} \!\!\!\!\!\! \beta^m, 
\end{equation}
and where ``more'' stands for the terms in the first sum that arise upon
taking $i=j$ or $i=k$. The structure constants in the Iwasawa
frame (with respect to the coframe in Equation~(\ref{coframe})) are
related to the structure constants $f^i{}_{jk}$ through 
\begin{equation}
  C^i{}_{jk} = \sum_{i',j',k'} f^{i'}{}_{j'k'}\cn^{-1}_{ii'} \cn_{jj'} \cn_{kk'}
\end{equation}
 and depend therefore
on the dynamical variables. Similarly, the potential
$V_{(p)}^\mathrm{magn}$ becomes 
\begin{equation}
  V_{(p)}^\mathrm{magn} = \frac{1}{2 \,(p+1)!}
  \!\! \sum_{i_1, i_2, \cdots, i_{p+1}} \!\!\!\!
  e^{-2 m_{i_{1} \cdots i_{p+1}}(\beta)} (\cf^h_{(p) \, i_1 \cdots i_{p+1}})^2,
\end{equation}
 where
the field strengths $\cf^h_{(p) \, i_1 \cdots i_{p+1}}$ reduce to
the ``$A C$'' terms in $dA$ and depend on the potentials and the
off-diagonal Iwasawa variables.


\subsubsection{The ultralocal Hamiltonian}

Let us now come back to the general, inhomogeneous case and
express the dynamics in the frame $\{dx^0, \psi^i\}$ where the
$\psi^i$'s form a ``generic'' non-holonomic frame in space, 
\begin{equation}
  d\psi^i = -\frac{1}{2} f^i{}_{jk}(x^m) \, \psi^j \wedge \psi^k.
\end{equation}
Here the $f^i{}_{jk}$'s are in general
space-dependent. In the non-holonomic frame, the exact Hamiltonian
takes the form 
\begin{equation}
  \ch = \ch^\mathrm{UL} + \ch^\mathrm{gradient},
\end{equation}
where the ultralocal part $\ch^\mathrm{UL}$ is given by
Equations~(\ref{calH1}) and~(\ref{formulaforR}) with the relevant
$f^i{}_{jk}$'s, and where $\ch^\mathrm{gradient}$ involves the spatial
gradients of $f^i{}_{jk}$, $\beta^m$, $\phi$ and $\cn_{ij}$.

The first part of the BKL conjecture states that one can drop
$\ch^\mathrm{gradient}$ asymptotically; namely, the dynamics of a
generic solution of the Einstein--$p$-form-dilaton equations (not
necessarily spatially homogeneous) is asymptotically determined,
as one goes to the spatial singularity, by the ultralocal
Hamiltonian 
\begin{equation}
  H^\mathrm{UL} = \int d^dx \, \ch^\mathrm{UL},
\end{equation}
provided that the phase space constants
$f^{i}{}_{jk}(x^m) = - f^{i}{}_{kj}(x^m)$ are such that all
exponentials in the above potentials
do appear. In other words, the $f$'s must be chosen such that none
of the coefficients of the exponentials, which involve $f$ and the
fields, identically vanishes -- as would be the case, for example,
if $f^{i}{}_{jk} = 0$ since then the potentials $V_G$ and
$V_{(p)}^\mathrm{magn}$ are equal to zero. This is always possible
because the $f^{i}{}_{jk}$, even though independent of the
dynamical variables, may in fact depend on $x$ and so are not
required to fulfill relations ``$ff = 0$'' analogous to the Bianchi
identity since one has instead ``$\partial f + ff = 0$''.

\subsubsection*{Comments}

\begin{enumerate}
\item As we shall see, the conditions on the $f$'s (that all
  exponentials in the potential should be present) can be considerably
  weakened. It is necessary that only the relevant exponentials (in
  the sense defined in Section~\ref{section:DynamicsBilliardHyp}) be
  present. Thus, one can correctly capture the asymptotic BKL
  behavior of a generic solution with fewer exponentials. In the case
  of eleven-dimensional supergravity the spatial curvature is
  asymptotically negligible with respect to the electromagnetic terms
  and one can in fact take a holonomic frame for which $f^{i}{}_{jk} =
  0$ (and hence also $ C^{i}{}_{jk} = 0$).
\item The actual values of the $f^{i}{}_{jk}$ (provided they
  fulfill the criterion given above or rather its weaker form just
  mentioned) turn out to be irrelevant in the BKL-limit because they
  can be absorbed through redefinitions. This is for instance why the
  Bianchi~VIII and~IX models, even though they correspond to different
  groups, can both be used to describe the BKL behavior in four
  spacetime dimensions.
\end{enumerate}


\subsection{Dynamics as a billiard in hyperbolic space}
\label{section:DynamicsBilliardHyp}

The second step in the BKL-limit \index{BKL-limit} is to take the
sharp wall limit \index{billiard wall}
of the potentials.\epubtkFootnote{In this article we will exclusively
  restrict ourselves to considerations involving the sharp wall
  limit. However, in recent work \cite{Damour:2007nb} it was
  argued that in order to have a rigorous treatment of the dynamics
  close to the singularity also in the chaotic case, it is necessary
  to go beyond the sharp wall limit. This implies that one should
  retain the exponential structure of the dominant walls.} This leads
to the billiard \index{cosmological billiard|bb} picture. It is
crucial here that the coefficients in front of the dominant walls are
all positive. Again, just as for the first step, this limit has not
been fully justified. Only heuristic, albeit convincing, arguments
have been put forward.

The idea is that as one goes to the singularity, the exponential
potentials get sharper and sharper and can be replaced in the
limit by the corresponding $\Theta_\infty$-function, denoted for
short $\Theta$ and defined by $\Theta(x) = 0$ for $x<0$ and
$\Theta(x) = + \infty$ for $x>0$. Taking into account the facts
that $a \Theta(x) = \Theta(x)$ for all $a>0$, as well as that some
walls can be neglected, one finds that the Hamiltonian becomes in
the sharp wall limit 
\begin{equation}
  H = \int d^d x \, \ch^\mathrm{sharp},
\end{equation}
with
\begin{eqnarray}
  \ch^\mathrm{sharp} &=&
  K + \sum_{i<j} \Theta \left(-2 s_{ji}(\beta)\right) + \!\!\!\!
  \sum_{i\not= j,i \not= k, j \not= k} \!\!\!\!
  \Theta(-2\alpha_{ijk}(\beta))
  \nonumber
  \\
  && + \!\!\!\! \sum_{i_1< i_2< \cdots< i_p} \!\!\!\!
  \Theta(-2 e_{i_1 \cdots i_p}(\beta)) + \!\!\!\!
  \sum_{i_1< i_2< \cdots< i_{p+1}} \!\!\!\!
  \Theta(-2 m_{i_{1} \cdots i_{p+1}}(\beta)),
  \label{sharp}
\end{eqnarray}
where $s_{ji}(\beta) = \beta^j -\beta^i$. See~\cite{DHNReview} for
more information.

The description of the motion of the scale factors (at each
spatial point) is easy to give in that limit. Because the
potential walls \index{billiard wall} are infinite (and positive), the motion is
constrained to the region where the arguments of all
$\Theta$-functions are negative, i.e., to 
\begin{equation}
  s_{ji}(\beta) \geq 0 \, (i<j),
  \qquad
  \alpha_{ijk}(\beta) \geq 0,
  \qquad
  e_{i_1 \cdots i_p}(\beta) \geq 0,
  \qquad
  m_{i_{1} \cdots i_{p+1}}(\beta) \geq 0.
\end{equation}
In that region, the motion is governed by the kinetic term $K$, i.e.,
is a geodesic for the metric in the space of the scale factors. Since
that metric is flat, this is a straight line. In addition, the
constraint $\ch=0$, which reduces to $K=0$ away from the potential
walls, forces the straight line to be null. We shall assume that the
time orientation in the space of the scale factors is such that the
straight line is future-oriented ($\g \rightarrow 0$ in the future).

It is easy to check that all the walls appearing in
Equation~(\ref{sharp}), collectively denoted $F_A(\beta) \equiv F_{A
  \mu} \beta^\mu =0$, are timelike hyperplanes. This is because the
squared norms of all the $F_A$'s are positive,
\begin{equation}
  (F_A \vert F_A) =
  \sum_i \left( \frac{\partial F_A}{\partial \beta^i} \right)^2 -
  \frac{1}{d-1} \left(\sum_i \frac{\partial F_A}{\partial \beta^i} \right)^2 +
  \left( \frac{\partial F_A}{\partial \phi} \right)^2 >0.
\end{equation}
Explicitly, one finds 
\begin{equation}
  \begin{array}{rcl}
    (s_{ji} \vert s_{ji} ) &=& 2,
    \\ [0.5 em]
    (\alpha_{ijk} \vert \alpha_{ijk}) &=& 2,
    \\ [0.25 em]
    (e_{i_1 \cdots i_p} \vert e_{i_1 \cdots i_p}) &=& \displaystyle
    \frac{p (d-p-1)}{d-1} + \frac{\left( \lambda^{(p)} \right)^2}{4},
    \\ [0.75 em]
    (m_{i_{1} \cdots i_{p+1}} \vert m_{i_{1} \cdots i_{p+1}}) &=& \displaystyle
    \frac{p (d-p-1)}{d-1} + \frac{\left( \lambda^{(p)} \right)^2}{4}.
  \end{array}
\end{equation}

Because the potential walls \index{billiard wall} are timelike, they have a non-empty
intersection with the forward light cone in the space of the scale
factors. When the null straight line representing the evolution
of the scale factors hits one of the walls, it gets reflected
according to the rule~\cite{DH1} 
\begin{equation}
  v^\mu \quad \rightarrow \quad
  v^\mu - 2 \frac{v^\nu F_{A \nu}}{(F_A \vert F_A)} F_A^\mu,
  \label{billiardreflectionrule}
\end{equation}
where $v$ is the velocity vector (tangent to the straight line). This
reflection preserves the time orientation since the hyperplanes are
timelike and hence belong to the orthochronous Lorentz group
$O^\uparrow(k,1)$ where $k = d-1$ or $d$ according to whether there is
no or one dilaton. The conditions $s_{ji} = 0$ define the ``symmetry''
or ``centrifugal'' walls, the conditions $\alpha_{ijk} = 0$ define the
``curvature'' or ``gravitational'' walls, the conditions $e_{i_1
  \cdots i_p}=0$ define the ``electric'' walls, while the conditions
$m_{i_{1} \cdots i_{p+1}} = 0$ define the ``magnetic'' walls.

The motion is thus a succession of future-oriented null straight
line segments interrupted by reflections 
\index{geometric reflection|bb} against the walls, where the motion
undergoes a reflection belonging to $O^\uparrow(k,1)$. Whether the
collisions eventually stop or continue forever is better visualized by
projecting the motion radially on the positive sheet of the unit
hyperboloid, as was done first in the pioneering work of Chitre and
Misner~\cite{Chitre, Misnerb} for pure gravity in four spacetime
dimensions. We recall that the positive sheet of the unit hyperboloid
$\sum (\beta^i)^2 - (\sum \beta^i)^2 + \phi^2 = -1$, $\sum \beta^i
>0$, provides a model of hyperbolic space (see,
e.g.,~\cite{Ratcliffe}). \index{hyperbolic space|bb}

The intersection of a timelike hyperplane with the unit hyperboloid
defines a hyperplane in hyperbolic space. The region in hyperbolic
space on the positive side of all hyperplanes is the allowed dynamical
region and is called the ``billiard table''. It is never compact in
the cases relevant to gravity, but it may or may not have finite
volume. The projection of the motion of the scale factors on the unit
hyperboloid is the same as the motion of a billiard ball in a
hyperbolic billiard: \index{cosmological billiard} geodesic arcs in
hyperbolic space within the billiard region, interrupted by collisions
against the bounding walls where the motion undergoes a specular
reflection.

When the volume of the billiard table is finite, the collisions with
the potential walls \index{billiard wall} never end (for generic
initial data) and the motion is chaotic. When, on the other hand, the
volume is infinite, generic initial data lead to a motion that
ultimately freely runs away to infinity. This is non-chaotic. For more
information, see~\cite{Ma69, Z84}. An interesting criterion for chaos
(equivalent to finite volume of hyperbolic billiard region) has been
given in~\cite{Ivashchuk:1999rm} in terms of illuminations of spheres
by point sources.

\subsubsection*{Comments}

\begin{enumerate}
\item The task of determining the billiard 
  \index{cosmological billiard} region is greatly simplified by the 
  observation that some
  walls are behind others and are thus not relevant. For instance, it
  is clear that if $\beta^2 - \beta^1 >0$ and $\beta^3 - \beta^2 >0$ ,
  then $\beta^3 - \beta^1>0$. Among the symmetry wall conditions, the
  only relevant ones are $\beta^{i+1} - \beta^i >0$, $ i=1, 2, \cdots
  , d-1$. Similarly, a wall of any given type can be written as a
  positive combination of the walls of the same type with smallest
  values of the indices $i$ of the $\beta$'s and the symmetry walls
  (e.g., the electric wall condition $\beta^2 > 0$ for a $1$-form with
  zero dilaton coupling can be written as $\beta^1 + (\beta^2 -
  \beta^1)>0$ and is thus a consequence of $\beta^1 >0$ and $\beta^2 -
  \beta^1>0$). Finally, one also verifies that in the presence of true
  $p$-forms ($0<p<d-1$), the gravitational walls are never relevant as
  they can be written as combinations of $p$-form walls with positive
  coefficients~\cite{DHRW02}.
\item It is interesting to determine the spatially homogeneous models
  that reproduce asymptotically the correct billiard limit. It is
  clear that in order to do so, homogeneous cosmological models need
  only contain the relevant walls. It is not necessary that they yield
  all the walls. Which homogeneity groups are acceptable depends on the
  system at hand. We list here a few examples. For vacuum gravity in
  four spacetime dimensions, the appropriate homogeneous models are
  the so-called Bianchi~VIII or IX models. For vacuum gravity in
  higher dimensions, the structure constants of the homogeneity group
  must fulfill the conditions of~\cite{DdRH88} \emph{and the metric
  must include off-diagonal components} (see also~\cite{dBPS03}). In
  the presence of a single $p$-form and no dilaton ($0<p<d-1$), the
  simplest (Abelian) homogeneity group can be taken~\cite{DH2}.
\end{enumerate}


\subsection{Rules for deriving the wall forms from the Lagrangian -- Summary}
\label{RulesForWalls}

We have recalled above that the generic behavior near a spacelike
singularity of the system with action~(\ref{keyaction}) can
be described at each spatial point in terms of a billiard \index{cosmological billiard} in
hyperbolic space. The action for the billiard ball reads, in the gauge
$N = \sqrt{\g}$,
\begin{equation}
  S = \int d x^{0} \left[ G_{\mu \nu} \frac{d \beta^\mu}{d x^{0}}
  \frac{d \beta^\nu}{d x^{0}} - V(\beta^\mu) \right],
\end{equation}
where we recall that $x^{0}\rightarrow \infty$ in the BKL-limit \index{BKL-limit}
(proper time $T \rightarrow 0^{+}$), and $G_{\mu \nu}$ is the metric
in the space of the scale factors,
\begin{equation}
  G_{\mu \nu} \, d\beta^\mu \, d\beta^\nu =
  \sum_{i=1}^d d\beta^i \, d\beta^i -
  \left(\sum_{i=1}^{d} d\beta^i \right) \left(\sum_{j=1}^{d} d\beta^j \right) +
  d \phi \, d \phi
\end{equation}
introduced in Equation~(\ref{metricscalefactors}) above. As
stressed there, this metric is flat and of Lorentzian signature.
Between two collisions, the motion is a free, geodesic motion. The
collisions with the walls \index{billiard wall|bb} are controlled by the potential
$V(\beta^\mu)$, which is a sum of sharp wall potentials. The walls
are hyperplanes and can be inferred from the Lagrangian. They are
as follows:

\begin{enumerate}
\item Gravity brings in the symmetry walls 
  \begin{equation}
    \beta^{i+1} - \beta^i = 0, 
    \label{symmetryW}
  \end{equation}
  with $i=1, 2, \cdots , d-1$, and the curvature wall 
  \begin{equation}
    2 \beta^1 + \beta^2 + \cdots + \beta^{d-2} = 0.
    \label{curvatureW}
  \end{equation}
\item Each $p$-form brings in an electric wall 
  \begin{equation}
    \beta^1 + \cdots + \beta^p + \frac{\lambda^{(p)}}{2} \phi=0 ,
    \label{electricW}
  \end{equation}
  and a magnetic wall 
  \begin{equation}
    \beta^1 + \cdots + \beta^{d-p-1} -\frac{\lambda^{(p)}}{2} \phi = 0.
    \label{magneticW}
  \end{equation}
\end{enumerate}

\noindent
We have written here only the (potentially) relevant walls. There are
other walls present in the potential, but because these are behind the
relevant walls, which are infinitely steep in the BKL-limit, they are
irrelevant. They are relevant, however, when trying to exhibit the
symmetry in a complete treatment where the BKL-limit is the zeroth
order term in a gradient expansion yet to be understood~\cite{DHN2}.

The scalar product dual to the scalar product in the space of the
scale factors is 
\begin{equation}
  (F\vert G) = \sum_i F_i G_i - \frac{1}{d-1}
  \biggl( \sum_i F_i \biggr) \biggl( \sum_j G_j \biggr) +
  F_\phi G_\phi
\end{equation}
for two linear forms $F= F_i \beta^i + F_\phi \phi$, $G= G_i \beta^i +
G_\phi \phi$.

These recipes are all that we shall need for investigating the
regularity properties of the billiards associated with the class
of actions Equation~(\ref{keyaction}).


\subsection{More on the free motion: The Kasner solution}

The free motion between two bounces is a straight line in the
space of the scale factors. In terms of the original metric
components, it takes the form of the Kasner solution with dilaton.
Indeed, the free motion is given by
\begin{displaymath}
  \beta^\mu = q^\mu x^{0} + \beta^\mu_0,
\end{displaymath}
where the ``velocities'' $q^\mu$ are subject to
\begin{displaymath}
  \sum_i (q^i)^2 - \left(\sum_i q^i\right)^2 + q_\phi^2 = 0,
\end{displaymath}
since the motion is lightlike by the Hamiltonian
constraint. \index{Hamiltonian constraint} The proper time $dT = -
\sqrt{\g}\, d x^{0}$ is then $T = B \exp(-K x^{0})$, with $K = \sum_i q^i$ and for
some constant $B$ (we assume, as before, that the singularity is at $T=0^+$).
Redefining then
\begin{displaymath}
  p^\mu = \frac{q^\mu}{\sum_i q^i} 
\end{displaymath}
yields the celebrated Kasner solution 
\begin{eqnarray}
  ds^2 &=& - dT^2 + \sum_i T^{2 p^i} \left(dx^i \right)^2,
  \\
  \phi &=& - p_\phi \ln T + A, 
\end{eqnarray}
subject to the constraints
\begin{equation}
  \sum_i p^i = 1,
  \qquad
  \sum_i (p^i)^2 + p_\phi^2 = 1,
\end{equation}
where $A$ is a constant of integration and where the coordinates $x^i$
have been suitably rescaled (if necessary).


\subsection{Chaos and billiard volume}

With our rules for writing down the billiard region, one can determine
in which case the volume of the billiard is finite and in which case
it is infinite. The finite-volume, chaotic case is also called
``mixmaster case'', a terminology introduced in four dimensions
in~\cite{Misner}.

The following results have been obtained:

\begin{itemize}
\item Pure gravity in $D \leq 10$ dimensions is chaotic, but ceases to
  be so for $D \geq 11$~\cite{Demaret1, Demaret2}.
\item The introduction of a dilaton removes chaos~\cite{BK1,
    AR01}. The gravitational four-derivative action in four dimensions,
  based on $R^2$, is dynamically equivalent to Einstein gravity coupled
  to a dilaton~\cite{Spindel:1992md}. Hence, chaos is removed also for
  this case.
\item $p$-form gauge fields ($0<p<d-1$) without scalar fields lead to
  a finite-volume billiard~\cite{DH2}.
\item When both $p$-forms and dilatons are included, the situation is
  more subtle as there is a competition between two opposing
  effects. One can show that if the dilaton couplings are in a
  ``subcritical'' open region that contains the origin -- i.e., ``not
  too big'' -- the billiard volume is infinite and the system is non
  chaotic. If the dilaton couplings are outside of that region, the
  billiard volume is finite and the system is chaotic~\cite{DHRW02}.
\end{itemize}


\subsection{A note on the constraints}

We have focused in the above presentation on the dynamical equations
of motion. The constraints were only briefly mentioned, with no
discussion, except for the Hamiltonian constraint. 
\index{Hamiltonian constraint} This is legitimate
because the constraints are first class and hence preserved by the
Hamiltonian evolution. Thus, they need only be imposed at some
``initial'' time. Once this is done, one does not need to worry about
them any more. Furthermore the momentum constraints and Gauss' law
constraints are differential equations relating the initial data at
different spatial points. This means that they do not constrain the
dynamical variables at a given point but involve also their gradients
-- contrary to the Hamiltonian constraint which becomes
ultralocal. Consequently, at any given point, one can freely choose
the initial data on the undifferentiated dynamical variables and then
use these data as (part of) the appropriate boundary data necessary to
integrate the constraints throughout space. This is why one can assert
that all the walls described above are generically present even when
the constraints are satisfied.

The situation is different in homogeneous cosmologies where the
symmetry relates the values of the fields at all spatial points.
The momentum and Gauss' law constraints become then algebraic
equations and might remove some relevant walls. But this feature
(removal of walls by the momentum and Gauss' law constraints) is
specific to some homogeneous cosmologies and does not hold in the
generic case where spatial gradients are non-zero.

A final comment: How the spatial diffeomorphism constraints and
Gauss' law fit in the conjectured infinite-dimensional symmetry is
a point that is still poorly understood. See, however,~\cite{Constraints} for recent progress in this direction.


\subsection{On the validity of the BKL conjecture -- A status report}

Providing a complete rigorous justification of the above
description of the behavior of the gravitational field in the
vicinity of a spacelike singularity is a formidable task that has
not been pushed to completion yet. The task is formidable because
the Einstein equations form a complicated nonlinear system of
partial differential equations. We shall assume throughout our
review that the BKL description is correct, based on the original
convincing arguments put forward by BKL themselves~\cite{BKL} and
the subsequent fruitful investigations that have shed further
important light on the validity of the conjecture. The billiard
description will thus be taken for granted.

For completeness, we provide in this section a short guide to
the work that has been accumulated since the late 1960's to
consolidate the BKL phenomenon.

As we have indicated, there are two aspects to the BKL conjecture:

\begin{enumerate}
\item The first part of the conjecture states that spatial points
  decouple as one goes to a spacelike singularity in the sense that
  the evolution can be described by a collection of systems of
  ordinary differential equations with respect to time, one such
  system at each spatial point. (\emph{``A spacelike singularity is
    local.''})
  \label{aspect_1}
\item The second part of the conjecture states that the
  system of ordinary differential equations with respect to time
  describing the asymptotic dynamics at any given spatial point can
  be asymptotically replaced by the billiard equations. If the
  matter content is such that the billiard table has infinite
  volume, the asymptotic behavior at each point is given by a
  (generalized) Kasner solution (\emph{``Kasner-like spacelike
    singularities}''). If, on the other hand, the matter content is
  such that the billiard table has finite volume, the asymptotic
  behavior at each point is a chaotic, infinite, oscillatory
  succession of Kasner epochs. (\emph{``Oscillatory, or mixmaster,
    spacelike singularities.''})
  \label{aspect_2}
\end{enumerate}

\noindent
A third element of the original conjecture was that the matter
could be neglected asymptotically. While generically true in four
spacetime dimensions (the exception being a massless scalar field,
equivalent to a fluid with the stiff equation of state $p=\rho$),
this aspect of the conjecture does not remain valid in higher
dimensions where the $p$-form fields might add relevant walls that
could change the qualitative asymptotic behavior. We shall thus
focus here only on Aspects~\ref{aspect_1} and~\ref{aspect_2}.

\begin{itemize}
\item In the Kasner-like case, the mathematical situation is
  easier to handle since the conjectured asymptotic behavior of the
  fields is then monotone and known in closed form. There exist
  theorems validating (generically) this conjectured asymptotic
  behavior, starting from the pioneering work of~\cite{AR01} (where
  the singularities with this behavior are called ``quiescent''),
  which was extended later in~\cite{DHRW02} to cover more general
  matter contents. See also~\cite{Berger:1998vxa,Isenberg:2002jg} for
  related work.
\item The situation is much more complicated in the oscillatory
  case, where only partial results exist. However, even though as
  yet incomplete, the mathematical and numerical studies of the BKL
  analysis has provided overwhelming support for its validity. Most
  work has been done in four dimensions.

  The first attempts to demonstrate that spacelike singularities are
  local were done in the simpler context of solutions with
  isometries. It is only recently that general solutions without
  symmetries have been treated, but this has been found to be
  possible only numerically so far~\cite{Gar04}. The literature on
  this subject is vast and we refer to~\cite{Andersson06, Gar04,
    Rendall} for points of entry into it. Let us note that an important
  element in the analysis has been a more precise reformulation of what
  is meant by ``local''. This has been achieved in~\cite{UEWE03}, where
  a precise definition involving a judicious choice of scale invariant
  variables has been proposed and given the illustrative name of {\it
    ``asymptotic silence''} -- the singularities being called {\it
    ``silent singularities''} since propagation of information is
  asymptotically eliminated.

  If one accepts that generic spacelike singularities are silent,
  one can investigate the system of ordinary differential equations
  that arise in the local limit. In four dimensions, this system is
  the same as the system of ordinary differential equations
  describing the dynamics of spatially homogeneous cosmologies of
  Bianchi type~IX. It has been effectively shown analytically
  in~\cite{Ringstrom:2000mk} that the Bianchi~IX evolution equations can
  indeed be replaced, in the generic case, by the billiard equations
  (with only the dominant, sharp walls) that produce the mixmaster
  behavior. This validates the second element in the BKL conjecture in
  four dimensions.
\end{itemize}

The connection between the billiard variables and the scale
invariant variables has been investigated recently in the
interesting works~\cite{Heinzle:2007kv, Uggla:2007bx}.

Finally, taking for granted the BKL conjecture, one might
analyze the chaotic properties of the billiard map (when the
volume is finite). Papers exploring this issue are~\cite{Chernoff83,
  Cornish:1996hx, KLKSS85, LLK71} (four dimensions)
and~\cite{Elskens:1987gj} (five dimensions). 

Let us finally mention the interesting recent
paper~\cite{Damour:2007nb}, in which a more precise formulation
of the BKL conjecture, aimed towards the chaotic case, is
presented. In particular, the main result of this work is an extension
of the Fuchsian techniques, employed, e.g., in~\cite{DHRW02}, which
are applicable also for systems exhibiting chaotic
dynamics. Furthermore, \cite{Damour:2007nb} examines the geometric
structure which is preserved close to the singularity, and it is shown
that this structure has a mathematical description in terms of a so
called ``partially framed flag''.

\newpage


\section{Hyperbolic Coxeter Groups}
\label{section:Coxeter}
\setcounter{equation}{0}

In this section, we develop the theory of Coxeter groups with a
particular emphasis on the hyperbolic case. The importance of
Coxeter groups for the BKL analysis stems from the fact that in
the case of the gravitational theories that have been studied most
(pure gravity, supergravities), the group generated by the
reflections \index{geometric reflection} in the billiard walls
\index{billiard wall} is a Coxeter group. This
follows, in turn, from the regularity of the corresponding
billiards, whose walls intersect at angles that are integer
submultiples of $\pi$.


\subsection[Preliminary example: The BKL billiard (vacuum $D=4$ gravity)]%
           {Preliminary example: The BKL billiard (vacuum \boldmath $D=4$ gravity)}

To illustrate the regularity of the gravitational billiards and
motivate the mathematical developments through an explicit
example, we first compute in detail the billiard \index{cosmological billiard}characterizing
vacuum, $D=4$ gravity. Since this corresponds to the case
originally considered by BKL, we call it the ``BKL billiard''. We show
in detail that the billiard reflections in this case are governed by
the ``extended modular group'' $PGL(2, \mbb{Z})$, which, as we shall
see, is isomorphic to the hyperbolic Coxeter group $A_1^{++}$. \index{$A_1^{++}$}


\subsubsection{Billiard reflections}
\label{section:BKLbilliard}

There are three scale factors so that after radial projection on
the unit hyperboloid, we get a billiard in two-dimensional
hyperbolic space. \index{hyperbolic space} The billiard region is
defined by the following relevant wall inequalities,
\begin{equation}
  \beta^2 - \beta^1 >0,
  \qquad
  \beta^3 - \beta^2 >0
\end{equation}
(symmetry walls) and 
\begin{equation}
  2 \beta^1 >0
\end{equation}
(curvature wall). The remarkable properties of this region
from our point of view are:

\begin{itemize}
\item It is a triangle (i.e., a simplex in two dimensions) because even
  though we had to begin with 6 walls (3 symmetry walls and 3
  curvature walls), only 3 of them are relevant.
\item The walls intersect at angles that are integer submultiples of
  $\pi$, i.e., of the form
  \begin{equation}
    \frac{\pi}{n},
  \end{equation}
  where $n$ is an integer. The symmetry walls intersect
  indeed at sixty degrees ($n = 3$) since the scalar product of the
  corresponding linear forms (of norm squared equal to $2$) is $-1$,
  while the gravitational wall makes angles of zero ($n = \infty$,
  scalar product $= -2$) and ninety ($n=2$, scalar product $=0$)
  degrees with the symmetry walls.
\end{itemize}

\noindent
These angles are captured in the matrix $A=(A_{ij})_{i,j=1, 2,3}$ of
scalar products, 
\begin{equation}
  A_{ij} = \left(\alpha_i \vert \alpha_j\right),
\end{equation}
which reads explicitly 
\begin{equation}
  A = \left(
  \begin{array}{@{}r@{\quad}r@{\quad}r@{}}
    2 & -2 & 0 \\
    -2 & 2 & -1 \\
    0 & -1 & 2
  \end{array}
  \right).
\end{equation}
Recall from the previous section that the scalar product of two linear
forms $F = F_i \beta^i$ and $G = G_i \beta^i$ is, in a three-dimensional
scale factor space, 
\begin{equation}
  (F \vert G) = \sum_i F_i G_i - \frac{1}{2}
  \left(\sum_i F_i\right) \left(\sum_i G_i \right),
\end{equation}
where we have taken $\alpha_1(\beta) \equiv 2 \beta^1$, $\alpha_2(\beta) \equiv \beta^2 -
\beta^1$ and $\alpha_3(\beta) \equiv \beta^3 - \beta^2$. The corresponding
billiard region is drawn in Figure~\ref{figure:BKLBilliard}.

\epubtkImage{BKLBilliard.png}{%
  \begin{figure}[htbp]
    \centerline{\includegraphics[width=50mm]{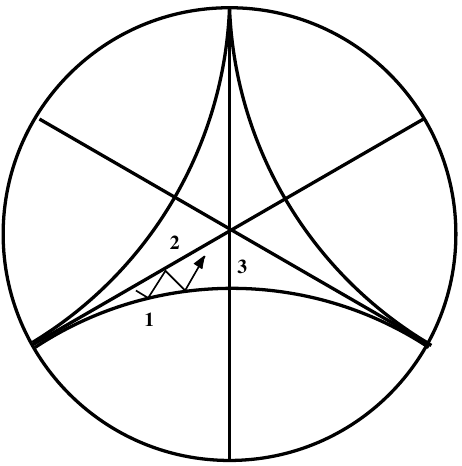}}
    \caption{The BKL billiard of pure four-dimensional
      gravity. The figure represents the billiard region projected
      onto the hyperbolic plane. The particle geodesic is confined
      to the fundamental region enclosed by the three walls
      $\al_1(\be)=2\be^{1}=0, \al_2(\be)=\be^2-\be^1=0$ and
      $\al_3(\be)=\be^{3}-\be^2=0$, as indicated by the numbering in
      the figure. The two symmetry walls $\al_2(\be)=0$ and
      $\al_3(\be)=0$ intersect at an angle of $\pi/3$, while the
      gravity wall $\al_1(\be)=0$ intersects, respectively, at
      angles $0$ and $\pi/2$ with the symmetry walls $\al_2(\be)=0$
      and $\al_3(\be)=0$. The particle has no direction of escape so
      the dynamics is chaotic.}
    \label{figure:BKLBilliard}
  \end{figure}}

Because the angles between the reflecting planes are integer
submultiples of $\pi$, the reflections \index{geometric reflection} in the walls bounding the
billiard region\epubtkFootnote{\label{footnote:reflections}$s_i$ is the reflection with respect
  to the hyperplane defined by $\alpha_i=0$, because it preserves the scalar product,
fixes the plane orthogonal to $\alpha_i$ and maps $\alpha_i$ on
$-\alpha_i$. Note that we are here being deliberately careless about
notation in order not to obscure the main point, namely that the
billiard reflections are elements of a Coxeter group. To be precise,
the linear forms $\alpha_i(\beta)$, $ i=1, 2, 3, $ really represent the
\emph{values} of the linear maps $\alpha_i : \beta \quad \rightarrow\quad
\alpha_i(\beta)\in\mathbb{R}$. The billiard ball moves in the space
of scale factors, say $\mathcal{M}_{\beta}$ ($\beta$-space),
\index{scale-factor space} and hence the maps $\alpha_i$, which define
the walls, belong to the dual space $\mathcal{M}_{\beta}^{\star}$ of
linear forms acting on $\mathcal{M}_{\beta}$. In order to be
compatible with the treatment in Section~\ref{section:DynamicsBilliardHyp}
(cf.~Equation~(\ref{billiardreflectionrule})),
Equation~(\ref{BKLbilliardReflection}) -- even though written here as a
reflection in the space $\mathcal{M}_{\beta}^{\star}$ -- really
corresponds to a geometric reflection in the space
$\mathcal{M}_{\beta}$, in which the particle moves. This will be
carefully explained in Section~\ref{section:CSAdynamics}
(cf.~Equations~(\ref{Cartanreflection2}) and~(\ref{Geometricreflection})),
after the necessary mathematical background has been introduced.},
\begin{equation}
  s_i(\ga) =
  \ga - 2 \frac{(\ga \vert \alpha_i)}{(\alpha_i \vert \alpha_i)} \alpha_i =
  \ga - (\ga \vert \alpha_i) \alpha_i,
  \label{BKLbilliardReflection}
\end{equation}
obey the following relations, 
\begin{equation}
  s_1 s_3 = s_3 s_1
  \quad \leftrightarrow \quad
  (s_1 s_3)^2 = 1,
  \qquad
  (s_2 s_3)^3 = 1.
\end{equation}
The product $s_1 s_3$ is a rotation by $2 \pi /2
= \pi $ and hence squares to one; the product $s_2 s_3$ is a
rotation by $2 \pi /3 $ and hence its cube is equal to one. There
is no power of the product $s_1 s_2$ that is equal to one,
something that one conventionally writes as 
\begin{equation}
 (s_1 s_2)^\infty = 1 .
\end{equation}

The group generated by the reflections $s_1$, $s_2$ and $s_3$ is
denoted $A_1^{++}$, \index{$A_1^{++}$} for reasons that will become
clear in the following, and coincides with the arithmetic group
$PGL(2,\mathbb{Z})$, as we will now show (see
also~\cite{FeingoldFrenkel, Kac, Humphreys}).


\subsubsection[On the group $PGL(2,\mathbb{Z})$]%
              {On the group \boldmath $PGL(2,\mathbb{Z})$}
\label{section:PGL2Z}

The group $PGL(2,\mathbb{Z})$ is defined as the group of $2 \times
2$ matrices $C$ with integer entries and determinant equal to $\pm
1$, with the identification of $C$ and $- C$, 
\begin{equation}
  PGL(2,\mathbb{Z}) = \frac{GL(2,\mathbb{Z})}{\mbb{Z}_2}.
\end{equation}
Note that although elements of the real general linear group $GL(2,\mbb{R})$
have (non-vanishing) unrestricted determinants, the discrete subgroup
$GL(2,\mbb{Z})\subset GL(2,\mbb{R})$ only allows for $\det C =\pm 1$
in order for the inverse $C^{-1}$ to also be an element of $GL(2,\mbb{Z})$.

There are two interesting realisations of $PGL(2,\mathbb{Z})$ in
terms of transformations in two dimensions:

\begin{itemize}
\item One can view $PGL(2,\mathbb{Z})$ as the group of fractional
  transformations of the complex plane 
  \begin{equation}
    C: z \rightarrow z' = \frac{az + b}{cz + d},
    \qquad
    a,b,c,d \in \mathbb{Z},
    \label{fraction}
  \end{equation}
  with 
  \begin{equation}
    ad - cb = \pm 1.
  \end{equation}
  Note that one gets the same transformation if $C$ is replaced by $-C$,
  as one should. It is an easy exercise to verify that the action of
  $PGL(2,\mbb{Z})$ when defined in this way maps the complex upper
  half-plane,
  \begin{equation}
    \mbb{H}=\{z\in\mbb{C}\, |\, \Im z> 0 \},
  \end{equation}
  onto itself whenever the determinant $ad-bc$ of $C$ is equal to $+1$.
  This is not the case, however, when $\det C = -1$.
\item For this reason, it is convenient to consider alternatively the following
  action of $PGL(2,\mathbb{Z})$, 
  \begin{eqnarray}
    z \rightarrow z' = \frac{az + b}{cz + d},
    & \qquad &
    \mbox{if } ad - cb = 1,
    \nonumber
    \\
    & \mbox{or} &
    \label{PGL2Z}
    \\
    z \rightarrow z' = \frac{a\bar{z} + b}{c\bar{z} + d},
    & \qquad &
    \mbox{if } ad - cb = - 1,
    \nonumber 
  \end{eqnarray}
  ($a,b,c,d \in \mathbb{Z}$), which does map the complex upper-half plane onto
  itself, i.e., which is such that $\Im z' >0$ whenever $\Im z >0$.

  The transformation~(\ref{PGL2Z}) is the composition of the
  identity with the transformation~(\ref{fraction}) when $\det C =
  1$, and of the complex conjugation transformation, $f: z
  \rightarrow \bar{z}$ with the transformation~(\ref{fraction}) when
  $\det C = -1$. Because the coefficients $a$, $b$, $c$, and $d$ are
  real, $f$ commutes with $C$ and furthermore the map~(\ref{fraction})
  $ \rightarrow$ (\ref{PGL2Z}) is a group
  isomorphism, so that we can indeed either view the group
  $PGL(2,\mathbb{Z})$ as the group of fractional
  transformations~(\ref{fraction}), or as the group of
  transformations~(\ref{PGL2Z}).
\end{itemize}

\noindent
An important subgroup of the group $PGL(2,\mathbb{Z})$ is the
group $PSL(2,\mathbb{Z})$ for which $ad - cb = 1$, also called
the ``modular group''. The translation $T: z \rightarrow z+1$ and the
inversion $S: z \rightarrow -1/z$ are examples of modular
transformations, 
\begin{equation}
  T = \left(
    \begin{array}{@{}r@{\quad}r@{}}
      1 & 1 \\
      0 & 1
    \end{array}
  \right),
  \qquad
  S = \left(
    \begin{array}{@{}r@{\quad}r@{}}
      0 & -1 \\
      1 & 0
    \end{array}
  \right).
\end{equation}
It is a classical result that
any modular transformation can be written as the product 
\begin{equation}
  T^{m_1} S T^{m_2} S \cdots S T^{m_k},
\end{equation}
but the representation is not unique~\cite{Ap97}.

Let $s_1$, $s_2$ and $s_3$ be the $PGL(2,\mathbb{Z})$-transformations
\begin{equation}
  \begin{array}{rcl}
    s_1: z &\quad \rightarrow& \quad -\bar{z},
    \\ [0.25 em]
    s_2: z &\quad \rightarrow& \quad 1-\bar{z},
    \\ [0.25 em]
    s_3: z &\quad \rightarrow& \quad \displaystyle \frac{1}{\bar{z}}, 
  \end{array}
\end{equation}
to which there correspond the matrices 
\begin{equation}
 s_1 = \left(
    \begin{array}{@{}r@{\quad}r@{}}
      1 & 0 \\
      0 & -1
    \end{array}
  \right),
  \qquad
  s_2 = \left(
    \begin{array}{@{}r@{\quad}r@{}}
      1 & - 1 \\
      0 & -1
    \end{array}
  \right),
  \qquad
  s_3 = \left(
    \begin{array}{@{}r@{\quad}r@{}}
      0 & 1 \\
      1 & 0
    \end{array}
  \right)
\end{equation}
The $s_i$'s are reflections in the straight lines $x=0$, $x= 1/2$ and
the unit circle $\vert z \vert = 1$, respectively. These are in fact
just the transformations of hyperbolic space $s_1$, $s_2$ and $s_3$
described in Section~\ref{section:BKLbilliard}, since the reflection
lines intersect at $0$, $90$ and $60$ degrees, respectively.

One easily verifies that $T = s_2 s_1$ and that $S=s_1 s_3 = s_3
s_1$. Since any transformation of $PGL(2,\mathbb{Z})$ not in
$PSL(2,\mathbb{Z})$ can be written as a transformation of
$PSL(2,\mathbb{Z})$ times, say, $s_1$ and since any transformation
of $PSL(2,\mathbb{Z})$ can be written as a product of $S$'s and
$T$'s, it follows that the group generated by the 3 reflections
$s_1$, $s_2$ and $s_3$ coincides with $PGL(2,\mathbb{Z})$, as
announced above. (Strictly speaking, $PGL(2,\mathbb{Z})$ could be
a quotient of that group by some invariant subgroup, but one may
verify that the kernel of the homomorphism is trivial (see
Section~\ref{PositiveAndNegative} below).) The fundamental domains for
$PGL(2,\mathbb{Z})$ and $PSL(2,\mathbb{Z})$ are drawn in
Figure~\ref{figure:PGL2ZPSL2Z}. The equivalence between
$PGL(2,\mbb{Z})$ and the Coxeter group $A_1^{++}$ \index{$A_1^{++}$}
has been discussed previously in~\cite{FeingoldFrenkel, Kac,
  Humphreys}.

\epubtkImage{PGL2ZPSL2Z.png}{%
  \begin{figure}[htbp]
    \centerline{\includegraphics[width=150mm]{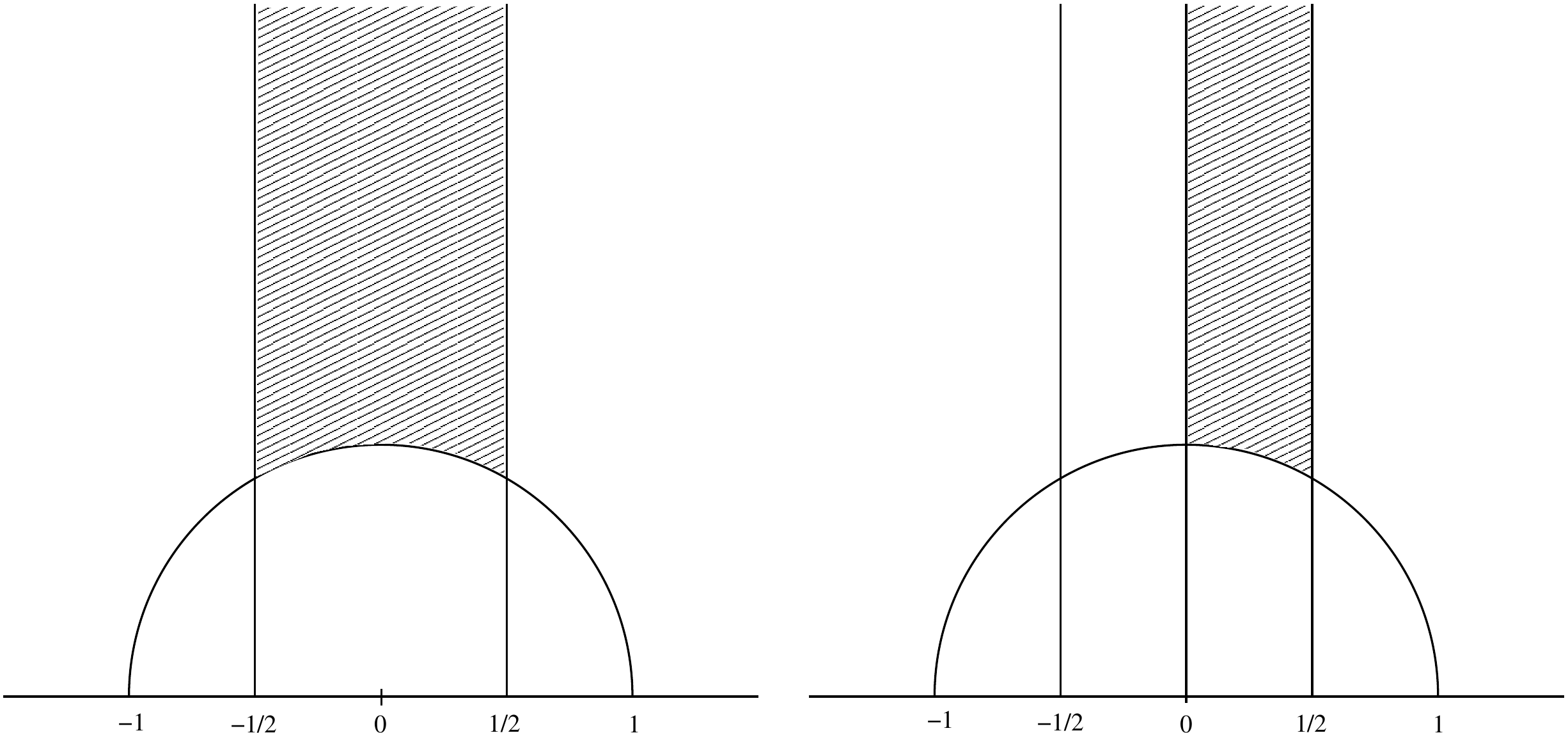}}
    \caption{The figure on the left hand side displays the action of
      the modular group $PSL(2, \mbb{Z})$ on the complex upper half plane
      $\mbb{H}=\{z\in\mbb{C}\, |\, \Im z > 0 \}$. The two generators of
      $PSL(2, \mbb{Z})$ are $S$ and $T$, acting as follows on the
      coordinate $z\in \mbb{H} : S(z)=-1/z; \, T(z)=z+1$, i.e., as an
      inversion and a translation, respectively. The shaded area indicates
      the fundamental domain $\mc{D}_{PSL(2, \mbb{Z})}=\{z\in \mbb{H}\,
      |\, -1/2 \leq \Re z\leq 1/2;\, |z|\geq 1\, \}$ for the action of
      $PSL(2, \mbb{Z})$ on $\mbb{H}$. The figure on the right hand side
      displays the action of the ``extended modular group'' $PGL(2,
      \mbb{Z})$ on $\mbb{H}$. The generators of $PGL(2, \mbb{Z})$ are
      obtained by augmenting the generators of $PSL(2, \mbb{Z})$ with the
      generator $s_1$, acting as $s_1(z)=-\bar{z}$ on $\mbb{H}$. The
      additional two generators of $PGL(2, \mbb{Z})$ then become:
      $s_2\equiv s_1\circ T; \, s_3\equiv s_1 \circ S$, and their actions
      on $\mbb{H}$ are $s_2(z)=1-\bar{z};\, s_3(z)=1/\bar{z}$. The new
      generator $s_1$ corresponds to a reflection in the line $\Re z=0$,
      the generator $s_2$ is in turn a reflection in the line $\Re z=1/2$,
      while the generator $s_3$ is a reflection in the unit circle
      $|z|=1$. The fundamental domain of $PGL(2, \mbb{Z})$ is
      $\mc{D}_{PGL(2, \mbb{Z})}=\{ z\in \mbb{H}\, |\, 0\leq \Re z\leq
      1/2; \, |z|\geq 1\, \}$, corresponding to half the fundamental
      domain of $PSL(2, \mbb{Z})$. The ``walls'' $\Re z=0, \Re z=1/2$ and
      $|z|=1$ correspond, respectively, to the gravity wall
      $\al_1(\be)=0$, the symmetry wall $\al_2(\be)=0$ and the symmetry
      wall $\al_3(\be)=0$ of Figure~\ref{figure:BKLBilliard}.}
    \label{figure:PGL2ZPSL2Z}
  \end{figure}}


\subsection{Coxeter groups -- The general theory}
\index{Coxeter group|bb}

We have just shown that the billiard group in the case of pure
gravity in four spacetime dimensions is the group
$PGL(2,\mathbb{Z})$. This group is generated by reflections and
is a particular example of a Coxeter group. Furthermore, as we
shall explain below, this Coxeter group turns out to be the Weyl
group of the (hyperbolic) Kac--Moody algebra $A_1^{++}$. Our first
encounter with Lorentzian Kac--Moody algebras in more general
gravitational theories will also be through their Weyl groups,
which are, exactly as in the four-dimensional case just described,
particular instances of (non-Euclidean) Coxeter groups, and which
arise as the groups of billiard reflections.

For this reason, we start by developing here some aspects of the
theory of Coxeter groups. An excellent reference on the subject
is~\cite{Humphreys}, to which we refer for more details and
information. We consider Kac--Moody algebras in
Section~\ref{section:KacMoody}.


\subsubsection{Examples}

Coxeter groups generalize the familiar notion of reflection groups in
Euclidean space. Before we present the basic definition, let us briefly
discuss some more illuminating examples.

\subsubsection*{The dihedral group \boldmath $I_2(3)\equiv A_2$}

Consider the dihedral group $I_2(3)$ of order 6 of symmetries of
the equilateral triangle in the Euclidean plane.

\epubtkImage{EquilateralNew.png}{%
  \begin{figure}[htbp]
    \centerline{\includegraphics[width=60mm]{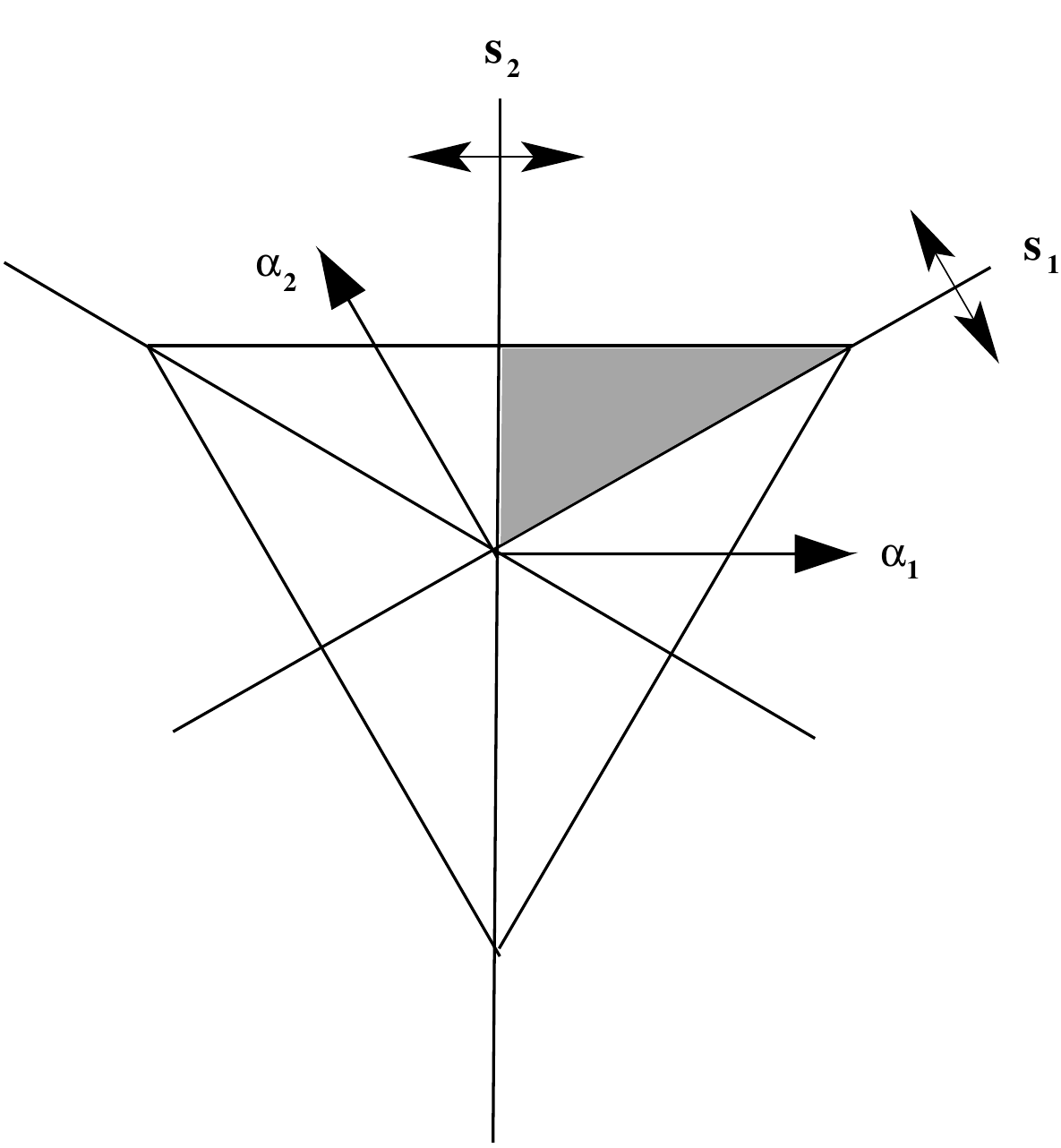}}
    \caption{The equilateral triangle with its 3 axes of
      symmetries. The reflections $s_1$ and $s_2$ generate the entire
      symmetry group. We have pictured the vectors $\alpha_1$ and
      $\alpha_2$ orthogonal to the axes of reflection and chosen to make
      an obtuse angle. The shaded region $\{w \vert (w \vert \alpha_1)
      \geq 0 \} \cap \{w \vert (w \vert \alpha_2) \geq 0 \}$ is a
      fundamental domain for the action of the group on the triangle. Note
      that the fundamental domain for the action of the group on the
      entire Euclidean plane extends indefinitely beyond the triangle but
      is, of course, still bounded by the two walls orthogonal to $\al_1$
      and $\al_2$.}
    \label{figure:equilateral}
  \end{figure}}

This group contains the identity, three reflections $s_1$, $s_2$
and $s_3$ about the three medians, the rotation $R_1$ of $2 \pi /
3$ about the origin and the rotation $R_2$ of $4 \pi/3$ about the
origin (see Figure~\ref{figure:equilateral}),
\begin{equation}
  I_2(3) = \{ 1, s_1, s_2, s_3, R_1, R_2 \}.
\end{equation}
The reflections act as follows\epubtkFootnote{Note that the discussion
  in Footnote~\ref{footnote:reflections} applies also here.},
\begin{equation}
  s_i(\ga) =
  \ga - 2 \frac{(\ga \vert \alpha_i)}{(\alpha_i \vert \alpha_i)} \alpha_i,
\end{equation}
where $(\ \vert \ )$ is here the Euclidean scalar product and where
$\alpha_i$ is a vector orthogonal to the hyperplane (here, line)
of reflection.

Now, all elements of the dihedral group $I_2(3)$ can be written as
products of the two reflections $s_1$ and $s_2$:
\begin{equation}
  1 = s_1^0,
  \qquad
  s_1 = s_1,
  \qquad
  s_2 = s_2,
  \qquad
  R_1 = s_1 s_2,
  \qquad
  R_2 = s_2 s_1,
  \qquad
  s_3 = s_1 s_2 s_1.
  \label{writing}
\end{equation}
Hence, the dihedral group $I_2(3)$ is generated by $s_1$ and
$s_2$. The writing Equation~(\ref{writing}) is not unique because
$s_1$ and $s_2$ are subject to the following relations,
\begin{equation}
  s_1^2 = 1,
  \qquad
  s_2^2 = 1,
  \qquad
  (s_1 s_2)^3 = 1.
  \label{relations}
\end{equation}
The first two relations merely follow from the fact
that $s_1$ and $s_2$ are reflections, while the third relation is
a consequence of the property that the product $s_1 s_2$ is a
rotation by an angle of $2 \pi / 3$. This follows, in turn, from
the fact that the hyperplanes (lines) of reflection make an angle
of $\pi/3$. There is no other relation between the generators
$s_1$ and $s_2$ because any product of them can be reduced, using
the relations Equation~(\ref{relations}), to one of the 6 elements in
Equation~(\ref{writing}), and these are independent.

The dihedral group $I_2(3)$ is also denoted $A_2$ because it is the
Weyl group of the simple Lie algebra $A_2$ (see
Section~\ref{section:KacMoody}). It is isomorphic to the permutation
group $S_3$ of three objects.

\subsubsection*{The infinite dihedral group \boldmath $I_2(\infty) \equiv A_1^+$}

Consider now the group of isometries of the Euclidean
line containing the symmetries about the points with integer or
half-integer values of $x$ ($x$ is a coordinate along the line) as
well as the translations by an integer. This is clearly an
infinite group. It is generated by the two reflections $s_1$
about the origin and $s_2$ about the point with coordinate $1/2$,
\begin{equation}
  s_1 (x) = - x,
  \qquad
  s_2 (x) = - (x-1).
\end{equation}
The product $s_2 s_1$ is a translation by $+1$
while the product $s_1 s_2$ is a translation by $-1$, so no power
of $s_1 s_2$ or $s_2 s_1$ gives the identity. All the powers
$(s_2 s_1)^k$ and $(s_1 s_2)^j$ are distinct (translations by $+k$
and $-j$, respectively). The only relations between the generators
are
\begin{equation}
  s_1^2 = 1 = s_2^2.
\end{equation}
This infinite dihedral group $I_2(\infty)$ is also denoted by $A_1^+$
because it is the Weyl group of the affine Kac--Moody algebra
$A_1^+$.


\subsubsection{Definition}

A Coxeter group \index{Coxeter group}${\mf{C}}$ is a group generated by a finite number
of elements $s_i$ ($i = 1, \cdots, n$) subject to relations that
take the form
\begin{equation}
  s_i^2 = 1 
  \label{rel1}
\end{equation}
and
\begin{equation}
  (s_i s_j)^{m_{ij}} = 1,
  \label{rel2}
\end{equation}
where the integers $m_{ij}$ associated
with the pairs $(i,j)$ fulfill
\begin{equation}
  \begin{array}{rcl}
    m_{ij}& =& m_{ji},
    \\
    m_{ij} &\geq& 2 \qquad (i \not= j).
  \end{array}
\end{equation}
Note that Equation~(\ref{rel1}) is a particular case of
Equation~(\ref{rel2}) with $m_{ii} = 1$. If there is no power of $s_i
s_j$ that gives the identity, as in our second example, we set, by
convention, $m_{ij} = \infty$. The generators $s_i$ are called
``reflections'' because of Equation~(\ref{rel1}), even though we have
not developed yet a geometric realisation of the group. This will be
done in Section~\ref{section:geometric} below.

The number $n$ of generators is called the rank of the Coxeter
group. The Coxeter group \index{Coxeter group} is completely specified by the integers
$m_{ij}$. It is useful to draw the set $\{m_{ij} \}$ pictorially
in a diagram $\Gamma$, called a Coxeter graph. \index{Coxeter graph|bb} With each
reflection $s_i$, one associates a node. Thus there are $n$ nodes
in the diagram. If $m_{ij} >2$, one draws a line between the node
$i$ and the node $j$ and writes $m_{ij}$ over the line, except if
$m_{ij}$ is equal to 3, in which case one writes nothing. The
default value is thus ``3''. When there is no line between $i$ and
$j$ ($i \not=j$), the exponent $m_{ij}$ is equal to 2. We have
drawn the Coxeter graphs for the Coxeter groups $I_2(3)$, $I_2(m)$
and for the Coxeter group $H_3$ of symmetries of the icosahedron.

\epubtkImage{A2.png}{%
  \begin{figure}[htbp]
    \centerline{\includegraphics[width=22mm]{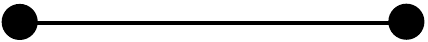}}
    \caption{The Coxeter graph of the symmetry group $I_2(3)
      \equiv A_2$ of the equilateral triangle.}
    \label{figure:A2}
  \end{figure}}

\epubtkImage{I2m.png}{%
  \begin{figure}[htbp]
    \centerline{\includegraphics[width=15mm]{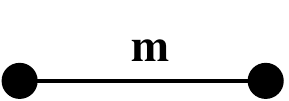}}
    \caption{The Coxeter graph of the dihedral group $I_2(m)$.}
    \label{figure:I2m}
  \end{figure}}

\epubtkImage{H3.png}{%
  \begin{figure}[htbp]
    \centerline{\includegraphics[width=30mm]{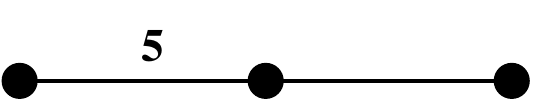}}
    \caption{The Coxeter graph of the symmetry group $H_3$ of the
      regular icosahedron.} \label{figure:H3}
  \end{figure}}

Note that if $m_{ij}= 2$, the generators $s_i$ and $s_j$ commute,
$s_i s_j = s_j s_i$. Thus, a Coxeter group $\mf{C}$ is the
direct product of the Coxeter subgroups associated with the
connected components of its Coxeter graph. For that reason, we
can restrict the analysis to Coxeter groups associated with
connected (also called irreducible) Coxeter graphs.

The Coxeter group may be finite or infinite as the previous
examples show.

\subsubsection*{Another example: \boldmath $C_2^+$}

It should be stressed that the Coxeter group can be infinite even
if none of the Coxeter exponent is infinite. Consider for
instance the group of isometries of the Euclidean plane generated
by reflections in the following three straight lines: (i) the
$x$-axis ($s_1$), (ii) the straight line joining the points
$(1,0)$ and $(0,1)$ ($s_2$), and (iii) the $y$-axis ($s_3$). The
Coxeter exponents are finite and equal to $4$ ($m_{12}= m_{21}=
m_{23}= m_{32}=4$) and $2$ ($m_{13}= m_{31}=2$). The Coxeter
graph is given in Figure~\ref{figure:C2pCoxeter}. The Coxeter group is the symmetry
group of the regular paving of the plane by squares and contains
translations. Indeed, the product $s_2 s_1 s_2$ is a reflection in
the line parallel to the $y$-axis going through $(1,0)$ and thus
the product $t = s_2 s_1 s_2 s_3$ is a translation by $+2$ in the
$x$-direction. All powers of $t$ are distinct; the group is
infinite. This Coxeter group is of affine type and is called $C_2^+$
(which coincides with $B_2^+$).

\epubtkImage{C2pCoxeter.png}{%
  \begin{figure}[htbp]
    \centerline{\includegraphics[width=50mm]{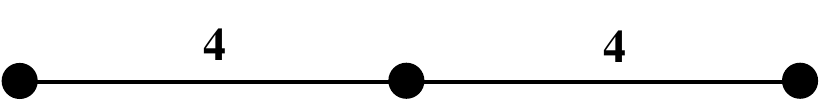}}
    \caption{The Coxeter graph of the affine Coxeter group
      $C_2^{+}$ corresponding to the group of isometries of the
      Euclidean plane.}
    \label{figure:C2pCoxeter}
  \end{figure}}

\subsubsection*{The isomorphism problem}

The Coxeter presentation of a given Coxeter group may
not be unique. Consider for instance the group $I_2(6)$ of order
12 of symmetries of the regular hexagon, generated by two
reflections $s_1$ and $s_2$ with
\begin{displaymath}
  s_1^2 = s_2^2 = 1,
  \qquad
  (s_1 s_2)^6 = 1.
\end{displaymath}
This group is isomorphic with the rank 3
(reducible) Coxeter group $I_2(3) \times \mathbb{Z}_2$, with
presentation
\begin{displaymath}
  r_1^2 = r_2^2 = r_3^2 = 1,
  \qquad
  (r_1 r_2)^3 = 1,
  \qquad
  (r_1 r_3)^2 = 1,
  \qquad
  (r_2 r_3)^2 = 1,
\end{displaymath}
the isomorphism being given by $f(r_1) = s_1$, $f(r_2) = s_1 s_2 s_1 s_2
s_1$, $f(r_3) = (s_1 s_2)^3$. The question of determining all such
isomorphisms between Coxeter groups is known as the ``isomorphism
problem of Coxeter groups''. This is a difficult problem whose general
solution is not yet known~\cite{Bahls}.


\subsubsection{The length function}
\label{section:LengthFunction}

An important concept in the theory of Coxeter groups is that of the
length of an element. The length of $w \in {\mf{C}}$ is by definition
the number of generators that appear in a minimal representation of
$w$ as a product of generators. Thus, if $w = s_{i_1} s_{i_2} \cdots
s_{i_l}$ and if there is no way to write $w$ as a product of less than
$l$ generators, one says that $w$ has length $l$.

For instance, for the dihedral group $I_2(3)$, the identity has
length zero, the generators $s_1$ and $s_2$ have length one, the
two non-trivial rotations have length two, and the third
reflection $s_3$ has length three. Note that the rotations have
representations involving two and four (and even a higher number
of) generators since for instance $s_1 s_2 = s_2 s_1 s_2 s_1$, but
the length is associated with the representations involving as 
few generators as possible. There might be more than one such
representation as it occurs for $s_3 = s_1 s_2 s_1 = s_2 s_1 s_2$.
Both involve three generators and define the length of $s_3$ to be
three.

Let $w$ be an element of length $l$. The length of $ws_i$ (where
$s_i$ is one of the generators) differs from the length of $w$ by
an odd (positive or negative) integer since the relations among
the generators always involve an even number of reflections. In
fact, $l(ws_i)$ is equal to $l+ 1$ or $l-1$ since $l(ws_i) \leq
l(w) + 1$ and $l(w \equiv w s_i s_i) \leq l(w s_i) +1$. Thus, in
$ws_i$, there can be at most one simplification (i.e., at most two
elements that can be removed using the relations).


\subsubsection{Geometric realization}
\label{section:geometric}

We now construct a geometric realisation for any given Coxeter
group. This enables one to view the Coxeter group \index{Coxeter group} as a group of
linear transformations acting in a vector space of dimension $n$, 
equipped with a scalar product preserved by the group.

To each generator $s_i$, associate a vector $\alpha_i$ of a basis
$\{ \alpha_1 , \cdots , \alpha_n \}$ of an $n$-dimensional vector
space $V$. Introduce a scalar product defined as follows,
\begin{equation}
  B(\alpha_i, \alpha_j) =
  - \cos \left( \frac{\pi}{m_{ij}} \right),
  \label{scalar}
\end{equation}
on the basis vectors and extend it to $V$ by linearity. Note that for
$i=j$, $m_{ii}=1$ implies $B(\alpha_i, \alpha_i)=1$ for all $i$. In
the case of
the dihedral group $A_2$, this scalar product is just the
Euclidean scalar product in the two-dimensional plane where the
equilateral triangle lies, as can be seen by taking the two
vectors $\alpha_1$ and $\alpha_2$ respectively orthogonal to the
first and second lines of reflection in Figure~\ref{figure:equilateral} and
oriented as indicated. But in general, the scalar
product~(\ref{scalar}) might not be of Euclidean signature and
might even be degenerate. This is the case for the infinite
dihedral group $I_2(\infty)$, for which the matrix $B$ reads
\begin{equation}
  B = \left(
    \begin{array}{@{}r@{\quad}r@{}}
      1 & -1 \\
      -1 & 1
    \end{array}
  \right)
\end{equation}
and has zero
determinant. We shall occasionally use matrix notations for the
scalar product, $B(\alpha,\ga) \equiv \alpha^T B \ga$.

However, the basis vectors are always all spacelike since they
have norm squared equal to $1$. For each $i$, the vector space
$V$ splits then as a direct sum
\begin{equation}
  V = \mbb{R} \alpha_i \oplus H_i,
\end{equation}
where $H_i$ is the hyperplane
orthogonal to $\alpha_i$ ($\delta \in H_i$ iff $B(\delta,
\alpha_i) = 0$). One defines the geometric reflection $\sigma_i$
as 
\begin{equation}
  \sigma_i(\ga) = \ga - 2 B(\ga, \alpha_i) \alpha_i.
\end{equation}
It is clear that $\sigma_i$
fixes $H_i$ pointwise and reverses $\alpha_i$. It is also clear
that $\sigma_i^2 = 1$ and that $\sigma_i$ preserves $B$,
\begin{equation}
  B \left( \sigma_i (\ga), \sigma_i (\ga') \right) = B(\ga, \ga'). 
\end{equation}
Note that in the
particular case of $A_2$, we recover in this way the reflections
$s_1$ and $s_2$.

We now verify that the $\sigma_i$'s also fulfill the relations
$(\sigma_i \sigma_j)^{m_{ij}} = 1$. To that end we consider the
plane $\Pi$ spanned by $\alpha_i$ and $\alpha_j$. This plane is
left invariant under $\sigma_i$ and $\sigma_j$. Two possibilities may
occur:

\begin{enumerate}
\item The induced scalar product on $\Pi$ is nondegenerate
  and in fact positive definite, or
  \label{case_1}
\item the induced scalar product is positive semi-definite, i.e.,
  there is a null direction orthogonal to any other direction.
  \label{case_2}
\end{enumerate}

\noindent
The second case occurs only when $m_{ij} = \infty$. The null direction
is given by $\lambda = \alpha_i + \alpha_j$.

\begin{itemize}
\item In Case~\ref{case_1}, $V$ splits as $\Pi \oplus \Pi^\perp$ and
  $(\sigma_i \sigma_j)^{m_{ij}}$ is clearly the identity on
  $\Pi^\perp$ since both $\sigma_i$ and $\sigma_j$ leave $\Pi^\perp$
  pointwise invariant. One needs only to investigate $(\sigma_i
  \sigma_j)^{m_{ij}}$ on $\Pi$, where the metric is positive
  definite. To that end we note that the reflections $\sigma_i$ and
  $\sigma_j$ are, on $\Pi$, standard Euclidean reflections in the
  lines orthogonal to $\alpha_i$ and $\alpha_j$, respectively. These
  lines make an angle of $\pi/m_{ij}$ and hence the product $\sigma_i
  \sigma_j$ is a rotation by an angle of $2 \pi / m_{ij}$. It follows
  that $(\sigma_i \sigma_j)^{m_{ij}} = 1$ also on $\Pi$.
\item In Case~\ref{case_2}, $m_{ij}$ is infinite and we must show that no
  power of the product $\sigma_i \sigma_j$ gives the identity. This is
  done by exhibiting a vector $\ga$ for which $(\sigma_i \sigma_j)^k
  (\ga) \not= \ga $ for all integers $k$ different from zero. Take for
  instance $\alpha_i$. Since one has $(\sigma_i \sigma_j) (\alpha_i) =
  \alpha_i + 2 \lambda$ and $(\sigma_i \sigma_j) (\lambda) = \lambda$,
  it follows that $(\sigma_i \sigma_j)^k (\alpha_i) = \alpha_i + 2 k
  \lambda \not= \alpha_i $ unless $k = 0$.
\end{itemize}

\noindent
As the defining relations are preserved, we can conclude that the
map $f$ from the Coxeter group generated by the $s_i$'s to the
geometric group generated by the $\sigma_i$'s defined on the
generators by $f(s_i) = \sigma_i$ is a group homomorphism. We
will show below that its kernel is the identity so that it is in
fact an isomorphism.

Finally, we note that if the Coxeter graph \index{Coxeter graph} is irreducible, as we
assume, then the matrix $B_{ij}$ is \emph{indecomposable}. A
matrix $A_{ij}$ is called \emph{decomposable} if after reordering
of its indices, it decomposes as a non-trivial direct sum, i.e.,
if one can slit the indices $i,j$ in two sets $J$ and $\Lambda$
such that $A_{ij} = 0$ whenever $i \in J, j \in \Lambda$ or $i \in
\Lambda, j \in J$. The indecomposability of $B$ follows from the
fact that if it were decomposable, the corresponding Coxeter graph
would be disconnected as no line would join a point in the set
$\Lambda$ to a point in the set $J$.


\subsubsection{Positive and negative roots}
\label{PositiveAndNegative}

A \emph{root} \index{root|bb} is any vector in the space $V$ of the geometric
realisation that can be obtained from one of the basis vectors
$\alpha_i$ by acting with an element $w$ of the Coxeter group
(more precisely, with its image $f(w)$ under the above
homomorphism, but we shall drop ``$f$'' for notational simplicity).
Any root $\alpha$ can be expanded in terms of the $\alpha_i$'s,
\begin{equation}
  \alpha = \sum_i c_i \alpha_i.
  \label{expansion}
\end{equation}
If the coefficients $c_i$ are all non-negative, we say that the root
$\alpha$ is positive and we write $\alpha >0$. If the coefficients
$c_i$ are all non-positive, we say that the root $\alpha$ is negative
and we write $\alpha <0$. Note that we use strict inequalities here because if $c_i=0$ for all $i$, then $\alpha$ is not a root. In particular, the $\alpha_i$'s themselves
are positive roots, called also ``simple'' roots. (Note that the
  simple roots considered here differ by normalization factors from
  the simple roots of Kac--Moody algebras, as we shall discuss below.)
We claim that roots are either positive or negative (there is no root
with some $c_i$'s in Equation~(\ref{expansion}) $>0$ and some other
$c_i$'s $<0$). The claim follows from the fact that the image of a
simple root by an arbitrary element $w$ of the Coxeter group is
necessarily either positive or negative.

This, in turn, is the result of the following theorem, which
provides a useful criterion to tell whether the length $l(ws_i)$
of $ws_i$ is equal to $l(w) + 1$ or $l(w) -1$.

\begin{theorem}
  $l(ws_i) = l(w) + 1$ if and only if $w(\alpha_i) >0$.

  \noindent The proof is given in~\cite{Humphreys}, page~111.
\end{theorem}

\noindent
It easily follows from this theorem that $l(ws_i) =
l(w) - 1$ if and only if $w(\alpha_i) <0$. Indeed, $l(ws_i) =
l(w) - 1$ is equivalent to $l(w) = l(ws_i) + 1$, i.e., $l((ws_i
)s_i) = l(ws_i) + 1$ and thus, by the theorem, $ws_i (\alpha_i)>
0$. But since $s_i(\alpha_i) = - \alpha_i$, this is equivalent to
$w(\alpha_i) <0$.

We have seen in Section~\ref{section:LengthFunction} that there are
only two possibilities for the length $l(w s_i)$. It is either equal
to $l(w) + 1$ or to $l(w) - 1$. From the theorem just seen, the root
$w(\alpha_i)$ is positive in the first case and negative in the
second. Since any root is the Coxeter image of one of the simple roots
$\alpha_i$, i.e., can be written as $w(\alpha_i)$ for some $w$ and
$\alpha_i$, we can conclude that the roots are either positive or
negative; there is no alternative.

The theorem can be used to provide a geometric interpretation of
the length function. One can show~\cite{Humphreys} that $l(w)$ is
equal to the number of positive roots sent by $w$ to negative
roots. In particular, the fundamental reflection $s$ associated
with the simple root $\alpha_s$ maps $\alpha_s$ to its negative
and permutes the remaining positive roots.

Note that the theorem implies also that the kernel of the
homomorphism that appears in the geometric realisation of the
Coxeter group is trivial. Indeed, assume $f(w) = 1$ where $w$ is
an element of the Coxeter group that is not the identity. It is
clear that there exists one group generator $s_i$ such that $l(w
s_i) = l(w) - 1$. Take for instance the last generator occurring
in a reduced expression of $w$. For this generator, one has
$w(\alpha_i) < 0$, which is in contradiction with the assumption
$f(w) = 1$.

Because $f$ is an isomorphism, we shall from now on identify the
Coxeter group with its geometric realisation and make no
distinction between $s_i$ and $\sigma_i$.


\subsubsection{Fundamental domain}

In order to describe the action of the Coxeter group, it is useful
to introduce the concept of fundamental domain. Consider first
the case of the symmetry group $A_2$ of the equilateral triangle.
The shaded region ${\cal F}$ in Figure~\ref{figure:A2} contains the
vectors $\ga$ such that $B(\alpha_1, \ga) \geq 0$ and
$B(\alpha_2, \ga) \geq 0$. It has the following important
property: Any orbit of the group $A_2$ intersects ${\cal F}$ once
and only once. It is called for this reason a ``fundamental
domain''. We shall extend this concept to all Coxeter groups.
However, when the scalar product $B$ is not positive definite,
there are inequivalent types of vectors and the concept of
fundamental domain can be generalized a priori in different ways,
depending on which region one wants to cover. (The entire
space? Only the timelike vectors? Another region?) The useful
generalization turns out not to lead to a fundamental domain of
the action of the Coxeter group on the entire vector space $V$,
but rather to a fundamental domain of the action of the Coxeter
group on the so-called Tits cone $\mc{X}$, which is such that the
inequalities $B(\alpha_i, \ga) \geq 0$ continue to play the
central role.

We assume that the scalar product is nondegenerate. Define for
each simple root $\alpha_i$ the open half-space
\begin{equation}
  A_i = \{ \ga \in V \, \vert \, B(\alpha_i, \ga) >0 \}.
\end{equation}
We define ${\cal E}$ to be the intersection
of all $A_i$,
\begin{equation}
  {\cal E} = \bigcap_i A_i.
\label{definitioncalE}
\end{equation}
This is a convex open cone, which is non-empty because the metric is
nondegenerate. Indeed, as $B$ is nondegenerate, one can, by a change
of basis, assume for simplicity that the bounding hyperplanes
$B(\alpha_i, \ga) = 0$ are the coordinate hyperplanes $x_i =
0$. ${\cal E}$ is then the region $x_i >0$ (with appropriate
orientation of the coordinates) and ${\cal F}$ is $x_i \geq 0$. The
closure
\begin{equation}
  \begin{array}{rcl}
    {\cal F} &=& \bar{{\cal E}} = \bigcap_i \bar{A}_i,
    \\ [0.5 em]
    \bar{A}_i & = & \{ \ga \in V \vert B(\alpha_i, \ga) \geq 0 \}
  \end{array}
\end{equation}
is then a closed convex cone\epubtkFootnote{Note that in
the case of the infinite dihedral group $I_2(\infty)$, for which
$B$ is degenerate, the definition does not give anything of
interest since ${\cal E} = \emptyset$. When $B$ is degenerate, the
formalism developed here can nevertheless be carried through but
one must go to the dual space $V^*$~\cite{Humphreys}.}.

We next consider the union of the images of ${\cal F}$ under the
Coxeter group,
\begin{equation}
  \mc{X} = \bigcup_{w \in \mf{C}} w({\cal F}).
\end{equation}
One can show~\cite{Humphreys} that this is also a convex cone, called
the Tits cone. Furthermore, ${\cal F}$ is a fundamental domain for the
action of the Coxeter group on the Tits cone; the orbit of any point
in $\mc{X}$ intersects ${\cal F}$ once and only
once~\cite{Humphreys}. The Tits cone does not coincide in general with
the full space $V$ and is discussed below in particular cases.


\subsection{Finite Coxeter groups}
\index{Coxeter group}

An important class of Coxeter groups are the finite ones, like
$I_2(3)$ above. One can show that a Coxeter group is finite if and
only if the scalar product defined by Equation~(\ref{scalar}) on $V$
is Euclidean~\cite{Humphreys}. Finite Coxeter groups coincide with
finite reflection groups in Euclidean space (through hyperplanes that
all contain the origin) and are discrete subgroups of $O(n)$. The
classification of finite Coxeter groups is known and is given in
Table~\ref{table:finite} for completeness. For finite Coxeter groups,
one has the important result that the Tits cone coincides with the
entire space $V$~\cite{Humphreys}.

\begin{table}
  \caption{Finite Coxeter groups.}
  \label{table:finite}
  \renewcommand{\arraystretch}{1.2}
  \vspace{0.5 em}
  \centering
  \begin{tabular}{m{30mm}|m{70mm}}
    \toprule
    Name & Coxeter graph \\
    \midrule
    $A_n$            & \includegraphics[width=55mm]{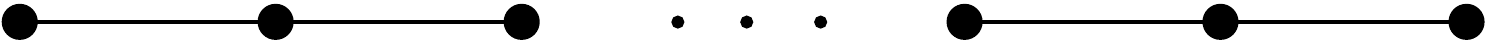} \\ [1 em]
    $B_n \equiv C_n$ & \includegraphics[width=55mm]{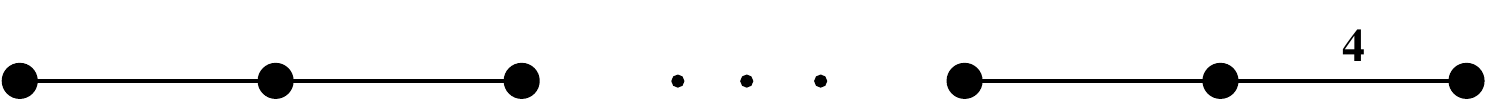} \\ [1 em]
    $D_n$            & \includegraphics[width=55mm]{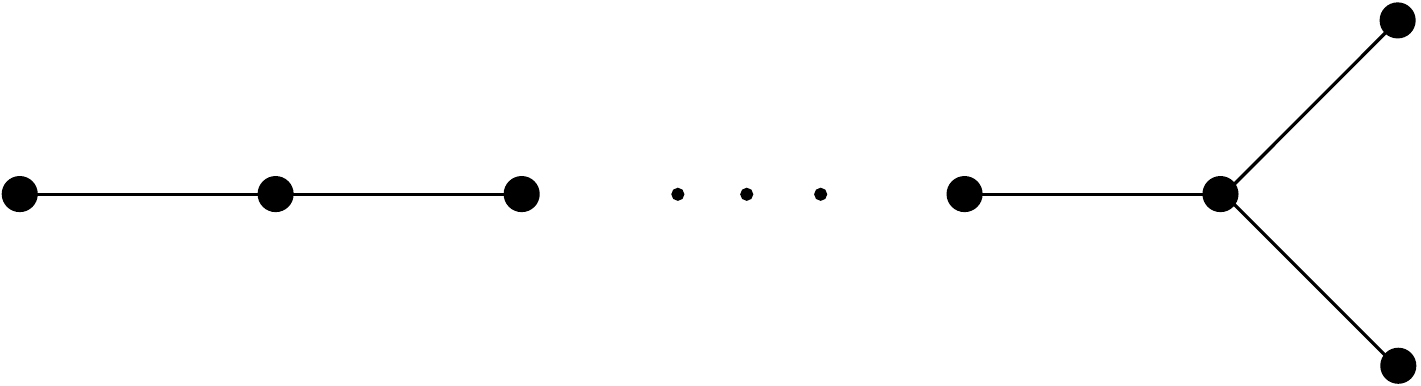} \\ [1 em]
    $I_2(m)$         & \includegraphics[width=13mm]{I2m} \\ [1 em]
    $F_4$            & \includegraphics[width=35mm]{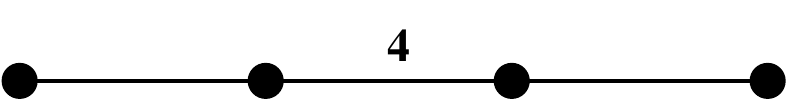} \\ [1 em]
    $E_6$            & \includegraphics[width=45mm]{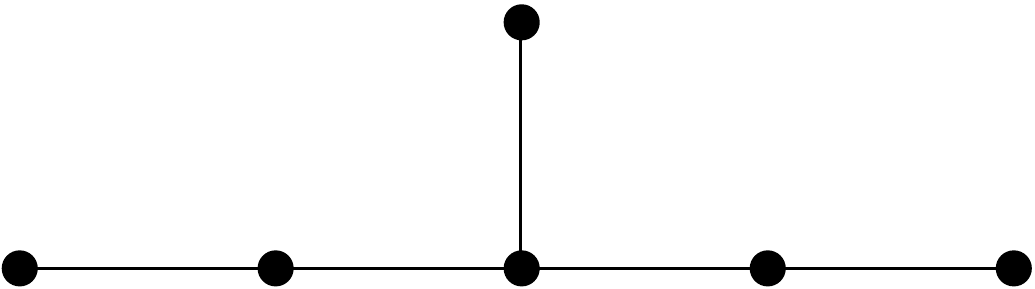} \\ [1 em]
    $E_7$            & \includegraphics[width=50mm]{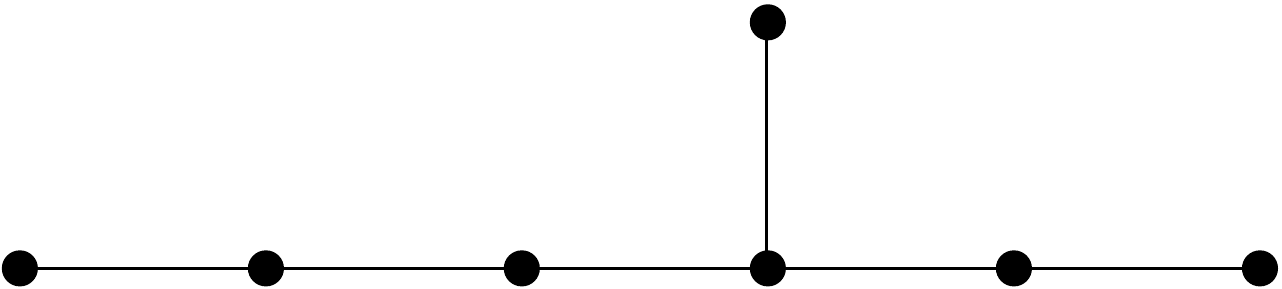} \\ [1 em]
    $E_8$            & \includegraphics[width=60mm]{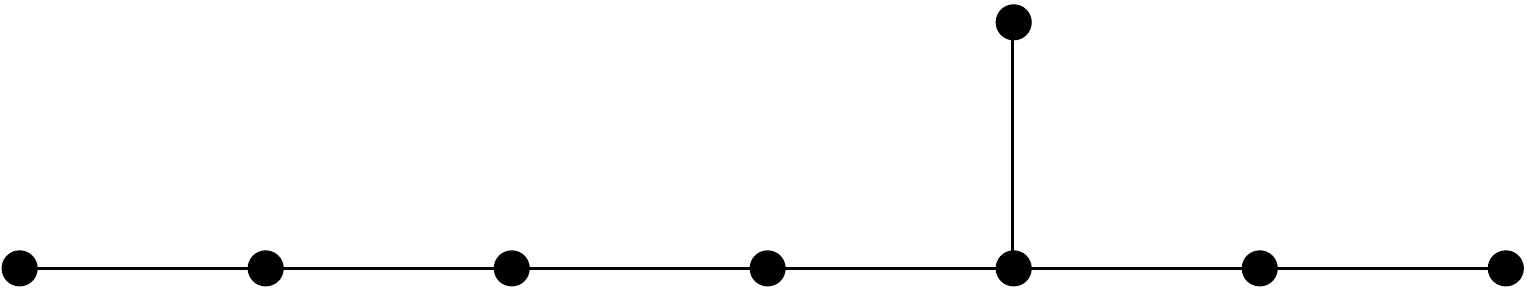} \\ [1 em]
    $H_3$            & \includegraphics[width=23mm]{H3} \\ [1 em]
    $H_4$            & \includegraphics[width=35mm]{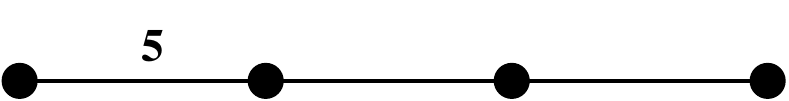} \\
    \bottomrule
  \end{tabular}
  \renewcommand{\arraystretch}{1.0}
\end{table}


\subsection{Affine Coxeter groups}
\label{AffineCox}
\index{Coxeter group}

Affine Coxeter groups are by definition such
that the bilinear form $B$ is positive semi-definite but not
positive definite. The radical $V^\perp$ (defined as the subspace
of vectors $x$ for which $B(x, y) \equiv x^T B y = 0$ for all $y$)
is then one-dimensional (in the irreducible case). Indeed, since
$B$ is positive semi-definite, its radical coincides with the set
$N$ of vectors such that $\lambda^T B \lambda = 0$ as can easily
be seen by going to a basis in which $B$ is diagonal (the
eigenvalues of $B$ are non-negative). Furthermore, $N$ is at least
one-dimensional since $B$ is not positive definite (one of
the eigenvalues is zero). Let $\mu$ be a vector in $V^\perp \equiv N$.
Let $\nu$ be the vector whose components are the absolute values
of those of $\mu$, $\nu_i = \vert \mu_i \vert$. Because $B_{ij}
\leq 0$ for $i \not= j$ (see definition of $B$ in Equation~(\ref{scalar})), one has
\begin{displaymath}
  0 \leq \nu^T B \nu \leq \mu^T B \mu = 0
\end{displaymath}
and thus the vector
$\nu$ belongs also to $V^\perp$. All the components of $\nu$ are
strictly positive, $\nu_i >0$. Indeed, let $J$ be the set of
indices for which $\nu_j > 0$ and $I$ the set of indices for
which $\nu_i = 0$. From $\sum_j B_{kj} \nu_j =0$ ($\nu \in
V^\perp$) one gets, by taking $k$ in $I$, that $B_{ij} =0$ for all
$i \in I$, $j \in J$, contrary to the assumption that the Coxeter
system is irreducible ($B$ is indecomposable). Hence, none of the
components of any zero eigenvector $\mu$ can be zero. If $V^\perp$
were more than one-dimensional, one could easily construct a zero
eigenvector of $B$ with at least one component equal to zero.
Hence, the eigenspace $V^\perp$ of zero eigenvectors is
one-dimensional.

Affine Coxeter groups can be identified with the groups generated
by affine reflections in Euclidean space (i.e., reflections
through hyperplanes that may not contain the origin, so that the
group contains translations) and have also been completely
classified~\cite{Humphreys}. The translation subgroup of an
affine Coxeter group $\mf{C}$ is an invariant subgroup and
the quotient $\mf{C}_0$ is finite; the affine Coxeter group
$\mf{C}$ is equal to the semi-direct product of its
translation subgroup by $\mf{C}_0$. We list all the affine
Coxeter groups in Table~\ref{table:affine}.

\begin{table}
  \caption{Affine Coxeter groups.}
  \label{table:affine}
  \renewcommand{\arraystretch}{1.2}
  \vspace{0.5 em}
  \centering
  \begin{tabular}{m{30mm}|m{70mm}}
    \toprule
    Name & Coxeter graph \\
    \midrule
    $A_1^+$         & \includegraphics[width=20mm]{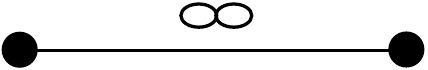} \\ [1 em]
    $A_n^+\, (n>1)$ & \includegraphics[width=55mm]{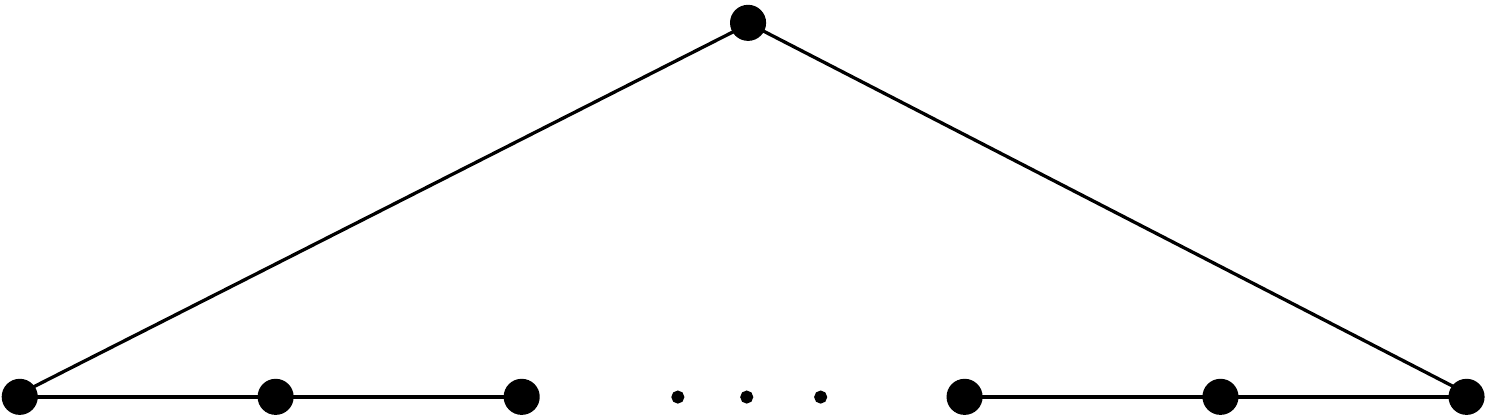} \\ [1 em]
    $B_n^+\, (n>2)$ & \includegraphics[width=55mm]{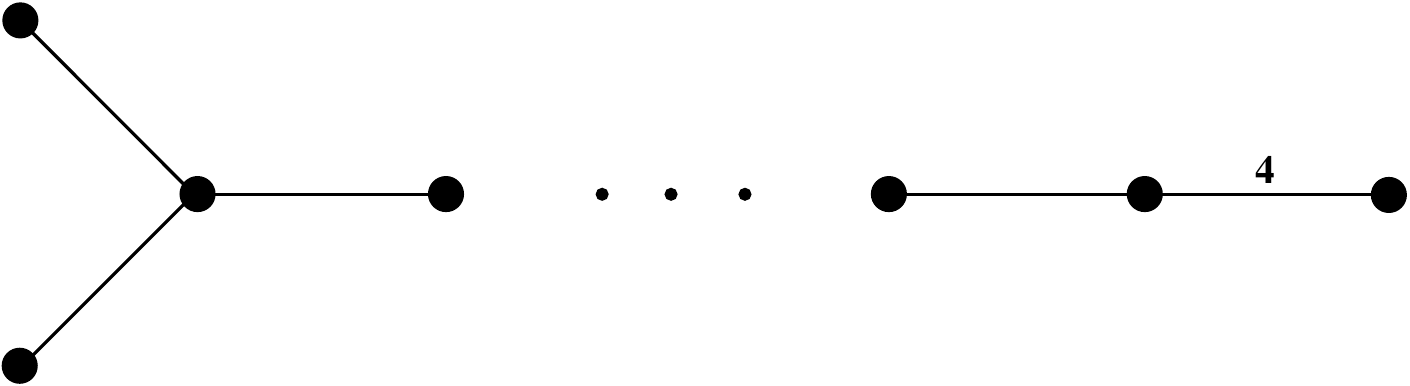} \\ [1 em]
    $C_n^+$         & \includegraphics[width=60mm]{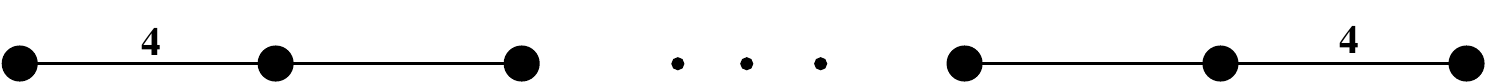} \\ [1 em]
    $D_n^+$         & \includegraphics[width=55mm]{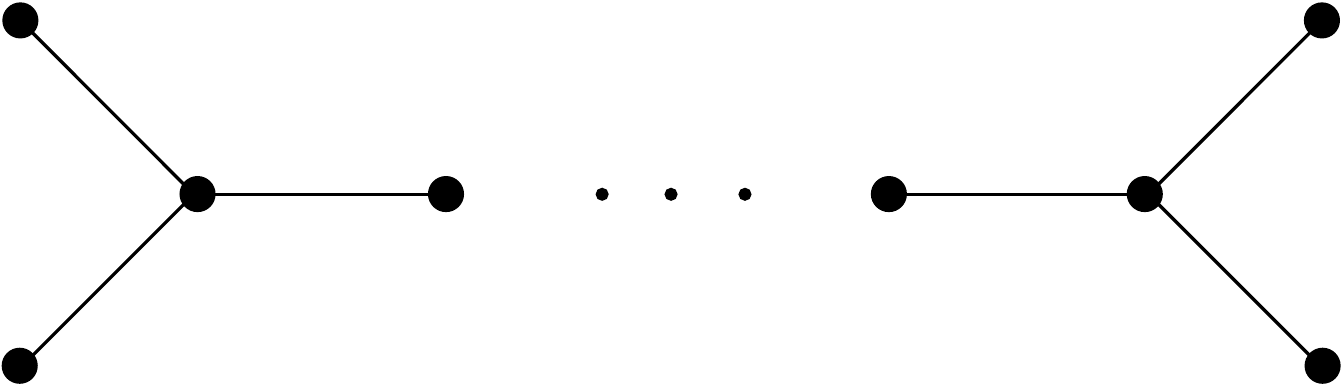} \\ [1 em]
    $G_2^+$         & \includegraphics[width=20mm]{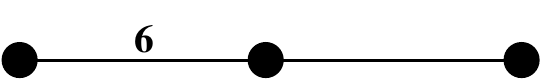} \\ [1 em]
    $F_4^+$         & \includegraphics[width=40mm]{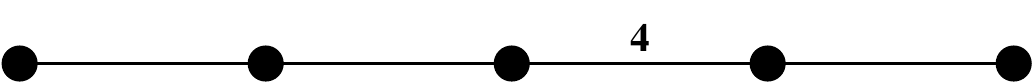} \\ [1 em]
    $E_6^+$         & \includegraphics[width=40mm]{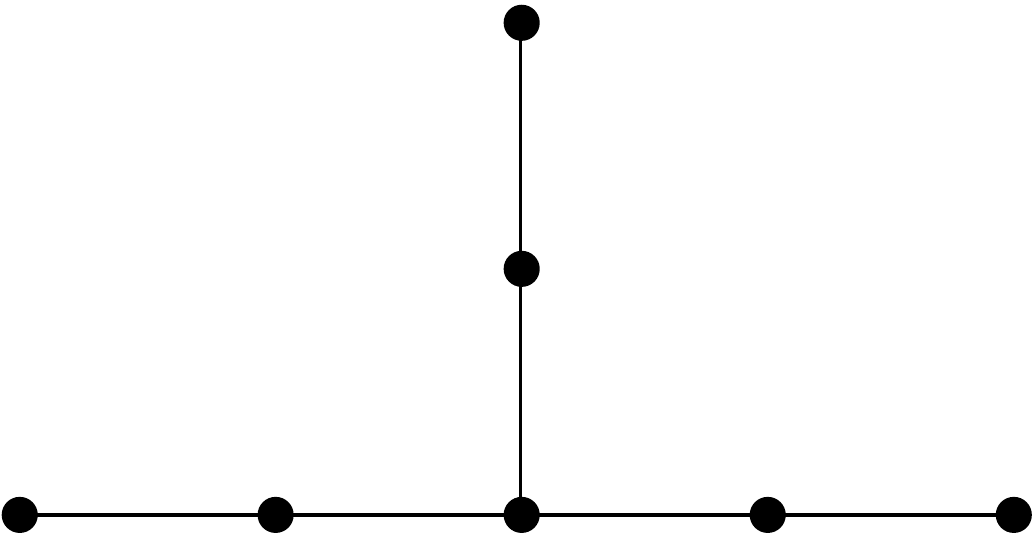} \\ [1 em]
    $E_7^+$         & \includegraphics[width=60mm]{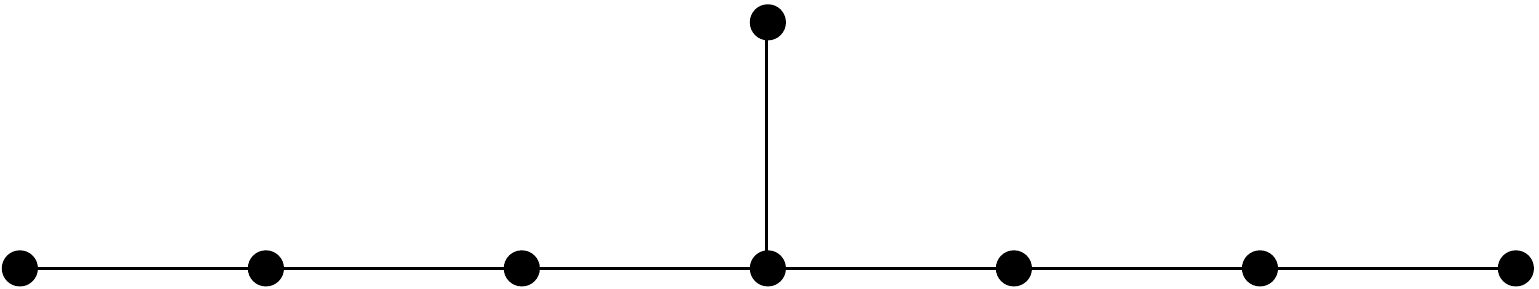} \\ [1 em]
    $E_8^+$         & \includegraphics[width=65mm]{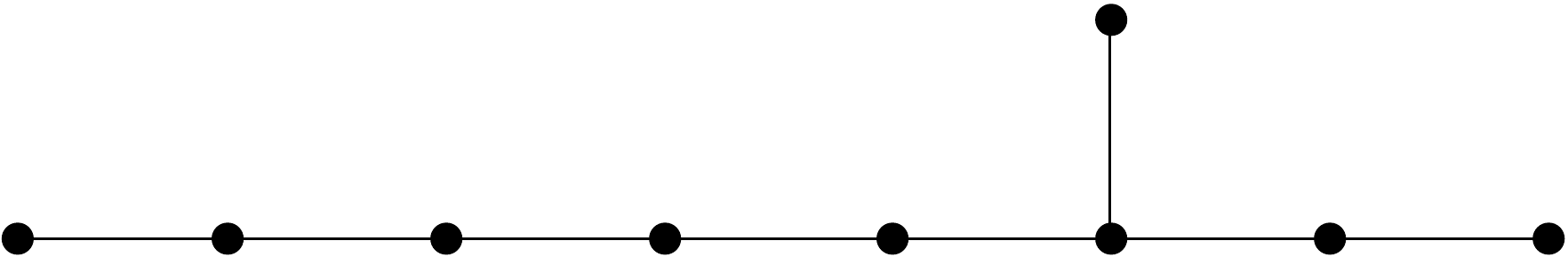} \\
    \bottomrule
  \end{tabular}
  \renewcommand{\arraystretch}{1.0}
\end{table}


\subsection{Lorentzian and hyperbolic Coxeter groups}
\label{section:Hyperbolic}
\index{Coxeter group}

Coxeter groups that are neither of
finite nor of affine type are said to be of indefinite type. An
important property of Coxeter groups of indefinite type is the
following. There exists a positive vector $(c_i)$ such that
$\sum_{j} B_{ij} c_j$ is negative~\cite{Kac}. A vector is said to
be positive (respectively, negative) if all its components are
strictly positive (respectively, strictly negative). This is
denoted $c_i>0$ (respectively, $c_i <0$). Note that a vector may
be neither positive nor negative, if some of its components are
positive while some others are negative. Note also that these
concepts refer to a specific basis. This property is demonstrated
in Appendix~\ref{appendix_1}.

We assume, as already stated, that the scalar product $B$ is
nondegenerate. Let $\{ \omega_i \}$ be the basis dual to the basis $\{
\alpha_i\}$ in the scalar product $B$,
\begin{equation}
  B(\alpha_i, \omega_j) = \delta_{ij}.
  \label{fund1}
\end{equation}
The $\omega_i$'s are called 
``fundamental weights''. (The fundamental weights are really
defined by Equation~(\ref{fund1}) up to normalization, as we will see
in Section~\ref{section:CrystalCoxeterGroups} on crystallographic
Coxeter groups. They thus differ from the solutions of
Equation~(\ref{fund1}) only by a positive multiplicative factor,
irrelevant for the present discussion.)

Consider the vector $v = \sum_i c_i \alpha_i$, where the vector
$c_i$ is such that $c_i> 0$ and $\sum_{j} B_{ij} c_j <0$. This
vector exists since we assume the Coxeter group to be of
indefinite type. Let $\Sigma$ be the hyperplane orthogonal to $v$.
Because $c_i>0$, the vectors $\omega_i$'s all lie on the positive
side of $\Sigma$, $B(v, \omega_i) = c_i >0$. By contrast, the
vectors $\alpha_i$'s all lie on the negative side of $\Sigma$
since $B(\alpha_i, v) = \sum_j B_{ij} c_j <0$. Furthermore, $v$
has negative norm squared, $B(v,v) = \sum_i c_i (\sum_j B_{ij}
c_j) <0$. Thus, in the case of Coxeter groups of indefinite type
(with a nondegenerate metric), one can choose a hyperplane such
that the positive roots \index{root} lie on one side of it and the fundamental
weights on the other side. The converse is true for Coxeter group
of finite type: In that case, there exists $c_i> 0$ such that
$\sum_j B_{ij} c_j$ is positive, implying that the positive roots
and the fundamental weights are on the same side of the hyperplane
$\Sigma$.

We now consider a particular subclass of Coxeter groups of
indefinite type, called Lorentzian Coxeter groups. These are
Coxeter groups such that the scalar product $B$ is of Lorentzian
signature $(n-1,1)$. They are discrete subgroups of the
orthochronous Lorentz group $O^+(n-1,1)$ preserving the time
orientation. Since the $\alpha_i$ are spacelike, the reflection
hyperplanes are timelike and thus the generating reflections $s_i$
preserve the time orientation. The hyperplane $\Sigma$ from the
previous paragraph is spacelike. In this section, we shall adopt
Lorentzian coordinates so that $\Sigma$ has equation $x^0 = 0$ and we
shall choose the time orientation so that the positive roots have a
negative time component. The fundamental weights have then a positive
time component. This choice is purely conventional and is made here
for convenience. Depending on the circumstances, the other time
orientation might be more useful and will sometimes be adopted later
(see for instance Section~\ref{HyperbolicKacMoodyAlgebras}).

Turn now to the cone ${\cal E}$ defined by Equation~(\ref{definitioncalE}).
This cone is clearly given by
\begin{equation}
  {\cal E} = \{ \lambda \in V \vert \, \forall \alpha_i
  \quad B(\lambda, \alpha_i) > 0 \} =
  \left\{ \sum d_i \omega_i \vert d_i >0 \right\}.
\end{equation}
Similarly, its closure ${\cal F}$ is given by
\begin{equation}
  {\cal F} = \{ \lambda \in V \vert \, \forall \alpha_i
  \quad B(\lambda, \alpha_i) \geq 0 \} =
  \left\{\sum d_i \omega_i \vert d_i \geq 0 \right\}.
  \label{coneFFundamental}
\end{equation}
The cone ${\cal F}$ is
thus the convex hull of the vectors $\omega_i$, which are on the
boundary of ${\cal F}$.

By definition, a hyperbolic Coxeter group is a Lorentzian Coxeter
group such that the vectors in ${\cal E}$ are all timelike,
$B(\lambda, \lambda) < 0$ for all $\lambda \in {\cal E}$.
Hyperbolic Coxeter groups are precisely the groups that emerge in
the gravitational billiards of physical interest. The
hyperbolicity condition forces $B(\lambda, \lambda) \leq 0$ for
all $\lambda \in {\cal F}$, and in particular, $B(\omega_i,
\omega_i) \leq 0$: The fundamental weights are timelike or null.
The cone ${\cal F}$ then lies within the light cone. This does not
occur for generic (non-hyperbolic) Lorentzian algebras.

The following theorem enables one to decide whether a Coxeter
group is hyperbolic by mere inspection of its Coxeter graph. \index{Coxeter graph}

\begin{theorem}
  Let $\mf{C}$ be a Coxeter group with
  irreducible Coxeter graph $\Gamma$. The Coxeter group is
  hyperbolic if and only if the following two conditions hold:
  
  \begin{itemize}
  \item The bilinear form $B$ is nondegenerate but not positive
    definite.
  \item For each $i$, the Coxeter graph obtained by removing the node
    $i$ from $\Gamma$ is of finite or affine type.
  \end{itemize}

  (Note: By removing a node, one might get a non-irreducible diagram
  even if the original diagram is connected. A reducible diagram defines
  a Coxeter group of finite type if and only if each irreducible
  component is of finite type, and a Coxeter group of affine type if and
  only if each irreducible component is of finite or affine type with at
  least one component of affine type.)

  \begin{newproof}  
    \begin{itemize}
    \item It is clear that if a Coxeter group is hyperbolic,
      then its bilinear form fulfills the first condition. Let $\omega_i$
      be one of the vectors of the dual basis. The vectors $\alpha_j$ with
      $j \not=i$ form a basis of the hyperplane $\Pi_i$ orthogonal to
      $\omega_i$. Because $\omega_i$ is non-spacelike (the group is
      hyperbolic), the hyperplane $\Pi_i$ is spacelike or null. The
      Coxeter graph defined by the $\alpha_j$ with $j \not=i$ (i.e., by
      removing the node $\alpha_i$) is thus of finite or affine type.
    \item Conversely, assume that the two conditions of the theorem
      hold. From the first condition, it follows that the set $N = \{
      \lambda \in V \, \vert\, B(\lambda, \lambda) <0 \}$ is non-empty. Let
      $\Pi_i$ be the hyperplane spanned by the $\alpha_j$ with $j \not=
      i$, i.e., orthogonal to $\omega_i$. From the second condition, it
      follows that the intersection of $N$ with each $\Pi_i$ is empty.
      Accordingly, each connected component of $N$ lies in one of the
      connected components of the complement of $\bigcup_i \Pi_i$, namely,
      is on a definite (positive or negative) side of each of the
      hyperplanes $\Pi_i$. These sets are of the form $\sum_i c_i
      \alpha_i$ with $c_i >0$ for some $i$'s (fixed throughout the set)
      and $c_i <0$ for the others. This forces the signature of $B$ to
      be Lorentzian since otherwise there would be at least a
      two-dimensional subspace $Z$ of $V$ such that $Z \setminus \{0\}
      \subset N$. Because $Z \setminus \{0\}$ is connected, it must lie
      in one of the subsets just described. But this is impossible since
      if $\lambda \in Z \setminus \{0\}$, then $-\lambda \in Z \setminus
      \{0\}$.
    \end{itemize}
    
    We now show that ${\cal E} \subset N$. Because the signature of
    $B$ is Lorentzian, $N$ is the inside of the standard light cone
    and has two components, the ``future'' component and the ``past''
    component. From the second condition of the theorem, each
    $\omega_i$ lies on or inside the light cone since the orthogonal
    hyperplane is non-timelike. Furthermore, all the $\omega_i$'s
    are future pointing, which implies that the cone ${\cal E}$ lies
    in $N$, as had to be shown (a positive sum of future pointing non
    spacelike vectors is non-spacelike). This concludes the proof of
    the theorem.
  \end{newproof}
\end{theorem}
 
\noindent
In particular, this theorem is useful for determining all
hyperbolic Coxeter groups once one knows the list of all finite
and affine ones. To illustrate its power, consider the Coxeter
diagram of Figure~\ref{figure:A7pp}, with 8 nodes on the loop and
one extra node attached to it (we shall see later that it is
called $A_7^{++}$).

\epubtkImage{A7pp.png}{%
  \begin{figure}[htbp]
    \centerline{\includegraphics[width=120mm]{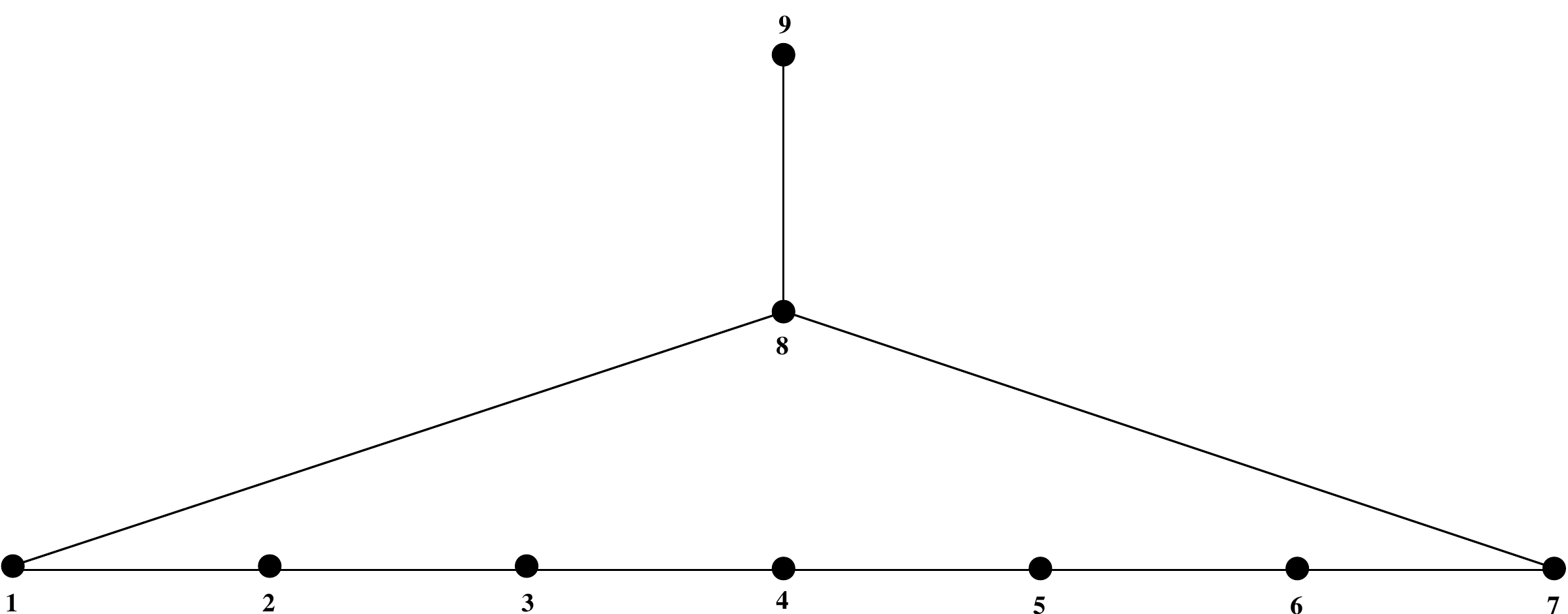}}
    \caption{The Coxeter graph of the group $A_7^{++}$.}
    \label{figure:A7pp}
  \end{figure}}

The bilinear form is given by 
\begin{equation}
 \frac{1}{2}
 \left(
   \begin{array}{@{}r@{\quad}r@{\quad}r@{\quad}r@{\quad}r@{\quad}r@{\quad}r@{\quad}r@{\quad}r@{}}
     2 & -1 & 0 & 0 & 0 & 0 & 0 & -1 & 0 \\
     -1 & 2 & - 1 & 0 & 0 & 0 & 0 & 0 & 0 \\
     0 & -1 & 2 & -1 & 0 & 0 & 0 & 0 & 0 \\
     0 & 0 & -1 & 2 & -1 & 0 & 0 & 0 & 0 \\
     0 & 0 & 0 & -1 & 2 & -1 & 0 & 0 & 0 \\
     0 & 0 & 0 & 0 & -1 & 2 & -1 & 0 & 0 \\
     0& 0 & 0 & 0 & 0 & -1 & 2 & -1 & 0 \\
     -1 & 0& 0 & 0 & 0 & 0 & -1 & 2 & -1 \\
     0 & 0 & 0& 0 & 0 & 0 & 0 & -1 & 2
   \end{array}
 \right).
\end{equation}
 and is of Lorentzian signature. If one removes the node
labelled 9, one gets the affine diagram $A_7^{+}$ (see
Figure~\ref{figure:A7p}). If one removes the node labelled 8, one
gets the finite diagram of the direct product group $A_1 \times A_7$
(see Figure~\ref{figure:A7A1}). Deleting the nodes labelled 1 or
7 yields the finite diagram of $A_8$ (see
Figure~\ref{figure:A8}). Removing the nodes labelled 2 or 6 gives
the finite diagram of $D_8$ (see Figure~\ref{figure:D8}). If one
removes the nodes labelled 3 or 5, one obtains the finite diagram
of $E_8$ (see Figure~\ref{figure:E8}). Finally, deleting the node
labelled 4 yields the affine diagram of $E_7^+$ (see
Figure~\ref{figure:E7pBis}). Hence, the Coxeter group is hyperbolic.

\epubtkImage{A7p.png}{%
  \begin{figure}[htbp]
    \centerline{\includegraphics[width=120mm]{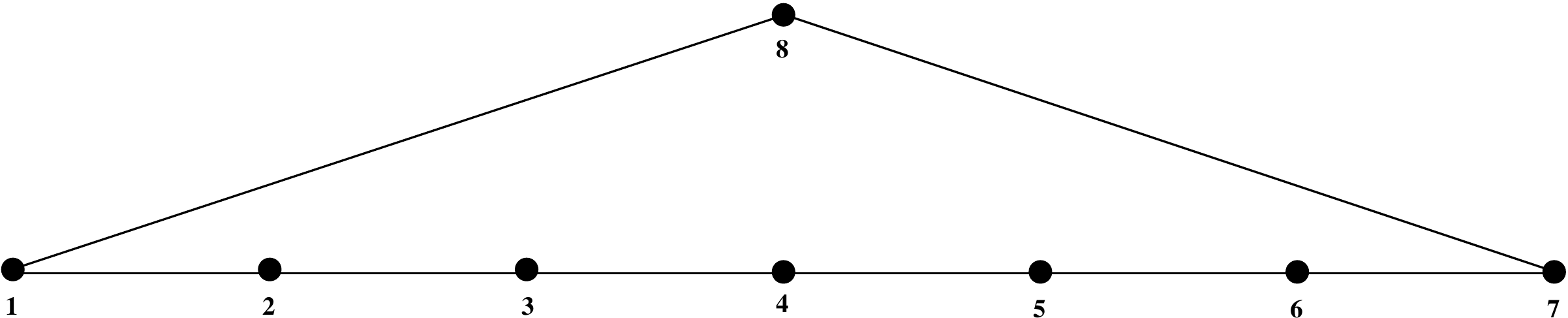}}
    \caption{The Coxeter graph of $A_7^{+}$.}
    \label{figure:A7p}
  \end{figure}}

\epubtkImage{A7A1.png}{%
  \begin{figure}[htbp]
    \centerline{\includegraphics[width=120mm]{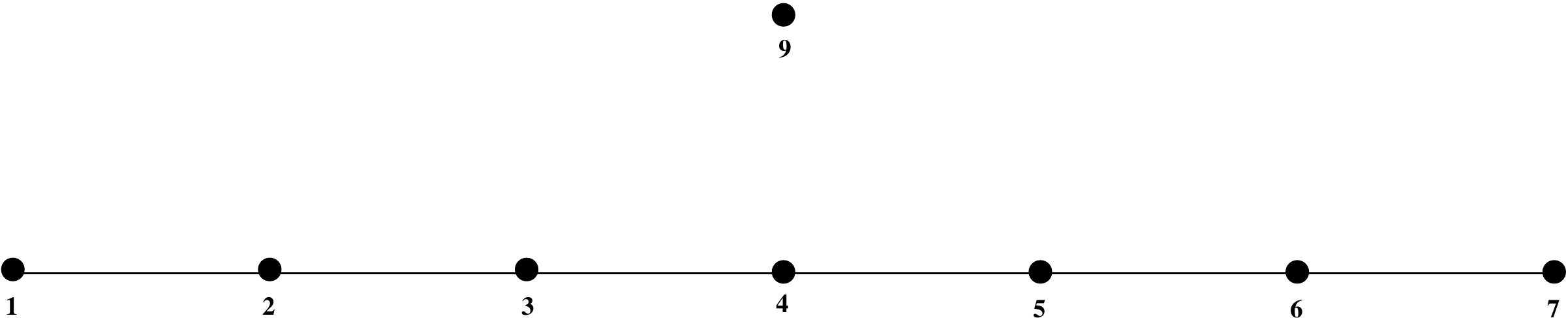}}
    \caption{The Coxeter graph of $A_7 \times A_1$.}
    \label{figure:A7A1}
  \end{figure}}

\epubtkImage{A8.png}{%
  \begin{figure}[htbp]
    \centerline{\includegraphics[width=120mm]{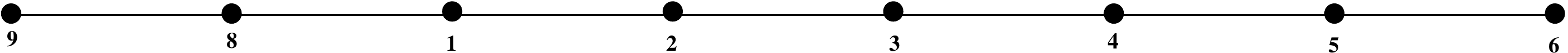}}
    \caption{The Coxeter graph of $A_8$.}
    \label{figure:A8}
  \end{figure}}

\epubtkImage{D8.png}{%
  \begin{figure}[htbp]
    \centerline{\includegraphics[width=120mm]{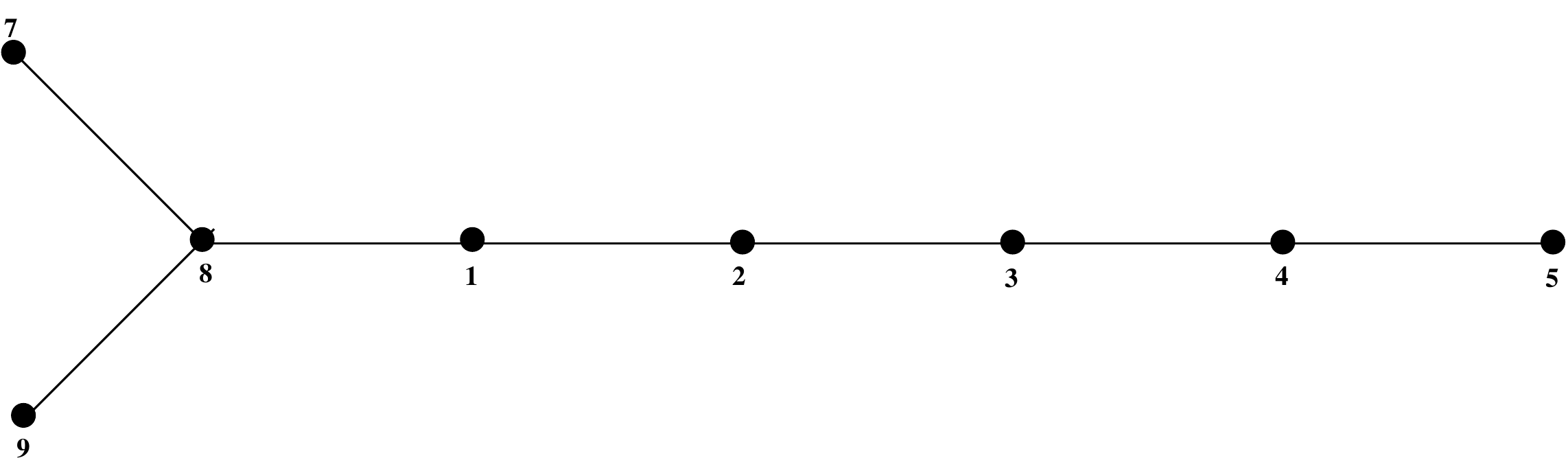}}
    \caption{The Coxeter graph of $D_8$.}
    \label{figure:D8}
  \end{figure}}

\epubtkImage{E8.png}{%
  \begin{figure}[htbp]
    \centerline{\includegraphics[width=120mm]{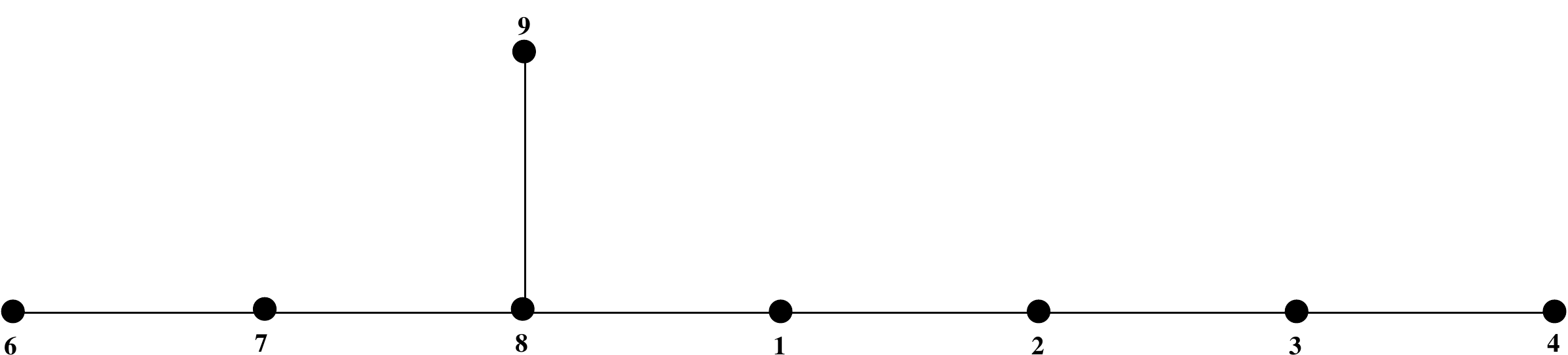}}
    \caption{The Coxeter graph of $E_8$.}
    \label{figure:E8}
  \end{figure}}

\epubtkImage{E7pBis.png}{%
  \begin{figure}[htbp]
    \centerline{\includegraphics[width=120mm]{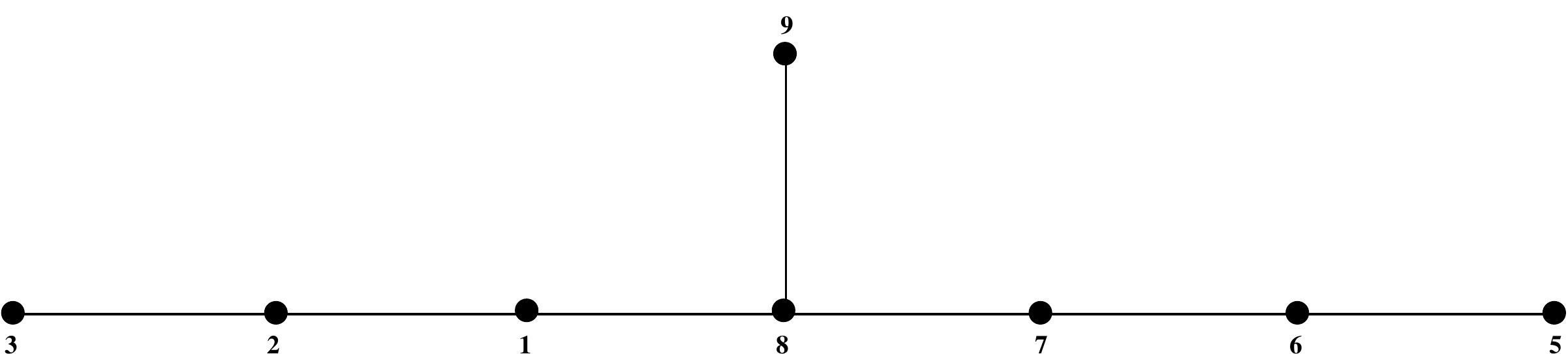}}
    \caption{The Coxeter graph of $E_7^+$.}
    \label{figure:E7pBis}
  \end{figure}}

Consider now the same diagram, with one more node in the loop
($A_8^{++}$). In that case, if one removes one of the middle
nodes 4 or 5, one gets the Coxeter group $E_7^{++}$, which is
neither finite nor affine. Hence, $A_8^{++}$ is not hyperbolic.

Using the two conditions in the theorem, one can in fact provide the
list of all irreducible hyperbolic Coxeter groups. The striking fact
about this classification is that hyperbolic Coxeter groups exist only
in ranks $3 \leq n \leq 10$, and, moreover, for $4\leq n\leq 10$ there
is only a finite number. In the $n=3$ case, on the other hand, there
exists an infinite class of hyperbolic Coxeter groups. In
Figure~\ref{figure:HyperbolicCGRank3} we give a general form of the
Coxeter graphs corresponding to all rank~3 hyperbolic Coxeter groups,
and in
Tables~\ref{table:HyperbolicCGRank4}\,--\,\ref{table:HyperbolicCGRank10}
we give the complete classification for $4\leq n\leq 10$.

Note that the inverse metric $(B^{-1})_{ij}$, which gives the
scalar products of the fundamental weights, has only negative
entries in the hyperbolic case since the scalar product of two
future-pointing non-spacelike vectors is strictly negative (it is
zero only when the vectors are both null and parallel, which does
not occur here).

One can also show~\cite{Kac, Humphreys} that in the hyperbolic
case, the Tits cone $\mc{X}$ coincides with the future light cone.
(In fact, it coincides with either the future light cone or the
past light cone. We assume that the time orientation in $V$ has
been chosen as in the proof of the theorem, so that the Tits cone
coincides with the future light cone.) This is at the origin of an
interesting connection with discrete reflection groups in
hyperbolic space \index{hyperbolic space} (which justifies the terminology). One may
realize hyperbolic space $\mc{H}_{n-1}$ as the upper sheet of the
hyperboloid $B(\lambda, \lambda) = -1 $ in $V$. Since the Coxeter
group is a subgroup of $O^+(n-1,1)$, it leaves this sheet
invariant and defines a group of reflections \index{geometric reflection} in $\mc{H}_{n-1}$. The
fundamental reflections are reflections through the hyperplanes in
hyperbolic space obtained by taking the intersection of the
Minkowskian hyperplanes $B(\alpha_i, \lambda) = 0$ with hyperbolic
space. These hyperplanes bound the fundamental region, which is
the domain to the positive side of each of these hyperplanes. The
fundamental region is a simplex with vertices $\bar{\omega}_i$,
where $\bar{\omega}_i$ are the intersection points of the lines
$\mbb{R} \omega_i$ with hyperbolic space. This intersection is at
infinity in hyperbolic space if $\omega_i$ is lightlike. The
fundamental region has finite volume but is compact only if the
$\omega_i$ are timelike.

Thus, we see that the hyperbolic Coxeter groups are the reflection
groups in hyperbolic space with a fundamental domain which (i) is
a simplex, and which (ii) has finite volume. The fact that the
fundamental domain is a simplex ($n$ vectors in $\mc{H}_{n-1}$) follows
from our geometric construction where it is assumed that the $n$
vectors $\alpha_i$ form a basis of $V$.

The group $PGL(2,\mathbb{Z})$ relevant to pure gravity in four
dimensions is easily verified to be hyperbolic.

For general information, we point out the following facts:

\begin{itemize}
\item Compact hyperbolic Coxeter groups (i.e., hyperbolic Coxeter
  groups with a compact fundamental region) exist only for ranks 3, 4
  and 5, i.e., in two, three and four-dimensional hyperbolic space. All
  hyperbolic Coxeter groups of rank $> 5$ have a fundamental region
  with at least one vertex at infinity. The hyperbolic Coxeter groups
  appearing in gravitational theories are always of the noncompact
  type.
\item There exist reflection groups in hyperbolic space whose
  fundamental domains are not simplices. Amazingly enough, these exist
  only in hyperbolic spaces of dimension $\leq 995$. If one imposes
  that the fundamental domain be compact, these exist only in
  hyperbolic spaces of dimension $\leq 29$. The bound can probably be
  improved~\cite{Vinberg}.
\item Non-hyperbolic Lorentzian Coxeter groups are associated through
  the above construction with infinite-volume fundamental regions
  since some of the vectors $\omega_i$ are spacelike, which imply that
  the corresponding reflection hyperplanes intersect beyond hyperbolic
  infinity.
\end{itemize}

\epubtkImage{HyperbolicCGRank3.png}{%
  \begin{figure}[htbp]
    \centerline{\includegraphics[width=50mm]{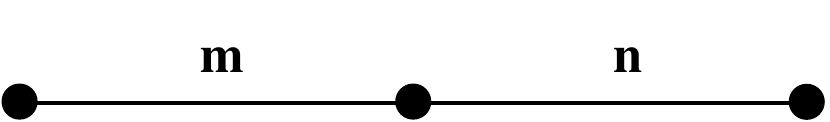}}
    \caption{This Coxeter graph corresponds to hyperbolic Coxeter
      groups for all values of $m$ and $n$ for which the associated
      bilinear form $B$ is not of positive definite or positive
      semidefinite type. This therefore gives rise to an infinite class of
      rank~3 hyperbolic Coxeter groups.}
    \label{figure:HyperbolicCGRank3}
  \end{figure}}

\begin{table}
  \caption{Hyperbolic Coxeter groups of rank 4.}
  \label{table:HyperbolicCGRank4}
  \renewcommand{\arraystretch}{2.0}
  \vspace{0.5 em}
  \centering
  \begin{tabular}{m{46mm}m{45mm}m{46mm}}
    \toprule
    & & \\ [-1.8 em]
    \includegraphics[width=45mm]{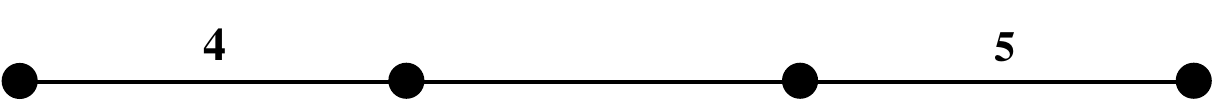} &
    \includegraphics[width=45mm]{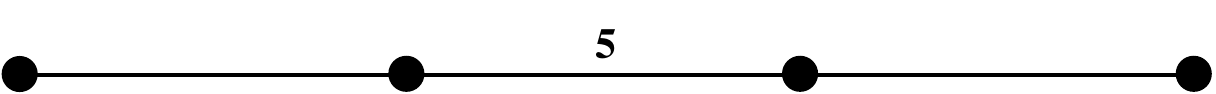} &
    \includegraphics[width=45mm]{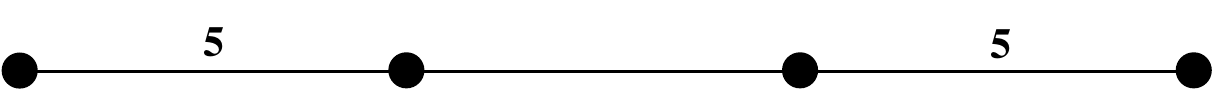} \\
    & & \\ [-1.8 em]
    \includegraphics[width=45mm]{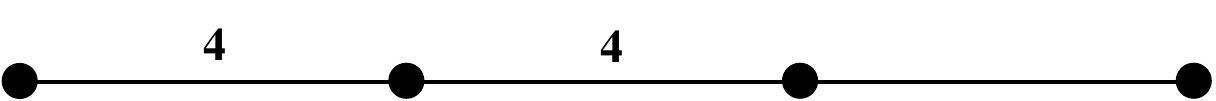} &
    \includegraphics[width=45mm]{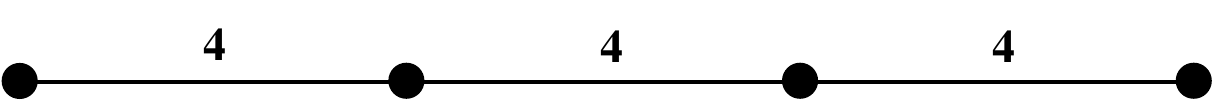} &
    \includegraphics[width=45mm]{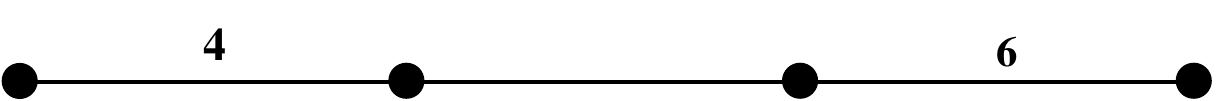} \\
    & & \\ [-1.8 em]
    \includegraphics[width=45mm]{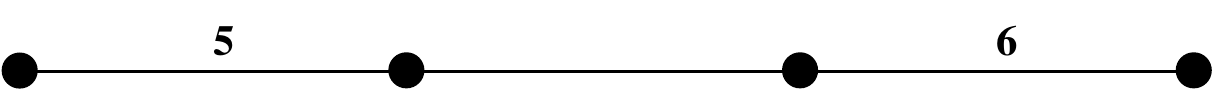} &
    \includegraphics[width=45mm]{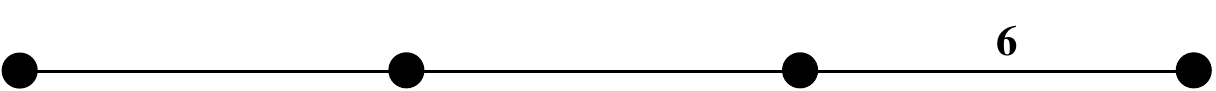} &
    \includegraphics[width=45mm]{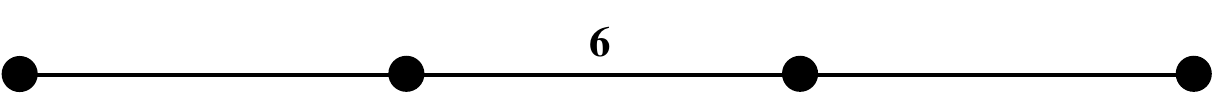} \\
    & & \\ [-1.8 em]
    \includegraphics[width=45mm]{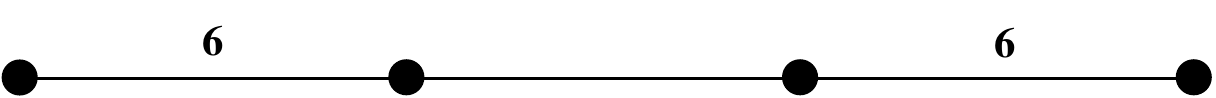} &
    \includegraphics[width=35mm]{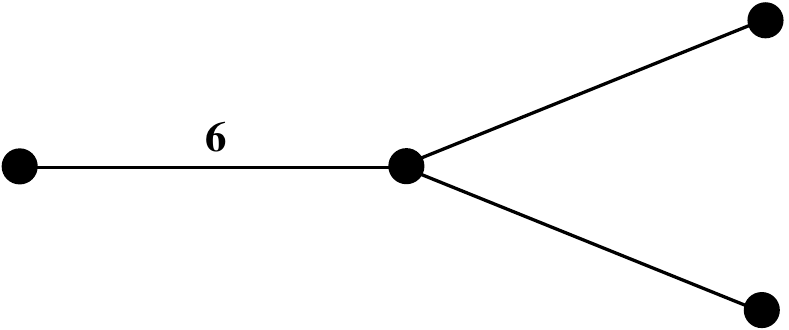} &
    \includegraphics[width=35mm]{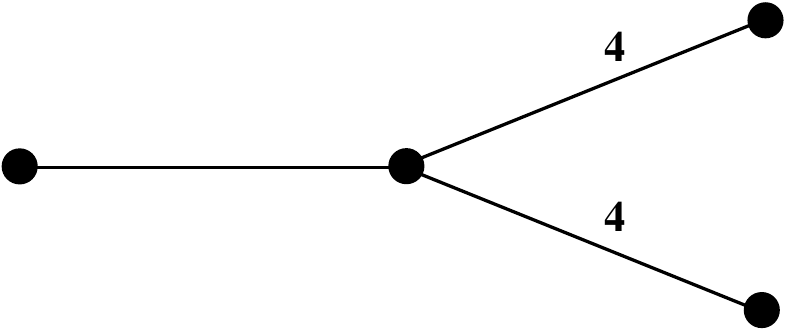} \\
    & & \\ [-1.8 em]
    \includegraphics[width=35mm]{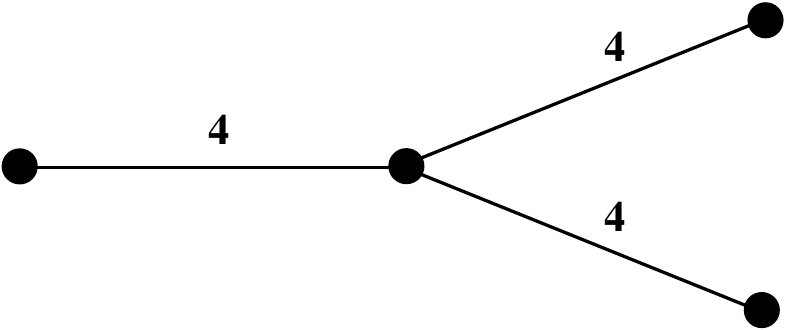} &
    \includegraphics[width=35mm]{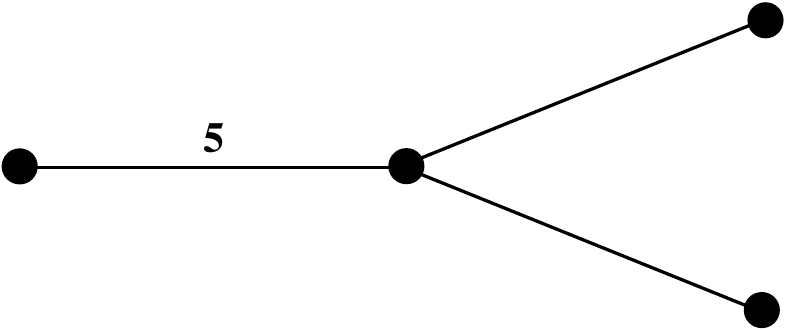} &
    \includegraphics[width=35mm]{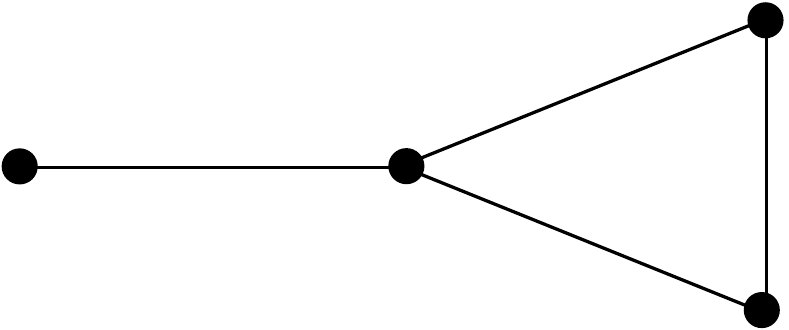} \\
    & & \\ [-1.8 em]
    \includegraphics[width=35mm]{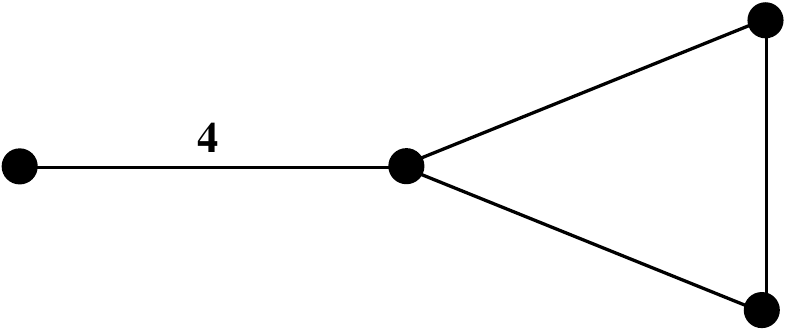} &
    \includegraphics[width=35mm]{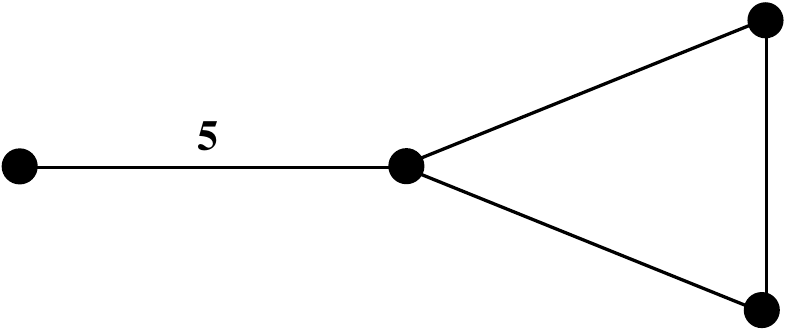} &
    \includegraphics[width=35mm]{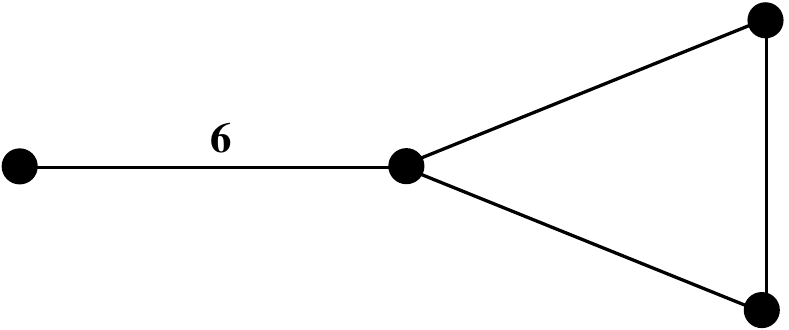} \\
    \bottomrule
  \end{tabular}
 \begin{tabular}{m{26mm}m{26mm}m{25mm}m{25mm}m{26mm}}
    & & & & \\ [-1.8 em]
    \includegraphics[width=20mm]{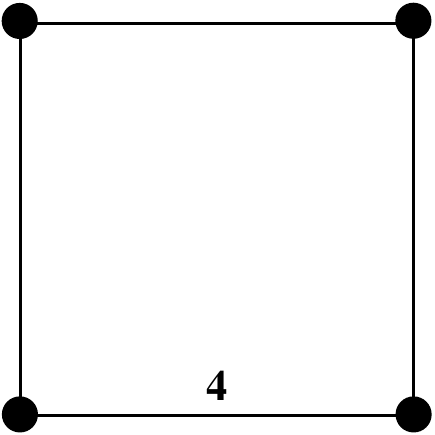} &
    \includegraphics[width=20mm]{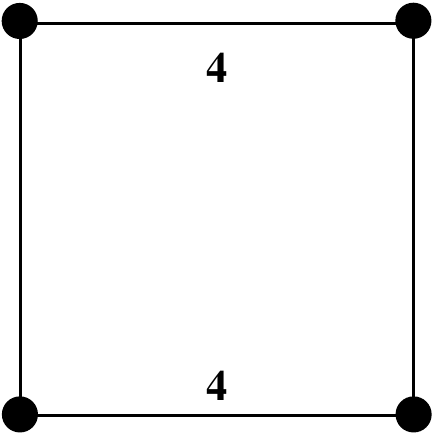} &
    \includegraphics[width=20mm]{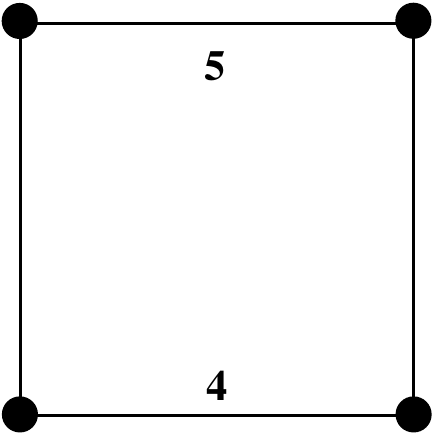} &
    \includegraphics[width=20mm]{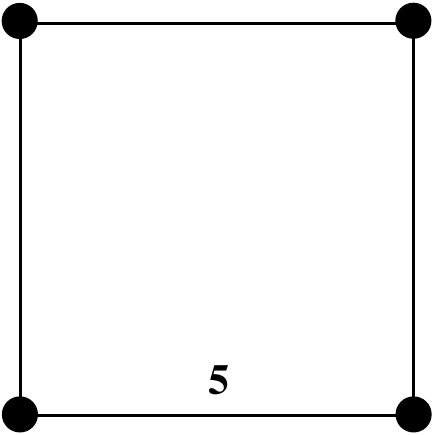} &
    \includegraphics[width=20mm]{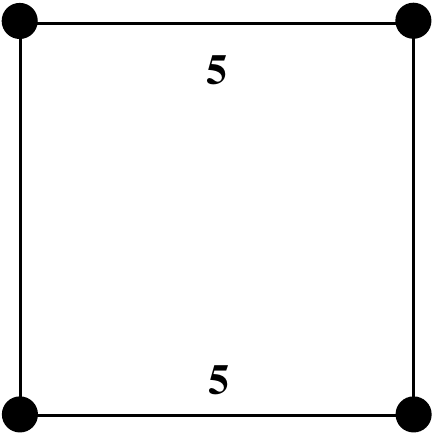} \\
    & & & & \\ [-1.8 em]
    \includegraphics[width=20mm]{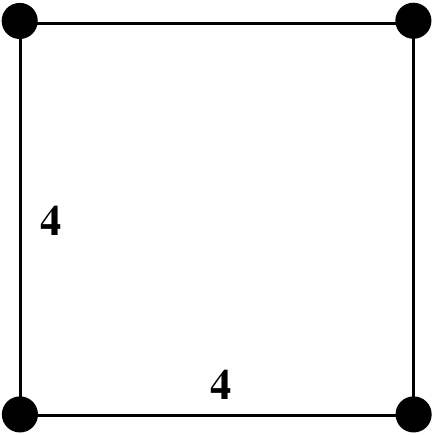} &
    \includegraphics[width=20mm]{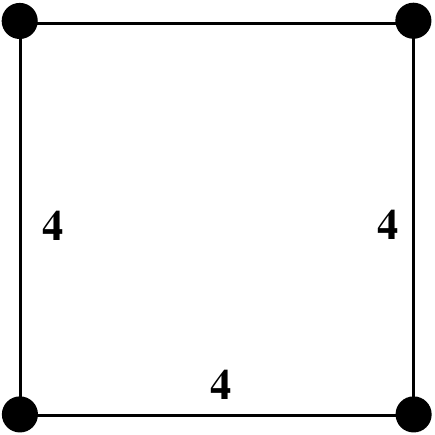} &
    \includegraphics[width=20mm]{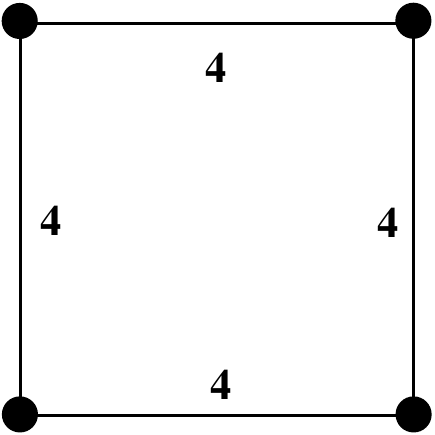} &
    \includegraphics[width=20mm]{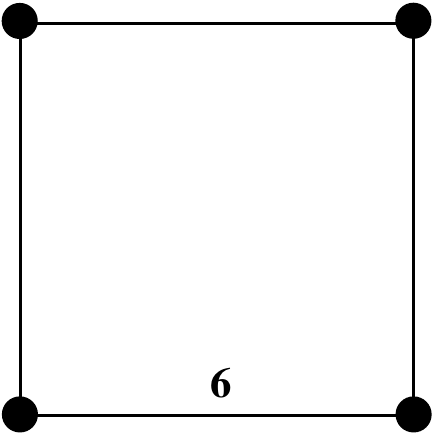} &
    \includegraphics[width=20mm]{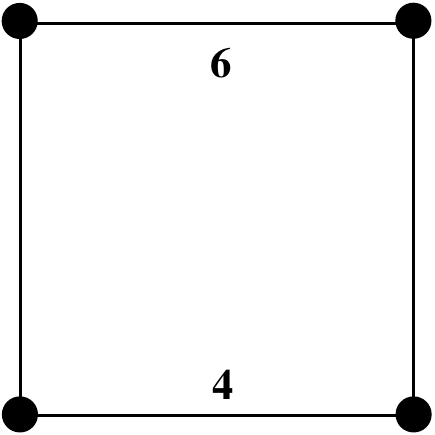} \\
    & & & & \\ [-1.8 em]
    \includegraphics[width=20mm]{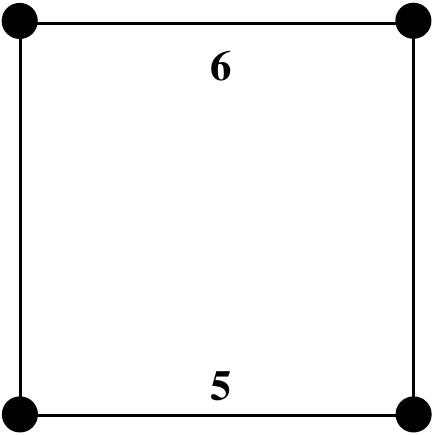} &
    \includegraphics[width=20mm]{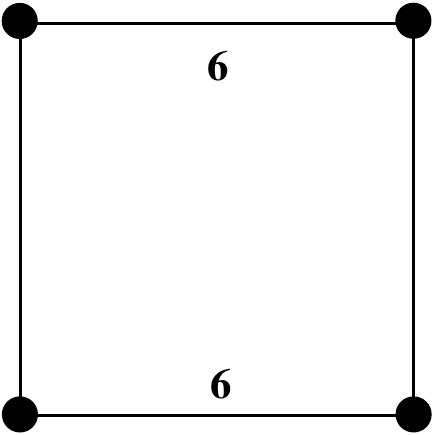} &
    \includegraphics[width=20mm]{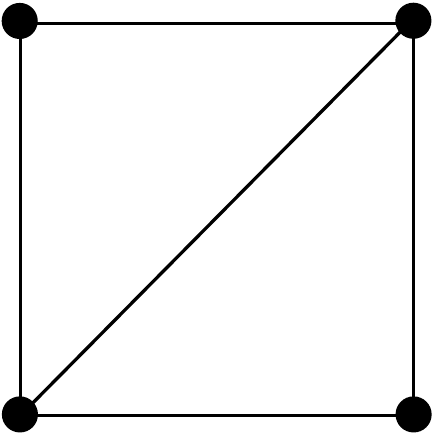} &
    \includegraphics[width=23mm]{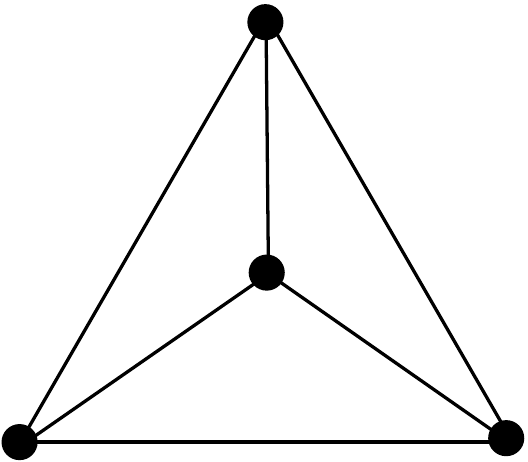} \\
    \bottomrule
  \end{tabular}
  \renewcommand{\arraystretch}{1.0}
\end{table}

\begin{table}
  \caption{Hyperbolic Coxeter groups of rank 5.}
  \label{table:HyperbolicCGRank5}
  \renewcommand{\arraystretch}{2.0}
  \vspace{0.5 em}
  \centering
  \begin{tabular}{m{70mm}m{70mm}}
    \toprule
    & \\ [-2.0 em]
    \includegraphics[width=65mm]{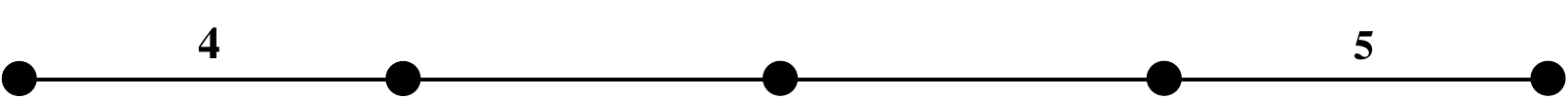} &
    \includegraphics[width=65mm]{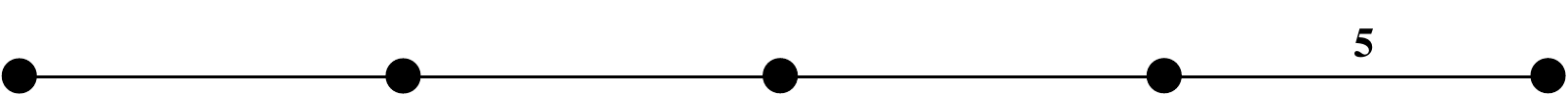} \\
    & \\ [-2.0 em]
    \includegraphics[width=65mm]{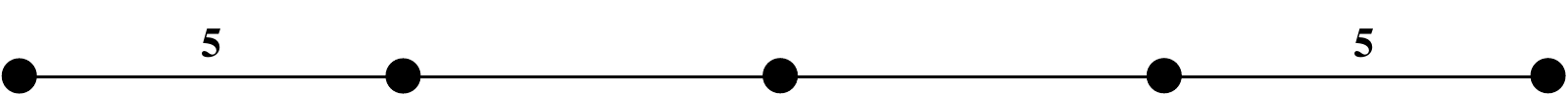} &
    \includegraphics[width=65mm]{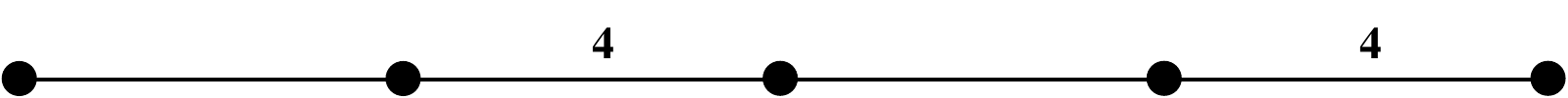} \\
    \bottomrule
  \end{tabular}
  \begin{tabular}{m{45mm}m{46mm}m{45mm}}
    & & \\ [-2.0 em]
    \includegraphics[width=46mm]{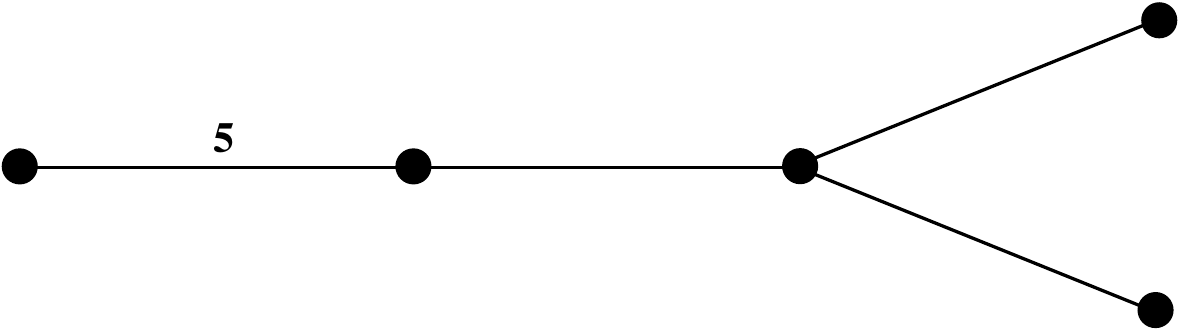} &
    \includegraphics[width=46mm]{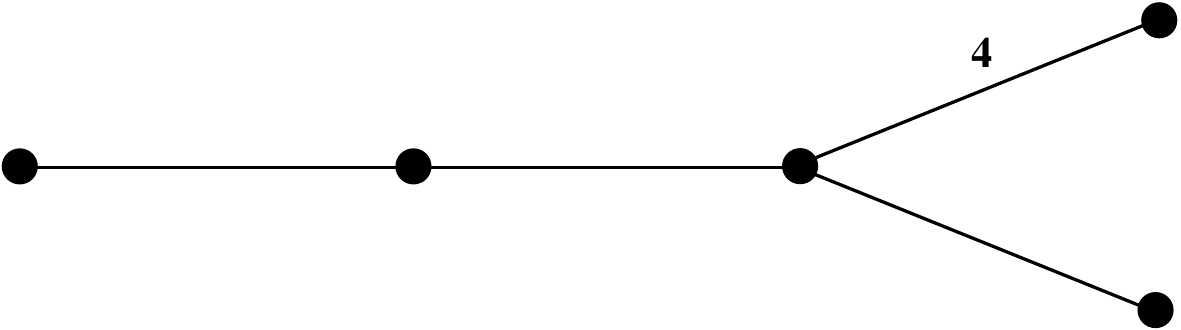}&
    \includegraphics[width=46mm]{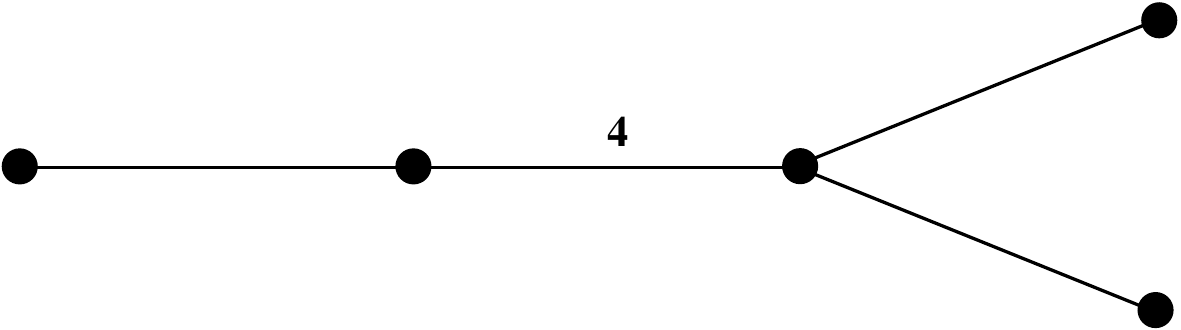} \\
    & & \\ [-2.0 em]
    \includegraphics[width=46mm]{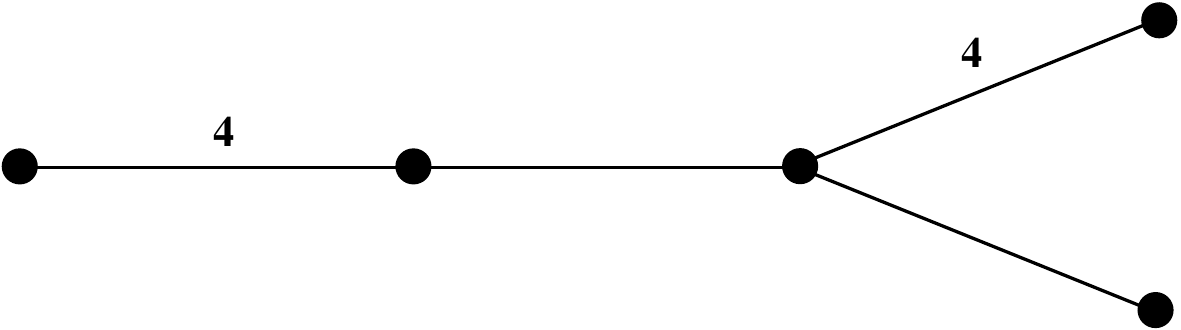}&
    \includegraphics[width=46mm]{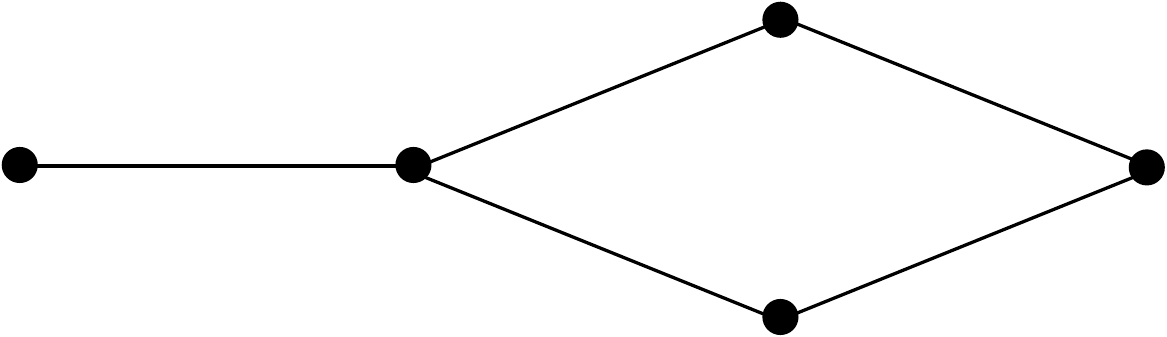} &
    \includegraphics[width=46mm]{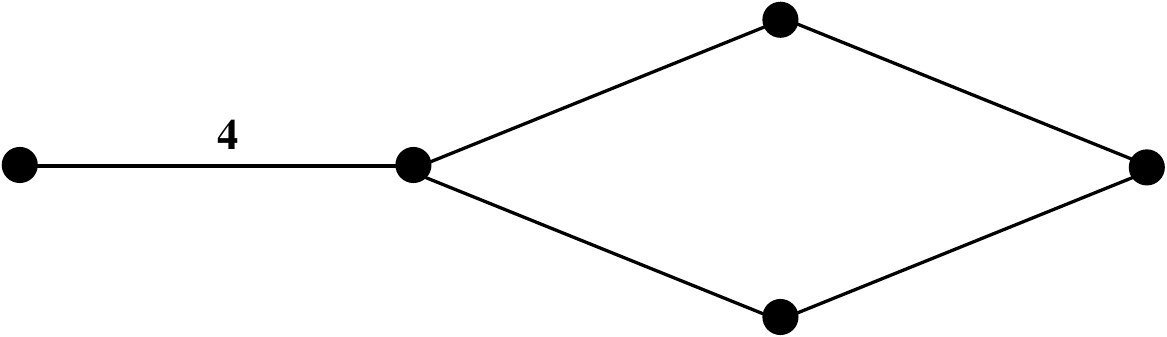} \\
    \bottomrule
  \end{tabular}
  \begin{tabular}{m{33mm}m{33mm}m{32mm}m{33mm}}
    & & & \\ [-2.0 em]
    \includegraphics[width=30mm]{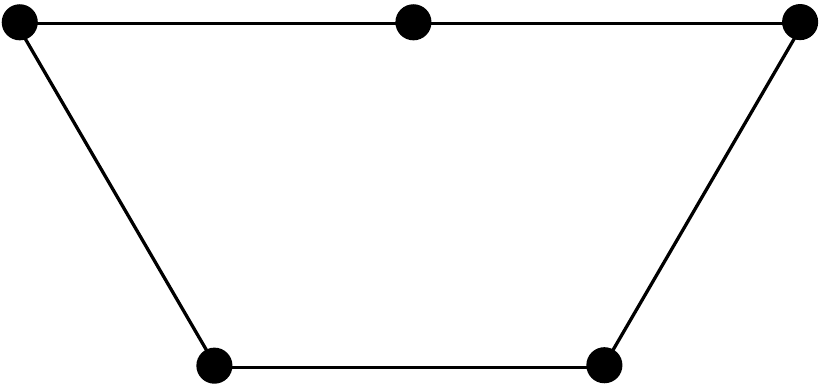} &
    \includegraphics[width=30mm]{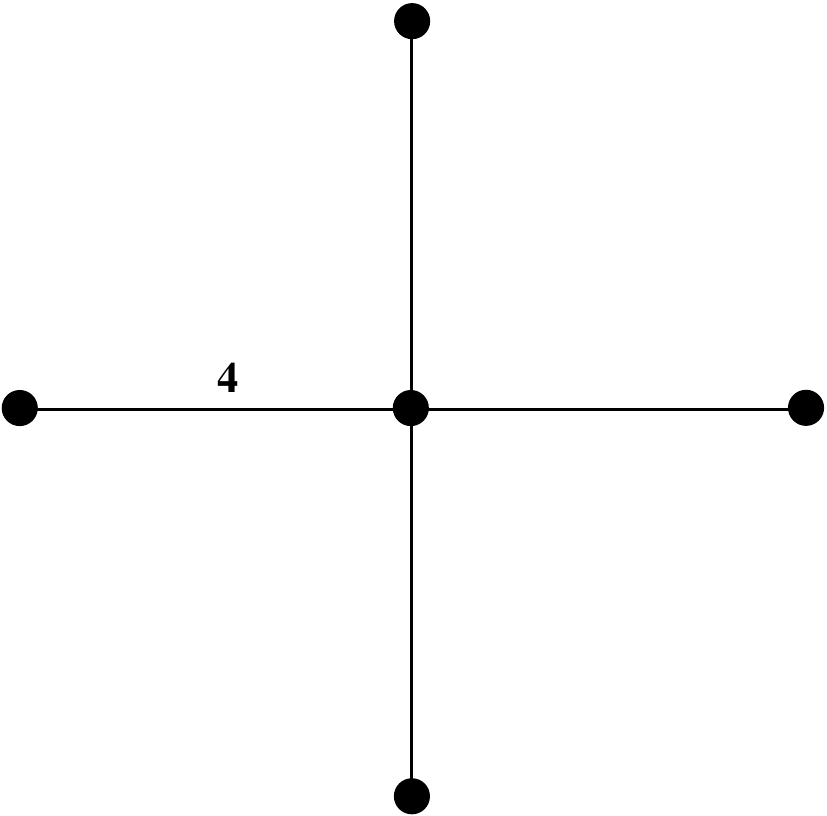} &
    \includegraphics[width=17mm]{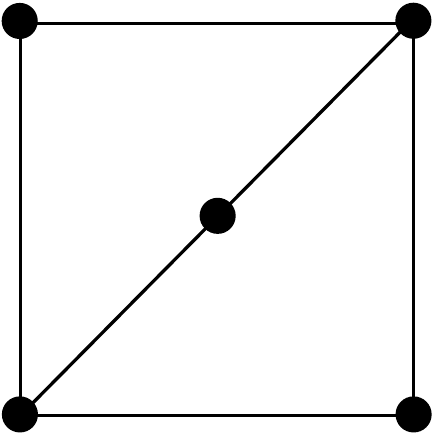} &
    \includegraphics[width=30mm]{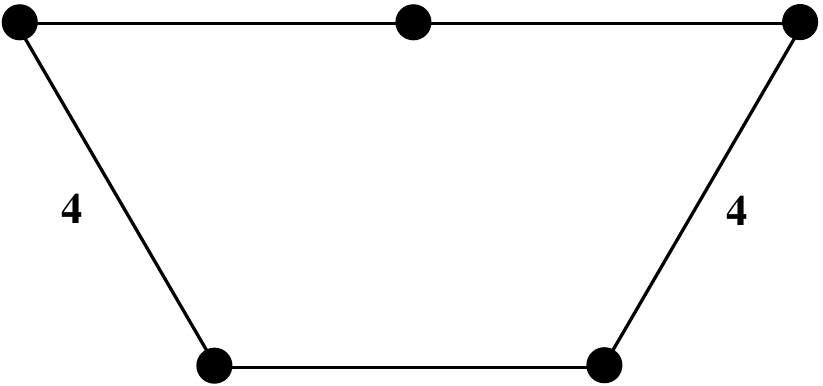} \\
    \bottomrule
  \end{tabular}
  \renewcommand{\arraystretch}{1.0}
\end{table}

\begin{table}
  \caption{Hyperbolic Coxeter groups of rank 6.}
  \label{table:HyperbolicCGRank6}
  \renewcommand{\arraystretch}{2.0}
  \vspace{0.5 em}
  \centering
  \begin{tabular}{m{70mm}m{70mm}}
    \toprule
    & \\ [-1.8 em]
    \includegraphics[width=70mm]{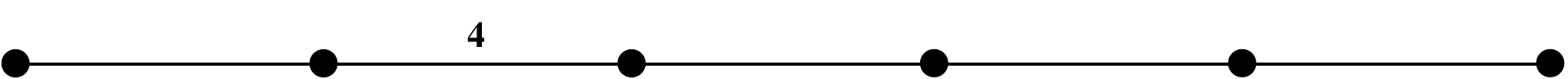} &
    \includegraphics[width=70mm]{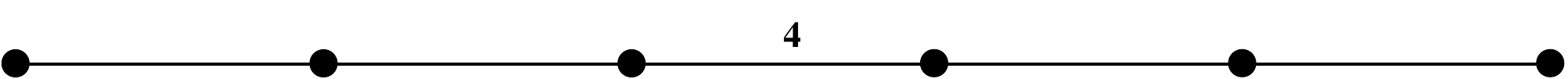} \\
    & \\ [-1.8 em]
    \includegraphics[width=70mm]{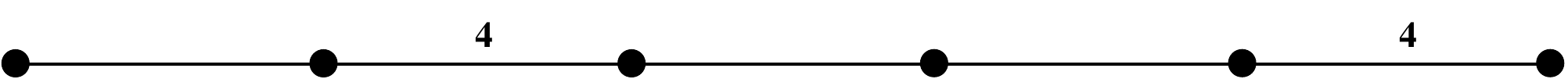} &
    \includegraphics[width=60mm]{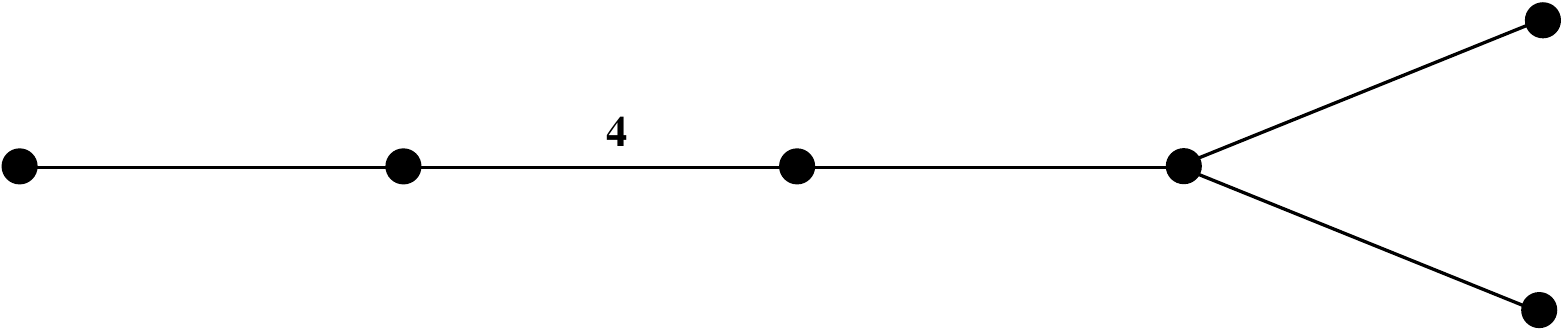} \\
    & \\ [-1.8 em]
    \includegraphics[width=60mm]{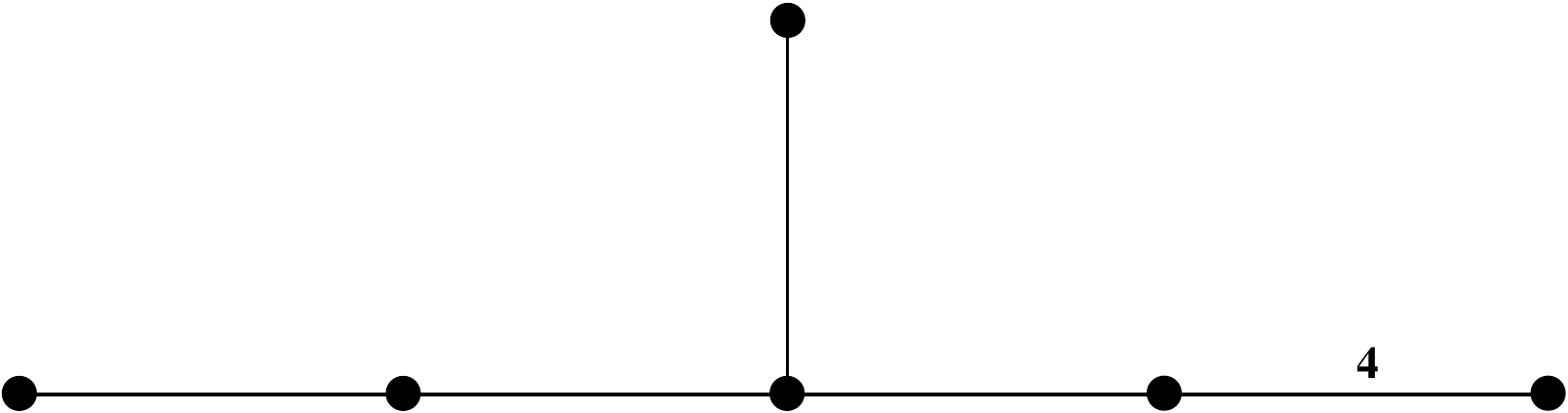} &
    \includegraphics[width=60mm]{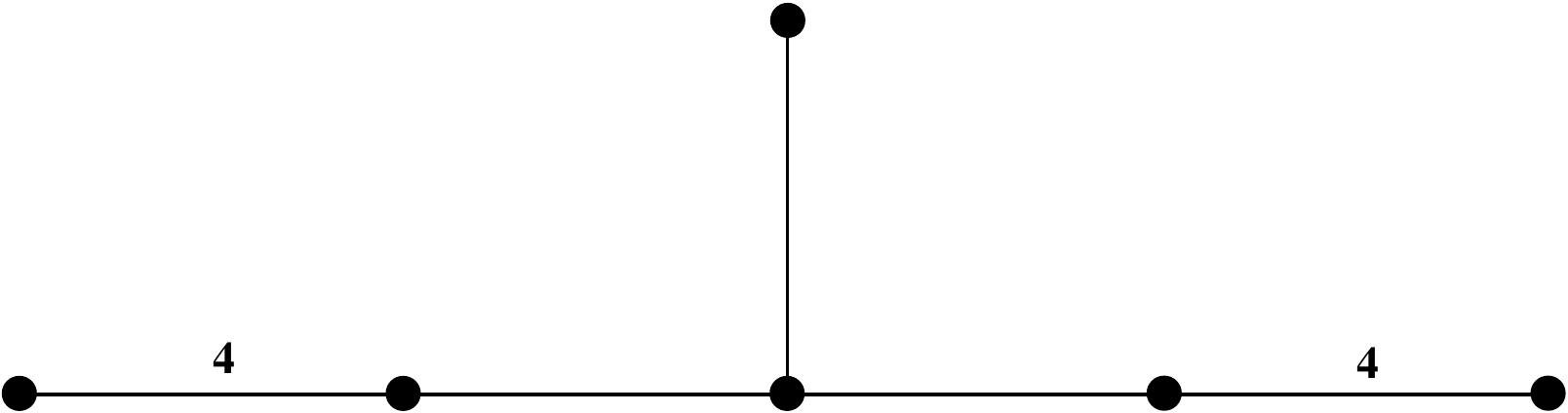} \\
    \bottomrule
  \end{tabular}
  \begin{tabular}{m{45mm}m{46mm}m{45mm}}
    & & \\ [-2.0 em]
    \includegraphics[width=45mm]{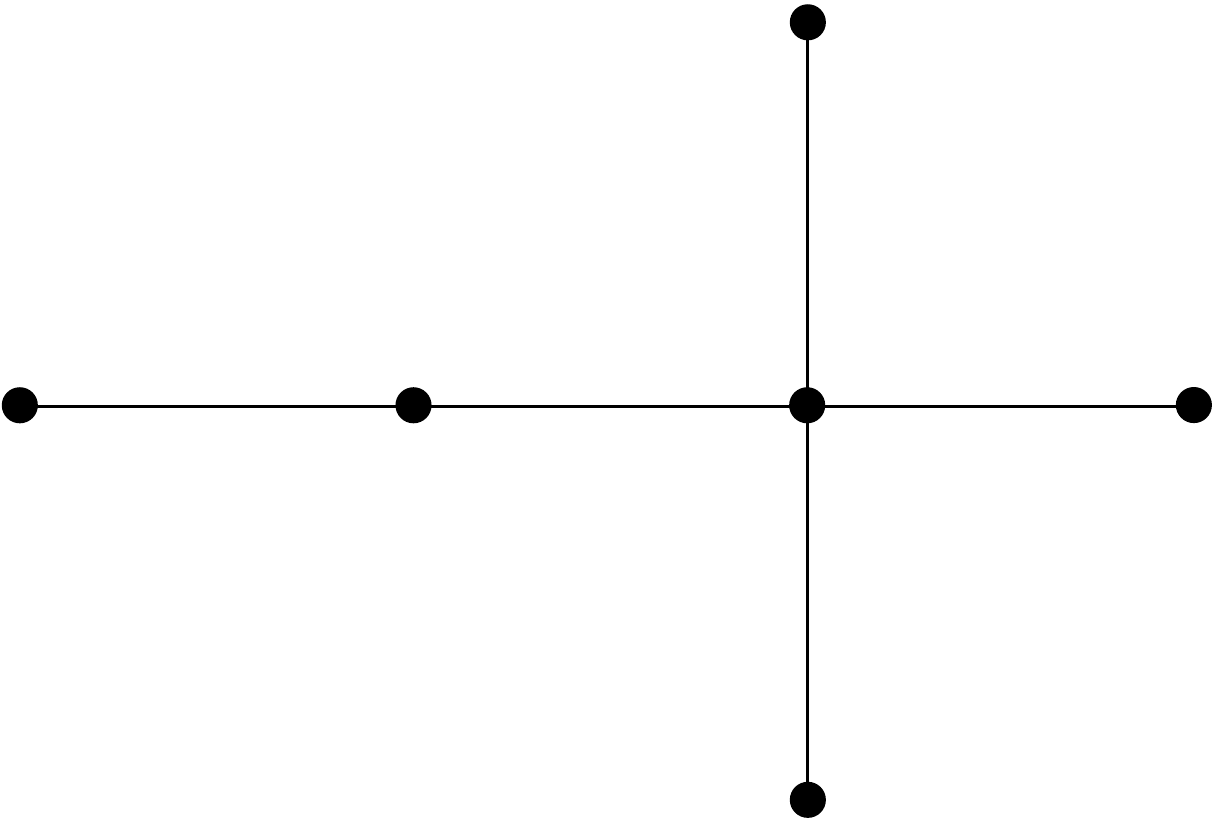} &
    \includegraphics[width=45mm]{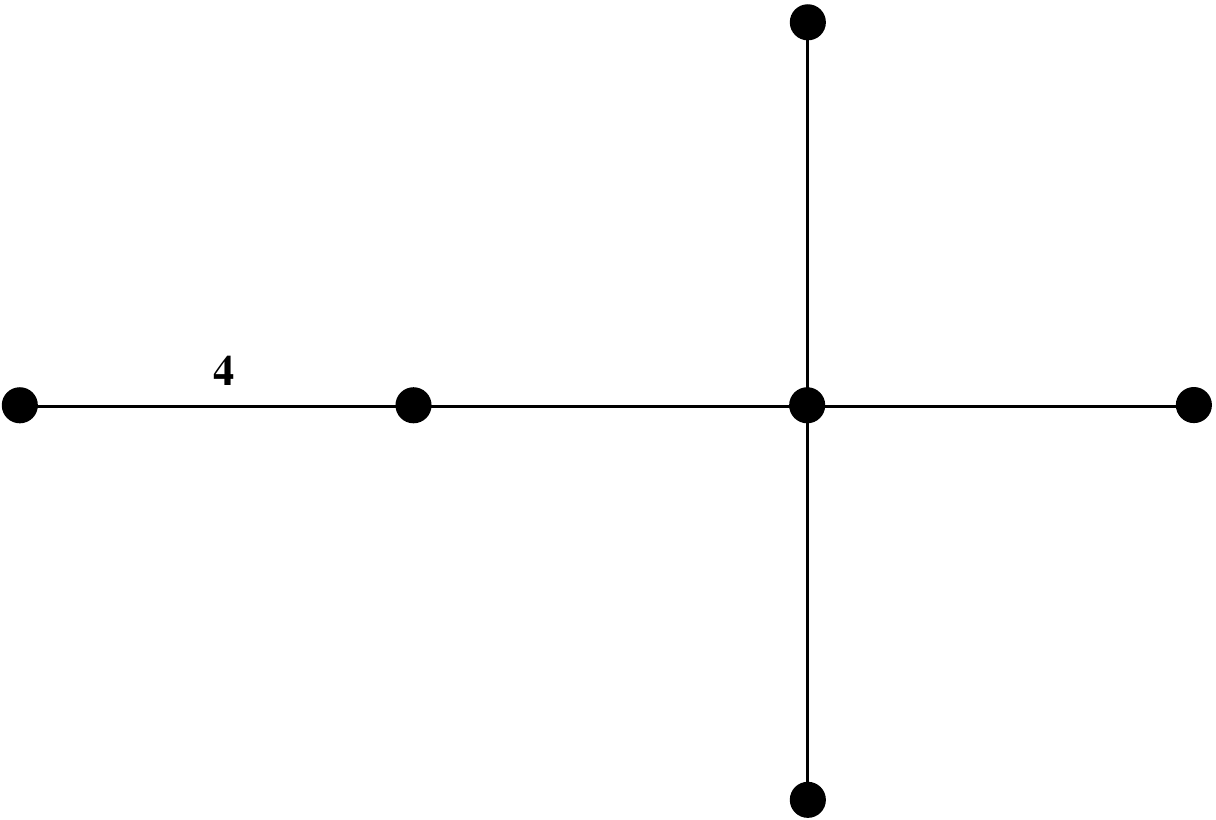} &
    \includegraphics[width=30mm]{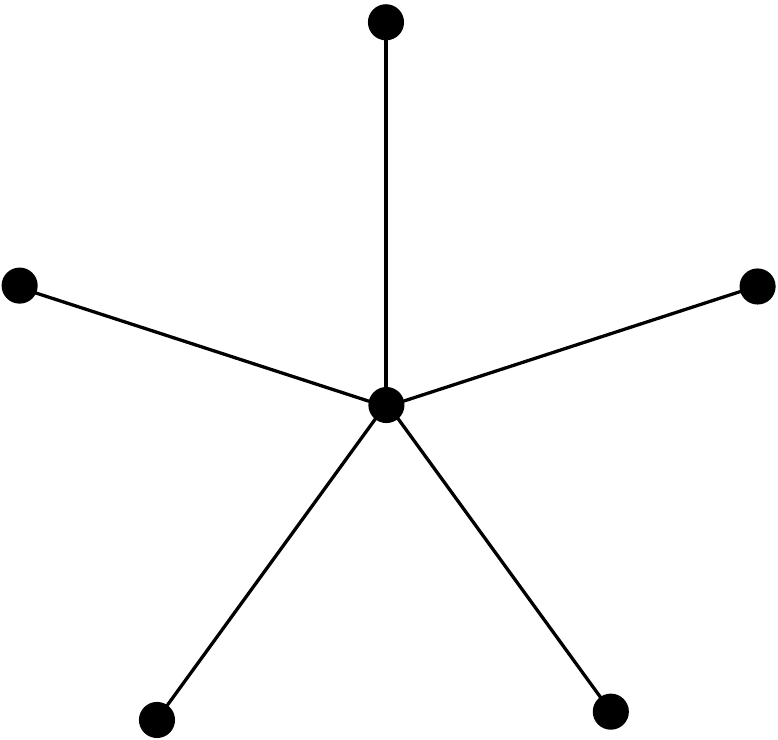} \\
    & & \\ [-1.8 em]
    \includegraphics[width=40mm]{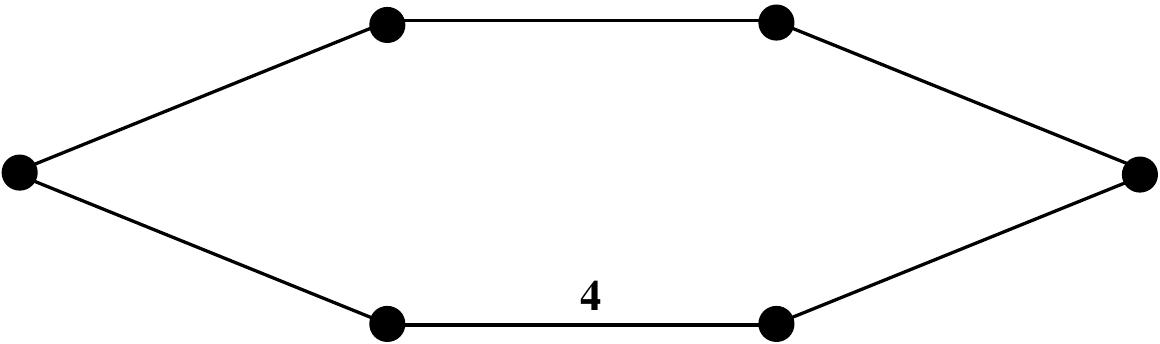} &
    \includegraphics[width=40mm]{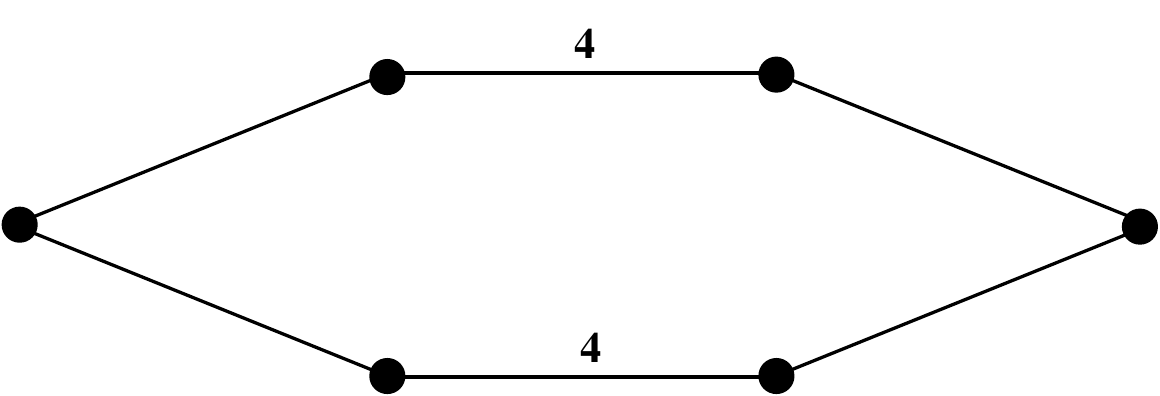} &
    \includegraphics[width=40mm]{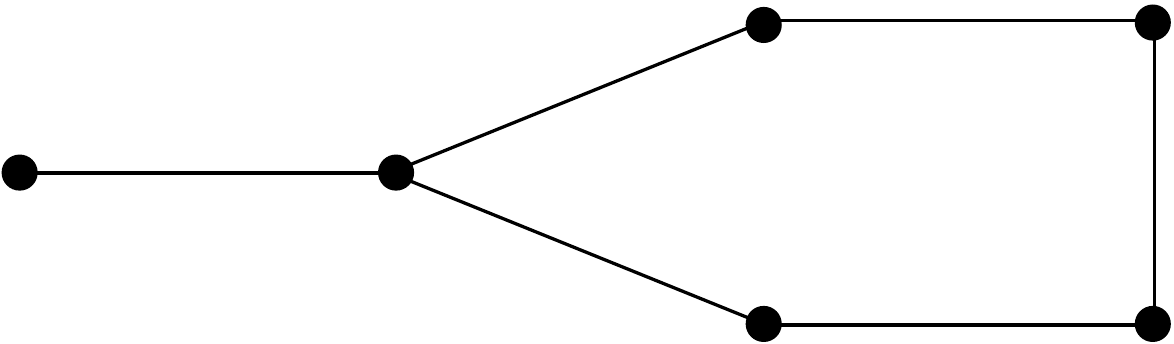}\\
    \bottomrule
  \end{tabular}
  \renewcommand{\arraystretch}{1.0}
\end{table}

\begin{table}
  \caption{Hyperbolic Coxeter groups of rank 7.}
  \label{table:HyperbolicCGRank7}
  \renewcommand{\arraystretch}{2.0}
  \vspace{0.5 em}
  \centering
  \begin{tabular}{m{75mm}}
    \toprule
    \\ [-2.0 em]
    \includegraphics[width=75mm]{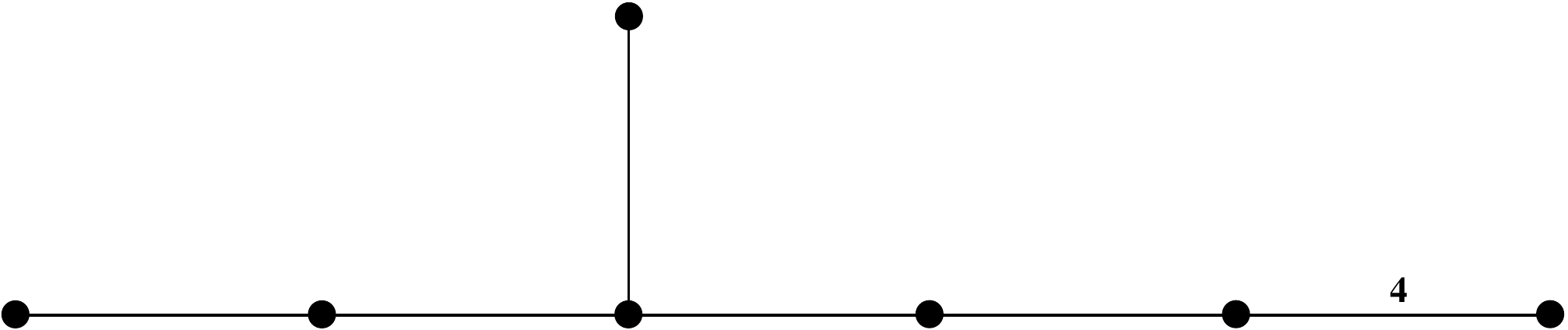} \\  
    \\ [-2.0 em]
    \includegraphics[width=60mm]{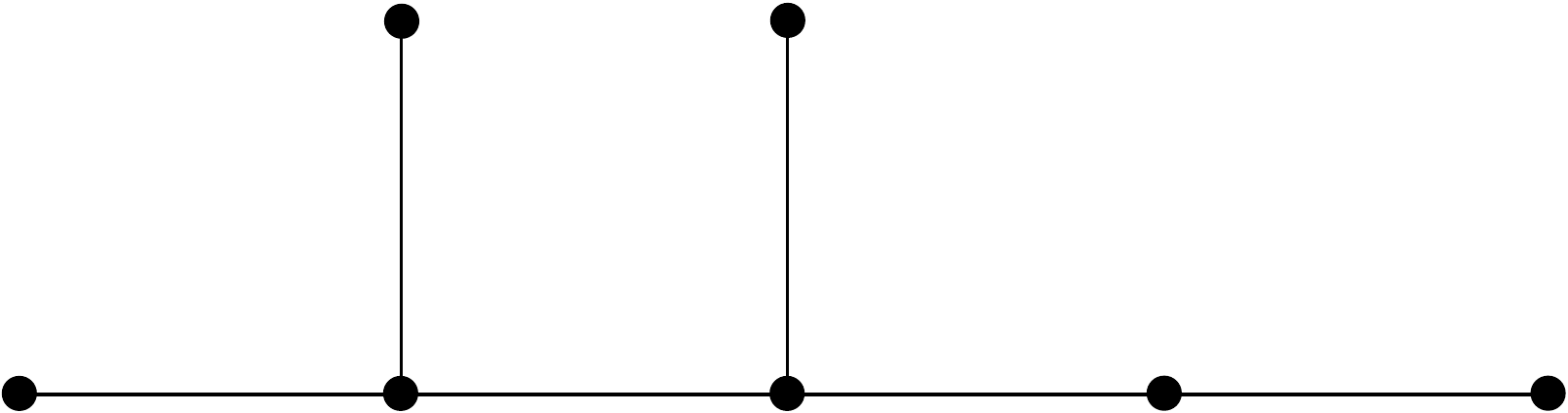} \\
    \\ [-2.0 em]
    \includegraphics[width=60mm]{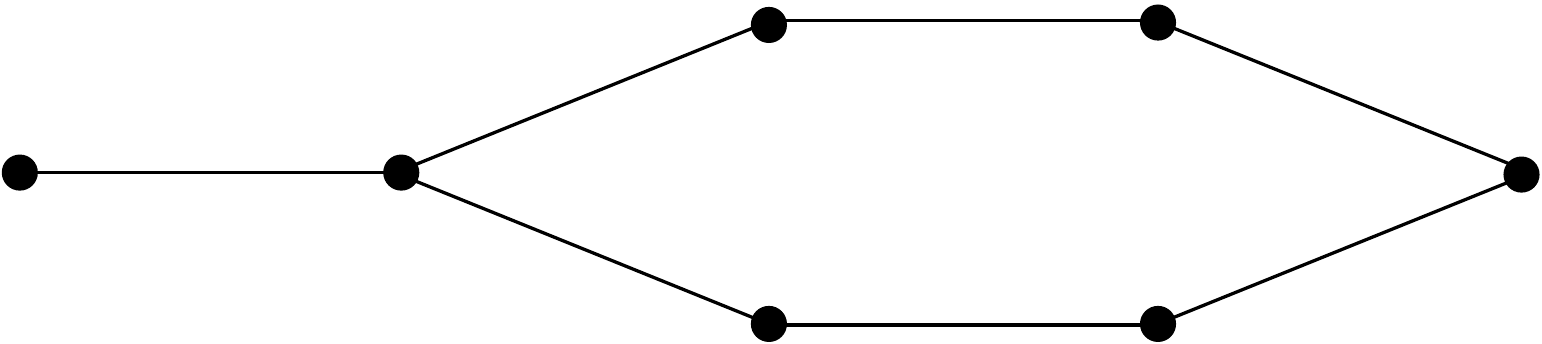} \\
    \bottomrule
  \end{tabular}
  \renewcommand{\arraystretch}{1.0}
\end{table}

\begin{table}
  \caption{Hyperbolic Coxeter groups of rank 8.}
  \label{table:HyperbolicCGRank8}
  \renewcommand{\arraystretch}{2.0}
  \vspace{0.5 em}
  \centering
  \begin{tabular}{m{70mm}m{60mm}}
    \toprule
    & \\ [-2.0 em]
    \includegraphics[width=70mm]{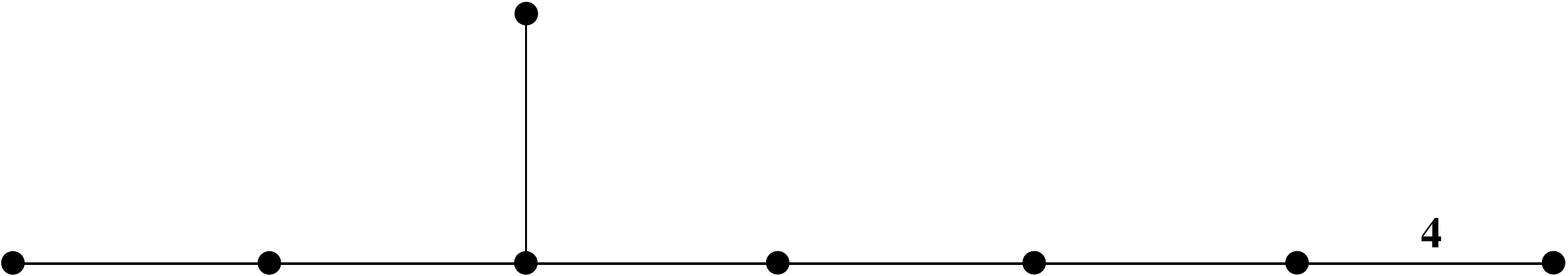} &
    \includegraphics[width=60mm]{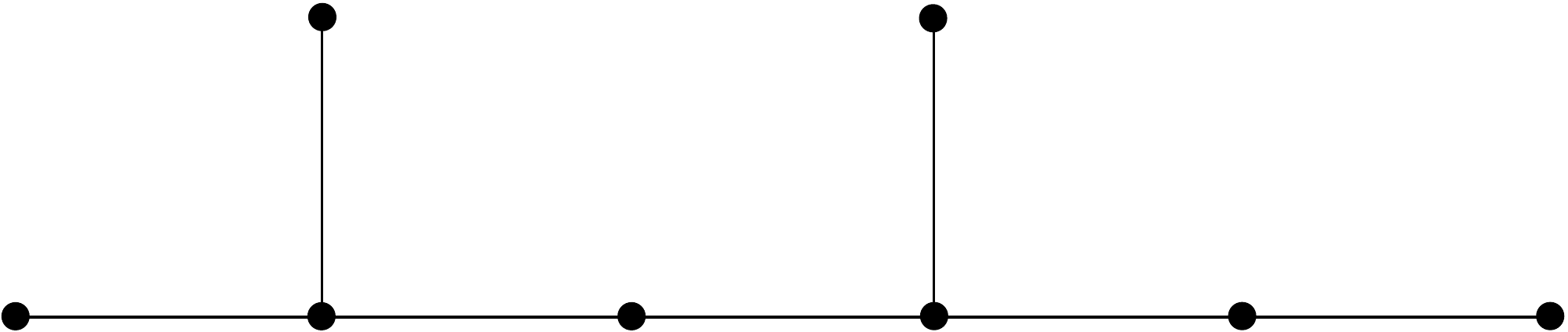} \\
    & \\ [-2.0 em]
    \includegraphics[width=60mm]{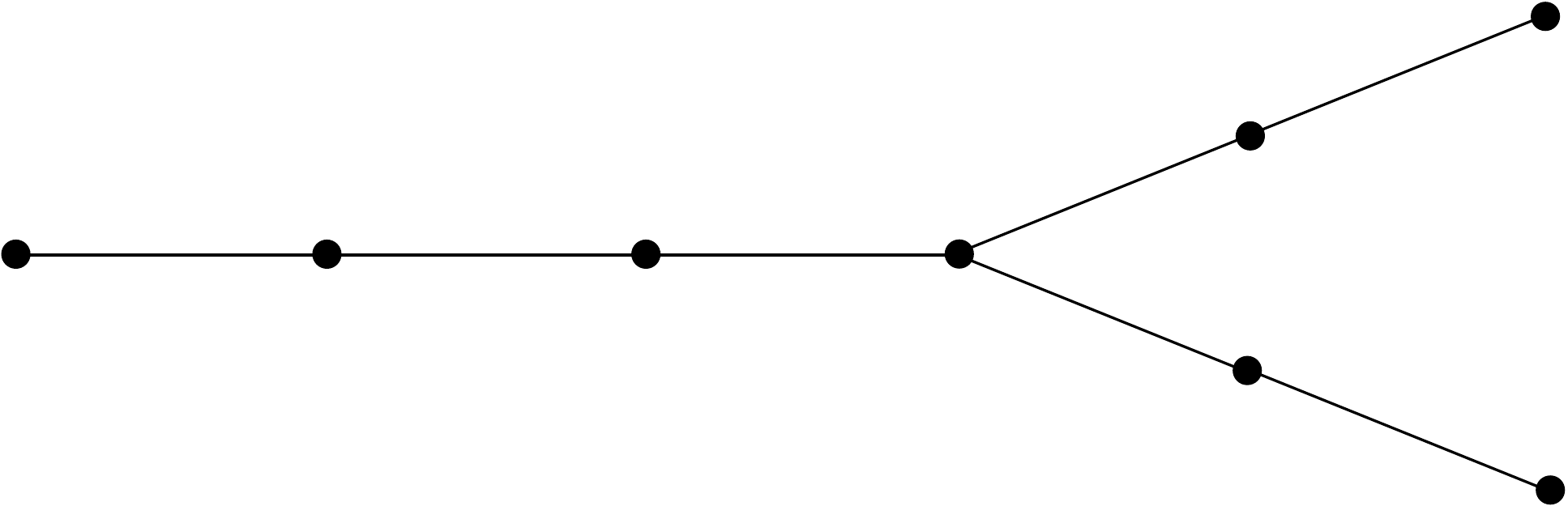} &
    \includegraphics[width=50mm]{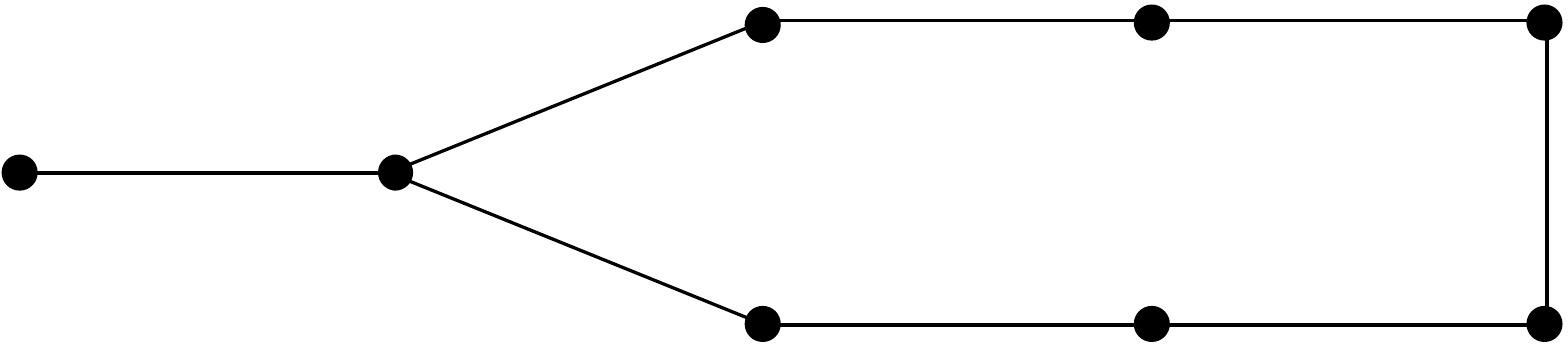} \\
    \bottomrule
  \end{tabular}
  \renewcommand{\arraystretch}{1.0}
\end{table}

\begin{table}
  \caption{Hyperbolic Coxeter groups of rank 9.}
  \label{table:HyperbolicCGRank9}
  \renewcommand{\arraystretch}{2.0}
  \vspace{0.5 em}
  \centering
  \begin{tabular}{m{100mm}}
    \toprule
    \\ [-2.0 em]
    \includegraphics[width=100mm]{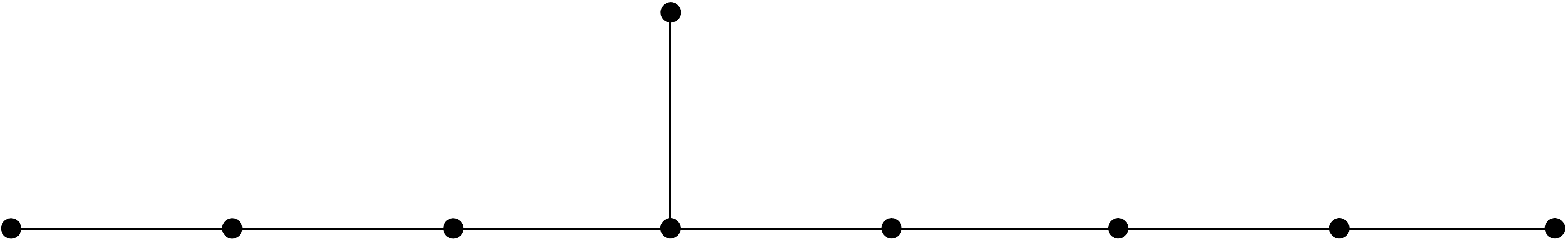} \\
    \\ [-2.0 em]
    \includegraphics[width=100mm]{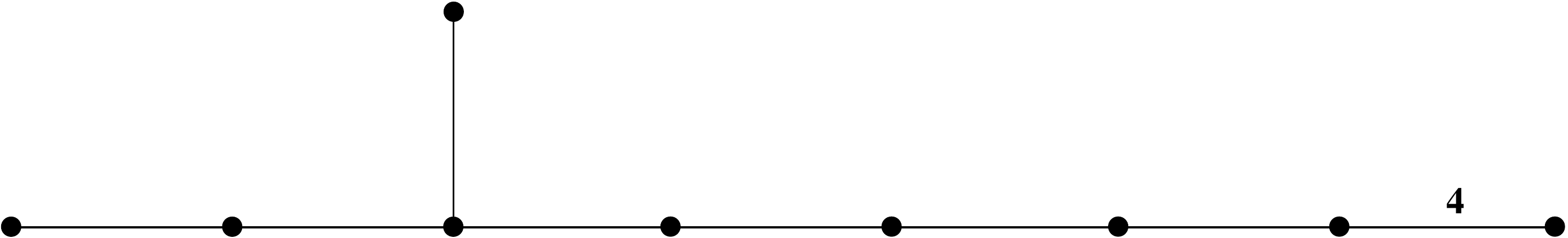} \\
    \\ [-2.0 em]
    \includegraphics[width=95mm]{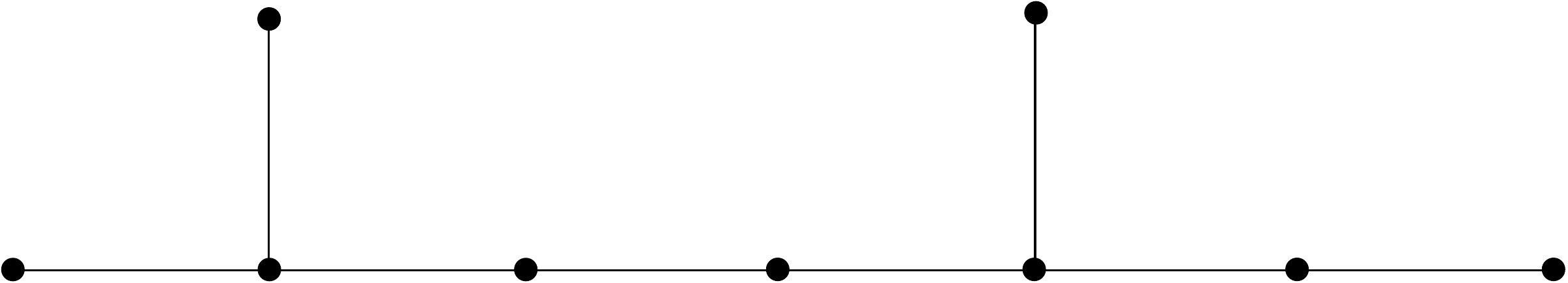} \\
    \\ [-2.0 em]
    \includegraphics[width=80mm]{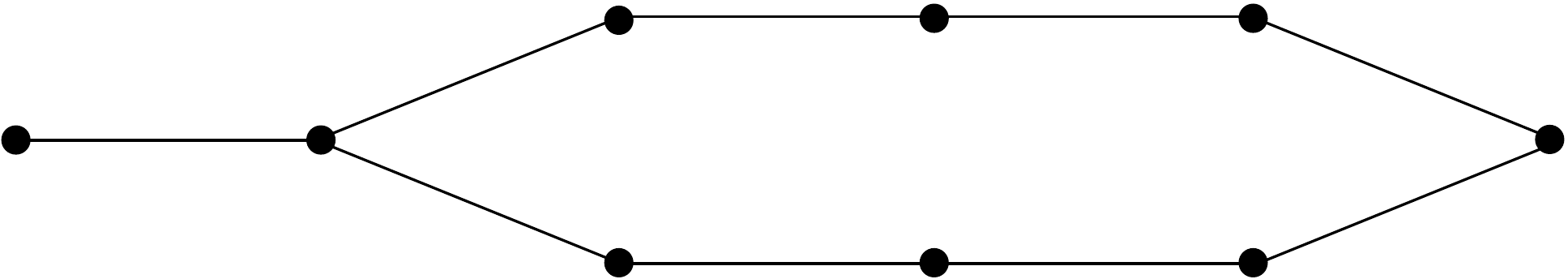} \\
    \bottomrule
  \end{tabular}
  \renewcommand{\arraystretch}{1.0}
\end{table}

\begin{table}
  \caption{Hyperbolic Coxeter groups of rank 10.}
  \label{table:HyperbolicCGRank10}
  \renewcommand{\arraystretch}{2.0}
  \vspace{0.5 em}
  \centering
  \begin{tabular}{m{110mm}}
    \toprule
    \\ [-2.0 em]
    \includegraphics[width=110mm]{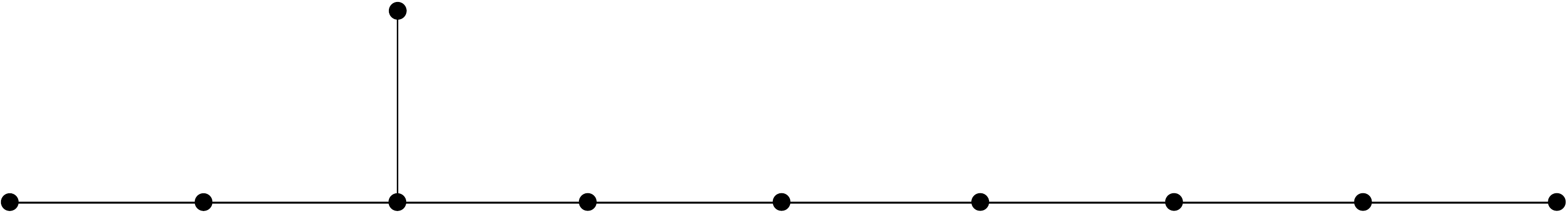} \\  
    \\ [-2.0 em]
    \includegraphics[width=110mm]{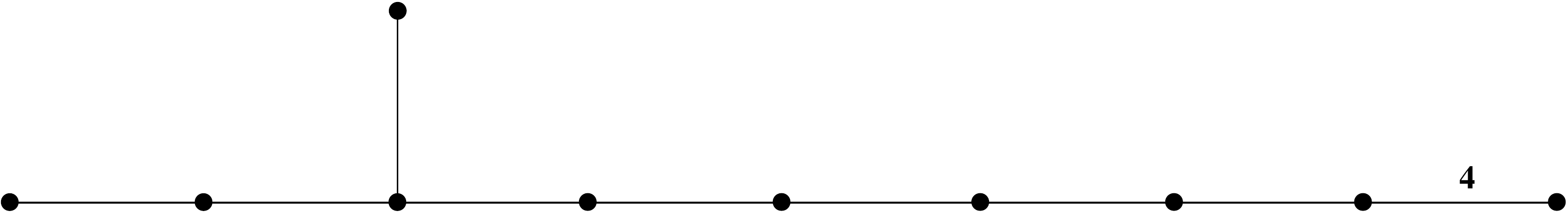} \\
    \\ [-2.0 em]
    \includegraphics[width=110mm]{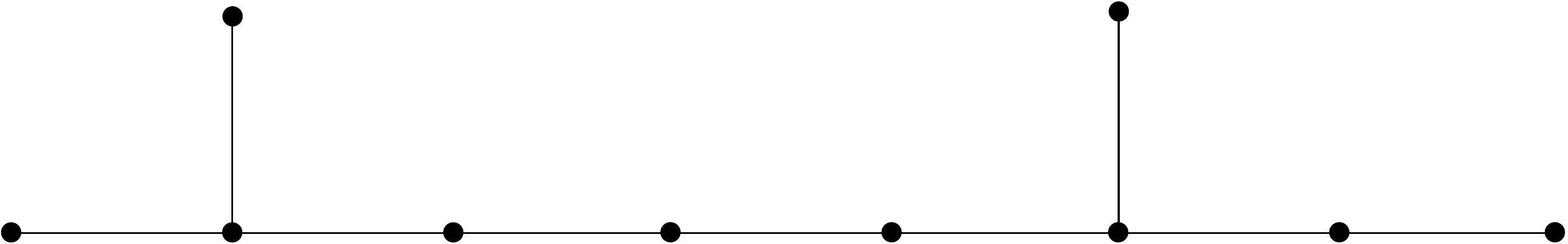} \\
    \bottomrule
  \end{tabular}
  \renewcommand{\arraystretch}{1.0}
\end{table}


\subsection{Crystallographic Coxeter groups}
\label{section:CrystalCoxeterGroups}

Among the Coxeter groups, only those that are crystallographic
correspond to Weyl groups of Kac--Moody algebras. Therefore we now
introduce this important concept. By definition, a Coxeter group is
crystallographic if it stabilizes a lattice in $V$. This lattice need
not be the lattice generated by the $\alpha_i$'s. As discussed
in~\cite{Humphreys}, a Coxeter group is crystallographic if and only
if two conditions are satisfied: (i) The integers $m_{ij}$ ($i \not=
j$) are restricted to be in the set $\{2,3,4,6, \infty \}$, and (ii)
for any closed circuit in the Coxeter graph of $\mf{C}$, the number of
edges labelled 4 or 6 is even.

Given a crystallographic Coxeter group, it is easy to exhibit a
lattice $L$ stabilized by it. We can construct a basis for that
lattice as follows. The basis vectors $\mu_i$ of the lattice are
multiples of the original simple roots, $\mu_i = c_i \alpha_i$ for
some scalars $c_i$ which we determine by applying the following rules:

\begin{itemize}
\item $m_{ij} = 3 \Rightarrow c_i = c_j $.
\item $m_{ij} = 4 \Rightarrow c_i = \sqrt{2} c_j$ or $c_j = \sqrt{2} c_i$.
\item $m_{ij} = 6 \Rightarrow c_i = \sqrt{3} c_j$ or $c_j = \sqrt{3} c_i$.
\item $m_{ij} = \infty \Rightarrow c_i = c_j$.
\end{itemize}

\noindent
One easily verifies that $\sigma_i(\mu_j) = \mu_j -
d_{ij} \mu_i $ for some integers $d_{ij}$. Hence $L$ is indeed
stabilized. The integers $d_{ij}$ are equal to $2 \frac{B(\mu_i ,
\mu_j)}{B(\mu_i, \mu_i)}$.

The rules are consistent as can be seen by starting from an
arbitrary node, say $\alpha_1$, for which one takes $c_1 = 1$. One
then proceeds to the next nodes in the (connected) Coxeter graph
by applying the above rules. If there is no closed circuit, there
is no consistency problem since there is only one way to proceed
from $\alpha_1$ to any given node. If there are closed circuits,
one must make sure that one comes back to the same vector after
one turn around any circuit. This can be arranged if the number
of steps where one multiplies or divides by $\sqrt{2}$
(respectively, $\sqrt{3}$) is even.

Our construction shows that the lattice $L$ is not unique. If
there are only two different lengths for the lattice vectors
$\mu_i$, it is convenient to normalize the lengths so that the
longest lattice vectors have length squared equal to two. This
choice simplifies the factors $2 \frac{B(\mu_i , \mu_j)}{B(\mu_i,
\mu_i)}$.

The rank 10 hyperbolic Coxeter groups are all crystallographic.
The lattices preserved by $E_{10}$ \index{$E_{10}$} and $DE_{10}$
\index{$DE_{10}$} are unique up to
an overall rescaling because the non-trivial $m_{ij}$ ($i \not=
j$) are all equal to $3$ and there is no choice in the ratios
$c_i/c_j$, all equal to one (first rule above). The Coxeter group
$BE_{10}$ \index{$BE_{10}$} preserves two (dual) lattices.

\subsubsection*{On the normalization of roots and weights in the
  crystallographic case}

Since the vectors $\mu_i$ and $\alpha_i$ are proportional, they
generate identical reflections. Even though they do not
necessarily have length squared equal to unity, the vectors
$\mu_i$ are more convenient to work with because the lattice
preserved by the Coxeter group is simply the lattice $\sum_i
\mathbb{Z} \mu_i$ of points with integer coordinates in the
basis $\{\mu_i\}$. For this reason, we shall call from now on
``simple roots'' the vectors $\mu_i$ and, to follow common
practice, \emph{will sometimes even rename them} $\alpha_i$. Thus,
in the crystallographic case, the (redefined) simple roots \index{root} are
appropriately normalized to the lattice structure. It turns out
that it is with this normalization that simple roots of Coxeter
groups correspond to simple roots of Kac--Moody algebras defined in
the Section~\ref{section:fundamentalweights}. A \emph{root} is any
point on the root lattice that is in the Coxeter orbit of some (redefined)
simple root. It is these roots that coincide with the (real)
roots of Kac--Moody algebras.

It is also useful to rescale the fundamental weights. The
rescaled fundamental weights, of course proportional to
$\omega_i$, are denoted $\Lambda_i$. The convenient normalization
is such that 
\begin{equation}
 (\Lambda_i \vert \mu_j) = \frac{(\mu_j \vert
\mu_j)}{2}\delta_{ij}.
\end{equation}
With this normalization, they coincide
with the fundamental weights of Kac--Moody algebras, to be considered
in Section~\ref{section:KacMoody}.

\newpage


\section{Lorentzian Kac--Moody Algebras}
\label{section:KacMoody}
\setcounter{equation}{0}

The explicit appearance of infinite crystallographic Coxeter
groups in the billiard limit suggests that gravitational theories
might be invariant under a huge symmetry described by Lorentzian
Kac--Moody algebras (defined in Section~\ref{section:definitions}). 
Indeed, there is an intimate connection
between crystallographic Coxeter groups and Kac--Moody algebras.
This connection might be familiar in the finite case. For
instance, it is well known that the finite symmetry group $A_2$ of
the equilateral triangle (isomorphic to the group of permutations
of 3 objects) and the corresponding hexagonal pattern of roots are
related to the finite-dimensional Lie algebra $\mf{sl}(3,\mbb{R})$ (or
$\mf{su}(3)$). The group $A_2$ is in fact the Weyl group of
$\mf{sl}(3,\mbb{R})$ (see Section~\ref{section:weylgroup}).

This connection is not peculiar to the Coxeter group $A_2$ but is
generally valid: Any crystallographic Coxeter group \index{Coxeter group} is the Weyl
group of a Kac--Moody algebra traditionally denoted in the same
way (see Section~\ref{section:weylgroup}). 
This is the reason why it is expected that the Coxeter groups
might signal a bigger symmetry structure. And indeed, there are
indications that this is so since, as we shall discuss in Section~\ref{section:sigmamodels}, an
attempt to reformulate the gravitational Lagrangians in a way that
makes the conjectured symmetry manifest yields intriguing results.

The purpose of this section is to develop the mathematical
concepts underlying Kac--Moody algebras and to explain the
connection between Coxeter groups and Kac--Moody algebras. How
this is relevant to gravitational theories will be discussed in
Section~\ref{section:KMBilliardsI}.


\subsection{Definitions}
\label{section:definitions}

An $n \times n$ matrix A is called
a ``generalized Cartan matrix'' \index{Cartan matrix|bb} (or just ``Cartan matrix'' for short)
if it satisfies the following conditions\epubtkFootnote{We are employing the convention of Kac~\cite{Kac} for the Cartan matrix. There exists an alternative definition of Kac--Moody algebras in the literature, in which the transposed matrix $A^T$ is used instead.}:
\begin{eqnarray}
  A_{ii} &=& 2
  \qquad
  \forall i = 1, \cdots, n,
  \\
  A_{ij} &\in& \mathbb{Z}_-
  \qquad
  (i \not= j),\\ 
  A_{ij} &=& 0 \quad \Rightarrow \quad A_{ji} = 0,
\end{eqnarray}
where $\mathbb{Z}_-$ denotes the non-positive integers. One can
encode the Cartan matrix in terms of a Dynkin diagram, \index{Dynkin diagram|bb} which is
obtained as follows:

\begin{enumerate}
\item For each $i = 1, \cdots, n$, one associates a node in the diagram.
\item One draws a line between the node $i$ and the node $j$ if
  $A_{ij} \not=0$; if $A_{ij} = 0$ ($= A_{ji}$), one draws no line
  between $i$ and $j$.
\item One writes the pair $(A_{ij}, A_{ji})$ over the line joining $i$
  to $j$. When the products $A_{ij} \cdot A_{ji}$ are all $\leq 4$
  (which is the only situation we shall meet in practice), this third
  rule can be replaced by the following rules:
  \begin{enumerate}
  \item one draws a number of lines between $i$ and $j$ equal to
    $\max (\vert A_{ij} \vert, \vert A_{ji} \vert)$;
  \item one draws an arrow from $j$ to $i$ if $\vert A_{ij} \vert >
    \vert A_{ji} \vert.$
  \end{enumerate}
\end{enumerate}

\epubtkImage{Mix.png}{%
  \begin{figure}[htbp]
    \centerline{\includegraphics[width=30mm]{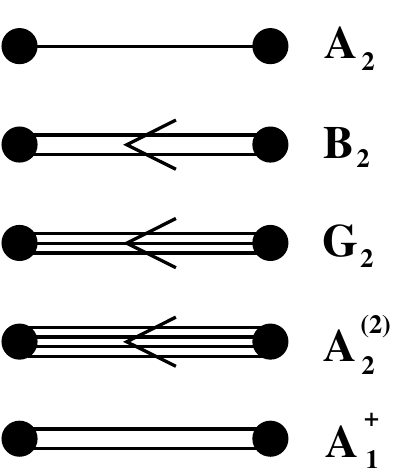}}
    \caption{The Dynkin diagrams corresponding to the finite Lie
      algebras $A_2, B_2$ and $G_2$ and to the affine Kac--Moody algebras
      $A_{2}^{(2)}$ and $A_1^{+}$.}
    \label{figure:Mix}
  \end{figure}}

So, for instance, the Dynkin diagrams in Figure~\ref{figure:Mix} correspond to the
Cartan matrices
\begin{equation}
  A[A_2] = \left(
    \begin{array}{@{}r@{\quad}r@{}}
      2 & -1 \\
      -1 & 2 
    \end{array}
  \right), 
  \label{AA22}
\end{equation}
\begin{equation}
  A[B_2] = \left(
    \begin{array}{@{}r@{\quad}r@{}}
      2 & -2 \\
      -1 & 2
    \end{array}
  \right), 
  \label{BBB222}
\end{equation}
\begin{equation}
  A[G_2] = \left(
    \begin{array}{@{}r@{\quad}r@{}}
      2 & -3 \\
      -1 & 2
    \end{array}
  \right),
\end{equation}
\begin{equation}
  A[A_2^{(2)}] = \left(
    \begin{array}{@{}r@{\quad}r@{}}
      2 & -4 \\
      -1 & 2
    \end{array}
  \right),
\end{equation}
\begin{equation}
  A[A_1^{+}] = \left(
    \begin{array}{@{}r@{\quad}r@{}}
      2 & -2 \\
      -2 & 2
    \end{array}
  \right), 
  \label{A+1A+1}
\end{equation}
respectively. If the Dynkin diagram \index{Dynkin diagram} is
connected, the matrix $A$ is indecomposable. This is what shall be
assumed in the following.

Although this is not necessary for developing the general theory,
we shall impose two restrictions on the Cartan matrix. The first
one is that $\det A \not=0$; the second one is that $A$ is
symmetrizable. The restriction $\det A \not=0$ excludes the
important class of affine algebras and will be lifted below. We
impose it at first because the technical definition of the
Kac--Moody algebra \index{Kac--Moody algebra} when $\det A =0$ is then
slightly more involved.

The second restriction imposes that there exists an invertible
diagonal matrix $D$ with positive elements $\epsilon_i$ and a
symmetric matrix $S$ such that
\begin{equation}
  A = D S.
\end{equation}
The matrix $S$ is called
a symmetrization of $A$ and is unique up to an overall positive
factor because $A$ is indecomposable. To prove this, choose the first
(diagonal) element $\epsilon_1>0$ of $D$ arbitrarily. Since $A$
is indecomposable, there exists a nonempty set $J_1$ of indices
$j$ such that $A_{1j} \not=0$. One has $A_{1j} = \epsilon_1
S_{1j}$ and $A_{j1} = \epsilon_j S_{j1}$. This fixes the
$\epsilon_j$'s $>0$ in terms of $\epsilon_1$ since $S_{1j} =
S_{j1}$. If not all the elements $\epsilon_j$ are determined at
this first step, we pursue the same construction with the elements
$A_{jk} = \epsilon_j S_{jk}$ and $A_{kj} = \epsilon_k S_{kj} =
\epsilon_k S_{kj}$ with $j \in J_1$ and, more generally, at step
$p$, with $j \in J_1 \cap J_2 \cdots \cap J_p$. As the matrix $A$
is assumed to be indecomposable, all the elements $\epsilon_i$ of
$D$ and $S_{ij}$ of $S$ can be obtained, depending only on the
choice of $\epsilon_1$. One gets no contradicting values for the
$\epsilon_j$'s because the matrix $A$ is assumed to be
symmetrizable.

In the symmetrizable case, one can characterize the Cartan matrix
according to the signature of (any of) its symmetrization(s). One
says that $A$ is of finite type if $S$ is of Euclidean signature,
and that it is of Lorentzian type if $S$ is of Lorentzian
signature.

Given a Cartan matrix $A$ (with $\det A \not= 0$), one defines the
corresponding Kac--Moody algebra \index{Kac--Moody algebra|bb}
$\mf{g}=\mf{g}(A)$ as the algebra
generated by $3 n$ generators $h_i, e_i, f_i$ subject to the
following ``Chevalley--Serre'' relations (in addition to the Jacobi
identity and anti-symmetry of the Lie bracket),
\begin{equation}
  \begin{array}{rcll}
    [h_i, h_j] &=& 0,
    \\ [0 em]
    [h_i, e_j] &=& A_{ij} e_j
    & \qquad
    (\mbox{no summation on }j),
    \\ [0 em]
    [h_i, f_j] &=& - A_{ij} f_j
    & \qquad
    (\mbox{no summation on }j),
    \\ [0 em]
    [e_i, f_j] &=& \delta_{ij} h_j
    & \qquad
    (\mbox{no summation on }j),
  \end{array}
  \label{KM}
\end{equation}
\begin{equation}
  \ad_{e_i}^{1-A_{ij}} (e_j) = 0,
  \qquad
  \ad_{f_i}^{1-A_{ij}} (f_j) = 0,
  \qquad
  i \not= j.
  \label{Serre}
\end{equation}
The relations~(\ref{Serre}), called Serre relations, \index{Serre relations|bb} read
explicitly
\begin{equation}
  \underbrace{[e_i, [e_i, [e_i, \cdots ,[e_i, e_j]] \cdots ]}_{1-A_{ij}\mbox{ commutators}} = 0
\end{equation}
(and likewise for the $f_k$'s).

Any multicommutator can be reduced, using the Jacobi identity and
the above relations, to a multicommutator involving only the
$e_i$'s, or only the $f_i$'s. Hence, the Kac--Moody algebra splits as a
direct sum (``triangular decomposition'')
\begin{equation}
  \mf{g} = \mf{n}_- \oplus \mf{h} \oplus \mf{n}_+, 
  \label{TriangularDecomposition}
\end{equation}
where $\mf{n}_-$ is the subalgebra involving the multicommutators
$[f_{i_1}, [f_{i_2}, \cdots,[f_{i_{k-1}}, f_{i_k}] \cdots] $,
$\mf{n}_+$ is the subalgebra involving the multicommutators
$[e_{i_1}, [e_{i_2}, \cdots,[e_{i_{k-1}}, e_{i_k}] \cdots] $ and
$\mf{h}$ is the Abelian subalgebra containing the $h_i$'s. This is
called the \emph{Cartan subalgebra} \index{Cartan subalgebra|bb} and
its dimension $n$ is the \emph{rank} of the Kac--Moody algebra
$\mf{g}$. It should be stressed that the direct sum
Equation~(\ref{TriangularDecomposition}) is a direct sum of
$\mf{n}_-$, $\mf{h}$ and $\mf{n}_+$ as vector spaces, not as
subalgebras (since these subalgebras do not commute).

A priori, the numbers of the multicommutators
\begin{displaymath}
  [f_{i_1}, [f_{i_2}, \cdots,[f_{i_{k-1}}, f_{i_k}] \cdots]
  \qquad \mbox{and} \qquad
  [e_{i_1}, [e_{i_2}, \cdots,[e_{i_{k-1}}, e_{i_k}] \cdots]
\end{displaymath}
are infinite, even after one has taken into account the Jacobi
identity. However, the Serre relations impose non-trivial
relations among them, which, in some cases, make the Kac--Moody
algebra finite-dimensional. Three worked examples are given in
Section~\ref{section:threeexamples} to illustrate the use of the Serre
relations. In fact, one can show~\cite{Kac} that the Kac--Moody
algebra is finite-dimensional if and only if the symmetrization $S$ of
$A$ is positive definite. In that case, the algebra is one of the
finite-dimensional simple Lie algebras given by the Cartan
classification. The list is given in Table~\ref{table:FiniteLieAlgebras}.

When the Cartan matrix $A$ is of Lorentzian signature the Kac--Moody
algebra $\mf{g}(A)$, constructed from $A$ using the Chevalley--Serre
relations, is called a \emph{Lorentzian Kac--Moody algebra}. This is
the case of main interest for the remainder of this paper.

\begin{table}
  \caption{Finite Lie algebras.}
  \label{table:FiniteLieAlgebras}
  \renewcommand{\arraystretch}{1.2}
  \vspace{0.5 em}
  \centering
  \begin{tabular}{m{30mm}|m{60mm}}
    \toprule
    Name & Dynkin diagram \\
    \midrule
    $A_n$ &  \includegraphics[width=55mm]{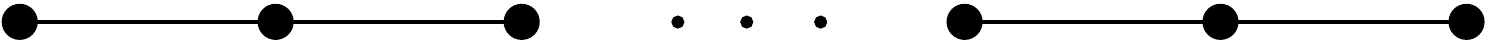} \\ [1 em]
    $B_n$ & \includegraphics[width=55mm]{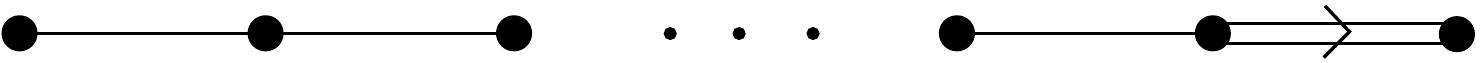} \\ [1 em]
    $C_n$ & \includegraphics[width=55mm]{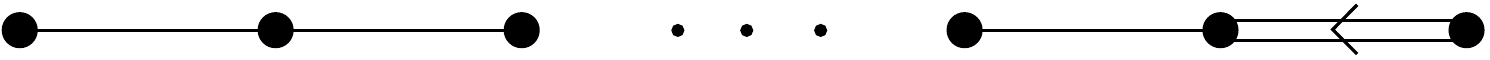} \\ [1 em]
    $D_n$ & \includegraphics[width=55mm]{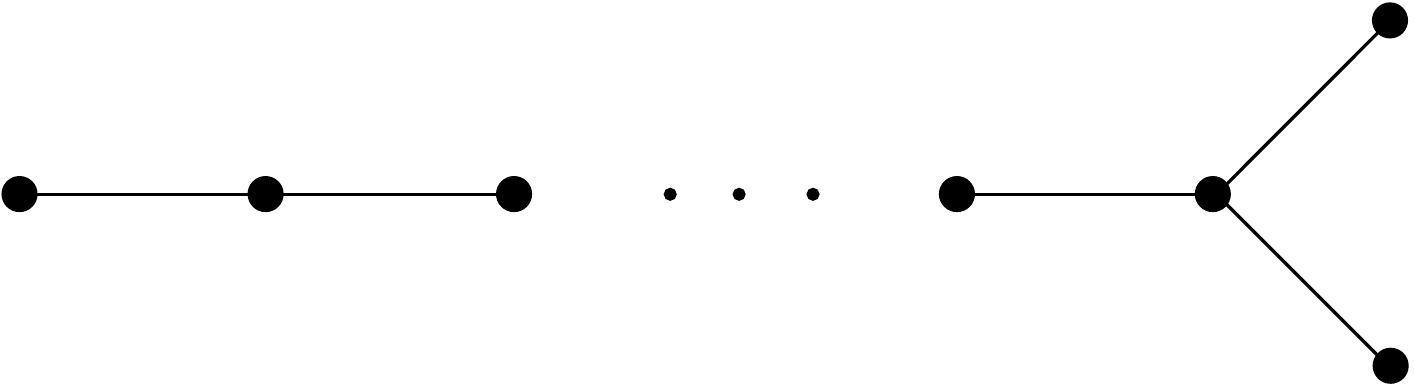} \\ [1 em]
    $G_2$ & \includegraphics[width=12mm]{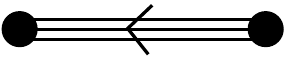} \\ [1 em]
    $F_4$ & \includegraphics[width=33mm]{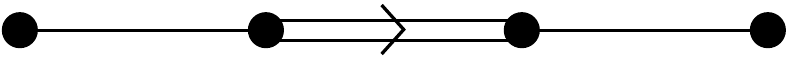} \\ [1 em]
    $E_6$ & \includegraphics[width=40mm]{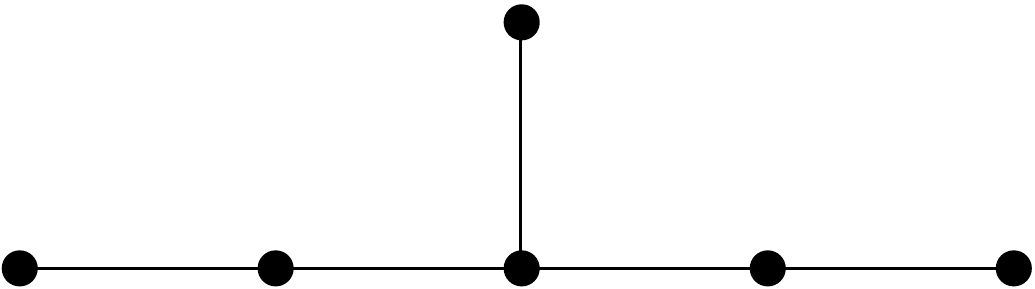} \\ [1 em]
    $E_7$ & \includegraphics[width=50mm]{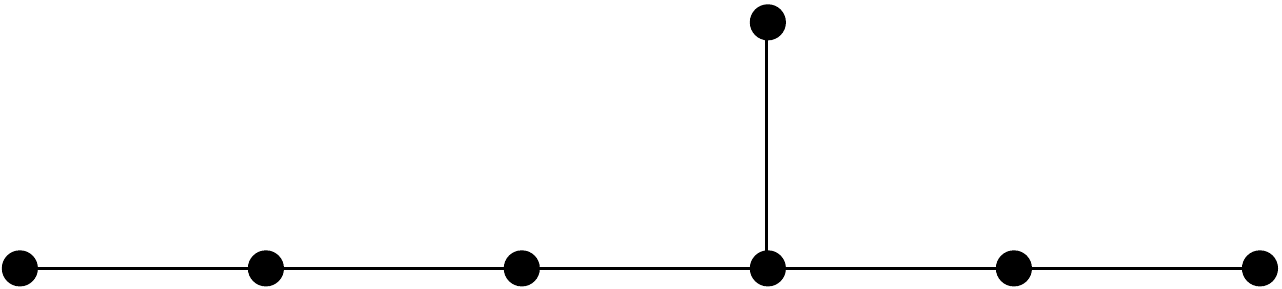} \\ [1 em]
    $E_8$ & \includegraphics[width=60mm]{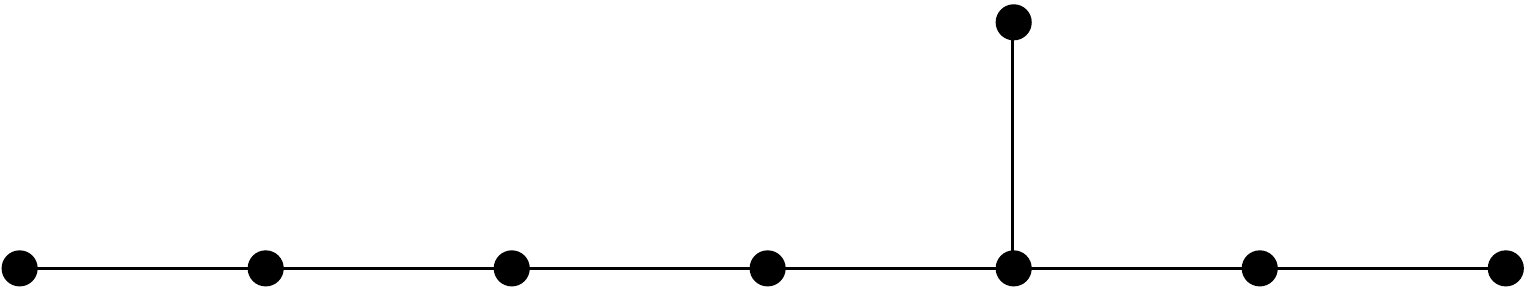} \\
    \bottomrule
  \end{tabular}
  \renewcommand{\arraystretch}{1.0}
\end{table}


\subsection{Roots}
\index{root|bb}

The adjoint action of the Cartan subalgebra on
$\mf{n}_+$ and $\mf{n}_-$ is diagonal. Explicitly,
\begin{equation}
  [h, e_i] = \alpha_i(h) e_i
  \qquad
  (\mbox{no summation on }i)
\end{equation}
for any element $h \in \mf{h}$, where $\alpha_i$ is the linear form
on $\mf{h}$ (i.e., the element of the dual $\mf{h}^*$)
defined by $\alpha_i (h_j) = A_{ji}$. The $\alpha_i$'s are called
the simple roots. Similarly,
\begin{equation}
  [h, [e_{i_1}, [e_{i_2}, \cdots,[e_{i_{k-1}}, e_{i_k}] \cdots]] =
  (\alpha_{i_1} + \alpha_{i_2} + \cdots \alpha_{i_k})(h) \,
  [e_{i_1}, [e_{i_2}, \cdots,[e_{i_{k-1}}, e_{i_k}] \cdots]
\end{equation}
and, if $[e_{i_1}, [e_{i_2}, \cdots,[e_{i_{k-1}}, e_{i_k}] \cdots]$ is
non-zero, one says that $\alpha_{i_1} + \alpha_{i_2} + \cdots
\alpha_{i_k}$ is a positive root. On the negative side, $\mf{n}_-$,
one has
\begin{equation}
  [h, [f_{i_1}, [f_{i_2}, \cdots,[f_{i_{k-1}}, f_{i_k}] \cdots] ] =
  - (\alpha_{i_1} + \alpha_{i_2} + \cdots \alpha_{i_k})(h) \,
  [f_{i_1}, [f_{i_2}, \cdots,[f_{i_{k-1}}, f_{i_k}] \cdots]
\end{equation}
and $-(\alpha_{i_1} + \alpha_{i_2} + \cdots \alpha_{i_k})(h)$ is called
a negative root when $[f_{i_1}, [f_{i_2}, \cdots,[f_{i_{k-1}},
f_{i_k}]$ is non-zero. This occurs if and only if $[e_{i_1},
[e_{i_2}, \cdots,[e_{i_{k-1}}, e_{i_k}] \cdots]$ is non-zero: $-
\al$ is a negative root if and only if $\al$ is a positive
root.

We see from the construction that the roots (linear forms $\alpha$
such that $[h, x] = \alpha(h) x$ has nonzero solutions $x$) are
either positive (linear combinations of the simple roots $\al_i$ with
integer non-negative coefficients) or negative (linear
combinations of the simple roots with integer non-positive
coefficients). The set of positive roots is denoted by $\Delta_+$;
that of negative roots by $\Delta_-$. The set of all roots is
$\Delta$, so we have $\Delta = \Delta_+ \cup \Delta_-$. The
simple roots are positive and form a basis of $\mf{h}^*$. One
sometimes denotes the $h_i$ by $\alpha_i^\vee$ (and thus,
$[\alpha_i^\vee, e_j] = A_{ij} e_j$ etc). Similarly, one also uses
the notation $\langle \cdot, \cdot \rangle$ for the standard
pairing between $\mf{h}$ and its dual $\mf{h}^*$, i.e., $\langle
\alpha, h \rangle = \alpha(h)$. In this notation the entries of the
Cartan matrix can be written as
\begin{equation}
  A_{ij}=\al_j(\al_i^{\vee}) = \left< \al_j, \al_i^{\vee} \right>.
\end{equation}

Finally, the root lattice \index{root lattice|bb} $Q$ is the set of linear combinations with
integer coefficients of the simple roots,
\begin{equation}
  Q = \sum_i \mathbb{Z} \alpha_i.
\end{equation}

All roots \index{root} belong to the root lattice, of course, but the converse
is not true: There are elements of $Q$ that are not roots.


\subsection{The Chevalley involution}

The symmetry between the positive and negative subalgebras $\mf{n}_+$ and
$\mf{n}_-$ of the Kac--Moody algebra can be rephrased formally
as follows: The Kac--Moody algebra is invariant under the Chevalley
involution $\tau$, defined on the generators as
\begin{equation}
  \tau (h_i) = - h_i,
  \qquad
  \tau(e_i) = - f_i,
  \qquad
  \tau(f_i) = - e_i.
\end{equation}
The Chevalley involution \index{Chevalley involution|bb} is in fact an algebra automorphism that
exchanges the positive and negative sides of the algebra.

Finally, we quote the following useful theorem. \vspace{.2cm}

\begin{theorem}
  The Kac--Moody algebra $\mf{g}$ defined by the relations~(\ref{KM},
  \ref{Serre}) is simple.

  \noindent
  The proof may be found in Kac' book~\cite{Kac}, page~12.
\end{theorem}

\noindent
We note that invertibility and indecomposability of the Cartan matrix
$A$ are central ingredients in the proof. In
particular, the theorem does not hold in the affine case, for
which the Cartan matrix is degenerate and has non-trivial
ideals\epubtkFootnote{We recall that an ideal
$\mf{i}$ is a subalgebra such that $[\mf{i},\mf{g}] \subset
\mf{i}$. A simple algebra has no non-trivial ideals.} (see~\cite{Kac}
and Section~\ref{section:theAffineCase}).


\subsection{Three examples}
\label{section:threeexamples}

To get a feeling for how the Serre relations \index{Serre relations} work, we treat in detail
three examples.

\begin{itemize}

\item $A_2$: We start with $A_2$, the Cartan matrix of which is
  Equation~(\ref{AA22}). The defining relations are then:
  \begin{equation}
    \begin{array}{rcl@{\qquad}rcl@{\qquad}rcl}
      [h_1, h_2] &=& 0, &
      [h_1, e_1] &=& 2e_1, &
      [h_1, e_2] &=& - e_2,
      \\ [0.25 em]
      [h_1, f_1] &=& -2 f_1, &
      [h_1, f_2] &=& f_2, &
      [h_2, e_1] &=& - e_1,
      \\ [0.25 em]
      [h_2, e_2] &=& 2 e_2, &
      [h_2, f_1] &=& f_1, &
      [h_2, f_2] &=& -2 f_2,
      \\ [0.25 em]
      [e_1, [e_1, e_2]] &=& 0, &
      [e_2,[e_2, e_1]] &=& 0, &
      [f_1, [f_1, f_2]] &=& 0,
      \\ [0.25 em]
      [f_2, [f_2, f_1]] &=& 0 &
      [e_i,f_j] &=& \delta_{ij} h_j.
    \end{array}
  \end{equation}
  The commutator $[e_1, e_2]$ is not killed by the defining
  relations and hence is not equal to zero (the defining relations
  are \emph{all} the relations). All the commutators with three (or
  more) $e$'s are however zero. A similar phenomenon occurs on the
  negative side. Hence, the algebra $A_2$ is eight-dimensional and
  one may take as basis $\{h_1, h_2, e_1, e_2, [e_1,e_2], f_1, f_2,
  [f_1, f_2] \}$. The vector $[e_1, e_2]$ corresponds to the
  positive root $\alpha_1 + \alpha_2$.
\item $B_2$: The algebra $B_2$, the Cartan matrix of which is
  Equation~(\ref{BBB222}), is defined by the same set of generators,
  but the Serre relations are now $[e_1, [e_1, [e_1, e_2]]] = 0$ and
  $[e_2, [e_2, e_1]] = 0$ (and similar relations for the $f$'s). The
  algebra is still finite-dimensional and contains, besides the
  generators, the commutators $[e_1, e_2]$, $[e_1, [e_1, e_2]]$, their
  negative counterparts $[f_1, f_2]$ and $[f_1, [f_1, f_2]]$, and
  nothing else. The triple commutator $[e_1, [e_1, [e_1, e_2]]]$
  vanishes by the Serre relations. The other triple commutator $[e_2,
  [e_1, [e_1, e_2]]]$ vanishes also by the Jacobi identity and the
  Serre relations,
  \begin{displaymath}
    [e_2, [e_1,[e_1, e_2]]] = [[e_2, e_1],[e_1, e_2]] +
    [e_1, [e_2, [e_1, e_2]]] = 0.
  \end{displaymath}
  (Each term on the right-hand side is zero: The first by antisymmetry
  of the bracket and the second because $[e_2, [e_1, e_2]] = - [e_2,
  [e_2, e_1]] = 0$.) The algebra is 10-dimensional and is isomorphic
  to $\mf{so}(3,2)$.
\item $A_1^{+}$: We now turn to $A_1^+$, the Cartan matrix of which is
  Equation~(\ref{A+1A+1}). This algebra is defined by the same set of
  generators as $A_2$, but with Serre relations given by
  \begin{equation}
    \begin{array}{rcl}
      [e_1, [e_1,[e_1, e_2]]] & = & 0,
      \\ [0.2 em]
      [e_2, [e_2,[e_2, e_1]]] & = & 0
    \end{array}
  \end{equation}
  (and similar relations for the $f$'s). This innocent-looking change in
  the Serre relations has dramatic consequences because the
  corresponding algebra is infinite-dimensional. (We analyze here the
  algebra generated by the $h$'s, $e$'s and $f$'s, which is in fact
  the derived Kac--Moody algebra -- see
  Section~\ref{section:theAffineCase} on affine Kac--Moody
  algebras. The derived algebra is already infinite-dimensional.) To
  see this, consider the $\mf{sl}(2, \mbb{R})$ current algebra, defined
  by 
  \begin{equation}
    [J_m^a, J_n^b] = f^{ab}{}_{c} J^c_{m+n} + m k^{ab} c \delta_{m+n,0},
    \label{affineKMsl2}
  \end{equation}
  where $a = 3, +, -$,
  $f^{ab}{}_{c}$ are the structure constants of
  $\mf{sl}(2, \mbb{R})$ and where $k^{ab}$ is the invariant metric on
  $\mf{sl}(2, \mbb{R})$ which we normalize here so that $k^{-+} = 1$. The
  subalgebra with $n=0$ is isomorphic to $\mf{sl}(2, \mbb{R})$,
  \begin{displaymath}
    [J_0^3 , J_0^+] = 2 J_0^+,
    \qquad
    [J_0^3, J_0^-] = - 2 J_0^-,
    \qquad
    [J_0^+, J_0^-] = J_0^3.
  \end{displaymath}
  The current algebra~(\ref{affineKMsl2}) is generated by $J_0^a$,
  $c$, $J_1^-$ and $J_{-1}^+$ since any element can be written as a multi-commutator
  involving them. The map
  \begin{equation}
    \begin{array}{rcl@{\qquad}rcl}
      h_1 &\rightarrow& J_0^3, &
      h_2 &\rightarrow& - J_0^3 + c,
      \\ [0.25 em]
      e_1 &\rightarrow& J_0^+, &
      e_2 &\rightarrow& J_1^-,
      \\ [0.25 em]
      f_1 &\rightarrow& J_0^-, &
      f_2 &\rightarrow& J_{-1}^+
    \end{array}
  \end{equation}
  preserves the defining relations of the Kac--Moody
  algebra and defines an isomorphism of the (derived) Kac--Moody
  algebra with the current algebra. The Kac--Moody algebra is
  therefore infinite-dimensional. One can construct non-vanishing
  infinite multi-commutators, in which $e_1$ and $e_2$ alternate:
  \begin{equation}
    \begin{array}{rcl@{\qquad}l}
      [e_1, [e_2, [e_1, \cdots, [e_1, e_2] \cdots ]]] &\sim& J^3_n &
      (n~e_1\mbox{'s and }n~e_2\mbox{'s}),
      \\ [0.25 em]
      [e_1, [e_2, [e_1, \cdots, [e_2, e_1] \cdots ]]] &\sim& J^+_n &
      (n+1~e_1\mbox{'s and }n~e_2\mbox{'s}),
      \\ [0.25 em]
      [e_2, [e_1, [e_2, \cdots, [e_1, e_2] \cdots ]]] &\sim& J^-_{n+1} &
      (n~e_1\mbox{'s and }n+1~e_2\mbox{'s}).
    \end{array}
  \end{equation}
  The Serre relations do not cut
  the chains of multi-commutators to a finite number.
\end{itemize}

\noindent
We see from these examples that the exact consequences of the
Serre relations might be intricate to derive explicitly. This is
one of the difficulties of the theory.


\subsection{The affine case}
\label{section:theAffineCase}

The affine case is characterized by the conditions that the Cartan
matrix has vanishing determinant, is symmetrizable and is such that
its symmetrization $S$ is positive semi-definite (only one zero
eigenvalue). As before, we also take the Cartan matrix to be
indecomposable. By a reasoning analogous to what we did in
Section~\ref{AffineCox} above, one can show that the radical of $S$ is
one-dimensional and that the ranks of $S$ and $A$ are equal to $n-1$.

One defines the corresponding Kac--Moody algebras in terms of $3n
+1$ generators, which are the same generators $h_i,e_i,f_i$
subject to the same conditions~(\ref{KM}, \ref{Serre})
as above, plus one extra generator $\eta$ which can be taken to
fulfill
\begin{equation}
  [\eta,h_i] = 0,
  \qquad
  [\eta, e_i] = \delta_{1i} e_1,
  \qquad
  [\eta, f_i] = -\delta_{1i} f_1.
\end{equation}
The algebra admits the same triangular decomposition as above,
\begin{equation}
  \mf{g} = \mf{n}_- \oplus \mf{h} \oplus \mf{n}_+,
\end{equation}
but now the Cartan subalgebra \index{Cartan subalgebra}
$\mf{h}$ has dimension $n+1$ (it contains the extra generator
$\eta$).

Because the matrix $A_{ij}$ has vanishing determinant, one can
find $a_i$ such that $\sum_i a_i A_{ij} = 0$. The element $c =
\sum_i a_i h_i$ is in the center of the algebra. In fact, the
center of the Kac--Moody algebra is one-dimensional and coincides
with $\mathbb{C} c$~\cite{Kac}. The derived algebra $\mf{g}'=
[\mf{g}, \mf{g}]$ is the subalgebra generated by $h_i, e_i, f_i$
and has codimension one (it does not contain $\eta$). One has
\begin{equation}
  \mf{g} = \mf{g}' \oplus \mathbb{C}\eta
\end{equation}
(direct sum of vector spaces, not as algebras). The
only proper ideals of the affine Kac--Moody algebra $\mf{g}$ are
$\mf{g}'$ and $\mathbb{C} c$.

Affine Kac--Moody algebras appear in the BKL context as subalgebras
of the relevant Lorentzian Kac--Moody algebras. Their complete
list is known and is given in Tables~\ref{table:affineKM1}
and~\ref{table:affineKM2}.

\begin{table}
  \caption{Untwisted affine Kac--Moody algebras.}
  \label{table:affineKM1}
  \renewcommand{\arraystretch}{1.2}
  \vspace{0.5 em}
  \centering
  \begin{tabular}{m{30mm}|m{70mm}}
    \toprule
    Name & Dynkin diagram \\
    \midrule
    $A_1^+$ &          \includegraphics[width=17mm]{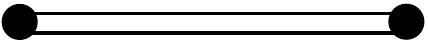} \\ [1 em]
    $A_n^+\, (n>1)$ & \includegraphics[width=57mm]{Anp} \\ [1 em]
    $B_n^+$ &          \includegraphics[width=57mm]{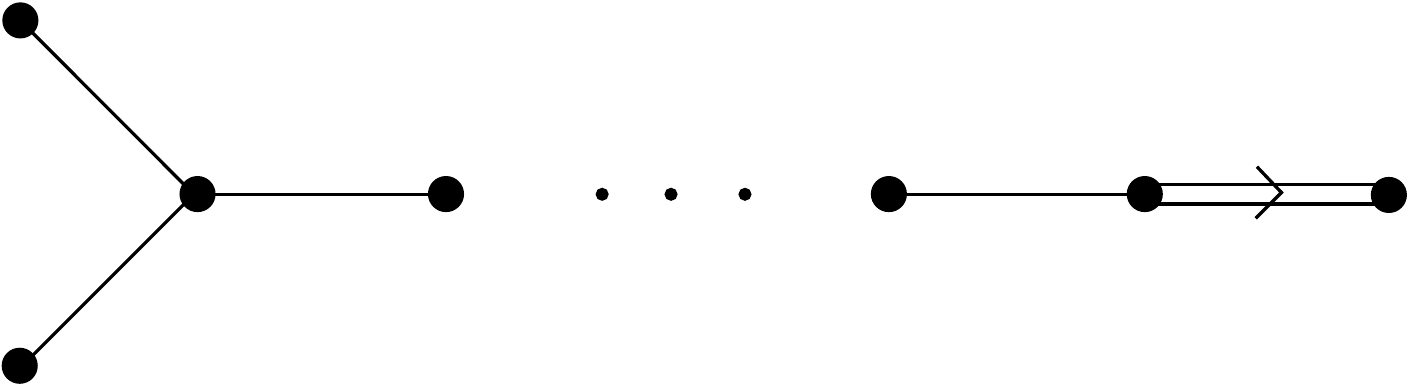} \\ [1 em]
    $C_n^+$ &          \includegraphics[width=63mm]{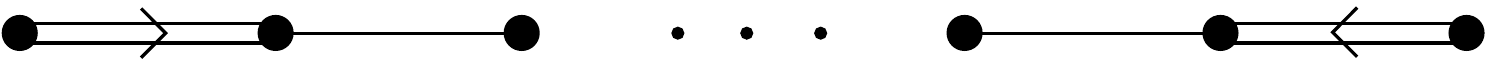} \\ [1 em]
    $D_n^+$ &          \includegraphics[width=55mm]{Dnp} \\ [1 em]
    $G_2^+$ &          \includegraphics[width=22mm]{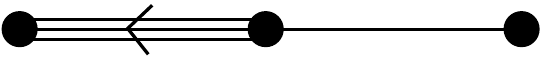} \\ [1 em]
    $F_4^+$ &          \includegraphics[width=43mm]{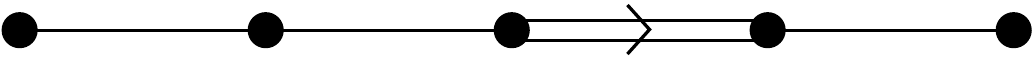} \\ [1 em]
    $E_6^+$ &          \includegraphics[width=40mm]{E6p} \\ [1 em]
    $E_7^+$ &          \includegraphics[width=60mm]{E72p} \\ [1 em]
    $E_8^+$ &          \includegraphics[width=70mm]{E82p} \\
    \bottomrule
  \end{tabular}
  \renewcommand{\arraystretch}{1.0}
\end{table}

\begin{table}
  \caption[Twisted affine Kac--Moody algebras.]{Twisted affine
  Kac--Moody algebras. We use the notation of Kac~\cite{Kac}.}
  \label{table:affineKM2}
  \renewcommand{\arraystretch}{1.2}
  \vspace{0.5 em}
  \centering
  \begin{tabular}{m{30mm}|m{90mm}}
    \toprule
    Name & Dynkin diagram \\
    \midrule
    $A_2^{(2)}$ &                    \includegraphics[width=17mm]{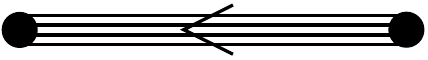} \\ [1 em]
    $A_{2n}^{(2)}\,\, (n\geq 2)$ & \includegraphics[width=90mm]{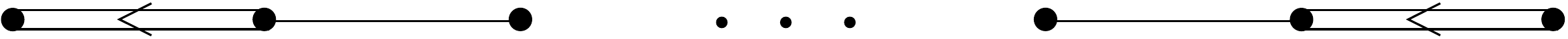} \\ [1 em]
    $A_{2n-1}^{(2)}\, (n\geq 3)$ &  \includegraphics[width=90mm]{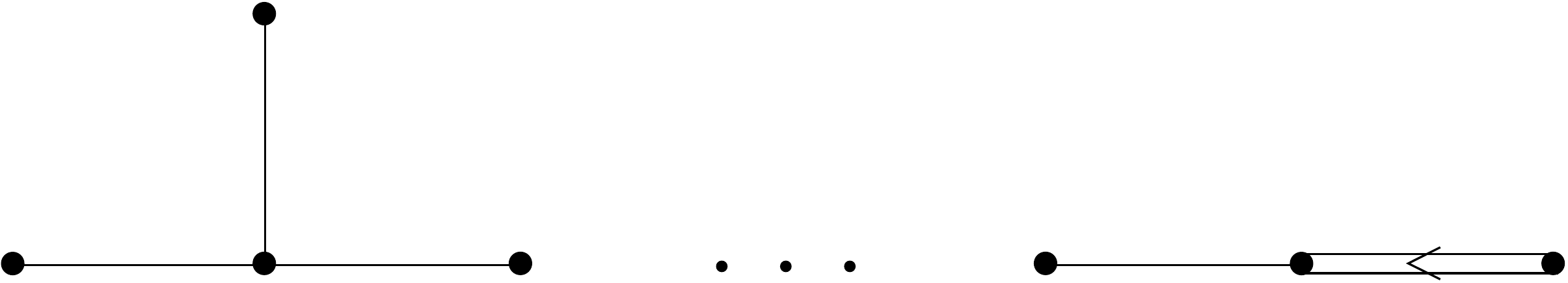} \\ [1 em]
    $D_{n+1}^{(2)}$ &                \includegraphics[width=90mm]{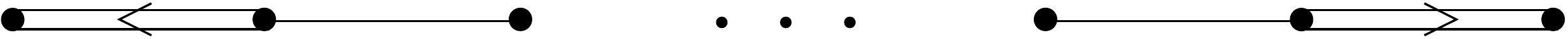} \\ [1 em]
    $E_6^{(2)}$ &                    \includegraphics[width=68mm]{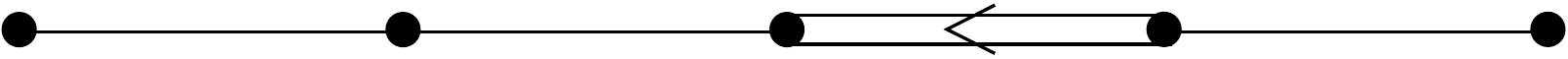} \\ [1 em]
    $D_4^{(3)}$ &                    \includegraphics[width=34mm]{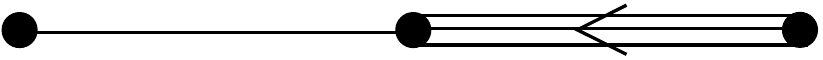} \\
    \bottomrule
  \end{tabular}
  \renewcommand{\arraystretch}{1.0}
\end{table}


\subsection{The invariant bilinear form}


\subsubsection{Definition}
\label{section:bilinearformdefinition}

To proceed, we assume, as announced above, that the Cartan matrix
is invertible and symmetrizable since these are the only cases
encountered in the billiards. Under these assumptions, an
invertible, invariant bilinear form is easily defined on the
algebra. We denote by $\epsilon_i$ the diagonal elements of $D$,
\begin{equation}
  A = DS, \qquad
  D = \diag (\epsilon_1, \epsilon_2 \cdots, \epsilon_n).
\end{equation}
First, one defines an invertible
bilinear form in the dual $\mf{h}^*$ of the Cartan subalgebra.
This is done by simply setting 
\begin{equation}
  (\alpha_i \vert \alpha_j) = S_{ij}
  \label{scalar roots}
\end{equation}
for the simple roots. It follows from $A_{ii} = 2$ that 
\begin{equation}
  \epsilon_i = \frac{2}{(\alpha_i \vert \alpha_i)} 
  \label{32}
\end{equation}
and thus the Cartan matrix can be written as
\begin{equation}
  A_{ij} = 2 \frac{(\alpha_i \vert \alpha_j)}{(\alpha_i \vert \alpha_i)}.
\end{equation}
It is customary to fix the normalization of
$S$ so that the longest roots have $(\alpha_i \vert \alpha_i) =
2$. As we shall now see, the definition (\ref{scalar roots}) leads
to an invariant bilinear form on the Kac--Moody algebra.

Since the bilinear form $( \cdot \vert \cdot)$ is nondegenerate
on $\mf{h}^*$, one has an isomorphism $\mu : \mf{h}^* \rightarrow
\mf{h}$ defined by 
\begin{equation}
  \langle \alpha, \mu(\ga) \rangle = (\alpha \vert \ga).
  \label{34}
\end{equation}
This isomorphism induces a bilinear form on the Cartan subalgebra,
\index{Cartan subalgebra}
also denoted by $( \cdot \vert \cdot)$. The inverse isomorphism is
denoted by $\nu$ and is such that 
\begin{equation}
  \langle \nu(h), h' \rangle = (h \vert h'),
  \qquad
  h, h' \in \mf{h}. 
  \label{35}
\end{equation}
Since the Cartan elements $h_i \equiv \alpha_i^\vee$ obey 
\begin{equation}
  \langle \alpha_i, \alpha_j^\vee \rangle = A_{ji}, 
\end{equation}
it is clear from the definitions that 
\begin{equation}
  \nu(h_i) \equiv \nu (\alpha_i^\vee) =
  \epsilon_i \alpha_i
  \quad \Leftrightarrow \quad
  h_i \equiv \alpha_i^\vee =
  \frac{2 \mu(\alpha_i)}{(\alpha_i \vert \alpha_i)},
  \label{37}
\end{equation}
and thus also 
\begin{equation}
  (h_i \vert h_j) = \epsilon_i \epsilon_j S_{ij}.
\end{equation}

The bilinear form $( \cdot \vert \cdot)$ can be uniquely extended
from the Cartan subalgebra to the entire algebra by requiring that
it is \emph{invariant}, i.e., that it fulfills 
\begin{equation}
  ([x,y] \vert z) = (x \vert [y,z]) \quad \forall\, x,y,z\in \mf{g}.
\end{equation}
For instance, for the $e_i$'s and $f_i$'s one finds 
\begin{equation}
  (h_i \vert e_j) A_{kj} = (h_i \vert [h_k, e_j]) =
  ([h_i, h_k] \vert e_j) = 0 \quad \Rightarrow \quad
  (h_i \vert e_j) = 0,
\end{equation}
and similarly
\begin{equation}
  (h_i \vert f_j) = 0.
\end{equation}
In the same way we have
\begin{equation}
  A_{ij} (e_j \vert f_k) = ([h_i, e_j]\vert f_k) =
  (h_i \vert [e_j, f_k]) = (h_i \vert h_j) \delta_{jk} =
  A_{ij}\epsilon_j \delta_{jk},
\end{equation}
and thus
\begin{equation}
  (e_i \vert f_j) = \epsilon_i \delta_{ij}.
\end{equation}
Quite generally, if $e_\alpha$ and $e_\ga$ are root vectors
corresponding respectively to the roots $\alpha$ and $\ga$,
\begin{displaymath}
  [h, e_\alpha] = \alpha(h) e_\alpha,
  \qquad
  [h, e_\ga] = \ga(h) e_\ga,
\end{displaymath}
then $(e_\alpha \vert e_\ga) = 0$ unless $\ga = -\alpha$. Indeed, one
has
\begin{displaymath}
  \alpha(h) (e_\alpha \vert e_\ga) = ([h, e_\alpha] \vert e_\ga) =
  - (e_\alpha \vert [h, e_\ga]) = -\ga(h) (e_\alpha \vert e_\ga),
\end{displaymath}
and thus
\begin{equation}
  (e_\alpha \vert e_\ga) = 0
  \qquad
  \mbox{if } \alpha + \ga \not= 0.
\end{equation}
It is proven in~\cite{Kac} that the invariance condition on
the bilinear form defines it indeed consistently and that it is
nondegenerate. Furthermore, one finds the relations 
\begin{equation}
  [h,x] = \alpha(h) x,
  \qquad
  [h,y] = - \alpha(h)y \quad \Rightarrow \quad
  [x, y] = (x \vert y) \mu(\alpha).
\end{equation}


\subsubsection{Real and imaginary roots}
\index{root}

Consider the restriction $( \cdot | \cdot )_{\mbb{R}}$ of the bilinear
form to the real vector space $\mf{h}^{\star}_{\mbb{R}}$ obtained by
taking the real span of the simple roots,
\begin{equation}
  \mf{h}^{\star}_{\mbb{R}}=\sum_{i}\mbb{R}\al_i.
\end{equation}
This defines a scalar product with a definite signature. As we have
mentioned, the signature is Euclidean if and only if the algebra
is finite-dimensional~\cite{Kac}. In that case, all roots -- and
not just the simple ones -- are spacelike, i.e., such that
$(\alpha \vert \alpha) >0$.

When the algebra is infinite-dimensional, the invariant scalar
product does not have Euclidean signature. The spacelike roots
are called ``real roots'', the non-spacelike ones are called
``imaginary roots''~\cite{Kac}. While the real roots are
nondegenerate (i.e., the corresponding eigenspaces, called ``root
spaces'', are one-dimensional), this is not so for imaginary roots.
In fact, it is a challenge to understand the degeneracy of
imaginary roots for general indefinite Kac--Moody algebras, and, in
particular, for Lorentzian Kac--Moody algebras.

Another characteristic feature of real roots, familiar from
standard finite-dimensional Lie algebra theory, is that if
$\alpha$ is a (real) root, no multiple of $\alpha$ is a root
except $\pm \alpha$. This is not so for imaginary roots, where $2
\alpha$ (or other non-trivial multiples of $\alpha$) can be a root
even if $\alpha$ is. We shall provide explicit examples below.

Finally, while there are at most two different root lengths in
the finite-dimensional case, this is no longer true even for real
roots in the case of infinite-dimensional Kac--Moody
algebras\epubtkFootnote{Imaginary roots may have arbitrarily negative
length squared in general.}. When all the real roots have the same
length, one says that the algebra is ``simply-laced''. Note that
the imaginary roots (if any) do not have the same length, except in
the affine case where they all have length squared equal to zero.


\subsubsection{Fundamental weights and the Weyl vector}
\label{section:fundamentalweights}

The fundamental weights $\{\Lambda_i\}$ of the Kac--Moody algebra
are vectors in the dual space $\mf{h}^*$ of the Cartan
subalgebra defined by 
\begin{equation}
  \langle \Lambda_i, \alpha_j^\vee \rangle = \delta_{ij}.
\end{equation}
This implies 
\begin{equation}
  (\Lambda_i \vert \alpha_j) = \frac{\delta_{ij}}{\epsilon_j}.
\end{equation}

The Weyl vector $\rho \in \mf{h}^*$ is defined by 
\begin{equation}
  (\rho \vert \alpha_j) = \frac{1}{\epsilon_j}
\end{equation}
and is thus equal to
\begin{equation}
  \rho = \sum_i \Lambda_i.
\end{equation}


\subsubsection{The generalized Casimir operator}

From the invariant bilinear form, one can construct a generalized Casimir
operator as follows.

We denote the eigenspace associated with $\alpha$ by
$\frak{g}_\alpha$. This is called the ``root space'' of $\al$ and is defined as
\begin{equation}
  \mf{g}_{\al}=\{x\in \mf{g}\, |\, [x,h]=\al(h)x,
  \quad \forall \, h\in \mf{h}\}.
\end{equation}
A representation of the Kac--Moody algebra is
called restricted if for every vector $v$ of the representation
subspace $V$, one has $\frak{g}_\alpha\cdot v = 0$ for all but a
finite number of positive roots $\alpha$.

Let $\{e_\alpha^K\}$ be a basis of $\frak{g}_\alpha$ and let
$\{e_{- \alpha}^K\}$ be the basis of $\frak{g}_{-\alpha}$ dual to
$\{e_\alpha^K\}$ in the $B$-metric, 
\begin{equation}
  (e_{\alpha}^K \vert e_{-\alpha}^L) = \delta^{KL} .
\end{equation}
Similarly, let $\{u_i \}$ be a basis of $\mf{h}$ and $\{ u^i \}$ the
dual basis of $\mf{h}$ with respect to the bilinear form $(\cdot \vert \cdot)$,
\begin{equation}
  (u_i \vert u^j) = \delta_i^j.
\end{equation}
We set 
\begin{equation}
  \Omega = 2 \mu(\rho) + \sum_i u^i u_i +
  2 \!\! \sum_{\alpha \in \Delta_+} \sum_K e^K_{- \alpha} e^K_\alpha,
  \label{Casimir}
\end{equation}
where $\rho$ is the Weyl vector. Recall from
Section~\ref{section:bilinearformdefinition} that $\mu$ is an
isomorphism from $\mathfrak{h}^{\star}$ to $\mathfrak{h}$, so, since
$\rho\in\mathfrak{h}^{\star}$, the expression $\mu(\rho)$ belongs to
$\mathfrak{h}$ as required. When acting on any vector of a restricted
representation, $\Omega$ is well-defined since only a finite number of
terms are different from zero.

It is proven in~\cite{Kac} that $\Omega$ commutes with all the
operators of any restricted representation. For that reason, it
is known as the (generalized) Casimir operator. It is quadratic
in the generators\epubtkFootnote{The generalized Casimir operator $\Omega$
is the only known polynomial element of the center $Z$ of the universal
enveloping algebra $U(\mf{g})$ of an indefinite Kac--Moody algebra $\mf{g}$.
However, Kac~\cite{Kac1984} has proven the existence of higher order
non-polynomial Casimir operators which are elements of the center $Z_{\mf{F}}$
of a suitable completion $U_{\mf{F}}(\mf{g})$ of the universal enveloping
algebra of $\mf{g}$. Recently, an explicit physics-inspired construction
was made, following~\cite{Kac1984}, for affine $\mf{g}$ in
terms of Wilson loops for WZW-models~\cite{Alekseev:2007in}.}.

\subsubsection*{Note}

This definition -- and, in particular, the
presence of the linear term $\mu(\rho)$ -- might seem a bit
strange at first sight. To appreciate it, turn to a
finite-dimensional simple Lie algebra. In the above notations, the
usual expression for the quadratic Casimir operator reads
\begin{equation}
  \Omega_{\mbox{finite}} = \sum_A \kappa^{AB} T_A T_B =
  \sum_i u^i u_i + \!\! \sum_{\alpha \in \Delta_+}
  (e_{- \alpha} e_\alpha + e_{\alpha} e_{-\alpha})
\end{equation}
(without degeneracy index $K$ since the roots are nondegenerate in
the finite-dimensional case). Here, $\kappa^{AB}$ is the Killing
metric and $\{T_A\}$ a basis of the Lie algebra. This expression is
not ``normal-ordered'' because there are, in the last term, lowering
operators standing on the right. We thus replace the last term by
\begin{eqnarray}
  \sum_{\alpha \in \Delta_+} e_{ \alpha} e_{-\alpha} &=&
  \sum_{\alpha \in \Delta_+} e_{- \alpha} e_\alpha +
  \sum_{\alpha \in \Delta_+} [e_{ \alpha}, e_{-\alpha}]
  \nonumber
  \\
  &=& \sum_{\alpha \in \Delta_+} e_{- \alpha} e_\alpha +
  \sum_{\alpha \in \Delta_+} \mu(\alpha).
\end{eqnarray}
Using the fact that in a finite-dimensional Lie algebra, $\rho = (1/2)
\sum_{\alpha \in \Delta_+} \alpha$, (see, e.g.,~\cite{Fuchs:1997jv})
one sees that the Casimir operator can be rewritten in ``normal
ordered'' form as in Equation~(\ref{Casimir}). The advantage of the
normal-ordered form is that it makes sense also for
infinite-dimensional Kac--Moody algebras in the case of restricted
representations.


\subsection{The Weyl group}
\label{section:weylgroup}

The Weyl group $\mf{W}[\mf{g}]$ of a Kac--Moody algebra $\mf{g}$ is
a discrete group of transformations acting on $\mf{h}^*$. It is
defined as follows. One associates a ``fundamental Weyl
reflection'' $r_i\in\mf{W}[\mf{g}]$ to each simple root through the
formula 
\begin{equation}
  r_i(\lambda) = \lambda -
  2 \frac{(\lambda \vert \alpha_i)}{(\alpha_i \vert \alpha_i)} \alpha_i.
  \label{fundamentalWeyl}
\end{equation}
The Weyl group is just the group
generated by the fundamental Weyl reflections. In particular, 
\begin{equation}
  r_i(\alpha_j) = \alpha_j - A_{ij} \alpha_i
  \qquad
  (\mbox{no summation on }i). 
  \label{fundamentalWeylbis}
\end{equation}

The Weyl group enjoys a number of interesting properties~\cite{Kac}:

\begin{itemize}
\item It preserves the scalar product on $\mf{h}^*$.
\item It preserves the root lattice and hence is crystallographic.
\item Two roots that are in the same orbit have identical
  multiplicities.
\item Any real root has in its orbit (at least) one simple root and
  hence, is nondegenerate.

\item The Weyl group is a Coxeter group. \index{Coxeter group} The
  connection between the Coxeter exponents and the Cartan integers
  $A_{ij}$ is given in Table~\ref{table:CatanCoxeter} ($i \not=j$).
\end{itemize}

\begin{table}
  \caption{Cartan integers and Coxeter exponents.}
  \renewcommand{\arraystretch}{1.2}
  \vspace{0.5 em}
  \centering
  \begin{tabular}{l|c}
    \toprule
    $ A_{ij} A_{ji} $ & $ m_{ij} $ \\
    \midrule
     0       &  2  \\
     1       &  3  \\
     2       &  4  \\
     3       &  6  \\
    $\geq$ 4 & $\infty$ \\
    \bottomrule
  \end{tabular}
  \label{table:CatanCoxeter}
  \renewcommand{\arraystretch}{1.0}
\end{table}

\noindent
This close relationship between Coxeter groups and Kac--Moody algebras
is the reason for denoting both with the same notation (for instance,
$A_n$ denotes at the same time the Coxeter group with Coxeter graph 
\index{Coxeter graph} of type $A_n$ and the Kac--Moody algebra with
Dynkin diagram $A_n$).

Note that different Kac--Moody algebras may have the same Weyl
group. This is in fact already true for finite-dimensional Lie
algebras, where dual algebras (obtained by reversing the arrows in the
Dynkin diagram) have the same Weyl group. This property can be seen
from the fact that the Coxeter exponents are related to the
duality-invariant product $A_{ij}A_{ji}$. But, on top of this, one
sees that whenever the product $A_{ij} A_{ji} $ exceeds four, which
occurs only in the infinite-dimensional case, the Coxeter exponent
$m_{ij}$ is equal to infinity, independently of the exact value of
$A_{ij} A_{ji} $. Information is thus clearly lost. For example, the
Cartan matrices 
\begin{equation}
  \left(
    \begin{array}{@{}r@{\quad}r@{\quad}r@{}}
      2 & -2 & - 2 \\
      -2 & 2 & - 2 \\
      -2 & -2 & 2 
    \end{array}
  \right),
  \qquad
  \left(
    \begin{array}{@{}r@{\quad}r@{\quad}r@{}}
      2 & -9 & -8 \\
      -4 & 2 & -5 \\
      -3 & -7 & 2
    \end{array}
  \right)
\end{equation}
lead to the same Weyl group, even though the corresponding Kac--Moody
algebras are not isomorphic or even dual to each other.

Because the Weyl groups are (crystallographic) Coxeter groups, we can
use the theory of Coxeter groups to analyze them. In the Kac--Moody
context, the fundamental region is called ``the fundamental Weyl
chamber''.

We also note that by (standard vector space) duality, one can define
the action of the Weyl group in the Cartan subalgebra 
\index{Cartan subalgebra} $\mf{h}$, such that 
\begin{equation}
  \langle \ga, r_i^\vee (h) \rangle =
  \langle r_i(\ga), h \rangle
  \qquad
  \mbox{for } \ga\in\mf{h}^{\star} \mbox{ and } h\in\mf{h}.
\end{equation}
One has using Equations~(\ref{32}, \ref{34}, \ref{35}, \ref{37}),
\begin{equation}
  r_i^\vee (h) = h - \langle\alpha_i , h \rangle h_i = h - 2
  \frac{(h \vert h_i)}{(h_i \vert h_i)} h_i.
  \label{4.61xx}
\end{equation}

Finally, we leave it to the reader to verify that when the
products $A_{ij} A_{ji} $ are all $\leq 4$, then the geometric
action of the Coxeter group considered in
Section~\ref{section:geometric} and the geometric action of the Weyl
group considered here coincide. The (real) roots and the fundamental
weights differ only in the normalization and, once this is taken into
account, the metrics coincide. This is not the case when some products
$A_{ij} A_{ji} $ exceed 4. It should be also pointed out that the
imaginary roots of the Kac--Moody algebras do not have immediate
analogs on the Coxeter side.

\subsubsection*{Examples}

\begin{itemize}
\item Consider the Cartan matrices
  \begin{displaymath}
    A^{\prime} = \left(
      \begin{array}{@{}r@{\quad}r@{\quad}r@{}}
        2 & -2 & 0 \\
        -2 & 2 & -1 \\
        0 & -1 & 2 
      \end{array}
    \right),
    \qquad
    A^{\prime\prime} = \left(
      \begin{array}{@{}r@{\quad}r@{\quad}r@{}}
        2 & -4 & 0 \\
        -1 & 2 & -1 \\
        0 & -1 & 2 
      \end{array}
    \right)
  \end{displaymath}
  As the first (respectively, second) Cartan matrix defines the Lie
  algebra $A_1^{++}$ \index{$A_1^{++}$|bb} (respectively $A_2^{(2)+}$)
  introduced below in Section~\ref{section:Overextensions}, we also
  write it as $A^{\prime}\equiv A[A_1^{++}]$ (respectively,
  $A^{\prime\prime}\equiv A[A_2^{(2)+}]$). We denote the associated
  sets of simple roots by
  $\{\al_1^{\prime},\al_2^{\prime},\al_3^{\prime}\}$ and
  $\{\al_1^{\pp},\al_2^{\pp},\al_3^{\pp}\}$, respectively. In both
  cases, the Coxeter exponents are $m_{12} = \infty$, $m_{13} = 2$,
  $m_{23} = 3$ and the metric $B_{ij}$ of the geometric Coxeter
  construction is
    \begin{displaymath}
    A^{\prime} = \left(
      \begin{array}{@{}r@{\quad}r@{\quad}r@{}}
        1 & -1 & 0 \\
        -1 & 1 & -\frac{1}{2} \\
        0 & -\frac{1}{2} & 1 
      \end{array}
    \right).
  \end{displaymath}
  We associate the simple roots $\{\al_1,\al_2,\al_3\}$ with the
  geometric realisation of the Coxeter group $\mf{B}$ defined by the
  matrix $B$. These roots may a priori differ by normalizations from
  the simple roots of the Kac--Moody algebras described by the Cartan
  matrices $A^{\p}$ and $A^{\pp}$.

  Choosing the longest Kac--Moody roots to have squared length equal
  to two yields the scalar products
  \begin{displaymath}
    S^{\prime} = \left(
      \begin{array}{@{}r@{\quad}r@{\quad}r@{}}
        2 & -2 & 0 \\
        -2 & 2 & -1 \\
        0 & -1 & 2 
      \end{array}
    \right),
    \qquad
    S^{\pp} = \left(
      \begin{array}{@{}r@{\quad}r@{\quad}r@{}}
        \frac{1}{2} & -1 & 0 \\
        -1 & 2 & -1 \\
        0 & -1 & 2 
      \end{array}
    \right).
  \end{displaymath}
  Recall now from Section~\ref{section:Coxeter} that the fundamental
  reflections $\si_i\in\mf{B}$ have the following geometric
  realisation
  \begin{equation}
    \si_i(\al_j)=\al_j-2B_{ij}\al_i
    \qquad
    (i=1,2,3),
  \end{equation}
  which in this case becomes 
  \begin{displaymath}
    \begin{array}{l@{\quad}l@{\qquad}l@{\qquad}l}
      \si_1: &
      \alpha_1 \rightarrow - \alpha_1, &
      \alpha_2 \rightarrow \alpha_2 + 2 \alpha_1, &
      \alpha_3 \rightarrow \alpha_3,
      \\ [0.2 em]
      \si_2: &
      \alpha_1 \rightarrow \alpha_1 + 2 \alpha_2, &
      \alpha_2 \rightarrow - \alpha_2, &
      \alpha_3 \rightarrow \alpha_3 + \alpha_2,
      \\ [0.2 em]
      \si_3: &
      \alpha_1 \rightarrow \alpha_1, &
      \alpha_2 \rightarrow \alpha_2 + \alpha_3, &
      \alpha_3 \rightarrow - \alpha_3.
    \end{array}
  \end{displaymath}
  We now want to compare this geometric realisation of $\mf{B}$ with
  the action of the Weyl groups of $A^{\p}$ and $A^{\pp}$ on the
  corresponding simple roots $\al_i^{\p}$ and $\al_i^{\pp}$. According
  to Equation~(\ref{fundamentalWeylbis}), the Weyl group
  $\mf{W}[A_1^{++}]$ acts as follows on the roots $\alpha'_i$
  \begin{displaymath}
    \begin{array}{l@{\quad}l@{\qquad}l@{\qquad}l}
      r'_1: &
      \alpha'_1 \rightarrow - \alpha'_1, &
      \alpha'_2 \rightarrow \alpha'_2 + 2 \alpha'_1, &
      \alpha'_3 \rightarrow \alpha'_3,
      \\ [0.2 em]
      r'_2: &
      \alpha'_1 \rightarrow \alpha'_1 + 2 \alpha'_2, &
      \alpha'_2 \rightarrow -\alpha'_2, &
      \alpha'_3 \rightarrow \alpha'_3 + \alpha'_2,
      \\ [0.2 em]
      r'_3: &
      \alpha'_1 \rightarrow \alpha'_1, &
      \alpha'_2 \rightarrow \alpha'_2 + \alpha'_3, &
      \alpha'_3 \rightarrow - \alpha'_3,
    \end{array}
  \end{displaymath}
  while the Weyl group $\mf{W}[A_2^{(2)+}]$ acts as 
  \begin{displaymath}
    \begin{array}{l@{\quad}l@{\qquad}l@{\qquad}l}
      r''_1: &
      \alpha''_1 \rightarrow - \alpha''_1, &
      \alpha''_2 \rightarrow \alpha''_2 + 4 \alpha''_1, &
      \alpha''_3 \rightarrow \alpha''_3,
      \\ [0.2 em]
      r''_2: &
      \alpha''_1 \rightarrow \alpha''_1 + \alpha''_2, &
      \alpha''_2 \rightarrow -\alpha''_2, &
      \alpha''_3 \rightarrow \alpha''_3 + \alpha''_2,
      \\ [0.2 em]
      r''_3: &
      \alpha''_1 \rightarrow \alpha''_1, &
      \alpha''_2 \rightarrow \alpha''_2 + \alpha''_3, &
      \alpha''_3 \rightarrow - \alpha''_3.
    \end{array}
  \end{displaymath}
  We see that the reflections coincide, $\si_1 = r'_1 = r''_1$, $\si_2 =
  r'_2 = r''_2$, $\si_3 = r'_3 = r''_3$, as well as the scalar products,
  provided that we set $2 \alpha''_1 = \alpha'_1$, $\alpha''_2 =
  \alpha'_2$, $\alpha'_3 = \alpha_3$ and $\alpha'_i = \sqrt{2} \alpha_i
  $. The Coxeter group $\mf{B}$ generated by the reflections thus
  preserves the lattices
  \begin{equation}
    Q^{\p}=\sum_{i}\mathbb{Z} \alpha'_i
    \qquad \mbox{and} \qquad
    Q^{\pp}=\sum_{i}\mathbb{Z} \alpha''_i,
  \end{equation}
  showing explicitly that, in the present case, the lattices preserved
  by a Coxeter group are not unique -- and might not even be dual to
  each other.

  It follows, of course, that the Weyl groups of the Kac--Moody algebras
  $A_1^{++}$ and $A_1^{(2)+}$ are the same,
  \begin{equation}
    \mf{W}[A_1^{++}]=\mf{W}[A_{2}^{(2)+}]=\mf{B}.
  \end{equation}
\item Consider now the Cartan matrix
  \begin{displaymath}
    A^{\ppp} = \left(
      \begin{array}{@{}r@{\quad}r@{\quad}r@{}}
        2 & -6 & 0 \\
        -1 & 2 & -1 \\
        0 & -1 & 2
      \end{array}
    \right),
  \end{displaymath}
  and its symmetrization
  \begin{displaymath}
    S^{\ppp} = \left(
      \begin{array}{@{}r@{\quad}r@{\quad}r@{}}
        \frac{1}{3} & -1 & 0 \\
        -1 & 2 & -1 \\
        0 & -1 & 2 
      \end{array}
    \right),
  \end{displaymath}
  The Weyl group $\mf{W}[A^{\ppp}]$ of the corresponding Kac--Moody
  algebra is isomorphic to the Coxeter group $\mf{B}$ above since,
  according to the rules, the Coxeter exponents are identical. But the
  action is now
  \begin{displaymath}
    \begin{array}{l@{\quad}l@{\qquad}l@{\qquad}l}
      r'''_1: &
      \alpha'''_1 \rightarrow - \alpha'''_1, &
      \alpha'''_2 \rightarrow \alpha'''_2 + 6 \alpha'''_1, &
      \alpha'''_3 \rightarrow \alpha'''_3
      \\ [0.2 em]
      r'''_2: &
      \alpha'''_1 \rightarrow \alpha'''_1 + \alpha'''_2, &
      \alpha'''_2 \rightarrow -\alpha'''_2 &
      \alpha'''_3 \rightarrow \alpha'''_3 + \alpha'''_2
      \\ [0.2 em]
      r'''_3: &
      \alpha'''_1 \rightarrow \alpha'''_1, &
      \alpha'''_2 \rightarrow \alpha'''_2 + \alpha'''_3, &
      \alpha'''_3 \rightarrow - \alpha'''_3
    \end{array}
  \end{displaymath}
  and cannot be made to coincide with the previous action by rescalings
  of the $\alpha'''_i$'s. One can easily convince oneself of the
  inequivalence by computing the eigenvalues of the matrices $S'$, $S''$
  and $S'''$ with respect to $B$.
\end{itemize}


\subsection{Hyperbolic Kac--Moody algebras}
\label{HyperbolicKacMoodyAlgebras}

Hyperbolic Kac--Moody algebras are by definition Lorentzian
Kac--Moody algebras with the property that removing any node from
their Dynkin diagram \index{Dynkin diagram} leaves one with a Dynkin
diagram of affine or finite type. The Weyl group of hyperbolic
Kac--Moody algebras is a crystallographic hyperbolic Coxeter group (as
defined in Section~\ref{section:Hyperbolic}). Conversely, any
crystallographic hyperbolic Coxeter group is the Weyl group of at
least one hyperbolic Kac--Moody algebra. Indeed, consider one of
the lattices preserved by the Coxeter group as constructed in
Section~\ref{section:CrystalCoxeterGroups}. The
matrix with entries equal to the $d_{ij}$ of that
section is the Cartan matrix of a Kac--Moody algebra that has
this given Coxeter group as Weyl group.

The hyperbolic Kac--Moody algebras have been classified
in~\cite{Saclioglu89} and exist only up to rank 10 (see
also~\cite{deBuyl:2004md}). In rank 10, there are four possibilities,
known as $E_{10} \equiv E_8^{++}$, $BE_{10} \equiv B_8^{++}$, $DE_{10}
\equiv D_8^{++}$ and $CE_{10} \equiv A_{15}^{(2)+}$, $BE_{10}$ and
$CE_{10}$ being dual to each other and possessing the same Weyl group
(the notation will be explained below).


\subsubsection[The fundamental domain $\mathcal{F}$]%
              {The fundamental domain \boldmath $\mathcal{F}$}

For a hyperbolic Kac--Moody algebra, the fundamental weights
$\Lambda_i$ are timelike or null and lie within the (say) past
lightcone. Similarly, the fundamental Weyl chamber \index{Weyl chamber|bb} $\mathcal{F}$
defined by $\{ v \in \mathcal{F} \Leftrightarrow (v \vert
\alpha_i) \geq 0 \}$ also lies within the past lightcone and is a
fundamental region for the action of the Weyl group on the Tits
cone, which coincides in fact with the past light cone. All these
properties carries over from our discussion of hyperbolic
Coxeter groups in Section~\ref{section:Coxeter}.

The positive imaginary roots $\alpha_K$ of the algebra fulfill
$(\alpha_K \vert \Lambda_i) \geq 0$ (with, for any $K$, strict
inequality for at least one $i$) and hence, since they are non-spacelike,
must lie in the \emph{future} light cone. Recall indeed
that the scalar product of two non-spacelike vectors with the same
time orientation is non-positive. For this reason, it is also of
interest to consider the action of the Weyl group on the future
lightcone, obtained from the action on the past lightcone by
mere changes of signs. A fundamental region is clearly given by
$- \mathcal{F}$. Any imaginary root is Weyl-conjugated to one
that lies in $- \mathcal{F}$.


\subsubsection{Roots and the root lattice}
\index{root}
\index{root lattice}

We have mentioned that not all points on the root lattice $Q$ of a
Kac--Moody algebras are actually roots. For hyperbolic algebras,
there exists a simple criterion which enables one to determine
whether a point on the root lattice is a root or not. We give it first
in the case where all simple roots have equal length squared
(assumed equal to two).

\begin{theorem}
  Consider a hyperbolic Kac--Moody algebra such that $(\alpha_i \vert
  \alpha_i) = 2$ for all simple roots $\alpha_i$. Then, any point
  $\alpha$ on the root lattice $Q$ with $(\alpha \vert \alpha) \leq 2$
  is a root (note that $(\alpha \vert \alpha)$ is even). In
  particular, the set of real roots is the set of points on the root
  lattice with $(\alpha \vert \alpha) = 2$, while the set of imaginary
  roots is the set of points on the root lattice (minus the origin)
  with $(\alpha \vert \alpha) \leq 0$.

  \noindent
  For a proof, see~\cite{Kac}, Chapter~5.
\end{theorem}

\noindent
The version of this theorem applicable to Kac--Moody algebras with
different simple root lengths is the following.

\begin{theorem}
  Consider a hyperbolic algebra with root lattice $Q$. Let $a$ be the
  smallest length squared of the simple roots, $a = \min_{i} (\alpha_i
  \vert \alpha_i)$. Then we have:
  
  \begin{itemize}
  \item The set of all short real roots is $\{\alpha \in Q \, |\,
    (\alpha \vert \alpha) = a \}$.
  \item The set of all real roots is
    \begin{displaymath}
      \left\{ \alpha = \sum_i k_i \alpha_i
      \in Q \, |\,  (\alpha \vert \alpha) > 0 \mbox{ and } k_i
      \frac{(\alpha_i \vert \alpha_i)}{(\alpha \vert \alpha)} \in
      \mathbb{Z} \; \forall i \right\}.
    \end{displaymath} 
  \item The set of all imaginary roots is the set of points on the root
    lattice (minus the origin) with $(\alpha \vert \alpha) \leq 0$.
  \end{itemize}

  \noindent
  For a proof, we refer again to~\cite{Kac}, Chapter~5.
\end{theorem}

\noindent
We shall illustrate these theorems in the examples below. Note
that it follows in particular from the theorems that if $\alpha$
is an imaginary root, all its integer multiples are also imaginary
roots.


\subsubsection{Examples}

We discuss here briefly two examples, namely $A_1^{++}$, \index{$A_1^{++}$} for which all
simple roots have equal length, and $A_2^{(2)+}$, with respective
Dynkin diagrams shown in Figures~\ref{figure:A1pp(nonumbers)}
and~\ref{figure:A2(2)p(nonumbers)}.

\epubtkImage{A1pp_nonumbers.png}{%
  \begin{figure}[htbp]
    \centerline{\includegraphics[width=40mm]{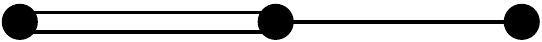}}
    \caption{The Dynkin diagram of the hyperbolic Kac--Moody
      algebra $A_1^{++}$. This algebra is obtained through a
      standard overextension of the finite Lie algebra $A_1$.}
    \label{figure:A1pp(nonumbers)}
  \end{figure}}

\epubtkImage{A2_2p_nonumbers.png}{%
  \begin{figure}[htbp]
    \centerline{\includegraphics[width=40mm]{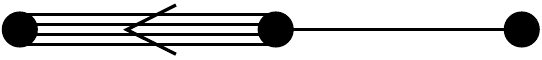}}
    \caption{The Dynkin diagram of the hyperbolic Kac--Moody
      algebra $A_2^{(2)+}$. This algebra is obtained through a
      Lorentzian extension of the twisted affine Kac--Moody algebra
      $A_{2}^{(2)}$.}
    \label{figure:A2(2)p(nonumbers)}
  \end{figure}}

\subsubsection*{The Kac--Moody Algebra \boldmath $A_1^{++}$}
\index{$A_1^{++}$}

This is the algebra associated with vacuum four-dimensional Einstein gravity
and the BKL billiard. We encountered its Weyl group
$PGL(2,\mathbb{Z})$ already in Section~\ref{section:BKLbilliard}.
The algebra is also denoted $AE_3$, or $H_3$. The Cartan matrix
\index{Cartan matrix} is
\begin{equation}
  \left(
    \begin{array}{@{}r@{\quad}r@{\quad}r@{}}
      2 & -2 & 0 \\
      -2 & 2 & -1 \\
      0 & -1 & 2 
    \end{array}
  \right).
\end{equation}
As it follows from our analysis in Section~\ref{section:BKLbilliard},
the simple roots may be identified with the following linear forms
$\alpha_i(\beta)$ in the three-dimensional space of the
$\beta^{i}$'s, 
\begin{equation}
  \alpha_1(\beta) = 2 \beta^1,
  \qquad
  \alpha_2(\beta) = \beta^2 - \beta^1,
  \qquad
  \alpha_3(\beta) = \beta^3 - \beta^2
\end{equation}
with scalar product 
\begin{equation}
 (F \vert G) = \sum_i F_i G_i - \frac{1}{2}
 \left(\sum_i F_i\right) \left(\sum_i G_i \right) 
 \label{scalarproductA1++}
\end{equation}
for two linear forms $F = F_i \beta^i$ and $G = G_i \beta^i$. It is
sometimes convenient to analyze the root system in terms of an
``affine'' level $\ell$ that counts the number of times the root
$\alpha_3$ occurs: The root $k \alpha_1 + m \alpha_2 + \ell \alpha_3$
has by definition level $\ell$ \epubtkFootnote{We discuss in detail a
  different kind of level decomposition in
  Section~\ref{section:LevelDecomposition}.}. We shall consider here
only positive roots for which $k, m , \ell \geq 0$.

Applying the first theorem, one easily verifies that the only
positive roots at level zero are the roots $k \alpha_1 + m
\alpha_2$, $\vert k - m \vert \leq 1$ ($k, m \geq 0$) of the
affine subalgebra $A_1^+$. When $k=m$, the root is imaginary and
has length squared equal to zero. When $\vert k - m \vert = 1$,
the root is real and has length squared equal to two.

Similarly, the only roots at level one are $(m+a) \alpha_1 + m
\alpha_2 + \alpha_3$ with $a^2 \leq m$, i.e., $-[\sqrt{m}] \leq a
\leq [\sqrt{m}]$. Whenever $\sqrt{m}$ is an integer, the roots $(m
\pm \sqrt{m}) \alpha_1 + m \alpha_2 + \alpha_3 $ have squared
length equal to two and are real. The roots $(m+a) \alpha_1 + m
\alpha_2 + \alpha_3$ with $a^2 < m$ are imaginary and have squared
length equal to $2(a^2 +1 -m) \leq 0$. In particular, the root
$m(\alpha_1 + \alpha_2) + \alpha_3$ has length squared equal to
$2(1-m)$. Of all the roots at level one with $m>1$, these are the
only ones that are in the fundamental domain $- {\cal F}$ (i.e.,
that fulfill $(\beta \vert \alpha_i) \leq 0$). When $m = 1$, none
of the level-1 roots is in $- {\cal F}$ and is either in the Weyl
orbit of $\alpha_1 + \alpha_2$, or in the Weyl orbit of
$\alpha_3$.

We leave it to the reader to verify that the roots at level two
that are in the fundamental domain $- {\cal F}$ take the form
$(m-1) \alpha_1 + m \alpha_2 + 2 \alpha_3$ and $m (\alpha_1 +
\alpha_2) + 2 \alpha_3$ with $m \geq 4$. Further information on
the roots of $A_1^{++}$ may be found in~\cite{Kac}, Chapter~11,
page~215.

\subsubsection*{The Kac--Moody Algebra \boldmath $A_2^{(2)+}$}

This is the algebra associated with the Einstein--Maxwell theory (see
Section~\ref{section:KMBilliardsII}). The notation will be
explained in Section~\ref{section:Overextensions}.
The Cartan matrix \index{Cartan matrix} is 
\begin{equation}
  \left(
    \begin{array}{@{}r@{\quad}r@{\quad}r@{}}
      2 & -4& 0 \\
      -1 & 2 & -1 \\
      0 & -1 & 2 
    \end{array}
  \right),
\end{equation}
and there are now two lengths for the
simple roots. The scalar products are 
\begin{equation}
  (\alpha_1 \vert \alpha_1) = \frac{1}{2},
  \qquad
  (\alpha_1 \vert \alpha_2) = - 1 = (\alpha_2 \vert \alpha_1),
  \qquad
  (\alpha_2 \vert \alpha_2) = 2.
\end{equation}
One may realize the simple roots as the linear forms 
\begin{equation}
  \alpha_1(\beta) = \beta^1,
  \qquad
  \alpha_2(\beta) = \beta^2 - \beta^1,
  \qquad
  \alpha_3(\beta) = \beta^3 - \beta^2
\end{equation}
in the three-dimensional space of the $\beta^i$'s with scalar product
Equation~(\ref{scalarproductA1++}).

The real roots, which are Weyl conjugate to one of the simple
roots $\alpha_1$ or $\alpha_2$ ($\alpha_3$ is in the same Weyl
orbit as $\alpha_2$), divide into long and short real roots. The
long real roots are the vectors on the root lattice with squared
length equal to two that fulfill the extra condition in the
theorem. This condition expresses here that the coefficient of
$\alpha_1$ should be a multiple of $4$. The short real roots are
the vectors on the root lattice with length squared equal to
one-half. The imaginary roots are all the vectors on the root
lattice with length squared $\leq 0$.

We define again the level $\ell$ as counting the number of times
the root $\alpha_3$ occurs. The positive roots at level zero are
the positive roots of the twisted affine algebra $A_2^{(2)}$,
namely, $\alpha_1$ and $(2m + a) \alpha_1 + m \alpha_2$, $m = 1,
2, 3, \cdots$, with $a = -2, -1, 0, 1, 2$ for $m$ odd and $a= -1,
0, 1$ for $m$ odd. Although belonging to the root lattice and of
length squared equal to two, the vectors $(2m \pm 2) \alpha_1 + m
\alpha_2$ are not long real roots when $m$ is even because they
fail to satisfy the condition that the coefficient $(2m \pm 2) $
of $\alpha_1$ is a multiple of $4$. The roots at level zero are
all real, except when $a=0$, in which case the roots $m(2 \alpha_1
+ \alpha_2)$ have zero norm.

To get the long real roots at level one, we first determine the
vectors $\alpha = \alpha_3 + k \alpha_1 + m \alpha_2$ of squared
length equal to two. The condition $(\alpha \vert \alpha) = 2$
easily leads to $m = p^2$ for some integer $p \geq 0$ and $k = 2
p^2 \pm 2 p= 2p (p\pm 1)$. Since $k$ is automatically a multiple
of $4$ for all $p = 0, 1, 2 , 3 , \cdots $, the corresponding
vectors are all long real roots. Similarly, the short real roots
at level one are found to be $(2 p^2 + 1) \alpha_1 + (p^2 + p +
1) \alpha_2 + \alpha_3$ and $(2 p^2 + 4 p + 3)\alpha_1 + (p^2 + p
+ 1) \alpha_2 + \alpha_3$ for $p$ a non-negative integer.

Finally, the imaginary roots at level one in the fundamental
domain $- {\cal F}$ read $(2 m -1) \alpha_1 + m \alpha_2 +
\alpha_3$ and $2m\alpha_1 + m \alpha_2 + \alpha_3$ where $m$ is an
integer greater than or equal to $2$. The first roots have
length squared equal to $- 2 m + \frac{5}{2}$, the second have
length squared equal to $- 2 m + 2$.


\subsection{Overextensions of finite-dimensional Lie algebras}
\label{section:Overextensions}

An interesting class of Lorentzian Kac--Moody algebras can be
constructed by adding simple roots to finite-dimensional simple Lie
algebras in a particular way which will be described below. These are
called ``overextensions''. \index{overextension|bb}

In this section, we let ${\mf{g}}$ be a complex, finite-dimensional,
simple Lie algebra of rank $r$, with simple roots $\alpha_1, \cdots,
\alpha_r$. As stated above, normalize the roots so that the long roots
have length squared equal to $2$ (the short roots, if any, have then
length squared equal to $1$ (or $2/3$ for $G_2$)). The roots of
simply-laced algebras are regarded as long roots.

Let $\alpha = \sum_i n_i \alpha_i$, $n_i \geq 0$ be a positive
root. One defines the \emph{height} of $\alpha$ as 
\begin{equation}
  \htx (\alpha) = \sum_i n_i.
\end{equation}
Among the roots of ${\mf{g}}$, there is a unique one that
has highest height, called the highest root. We denote it by
$\theta$. It is long and it fulfills the property that $(\theta
\vert \alpha_i) \geq 0$ for all simple roots $\alpha_i$, and 
\begin{equation}
  2 \frac{(\alpha_i \vert \theta)}{(\theta \vert \theta)} \in \mathbb{Z},
  \qquad
  2 \frac{(\theta \vert \alpha_i)}{(\alpha_i \vert \alpha_i)} \in \mathbb{Z} 
\end{equation}
(see, e.g.,~\cite{Fuchs:1997jv}). We
denote by $V$ the $r$-dimensional Euclidean vector space spanned
by $\alpha_i$ ($i= 1, \cdots, r$). Let $M_2$ be the two-dimensional
Minkowski space with basis vectors $u$ and $v$ so that $(u \vert
u) = (v \vert v) = 0$ and $(u \vert v) = 1$. The metric in the
space $V \oplus M_2$ has clearly Minkowskian signature $(-, +, +,
\cdots, +)$ so that any Kac--Moody algebra whose simple roots span
$V \oplus M_2$ is necessarily Lorentzian.


\subsubsection{Untwisted overextensions} 

The standard overextensions ${\mf{g}}^{++}$ are obtained by adding
to the original roots of ${\mf{g}}$ the roots
\begin{displaymath}
  \alpha_0 = u - \theta,
  \qquad
  \alpha_{-1} = -u - v.
\end{displaymath}
The matrix $A_{ij} = 2 \frac{(\alpha_i \vert \alpha_j)}{(\alpha_i
  \vert \alpha_i)}$ where $i, j = -1, 0, 1, \cdots, r$ is a
(generalized) Cartan matrix and defines indeed a Kac--Moody algebra.

The root $\alpha_0$ is called the affine root and the algebra
${\mf{g}}^+$ (${\mf{g}}^{(1)}$ in Kac's notations~\cite{Kac}) with
roots $\alpha_0, \alpha_1, \cdots, \alpha_r$ is the untwisted
affine extension of ${\mf{g}}$. The root $\alpha_{-1}$ is known as
the overextended root. One clearly has rank$({\mf{g}}^{++}) =$
rank$({\mf{g}}) +2$. The overextended root has vanishing scalar
product with all other simple roots except $\alpha_0$. One has
explicitly $(\alpha_{-1} \vert \alpha_{-1}) = 2 = (\alpha_0 \vert
\alpha_0)$ and $(\alpha_{-1} \vert \alpha_0) = -1$, which shows
that the overextended root is attached to the affine root (and
only to the affine root) with a single link.

Of these Lorentzian algebras, the following ones are hyperbolic:

\begin{itemize}
\item $A_k^{++}$ ($k \leq 7$),
\item $B_k^{++}$ ($k \leq 8$),
\item $C_k^{++}$ ($k \leq 4$),
\item $D_k^{++}$ ($k \leq 8$),
\item $G_2^{++}$,
\item $F_4^{++}$,
\item $E_k^{++}$ ($k=6,7,8$).
\end{itemize}

The algebras $B_8^{++}$, $D_8^{++}$ and $E_8^{++}$
are also denoted $BE_{10}$, $DE_{10}$ and $E_{10}$, respectively.

\subsubsection*{A special property of \boldmath $E_{10}$}
\index{$E_{10}$|bb}

Of these maximal rank hyperbolic algebras, $E_{10}$ plays a very
special role. Indeed, one can verify that the determinant of its
Cartan matrix is equal to $-1$. It follows that the lattice of
$E_{10}$ is self-dual, i.e., that the fundamental weights belong
to the root lattice of $E_{10}$. In view of the above theorem on
roots of hyperbolic algebras and of the hyperboliticity of
$E_{10}$, the fundamental weights of $E_{10}$ are actually
(imaginary) roots since they are non-spacelike. The root lattice
of $E_{10}$ is the only Lorentzian, even, self-dual lattice in 10
dimensions (these lattices exist only in 2 mod 8 dimensions).


\subsubsection{Root systems in Euclidean space}
\label{section:RootSystemsEuclidean}

In order to describe the ``twisted'' overextensions, \index{overextension} we need to
introduce the concept of a ``root system''.

A \emph{root system} in a real Euclidean space
$V$ is by definition a finite subset $\Delta$ of nonzero elements
of $V$ obeying the following two conditions: 
\begin{eqnarray}
  && \Delta \mbox{ spans } V,
  \\
  && \forall \alpha,\,\beta \in \Delta:
  \qquad
  \left\{
    \begin{array}{l}
      A_{\alpha,\beta}=2\frac{(\alpha\vert\beta)}{(\beta\vert\beta)}\in \ZZ,
      \\
      \beta - A_{\beta,\alpha}\,\alpha\in \Delta.
    \end{array}
  \right.
  \label{4.72}
\end{eqnarray}
The elements of $\Delta$ are called the \emph{roots}. From the
definition one can prove the following properties~\cite{Helgason}:

\begin{enumerate}
\item If $\alpha \in \Delta$, then $-\alpha \in \Delta$.
\item If $\alpha \in \Delta$, then the only elements of $\Delta$
  proportional to $\alpha$ are $\pm \frac 12\alpha$, $\pm \alpha$,
  $\pm 2\alpha$. If only $\pm\alpha$ occurs (for all roots $\alpha$),
  the root system is said to be {\sl reduced} (\emph{proper} in
  ``Araki terminology''~\cite{Araki}).
\item If $\alpha$, $\beta\in \Delta$, then $0\leq A_{\alpha,\beta}\,
  A_{\beta,\alpha}\leq 4$, i.e., $A_{\alpha,\beta}=0,\,\pm 1,\,\pm
  2,\,\pm 3,\,\pm 4$; the last occurrence appearing only for
  $\beta=\pm 2\alpha$, i.e., for nonreduced systems. (The proof of
  this point requires the use of the Schwarz inequality.)
\item If $\alpha$, $\beta\in \Delta$ are not proportional to each
  other and $(\alpha\vert\alpha)\leq (\beta\vert\beta)$ then
  $A_{\alpha,\beta}=0,\,\pm1$. Moreover if $(\alpha\vert\beta)\not =
  0$, then $(\beta\vert\beta)=(\alpha\vert\alpha),\, 2\,
  (\alpha\vert\alpha),\mbox{ or } 3\, (\alpha\vert\alpha)$.
\item If $\alpha$, $\beta\in \Delta$, but $\alpha-\beta \not\in \Delta
  \cup 0$, then $(\alpha\vert\beta) \leq 0$ and, as a consequence, if
  $\alpha$, $\beta\in \Delta$ but $\alpha\pm\beta \not\in \Delta
  \cup 0$ then  $(\alpha\vert\beta) = 0$. That $(\alpha \vert
  \beta) \leq 0$ can be seen as follows. Clearly, $\alpha$ and
  $\beta$ can be assumed to be linearly
  independent\epubtkFootnote{If they were not, one would have by the
    second point above $\beta = \pm \frac 12\alpha$, $\beta = \pm
    \alpha$ or $\beta = \pm 2\alpha$. If the minus sign holds, then
    $(\alpha \vert \beta)$ is automatically  $<0$ and there is nothing
    to be proven. So we only need to consider the cases $\beta = +
    \frac 12\alpha$, $\beta = + \alpha$ or $\beta = + 2\alpha$. In
    the first case, $\alpha - \beta = \beta \in \Delta$, in the second
    case $\alpha - \beta = 0$, and in the last case $\alpha - \beta =
    - \alpha \in \Delta$ so these three cases are in fact excluded by
    the assumption. We can therefore assume $\alpha$ and $\beta$ to be
    linearly independent.}. Now, assume $(\alpha \vert \beta) >0$. By
  the previous point, $A_{\alpha, \beta} = 1$ or $A_{\beta, \alpha}
  = 1$. But then either $\alpha - A_{\alpha, \beta} \beta = \alpha -
  \beta \in \Delta$ or $-(\beta - A_{\beta,\alpha} \alpha) = \alpha
  - \beta \in \Delta$ by (\ref{4.72}), contrary to the
  assumption. This proves that $(\alpha \vert \beta) \leq 0$.
\end{enumerate}

Since $\Delta$ spans the vector space $V$, one can chose a basis
$\{\alpha_i\}$ of elements of $V$ within $\Delta$. This can
furthermore be achieved in such a way the $\alpha_i$ enjoy the
standard properties of simple roots of Lie algebras so that in
particular the concepts of positive, negative and highest roots
can be introduced~\cite{Helgason}.

All the abstract root systems in Euclidean space have been
classified (see, e.g.,~\cite{Helgason}) with the following results:

\begin{itemize}
\item The most general root system is obtained by taking a union
  of irreducible root systems. An irreducible root system is one
  that cannot be decomposed into two disjoint nonempty orthogonal
  subsets.
\item The irreducible reduced root systems are simply the root
  systems of finite-dimensional simple Lie algebras ($A_n$ with $n
  \geq 1$, $B_n$ with $n \geq 3$, $C_n$ with $n \geq 2$, $D_n$ with
  $n \geq 4$, $G_2$, $F_4$, $E_6$, $E_7$ and $E_8$).
\item Irreducible nonreduced root systems are all given by the
  so-called $(BC)_{n}$-systems. A $(BC)_{n}$-system is obtained by
  combining the root system of the algebra $B_n$ with the root
  system of the algebra $C_n$ in such a way that the long roots of
  $B_n$ are the short roots of $C_n$. There are in that case three
  different root lengths. Explicitly $\Delta$ is given by the $n$
  unit vectors $\vec e_{k }$ and their opposite $-\vec e_{k }$ along
  the Cartesian axis of an $n$-dimensional Euclidean space, the $2n$
  vectors $\pm 2 \,\vec e_{k }$ obtained by multiplying the previous
  vectors by 2 and the $2n(n-1)$ diagonal vectors $\pm \vec e_{k
  }\pm \vec e_{k' }$, with $k\not = k'$ and $k,k'= 1,\ldots, n$. The
  $n=3$ case is pictured in Figure~\ref{figure:BC23}. The Dynkin
  diagram of $(BC)_r$ is the Dynkin diagram of $B_r$ with a double
  circle $\bigcirc\!\!\!\!\!\!\:\circ\hspace{2.0pt}$ over the simple
  short root, say $\alpha_1$, to indicate that $2 \alpha_1$ is also
  a root.
\end{itemize}

\epubtkImage{BC23.png}{%
  \begin{figure}[htbp]
    \centerline{\includegraphics[width=120mm]{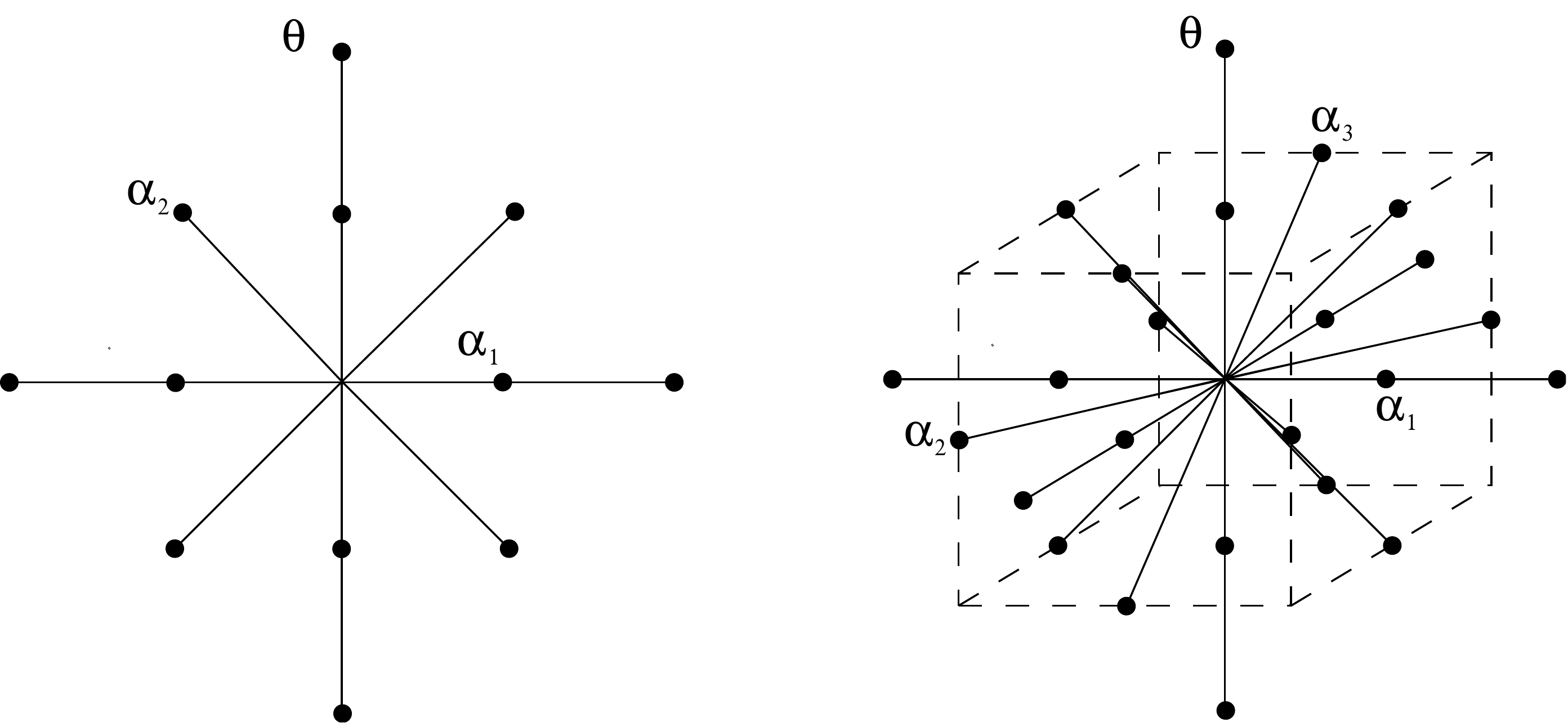}}
    \caption{The nonreduced $(BC)_{2}$- and $(BC)_{3}$-root
      systems. In each case, the highest root $\theta$ is displayed.}
    \label{figure:BC23}
  \end{figure}}

It is sometimes convenient to rescale the roots by the factor
$(1/\sqrt{2})$ so that the highest root $\theta= 2(\alpha_1 +
\alpha_2 + \cdots + \alpha_r)$~\cite{Helgason} of the
$(BC)$-system has length 2 instead of 4.


\subsubsection{Twisted overextensions}

We follow closely~\cite{HenneauxJulia}. Twisted affine algebras
are related to either the $(BC)$-root systems or to extensions by
the highest short root (see~\cite{Kac}, Proposition~6.4).

\subsubsection*{Twisted overextensions associated with the \boldmath
  $(BC)$-root systems}

These are the overextensions \index{overextension} relevant for
some of the gravitational billiards. The construction proceeds as
for the untwisted overextensions, but the starting point is now
the $(BC)_r$ root system with rescaled roots. The highest root has
length squared equal to 2 and has non-vanishing scalar product
only with $\alpha_r$ ($(\alpha_r \vert \theta) = 1$). The
overextension procedure (defined by the same formulas as in the
untwisted case) yields the algebra $(BC)_r^{++}$, also denoted
$A_{2 r}^{(2)+}$.

There is an alternative overextension $A_{2 r}^{(2)'+}$ that can
be defined by starting this time with the algebra $C_r$ but taking
\emph{one-half} the highest root of $C_r$ to make the extension
(see~\cite{Kac}, formula in Paragraph~6.4, bottom of page~84). The
formulas for $\alpha_0$ and $\alpha_{-1}$ are $2 \alpha_0 = u -
\theta$ and $2 \alpha_{-1} = -u -v $ (where $\theta$ is now the
highest root of $C_r$). The Dynkin diagram of $A_{2 r}^{(2)'+}$
is dual to that of $A_{2 r}^{(2)+}$. (Duality amounts to reversing
the arrows in the Dynkin diagram, i.e., replacing the
(generalized) Cartan matrix by its transpose.)

The algebras $A_{2 r}^{(2)+}$ and $A_{2 r}^{(2)'+}$ have rank
$r+2$ and are hyperbolic for $r \leq 4$. The intermediate affine
algebras are in all cases the twisted affine algebras
$A_{2r}^{(2)}$. We shall see in Section~\ref{section:KMBilliardsII}
that by coupling to three-dimensional gravity a coset model
$\mc{G}/\mc{K}(\mc{G})$, where the so-called restricted root system
(see Section~\ref{section:FiniteRealLieAlgebras}) of the
(real) Lie algebra $\mf{g}$ of the Lie group $\mc{G}$ is of
$(BC)_r$-type, one can realize all the $A_{2r}^{(2)+}$ algebras.

\subsubsection*{Twisted overextensions associated with the highest short
  root}

We denote by $\theta_s$ the unique short
root of heighest weight. It exists only for non-simply laced
algebras and has length $1$ (or $2/3$ for $G_2$). The twisted
overextensions \index{overextension} are defined as the standard overextensions but one
uses instead the highest short root $\theta_s$. The formulas for
the affine and overextended roots are
\begin{displaymath}
  \alpha_0 = u - \theta_s,
  \qquad
  \alpha_{-1} = - u - \frac{1}{2} v,
  \qquad
  ({\mf{g}} = B_r, C_r, F_4)
\end{displaymath}
or
\begin{displaymath}
  \alpha_0 = u - \theta_s,
  \qquad
  \alpha_{-1} = - u - \frac{1}{3}v,
  \qquad
  ({\mf{g}} = G_2).
\end{displaymath}
(We choose the overextended root to
have the same length as the affine root and to be attached to it
with a single link. This choice is motivated by considerations of
simplicity and yields the fourth rank ten hyperbolic algebra when
${\mf{g}} = C_8$.)

The affine extensions generated by $\alpha_0, \cdots, \alpha_r$
are respectively the twisted affine algebras $D_{r+1}^{(2)}$
(${\mf{g}} = B_r$), $A_{2r-1}^{(2)}$ ($\mf{g} = C_r$),
$E_6^{(2)}$ (${\mf{g}} = F_4$) and $D_4^{(3)}$ (${\mf{g}} = G_2$).
These twisted affine algebras are related to external
automorphisms of $D_{r+1}$, $A_{2r-1}$, $E_6$ and $D_4$,
respectively (the same holds for $A_{2r}^{(2)}$ above)~\cite{Kac}.
The corresponding twisted overextensions have the following
features.

\begin{itemize}
\item The overextensions $D_{r+1}^{(2)+}$ have
  rank $r + 2$ and are hyperbolic for $r \leq 4$.
\item The overextensions $A_{2r-1}^{(2) +}$ have rank $r + 2$ and are
  hyperbolic for $r \leq 8$. The last hyperbolic case, $r=8$,
  yields the algebra $A_{15}^{(2) +}$, also denoted $CE_{10}$. It is
  the fourth rank-10 hyperbolic algebra, besides $E_{10}$, $BE_{10}$
  and $DE_{10}$.
\item The overextensions $E_6^{(2)+}$ (rank 6) and
  $D_4^{(3)+}$ (rank 4) are hyperbolic.
\end{itemize}

\noindent
We list in Table~\ref{table:TwistedOverextensions} the Dynkin diagrams
of all twisted overextensions.

\begin{table}
  \caption{Twisted overextended Kac--Moody algebras.}
  \renewcommand{\arraystretch}{1.2}
  \vspace{0.5 em}
  \centering
  \begin{tabular}{m{30mm}|m{100mm}}
    \toprule
    Name & Dynkin diagram \\
    \midrule
    $A_2^{(2)+}$ &                       \includegraphics[width=28mm]{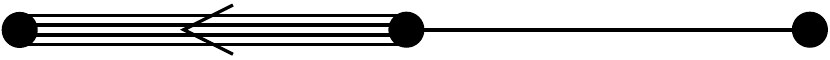} \\ [1 em]
    $A_2^{(2)^{\p}+}$ &                  \includegraphics[width=28mm]{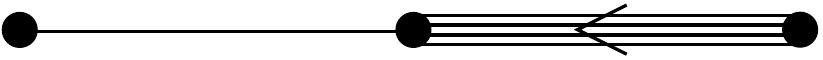} \\ [1 em]
    $A_{2n}^{(2)+}\,\, (n\geq 2)$ &      \includegraphics[width=100mm]{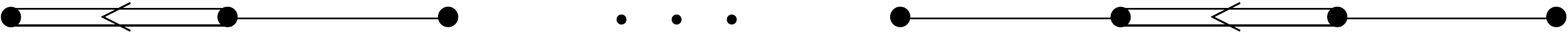} \\ [1 em]
    $A_{2n}^{(2)^{\p}+}\,\, (n\geq 2)$ & \includegraphics[width=100mm]{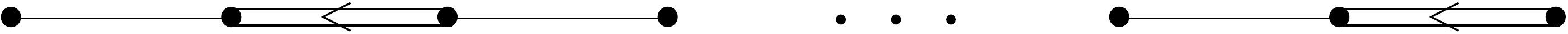} \\ [1 em]
    $A_{2n-1}^{(2)+}\, (n\geq 3)$ &      \includegraphics[width=100mm]{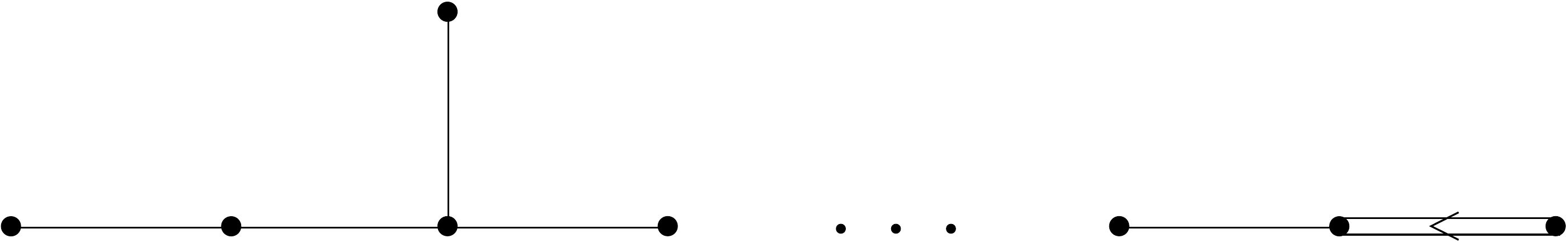} \\ [1 em]
    $D_{n+1}^{(2)+}$ &                   \includegraphics[width=100mm]{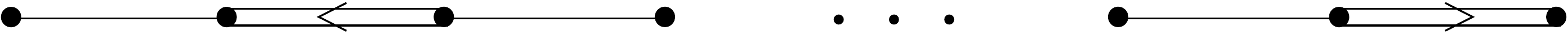} \\ [1 em]
    $E_6^{(2)+}$ &                       \includegraphics[width=73mm]{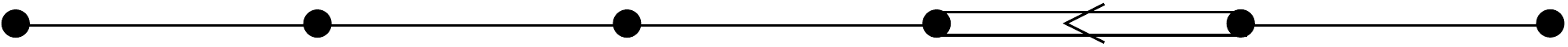} \\ [1 em]
    $D_4^{(3)+}$ &                       \includegraphics[width=45mm]{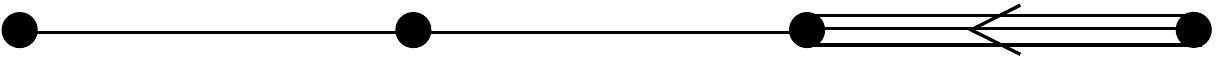} \\
    \bottomrule
  \end{tabular}
  \label{table:TwistedOverextensions}
  \renewcommand{\arraystretch}{1.0}
\end{table}

A satisfactory feature of the class of overextensions
(standard \emph{and} twisted) is that it is closed under duality.
For instance, $A_{2r-1}^{(2)+}$ is dual to $B_r^{++}$. In fact,
one could get the twisted overextensions associated with the
highest short root from the standard overextensions precisely by
requiring closure under duality. A similar feature already holds
for the affine algebras.

Note also that while not all hyperbolic Kac--Moody algebras are
symmetrizable, the ones that are obtained through the process of
overextension are.


\subsubsection{Algebras of Gaberdiel--Olive--West type}

One can further extend the overextended algebras \index{overextension}
to get ``triple extensions'' or ``very extended algebras''. This is
done by adding a further simple root attached with a single link to
the overextended root of Section~\ref{section:Overextensions}. For
instance, in the case of $E_{10}$, \index{$E_{10}$} one gets $E_{11}$
with the Dynkin diagram displayed in Figure~\ref{figure:E11}. These
algebras are Lorentzian, but not hyperbolic.

\epubtkImage{E11.png}{%
  \begin{figure}[t]
    \centerline{\includegraphics[width=110mm]{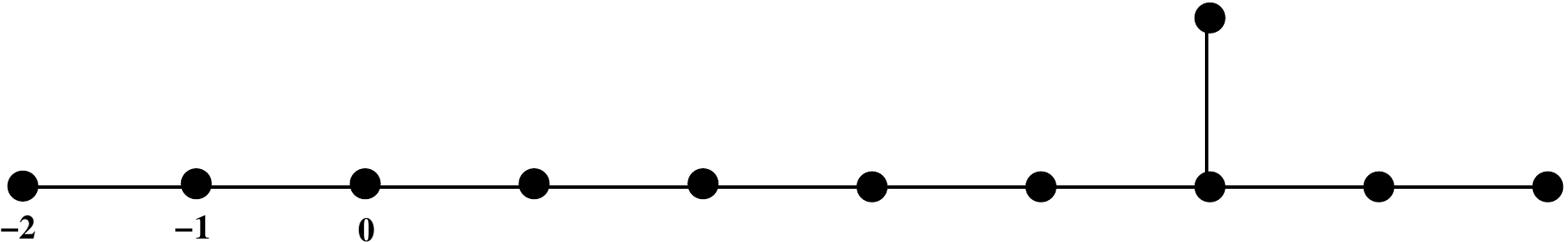}}
    \caption{The Dynkin diagram of $E_{11}$. Labels $0, -1$ and $-2$
      enumerate the nodes corresponding, respectively, to the affine root
      $\al_0$, the overextended root $\al_{-1}$ and the ``very extended''
      root $\al_{-2}$.}
    \label{figure:E11}
  \end{figure}}

The very extended algebras belong to a more general class of algebras
considered by Gaberdiel, Olive and West in~\cite{OliveWest}. These are
defined to be algebras with a connected Dynkin diagram that possesses
at least one node whose deletion yields a diagram with connected
components that are of finite type except for at most one of affine
type. For a hyperbolic algebra, the deletion of \emph{any} node should
fulfill this condition. The algebras of Gaberdiel, Olive and West are
Lorentzian if not of finite or affine
type~\cite{Ruuska:1991ne, OliveWest}. They include the overextensions \index{overextension}
of Section~\ref{section:Overextensions}. The untwisted and twisted
very extended algebras are clearly also of this type, since removing
the affine root gives a diagram with the requested properties.

Higher order extensions with special additional properties have been
investigated in~\cite{Forte:2005uy}.


\subsection{Regular subalgebras of Kac--Moody algebras}
\label{Regular}
\index{regular subalgebra|bb}

This section is based on~\cite{Henneaux:2006gp}.


\subsubsection{Definitions}
\label{definitions}

Let $\mf{g}$ be a Kac--Moody algebra, and let $\bar{\mathfrak{g}}$ be a
subalgebra of $\mathfrak{g}$ with triangular decomposition
$\bar{\mathfrak{g}}= \bar{\mf{n}}_- \oplus \bar{\mathfrak{h}} \oplus
\bar{\mf{n}}_+ $. We assume that $\bar{\mathfrak{g}}$ is canonically
embedded in $\mathfrak{g}$, i.e., that the Cartan subalgebra \index{Cartan subalgebra}
$\bar{\mathfrak{h}}$ of $\bar{\mathfrak{g}}$ is a subalgebra of
the Cartan subalgebra $\mathfrak{h}$ of $\mathfrak{g}$,
$\bar{\mathfrak{h}} \subset \mathfrak{h}$, so that
$\bar{\mathfrak{h}}= \bar{\mathfrak{g}} \cap \mathfrak{h}$. We
shall say that $\bar{\mathfrak{g}}$ is regularly embedded in
$\mathfrak{g}$ (and call it a ``regular subalgebra'') if and only if two
conditions are fulfilled: (i) The root generators of
$\bar{\mathfrak{g}}$ are root generators of $\mathfrak{g}$, and
(ii) the simple roots of $\bar{\mathfrak{g}}$ are real roots of
$\mathfrak{g}$. It follows that the Weyl group of
$\bar{\mathfrak{g}}$ is a subgroup of the Weyl group of
$\mathfrak{g}$ and that the root lattice of $\bar{\mathfrak{g}}$
is a sublattice of the root lattice of $\mathfrak{g}$.

The second condition is automatic in the finite-dimensional case
where there are only real roots. It must be separately imposed in
the general case. Consider for instance the rank 2 Kac--Moody
algebra $\mf{g}$ with Cartan matrix
\begin{displaymath}
  \left(
    \begin{array}{@{}r@{\quad}r@{}}
      2 & -3 \\
      -3 & 2
    \end{array}
  \right).
\end{displaymath}
Let 
\begin{eqnarray}
  x &=& \frac{1}{\sqrt{3}}[e_1,e_2],
  \\
  y &=& \frac{1}{\sqrt{3}}[f_1,f_2],
  \\
  z &=& -(h_1 + h_2).
\end{eqnarray}
It is easy to verify that $x,y,z$ define an $A_1$ subalgebra
of $\mf{g}$ since $[z, x] = 2 x$, $[z,y] = -2 y$ and $[x,y] =
z$. Moreover, the Cartan subalgebra of $A_1$ is a subalgebra of
the Cartan subalgebra of $\mf{g}$, and the step operators of
$A_1$ are step operators of $\mf{g}$. However, the simple
root $\alpha = \alpha_1 + \alpha_2$ of $A_1$ (which is an
$A_1$-real root since $A_1$ is finite-dimensional), is an
imaginary root of $\mf{g}$: $\alpha_1 + \alpha_2$ has norm
squared equal to $-2$. Even though the root lattice of $A_1$
(namely, $\{\pm \alpha\}$) is a sublattice of the root lattice of
$\mf{g}$, the reflection in $\alpha$ is not a Weyl reflection
of $\mf{g}$. According to our definition, this embedding of
$A_1$ in $\mf{g}$ is not a regular embedding.


\subsubsection[Examples -- Regular subalgebras of $E_{10}$]%
              {Examples -- Regular subalgebras of \boldmath $E_{10}$}
\label{section:examplesregularsubalgebra}
\index{regular subalgebra}

We shall describe some regular subalgebras of $E_{10}$.
The Dynkin diagram of $E_{10}$ is displayed in Figure~\ref{figure:E10c}.

\subsubsection*{\boldmath $A_9 \subset \mathcal{B} \subset E_{10}$}

A first, simple, example of a regular
embedding is the embedding of $A_9$ in $E_{10}$ which will be used
to define the level when trying to reformulate eleven-dimensional
supergravity as a nonlinear sigma model. This is not a maximal
embedding since one can find a proper subalgebra $\mathcal{B}$ of
$E_{10}$ that contains $A_9$. One may take for $\mathcal{B}$ the
Kac--Moody subalgebra of $E_{10}$ generated by the operators at
levels $0$ and $\pm 2$, which is a subalgebra of the algebra
containing all operators of even level\epubtkFootnote{We thank Axel
Kleinschmidt for an informative comment on this point.}. It is
regularly embedded in $E_{10}$. Its Dynkin diagram is shown in
Figure~\ref{figure:E7ppp}.

\epubtkImage{E10.png}{%
  \begin{figure}[t]
    \centerline{\includegraphics[width=100mm]{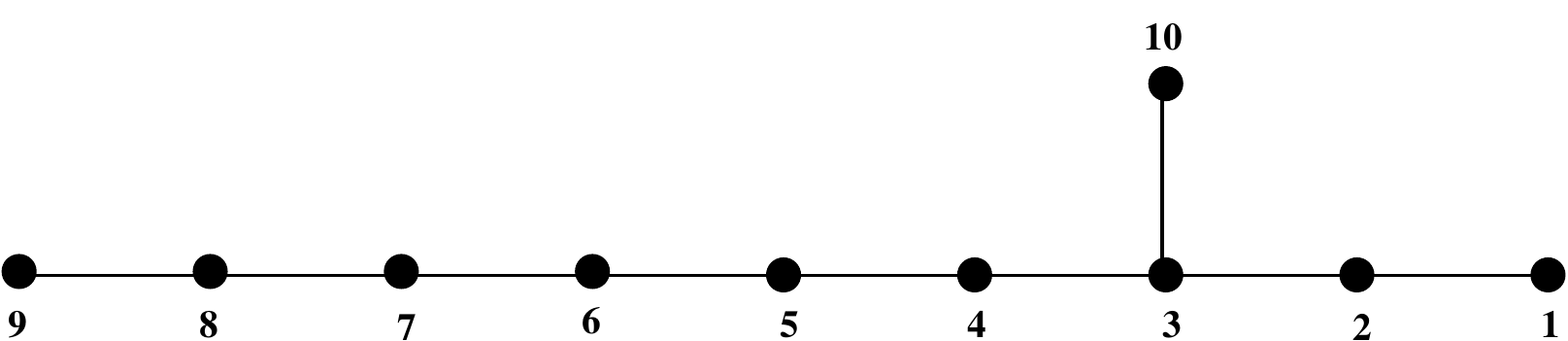}}
    \caption{The Dynkin diagram of $E_{10}$. \index{$E_{10}$|bb}
      Labels $1,\cdots, 7$ and $10$ enumerate the nodes corresponding
      the regular $E_{8}$ subalgebra discussed in the text.}
    \label{figure:E10c}
  \end{figure}}

\epubtkImage{E7ppp.png}{%
  \begin{figure}[htbp]
    \centerline{\includegraphics[width=100mm]{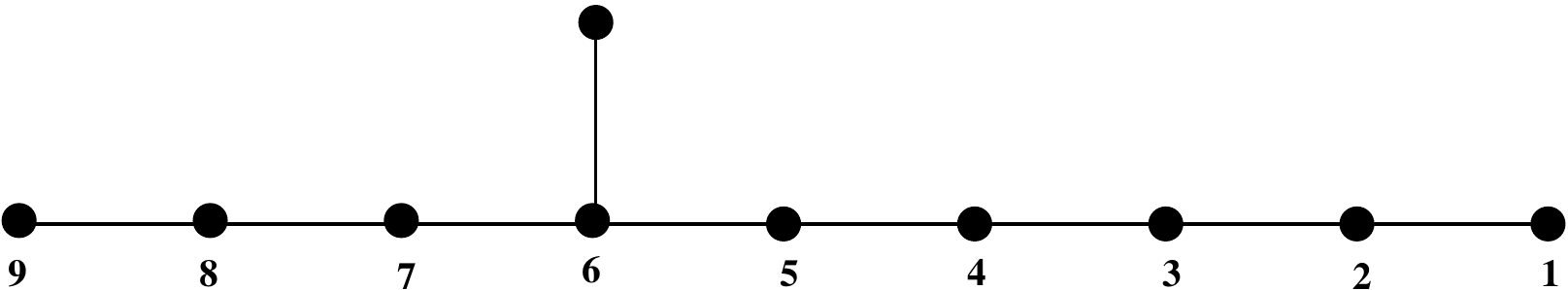}}
    \caption{The Dynkin diagram of $\mathcal{B} \equiv E^{+++}_7$. The
      root without number is the root denoted $\bar{\alpha}_{10}$ in the
      text.} \label{figure:E7ppp}
  \end{figure}}

In terms of the simple roots of $E_{10}$, the simple
roots of $\mathcal{B}$ are $\alpha_1$ through $\alpha_9$ and
$\bar{\alpha}_{10} = 2 \alpha_{10} + \alpha_1 + 2 \alpha_2 + 3
\alpha_3 + 2 \alpha_4 + \alpha_5$. The algebra $\mathcal{B}$ is
Lorentzian but not hyperbolic. It can be identified with the
``very extended'' algebra $E^{+++}_7$~\cite{OliveWest}.

\subsubsection*{\boldmath $DE_{10} \subset E_{10}$}
\index{$DE_{10}$|bb}

In~\cite{Dynkin:1957um}, Dynkin has given a method for finding all
maximal regular subalgebras \index{regular subalgebra} of
finite-dimensional simple Lie algebras. The method is based on using
the highest root and is not generalizable as such to general
Kac--Moody algebras for which there is no highest root. Nevertherless,
it is useful for constructing regular embeddings of overextensions
\index{overextension} of finite-dimensional simple Lie algebras. We
illustrate this point in the case of $E_8$ and its overextension
$E_{10}\equiv E_{8}^{++}$. In the notation of
Figure~\ref{figure:E10c}, the simple roots of $E_{8}$ (which is
regularly embedded in $E_{10}$) are $\alpha_1, \cdots , \alpha_7$ and
$\alpha_{10}$.

Applying Dynkin's procedure to $E_8$, one easily finds that $D_8$
can be regularly embedded in $E_8$. The simple roots of $D_8
\subset E_8$ are $\alpha_2, \alpha_3, \alpha_4, \alpha_5,
\alpha_6, \alpha_7$, $\alpha_{10}$ and $\beta \equiv - \theta_{E_8}$,
where 
\begin{equation}
\theta_{E_8} = 3 \alpha_{10} + 6 \alpha_3 + 4 \alpha_2 + 2
\alpha_1 + 5 \alpha_4 + 4 \alpha_5 + 3 \alpha_6 + 2 \alpha_7 
\end{equation}
is the highest root of $E_8$.
One can replace this embedding, in which a simple root of $D_8$,
namely $\beta$, is a negative root of $E_8$ (and the corresponding
raising operator of $D_8$ is a lowering operator for $E_8$), by an
equivalent one in which all simple roots of $D_8$ are positive
roots of $E_8$.

This is done as follows. It is reasonable to guess that the
searched-for Weyl element that maps the ``old'' $D_8$ on the ``new''
$D_8$ is some product of the Weyl reflections in the four
$E_8$-roots orthogonal to the simple roots $\alpha_3$, $\alpha_4$,
$\alpha_5$, $\alpha_6$ and $\alpha_7$, expected to be shared (as
simple roots) by $E_8$, the old $D_8$ and the new $D_8$ -- and
therefore to be invariant under the searched-for Weyl element.
This guess turns out to be correct: Under the action of the
product of the commuting $E_8$-Weyl reflections in the $E_8$-roots
$\mu_1 = 2 \alpha_1 + 3 \alpha_2 + 5 \alpha_3 + 4 \alpha_4 + 3
\alpha_5 + 2 \alpha_6 + \alpha_7+ 3 \alpha_{10}$ and $\mu_2= 2
\alpha_1 + 4 \alpha_2 + 5 \alpha_3 + 4 \alpha_4 + 3 \alpha_5 + 2
\alpha_6 + \alpha_7+ 2 \alpha_{10}$, the set of $D_8$-roots
$\{\alpha_2,\alpha_3,\alpha_4,\alpha_5,\alpha_6, \alpha_7,
\alpha_{10}, \beta \}$ is mapped on the equivalent set of positive
roots $\{\alpha_{10},\alpha_3,\alpha_4,\alpha_5,\alpha_6,
\alpha_7, \alpha_{2}, \bar{\beta} \}$, where 
\begin{equation}
  \bar{\beta} =
  2 \alpha_1 + 3 \alpha_2 + 4 \alpha_3 + 3 \alpha_4 + 2 \alpha_5 +
  \alpha_6 + 2 \alpha_{10}.
\end{equation}
In this equivalent embedding, all
raising operators of $D_8$ are also raising operators of $E_8$.
What is more, the highest root of $D_8$, 
\begin{equation}
  \theta_{D_8} =
  \alpha_{10} + 2 \alpha_3 + 2\alpha_4 + 2 \alpha_5 + 2 \alpha_6 +
  2 \alpha_7 + \alpha_{2}+ \bar{\beta}
\end{equation}
is equal to the highest root
of $E_8$. Because of this, the affine root $\alpha_8$ of the
untwisted affine extension $E_8^+$ can be identified with the
affine root of $D_8^+$, and the overextended root $\alpha_9$ can
also be taken to be the same. Hence, $DE_{10}$ can be regularly
embedded in $E_{10}$ (see Figure~\ref{figure:DE10}).

\epubtkImage{DE10.png}{%
  \begin{figure}[htbp]
    \centerline{\includegraphics[width=90mm]{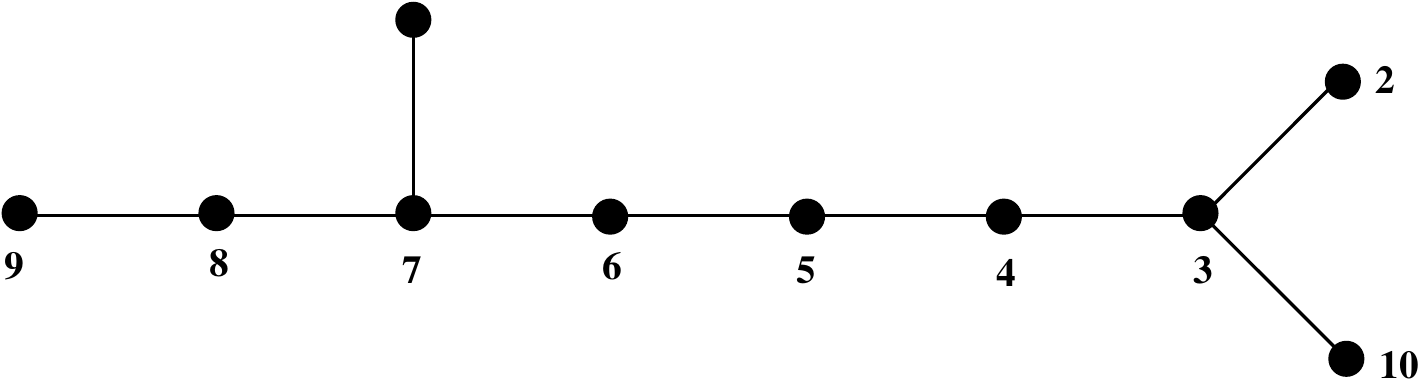}}
    \caption{$DE_{10}\equiv D_{8}^{++}$ regularly embedded in
      $E_{10}$. Labels $2,\cdots, 10$ represent the simple roots
      $\al_2,\cdots, \al_{10}$ of $E_{10}$ and the unlabeled node
      corresponds to the positive root $\bar{\be}=2\al_1+
      3\al_2+4\al_3+3\al_4+2\al_5+\al_6+2\al_{10}$.}
    \label{figure:DE10}
  \end{figure}}

The embedding just described is in fact relevant to string theory
and has been discussed from various points of view in previous
papers~\cite{E10andIIA,Brown:2004ar}. By dimensional reduction of
the bosonic sector of eleven-dimensional supergravity on a circle,
one gets, after dropping the Kaluza--Klein vector and the 3-form,
the bosonic sector of pure $\mc{N}=1$ ten-dimensional supergravity. The
simple roots of $DE_{10}$ \index{$DE_{10}$} are the symmetry walls and the electric
and magnetic walls of the 2-form and coincide with the positive
roots given above~\cite{ArithmeticalChaos}. A similar construction
shows that $A_8^{++}$ can be regularly embedded in $E_{10}$, and that
$DE_{10}$ can be regularly embedded in $BE_{10}\equiv
B_8^{++}$. \index{$BE_{10}$} See~\cite{Hillmann:2006ic} for a recent discussion of
$DE_{10}$ in the context of Type~I supergravity.


\subsubsection{Further properties}

As we have just seen, the
raising operators of $\bar{\mathfrak{g}}$ might be raising or
lowering operators of $\mathfrak{g}$. We shall consider here only
the case when the positive (respectively, negative) root generators
of $\bar{\mathfrak{g}}$ are also positive (respectively, negative)
root generators of $\mathfrak{g}$, so that $\bar{\mf{n}}_- =
\mf{n}_- \cap \bar{\mathfrak{g}}$ and $\bar{\mf{n}}_+ = \mf{n}_+
\cap \bar{\mathfrak{g}}$ (``positive regular embeddings''). This
will always be assumed from now on.

In the finite-dimensional case, there is a useful criterion to
determine regular algebras from subsets of roots. This criterion,
which does not use the highest root, has been generalized to
Kac--Moody algebras in~\cite{Feingold}. It covers also non-maximal
regular subalgebras \index{regular subalgebra} and goes as follows:

\begin{theorem}
  Let $\Phi^+_\mathrm{real}$ be the
  set of positive real roots of a Kac--Moody algebra $\mf{g}$. Let
  $\ga_1, \cdots, \ga_n \in \Phi^+_\mathrm{real}$ be chosen such that none
  of the differences $\ga_i - \ga_j$ is a root of $\mf{g}$. Assume
  furthermore that the $\ga_i$'s are such that the matrix $C
  =[C_{ij}] = [2 \left( \ga_i \vert \ga_j \right) /\left(\ga_i \vert
    \ga_i \right)]$ has non-vanishing determinant. For each $1 \leq i
  \leq n$, choose non-zero root vectors $E_i$ and $F_i$ in the
  one-dimensional root spaces corresponding to the positive real
  roots $\ga_i$ and the negative real roots $-\ga_i$, respectively,
  and let $H_i = [E_i, F_i]$ be the corresponding element in the
  Cartan subalgebra of $\mf{g}$. Then, the (regular) subalgebra of
  $\mf{g}$ generated by $\{E_i, F_i, H_i\}$, $i= 1, \cdots, n$, is a
  Kac--Moody algebra with Cartan matrix \index{Cartan matrix} $[C_{ij}]$.
  
  \begin{newproof}
    The proof of this theorem is given in~\cite{Feingold}. Note that
    the Cartan integers $2 \f{\left( \ga_i \vert \ga_j \right)}{
      \left(\ga_i \vert \ga_i \right)}$ are indeed
    integers (because the $\ga_i$'s are positive real roots), which
    are non-positive (because $\ga_i - \ga_j$ is not a root), so that
    $[C_{ij}]$ is a Cartan matrix.
  \end{newproof}
\end{theorem}

\subsubsection*{Comments}

\begin{enumerate}
\item When the Cartan matrix is degenerate, the corresponding
  Kac--Moody algebra has nontrivial ideals~\cite{Kac}. Verifying that
  the Chevalley--Serre relations are fulfilled is not sufficient to
  guarantee that one gets the Kac--Moody algebra corresponding to the
  Cartan matrix $[C_{ij}]$ since there might be non-trivial
  quotients. Situations in which the algebra generated by the set
  $\{E_i, F_i, H_i\}$ is the quotient of the Kac--Moody algebra with
  Cartan matrix $[C_{ij}]$ by a non-trivial ideal were discussed
  in~\cite{Henneaux:2006gp}.
\item If the matrix $[C_{ij}]$ is decomposable, say $C=D\oplus E$ with
  $D$ and $E$ indecomposable, then the Kac--Moody algebra
  $\mathbb{KM}(C)$ generated by $C$ is the direct sum of the Kac--Moody
  algebra $\mathbb{KM}(D)$ generated by $D$ and the Kac--Moody algebra
  $\mathbb{KM}(E)$ generated by $E$. The subalgebras $\mathbb{KM}(D)$
  and $\mathbb{KM}(E)$ are ideals. If $C$ has non-vanishing
  determinant, then both $D$ and $E$ have non-vanishing
  determinant. Accordingly, $\mathbb{KM}(D)$ and $\mathbb{KM}(E)$ are
  simple~\cite{Kac} and hence, either occur faithfully or
  trivially. Because the generators $E_i$ are linearly independent,
  both $\mathbb{KM}(D)$ and $\mathbb{KM}(E)$ occur
  faithfully. Therefore, in the above theorem the only case that
  requires special treatment is when the Cartan matrix $C$ has
  vanishing determinant.
\end{enumerate}

\noindent As we have mentioned above, it is convenient to
universally normalize the Killing form of Kac--Moody algebras in
such a way that the long real roots have always the same squared
length, conveniently taken equal to two. It is then easily seen
that the Killing form of any regular Kac--Moody subalgebra of
$E_{10}$ coincides with the invariant form induced from the
Killing form of $E_{10}$ through the embedding since $E_{10}$ is
``simply laced''. This property does not hold for non-regular
embeddings as the example given in Section~\ref{section:definitions}
shows: The subalgebra $A_1$ considered there has an induced form
equal to minus the standard Killing form. 

\newpage


\section{Kac--Moody Billiards I -- The Case of Split Real Forms}
\label{section:KMBilliardsI}
\setcounter{equation}{0}

In this section we will begin to explore in more detail the
correspondence between Lorentzian Coxeter groups and the limiting
behavior of the dynamics of gravitational theories close to a
spacelike singularity.

We have seen in Section~\ref{section:BKL} that in the BKL-limit, 
\index{BKL-limit} the dynamics of gravitational theories is equivalent
to a billiard dynamics in a region of hyperbolic space. 
\index{hyperbolic space} In the generic
case, the billiard region has no particular feature. However, we
have seen in Section~\ref{section:Coxeter} that in the case
of pure gravity in four spacetime dimensions, the billiard region
has the remarkable property of being the fundamental domain of the
Coxeter group $PGL(2,\mathbb{Z})$ acting on two-dimensional
hyperbolic space.

This is not an accident. Indeed, this feature arises for all
gravitational theories whose toroidal dimensional reduction to three
dimensions exhibits hidden symmetries, in the sense that the reduced
theory can be reformulated as three-dimensional gravity coupled to a
nonlinear sigma-model based on $\mc{U}_3/ \mc{K}(\mc{U}_3)$, where
$\mc{K}(\mc{U}_3)$ is the maximal compact subgroup 
\index{maximal compact subgroup} of $\mc{U}_3$. The
``hidden'' symmetry group $\mc{U}_3$ is also called, by a
generalization of language, ``the U-duality
group''~\cite{Obers:1998fb}. This situation covers the cases of pure
gravity in any spacetime dimension, as well as all known supergravity
models. In all these cases, the billiard region is the fundamental
domain of a Lorentzian Coxeter group \index{Coxeter group} (``Coxeter
billiard''). Furthermore, the Coxeter group in question is
crystallographic and turns out to be the Weyl group of a Lorentzian
Kac--Moody algebra. The billiard table is then the fundamental Weyl
chamber of a Lorentzian Kac--Moody algebra~\cite{ArithmeticalChaos,
  HyperbolicKaluzaKlein} and the billiard is also called a ``Kac--Moody
billiard''. This enables one to reformulate the dynamics as a motion
in the Cartan subalgebra of the Lorentzian Kac--Moody algebra, hinting
at the potential -- and still conjectural at this stage -- existence
of a deeper, infinite-dimensional symmetry of the theory.

The purpose of this section is threefold:

\begin{enumerate}

\item First, we exhibit other theories besides pure gravity in four
  dimensions which also lead to a Coxeter billiard. We stress further
  how exceptional these theories are in the space of all theories
  described by the action Equation~(\ref{keyaction}).

\item Second, we show how to reformulate the dynamics as a motion in
  the Cartan subalgebra of a Lorentzian Kac--Moody algebra.

\item Finally, we connect the Lorentzian Kac--Moody algebra that
  appears in the BKL-limit to the ``hidden'' symmetry group $\mc{U}_3$
  in the simplest case when the real Lie algebra ${\mathfrak{u}}_3$ of
  the group $\mc{U}_3$ is the split real form of the corresponding
  complexified Lie algebra ${\mathfrak{u}}_3^{\mathbb{C}}$. (These
  concepts will be defined below.) The general case will be dealt with
  in Section~\ref{section:KMBilliardsII}, after we have recalled the
  most salient features of the theory of real forms in
  Section~\ref{section:FiniteRealLieAlgebras}.

\end{enumerate}


\subsection{More on Coxeter billiards}


\subsubsection[The Coxeter billiard of pure gravity in $D$ spacetime dimensions]%
              {The Coxeter billiard of pure gravity in \boldmath $D$ spacetime dimensions}
\index{cosmological billiard}

We start by providing other examples of
theories leading to regular billiards, focusing first on pure
gravity in any number of $D$ ($>3$) spacetime dimensions. In this
case, there are $d=D-1$ scale factors $\beta^i$ and the relevant
walls are the symmetry walls, Equation~(\ref{symmetryW}), 
\begin{equation}
  s_i(\beta) \equiv \beta^{i+1} - \beta^i= 0
  \qquad
  (i=1, 2, \cdots, d-1),
\end{equation}
and the curvature wall, Equation~(\ref{curvatureW}),
\begin{equation}
 r(\beta) \equiv 2 \beta^1 + \beta^2 + \cdots + \beta^{d-2}= 0.
\end{equation}
There are thus $d$ relevant walls, \index{billiard wall} which define
a simplex in ($d-1$)-dimensional hyperbolic
space $\mc{H}_{d-1}$. The scalar products of the linear forms defining these
walls are easily computed. One finds as non-vanishing products
\begin{equation}
  \begin{array}{rcl}
    (s_i \vert s_i) &=& \phantom{-}2
    \qquad
    (i = 1, \cdots, d-1),
    \\ [0.25 em]
    (r \vert r) &=& \phantom{-}2,
    \\ [0.25 em]
    (s_{i+1} \vert s_i) &=& -1
    \qquad
    (i = 2, \cdots, d-1)
    \\ [0.25 em]
    (r \vert s_1) &=& -1,
    \\ [0.25 em]
    (r \vert s_{d-2}) &=& -1.
  \end{array}
\end{equation}
The matrix of the scalar products of the wall forms is thus the Cartan
matrix of the (simply-laced) Lorentzian Kac--Moody algebra
$A_{d-2}^{++}$ with Dynkin diagram as in
Figure~\ref{figure:AnppNumbered}. The roots of the underlying
finite-dimensional algebra $A_{d-2}$ are given by $s_i$ ($i = 1,
\cdots, d-3$) and $r$. The affine root is $s_{d-2}$ and the
overextended root is $s_{d-1}$.

\epubtkImage{AnppNumbered.png}{%
  \begin{figure}[htbp]
    \centerline{\includegraphics[width=110mm]{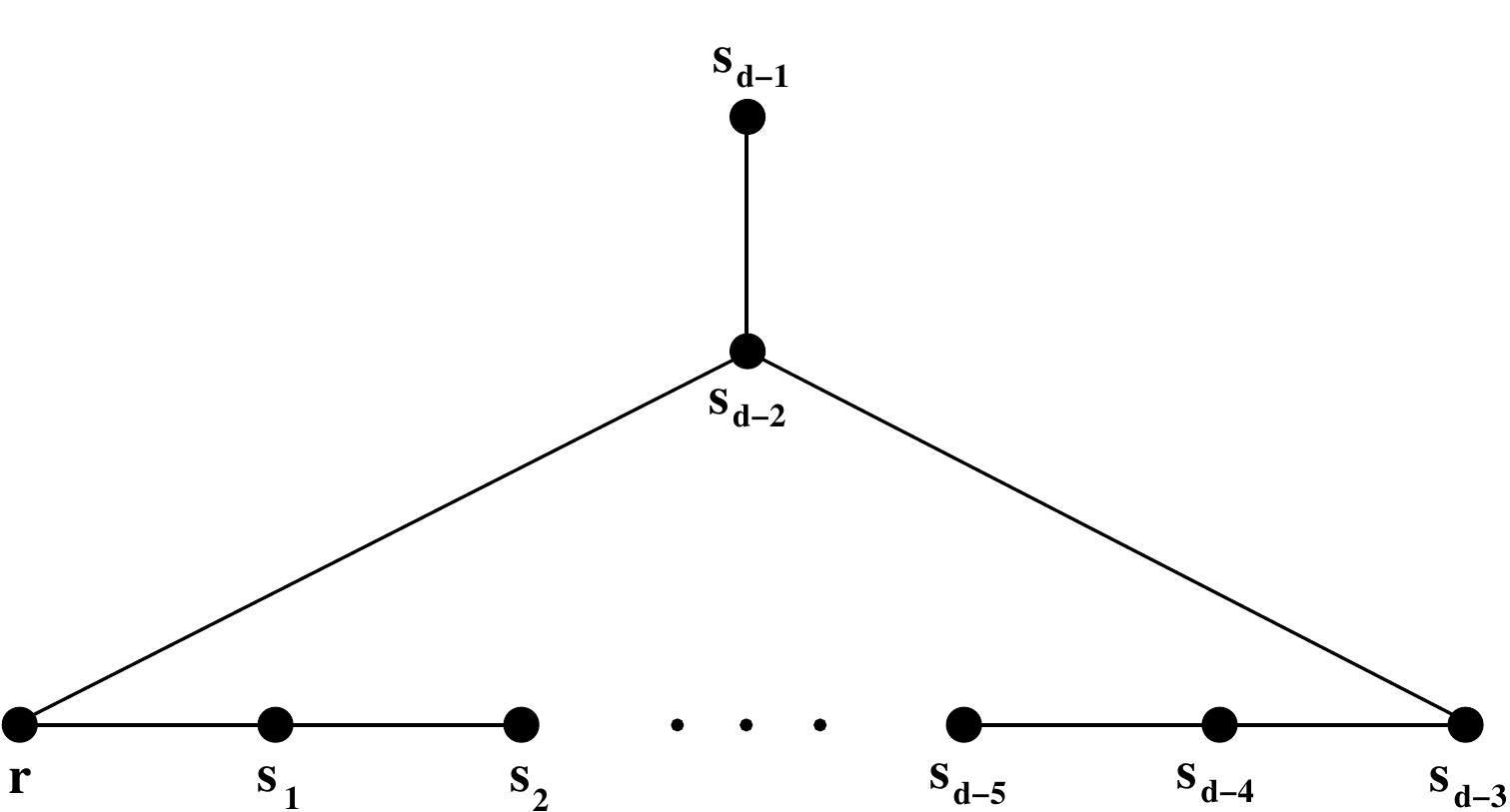}}
    \caption{The Dynkin diagram of the hyperbolic Kac--Moody algebra
      $A_{d-2}^{++}$ which controls the billiard dynamics of pure
      gravity in $D=d+1$ dimensions. The nodes $s_{1}, \cdots, s_{d-1}$
      represent the ``symmetry walls'' arising from the off-diagonal
      components of the spatial metric, and the node $r$ corresponds to
      a ``curvature wall'' coming from the spatial curvature. The
      horizontal line is the Dynkin diagram of the underlying
      $A_{d-2}$-subalgebra and the two topmost nodes, $s_{d-2}$ and
      $s_{d-1}$, give the affine- and overextension, respectively.}
    \label{figure:AnppNumbered}
  \end{figure}}

Accordingly, in the case of pure gravity in any number of
spacetime dimensions, one finds also that the billiard region is
regular. This provides new examples of Coxeter billiards, with
Coxeter groups $A_{d-2}^{++}$, which are also Kac--Moody billiards
since the Coxeter groups are the Weyl groups of the Kac--Moody
algebras $A_{d-2}^{++}$.


\subsubsection{The Coxeter billiard for the coupled gravity-3-Form system}

\subsubsection*{Coxeter polyhedra}

Let us review the conditions that must be fulfilled in
order to get a Kac--Moody billiard and let us emphasize how
restrictive these conditions are. The billiard region computed
from any theory coupled to gravity with $n$ dilatons in $D=d+1$
dimensions always defines a convex polyhedron in a $(d+n-1)$-dimensional
hyperbolic space $\mc{H}_{d+n-1}$. In the general case,
the dihedral angles between adjacent faces of $\mc{H}_{d+n-1}$ can
take arbitrary continuous values, which depend on the dilaton
couplings, the spacetime dimensions and the ranks of the $p$-forms
involved. However, only if the dihedral angles are integer
submultiples of $\pi$ (meaning of the form $\pi/k$ for $k\in
\mbb{Z}_{\geq 2}$) do the reflections in the faces of
$\mc{H}_{d+n-1}$ define a Coxeter group. \index{Coxeter group} In this special case the
polyhedron is called a \emph{Coxeter polyhedron}. This Coxeter
group is then a (discrete) subgroup of the isometry group of
$\mc{H}_{d+n-1}$.

In order for the billiard region to be identifiable with the
fundamental Weyl chamber \index{Weyl chamber} of a Kac--Moody algebra, the Coxeter
polyhedron should be a \emph{simplex}, i.e., bounded by $d+n$
walls in a $d+n-1$-dimensional space. In general, the Coxeter
polyhedron need not be a simplex.

There is one additional condition. The angle $\vartheta$ between
two adjacent faces $i$ and $j$ is given, in terms of the Coxeter
exponents, by 
\begin{equation}
  \vartheta= \f{\pi}{m_{ij}}.
  \label{anglebetweenadjacentfaces}
\end{equation}
Coxeter groups that correspond to Weyl groups of Kac--Moody algebras 
are the \emph{crystallographic} Coxeter groups for which $m_{ij}\in
\{2,3,4,6,\infty\}$. So, the requirement for a gravitational
theory to have a Kac--Moody algebraic description is not just that
the billiard region is a Coxeter simplex but also that the angles
between adjacent walls are such that the group of reflections in
these walls is crystallographic.

These conditions are very restrictive and hence gravitational
theories which can be mapped to a Kac--Moody algebra in the
BKL-limit \index{BKL-limit} are rare.

\subsubsection*{The Coxeter billiard of eleven-dimensional supergravity}

Consider for instance the action~(\ref{keyaction}) for
gravity coupled to a single three-form in $D=d+1$ spacetime
dimensions. We assume $D \geq 6$ since in lower dimensions the
3-form is equivalent to a scalar ($D = 5$) or has no degree of
freedom ($D<5$).

\begin{theorem}
  Whenever a $p$-form ($p \geq 1$) is present, the curvature wall is
  subdominant as it can be expressed as a linear combination with
  positive coefficients of the electric and magnetic walls of the
  $p$-forms. (These walls are all listed in
  Section~\ref{RulesForWalls}.)
  
  \begin{newproof}
    The dominant electric wall is (assuming the
    presence of a dilaton) 
    \begin{equation}
      e_{1\cdots p}(\be)\equiv
      \be^{1}+\be^{2}+\cdots +\be^{p}-\f{\la_p}{2}\phi = 0,
      \label{electricwallcondition}
    \end{equation}
    while one of the magnetic wall reads 
    \begin{equation}
      m_{1, p+1, \cdots , d-2} (\be) \equiv
      \be^{1}+\be^{p+1}+\cdots +\be^{d-2}+\f{\la_p}{2}\phi = 0,
    \end{equation}
    so that the dominant curvature wall is just the sum
    $e_{1\cdots p}(\be) + m_{1, p+1, \cdots , d-2} (\be)$.
  \end{newproof}
\end{theorem}

It follows that in the case of gravity coupled to a
single three-form in $D=d+1$ spacetime dimensions, the relevant walls
are the symmetry walls, Equation~(\ref{symmetryW}), 
\begin{equation}
  s_i(\beta) \equiv \beta^{i+1} - \beta^i= 0,
  \qquad
  i=1, 2, \cdots , d-1
\end{equation}
(as always) and the electric wall 
\begin{equation}
  e_{123}(\beta) \equiv \beta^1 + \beta^2 + \beta^{3} = 0
\end{equation}
($D \geq 8$) or the magnetic wall
\begin{equation}
  m_{1 \cdots D-5}(\beta) \equiv
  \beta^1 + \beta^2 + \cdots \beta^{D-5} = 0
\end{equation}
($D \leq 8$). Indeed, one can express the
magnetic walls as linear combinations with (in general
non-integer) positive coefficients of the electric walls for $D
\geq 8$ and vice versa for $D \leq 8$. Hence the billiard table
is always a simplex (this would not be true had one a dilaton and
various forms with different dilaton couplings).

However, it is only for $D=11$ that the billiard is a Coxeter
billiard. In all the other spacetime dimensions, the angle
between the relevant $p$-form wall and the symmetry wall that does
not intersect it orthogonally is not an integer submultiple of
$\pi$. More precisely, the angle between

\begin{itemize}
\item the magnetic wall $\beta^1$ and the symmetry wall
  $\beta^2 - \beta^1$ ($D=6$),
\item the magnetic wall $\beta^1 + \beta^2$ and the symmetry wall
  $\beta^3 - \beta^2$ ($D=7$), and
\item the electric wall $\beta^1 + \beta^2 + \beta^3$ and the symmetry
  wall $\beta^4 - \beta^3$ ($D \geq 8$),
\end{itemize}

is easily verified to be an integer submultiple of $\pi$ only for
$D=11$, for which it is equal to $\pi/3$.

From the point of view of the regularity of the billiard, the
spacetime dimension $D=11$ is thus privileged. This is of course
also the dimension privileged by supersymmetry. It is quite
intriguing that considerations \emph{a priori} quite different
(billiard regularity on the one hand, supersymmetry on the other
hand) lead to the same conclusion that the gravity-3-form
system is quite special in $D=11$ spacetime dimensions.

For completeness, we here present the wall system relevant for the
special case of $D=11$. We obtain ten dominant wall forms, which
we rename $\al_1, \cdots, \al_{10}$, 
\begin{equation}
  \begin{array}{rcl}
    \al_{m}(\be)&=& \be^{m+1}-\be^m
    \qquad
    (m=1, \cdots, 10),
    \\ [0.25 em]
    \al_{10}(\be)&=& \be^1+\be^2+\be^3.
  \end{array}
\end{equation}
Then, defining a new
collective index $i=(m, 10)$, we see that the scalar products
between these wall forms can be organized into the matrix 
\begin{equation}
  A_{ij}=2\f{(\al_{i}|\al_{j})}{(\al_{i}|\al_{i})}= \left(
    \begin{array}{@{}c@{\quad}c@{\quad}c@{\quad}c@{\quad}c@{\quad}c@{\quad}c@{\quad}c@{\quad}c@{\quad}c@{}}
      2 & -1 & 0 & 0 & 0 & 0 & 0 & 0 & 0 & 0 \\
      -1 & 2 & -1 & 0 & 0 & 0 & 0 & 0 & 0 & 0 \\
      0 & -1 & 2 & -1 & 0 & 0 & 0 & 0 & 0 & -1 \\
      0 & 0 & -1 & 2 & -1 & 0 & 0 & 0 & 0 & 0 \\
      0 & 0 & 0 & -1 & 2 & -1 & 0 & 0 & 0 & 0 \\
      0 & 0 & 0 & 0 & -1 & 2 & -1 & 0 & 0 & 0 \\
      0 & 0 & 0 & 0 & 0 & -1 & 2 & -1 & 0 & 0 \\
      0 & 0 & 0 & 0 & 0 & 0 & -1 & 2 & -1 & 0 \\
      0 & 0 & 0 & 0 & 0 & 0 & 0 & -1 & 2 & 0 \\
      0 & 0 & -1 & 0 & 0 & 0 & 0 & 0 & 0 & 2 \\
    \end{array}
  \right),
\end{equation}
which can be identified with the Cartan matrix \index{Cartan matrix}
of the hyperbolic Kac--Moody algebra \index{Kac--Moody algebra}
$E_{10}$ that we have encountered in
Section~\ref{section:examplesregularsubalgebra}. We again display the
corresponding Dynkin diagram \index{Dynkin diagram} in
Figure~\ref{figure:E10}, where we point out the explicit relation
between the simple roots and the walls of the Einstein--3-form
theory. It is clear that the nine dominant symmetry wall forms
correspond to the simple roots \index{root} $\al_{m}$ of the
subalgebra $\mf{sl}(10,\mbb{R})$. The enlargement to $E_{10}$ is due
to the tenth exceptional root realized here through the dominant
electric wall form $e_{123}$.

\epubtkImage{E10.png}{%
  \begin{figure}[htbp]
    \centerline{\includegraphics[width=100mm]{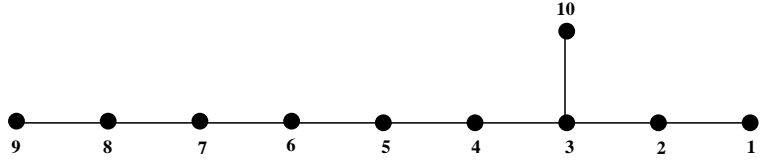}}
    \caption{The Dynkin diagram of $E_{10}$. Labels $m=1,\cdots, 9$
      enumerate the nodes corresponding to simple roots, $\al_{m}$, of the
      $\mf{sl}(10,\mathbb{R})$ subalgebra and the exceptional node,
      labeled ``$10$'', is associated to the electric wall
      $\al_{10}=e_{123}$.}
    \label{figure:E10}
  \end{figure}}


\subsection{Dynamics in the Cartan subalgebra}
\label{section:CSAdynamics}

We have just learned that, in some cases, the group of reflections 
\index{geometric reflection} that describe the (possibly chaotic)
dynamics in the BKL-limit \index{BKL-limit} is a Lorentzian Coxeter
group $\mf{C}$. In this section we fully exploit this algebraic fact
and show that whenever $\mf{C}$ is \emph{crystallographic}, the
dynamics takes place in the Cartan subalgebra $\mf{h}$ of the
Lorentzian Kac--Moody algebra  \index{Kac--Moody algebra} $\mf{g}$,
for which $\mf{C}$ is the Weyl group. Moreover, we show that the
``billiard table'' can be identified with the fundamental Weyl chamber
in $\mf{h}$.


\subsubsection{Billiard dynamics in the Cartan subalgebra}
\index{Cartan subalgebra}
\index{cosmological billiard}

\subsubsection*{Scale factor space and the wall system}

Let us first briefly review some of the salient features
encountered so far in the analysis. In the following we denote by
$\mc{M}_{\be}$ the Lorentzian ``scale factor''-space (or
$\be$-space) in which the billiard dynamics takes place. Recall
that the metric in $\mc{M}_\be$, induced by the Einstein--Hilbert
action, is a flat Lorentzian metric, whose explicit form in terms
of the (logarithmic) scale factors reads 
\begin{equation}
  G_{\mu\nu} \, d\be^{\mu} \, d\be^{\nu} =
  \sum_{i=1}^{d} d\be^{i} \, d\be^{i} -
  \left(\sum_{i=1}^{d}d\be^{i}\right)
  \left(\sum_{j=1}^{d}d\be^{j}\right) + d\phi \, d\phi, 
\label{metricinbetaspace}
\end{equation}
where $d$ counts the number of physical spatial dimensions (see
Section~\ref{RulesForWalls}). The role of all other ``off-diagonal''
variables in the theory is to interrupt the free-flight motion of the
particle, by adding walls \index{billiard wall} in $\mc{M}_{\be}$ that
confine the motion to a limited region of scale factor space, namely a
convex cone bounded by timelike hyperplanes. When projected onto the
unit hyperboloid, this region defines a simplex in hyperbolic space
\index{hyperbolic space} which we refer to as the ``billiard table''.

One has, in fact, more than just the walls. The theory provides
these walls with a specific normalization through the Lagrangian,
which is crucial for the connection to Kac--Moody algebras. Let us
therefore discuss in somewhat more detail the geometric properties of
the wall system. The metric, Equation~(\ref{metricinbetaspace}), in
scale factor space can be seen as an extension of a flat Euclidean
metric in Cartesian coordinates, and reflects the Lorentzian nature of
the vector space $\mc{M}_{\be}$. In this space we may identify a pair
of coordinates $(\be^{i}, \phi)$ with the components of a vector
$\be\in \mc{M}_{\be}$, with respect to a basis $\{\bar{u}_{\mu}\}$ of
$\mc{M}_{\be}$, such that
\begin{equation}
  \bar{u}_{\mu}\cdot \bar{u}_{\nu}=G_{\mu\nu}.
\end{equation}
The walls themselves are then defined by hyperplanes in this linear
space, i.e., as linear forms $\om=\om_{\mu}\underline{\si}^{\mu}$, for
which $\om=0$, where $\{\underline{\si}^{\mu}\}$ is the basis dual to
$\{\bar{u}^{\mu}\}$. The pairing $\om(\be)$ between a vector $\be\in
\mc{M}_\be$ and a form $\om\in\mc{M}_{\be}^{\star}$ is sometimes also
denoted by $\left< \om, \be \right>$, and for the two dual bases we
have, of course, 
\begin{equation}
  \left< \underline{\si}^{\mu}, \bar{u}_{\nu}\right> =\delta^{\mu}_{\nu}.
\end{equation}
We therefore find that the walls can be written as linear forms in the
scale factors:
\begin{equation}
  \om(\be)=\sum_{\mu, \nu}\om_{\mu}\be^{\nu}
  \left<\underline{\si}^{\mu}, \bar{u}_{\nu}\right>=
  \sum_{\mu}\om_{\mu}\be^{\mu}=
  \sum_{i=1}^{d}\om_{i}\be^{i}+\om_{\phi}\phi.
  \label{wallforms}
\end{equation}
We call $\om(\be)$ \emph{wall forms}. With this interpretation they
belong to the dual space $\mc{M}_{\be}^{\star}$, i.e., 
\begin{equation}
  \begin{array}{rcl}
    \mc{M}_{\be}^{\star} \, \ni\, \om :
    \mc{M}_{\be} &\quad \longrightarrow& \quad \mbb{R},
    \\ [0.25 em]
    \be &\quad \longmapsto& \quad \om(\be).
  \end{array}
  \label{dualspaces}
\end{equation}
From Equation~(\ref{dualspaces}) we may conclude that the walls
bounding the billiard are the hyperplanes $\om=0$ through the origin
in $\mc{M}_{\be}$ which are orthogonal to the vector with components
$\om^{\mu}=G^{\mu\nu}\om_{\nu}$.

It is important to note that it is the wall \index{billiard wall}
forms that the theory provides, as arguments of the exponentials in
the potential, and not just the hyperplanes on which these forms $\om$
vanish. The scalar products between the wall forms are computed using
the metric in the dual space $\mc{M}_{\be}^{\star}$, whose explicit
form was given in Section~\ref{RulesForWalls}, 
\begin{equation}
  (\om |\om^{\prime})\equiv G^{\mu\nu}\om_{\mu}\om_{\nu}=
  \sum_{i=1}^{d}\om_{i}\om_{i}^{\prime}-\f{1}{d-1}
  \left(\sum_{i=1}^{d}\om_{i}\right)
  \left(\sum_{j=1}^{d}\om_{j}^{\prime}\right)+
  \om_{\phi}\om_{\phi}^{\prime},
  \qquad
  \om, \om^{\p}\in\mc{M}_{\be}. 
  \label{inversemetric}
\end{equation}

\subsubsection*{Scale factor space and the Cartan subalgebra}

The crucial additional observation is that (for the ``interesting''
theories) the matrix $A$ associated with the relevant walls
$\om_A$, 
\begin{equation}
  A_{AB} = 2 \frac{(\om_A \vert \om_B)}{(\om_A \vert \om_A)}
\end{equation}
is a Cartan matrix, \index{Cartan matrix} i.e., besides having 2's on its
diagonal, which is rather obvious, it has as off-diagonal entries
non-positive integers (with the property $A_{AB} \not= 0
\Rightarrow A_{BA} \not=0$). This Cartan matrix is of course
symmetrizable since it derives from a scalar product.

For this reason, one can usefully identify the space of the scale
factors with the Cartan subalgebra \index{Cartan subalgebra} $\mf{h}$
of the Kac--Moody algebra $\mf{g}(A)$ defined by $A$. In that
identification, the wall forms become the simple roots, \index{root}
which span the vector space
$\mf{h}^{\star}=\spn\{\al_{1},\cdots,\al_{r}\}$ dual to the Cartan
subalgebra. The rank $r$ of the algebra is equal to the number of
scale factors $\beta^\mu$, including the dilaton(s) if any
($(\beta^\mu) \equiv (\be^{i}, \phi)$). This number is also equal to
the number of walls since we assume the billiard to be a simplex. So,
both $A$ and $\mu$ run from $1$ to $r$. The metric in $\mc{M}_\be$,
Equation~(\ref{metricinbetaspace}), can be identified with the
invariant bilinear form of $\mf{g}$, restricted to the Cartan
subalgebra $\mf{h}\subset \mf{g}$. The scale factors $\beta^\mu$ of
$\mc{M}_\be$ become then coordinates $h^{\mu}$ on the Cartan
subalgebra $\mf{h}\subset\mf{g}(A)$.

The Weyl group of a Kac--Moody algebra has been defined first in
the space $\mf{h}^{\star}$ as the group of reflections in the
walls orthogonal to the simple roots. Since the metric is non
degenerate, one can equivalently define by duality the Weyl group
in the Cartan algebra $\mf{h}$ itself (see
Section~\ref{section:weylgroup}). For each reflection $r_i$ on
$\mf{h}^{\star}$ we associate a dual reflection $r_i^{\vee}$ as
follows,
\begin{equation}
  r_i^{\vee}(\be)=\be-\left< \al_i, \be\right > \al_i^{\vee},
  \qquad
  \be, \al_i^{\vee}\in\mf{h},
  \label{Cartanreflection}
\end{equation}
which is the reflection relative to
the hyperplane $\al_i(\be)=\left< \al_i, \be \right> =0$. This
expression can be rewritten (see Equation~(\ref{4.61xx})), 
\begin{equation}
  r_i^{\vee}(\be)=
  \be-\f{2(\be|\al_i^{\vee})}{(\al_i^{\vee}|\al_i^{\vee})}\al_i^{\vee},
  \label{Cartanreflection2}
\end{equation}
or, in terms of the scale factor coordinates $\beta^\mu$, 
\begin{equation}
  \be^{\mu} \longrightarrow {\be^{\mu}}^{\p} =
  \be^{\mu}-\f{2(\be\vert \om^{\vee})}{(\om^\vee \vert \om^\vee)}\om^{\vee\mu}. 
  \label{Geometricreflection}
\end{equation}
This is precisely the billiard reflection
Equation~(\ref{billiardreflectionrule}) found in
Section~\ref{section:DynamicsBilliardHyp}.

Thus, we have the following correspondence:
\begin{equation}
  \begin{array}{rcl}
    \mc{M}_{\be} &\equiv& \mf{h},
    \\ [0.25 em]
    \mc{M}_{\be}^{\star} &\equiv& \mf{h}^{\star},
    \\ [0.25 em]
    \om_{A}(\be) &\equiv& \al_{A}(h),
    \\ [0.25 em]
    \mbox{billiard wall reflections} \index{geometric reflection|bb} &\equiv&
    \mbox{fundamental Weyl reflections}.
  \end{array}
  \label{identifications} 
\end{equation}
As we have also seen, the Kac--Moody algebra $\mf{g}(A)$ is Lorentzian
since the signature of the metric Equation~(\ref{metricinbetaspace})
is Lorentzian. This fact will be crucial in the analysis of subsequent
sections and is due to the presence of gravity, where conformal
rescalings of the metric define timelike directions in scale factor
space.

We thereby arrive at the following important
result~\cite{ArithmeticalChaos, HyperbolicKaluzaKlein, DHNReview}:

\begin{center}
  \begin{tabular}{c}
    \hline \hline \\
    \emph{The dynamics of (a restricted set of) theories coupled to
      gravity can in the BKL-limit be} \\
    \emph{mapped to a billiard \index{cosmological billiard} motion in
      the Cartan subalgebra
      $\mf{h}$ of a Lorentzian Kac--Moody algebra $\mf{g}$.} \\ [1 em]
    \hline \hline
  \end{tabular}
\end{center}


\subsubsection{The fundamental Weyl chamber and the billiard table}

Let $\mc{B}_{\mc{M}_{\be}}$ denote the region in scale factor
space to which the billiard motion is confined, 
\begin{equation}
  \mcBe =\{\be \in \mc{M}_\be\, | \, \om_{A}(\be)\geq 0\},
  \label{billiardtable}
\end{equation}
where the index $A$ runs over all
relevant walls. On the algebraic side, the fundamental Weyl
chamber in $\mf{h}$ is the closed convex (half) cone given by 
\begin{equation}
  \mc{W}_{\mf{h}}=\{h \in \mf{h}\, | \, \al_{A}(h) \geq 0;
  \, A=1,\cdots, \rank \mf{g} \}. 
  \label{WeylChamber}
\end{equation}
We see that the conditions $\al_{A}(h) \geq 0$ defining
$\mc{W}_{\mf{h}}$ are equivalent, upon examination of
Equation~(\ref{identifications}), to the conditions $\om_{A}(\be)\geq
0$ defining the billiard table $\mcBe$.

We may therefore make the crucial identification 
\begin{equation}
  \mc{W}_{\mf{h}}\equiv \mc{B}_{\mc{M}_\be},
  \label{WeylChamberandBilliardRegion}
\end{equation}
which means that the particle geodesic is confined to move within the
fundamental Weyl chamber \index{Weyl chamber} of $\mf{h}$. From the
billiard analysis in Section~\ref{section:BKL} we know that the
piecewise motion in scale-factor space \index{scale-factor space} is
controlled by geometric reflections \index{geometric reflection} with
respect to the walls $\om_{A}(\be)=0$. By comparing with the dominant
wall forms and using the correspondence in
Equation~(\ref{identifications}) we may further conclude that the
geometric reflections of the coordinates $\be^{\mu}(\tau)$ are
controlled by the Weyl group in the Cartan subalgebra of $\mf{g}(A)$.


\subsubsection{Hyperbolicity implies chaos}

We have learned that the BKL dynamics is chaotic if and only if
the billiard table is of finite volume when projected onto the
unit hyperboloid. From our discussion of hyperbolic Coxeter
groups in Section~\ref{section:Hyperbolic}, we see
that this feature is equivalent to hyperbolicity of the
corresponding Kac--Moody algebra. This leads to the crucial
statement~\cite{ArithmeticalChaos,HyperbolicKaluzaKlein,DHNReview}:

\begin{center}
  \begin{tabular}{c}
    \hline \hline \\
    \emph{If the billiard region of a gravitational system in the
      BKL-limit can be identified with the} \\
    \emph{fundamental Weyl chamber \index{Weyl chamber}
      of a hyperbolic Kac--Moody algebra,
      then the dynamics is chaotic.} \\ [1 em]
    \hline \hline
  \end{tabular}
\end{center}

As we have also discussed above, hyperbolicity can be rephrased in
terms of the fundamental weights $\Lambda_i$ defined as 
\begin{equation}
  \left<\Lambda_{j}, \al_i^{\vee}\right> =
  \f{2(\Lambda_j|\al_i)}{(\al_i|\al_i)} \equiv \delta_{ij},
  \qquad
  \al_i^{\vee}\in \mf{h},\, \Lambda_i\in\mf{h}^{\star}.
  \label{fundamentalweightcondition}
\end{equation}
Just as the fundamental
Weyl chamber in $\cH^\star$ can be expressed in terms of the
fundamental weights (see Equation~(\ref{coneFFundamental})), the
fundamental Weyl chamber in $\cH$ can be expressed in a similar
fashion in terms of the fundamental coweights: 
\begin{equation}
  \mc{W}_{\mf{h}}=\{\be\in\mf{h}\, |\, \be=
  \sum_i a_i \Lambda_i^{\vee}, \, a_i\in\mbb{R}_{\geq 0}\}.
\end{equation}
As we have seen (Sections~\ref{section:Hyperbolic}
and~\ref{HyperbolicKacMoodyAlgebras}), hyperbolicity holds if and
only if none of the fundamental weights are spacelike, 
\begin{equation}
  (\Lambda_{i}|\Lambda_{i})\leq 0,
  \label{timelikecondition}
\end{equation}
for all $ i\in \{1, \cdots, \rank \mf{g}\}$.

\subsubsection*{Example: Pure gravity in \boldmath $D=3+1$ and $A_{1}^{++}$}
\index{$A_1^{++}$}

Let us return once more to the example of pure
four-dimensional gravity, i.e., the original ``BKL billiard''. We
have already found in Section~\ref{section:Coxeter} that the
three dominant wall forms give rise to the Cartan matrix of the
hyperbolic Kac--Moody algebra
$A_1^{++}$~\cite{HyperbolicKaluzaKlein,DHNReview}. Since the algebra
is hyperbolic, this theory exhibits chaotic behavior. In this
example, we verify that the Weyl chamber is indeed contained
within the lightcone by computing explicitly the norms of the
fundamental weights.

It is convenient to first write the simple roots in the $\be$-basis as
follows>
\begin{equation}
  \begin{array}{rcl}
    \al_{1}&=& (2,0,0)
    \\ [0.25 em]
    \al_{2}&=& (-1,1,0)
    \\ [0.25 em]
    \al_{3}&=& (0,-1,1).
  \end{array}
  \label{simplerootsAE3}
\end{equation}

Since the Cartan matrix of $A_1^{++}$ is symmetric, the relations
defining the fundamental weights are
\begin{equation}
  (\al_{i}|\Lambda_{j})\equiv \delta_{ij}.
\end{equation}
By solving these equations for $\Lambda_i$ we deduce that the
fundamental weights are
\begin{equation}
  \begin{array}{rcl}
    \Lambda_{1}&=& \displaystyle
    -\f{3}{2}\al_{1}-2\al_{2}-\al_{3} =(-1,-1,-1),
    \\ [0.75 em]
    \Lambda_{2}&=& \displaystyle
    -2\al_{1}-2\al_{2}-2\al_{3} =(0,1,-1),
    \\ [0.5 em]
    \Lambda_{3}&=& \displaystyle
    -\al_{1}-\al_{2}=(-1,-1,0),
  \end{array}
  \label{FundamentalweightsAE3} 
\end{equation}
where in the last step we have
written the fundamental weights in the $\be$-basis. The norms may
now be computed with the metric in root space and become 
\begin{equation}
  (\Lambda_{1}|\Lambda_{1})=-\f{3}{2},
  \qquad
  (\Lambda_{2}|\Lambda_{2})=-2,
  \qquad
  (\Lambda_{3}|\Lambda_{3})=0. 
  \label{NormsFundamentalweightsAE3}
\end{equation}
We see that $\Lambda_{1}$ and $\Lambda_{2}$ are timelike and
that $\Lambda_{3}$ is lightlike. Thus, the Weyl chamber is indeed
contained inside the lightcone, the algebra is hyperbolic and the
billiard is of finite volume, in agreement with what we already
found \cite{HyperbolicKaluzaKlein}.


\subsection{Understanding the emerging Kac--Moody algebra}
\label{section:understanding}

We shall now relate the Kac--Moody algebra whose fundamental Weyl
chamber emerges in the BKL-limit \index{BKL-limit} to the U-duality group that
appears upon toroidal dimensional reduction to three spacetime
dimensions. We shall do this first in the case when
${\mathfrak{u}}_3$ is a split real form. By this we mean that the
real algebra ${\mathfrak{u}}_3$ possesses the same Chevalley--Serre
presentation as ${\mathfrak{u}}_3^{\mathbb{C}}$, but with
coefficients restricted to be real numbers. This restriction is
mathematically consistent because the coefficients appearing in
the Chevalley--Serre presentation are all reals (in fact,
integers).

The fact that the billiard structure is preserved under reduction
turns out to be very useful for understanding the emergence of
``overextended'' algebras in the BKL-limit. By computing the
billiard in three spacetime dimensions instead of in maximal
dimension, the relation to U-duality groups becomes particularly
transparent and the computation of the billiard follows a similar
pattern for all cases. We will see that if ${\mathfrak{u}}_3$ is
the algebra representing the internal symmetry of the
non-gravitational degrees of freedom in three dimensions, then the
billiard is controlled by the Weyl group of the overextended
algebra ${\mathfrak{u}}_3^{++}$. The analysis is somewhat more
involved when ${\mathfrak{u}}_3$ is non-split, and we postpone a
discussion of this until Section~\ref{section:KMBilliardsII}.


\subsubsection{Invariance under toroidal dimensional reduction}

It was shown in~\cite{InvarianceUnderCompactification} that the
structure of the billiard for any given theory is completely
unaffected by dimensional reduction on a torus. In this section we
illustrate this by an explicit example rather than in full
generality. We consider the case of reduction of
eleven-dimensional supergravity on a circle.

The compactification ansatz in the conventions
of~\cite{Cremmer:1997ct, InvarianceUnderCompactification} is
\begin{equation}
  \g_{MN}=\left(
    \begin{array}{@{}c@{\quad}c@{}}
      e^{-2(\f{4}{3\sqrt{2}}\hat{\varphi})} &
      e^{-2(\f{4}{3\sqrt{2}}\hat{\varphi})}\hat{\mc{A}}_{\nu}
      \\
      e^{-2(\f{4}{3\sqrt{2}}\hat{\varphi})}\hat{\mc{A}}_{\mu} &
      e^{-2(\f{-1}{6\sqrt{2}}\hat{\varphi})}\hat{\g}_{\mu\nu}+
      e^{-2(\f{4}{3\sqrt{2}}\hat{\varphi})}\hat{\mc{A}}_{\mu}\hat{\mc{A}}_{\nu}
    \end{array}
  \right),
  \label{compacificationansatz}
\end{equation}
where $\mu,\nu=0, 2, \cdots , 10$, i.e., the compactification is
performed along the first
spatial direction\epubtkFootnote{Taking the first spatial direction as
  compactification direction is convenient, for it does not change
  the conventions on the simple roots. More precisely, the
  Kaluza--Klein ansatz is compatible in that case with our Iwasawa
  decomposition~(\ref{IwasawaII}) \index{Iwasawa decomposition} of the
  spatial metric with $\cn$
  an upper triangular matrix. The (equivalent) choice of the tenth
  direction as compactification direction would correspond to a
  different (equivalent) choice of $\cn$.}. We will refer to the new
lower-dimensional fields $\hat{\varphi}$ and $\hat{\mc{A}}_{\mu}$
as the dilaton and the Kaluza--Klein (KK) vector, respectively.
Quite generally, hatted fields are low-dimensional fields. The
ten-dimensional Lagrangian becomes 
\begin{eqnarray}
  \mc{L}_{(10)}^{\mathrm{SUGRA}_{11}}&=&
  R_{(10)}\star \mathbf{1}-\star d\hat{\varphi}\wedge d\hat{\varphi}-
  \f{1}{2}e^{-2(\f{3}{2\sqrt{2}}\hat{\varphi})}\star
  \hat{\mc{F}}^{(2)}\wedge \hat{\mc{F}}^{(2)}
  \nonumber
  \\
  & & -\f{1}{2}e^{-2(\f{1}{2\sqrt{2}}\hat{\varphi})}\star\hat{F}^{(4)}
  \wedge\hat{F}^{(4)}-\f{1}{2}e^{-2(\f{-1}{\sqrt{2}}\hat{\varphi})}
  \star\hat{F}^{(3)}\wedge\hat{F}^{(3)},
  \label{compactifiedLagrangian}
\end{eqnarray}
where $\hat{\mc{F}}^{(2)}=d\hat{\mc{A}}^{(1)}$ and $\hat{F}^{(4)},
\hat{F}^{(3)}$ are the field strengths in ten dimensions
originating from the eleven-dimensional 3-form field strength
$F^{(4)}=dA^{(3)}$.

Examining the new form of the metric reveals that the role of the
scale factor $\be^1$, associated to the compactified dimension, is
now instead played by the ten-dimensional dilaton,
$\hat{\varphi}$. Explicitly we have 
\begin{equation}
  \be^{1}=\f{4}{3\sqrt{2}}\hat{\varphi}.
  \label{compactifiedscalefactor}
\end{equation}
The nine remaining
eleven-dimensional scale factors, $\be^2, \cdots, \be^{10}$, may in
turn be written in terms of the new scale factors,
$\hat{\be}^{a}$, associated to the ten-dimensional metric,
$\hat{\g}_{\mu\nu}$, and the dilaton in the following way:
\begin{equation}
  \be^{a}=\hat{\be}^{a}-\f{1}{6\sqrt{2}}\hat{\varphi}
  \qquad
  (a=2,\cdots, 10). 
  \label{tendimensionalscalefactors}
\end{equation}
We are interested in finding the dominant wall forms in terms of the new
scale factors $\hb_2,\cdots , \hb_{10}$ and $\hvp$. It is clear
that we will have eight ten-dimensional symmetry walls,
\begin{equation}
  \hat{s}_{\hm}(\hb)=\hb^{\hm+1}-\hb^{\hm}
  \qquad
  (\hm=2, \cdots , 9), 
  \label{tendimensionalsymmetrywalls}
\end{equation}
which correspond to
the eight simple roots of $\mf{sl}(9,\mathbb{R})$. Using
Equation~(\ref{compactifiedscalefactor}) and
Equation~(\ref{tendimensionalscalefactors}) one may also check that the
symmetry wall $\be^2-\be^1$, that was associated with the
compactified direction, gives rise to an electric wall of the
Kaluza--Klein vector, 
\begin{equation}
  \hat{e}^{\hat{\mc{A}}}_{2}(\hb)=\hb^2-\f{3}{2\sqrt{2}}\hvp.
  \label{KKelectricwall}
\end{equation}
The metric in the dual space gets modified in a natural way,
\begin{equation}
  (\hat{\al}_{k}|\ha_{l})=
  \sum_{i=2}^{10}\ha_{ki}\ha_{li}-\f{1}{8}
  \left(\sum_{i=2}^{10}\ha_{ki}\right)
  \left(\sum_{j=2}^{10}\ha_{lj}\right)+
  \ha_{k\hvp}\ha_{l\hvp},
  \label{10Dmetric}
\end{equation}
i.e., the dilaton contributes with a flat
spatial direction. Using this metric it is clear that
$\hat{e}^{\hat{\mc{A}}}_{2}$ has non-vanishing scalar product
only with the second symmetry wall $\hat{s}_{2}=\hb^{3}-\hb^{2}$,
$(\hat{e}^{\hat{\mc{A}}}_{2}|\hat{s}_{2})=-1$, and it follows that
the electric wall of the Kaluza--Klein vector plays the role of the
first simple root of $\mf{sl}(10,\mathbb{R})$, $\ha_{1}\equiv
\hat{e}^{\hat{\mc{A}}}_{2}$. The final wall form that completes
the set will correspond to the exceptional node labeled ``10'' in
Figure~\ref{figure:E10} and is now given by one of the electric
walls of the NS-NS 2-form $\hat{A}^{(2)}$, namely 
\begin{equation}
  \ha_{10}\equiv\hat{e}^{\hat{A}^{(2)}}_{23}(\hb)=
  \hb^2+\hb^3+\f{1}{\sqrt{2}}\hvp.
  \label{exceptionalroot}
\end{equation}
It is then easy to verify that this
wall form has non-vanishing scalar product only with the third
simple root $\ha_{3}=\hat{s}_{3}$,
$(\hat{e}^{\hat{A}^{(2)}}_{23}|\hat{s}_{3})=-1$, as desired.

We have thus shown that the $E_{10}$ structure is sufficiently
rigid to withstand compactification on a circle with the new
simple roots explicitly given by 
\begin{equation}
  \{\ha_{1}, \ha_{2}, \cdots, \ha_{9}, \ha_{10}\} =
  \{\hat{e}^{\hat{\mc{A}}}_{2}, \hat{s}_{2}, \cdots,
  \hat{s}_{9}, \hat{e}^{\hat{A}^{(2)}}_{23}\}.
  \label{newrootsofE10}
\end{equation}
This result is in fact true also for
the general case of compactification on tori, $T^{n}$. When
reaching the limiting case of three dimensions, all the non-gravity
wall forms correspond to the electric and magnetic walls of the
axionic scalars. We will discuss this case in detail below.

For non-toroidal reductions the above analysis is drastically
modified~\cite{Wesley2005bd,Wesley2006cd}. The topology of
the internal manifold affects the dominant wall system, and hence
the algebraic structure in the lower-dimensional theory is
modified. In many cases, the billiard of the effective
compactified theory is described by a (non-hyperbolic) regular
Lorentzian subalgebra of the original hyperbolic Kac--Moody
algebra~\cite{DPDW}.

The walls are also invariant under dualization of a $p$-form into
a $(D-p-2)$-form; this simply exchanges magnetic and electric
walls.


\subsubsection{Iwasawa decomposition for split real forms}
\index{Iwasawa decomposition|bb}

We will now exploit the invariance of the billiard under dimensional
reduction, by considering theories that -- when compactified on a
torus to three dimensions -- exhibit ``hidden'' internal global symmetries
$\mc{U}_3$. By this we mean that the three-dimensional reduced
theory is described, after dualization of all vectors to scalars,
by the sum of the Einstein--Hilbert Lagrangian coupled to the
Lagrangian for the nonlinear sigma model \index{nonlinear sigma model}
$\mc{U}_3/\mc{K}(\mc{U}_3)$. Here, $\mc{K}(\mc{U}_3)$ is the
maximal compact subgroup \index{maximal compact subgroup} defining the
``local symmetries''. In order to understand the connection between
the U-duality group $\mc{U}_3$ and the Kac--Moody algebras appearing
in the BKL-limit, we must first discuss some important features of the
Lie algebra $\mf{u}_3$.

Let $\mf{u}_3$ be a split real form, meaning that it can be
defined in terms of the Chevalley--Serre presentation of the
complexified Lie algebra $\mf{u}_3^{\mbb{C}}$ by simply
restricting all linear combinations of generators to the real
numbers $\mbb{R}$. Let $\mf{h}_3$ be the Cartan subalgebra 
\index{Cartan subalgebra} of $\mf{u}_3$ appearing in the
Chevalley--Serre presentation, spanned by the generators
$\al_1^{\vee}, \cdots, \al_n^{\vee}$. It is maximally noncompact (see
Section~\ref{section:FiniteRealLieAlgebras}). An \emph{Iwasawa
  decomposition} \index{Iwasawa decomposition} of $\mf{u}_3$ is a
direct sum of vector spaces of the following form,
\begin{equation}
  \mf{u}_3=\mf{k}_3\oplus \mf{h}_3\oplus \mf{n}_3,
\end{equation}
where $\mf{k}_3$ is the ``maximal
compact subalgebra'' of $\mf{u}_3$, and $\mf{n}_3$ is the nilpotent
subalgebra spanned by the positive root generators $E_{\al},\,
\forall\al\in \Delta_+$.

The corresponding Iwasawa decomposition at the group level enables
one to write uniquely any group element as a product of an element
of the maximally compact subgroup times an element in the subgroup
whose Lie algebra is $\mf{h}_3$ times an element in the subgroup
whose Lie algebra is $\mf{n}_3$. An arbitrary element $\mc{V}(x)$
of the coset $\mc{U}_3/\mc{K}(\mc{U}_3)$ is defined as the set of
equivalence classes of elements of the group modulo elements in
the maximally compact subgroup. Using the Iwasawa decomposition,
one can go to the ``Borel gauge'', where the elements in the coset
are obtained by exponentiating only generators belonging to the
\emph{Borel subalgebra}, \index{Borel subalgebra|bb}
\begin{equation}
  \mf{b}_3=\mf{h}_3\oplus \mf{n}_3\subset \mf{u}_3.
\end{equation}
In that gauge we have 
\begin{equation}
  \mc{V}(x)=
  \Exp \left[ \phi(x) \cdot \mf{h}_{3}\right]\,
  \Exp \left[\chi(x)\cdot \mf{n}_3\right],
  \label{elementofsplitcosetU3}
\end{equation}
where $\phi$ and $\chi$ are
(sets of) coordinates on the coset space \index{coset space} 
$\mc{U}_3/\mc{K}(\mc{U}_3)$. A Lagrangian based on this coset will
then take the generic form (see Section~\ref{section:sigmamodels}) 
\begin{equation}
  \mc{L}_{\mc{U}_3/\mc{K}(\mc{U}_3)}=
  \!\!\sum_{i=1}^{\dim \mf{h}_3}\!\!
  \pa_x \phi^{(i)}(x) \pa_x \phi^{(i)}(x)+
  \!\!\sum_{\al\in \Delta_+} e^{2\al(\phi)}\!\!
  \left[\pa_x\chi^{(\al)}(x)+\cdots\right]
  \left[\pa_x\chi^{(\al)}(x)+\cdots\right],
  \label{LagrangiansplitU3}
\end{equation}
where $x$ denotes coordinates in
spacetime and the ``ellipses'' denote correction terms that are of
no relevance for our present purposes. We refer to the fields
$\{\phi\}$ collectively as \emph{dilatons} and the fields
$\{\chi\}$ as \emph{axions}. There is one axion field
$\chi^{(\al)}$ for each positive root $\al\in\Delta_+$ and one
dilaton field $\phi^{(i)}$ for each Cartan generator
$\al_i^{\vee}\in\mf{h}_3$.

The Lagrangian~(\ref{LagrangiansplitU3}) coupled to the pure
three-dimensional Einstein--Hilbert term is the key to
understanding the appearance of the Lorentzian Coxeter group
$\mf{u}_3^{++}$ in the BKL-limit.


\subsubsection{Starting at the bottom -- Overextensions of
  finite-dimensional Lie algebras}
\label{section:3Dbilliard}

To make the point explicit, we will again limit our analysis to
the example of eleven-dimensional supergravity. Our starting point
is then the Lagrangian for this theory compactified on an
8-torus, $T^{8}$, to $D=2+1$ spacetime dimensions (after all form
fields have been dualized into scalars),
\begin{equation}
  \mc{L}^{\mathrm{SUGRA}_{11}}_{(3)}=R_{(3)}\star \mathbf{1}-
  \sum_{i=1}^{8}\star d\hvp^{(i)} \wedge d\hvp^{(i)}-
  \f{1}{2}\sum_{q=1}^{120}e^{2\al_{q}(\hvp)}\star
  (d\hat{\chi}^{(q)}+\cdots )\wedge (d \hat{\chi}^{(q)}+\cdots).
  \label{11Sugra3D}
\end{equation}
The second two terms in this Lagrangian correspond to the coset model
$\mc{E}_{8(8)}/(\mathrm{Spin}(16)/\mathbb{Z}_2)$, where $\mc{E}_{8(8)}$
denotes the group obtained by exponentiation of the split form
$E_{8(8)}$ of the complex Lie algebra $E_{8}$ and
$\mathrm{Spin}(16)/\mathbb{Z}_2$ is the maximal compact subgroup of
$\mc{E}_{8(8)}$~\cite{Cremmer:1978ds,Marcus:1983hb,
  Cremmer:1997ct}. The 8 dilatons $\hvp$ and the 120 axions
$\chi^{(q)}$ are coordinates on the coset space\epubtkFootnote{This
  structure of $\mc{E}_{8(8)}$ can be understood as follows. The
  248-dimensional Lie algebra $E_{8(8)}$ can be represented as
  ${\mathfrak{so}(16)}\oplus {\mathcal{S}_{16}}$ (direct sum of vector
  spaces), where ${\mathcal{S}_{16}}$ constitutes a 128-dimensional
  representation space of the group $\mathrm{Spin}(16)$, that transforms like
  Majorana--Weyl spinors. Using Dirac matrices $\Gamma_{a\
    \mu}^{\,\nu}$, the commutation relations read:\\[3mm]
\hphantom{$\displaystyle\left[ M_{ab},Q_{\mu}\right]\left[ Q_{\mu},Q_{\nu}\right]$}$\displaystyle\left[ M_{ab},M_{cd}\right] = \delta_{ac}M_{bd}+\delta_{bd}M_{ac}-\ad M_{bc}-\delta_{bc}M_{ad},$\\[3mm]
\hphantom{$\displaystyle\left[ M_{ab},M_{cd}\right]\left[ Q_{\mu},Q_{\nu}\right]$}$\displaystyle\left[ M_{ab},Q_{\mu}\right] = \frac 12 \Gamma_{[ab]\mu}^{\,\nu} Q_{\nu},$\\[3mm]
\hphantom{$\displaystyle\left[ M_{ab},M_{cd}\right]\left[ M_{ab},Q_{\mu}\right]$}$\displaystyle\left[ Q_{\mu},Q_{\nu}\right] = \Gamma^{[ab]}_{\mu\nu}M_{ab}.$\\[3mm]
  %
  %
  For more information about $E_{8(8)}$ see~\cite{Marcus:1983hb}, and
  for a general discussion of real forms of Lie algebras see
  Section~\ref{section:FiniteRealLieAlgebras}.}. \index{coset space}
Furthermore, the $\al_{q}(\hvp)$ are linear forms on the elements of the Cartan
subalgebra $h=\hvp^{i}\al^{\vee}_{i}$ and they correspond to the
positive roots of $E_{8(8)}$ \epubtkFootnote{In the following we write
  simply $E_8$ and it is understood that we refer to the split real
  form $E_{8(8)}$.}. As before, we do not write explicitly the
corrections to the curvatures $d\hat{\chi}$ that appear in the
compactification process. The entire set of positive roots can be
obtained by taking linear combinations of the seven simple roots of
$\mf{sl}(8,\mathbb{R})$ (we omit the ``hatted'' notation on the
roots since there is no longer any risk of confusion),
\begin{equation}
  \begin{array}{rcl@{\qquad}rcl}
    \al_{1}(\hvp)&=& \displaystyle
    \f{1}{\sqrt{2}}
    \left(\f{\sqrt{7}}{2}\hvp_2-\f{3}{2}\hvp_1\right), &
    \al_{2}(\hvp)&=& \displaystyle
    \f{1}{\sqrt{2}}
    \left(\f{2\sqrt{3}}{\sqrt{7}}\hvp_3-\f{4}{\sqrt{7}}\hvp_2\right),
    \\ [1.5 em]
    \al_{3}(\hvp)&=& \displaystyle
    \f{1}{\sqrt{2}}
    \left(\f{\sqrt{5}}{\sqrt{3}}\hvp_4-\f{\sqrt{7}}{\sqrt{3}}\hvp_3\right), &
    \al_{4}(\hvp)&=& \displaystyle
    \f{1}{\sqrt{2}}
    \left(\f{2\sqrt{2}}{\sqrt{5}}\hvp_5-\f{2\sqrt{3}}{\sqrt{5}}\hvp_4\right),
    \\ [1.5 em]
    \al_{5}(\hvp)&=& \displaystyle
    \f{1}{\sqrt{2}}
    \left(\f{\sqrt{3}}{\sqrt{2}}\hvp_6-\f{\sqrt{5}}{\sqrt{2}}\hvp_5 \right), &
    \al_{6}(\hvp)&=& \displaystyle
    \f{1}{\sqrt{2}}
    \left(\f{2}{\sqrt{3}}\hvp_6-\f{2\sqrt{2}}{\sqrt{3}}\hvp_5\right),
    \\ [1.5 em]
    \al_{7}(\hvp)&=& \displaystyle
    \f{1}{\sqrt{2}}\left(\hvp_8-\sqrt{3}\hvp_7\right),
  \end{array}
  \label{simplerootsA7} 
\end{equation}
and the exceptional root 
\begin{equation}
  \al_{10}(\hvp)=\f{1}{\sqrt{2}}
  \left(\hvp_1+\f{3}{\sqrt{7}}\hvp_2+\f{2\sqrt{3}}{\sqrt{7}}\hvp_3\right).
  \label{exceptionalrootE8}
\end{equation}
These correspond exactly to the
root vectors $\vec{b}_{i,i+1}$ and $\vec{a}_{123}$ as they appear
in the analysis of~\cite{Cremmer:1997ct}, except for the
additional factor of $\f{1}{\sqrt{2}}$ needed to compensate for
the fact that the aforementioned reference has an additional
factor of $2$ in the Killing form. Hence, using the Euclidean
metric $\delta_{ij}$ $(i,j=1,\cdots, 8)$ one may check that the
roots defined above indeed reproduce the Cartan matrix of $E_{8}$.

Next, we want to determine the billiard structure for this
Lagrangian. As was briefly mentioned before, in the reduction from
eleven to three dimensions all the non-gravity walls associated to the
eleven-dimensional 3-form $A^{(3)}$ have been transformed, in the
same spirit as for the example given above, into electric and
magnetic walls of the axionic scalars $\hat{\chi}$. Since the
terms involving the electric fields $\pa_{t}\hat{\chi}^{(i)}$
possess no spatial indices, the corresponding wall forms do not
contain any of the remaining scale factors $\hb^9, \hb^{10}$, and
are simply linear forms on the dilatons only. In fact the dominant
electric wall forms are just the simple roots of $E_{8}$, 
\begin{equation}
  \begin{array}{rcl}
    \hat{e}_{a}^{\hat{\chi}}(\hvp)&=&\al_{a}(\hvp)
    \qquad
    (a=1,\cdots, 7),
    \\ [0.25 em]
    \hat{e}_{10}^{\hat{\chi}}(\hvp)&=&\al_{10}(\hvp).
  \end{array}
  \label{dominantwall3D} 
\end{equation}
The magnetic wall forms naturally
come with one factor of $\hb$ since the magnetic field strength
$\pa_i\hat{\chi}$ carries one spatial index. The dominant magnetic
wall form is then given by 
\begin{equation}
  \hat{m}_{9}^{\hat{\chi}}(\hb, \hvp)=\hb^9-\theta(\hvp), 
  \label{dominantmagneticwall3D}
\end{equation}
where $\theta(\hvp)$ denotes the highest root of $E_{8}$ which
takes the following form in terms of the simple roots,
\begin{equation}
  \theta=
  2\al_{1}+4\al_{2}+6\al_{3}+5\al_{4}+4\al_{5}+3\al_{6}+2\al_{7}+3\al_{10}=
  \sqrt{2} \, \hvp_8. 
  \label{highestrootE8}
\end{equation}
Since we are in three dimensions there is no curvature wall and hence
the only wall associated to the Einstein--Hilbert term is the symmetry
wall
\begin{equation}
  \hat{s}_{9}=\hb^{10}-\hb^9, 
  \label{3Dsymmetrywall}
\end{equation}
coming from the three-dimensional metric $\hat{\g}_{\mu\nu}$
($\mu,\nu=0, 9, 10$). We have thus found all the dominant wall
forms in terms of the lower-dimensional variables.

The structure of the corresponding Lorentzian Kac--Moody algebra is
now easy to establish in view of our discussion of overextensions \index{overextension}
in Section~\ref{section:Overextensions}. The relevant walls listed above
are the simple roots of the (untwisted) overextension $E_8^{++}$.
Indeed, the relevant electric roots are the simple roots of $E_8$,
the magnetic root of Equation~(\ref{dominantmagneticwall3D}) provides the
affine extension, while the gravitational root of
Equation~(\ref{3Dsymmetrywall}) is the overextended root.

What we have found here in the case of eleven-dimensional supergravity
also holds for the other theories with U-duality algebra
$\mf{u}_3$ in 3 dimensions when $\mf{u}_3$ is a split real form.
The Coxeter group and the corresponding Kac--Moody algebra are
given by the untwisted overextension $\mf{u}_3^{++}$. This
overextension emerges as
follows~\cite{InvarianceUnderCompactification}:

\begin{itemize}
\item The dominant electric wall forms
  $\hat{e}^{\hat{\chi}}(\hvp)$ for the supergravity theory in
  question are in one-to-one correspondence with the simple roots
  of the associated U-duality algebra $\mf{u}_3$.
\item Adding the dominant magnetic wall form
  $\hat{m}^{\hat{\chi}}(\hb,\hvp)=\hb^9-\theta(\hvp)$ corresponds
  to an \emph{affine extension} $\mf{u}_3^{+}$ of $\mf{u}_3$.
\item Finally, completing the set of dominant wall forms with the
  symmetry wall $\hat{s}_{9}(\hb)=\hb^{10}-\hb^{9}$, which is the
  only gravitational wall form existing in three dimensions, is
  equivalent to an \emph{overextension} $\mf{u}_3^{++}$ of
  $\mf{u}_3$.
\end{itemize}

Thus we see that the appearance of overextended algebras in the
BKL-analysis of supergravity theories is a generic phenomenon
closely linked to hidden symmetries.


\subsection{Models associated with split real forms}
\label{section:classificationsplit}

In this section we give a complete list of all theories whose billiard
description can be given in terms of a Kac--Moody algebra that is the
untwisted overextension of a split real form of the associated
U-duality algebra (see Table~\ref{table:split}). These are precisely
the \emph{maximally oxidized} theories introduced
in~\cite{Breitenlohner:1987dg} and further examined
in~\cite{Cremmer:1999du}. These theories are completely classified by
their global symmetry groups $\mc{U}_3$ arising in three
dimensions. For the string-related theories the group $\mc{U}_3$ is
the (classical version of) the U-duality symmetry obtained by
combining the S- and T-dualities in three
dimensions~\cite{Obers:1998fb}. Thereof the notation $\mc{U}_3$ for
the global symmetry group in three dimensions. We extend the
classification to the non-split case in
Section~\ref{section:KMBilliardsII}.

Let us also note here that, as shown in~\cite{deBuyl:2003ub},
the billiard analysis sheds light on the problem of oxidation,
i.e., the problem of finding the maximum spacetime dimension in
which a theory with a given duality group in three dimensions can
be reformulated. More on this question can be found
in~\cite{Keurentjes:2002xc, Keurentjes:2002rc}.

\begin{landscape}
\begin{table}
  \caption[Complete list of theories that exhibit extended coset
  symmetries of split real Lie algebras upon compactification to three
  spacetime dimensions.]{We present here the complete list of theories
  that exhibit extended coset symmetries of split real Lie algebras
  upon compactification to three spacetime dimensions. In the leftmost
  column we give the coset space \index{coset space} which is relevant
  in each case. We also list the Kac--Moody algebras that underlie the
  gravitational dynamics in the BKL-limit. These appear as
  overextensions of the finite Lie algebras found in three
  dimensions. Finally we indicate which of these theories are related
  to string/M-theory.}
  \renewcommand{\arraystretch}{1.7}
  \vspace{0.5 em}
  \centering
  \begin{tabular}{l|>{\raggedright}p{114mm}|p{32mm}|p{28mm}}
    \toprule
    $\mc{U}_3/\mc{K}(\mc{U}_3)$ &
    Lagrangian in maximal dimension &
    Kac--Moody algebra &
    String/M-theory \\
    \midrule 
        $\f{SL(n+1, \mbb{R})}{SO(n+1)}$
      &
        ${\mc{L}}_{n+3} = R\star \mathbf{1}$
      &
        $AE_{n+2}\equiv A_n^{++}$
      &
        No
    \\
        $\f{SO(n,n+1)}{SO(n)\times SO(n+1)}$
      &
        ${\mc{L}}_{n+2} = R\star \mathbf{1}-\mbox{}$\hspace{0em}%
        $\star d\phi\wedge d\phi-\mbox{}$\hspace{0em}%
        $\f{1}{2} e^{2\f{\sqrt{2}}{\sqrt{n}}\phi}\star G^{(3)}\wedge G^{(3)}-\mbox{}$\hspace{0em}%
        $\f{1}{2} e^{\f{2}{\sqrt{n}}\phi} \star F^{(2)}\wedge F^{(2)}$,\hspace{2em}%
        $G^{(3)}=dB^{(2)}+\mbox{}$\hspace{0em}%
        $\f{1}{2}A^{(1)}\wedge A^{(1)}$,\hspace{2em}%
        $F^{(2)}=dA^{(1)}$
      &
        $BE_{n+2}\equiv B_n^{++}$
      &
        No
    \\
        $ \f{Sp(n)}{U(n)}$
      &
        ${\mc{L}}_4 = R\star \mathbf{1}-\mbox{}$\hspace{0em}%
        $\star d\vec\phi\wedge d\vec\phi-\mbox{}$\hspace{0em}%
        $\f{1}{2} \sum_{\alpha} e^{2 \vec \sigma_\alpha.\vec\phi} \star (d\chi^\alpha + \cdots)\wedge (d\chi^\alpha+ \cdots)-\mbox{}$\hspace{0em}%
        $\f{1}{2}\sum_{a=1}^{n-1} e^{\vec e_a.\vec\phi\sqrt{2}}\star dA^a_{(1)}\wedge dA^a_{(1)}$
      &
        $CE_{n+2}\equiv C_n^{++}$
      &
        No
    \\
        $\f{SO(n,n)}{SO(n)\times SO(n)}$
      &
        ${\mc{L}}_{n+2} = R\star \mathbf{1} -\mbox{}$\hspace{0em}%
        $\star d\phi\wedge d\phi -\mbox{}$\hspace{0em}%
        $\f{1}{2} \,e^{\f{4}{\sqrt{n}}\phi}\star dB^{(2)}\wedge dB^{(2)}$
      &
        $DE_{n+2} \equiv D_n^{++}$
      &
        \footnotesize type I ($n=8$) / bosonic string ($n=24$)
    \\
        $ \f{G_{2(2)}}{SU(8)}$
      &
        ${\mc{L}}_5 = R\star \mathbf{1} -\mbox{}$\hspace{0em}%
        $\f{1}{2}\star F^{(2)}\wedge F^{(2)} +\mbox{}$\hspace{0em}%
        $\f{1}{3\sqrt{3}} F^{(2)}\wedge F^{(2)}\wedge A^{(1)}$,\hspace{0em}%
        $F^{(2)}=dA^{(1)}$
      &
        $G_2^{++}$
      &
        No
    \\
        $\f{F_{4(4)}}{Sp(3)\times SU(3)}$
      &
        ${\mc{L}}_6 = R\star \mathbf{1} -\mbox{}$\hspace{0em}%
        $\star d\phi\wedge d\phi -\mbox{}$\hspace{0em}%
        $\f{1}{2} e^{2\phi}\star d\chi\wedge d\chi -\mbox{}$\hspace{0em}%
        $\f{1}{2} e^{-2\phi} \star H^{(3)}\wedge H^{(3)} -\mbox{}$\hspace{0em}%
        $\f{1}{2} \star G^{(3)} \wedge$ $G^{(3)} -\mbox{}$\hspace{0em}%
        $\f{1}{2} e^{\phi} \star F^{+}_{(2)}\wedge F^{+}_{(2)} -\mbox{}$\hspace{0em}%
        $\f{1}{2} e^{-\phi} \star F^{-}_{(2)}\wedge F^{-}_{(2)} -\mbox{}$\hspace{0em}%
        $\f{1}{\sqrt{2}}\chi \,H^{(3)}\wedge G^{(3)} -\mbox{}$\hspace{0em}%
        $\f{1}{2} A^{+}_{(1)}\wedge F^{+}_{(2)}\wedge$ $ H^{(3)} -\mbox{}$\hspace{0em}%
        $\f{1}{2} A^{+}_{(1)}\wedge F^{-}_{(2)}\wedge G^{(3)}$,\hspace{2em}%
        $F^{+}_{(2)}=dA^{+}_{(1)}+\mbox{}$\hspace{0em}%
        $\f{1}{\sqrt{2}}\chi dA^{-}_{(1)}$,\hspace{2em}%
        $F^{-}_{(2)}=dA^{-}_{(1)}$,\hspace{2em}%
        $H^{(3)}=dB^{(2)}+\f{1}{2}A^{-}_{(1)}\wedge dA^{-}_{(1)}$,\hspace{2em}%
        $G^{(3)}=dC^{(2)}-\mbox{}$\hspace{0em}%
        $\f{1}{\sqrt{2}}\chi H^{(3)} -\mbox{}$\hspace{0em}%
        $\f{1}{2}A^{+}_{(1)}\wedge dA^{-}_{(1)}$
      &
        $F_4^{++}$
      &
        No
    \\
        $ \f{E_{6(6)}}{Sp(4)/\mbb{Z}_2}$
      &
        ${\mc{L}}_8 = R\star \mathbf{1} -\mbox{}$\hspace{0em}%
        $\star d\phi\wedge d\phi -\mbox{}$\hspace{0em}%
        $\f{1}{2} e^{2\sqrt{2}\phi}\star d\chi\wedge d\chi -\mbox{}$\hspace{0em}%
        $\f{1}{2}e^{-\sqrt{2}\phi}\star G^{(4)}\wedge G^{(4)} +\mbox{}$\hspace{0em}%
        $\chi\,G^{(4)}\wedge G^{(4)}$,\hspace{2em}%
        $G^{(4)}=dC^{(3)}$
      &
        $E_{6}^{++}$
      &
        No
    \\
        $ \f{E_{7(7)}}{SU(8)/\mbb{Z}_2}$
      &
        $ {\mc{L}}_9 =  R\star \mathbf{1} -\mbox{}$\hspace{0em}%
        $\star d\phi\wedge d\phi -\mbox{}$\hspace{0em}%
        $\frac{1}{2} e^{\frac{2\sqrt{2}}{\sqrt{7}}\phi}\star dC^{(3)}\wedge dC^{(3)} -\mbox{}$\hspace{0em}%
        $\frac{1}{2} e^{-\frac{4\sqrt{2}}{\sqrt{7}}\phi}\star dA^{(1)}\wedge dA^{(1)} -\mbox{}$\hspace{0em}%
        $\frac{1}{2} dC^{(3)}\wedge dC^{(3)}\wedge A^{(1)}$
      &
        $E_7^{++}$
      &
        No
    \\
        $ \f{E_{8(8)}}{\mathrm{Spin}(16)/\mbb{Z}_2} $
      &
        $ {\mc{L}}_{11} = R\star \mathbf{1} -\mbox{}$\hspace{0em}%
        $\frac{1}{2}\star dC^{(3)} \wedge dC^{(3)}-\mbox{}$\hspace{0em}%
        $\frac{1}{6} dC^{(3)}\wedge dC^{(3)}\wedge C^{(3)} $
      &
        $E_{10}\equiv E_8^{++}$
      &
        \footnotesize M-theory, type IIA and type~IIB string theory
    \\
    \bottomrule
  \end{tabular}
  \label{table:split}
  \renewcommand{\arraystretch}{1.0}
\end{table}
\end{landscape}

\newpage


\section{Finite-Dimensional Real Lie Algebras}
\label{section:FiniteRealLieAlgebras}
\setcounter{equation}{0}

In this section we explain the basic theory of real forms of
finite-dimensional Lie algebras. This material is somewhat technical
and may therefore be skipped at a first reading. The theory of real
forms of Lie algebras is required for a complete understanding of
Section~\ref{section:KMBilliardsII}, which deals with the general case
of Kac--Moody billiards for non-split real forms. However, for the
benefit of the reader who wishes to proceed directly to the physical
applications, we present a brief summary of the main points in the
beginning of Section~\ref{section:KMBilliardsII}.

Our intention with the following presentation is to provide an
accessible reference on the subject, directed towards physicists. We
therefore consider this section to be somewhat of an entity of its
own, which can be read independently of the rest of the
paper. Consequently, we introduce Lie algebras in a rather different
manner compared to the presentation of Kac--Moody algebras in
Section~\ref{section:KacMoody}, emphasizing here more involved
features of the general structure theory of real Lie algebras rather
than relying entirely on the Chevalley--Serre basis and its
properties. In the subsequent section, the reader will then see these
two approaches merged, and used simultaneously to describe the
billiard structure of theories whose U-duality algebras in three
dimensions are given by arbitrary real forms.

We have adopted a rather detailed and explicit presentation. We
do not provide all proofs, however, referring the reader
to~\cite{Helgason, Knapp, Loos, Helminck} for more information
(including definitions of basic Lie algebra theory concepts).

There are two main approaches to the classification of real forms
of finite-dimensional Lie algebras. One focuses on the maximal
compact Cartan subalgebra and leads to Vogan diagrams. 
\index{Vogan diagram} The other focuses on the maximal noncompact
Cartan subalgebra and leads to Tits--Satake diagrams. 
\index{Tits--Satake diagram} It is this second approach that is of
direct use in the billiard analysis. However, we have chosen to
present here both approaches as they mutually enlighten each
other.


\subsection{Definitions}

Lie algebras are usually, in a first step at least,
considered as complex, i.e., as complex vector spaces, structured by
an antisymmetric internal bilinear product, the Lie bracket,
obeying the Jacobi identity. If $\{T_{\al}\}$ denotes a basis of such
a complex Lie algebra $\cG$ of dimension $n$ (over $\CC$), we may
also consider $\cG$ as a real vector space of double dimension
$2\,n$ (over $\RR$), a basis being given by $\{T_{\al},\, iT_{\al}\}$.
Conversely, if $\GO$ is a real Lie algebra, by extending the field
of scalars from $\RR$ to $\CC$, we obtain the complexification of
$\GO$, denoted by $\GC$, defined as: 
\begin{equation}
 \GC = \GO\otimes_{\RR}\CC.
\end{equation}
Note that $(\GC)^\RR=\GO\op i\GO$ and $ \dim_{\RR}(\GC)^\RR=
2 \, \dim_{\RR }(\GO)$. When a complex Lie algebra
$\cG$, considered as a real algebra, has a decomposition
\begin{equation}
 \GR = \GO\op i\,\GO,
\label{realform}
\end{equation}
with $\GO$ being a real Lie algebra, we say that $\GO$ is a real form of
the complex Lie algebra $\cG$. In other words, a real form of a
complex algebra exists if and only if we may choose a basis of the
complex algebra such that all the structure constants become real.
Note that while $\GR$ is a real space, multiplication by a complex
number is well defined on it since $\GO\op i\GO =
\GO\otimes_{\RR}\CC$. As we easily see from Equation~(\ref{realform}),
\begin{equation}
  \CC\times\GR \quad \rightarrow \quad \GR:
  (a+i\,b, X_{0}+i\, Y_{0}) \mapsto (a X_{0}-b Y_{0})+ i\,(a Y_{0} + b
  X_{0}),
  \label{Cmult},
\end{equation}
where $a, b\in \mbb{R}$ and $X_0, Y_0\in \mf{g}_0$.

The Killing form is defined by 
\begin{equation}
  B(X,Y) = \Tr(\ad X \ad Y).
\end{equation}
The Killing forms on $\GR$ and $\GC$ or $\GO$ are
related as follows. If we split an arbitrary generator $ Z$ of
$\cG$ according to Equation~(\ref{realform}) as
$ Z= X_{0}+i\, Y_{0}$, we may write: 
\begin{equation}
  B_{\GR}(Z,\,Z^\prime)=2 \Real B_{\GC}(Z,\,Z^\prime) =
  2 \left(B_{\GO}(X_0,\,X^\prime_0) - B_{\GO}(Y_0,\,Y^\prime_0)\right).
\end{equation}
Indeed, if $\ad_{\cG}{ Z}$ is a complex $n \times n$ matrix,
$\ad_{\GR}( X_{0}+i\, Y_{0})$ is a real $2n\times 2n$ matrix:
\begin{equation}
  \ad_{\GR}(X_{0}+i\, Y_{0})=
  \left(
    \begin{array}{@{}r@{\quad}r@{}}
      \ad_{\GO} X_{0} & -\ad_{\GO} Y_{0} \\
      \ad_{\GO} Y_{0} & \ad_{\GO} X_{0}
    \end{array}
  \right).
\end{equation}


\subsection[A preliminary example: $\mf{sl}(2, \CC)$]%
           {A preliminary example: \boldmath $\mf{sl}(2, \CC)$}
\label{section:SL2Example}

Before we proceed to develop the general theory of real forms, we
shall in this section discuss in detail some properties of the real
forms of $A_1=\mathfrak{sl}(2,\CC)$. This is a nice example, which
exhibits many properties that turn out not to be specific just to the
case at hand, but are, in fact, valid also in the general framework of
semi-simple Lie algebras. The main purpose of subsequent sections will
then be to show how to extend properties that are immediate in the
case of $\mathfrak{sl}(2,\CC)$, to general semi-simple Lie algebras.


\subsubsection[Real forms of $\mathfrak{sl}(2,\CC)$]%
              {Real forms of \boldmath $\mathfrak{sl}(2,\CC)$}

The complex Lie algebra $\mathfrak{sl}(2,\CC)$ can be represented
as the space of complex linear combinations of the three matrices

\begin{equation}
  \label{sl2R}
  h=\left(
    \begin{array}{@{}r@{\quad}r@{}}
      1 & 0 \\
      0 & -1
    \end{array}
  \right),
  \qquad
  e=\left(
    \begin{array}{@{}r@{\quad}r@{}}
      0 & 1 \\
      0 & 0
    \end{array}
  \right),
  \qquad
  f=\left(
    \begin{array}{@{}r@{\quad}r@{}}
      0 & 0 \\
      1 & 0 \end{array}
  \right)
\end{equation}
which satisfy the well known commutation relations 
\begin{equation}
  \label{sl2Rcr}
  \Lb{{h}}{e}=2\,e,
  \qquad
  \Lb{{h}}{{f}}=-2\,{f},
  \qquad
  \Lb{e}{{f}}= {h}.
\end{equation}
A crucial property of these commutation relations is that the
structure constants defined by the brackets are all real. Thus by
restricting the scalars in the linear combinations from the
complex to the real numbers, we still obtain closure for the Lie
bracket on real combinations of $ h, e$ and $f$, defining thereby
a real form of the complex Lie algebra $\mathfrak{sl}(2,\CC)$: the
real Lie algebra $\mathfrak{sl}(2,\RR)$ \epubtkFootnote{Actually, the
  structure constants are integers and thus allows for defining the
  arithmetic subgroup $SL(2,\ZZ)\subset SL(2, \mbb{R})$.}. As we have
indicated above, this real form of $\mathfrak{sl}(2,\CC)$ is called
the ``split real form''.

Another choice of $\mathfrak{sl}(2,\CC)$ generators that,
similarly, leads to a real Lie algebra consists in taking $i$
times the Pauli matrices $\sigma^x$, $\sigma^y$, $\sigma^z$, i.e.,
\begin{equation}
  \label{su2R}
  {\tau}^x=i(e+{f})=\left(
    \begin{array}{@{}r@{\quad}r@{}}
      0 & i \\
      i & 0
    \end{array}
  \right),
  \qquad
  {\tau}^y=(e-{f})=\left(
    \begin{array}{@{}r@{\quad}r@{}}
      0 & 1 \\
      -1 & 0
    \end{array}\right),
  \qquad
  {\tau}^z=i{h}=\left(
    \begin{array}{@{}r@{\quad}r@{}}
      i & 0 \\
      0 & -i
    \end{array}
  \right).
\end{equation}
The real linear combinations of these matrices form the
familiar $\mathfrak{su}(2)$ Lie algebra (a real Lie algebra, even
if some of the matrices using to represent it are complex). This
real Lie algebra is non-isomorphic (as a real algebra) to
$\mathfrak{sl}(2,\RR)$ as there is no real change of basis that
maps $\{h, e, f\}$ on a basis with the $\mathfrak{su}(2)$
commutation relations. Of course, the two algebras are isomorphic
over the complex numbers.


\subsubsection{Cartan subalgebras}
\index{Cartan subalgebra}

Let $\mathfrak{h}$ be a subalgebra of $\mathfrak{sl}(2,\RR)$. We
say that $\mf{h}$ is a \emph{Cartan subalgebra} of $\mathfrak{sl}(2,\RR)$
if it is a Cartan subalgebra of $\mathfrak{sl}(2,\CC)$ when the
real numbers are replaced by the complex numbers. Two Cartan
subalgebras $\mathfrak{h}_1$ and $\mathfrak{h}_2$ of
$\mathfrak{sl}(2,\RR)$ are said to be equivalent (as Cartan
subalgebras of $\mathfrak{sl}(2,\RR)$) if there is an automorphism
$a$ of $\mathfrak{sl}(2,\RR)$ such that $a(\mathfrak{h}_1) =
\mathfrak{h}_2$.

The subspace $\RR h$ constitutes clearly a Cartan subalgebra of
$\mathfrak{sl}(2,\RR)$. The adjoint action of $h$ is diagonal in
the basis $\{e,\,f,\,h\}$ and can be represented by the matrix
\begin{equation}
  \left(
    \begin{array}{@{}r@{\quad}r@{\quad}r@{}}
      2 & 0 & 0 \\
      0 & -2 & 0 \\
      0 & 0 & 0
    \end{array}
  \right).
\end{equation}
Another Cartan subalgebra of $\mathfrak{sl}(2,\RR)$ is given
by $\RR(e - {f})\equiv\RR{{\tau}^y}$, whose adjoint action with
respect to the same basis is represented by the matrix
\begin{equation}
  \left(
    \begin{array}{@{}r@{\quad}r@{\quad}r@{}}
      0 & 1 & 1 \\
      -2 & 0 & 0 \\
      -2 & 0 & 0
    \end{array}
  \right).
\end{equation}
Contrary to the matrix representing $\ad_h$, in addition to 0 this
matrix has two imaginary eigenvalues: $\pm 2\,i$. Thus, there can be
no automorphism $a$ of $\mathfrak{sl}(2,\RR)$ such that $\tau^y =
\lambda a(h)$, $\lambda \in \mathbb{R}$ since $\ad_{a(h)}$ has the
same eigenvalues as $\ad_h$, implying that the eigenvalues of $\lambda
\, \ad_{a(h)}$ are necessarily real ($\lambda \in \mathbb{R}$).

Consequently, even though they are equivalent over the complex
numbers since there is an automorphism in $SL(2,\CC)$ that
connects the complex Cartan subalgebras $\CC\,h$ and $\CC\,{
\tau}^y$, we obtain
\begin{equation}
  {\tau}^y=i \, \Ad
  \left( \Exp \left[i\frac \pi 4 (e + {f}) \right] \right){h},
  \qquad
  {h}=i\, \Ad \left( \Exp \left[\frac \pi 4 {\tau}^x \right] \right){\tau}^y.
\end{equation}
The real Cartan subalgebras generated by
$h$ and ${\tau}^y$ are non-isomorphic over the real numbers.


\subsubsection{The Killing form}

The Killing form of $SL(2,\RR)$ reads explicitly 
\begin{equation}
  B=\left(
    \begin{array}{@{}r@{\quad}r@{\quad}r@{}}
      0 & 4 & 0 \\
      4 & 0 & 0 \\
      0 & 0 & 8
    \end{array}
  \right)
\end{equation}
in the basis $\{e,\,f,\,h\}$. The Cartan subalgebra $\RR h$ is spacelike
while the Cartan subalgebra $\RR \tau^y$ is timelike. This is
another way to see that these are inequivalent since the
automorphisms of $\mathfrak{sl}(2,\RR)$ preserve the Killing form.
The group $\mathrm{Aut}[\mf{sl}(2, \mbb{R})]$ of automorphisms of
$\mathfrak{sl}(2,\RR)$ is $ SO(2,1)$,
while the subgroup $\mathrm{Int}[\mf{sl}(2, \mbb{R})]\subset
\mathrm{Aut}[\mf{sl}(2, \mbb{R})]$ of inner automorphisms is the
connected component $ SO(2,1)^+$ of $SO(2,1)$. All spacelike directions are
equivalent, as are all timelike directions, which shows that all
the Cartan subalgebras of $\mathfrak{sl}(2,\RR)$ can be obtained by
acting on these two inequivalent particular ones by
$\mathrm{Int}[\mathfrak{sl}(2,\RR)]$, i.e., the adjoint action of the
group $SL(2,\RR)$. The lightlike directions do not define Cartan
subalgebras because the adjoint action of a lighlike vector is
non-diagonalizable. In particular $\RR e$ and $\RR f$ are not Cartan
subalgebras even though they are Abelian.

By exponentiation of the generators $h$ and $\tau^y$, we obtain
two subgroups, denoted ${ \mc{A}}$ and ${\mc{K}}$: 
\begin{eqnarray}
  \mc{A}&=&\left\{ \Exp [t\,{h}] = \left(
      \begin{array}{@{}r@{\quad}r@{}}
        e^t & 0 \\
        0 & e^{-t}
      \end{array}
    \right) \vert t\in \RR\right\}\simeq\RR,
  \label{ncg}
  \\
  \mc{K}&=&\left\{ \Exp [t\,{\tau}^y] = \left(
      \begin{array}{@{}r@{\quad}r@{}}
        \cos(t) & \sin(t) \\
        -\sin(t)&\cos(t)
      \end{array}
    \right)\vert t\in [0,2\pi[\,\right\}\simeq \RR/\ZZ.
  \label{cg}
\end{eqnarray}
The subgroup defined by Equation~(\ref{ncg}) is noncompact, the one defined
by Equation~(\ref{cg}) is compact; consequently the
generator $h$ is also said to be noncompact while ${\tau}^y $ is
called compact.


\subsubsection[The compact real form $\mathfrak{su}(2)$]%
              {The compact real form \boldmath $\mathfrak{su}(2)$}

The Killing metric on the group $\mathfrak{su}(2)$ is negative
definite. In the basis $\{\tau^x, \tau^y, \tau^z\}$, it reads 
\begin{equation}
  B =\left(
    \begin{array}{@{}r@{\quad}r@{\quad}r@{}}
      -8 & 0 & 0 \\
      0 & -8 & 0 \\
      0 & 0 & -8
    \end{array}
  \right).
\end{equation}

The corresponding group obtained by exponentiation is $SU(2)$,
which is isomorphic to the 3-sphere and which is accordingly
compact. All directions in $\mathfrak{su}(2)$ are equivalent and
hence, all Cartan subalgebras are $SU(2)$ conjugate to
$\RR{\tau}^y$. Any generator provides by exponentiation a group
isomorphic to $\RR/\ZZ$ and is thus compact.

Accordingly, while $\mathfrak{sl}(2,\RR)$ admits both compact and
noncompact Cartan subalgebras, the Cartan subalgebras of
$\mathfrak{su}(2)$ are all compact. The real algebra
$\mathfrak{su}(2)$ is called the compact real form of
$\mathfrak{sl}(2,\CC)$. One often denotes the real forms by their
signature. Adopting Cartan's notation $A_1$ for
$\mathfrak{sl}(2,\CC)$, one has $\mathfrak{sl}(2,\RR) \equiv A_{1
\, (1)}$ and $\mathfrak{su}(2)\equiv A_{1 \, (-3)}$. We shall verify before that there are no other real
forms of $\mathfrak{sl}(2,\CC)$.


\subsubsection[$\mathfrak{su}(2)$ and $\mathfrak{sl}(2,\RR)$ compared
  and contrasted -- The Cartan involution]%
              {\boldmath $\mathfrak{su}(2)$ and $\mathfrak{sl}(2,\RR)$ compared
  and contrasted -- The Cartan involution}

Within $\mathfrak{sl}(2,\CC)$, one may express the basis vectors
of one of the real subalgebras $\mathfrak{su}(2)$ or
$\mathfrak{sl}(2,\RR)$ in terms of those of the other. We obtain,
using the notations $ t=( e-f)$ and $ x=(e+f)$: 
\begin{equation}
  \begin{array}{rcl@{\qquad}rcl}
    {x} &=& -i\,{\tau}^x, &
    {\tau}^x &=& i\,{ x},
    \\
    {h} &=& -i\,{\tau}^z, &
    {\tau}^z &=& i\,{h},
    \\
    {t} &=& {\tau}^y, &
    {\tau}^y &=& {t}.
  \end{array}
\end{equation}
Let us remark that, in terms of the generators of $\mathfrak{su}(2)$,
the noncompact generators $x$ and $h$ of $\mathfrak{sl}(2,\RR)$ are
purely imaginary but the compact one $t$ is real.

More precisely, if $\tau$ denotes the conjugation\epubtkFootnote{A
conjugation on a complex Lie algebra is an antilinear involution,
preserving the Lie algebra structure.} of $\mathfrak{sl}(2,\CC)$
that fixes $\{{\tau}^x,\, {\tau}^y,\, \tau^z\}$, we
obtain: 
\begin{equation}
  \tau({x})=-{x},
  \qquad
  \tau({t})=+{t},
  \qquad
  \tau({h})=-{h},
\end{equation}
or, more generally,
\begin{equation}
  \forall {X} \in \mathfrak{sl}(2,\CC):\tau({ X})=-{X}^\dagger.
\end{equation}
Conversely, if we denote by $\sigma$ the
conjugation of $\mathfrak{sl}(2,\CC)$ that fixes the previous
$\mathfrak{sl}(2,\RR)$ Cartan subalgebra in
$\mathfrak{sl}(2,\CC)$, we obtain the usual complex conjugation of
the matrices: 
\begin{equation}
 \sigma({ X})=\,\overline{\!X}.
\end{equation}

The two conjugations $\tau$ and $\sigma$ of $\mathfrak{sl}(2,\CC)$
associated with the real subalgebras $\mathfrak{su}(2)$ and
$\mathfrak{sl}(2,\RR)$ of $\mathfrak{sl}(2,\CC)$ commute with each
other. Each of them, trivially, fixes pointwise the algebra
defining it and globally the other algebra, where it constitutes
an involutive automorphism (``involution'').

The Killing form is neither positive definite nor negative
definite on $\mathfrak{sl}(2,\RR)$: The symmetric matrices have
positive norm squared, while the antisymmetric ones have negative
norm squared. Thus, by changing the relative sign of the
contributions associated with symmetric and antisymmetric
matrices, one can obtain a bilinear form which is definite.
Explicitly, the involution $\theta$ of $\mathfrak{sl}(2,\RR)$
defined by $\theta(X) = - X^t$ has the feature that 
\begin{equation}
  B^\theta (X, Y) = - B(X, \theta Y)
\end{equation}
is positive definite. An involution of a real Lie algebra with that
property is called a ``Cartan involution'' \index{Cartan involution} (see
Section~\ref{section:cartaninvolution} for the general definition).

The Cartan involution $\theta$ is just the restriction to
$\mathfrak{sl}(2,\RR)$ of the conjugation $\tau$ associated with
the compact real form $\mathfrak{su}(2)$ since for real matrices
$X^\dagger = X^t$. One says for that reason that the compact real
form $\mathfrak{su}(2)$ and the noncompact real form
$\mathfrak{sl}(2,\RR)$ are ``aligned''.

Using the Cartan involution $\theta$, one can split
$\mathfrak{sl}(2,\RR)$ as the direct sum 
\begin{equation}
  \mathfrak{sl}(2,\RR) = \cK \oplus \cP,
\end{equation}
where $\cK$ is the subspace of antisymmetric
matrices corresponding to the eigenvalue $+1$ of the Cartan
involution while $\cP$ is the subspace of symmetric matrices
corresponding to the eigenvalue $-1$. These are also eigenspaces
of $\tau$ and given explicitly by $\cK=\RR { t}$ and $\cP=\RR{
x}\oplus \RR{h}$. One has 
\begin{equation}
  \mathfrak{su}(2) = \cK \oplus i \cP,
\end{equation}
 i.e., the real form $\mathfrak{sl}(2,\RR)$ is obtained from
 the compact form $\mathfrak{su}(2)$ by inserting an ``$i$'' in
front of the generators in $\cP$.


\subsubsection{Concluding remarks}

Let us close these preliminaries with some remarks.

\begin{enumerate}
\item The conjugation $\tau$ allows to define a Hermitian form on
  $\mathfrak{sl}(2,\CC)$: 
  \begin{equation}
    {X}\bullet{ Y}= -\Tr({Y}\tau({ X})) .
  \end{equation}
\item Any element of the group $SL(2,\RR)$ can be written as a product
  of elements belonging to the subgroups $ \mc{K}$, $ \mc{A}$ and ${
    \mc{N}}=\Exp [\RR e]$ (Iwasawa decomposition), \index{Iwasawa decomposition}
  \begin{equation}
    \Exp [\theta\,{ t}] \, \Exp [a\,{h}] \, \Exp [n\,{ e}] =
    \left(
      \begin{array}{@{}r@{\quad}r@{}}
        e^a\cos\theta & n\,e^a\cos \theta+e^{-a}\sin \theta
        \\
        -e^{a}\sin \theta & e^{-a}\cos \theta-n\,e^{a}\sin \theta
      \end{array}
    \right).
  \end{equation}
\item Any element of $\cP$ is conjugated via $ \mc{K}$ to a multiple
  of $h$ ,
  \begin{equation}
    \rho(\cos \alpha\, {h}+\sin \alpha\, { x})=
    \left(
      \begin{array}{@{}r@{\quad}r@{}}
        \cos \f{\alpha}{2} & \sin \f{\alpha}{2}
        \\
        -\sin\f{\alpha}{2}&\cos \f{\alpha}{2}
      \end{array}
    \right)
    \rho\,{h} \left(
      \begin{array}{@{}r@{\quad}r@{}}
        \cos \f{\alpha}{2} & -\sin \f{\alpha}{2}
        \\
        \sin \f{\alpha}{2} & \cos \f{\alpha}{2}
      \end{array}
    \right),
  \end{equation}
  so, denoting by $\cA=\RR\,{h}$ the (maximal) noncompact Cartan
  subalgebra of $\mathfrak{sl}(2,\RR)$, we obtain 
  \begin{equation}
    \cP=\Ad ({\mc{K}})\cA.
  \end{equation}
\item Any element of $SL(2,\RR)$ can be written as the product of an
  element of $ \mc{K}$ and an element of $\Exp [\cP]$. Thus, as a
  consequence of the previous remark, we have $SL(2,\RR)={
    \mc{K}\,\mA\,\mK}$ (Cartan)\epubtkFootnote{This decomposition is
    just the ``standard'' decomposition of any $2+1$ Lorentz
    transformation, into the product of a rotation followed by a
    boost in a fixed direction and finally followed by yet another
    rotation.}.
\item When the Cartan subalgebra of $\mathfrak{sl}(2,\RR)$ is chosen
  to be $\RR\,{h}$, the root vectors are $e$ and $ { f}$. We obtain
  the compact element $ t$, generating a non-equivalent Cartan
  subalgebra  by taking the combination 
  \begin{equation}
    {t}=e+\theta (e).
    \label{sl2Caycomp}
  \end{equation}
  Similarly, the normalized root vectors associated with ${ t}$ are (up
  to a complex phase) ${ E}_{\pm 2 i}= \frac 12({h} \mp i {x})$: 
  \begin{equation}
    \label{sl2Rtcr}
    \Lb{{t}}{{E}_{2i}}=2i\,{E}_{2i},
    \qquad
    \Lb{{t}}{{E}_{-2i}}=-2i\,{E}_{-2i},
    \qquad
    \Lb{{{E}_{2i}}}{{E}_{-2i}}= i\,{t}.
  \end{equation}
  Note that both the real and imaginary components of ${E}_{\pm 2 i}$
  are noncompact. They allow to obtain the noncompact Cartan
  generators $h, x$ by taking the combinations
  \begin{equation}
    \cos \alpha\, {h}+\sin \alpha\, {x} =
    e^{i\alpha}{E}_{2i}+e^{-i\alpha}{E}_{-2i}.
    \label{sl2Caync}
  \end{equation}
\end{enumerate}


\subsection{The compact and split real forms of a semi-simple Lie algebra}
\label{section:CompactAndRealForms}

We shall consider here only semi-simple Lie algebras. Over the complex
numbers, Cartan subalgebras are ``unique''\epubtkFootnote{We say that
  an object is ``unique'' when it is unique up to an internal
  automorphism.}. These subalgebras may be defined as maximal Abelian
subalgebras $\cH$ such that the transformations in $\ad[\cH]$ are
simultaneously diagonalizable (over $\CC$). \index{Cartan subalgebra}
Diagonalizability is an
essential ingredient in the definition. There might indeed exist
Abelian subalgebras of dimension higher than the rank (= dimension of
Cartan subalgebras), but these would involve non-diagonalizable
elements and would not be Cartan subalgebras\epubtkFootnote{For
  example, for the split form $E_{8(8)}$ of $E_8$, the 8 level
  3-elements and the 28 level 2-elements form an Abelian subalgebra
  since there are no elements at levels $>3$ (the level is defined in
    Section~\ref{section:LevelDecomposition}). This Abelian subalgebra
  has dimension 36, which is clearly much greater than the rank (8).
  We thank Bernard Julia for a discussion on this example. Note that
  for subgroups of the unitary group, diagonalizability is
  automatic.}.

We denote the set of nonzero roots as $\Delta$. One may complete the
Chevalley generators into a full basis, the so-called \emph{Cartan--Weyl
  basis}, such that the following commutation relations hold:
\begin{eqnarray}
  \left[ H, E_{\al}\right]&=&\alpha( H)\, E_{\al},
  \label{comHEa}
  \\
  \left[ E_{\al}, E_{\be}\right]&=&\left\{
    \begin{array}{ll}
      N_{\alpha,\,\beta} E_{\alpha+\beta} &
      \qquad \mbox{if } \alpha+\beta \in \Delta,
      \\
      H_{\al} &
      \qquad \mbox{if } \alpha+\beta = 0,
      \\
      0 &
      \qquad \mbox{if } \alpha+\beta \not\in \Delta,
    \end{array}
  \right.
  \label{comEE}
\end{eqnarray}
where $ H_{\al}$ is defined by duality thanks to the Killing
form $B({ X},{ Y})=\Tr(\ad{ X}\,\ad{ Y})$, which is non-singular on
semi-simple Lie algebras: 
\begin{equation}
  \forall H\in\cH : \alpha( H)=B( H_{\al}, H),
\end{equation}
and the generators are normalized according to (see
Equation~(\ref{Bea})) 
\begin{equation}
  B( E_{\al},\, E_{\be})=\delta_{\alpha+\beta,0}.
\end{equation}
The generators $ E_{\al}$ associated with the roots $\alpha$ (where
$\alpha$ need not be a simple root) may be chosen such that the
structure constants $N_{\alpha,\,\beta}$ satisfy the relations 
\begin{eqnarray}
  N_{\alpha,\beta}&=&
  -N_{\beta,\alpha}=-N_{-\alpha,-\beta}=N_{\beta,-\alpha-\beta},
  \label{Nrel}
  \\
  N_{\alpha,\,\beta}^{2}&=&
  \frac 12 q(p+1) (\alpha\vert\alpha),
  \qquad
  p,\, q \in \NN_{0}, 
  \label{N2}
\end{eqnarray}
where the scalar product between roots is defined as
\begin{equation}
  (\alpha\vert\beta)=B( H_{\al}, H_{\beta}).
  \label{avb}
\end{equation}
The non-negative integers $p$ and $q$ are such that the string of all
vectors $\beta + n\,\alpha$ belongs to $\Delta$ for $-p\leq n\leq q$;
they also satisfy the equation
$p-q=2(\beta\vert\alpha)/(\alpha\vert\alpha)$. A standard result
states that for semi-simple Lie algebras
\begin{equation}
  \label{abQQ}
  (\alpha\vert\beta)=
  \sum_{\gamma\in\Delta}(\alpha\vert\gamma)(\gamma\vert\beta)\in \QQ,
\end{equation}
from which we notice that the roots are real when evaluated on an
$ H_{\beta}$-generator,
\begin{equation}
  \alpha(H_{\beta})=(\alpha\vert\beta).
\end{equation}

An important consequence of this discussion is that in
Equation~(\ref{comEE}), the structure constants of the commutations
relations may all be chosen real. Thus, if we restrict ourselves to
real scalars we obtain a real Lie algebra ${\mathfrak{s}}_{0}$, which
is called the \emph{split real form} because it contains the maximal
number of noncompact generators. This real form of $\cG$ reads
explicitly 
\begin{equation}
  \label{split}
  {\mathfrak{s}}_{0} =
  \bigoplus_{\aD}\RR H_{\al}\oplus \bigoplus_{\aD}\RR {E}_{\alpha}.
\end{equation}
The signature of the Killing form on
${\mathfrak{s}}_{0}$ (which is real) is easily computed. First,
it is positive definite on the real linear span $\cH_{0}$ of the
$ H_{\al}$ generators. Indeed, 
\begin{equation}
  B(H_{\al},\, H_{\al})=(\alpha\vert\alpha)= \!\!
  \sum_{\gamma\in\Delta}(\alpha\vert\gamma)^2>0.
\end{equation}
Second, the invariance of the Killing form fixes the
normalization of the $ E_{\al}$ generators to one, 
\begin{equation}
  B({E_{\al}}, {E}_{-\alpha}) = 1,
\end{equation}
since\epubtkFootnote{Quite generally, if $ X_\alpha$ is a vector in
  $\aga$ and $ Y_{-\alpha}$ is a vector in ${\bf g}_{-\alpha}$, then
  one has $[ X_\alpha, Y_{-\alpha}] = B( X_\alpha, Y_{-\alpha})
  H_{\alpha}$.}
\begin{equation}
  B(\Lb{ E_{\al}}{{E}_{-\alpha}}, H_{\al})=(\alpha\vert\alpha)=
  -B({E}_{-\alpha},\Lb{E_{\al}}{H_{\al}})=
  (\alpha\vert\alpha)B(E_{\al},{E}_{-\alpha}) 
  \label{Bea}.
\end{equation}
Moreover, one has
$B(\aga,\,\agb)=0$ if $\alpha+\beta\not= 0$. Indeed
$\ad[\aga]\,\ad[\agb]$ maps $\ag{\mu}$ into $\ag{\mu+\alpha+\beta}$,
i.e., in matrix terms $\ad[\aga]\,\ad[\agb]$ has zero elements on the
diagonal when $\alpha+\beta\not= 0$. Hence, the vectors ${ E_{\al}} +
{ E}_{-\alpha}$ are spacelike and orthogonal to the vectors
${ E_{\al}} - E_{-\alpha}$, which are timelike. This implies
that the signature of the Killing form is 
\begin{equation}
  \left(\left.\frac 12 (\dim \cS_{0}+\rank \cS_{0})\right\vert_{+},
  \left.\frac 12 (\dim \cS_{0}-\rank\cS_{0})\right\vert_{-}\right).
\end{equation}
The split real form $\cS_{0}$ of $\cG$ is ``unique''.

On the other hand, it is not difficult to check that the linear
span 
\begin{equation}
  \label{compact}
  \mf{c}_{0}=\bigoplus_{\aD}\RR(i\, H_{\al})\oplus
  \bigoplus_{\aD}\RR({ E}_{\alpha}-{ E}_{-\alpha})\oplus
  \bigoplus_{\aD}\RR\,i\,({ E}_{\alpha}+{ E}_{-\alpha})
\end{equation}
also defines a real Lie algebra. An important property of this real
form is that the Killing form is negative definite on it. Its
signature is 
\begin{equation}
  (0\vert_{+},\dim \mf{c}_{0}\vert_{-}).
\end{equation}
This is an immediate consequence of the previous discussion
and of the way $\mf{c}_{0}$ is constructed. Hence, this real Lie algebra
is compact\epubtkFootnote{An algebra is said to be compact if its group
of internal automorphisms is compact in the topological sense. A
classic theorem states that a semi-simple algebra is compact if
and only if its Killing form is negative definite.}. For this reason,
$\mf{c}_{0}$ is called the ``compact real form'' of $\cG$. It is also
``unique''.


\subsection{Classical decompositions}


\subsubsection{Real forms and conjugations}

The compact and split real Lie algebras constitute the two ends of
a string of real forms that can be inferred from a given complex
Lie algebra. As announced, this section is devoted to the
systematic classification of these various real forms.

If $\GO$ is a real form of $\cG$, it defines a conjugation on $\cG$.
Indeed we may express any ${ Z} \in \cG $ as ${ Z}={
X}_{0}+ i\,{ Y}_{0}$ with ${ X}_{0}\in\GO$ and $i\,{ Y}_{0}
\in i\,\GO$, and the conjugation of $\cG$ with respect to $\GO$ is given
by
\begin{equation}
  {Z}\mapsto \overline{Z}={ X}_{0}-i\,{ Y}_{0}.
\end{equation}
Using Equation~(\ref{Cmult}), it is immediate to verify that this
involutive map is antilinear:
$\overline{\lambda\,{Z}}=\overline{\lambda}\,\overline{{ Z}}$, where
$\overline{\lambda}$ is the complex conjugate of the complex number
$\lambda$.

Conversely, if $\sigma$ is a conjugation on $\cG$, the set
$\cG_{\sigma}$ of elements of $\cG$ fixed by $\sigma$ provides a
real form of $\cG$. Then $\sigma$ constitutes the conjugation of
$\cG$ with respect to $\cG_{\sigma}$. Thus, on $\cG$, real forms
and conjugations are in one-to-one correspondence. The strategy
used to classify and describe the real forms of a given complex
simple algebra consists of obtaining all the nonequivalent
possible conjugations it admits.


\subsubsection{The compact real form aligned with a given real form}

Let $\GO$ be a real form of the complex semi-simple Lie algebra
$\cG^\mathbb{C}= \GO \otimes_{\mathbb{R}} \mathbb{C}$. Consider a
compact real form $\mf{c}_{0}$ of $\cG^\mathbb{C}$ and the respective
conjugations $\tau$ and $\sigma$ associated with $\mf{c}_{0}$ and $\GO$.
It may or it may not be that $\tau$ and $\sigma$ commute. When
they do, $\tau$ leaves $\GO$ invariant,
\begin{displaymath}
  \tau(\GO) \subset \GO
\end{displaymath}
and, similarly, $\sigma$ leaves $\mf{c}_{0}$ invariant,
$$\sigma(\mf{c}_{0}) \subset \mf{c}_{0}.$$ In that case, one says that the real
form $\GO$ and the compact real form $\mf{c}_{0}$ are ``aligned''.

Alignment is not automatic. For instance, one can always de-align
a compact real form by applying an automorphism to it while
keeping $\GO$ unchanged. However, there is a theorem that states
that given a real form $\GO$ of the complex Lie algebra
$\cG^\mathbb{C}$, there is always a compact real form $\mf{c}_{0}$
associated with it~\cite{Helgason, Knapp}. As this result is
central to the classification of real forms, we provide a proof in
Appendix~\ref{appendix_2}, where we also prove the uniqueness of the
Cartan involution. 

We shall from now on always consider the compact real form aligned with
the real form under study.


\subsubsection{Cartan involution and Cartan decomposition}
\label{section:cartaninvolution}
\index{Cartan involution|bb}

A Cartan involution $\theta$ of a real Lie algebra $\GO$ is an
involutive automorphism such that the symmetric, bilinear form
$B^\theta$ defined by 
\begin{equation}
  B^\theta(X,Y) = - B(X, \theta Y)
\end{equation}
is positive definite. If the algebra $\GO$ is compact, a Cartan
involution is trivially given by the identity.

A Cartan involution $\theta$ of the real semi-simple Lie algebra
$\GO$ yields the direct sum decomposition (called Cartan
decomposition) 
\begin{equation}
  \GO=\KO\oplus\PO,
  \label{CartanDecomp}
\end{equation}
where $\KO$ and $\PO$ are the $\theta$-eigenspaces of eigenvalues
$+1$ and $-1$. Explicitly, the decomposition of a Lie algebra element
is given by 
\begin{equation}
  X=\frac 12 ( X + \theta[ X])+\frac 12 (X - \theta[ X]).
\end{equation}
The eigenspaces obey the commutation relations 
\begin{equation}
  \Lb{\KO}{\KO}\subset\KO,
  \qquad
  \Lb{\KO}{\PO}\subset\PO,
  \qquad
  \Lb{\PO}{\PO}\subset\KO, 
  \label{comKP}
\end{equation}
from which we
deduce that $B(\KO,\PO)=0$ because the mappings $\ad[\KO]\,\ad[\PO]$
map $\PO$ on $\KO$ and $\KO$ on $\PO$. Moreover $\theta[\KO]=+\KO$
and $\theta[\PO]=-\PO$, and hence $B^\theta(\KO,\,\PO)=0$. In
addition, since $B^\theta$ is positive definite, the Killing form $B$
is negative definite on $\KO$ (which is thus a compact subalgebra)
but is positive definite on $\PO$ (which is not a subalgebra).

Define in $\cG^\mathbb{C}$ the algebra $\mf{c}_{0}$ by 
\begin{equation}
  \mf{c}_{0}=\KO\oplus i\PO.
\end{equation}
 It is clear that $\mf{c}_{0}$ is also a real form of
$\cG^\mathbb{C}$ and is furthermore compact since the Killing form
restricted to it is negative definite. The conjugation $\tau$
that fixes $\mf{c}_{0}$ is such that $\tau(X) = X$ ($X \in \KO$), $\tau(i
Y) = i Y$ ($Y \in \PO$) and hence $\tau(Y) = - Y$ ($Y \in \PO$).
It leaves $\GO$ invariant, which shows that $\mf{c}_{0}$ is aligned with
$\GO$. One has 
\begin{equation}
  \mf{c}_{0}=\KO\oplus i\PO,
  \qquad
  \KO=\GO\cap\mf{c}_{0},
  \qquad \PO=\GO\cap i\,\mf{c}_{0}.
  \label{RelUG}
\end{equation}

Conversely, let $\mf{c}_{0}$ be a compact real form aligned with $\GO$
and $\tau$ the corresponding conjugation. The restriction
$\theta$ of $\tau$ to $\GO$ is a Cartan involution. \index{Cartan involution} Indeed, one
can decompose $\GO$ as in Equation~(\ref{CartanDecomp}), with
Equation~(\ref{comKP}) holding since $\theta$ is an involution of
$\GO$. Furthermore, one has also Equation~(\ref{RelUG}), which shows
that $\KO$ is compact and that $B_\theta$ is positive definite.

This shows, in view of the result invoked above that an aligned
compact real form always exists, that any real form possesses a
Cartan involution and a Cartan decomposition. If there are two
Cartan involutions, $\theta$ and $\theta'$, defined on a real
semi-simple Lie algebra, one can show that they are conjugated by
an internal automorphism~\cite{Helgason, Knapp}. It follows that
any real semi-simple Lie algebra possesses a ``unique'' Cartan
involution.

On the matrix algebra $\ad[\GO]$, the Cartan involution is nothing
else than minus the transposition with respect to the scalar
product $B^\theta$, 
\begin{equation}
  \ad\theta X=-(\ad X)^{T}. 
  \label{symtr}
\end{equation}
Indeed, remembering that the transpose of a
linear operator with respect to $B^\theta$ is defined by
$B^\theta(X, AY) = B^\theta (A^T X, Y)$, one gets 
\begin{eqnarray}
  B^\theta(\ad\theta X( Y),\, Z)&=&
  -B(\Lb{\theta X}{ Y},\,\theta Z)=
  B( Y,\,\Lb{\theta X}{\theta Z})
  \nonumber
  \\
  &=&B( Y,\,\theta\Lb{ X}{ Z})=
  -B^\theta( Y,\,\ad X( Z))=
  -B^\theta((\ad X)^{T}( Y),\, Z).
\end{eqnarray}
Since $B_\theta$ is positive definite, this implies, in
particular, that the operator $\ad Y$, with $ Y \in \PO$, is
diagonalizable over the real numbers since it is symmetric, $\ad Y
= (\ad Y )^T$.

An important consequence of this~\cite{Helgason, Knapp} is that
any real semi-simple Lie algebra can be realized as a real matrix
Lie algebra, closed under transposition. One can also
show~\cite{Helgason, Knapp} that the Cartan decomposition of the Lie
algebra of a semi-simple group can be lifted to the group via a
diffeomorphism between ${\KO}\times \PO \mapsto \mG = \mK \exp [\PO]$,
where $\mK$ is a closed subgroup with $\KO$ as Lie algebra. It is this
subgroup that contains all the topology of $\mG$.


\subsubsection{Restricted roots}

Let $\GO$ be a real semi-simple Lie algebra. It admits a Cartan
involution $\theta$ that allows to split it into eigenspaces $\KO$
of eigenvalue $+1$ and $\PO$ of eigenvalue $-1$. We may choose in
$\PO$ a maximal Abelian subalgebra $\AO$ (because the dimension
of $\PO$ is finite). The set $\{\ad H \vert H \in \AO\}$ is a
set of symmetric transformations that can be simultaneously
diagonalized on $\RR$. Accordingly we may decompose $\GO$ into a
direct sum of eigenspaces labelled by elements of the dual space
$\AO^{*}$: 
\begin{equation}
  \GO=\bigoplus_{\lambda}\,\agl,
  \qquad
  \agl=\{ X\in\GO\vert\forall H\in \AO:\ad H( X)=\lambda( H)\, X\}.
\end{equation}

One, obviously non-vanishing, subspace is $\ag{0}$, which contains
$\AO$. The other nontrivial subspaces define the \emph{restricted
root spaces} of $\GO$ with respect to $\AO$, of the pair
$(\GO,\,\AO)$. The $\lambda$ that label these subspaces
$\agl$ are the \emph{restricted roots} and their elements are
called  \emph{restricted root vectors}. The set of all $\lambda$ is called the \emph{restricted root system}. \index{restricted root system|bb} By construction the
different $\agl$ are mutually $B^\theta$-orthogonal. The Jacobi
identity implies that $\Lb{\agl}{\ag{\mu}}\subset
\ag{\lambda+\mu}$, while $\AO\subset \PO$ implies that $\theta
\agl =\ag{-\lambda}$. Thus if $\lambda$ is a restricted root, so
is $-\lambda$.

Let $\cM$ be the centralizer of $\AO$ in $\KO$. The space
$\ag{0}$ is given by 
\begin{equation}
  \ag{0} =\AO\oplus\cM.
\end{equation}
 If $\TO$ is a
maximal Abelian subalgebra of $\cM$, the subalgebra
$\HO=\AO\oplus\TO$ is a Cartan subalgebra of the real algebra
$\GO$ in the sense that its complexification $\HC$ is a Cartan
subalgebra of $\GC$. Accordingly we may consider the set of
nonzero roots $\Delta$ of $\GC$ with respect to $\HC$ and write
\begin{equation}
  \GC=\HC\bigoplus_{\aD}(\ag{\alpha})^\CC.
\end{equation}

The restricted root space $\agl$ is given by 
\begin{equation}
  \agl=\GO\cap \!\!\!\! \bigoplus_{\scriptsize
    \begin{array}{l}
      \aD
      \\
      \alpha\vert_{\AO}=\lambda
    \end{array}
  }\!\!\!(\ag{\alpha})^\CC
\end{equation}
and similarly 
\begin{equation}
  \MC=\TC \!\!\!\! \bigoplus_{\scriptsize
    \begin{array}{l}
      \aD
      \\
      \alpha\vert_{\AO}=0
    \end{array}
  }\!\!\!(\ag{\alpha})^\CC.
\end{equation}

Note that the multiplicities of the restricted roots $\lambda$ might
be nontrivial even though the roots $\alpha$ are nondegenerate,
because distinct roots $\alpha$ might yield the same restricted root
when restricted to $\AO$.

Let us denote by $\Sigma$ the subset of nonzero restricted roots
and by $V_{\Sigma}$ the subspace of $\AO^{*}$ that they span. One
can show~\cite{Helgason, Knapp} that $\Sigma$ is a root system as
defined in Section~\ref{section:KacMoody}. This root
system need not be reduced. As for all root systems, one can
choose a way to split the roots into positive and negative ones.
Let $\Sigma^{+}$ be the set of positive roots and
\begin{equation}
  \cN=\bigoplus_{\lambda\in\Sigma^{+}}\agl.
\end{equation}
As $\Sigma^{+}$ is finite, $\cN$ is a nilpotent subalgebra of $\GO$
and $\AO\oplus \cN$ is a solvable subalgebra.


\subsubsection[Iwasawa and $\mK\mA\mK$ decompositions]%
              {Iwasawa and \boldmath $\mK\mA\mK$ decompositions}
\label{section:Iwasawa}
\index{Iwasawa decomposition}

The Iwasawa decomposition provides a global factorization of any
semi-simple Lie group in terms of closed subgroups. It can be
viewed as the generalization of the Gram--Schmidt orthogonalization
process.

At the level of the Lie algebra, the Iwasawa decomposition theorem
states that 
\begin{equation}
  \GO=\KO\oplus\AO\oplus\cN.
\end{equation}
Indeed any element $ X$ of $\GO$ can be decomposed as 
\begin{equation}
  X= X_{0}+\sum_{\lambda} X_{\lambda}=
  X_{0}+\!\!\sum_{\lambda\in\Sigma^{+}}(X_{-\lambda}+\theta X_{-\lambda})+
  \!\!\sum_{\lambda\in\Sigma^{+}}( X_{\lambda}-\theta X_{-\lambda}).
\end{equation}
The first term $ X_{0}$ belongs to $\ag{0} =\AO\oplus\cM
\subset \AO\oplus\KO$, while the second term belongs to $\KO$, the
eigenspace subspace of $\theta$-eigenvalue +1. The third term
belongs to $\cN$ since $\theta X_{-\lambda} \in \ag{\lambda}$.
The sum is furthermore direct. This is because one has obviously
$\KO\cap\AO=0$ as well as $\AO\cap\cN = 0$. Moreover, $\KO\cap\cN$
also vanishes because $\theta \cN\cap\cN=0$ as a consequence of
$\theta\cN =\bigoplus_{\lambda\in\Sigma^{+}}\ag{-\lambda}$.

The Iwasawa decomposition \index{Iwasawa decomposition} of the Lie algebra differs from the
Cartan decomposition and is tilted with respect
to it, in the sense that $\cN$ is neither in $\KO$ nor in $\PO$.
One of its virtues is that it can be elevated from the Lie algebra
$\GO$ to the semi-simple Lie group $\mG$. Indeed, it can be
shown~\cite{Helgason, Knapp} that the map 
\begin{equation}
  (k,a,n) \in \mK \times \mA \times \mN \mapsto k\, a\, n \in \mG
\end{equation}
is a global diffeomorphism. Here, the subgroups $\mK$, $\mA$ and $\mN$
have respective Lie algebras $\KO$, $\AO$, $\cN$. This decomposition is
``unique''.

There is another useful decomposition of $\mG$ in terms of a product
of subgroups. Any two generators of $\PO$ are conjugate via
internal automorphisms of the compact subgroup $\mK$. As a
consequence writing an element $g\in \mG$ as a product
$g=k\,\Exp [\PO]$, we may write $ \mG = \mK\,\mA\,\mK$, which
constitutes the so-called $\mK\mA\mK$ decomposition of the group (also
sometimes called the Cartan decomposition of the group although it
is not the exponention of the Cartan decomposition of the
algebra). Here, however, the writing of an element of $ \mG$ as
product of elements of $ \mK$ and $ \mA$ is, in general, not unique.


\subsubsection[$\theta$-stable Cartan subalgebras]%
              {\boldmath $\theta$-stable Cartan subalgebras}

As in the previous sections, $\GO$ is a real form of the complex
semi-simple algebra $\cG$, $\sigma$ denotes the conjugation it
defines, $\tau$ the conjugation that commutes with $\sigma$, $\mf{c}_{0}$
the associated compact aligned real form of $\cG$ and $\theta$ the
Cartan involution. \index{Cartan involution} It is also useful to introduce the involution
of $\cG$ given by the product $\sigma \tau$ of the commuting
conjugations. We denote it also by $\theta$ since it reduces to
the Cartan involution when restricted to $\GO$. Contrary to the
conjugations $\sigma$ and $\tau$, $\theta$ is linear over the
complex numbers. Accordingly, if we complexify the Cartan
decomposition $\GO = \KO \oplus \PO$, to 
\begin{equation}
  \cG = \cK \oplus \cP
\end{equation}
 with $\cK = \KO \otimes_\mathbb{R} \mathbb{C} = \KO \oplus i
\KO$ and $\cP = \PO \otimes_\mathbb{R} \mathbb{C} = \PO \oplus i
\PO$, the involution $\theta$ fixes $\cK$ pointwise while
$\theta( X) = - X$ for $ X \in \cP$.

Let $\HO$ be a $\theta$-stable Cartan subalgebra of $\GO$, i.e., a
subalgebra such that (i) $\theta(\HO) \subset \HO$, and (ii) $\cH
\equiv \HO^\mathbb{C}$ is a Cartan subalgebra of the complex
algebra $\cG$. One can decompose $\HO$ into compact and
noncompact parts, 
\begin{equation}
  \HO = \TO \oplus \AO,
  \qquad
  \TO = \HO \cap \KO,
  \qquad
  \AO = \HO \cap \PO.
\end{equation}

We have seen that for real Lie algebras, the Cartan subalgebras
are not all conjugate to each other; in particular, even though
the dimensions of the Cartan subalgebras are all equal to the rank
of $\cG$, the dimensions of the compact and noncompact
subalgebras depend on the choice of $\HO$. For example, for $\mf{sl}(2,
\mathbb{R})$, one may take $\HO = \mathbb{R} t$, in which case
$\TO = 0$, $\AO = \HO$. Or one may take $\HO = \mathbb{R}
\tau^y$, in which case $\TO = \HO$, $\AO = 0$.

One says that the $\theta$-stable Cartan subalgebra $\HO$ is {\it
maximally compact} if the dimension of its compact part $\TO$ is
as large as possible; and that it is \emph{maximally noncompact}
if the dimension of its noncompact part $\AO$ is as large as
possible. The $\theta$-stable Cartan subalgebra $\HO = \TO \oplus
\AO$ used above to introduce restricted roots, where $\AO$ is a
maximal Abelian subspace of $\PO$ and $\TO$ a maximal Abelian
subspace of its centralizer $\cM$, is maximally noncompact. If
$\cM=0$, the Lie algebra $\GO$ constitutes a split real form of
$\GC$. The \emph{real rank} of $\GO$ is the dimension of its
maximally noncompact Cartan subalgebras (which can be shown to be
conjugate, as are the maximally compact ones~\cite{Knapp}).


\subsubsection{Real roots -- Compact and non-compact imaginary roots}

Consider a general $\theta$-stable Cartan subalgebra $\HO = \TO
\oplus \AO$, which need not be maximally compact or maximally non
compact. A consequence of Equation~(\ref{symtr}) is that the matrices
of the real linear transformations $\ad H$ are real symmetric
for $ H\in\AO$ and real antisymmetric for $ H\in\TO$.
Accordingly, the eigenvalues of $\ad H$ are real (and $\ad H$
can be diagonalized over the real numbers) when $ H\in\AO$, while
the eigenvalues of $\ad H$ are imaginary (and $\ad H$ cannot be
diagonalized over the real numbers although it can be diagonalized
over the complex numbers) when $ H\in\TO$.

Let $\alpha$ be a root of $\cG$, i.e., a non-zero eigenvalue of
$\ad \cH$ where $\cH$ is the complexification of the
$\theta$-stable Cartan subalgebra $\cH = \HO \otimes_{\mbb{R}} \mbb{C} = \HO
\oplus i \HO$. As the values of the roots acting on a given $ H$
are the eigenvalues of $\ad H$, we find that the roots are real
on $\AO$ and imaginary on $\TO$. One says that a root is {\em
real} if it takes real values on $\HO = \TO \oplus \AO$, i.e., if
it vanishes on $\TO$. It is \emph{imaginary} if it takes imaginary 
values on $\HO$, i.e., if it vanishes on $\AO$, and {\em
complex} otherwise. These notions of ``real'' and ``imaginary''
roots should not be confused with the concepts with similar
terminology introduced in Section~\ref{section:KacMoody} in the
context of non-finite-dimensional Kac--Moody algebras.

If $\HO$ is a $\theta$-stable Cartan subalgebra, its
complexification $\cH = \HO \otimes_{\mbb{R}} \mbb{C} = \HO \oplus i \HO$ is
stable under the involutive authormorphism $ \theta=\tau\,\sigma$.
One can extend the action of $\theta$ from $\cH$ to
$\cH^*$ by duality. Denoting this transformation by the same
symbol $\theta$, one has 
\begin{equation}
  \forall H \in \cH \mbox{ and }
  \forall \alpha \in \cH^*,
  \qquad \theta(\alpha)( H) = \alpha(\theta^{-1} (H)),
\end{equation}
or, since $\theta^2 = 1$, 
\begin{equation}
  \theta(\alpha)( H)=\alpha(\theta H).
\end{equation}

Let $ E_{\al}$ be a nonzero root vector associated with the root
$\alpha$ and consider the vector $\theta E_{\al}$. One has
\begin{equation}
  \label{thtEa}
  \Lb{ H}{\theta E_{\al} }=\theta \Lb{\theta H }{ E_{\al}}=
  {\alpha(\theta H )}\, \theta E_{\al} =
  \theta(\alpha) ( H)\, \theta E_{\al},
\end{equation}
i.e., $\theta(\aga)= \bag_{\theta(\alpha)}$ because the roots are
nondegenerate, i.e., all root spaces are one-dimensional.

Consider now an imaginary root $\alpha$. Then for all $h\in \HO$ and
$a\in \AO$ we have $\alpha (h + a)=\alpha (h)$, while $\theta(\alpha)
\, (h + a) =\alpha(\theta(h+ a))=\alpha(h- a) = \alpha (h)$;
accordingly $\alpha=\theta(\alpha) $. Moreover, as the roots are
nondegenerate, one has $\theta E_{\al} =\pm E_{\al}$. Writing $ E_{\al}$ as 
\begin{equation}
  E_{\al}= X_{\alpha}+i\, Y_{\alpha}
  \qquad
  \mbox{with } X_{\alpha},\, Y_{\alpha}\in \GO , 
\label{splitroot}
\end{equation}
it is easy to check that $ \theta E_{\al} =+ E_{\al}$ implies that $
X_{\alpha}$ and $ Y_{\alpha}$ belong to $\KO$, while both are in
$\PO$ if $ \theta E_{\al} =- E_{\al}$. Accordingly, $\aga$ is
completely contained either in $\cK=\KO\oplus i\,\KO$ or in
$\cP=\PO\oplus i\,\PO$. If $\aga\subset \cK$, the imaginary root is
said to be \emph{compact}, and if $\aga\subset \cP$ it is said to be
\emph{noncompact}.


\subsubsection{Jumps between Cartan subalgebras -- Cayley transformations}

Suppose that $\beta$ is an imaginary noncompact root. Consider a
$\beta$-root vector $ E_{\be}\in\agb\subset \cP$. If this root is
expressed according to Equation~(\ref{splitroot}), then its conjugate, with
respect to (the conjugation $\sigma$ defined by) $\GO$, is 
\begin{equation}
  \sigma E_{\be}= X_{\beta}-i\, Y_{\beta}
  \qquad
  \mbox{with } X_{\beta},\, Y_{\beta}\in \PO.
\end{equation}
It belongs to $\ag{-\beta}$ because (using $\forall H\in\HO: \, \sigma H= H $) 
\begin{equation}
  \Lb{H}{\sigma E_{\be}}=\sigma\Lb{\sigma H}{ E_{\be}}=
  \overline{\beta(\sigma H)}\, \sigma E_{\be}=-\beta( H)\, \sigma E_{\be}.
\end{equation}
Hereafter, we shall denote $\sigma E_{\be}$ by $\overline
E_{\be}$. The commutator 
\begin{equation}
  \Lb{ E_{\be}}{\overline E_{\be}}=B( E_{\be},\,\overline E_{\be})\, H_{\beta}
\end{equation}
belongs to $i\,\cK_{0}$ since $\sigma ([ E_{\be},\overline E_{\be}]) =
[\overline E_{\be}, E_{\be}] = - [ E_{\be},\overline E_{\be}]$ and can be written,
after a renormalization of the generators $ E_{\be}$, as 
\begin{equation}
  \Lb{ E_{\be}}{\overline E_{\be}} =
  \frac 2{(\beta\vert\beta)}\, H_{\beta}= H^\prime_{\beta}
  \qquad
  \in i\,\cK_{0}.
  \label{norE}
\end{equation}
Indeed as $ E_{\be}\in\cP$, we have $\overline E_{\be}\in\cP$ and thus
$\theta\overline E_{\be}=-\overline E_{\be}$. This implies
\begin{displaymath}
  B(E_{\be}\,,\overline E_{\be})=-B( E_{\be},
  \,\theta\overline E_{\be})=B^\theta( E_{\be},\overline E_{\be})>0.
\end{displaymath}
The three generators $\{ H_{\be^{\p}},\, E_{\be},\, \overline
E_{\be}\}$ therefore define an $\mathfrak{sl}(2,\CC)$ subalgebra: 
\begin{equation}
  \Lb{E_{\be}}{\overline E_{\be}}= H_{\be^{\p}},
  \qquad
  \Lb{ H_{\be^{\p}}}{ E_{\be}}=2\, E_{\be},
  \qquad
  \Lb{ H_{\be^{\p}}}{\overline E_{\be}}=-2\,\overline E_{\be}.
\end{equation}
We may change the basis and take 
\begin{equation}
  h= E_{\be}+\overline E_{\be},
  \qquad
  e=\frac i2( E_{\be}-\overline E_{\be}- H_{\be^{\p}}),
  \qquad
  f=\frac i2( E_{\be}-\overline E_{\be}+ H_{\be^{\p}}),
\end{equation}
whose elements belong to $\GO$ (since they are fixed by $\sigma$) and
satisfy the commutation relations~(\ref{sl2Rcr}) 
\begin{equation}
  \Lb{e}{f} = h,
  \qquad
  \Lb{h}{e}=2\, e,
  \qquad
  \Lb{h}{f}=-2\, f.
\end{equation}
The subspace 
\begin{equation}
  \HO'=\ker (\beta\vert\HO)\oplus \RR h
\end{equation}
constitutes a new real Cartan subalgebra whose intersection with $\PO$
has one more dimension.

Conversely, if $\beta$ is a real root then $\theta(\beta)=-\beta$.
Let $ E_{\be}$ be a root vector. Then $\overline E_{\be}$ is also in $\agb$
and hence proportional to $ E_{\be}$. By adjusting the phase of $ E_{\be}$,
we may assume that $ E_{\be}$ belongs to $\GO$. At the same time,
$\theta E_{\be}$, also in $\GO$, is an element of $\ag{-\beta}$.
Evidently, $B( E_{\be},\,\theta E_{\be}) = - B^\theta( E_{\be},\, E_{\be})$ is
negative. Introducing $ H_{\be^{\p}}=2/(\beta\vert\beta)H_{\be}$ (which is
in $\PO$), we obtain the $\mathfrak{sl}(2,\RR)$ commutation
relations 
\begin{equation}
  \Lb{E_{\be}}{\theta E_{\be}}=- H_{\be^{\p}}\,\in\PO,
  \qquad
  \Lb{ H_{\be^{\p}}}{ E_{\be}}=2\, E_{\be},
  \qquad
  \Lb{ H_{\be^{\p}}}{\theta E_{\be}}=-2\,\theta E_{\be}.
\end{equation}
Defining the compact generator $ E_{\be}+\theta E_{\be} $, which
obviously belongs to $\GO$, we may build a new Cartan subalgebra of
$\GO$:
\begin{equation}
  \HO'=\ker (\beta\vert\HO)\oplus \RR\, ( E_{\be}+\theta E_{\be}),
\end{equation}
whose noncompact subspace is now one dimension less than
previously.

These two kinds of transformations -- called \emph{Cayley
transformations} -- allow, starting from a $\theta$-stable Cartan
subalgebra, to transform it into new ones with an increasing
number of noncompact dimensions, as long as noncompact imaginary
roots remain; or with an increasing number of compact dimensions,
as long as real roots remain. Exploring the algebra in this way,
we obtain all the Cartan subalgebras up to conjugacy. One can
prove that the endpoints are maximally noncompact and maximally
compact, respectively.

\begin{theorem}
  Let $\HO$ be a $\theta$ stable Cartan subalgebra of $\GO$. Then
  there are no noncompact imaginary roots if and only if $\HO$ is
  maximally noncompact, and there are no real roots if and only if
  $\HO$ is maximally compact~\cite{Knapp}.
\end{theorem}

\noindent
For a proof of this, note that we have already proven that if there are
imaginary noncompact (respectively, real) roots, then $\HO$ is
not maximally noncompact (respectively, compact). The converse is
demonstrated in~\cite{Knapp}.


\subsection{Vogan diagrams}

Let $\GO$ be a real semi-simple Lie algebra, $\cG$ its
complexification, $\theta$ a Cartan involution 
\index{Cartan involution} leading to the Cartan decomposition 
\begin{equation}
  \GO=\KO\oplus\PO,
\end{equation}
and $\HO$ a Cartan
$\theta$-stable subalgebra of $\GO$. Using, if necessary,
successive Cayley transformations, we may build a maximally
compact $\theta$-stable Cartan subalgebra $\HO=\TO\oplus\AO$, with
complexification $\cH=\cT\oplus \cA$. As usual we denote by
$\Delta$ the set of (nonzero) roots of $\cG$ with respect to
$\cH$. This set does not contain any real root, the compact
dimension being assumed to be maximal.

From $\Delta$ we may define a positive subset $\Delta^{+}$ by choosing
the first set of indices from a basis of $i\,\TO$, and then the next
set from a basis of $\AO$. Since there are no real roots, the roots in
$\Delta^{+}$ have at least one non-vanishing component along $i \,
\TO$, and the first non-zero one of these components is strictly
positive. Since $\theta =+1$ on $\TO$, and since there are no real
roots: $\theta\Delta^{+}=\Delta^{+}$. Thus $\theta$ permutes the
simple roots, fixes the imaginary roots and exchanges in 2-tuples
the complex roots: it constitutes an involutive automorphism of the
Dynkin diagram of $\cG$.

A Vogan diagram \index{Vogan diagram|bb} is associated to the triple
$(\GO,\HO,\Delta^{+})$ as follows. It corresponds to the standard
Dynkin diagram \index{Dynkin diagram} of $\Delta^{+}$, with additional
information: the 2-element orbits under $\theta$ are exhibited by
joining the correponding simple roots by a double arrow and the
1-element orbit is painted in black (respectively, not painted), if
the corresponding imaginary simple root is noncompact (respectively,
compact).


\subsubsection[Illustration -- The $\mathfrak{sl}(5,\CC)$ case]%
              {Illustration -- The \boldmath $\mathfrak{sl}(5,\CC)$ case}

The complex Lie algebra $\mf{sl}(5, \CC)$ can be represented as the
algebra of traceless $5\times 5$ complex matrices, the Lie bracket
being the usual commutator. It has dimension 24. In principle, in order to
compute the Killing form, one needs to handle the $24\times 24$
matrices of the adjoint representation. Fortunately, the
uniqueness (up to an overall factor) of the bi-invariant quadratic
form on a simple Lie algebra leads to the useful relation 
\begin{equation}
  B (X, Y)= \Tr(\ad X\, \ad Y)=10\, \Tr( X Y).
\end{equation}
 The coefficient
$10$ appearing in this relation is known as the Coxeter index of
$\mathfrak{sl}(5,\CC)$.

A Cartan--Weyl basis is obtained by taking the 20 nilpotent
generators $K^p{}_q$ (with $p\neq q$) corresponding to matrices, all
elements of which are zero except the one located at the
intersection of row $p$ and column $q$, which is equal to 1, 
\begin{equation}
  (K^p{}_q)^\alpha_\beta=\delta ^{\alpha\,p}\delta_{\beta\,q}
  \label{GLgenerators}
\end{equation}
and the four diagonal ones,
\begin{equation}
  \begin{array}{rcl@{\qquad}rcl}
    H_1 &=& \left(
      \begin{array}{@{}r@{\quad}r@{\quad}r@{\quad}r@{\quad}r@{}}
        1&0&0&0&0 \\
        0&-1&0&0&0 \\
        0&0&0&0&0 \\
        0&0&0&0&0 \\
        0&0&0&0&0
      \end{array}
    \right), &
    H_2 &=& \left(
      \begin{array}{@{}r@{\quad}r@{\quad}r@{\quad}r@{\quad}r@{}}
        0&0&0&0&0 \\
        0&1&0&0&0 \\
        0&0&-1&0&0 \\
        0&0&0&0&0 \\
        0&0&0&0&0
      \end{array}
    \right),
    \\ [3.5 em]
    H_3 &=& \left(
      \begin{array}{@{}r@{\quad}r@{\quad}r@{\quad}r@{\quad}r@{}}
        0&0&0&0&0 \\
        0&0&0&0&0 \\
        0&0&1&0&0 \\
        0&0&0&-1&0 \\
        0&0&0&0&0
      \end{array}
    \right), &
    H_4 &=& \left(
      \begin{array}{@{}r@{\quad}r@{\quad}r@{\quad}r@{\quad}r@{}}
        0&0&0&0&0 \\
        0&0&0&0&0 \\
        0&0&0&0&0 \\
        0&0&0&1&0 \\
        0&0&0&0&-1
      \end{array}
    \right),
  \end{array}
  \label{H14}
\end{equation}
which constitute a Cartan subalgebra $\cH$.

The root space is easily described by introducing the five linear
forms $\epsilon_p$, acting on diagonal matrices $d=\diag (d_1,\ldots,d_5)$ as follows: 
\begin{equation}
  \epsilon_p(d)=d_p.
  \label{e5}
\end{equation}
 In terms of these, the dual space $\cH^\star$ of
the Cartan subalgebra may be identified with the subspace 
\begin{equation}
  \left\{\epsilon=\sum_p A^p\, \epsilon_p\, \big|\, \sum_p A^p=0 \right\}.
\end{equation}
The 20 matrices ${K^p}_q$ are root vectors,
\begin{equation}
  \Lb{H_k}{{K^p}_q}=(\epsilon_p[H_k]-\epsilon_q[H_k]){K^p}_q,
\end{equation}
i.e., $K^p{}_q$ is a root vector associated to the root
$\epsilon_p-\epsilon_q$.

\subsubsection*{\boldmath $\mathfrak{sl}(5,\RR)$ and $\mathfrak{su}(5)$}

By restricting ourselves to real combinations of these generators
we obtain the real Lie algebra $\mathfrak{sl}(5,\RR)$. The
conjugation $\eta$ that it defines on $\mathfrak{sl}(5,\CC)$ is just
the usual complex conjugation. This $\mathfrak{sl}(5,\RR)$
constitutes the split real form $\mf{s}_0$ of $\mathfrak{sl}(5,\CC)$.
Applying the construction given in Equation~(\ref{compact}) to the
generators of $\mathfrak{sl}(5,\RR)$, we obtain the set of
antihermitian matrices 
\begin{equation}
  i\,H_k,
  \qquad
  K^p{}_q-{K^q}_p,
  \qquad
  i( K^p{}_q+{K^q}_p)
  \qquad
  (p>q),
\end{equation}
defining a basis of the real subalgebra $\mathfrak{su}(5)$. This is
the compact real form $\mf{c}_0$ of $\mf{sl}(5, \CC)$. The conjugation
associated to this algebra (denoted by $\tau$) is minus the Hermitian
conjugation, 
\begin{equation}
  \tau( X)=- X^\dagger.
\end{equation}
Since $[\eta, \tau] = 0$, $\tau$ induces a Cartan involution $\theta$
on $\mathfrak{sl}(5,\RR)$, providing a Euclidean form on the previous
$\mathfrak{sl}(5,\RR)$ subalgebra
\begin{equation}
  B^\theta( X,\, Y)=10\, \Tr( X Y^t),
\end{equation}
which can be extended to a Hermitian form on $ \mathfrak{sl}(5,\CC)$, 
\begin{equation}
  B^\theta( X,\, Y)=10\, \Tr( X Y^\dagger).
\end{equation}
Note that the generators $i\,H_k$ and $i( K^p{}_q+{K^q}_p)$ are real
generators (although described by complex matrices) since, e.g.,
$(i\,H_k)^\dagger=-i\,H_k^\dagger$, i.e., $\tau(i\,H_k) = i\,H_k$.

\subsubsection*{The other real forms}

The real forms of $\mathfrak{sl}(5,\CC)$ that are not
isomorphic to $\mf{sl}(5, \RR)$ or $\mf{su}(5)$ are isomorphic either
to $\mathfrak{su}(3,2)$ or $\mathfrak{su}(4,1)$. In terms of matrices
these algebras can be represented as 
\begin{equation}
  \begin{array}{rcl@{\qquad}l}
    X &=& \left(
      \begin{array}{@{}r@{\quad}r@{}}
        A&\Gamma \\
        \Gamma^\dagger & B
      \end{array}
    \right)
    &
    \mbox{where } A = -A^\dagger\in\CC^{p\times p},
    \quad
    B =-{B}^\dagger\in\CC^{q\times q},
    \\ [1 em]
    \Tr {A}+ \Tr {B} &=& 0,
    \quad
    {\Gamma} \in\CC^{p\times q}
    &
    \mbox{with } p=3 \mbox{ (respectively 4)} \mbox{ and } q=2
    \mbox{ (respectively 1)}.
  \end{array}
  \label{natsu}
\end{equation}
We shall call these ways of describing $\mathfrak{su}(p,q)$ the
``natural'' descriptions of $\mathfrak{su}(p,q)$. Introducing the
diagonal matrix 
\begin{equation}
  \label{supq}
  I_{p,\,q}=\left(\begin{array}{@{}l@{\quad}l@{}}
      Id^{p\times p}& \\
      &-Id^{q\times q}
    \end{array}\right),
\end{equation}
the conjugations defined by these subalgebras are given by:
\begin{equation}
\label{sn} \sigma_{p,\,q}( X)=-I_{p,\,q} X^\dagger
I_{p,\,q}.
\end{equation}

\subsubsection*{Vogan diagrams}

The Dynkin diagram of $\mathfrak{sl}(5,\CC)$ is of
$A_4$ type (see Figure~\ref{figure:Dynkinsl5C}).

\epubtkImage{Dynkinsl5C.png}{%
  \begin{figure}[htbp]
    \centerline{\includegraphics[width=50mm]{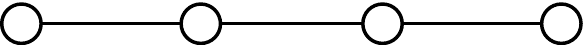}}
    \caption{The $A_{4}$ Dynkin diagram.}
    \label{figure:Dynkinsl5C}
  \end{figure}}

Let us first consider an $\mathfrak{su}(3,2)$ subalgebra. Diagonal
matrices define a Cartan subalgebra whose all elements are
compact. Accordingly all associated roots are imaginary. If we define
the positive roots using the natural ordering
$\epsilon_1>\epsilon_2>\epsilon_3>\epsilon_4>\epsilon_5$, the simple
roots $\alpha_1=\epsilon_1-\epsilon_2$,
$\alpha_2=\epsilon_2-\epsilon_3$, $\alpha_4=\epsilon_4-\epsilon_5$ are
compact but $\alpha_3=\epsilon_3-\epsilon_4$ is noncompact. The
corresponding Vogan diagram \index{Vogan diagram} is displayed in
Figure~\ref{figure:Vogsu32I}.

\epubtkImage{Vogsu32I.png}{%
  \begin{figure}[htbp]
    \centerline{\includegraphics[width=50mm]{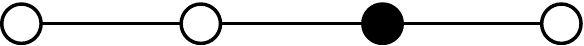}}
    \caption{A Vogan diagram associated to $\mathfrak{su}(3,2)$.}
    \label{figure:Vogsu32I}
  \end{figure}}

However, if instead of the natural order we define positive roots by
the rule $\epsilon_1>\epsilon_2>\epsilon_4>\epsilon_5>\epsilon_3$, the
simple positive roots are $\tilde\alpha_1=\epsilon_1-\epsilon_2$ and
$\tilde\alpha_3=\epsilon_4-\epsilon_5$ which are compact, and
$\tilde\alpha_2=\epsilon_2-\epsilon_4$ and
$\tilde\alpha_4=\epsilon_5-\epsilon_3$ which are noncompact. The
associated Vogan diagram is shown in Figure~\ref{figure:Vogsu32II}.

\epubtkImage{Vogsu32II.png}{%
  \begin{figure}[htbp]
    \centerline{\includegraphics[width=50mm]{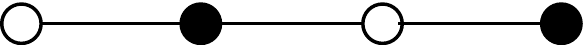}}
    \caption{Another Vogan diagram associated to $\mathfrak{su}(3,2)$.}
    \label{figure:Vogsu32II}
  \end{figure}}

Alternatively, the choice of order
$\epsilon_1>\epsilon_5>\epsilon_3>\epsilon_4>\epsilon_2$ leads to the
diagram in Figure~\ref{figure:Vogsu32III}.

\epubtkImage{Vogsu32III.png}{%
  \begin{figure}[htbp]
    \centerline{\includegraphics[width=50mm]{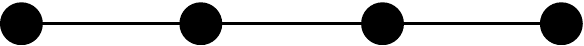}}
    \caption{Yet another Vogan diagram associated to $\mathfrak{su}(3,2)$.}
    \label{figure:Vogsu32III}
  \end{figure}}

There remain seven other possibilities, all describing the same
subalgebra $\mathfrak{su}(3,2)$. These are displayed in
Figure~\ref{figure:Vogsu32IV}.

\epubtkImage{Vogsu32IV.png}{%
  \begin{figure}[htbp]
    \centerline{\includegraphics[width=80mm]{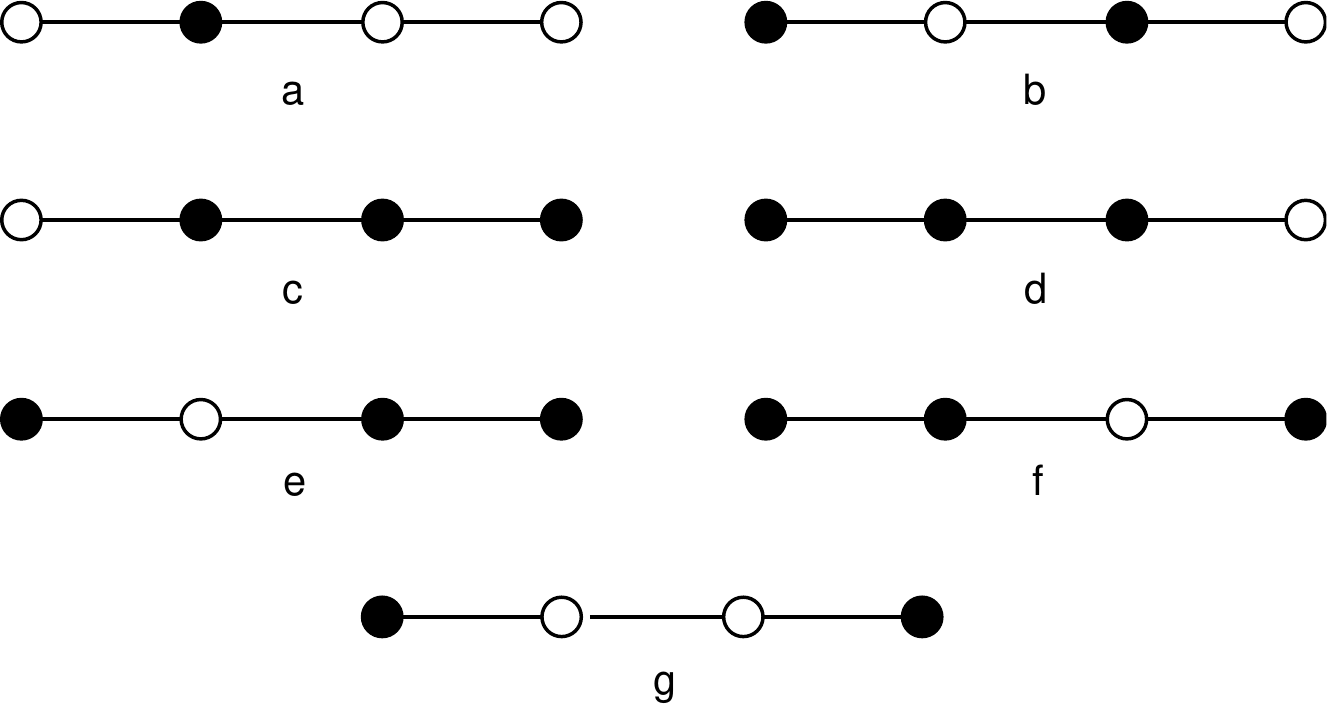}}
    \caption{The remaining Vogan diagrams associated to
      $\mathfrak{su}(3,2)$.}
    \label{figure:Vogsu32IV}
  \end{figure}}

In a similar way, we obtain four different Vogan diagrams for
$\mathfrak{su}(4,1)$, displayed in Figure~\ref{figure:Vogsu41}.

\epubtkImage{Vogsu41.png}{%
  \begin{figure}[htbp]
    \centerline{\includegraphics[width=80mm]{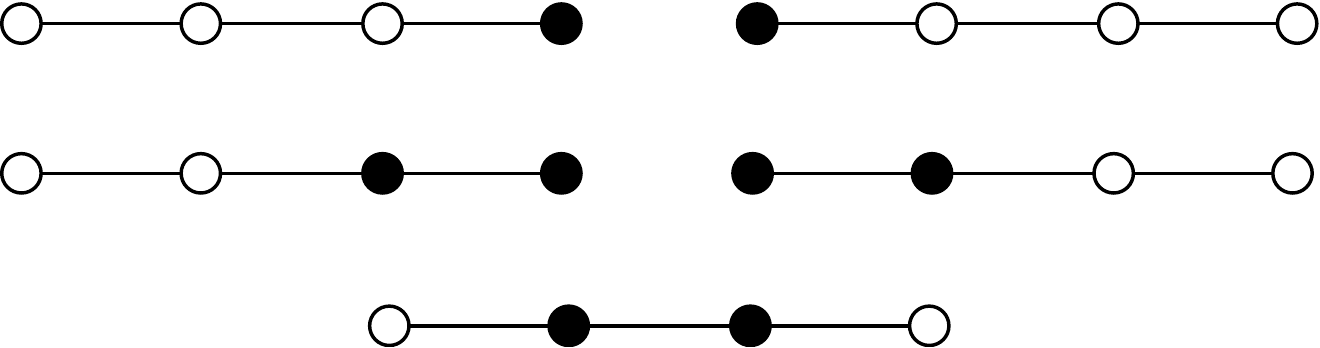}}
    \caption{The four Vogan diagrams associated to $\mathfrak{su}(4,1)$. }
    \label{figure:Vogsu41}
  \end{figure}}

Finally we have two non-isomorphic Vogan diagrams associated with
$\mathfrak{su}(5)$ and $\mathfrak{sl}(5,\RR)$. These are shown in
Figure~\ref{figure:Vogsu5sl5}.

\epubtkImage{Vogsu5sl5.png}{%
  \begin{figure}[htbp]
    \centerline{\includegraphics[width=100mm]{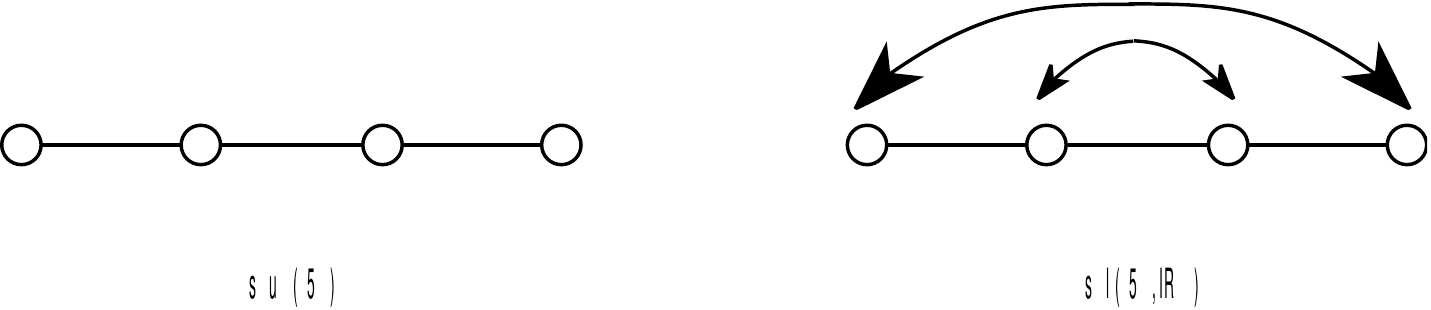}}
    \caption{The Vogan diagrams for $\mathfrak{su}(5)$
      and $\mathfrak{sl}(5,\RR)$.}
    \label{figure:Vogsu5sl5}
  \end{figure}}


\subsubsection{The Borel and de Siebenthal theorem}

As we just saw, the same real Lie algebra may yield different Vogan
diagrams only by changing the definition of positive roots. But
fortunately, a theorem of Borel and de Siebenthal tells us that we may
always choose the simple roots such that at most one of them is
noncompact~\cite{Knapp}. In other words, we may always assume that a
Vogan diagram possesses at most one black dot.

Furthermore, assume that the automorphism associated with the
Vogan diagram is the identity (no complex roots). Let
$\{\alpha_{p}\}$ be the basis of simple roots and $\{\Lambda_{q}\}$
its dual basis, i.e., $(\Lambda_{q}\vert\alpha_{p})=\delta_{p\,q}$.
Then the single painted simple root $\alpha_{p}$ may be chosen so
that there is no $q$ with $(\Lambda_{p}-\Lambda_{q}\vert
\Lambda_{q})>0$. This remark, which limits the possible simple root
that can be painted, is particularly helpful when analyzing the
real forms of the exceptional groups. For instance, from the Dynkin
diagram of $E_{8}$ (see Figure~\ref{figure:DynkinE8}), it is easy
to compute the dual basis and the matrix of scalar products
$B_{p\,q}=(\Lambda_{p}-\Lambda_{q}\vert \Lambda_{q})$.

\epubtkImage{DynkinE8.png}{%
  \begin{figure}[htbp]
    \centerline{\includegraphics[width=90mm]{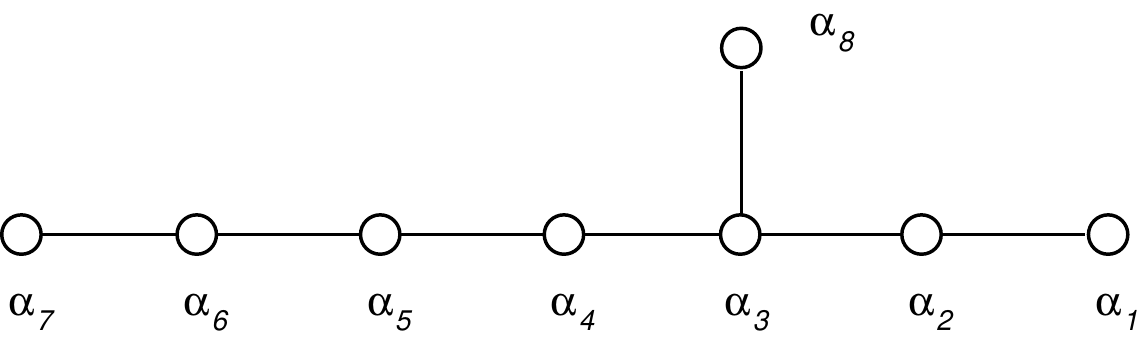}}
    \caption{The Dynkin diagram of $E_{8}$. Seen as a
      Vogan diagram, it corresponds to the maximally compact form of
      $E_{8}$.}
    \label{figure:DynkinE8}
  \end{figure}}

We obtain
\begin{equation}
  (B_{p\,q})=\left(
    \begin{array}{@{}r@{\quad}r@{\quad}r@{\quad}r@{\quad}r@{\quad}r@{\quad}r@{\quad}r@{}}
      -0 & -7 & -20 & -12 & -6 & -2 & -0 & -3 \\
      -3 & -0 & -10 & -4 & -0 & -2 & -2 & -2 \\
      -6 & -6 & -0 & -4 & -6 & -6 & -4 & -7 \\
      -4 &-2 & -6 & -0 & -3 & -4 & -3 & -4 \\
      -2 & -2 & -12 & -5 & -0 & -2 & -2 & -1 \\
      -0 & -6 & -18 & -10 & -4 & -0 & -1 & -2 \\
      -2 & -10 & -24 & -15 & -8 & -3 & -0 & -5 \\
      -1 & -4 & -15 & -8 & -3 & -0 & -1 & -0
    \end{array} \right),
\end{equation}
from which we see that there exist, besides the compact real form,
only two other non-isomorphic real forms of $E_{8}$, described by the
Vogan diagrams in Figure~\ref{figure:VoganE8}\epubtkFootnote{In the
  notation $H_{r(\sigma)}$ for a real form of the simple complex Lie
  algebra $H_{r}$, with the integer $\sigma$ referring to the
  signature.}.

\epubtkImage{VoganE8.png}{%
  \begin{figure}[htbp]
    \centerline{\includegraphics[width=80mm]{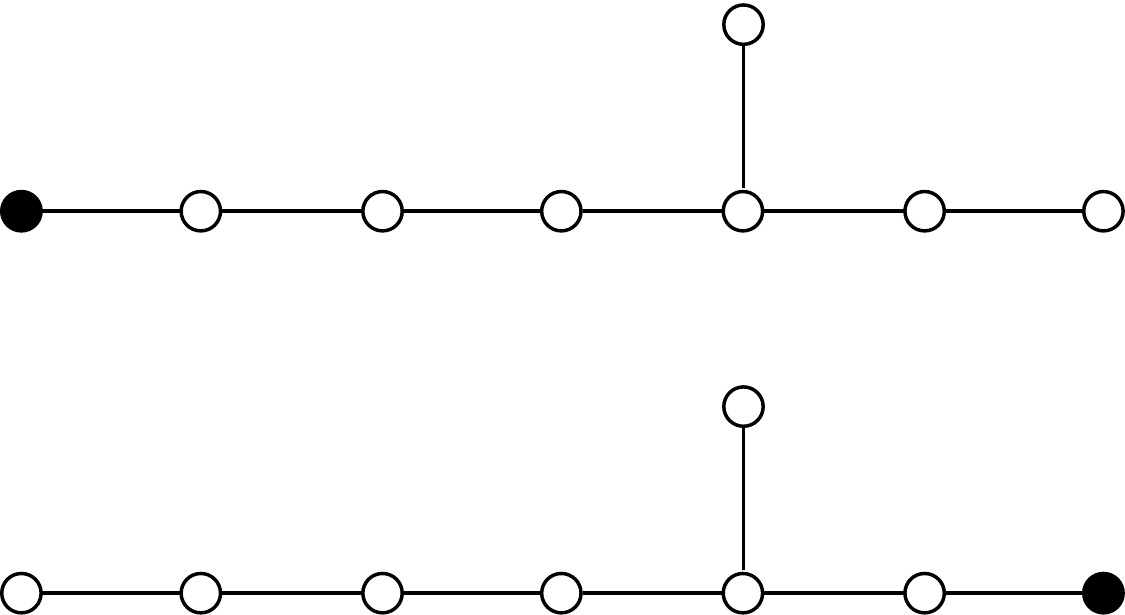}}
    \caption{Vogan diagrams of the two different noncompact real
      forms of $E_{8}$: $E_{8(-24)}$ and $E_{8(8)}$. The lower one corresponds
      to the split real form.}
    \label{figure:VoganE8}
  \end{figure}}


\subsubsection[Cayley transformations in $\mathfrak{su}(3,2)$]%
              {Cayley transformations in \boldmath $\mathfrak{su}(3,2)$}
\label{section:CTsu32}

Let us now illustrate the Cayley transformations. For this purpose,
consider again $\mathfrak{su}(3,2)$ with the imaginary diagonal
matrices as Cartan subalgebra and the natural ordering of the
$\epsilon_{k}$ defining the positive roots. As we have seen,
$\alpha_{3}=\epsilon_{3}-\epsilon_{4}$ is an imaginary noncompact
root. The associated $\mathfrak{sl}(2,\CC)$ generators are 
\begin{equation}
  E_{\alpha_{3}}=K^{3}_{4},
  \qquad
  \overline{ E_{\alpha_{3}}}=\sigma K^{3}_{4}=K^{4}_{3},
  \qquad
  i\,H_{3}.
\end{equation}
From the action of
$\alpha_{3}$ on the Cartan subalgebra $D=\spn \{i\,H_{k},\
k=1,\ldots,4\}$, we may check that 
\begin{equation}
  \ker (\alpha_{3}\vert D)=
  \spn \{ i\,H_{1},\, i(2\,H_{2}+H_{3}),\,i(2\,H_{4}+H_{3})\}, 
  \label{ka3}
\end{equation}
 and that $H^\prime=
( E_{\alpha_{3}}+\overline{ E_{\alpha_{3}}})=
(K^{3}_{4}+K^{4}_{3})$ is such that $\theta H^\prime=-H^\prime$
and $\sigma H^\prime= H^\prime$. Moreover $H^\prime$ commutes with
$\ker (\alpha_{3}\vert D)$. Thus 
\begin{equation}
  C=\ker (\alpha_{3}\vert D)\oplus \RR\,H^\prime 
  \label{CCSA}
\end{equation}
constitutes a $\theta$-stable Cartan subalgebra with one
noncompact dimension $H^\prime$. Indeed, we have
$B(H^\prime,H^\prime)=20$. If we compute the roots with respect to
this new Cartan subalgebra, we obtain twelve complex roots
(expressed in terms of their components in the basis dual to the
one implicitly defined by Equations~(\ref{ka3}) and~(\ref{CCSA}),
\begin{equation}
  \pm(i,-3i,i,\pm 1),
  \qquad
  \pm (0,i,-3i,\pm 1),
  \qquad
  \pm(i,i,-i,\pm 1),
\end{equation}
six imaginary roots 
\begin{equation}
  \pm i(2,-2,0,0),
  \qquad
  \pm i(1,-2,-2,0),
  \qquad
  \pm i(1,0,2,0),
\end{equation}
and a pair of real roots $\pm (0,0,0,2)$.

Let us first consider the Cayley transformation obtained using,
for instance, the real root $(0,0,0,2)$. An associated root
vector, belonging to $\GO$, reads 
\begin{equation}
  {E}=\left(
    \begin{array}{@{}r@{\quad}r@{\quad}r@{\quad}r@{\quad}r@{}}
      0&0&0&0&0 \\
      0&0&0&0&0 \\
      0&0&\frac i 2&-\frac i 2&0 \\
      0&0&\frac i 2&-\frac i 2&0 \\
      0&0&0&0&0 \\
    \end{array}
  \right).
\end{equation}
The corresponding compact Cartan generator is
\begin{equation}
  {h}=\left(
    \begin{array}{@{}r@{\quad}r@{\quad}r@{\quad}r@{\quad}r@{}}
      0&0&0&0&0 \\
      0&0&0&0&0 \\
      0&0& i &0&0 \\
      0&0&0&- i&0 \\
      0&0&0&0&0 \\
    \end{array}
  \right),
\end{equation}
which, together with the three generators in
Equation~(\ref{ka3}), provide a compact Cartan subalgebra of
$\mathfrak{su}(3,2)$.

If we consider instead the imaginary roots, we find for instance
that $K^{5}_{2}=-\tilde\theta K^{5}_{2}$ is a noncompact complex
root vector corresponding to the root $\beta =i(1,-2,-2,0)$. It
provides the noncompact generator $K^{2}_{5}+K^{5}_{2}$ which,
together with 
\begin{equation}
  \ker (\beta\vert C)=
  \spn \{i(2\,H_{1}+2\,H_{2}+H_{3}),\,i(2\,H_{1}+H_{3}+2\,H_{4}),\,
  K^{3}_{4}+K^{4}_{3}\},
\end{equation}
generates a maximally noncompact Cartan
subalgebra of $\mathfrak{su}(3,2)$. A similar construction can be
done using, for instance, the roots $\pm i(1,0,2,0)$, but not with
the roots $\pm i(2,-2,0,0)$ as their corresponding root vectors
$K^{1}_{2}$ and $K^{2}_{1}$ are fixed by $\tilde\theta$ and thus
are compact.


\subsubsection{Reconstruction}

We have seen that every real Lie algebra leads to a Vogan diagram. \index{Vogan diagram}
Conversely, every Vogan diagram defines a real Lie algebra. We
shall sketch the reconstruction of the real Lie algebras from the
Vogan diagrams here, referring the reader to~\cite{Knapp} for more
detailed information.

Given a Vogan diagram, the reconstruction of the associated real
Lie algebra proceeds as follows. From the diagram, which is a
Dynkin diagram with extra information, we may first construct the
associated complex Lie algebra, select one of its Cartan
subalgebras and build the corresponding root system. Then we may
define a compact real subalgebra according to Equation~(\ref{compact}).

We know the action of $\theta$ on the simple roots. This implies
that the set $\Delta$ of all roots is invariant under $\theta$.
This is proven inductively on the level of the roots, starting
from the simple roots (level 1). Suppose we have proven that the
image under $\theta$ of all the positive roots, up to level $n$
are in $\Delta$. If $\gamma$ is a root of level $n+1$, choose a
simple root $\alpha$ such that $(\gamma\vert\alpha)>0$. Then the
Weyl transformed root $s_{\alpha}\gamma = \beta$ is also a
positive root, but of smaller level. Since $\theta (\alpha)$ and
$\theta (\beta)$ are then known to be in $\Delta$, and since the
involution acts as an isometry, $\theta
(\gamma)=s_{\theta(\alpha)}(\theta(\beta))$ is also in $\Delta$.

One can transfer by duality the action of $\theta$ on $\cH^*$ to
the Cartan subalgebra $\cH$, and then define its action on the
root vectors associated to the simple roots according to the
rules 
\begin{equation}
  \theta E_{\al}=\left\{
    \begin{array}{l@{\qquad}l}
      E_{\al} &
      \mbox{if }\alpha \mbox{ is unpainted and invariant},
      \\
      - E_{\al} &
      \mbox{if } \alpha \mbox{ is painted and invariant},
      \\
      - E_{\theta[\alpha]} &
      \mbox{if } \alpha \mbox{ belongs to a 2-cycle}.
    \end{array}
  \right.
\end{equation}
These rules extend $\theta$ to an involution of $\cG$.

This involution is such that $\theta E_{\al}=a_{\alpha}\,
E_{\theta[\alpha]}$, with $a_{\alpha}=\pm 1$ \epubtkFootnote{The
  coefficients $a_{\al}$ are determined from the commutation relations
  as follows: $N_{\al, \be}a_{\al+\be}=N_{\theta[\al], \theta[\be]}
  a_{\al} a_{\be}$. Moreover, because $\theta$ is an automorphism of
  the root lattice we have $N^{2}_{\al,\be}=N^{2}_{\theta[\al],
    \theta[\be]},$ and so if $a_{\al}$ and $a_{\be}$ are equal to $\pm
  1$, then so is $a_{\al +\be}$. But since this is true for the simple
  roots it remains true for all roots.}. Furthermore it globally fixes
$\mf{c}_{0}$, $\theta\mf{c}_{0}=\mf{c}_{0}$. Let $\cK$ and $\cP$ be
the $+1$ or $-1$ eigenspaces of $\theta$ in $\cG=\cK\oplus\cP$. Define
$\KO=\mf{c}_{0}\cap\cK$ and $\PO=i(\mf{c}_{0}\cap \cP)$ so that
$\mf{c}_{0}=\KO\oplus i\,\PO$. Set 
\begin{equation}
  \GO=\KO\oplus\PO.
\end{equation}
Using $\theta \mf{c}_{0}=\mf{c}_{0}$, one then verifies that $\GO$
constitutes the desired real form of $\cG$~\cite{Knapp}.


\subsubsection[Illustrations: $\mathfrak{sl}(4,\RR)$ versus $\mathfrak{sl}(2,\HH)$]%
              {Illustrations: \boldmath $\mathfrak{sl}(4,\RR)$ versus $\mathfrak{sl}(2,\HH)$}

We shall exemplify the reconstruction of real algebras from Vogan
diagrams by considering two examples of real forms of
$\mathfrak{sl}(4,\CC)$. The diagrams are shown in
Figure~\ref{figure:Vogsl4Rsl2H}.

\epubtkImage{Vogsl4Rsl2H.png}{%
  \begin{figure}[htbp]
    \centerline{\includegraphics[width=90mm]{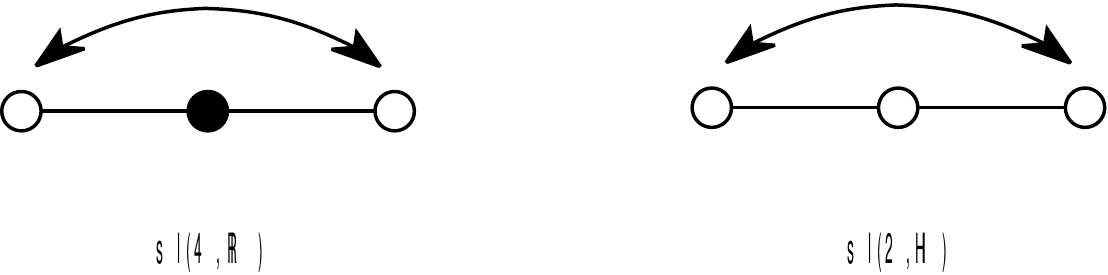}}
    \caption{The Vogan diagrams associated to a $\mathfrak{sl}(4\RR)$
      and $\mathfrak{sl}(2\HH)$ subalgebra.}
    \label{figure:Vogsl4Rsl2H}
  \end{figure}}

The $\theta$ involutions they describe are (the upper signs
correspond to the left-hand side diagram, the lower signs to the
right-hand side diagram): 
\begin{equation}
  \begin{array}{rcl@{\qquad}rcl@{\qquad}rcl}
    \theta H_{\alpha_{1}} &=& H_{\alpha_{3}}, &
    H_{\alpha_{2}} &=& H_{\alpha_{2}}, &
    \theta H_{\alpha_{3}} &=& H_{\alpha_{1}}
    \\ [0.25 em]
    \theta E_{\alpha_{1}} &=& E_{\alpha_{3}}, &
    \theta E_{\alpha_{2}} &=& \mp E_{\alpha_{2}}, &
    \theta E_{\alpha_{3}} &=& E_{\alpha_{1}}.
  \end{array}
\end{equation}
Using the commutations relations 
\begin{equation}
  \begin{array}{rcl}
    \Lb{E_{\alpha_{1}}}{E_{\alpha_{2}}} &=& E_{\alpha_{1}+\alpha_{2}},
    \\ [0.25 em]
    \Lb{E_{\alpha_{2}}}{E_{\alpha_{3}}} &=& E_{\alpha_{2}+\alpha_{3}},
    \\ [0.25 em]
    \Lb{E_{\alpha_{1}+\alpha_{2}}}{ E_{\alpha_{3}}} &=&
    E_{\alpha_{1}+\alpha_{2}+\alpha_{3}}=
    \Lb{ E_{\alpha_{1}}}{ E_{\alpha_{2}+\alpha_{3}}}
  \end{array}
\end{equation}
we obtain 
\begin{equation}
  \theta E_{\alpha_{1}+\alpha_{2}}=\pm E_{\alpha_{2}+\alpha_{3}},
  \qquad
  \theta E_{\alpha_{2}+\alpha_{3}}=\pm E_{\alpha_{1}+\alpha_{2}},
  \qquad
  \theta E_{\alpha_{1}+\alpha_{2}+\alpha_{3}}=
  \mp E_{\alpha_{1}+\alpha_{2}+\alpha_{3}}.
\end{equation}
Let us consider the left-hand side diagram. The corresponding $+1$
$\theta$-eigenspace $\cK$ has the following realisation,
\begin{equation}
 \cK = \spn \left\{ H_{\alpha_{1}}+ H_{\alpha_{3}},\, H_{\alpha_{2}},\,
   E_{\alpha_{1}}+ E_{\alpha_{3}},\, E_{-\alpha_{1}}+ E_{-\alpha_{3}},\,
   E_{\alpha_{1}+\alpha_{2}}+ E_{\alpha_{2}+\alpha{3}},\,
   E_{-\alpha_{1}-\alpha_{2}}+ E_{-\alpha_{2}-\alpha{3}}\right\},
\end{equation}
and the $-1$ $\theta$-eigenspace $\cP$ is given by
\begin{eqnarray}
  \cP&=&\spn \{ H_{\alpha_{1}}- H_{\alpha_{3}},\,
  E_{\pm\alpha_{2}},\, E_{\alpha_{1}}- E_{\alpha_{3}},\,
  E_{-\alpha_{1}}- E_{-\alpha_{3}},
  \nonumber
  \\
  & & \qquad ~\,
  E_{\alpha_{1}+\alpha_{2}}- E_{\alpha_{2}+\alpha_{3}},\,
  E_{-\alpha_{1}-\alpha_{2}}- E_{-\alpha_{2}-\alpha_{3}},\,
  E_{\pm(\alpha_{1}+\alpha_{2}+\alpha_{3})}\}.
\end{eqnarray}
The intersection $\mf{c}_{0}\cap\cK$ then leads to the
$\mathfrak{so}(4,\RR)=\mathfrak{so}(3,\RR)\oplus
\mathfrak{so}(3,\RR)$ algebra
\begin{eqnarray}
  \KO&=&\spn \{i( H_{\alpha_{1}}+ H_{\alpha_{3}}),\,
  (E_{\alpha_{1}}+ E_{\alpha_{3}}-E_{-\alpha_{1}}- E_{-\alpha_{3}}),\,
  i( E_{\alpha_{1}}+ E_{\alpha_{3}}+ E_{-\alpha_{1}}+ E_{-\alpha_{3}})\}
  \nonumber
  \\
  & &\oplus\, \spn \big\{i( H_{\alpha_{1}}+2\, H_{\alpha_{2}}+ H_{\alpha_{3}}),\,
  ( E_{\alpha_{1}+\alpha_{2}}+ E_{\alpha_{2}+\alpha_{3}}-
  E_{-(\alpha_{1}+\alpha_{2})}- E_{-(\alpha_{2}+\alpha_{3})}),
  \nonumber
  \\
  & & \qquad \qquad
  i( E_{\alpha_{1}+\alpha_{2}}+ E_{\alpha_{2}+\alpha_{3}}+
  E_{-(\alpha_{1}+\alpha_{2})}+ E_{-(\alpha_{2}+\alpha_{3})})\},
\end{eqnarray}
and the remaining noncompact generator subspace
$\PO=i(\mf{c}_{0}\cap\cP)$ becomes
\begin{eqnarray}
  \PO&=& \spn \{H_{\alpha_{1}}- H_{\alpha_{3}},\,
  (E_{\alpha_{1}}- E_{\alpha_{3}}+ E_{-\alpha_{1}}- E_{-\alpha_{3}}),\,
  i( E_{\alpha_{1}}- E_{\alpha_{3}}- E_{-\alpha_{1}}+ E_{-\alpha_{3}}),
  \nonumber
  \\
  & & \qquad ~\;
  (E_{\alpha_{1}+\alpha_{2}}- E_{\alpha_{2}+\alpha_{3}}+
  E_{-(\alpha_{1}+\alpha_{2})}- E_{-(\alpha_{2}+\alpha_{3})}),
  \nonumber
  \\
  & & \qquad ~\;
  i(E_{\alpha_{1}+\alpha_{2}}- E_{\alpha_{2}+\alpha_{3}}-
  E_{-(\alpha_{1}+\alpha_{2})}+ E_{-(\alpha_{2}+\alpha_{3})}),\,
  E_{\alpha_{2}}+ E_{-\alpha_{2}},\,i( E_{\alpha_{2}}- E_{-\alpha_{2}}),
  \nonumber
  \\
  & & \qquad ~\;
  E_{\alpha_{1}+\alpha_{2}+\alpha_{3}}+ E_{-(\alpha_{1}+\alpha_{2}+\alpha_{3})},\,
  i(E_{\alpha_{1}+\alpha_{2}+\alpha_{3}}-
  E_{-(\alpha_{1}+\alpha_{2}+\alpha_{3})})\}.
\end{eqnarray}

Doing the same exercise for the second diagram, we obtain the real
algebra $\mathfrak{sl}(2,\HH)$ with
$\KO=\mathfrak{so}(5,\RR)=\mathfrak{sp}(4,\RR)$, which is a
10-parameter compact subalgebra, and $\PO$ given by
\begin{eqnarray}
  \PO&=&\spn \{ H_{\alpha_{1}}- H_{\alpha_{3}},\,
  (E_{\alpha_{1}}- E_{\alpha_{3}}+E_{-\alpha_{1}}- E_{-\alpha_{3}}),\,
  i( E_{\alpha_{1}}- E_{\alpha_{3}}-E_{-\alpha_{1}}+ E_{-\alpha_{3}}),
  \nonumber
  \\
  & & \qquad ~\;
  (E_{\alpha_{1}+\alpha_{2}}+ E_{\alpha_{2}+\alpha_{3}}+
  E_{-(\alpha_{1}+\alpha_{2})}+ E_{-(\alpha_{2}+\alpha_{3})}),
  \nonumber
  \\
  & & \qquad ~\;
  i( E_{\alpha_{1}+\alpha_{2}}+ E_{\alpha_{2}+\alpha_{3}}-
  E_{-(\alpha_{1}+\alpha_{2})}- E_{-(\alpha_{2}+\alpha_{3})})\}.
\end{eqnarray}


\subsubsection{A pictorial summary -- All real simple Lie algebras
  (Vogan diagrams)}
\index{Vogan diagram}

The following tables provide all real simple Lie algebras and the
corresponding Vogan diagrams. The restrictions imposed on some of
the Lie algebra parameters eliminate the consideration of
isomorphic algebras. See~\cite{Knapp} for the derivation.


\begin{table}[htbp]
  \caption{Vogan diagrams ($A_{n}$ series)}
  \vspace{0.5 em}
  \centering
\begin{tabular}{c|c|c}
\toprule
$A_{n}$ series, $n\geq 1$ & Vogan diagram & Maximal compact subalgebra \\
\midrule
\parbox[t][10mm][t]{35mm}{
\centering $\mathfrak{su}(n+1)$}
&
\invisible{5}{3}
\parbox[t][10mm][t]{45mm}{
\centering \includegraphics[width=40mm]{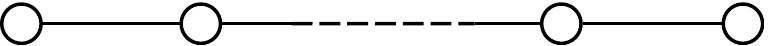}\\
\scriptsize No painted root}
&
$\mathfrak{su}(n+1)$ \\
\midrule
$\mathfrak{su}(p,q)$
&
\invisible{10}{3}
\parbox[t][15mm][t]{45mm}{
\centering \includegraphics[width=40mm]{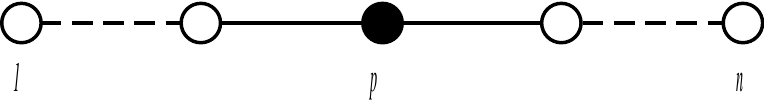}\\
\scriptsize Only the $p^{th}$ root is painted}
&
$\mathfrak{su}(p)\oplus \mathfrak{su}(q)\oplus u(1)$\\
\midrule
$\mathfrak{sl}(2n,\RR)$
&
\invisible{17}{3}
\parbox[t][20mm][t]{45mm}{
\centering \includegraphics[width=40mm]{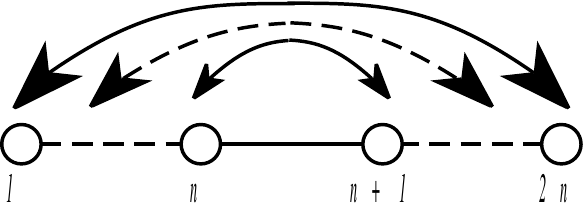}\\
\scriptsize Odd number of roots}
&
$\mathfrak{so}(2n)$\\
\midrule
$\mathfrak{sl}(2n+1,\RR)$
&
\invisible{17}{3}
\parbox[t][20mm][t]{45mm}{
\centering \includegraphics[width=40mm]{Vogslnplus1R}\\
\scriptsize Even number of (unpainted) roots}
& $\mathfrak{so}(2n+1)$ \\
\midrule
$\mathfrak{sl}(n+1,\HH)$
&
\invisible{17}{3}
\parbox[t][20mm][t]{45mm}{
\centering \includegraphics[width=45mm]{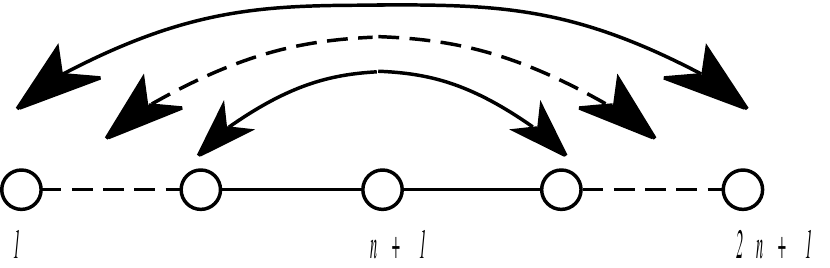}\\
\scriptsize Odd number of (unpainted) roots}
& $\mathfrak{sp}(n+1)$\\ 
\bottomrule
\end{tabular}
\end{table}

\begin{table}[htbp]
  \caption{Vogan diagrams ($B_{n}$ series)}
  \vspace{0.5 em}
  \centering
\begin{tabular}{c|c|c}
\toprule
$B_{n}$ series, $n\geq 2$ & Vogan diagram & Maximal compact subalgebra\\
\midrule
$\mathfrak{so}(2n+1)$
&
\invisible{5}{3}
\parbox[t][10mm][t]{45mm}{
\centering \includegraphics[width=40mm]{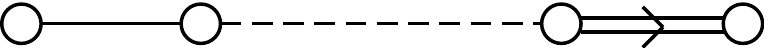}\\
\scriptsize No painted root}
&
$\mathfrak{so}(2n+1)$ \\
\midrule
\invisible{10}{3}
\parbox[t][15mm][t]{35mm}{
\centering $\mathfrak{so}(p,q)$\\
\scriptsize $p\leq n-\frac12 ,\, q=2n +1-p$}
&
\invisible{10}{3}
\parbox[t][20mm][t]{45mm}{
\includegraphics[width=45mm]{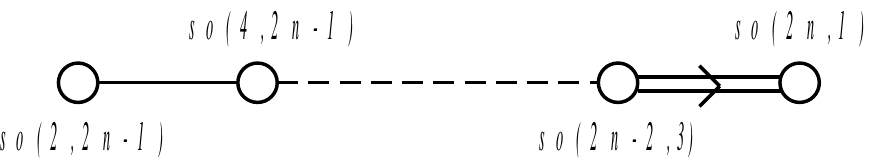}\\
\scriptsize Each of the roots, once painted, leads to the algebra
mentioned under it.}
&
$\mathfrak{so}(p)\oplus \mathfrak{so}(q)$ \\
\bottomrule
\end{tabular}
\end{table}

\begin{table}[htbp]
  \caption{Vogan diagrams ($C_{n}$ series)}
  \vspace{0.5 em}
  \centering
\begin{tabular}{c|c|c}
\toprule
$C_{n}$ series, $n\geq 3$ & Vogan diagram & Maximal compact subalgebra\\
\midrule
$\mathfrak{sp}(n)$
&
\invisible{5}{3}
\parbox[t][10mm][t]{45mm}{
\centering \includegraphics[width=40mm]{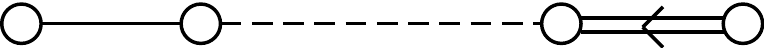}\\
\scriptsize No painted root,}
&
$\mathfrak{sp}(n)$ \\
\midrule
\parbox[t][20mm][t]{35mm}{
\centering $\mathfrak{sp}(p,\,q)$\\
{\scriptsize $0<p\leq\frac n2,\ q=n-p$}\\
 $\mathfrak{sp}(n,\RR)$}
&
\invisible{10}{3}
\parbox[t][20mm][t]{45mm}{
\includegraphics[width=45mm]{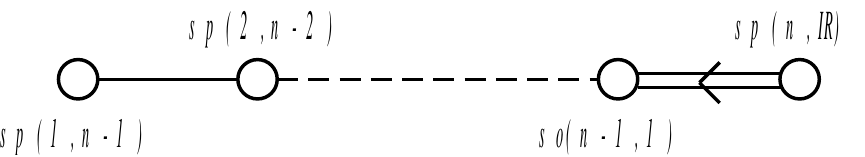}\\
\scriptsize Each of the roots, once painted, corresponds to the
algebra mentioned near it.}
&
\parbox[t][20mm][t]{30mm}{
\centering $\mathfrak{sp}(p)\oplus \mathfrak{sp}(q)$\\
\strut\\
$\mf{u}(n)$}\\
\bottomrule
\end{tabular}
\end{table}

\begin{table}[htbp]
  \caption{Vogan diagrams ($D_{n}$ series)}
  \vspace{0.5 em}
  \centering
\begin{tabular}{c|c|c}
\toprule
$D_{n}$ series, $n\geq 4$ & Vogan diagram & Maximal compact subalgebra\\
\midrule
\parbox[t][10mm][t]{35mm}{
\centering $\mathfrak{so}(2n)$}
&
\invisible{20}{3}
\parbox[t][15mm][t]{50mm}{
\centering \includegraphics[width=40mm]{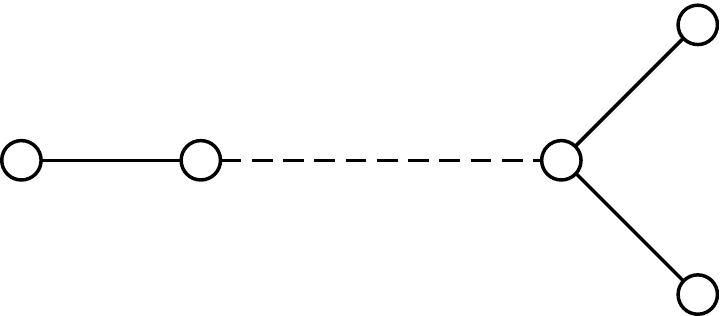}\\
\scriptsize No painted root}
&
$\mathfrak{so}(2n)$ \\
\midrule
\parbox[t][15mm][t]{35mm}{
\centering $\mathfrak{so}(2p,2q)$ \\
{\scriptsize $ 0<p \leq \frac n2,\, q=n-p $}\\
$\mf{so}^*(2n)$}
&
\invisible{23}{3}
\parbox[t][20mm][t]{50mm}{
\includegraphics[width=50mm]{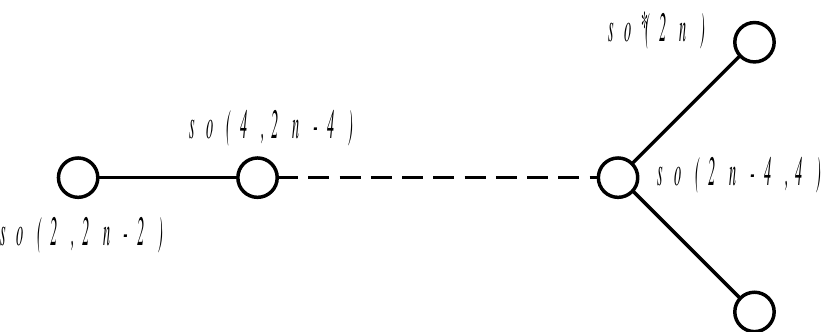}\\
\scriptsize Each of the roots, once painted, corresponds to the
algebra mentioned near it.}
&
\parbox[t][20mm][t]{30mm}{
\centering $\mathfrak{so}(2p)\oplus \mathfrak{so}(2q)$\\
\strut \\
$\mf{u}(n)$}\\
\midrule
\parbox[t][20mm][t]{35mm}{
\centering $\mathfrak{so}(2p+1,2q+1)$ \\
\scriptsize $ 0<p \leq\frac {n-1}2$,\\
$q=n-p-1 $}
&
\invisible{20}{3}
\parbox[t][20mm][t]{50mm}{
\includegraphics[width=50mm]{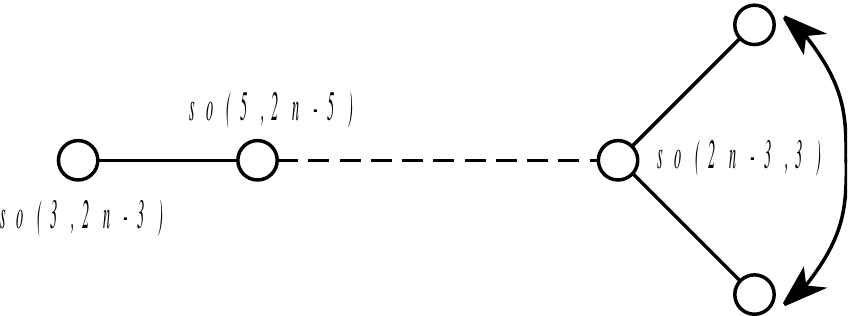}\\
\scriptsize Each of the roots, once painted, corresponds to the
algebra mentioned near it.\\ No root painted corresponds to\\
$\mathfrak{so}(1,2n-1)$.}
&
$\mathfrak{so}(2p+1)\oplus \mathfrak{so}(2q+1)$\\
\bottomrule
\end{tabular}
\end{table}

\begin{table}[htbp]
  \caption{Vogan diagrams ($G_{2}$ series)}
  \vspace{0.5 em}
  \centering
\begin{tabular}{c|c|c}
\toprule
$G_2$ & Vogan diagram & Maximal compact subalgebra \\
\midrule
\parbox[t][15mm][t]{35mm}{
\centering $G_{2}$
}
&
\invisible{7}{3}
\parbox[t][15mm][t]{50mm}{
\centering \includegraphics[width=15mm]{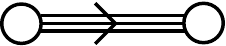}\\
\scriptsize No painted root, providing the real compact form}
&
$G_{2}$ \\
\midrule
$G_{2(2)}$
&
\invisible{8}{3}
\centering \includegraphics[width=15mm]{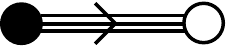}
&
$\mathfrak{su}(2)\oplus \mathfrak{su}(2)$\\
\bottomrule
\end{tabular}
\end{table}

\begin{table}[htbp]
  \caption{Vogan diagrams ($F_{4}$ series)}
  \vspace{0.5 em}
  \centering
\begin{tabular}{c|c|c}
\toprule
$F_4$ series & Vogan diagram & Maximal compact subalgebra\\
\midrule
\parbox[t][15mm][t]{35mm}{
\centering $F_{4}$
}
&
\invisible{8}{3}
\parbox[t][10mm][t]{50mm}{
\centering \includegraphics[width=40mm]{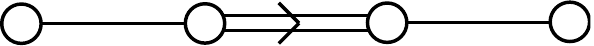}\\
\scriptsize No painted root, providing the real compact form}
&
$F_4$ \\
\midrule
$F_{4(4)}$
&\invisible{8}{3}
\centering \includegraphics[width=40mm]{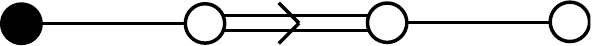}
&
$\mathfrak{sp}(3)\oplus \mathfrak{su}(2)$\\
\midrule
$F_{4(-20)}$
&
\invisible{8}{3}
\centering \includegraphics[width=40mm]{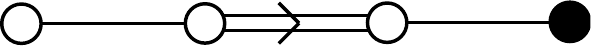}
&
$\mathfrak{so}(9)$\\
\bottomrule
\end{tabular}
\end{table}

\begin{table}[htbp]
  \caption{Vogan diagrams ($E_{6}$ series)}
  \vspace{0.5 em}
  \centering
\begin{tabular}{c|c|c}
\toprule
$E_{6}$ & Vogan diagram & Maximal compact subalgebra\\
\midrule
\parbox[t][15mm][t]{35mm}{
\centering $E_{6}$
}
&\invisible{15}{3}
\parbox[t][20mm][t]{45mm}{
\centering \includegraphics[width=40mm]{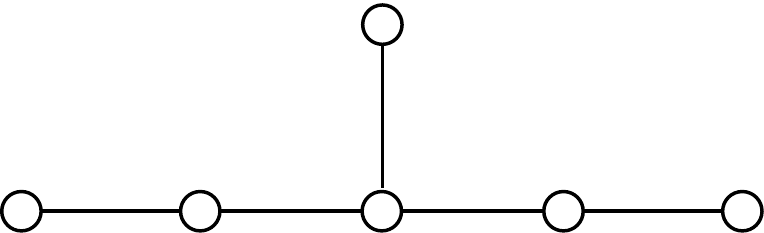}\\
\scriptsize No painted root, providing the real compact form}
&
$E_{6}$ \\
\midrule
$E_{6(6)}$
&
\invisible{23}{3}
\centering \includegraphics[width=40mm]{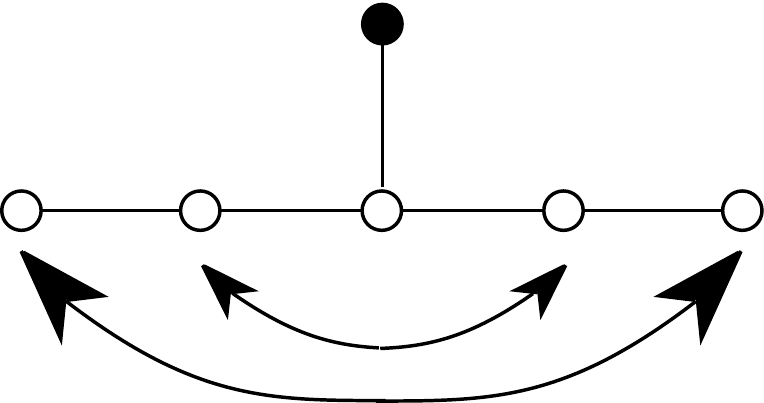}
&
$\mathfrak{sp}(4)$\\
\midrule
$E_{6(2)}$
&
\invisible{15}{3}
\centering \includegraphics[width=40mm]{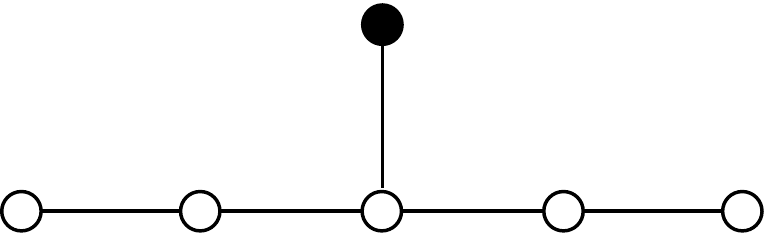}
&
$\mathfrak{su}(6)\oplus \mathfrak{su}(2)$\\
\midrule
$E_{6(-14)}$
&
\invisible{15}{3}
\centering \includegraphics[width=40mm]{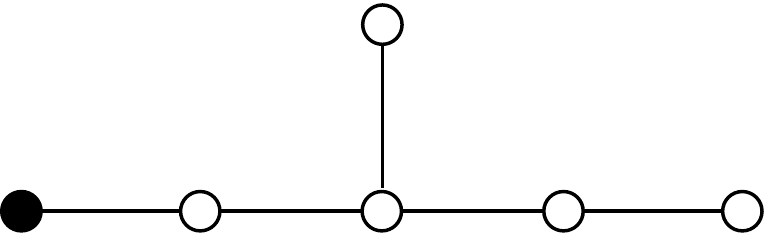}
&
$\mathfrak{su}(10)\oplus \mf{u}(1)$\\
\midrule
$E_{6(-26)}$
&
\invisible{23}{3}
\centering \includegraphics[width=40mm]{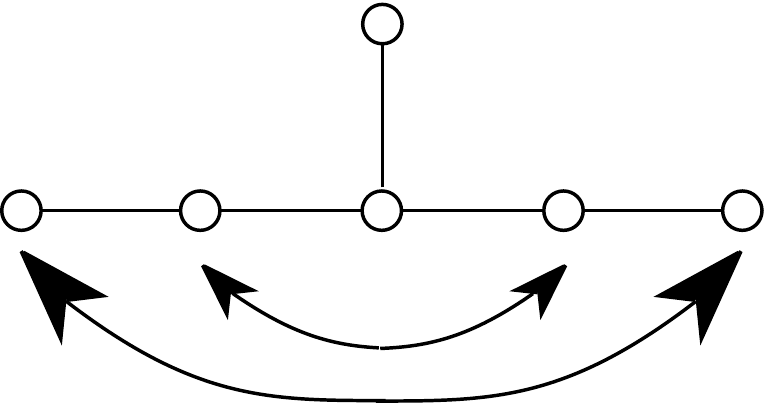}
&
$F_{4}$\\
\bottomrule
\end{tabular}
\end{table}

\begin{table}[htbp]
  \caption{Vogan diagrams ($E_{7}$ series)}
  \vspace{0.5 em}
  \centering
\begin{tabular}{c|c|c}
\toprule
$E_{7}$ & Vogan diagram & Maximal compact subalgebra\\
\midrule
\parbox[t][15mm][t]{35mm}{
\centering $E_{7}$
}
&
\invisible{12}{3}
\parbox[t][18mm][t]{45mm}{
\centering \includegraphics[width=40mm]{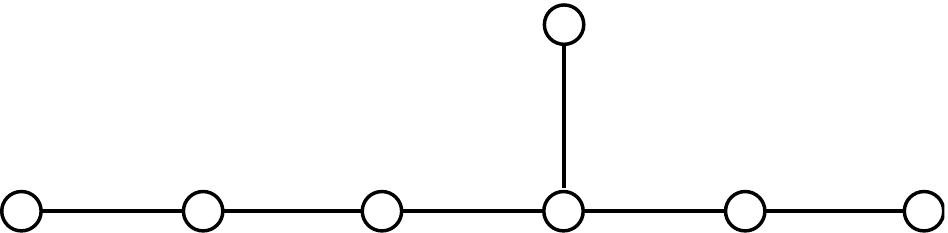}\\
\scriptsize No painted root, providing the real compact form}
&
$E_{7}$\\ 
\midrule
$E_{7(7)}$
&
\invisible{12}{3}
\centering \includegraphics[width=40mm]{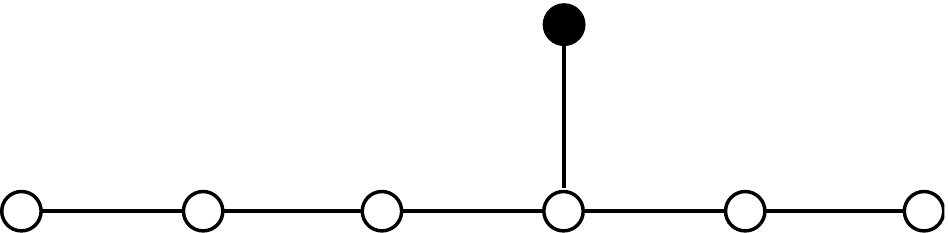}
&
$\mathfrak{su}(8)$\\
\midrule
$E_{7(43)}$
&
\invisible{12}{3}
\centering \includegraphics[width=40mm]{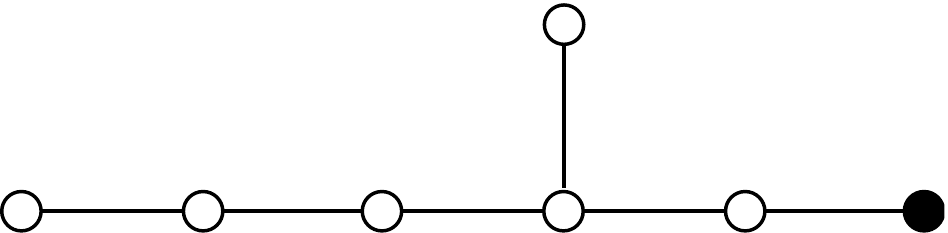}
&
$\mathfrak{so}(12)\oplus \mathfrak{su}(2)$\\
\midrule
$E_{7(-25)}$
&
\invisible{12}{3}
\centering \includegraphics[width=40mm]{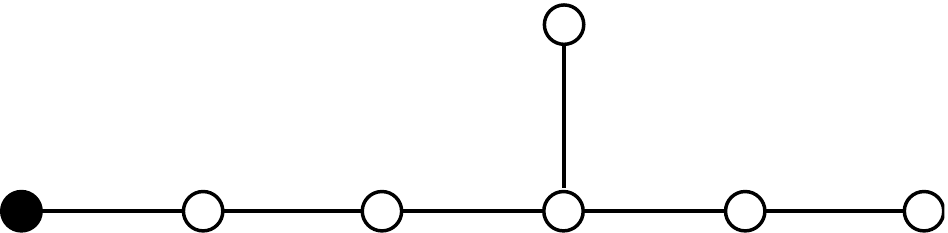}
&
$ E_{6}\oplus \mf{u}(1)$\\
\bottomrule
\end{tabular}
\end{table}

\begin{table}[htbp]
  \caption{Vogan diagrams ($E_{8}$ series)}
  \vspace{0.5 em}
  \centering
\begin{tabular}{c|c|c}
\toprule
$E_{8}$ & Vogan diagram & Maximal compact subalgebra \\
\midrule
\parbox[t][15mm][t]{35mm}{
\centering $E_{8}$
}
&
\invisible{10}{3}
\parbox[t][18mm][t]{45mm}{
\centering \includegraphics[width=40mm]{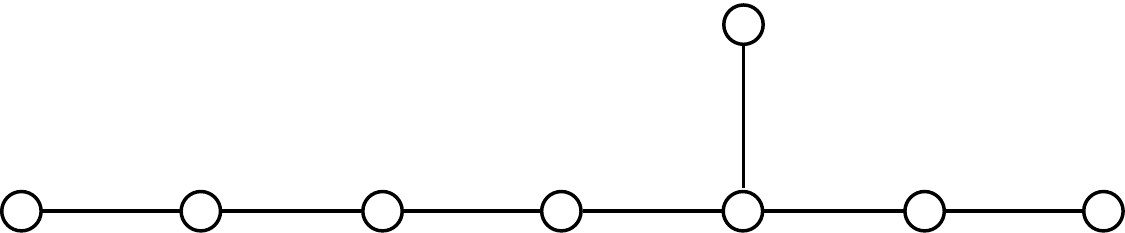}\\
\scriptsize No painted root, providing the real compact form}
&
$E_{8}$ \\
\midrule
$E_{8(8)}$
&
\invisible{12}{3}
\centering \includegraphics[width=40mm]{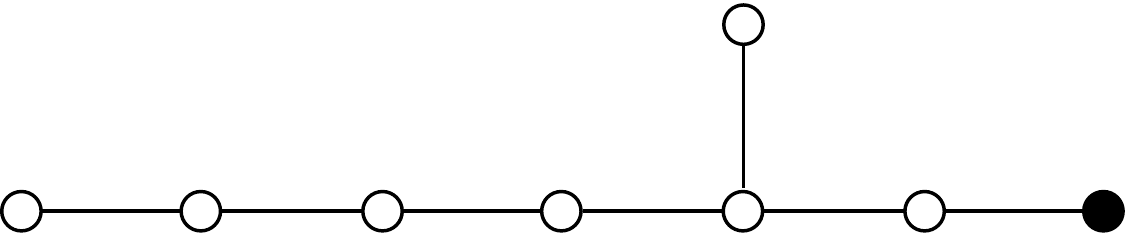}
&
$\mathfrak{so}(16)$\\
\midrule
$E_{8(-24)}$
&
\invisible{12}{3}
\centering \includegraphics[width=40mm]{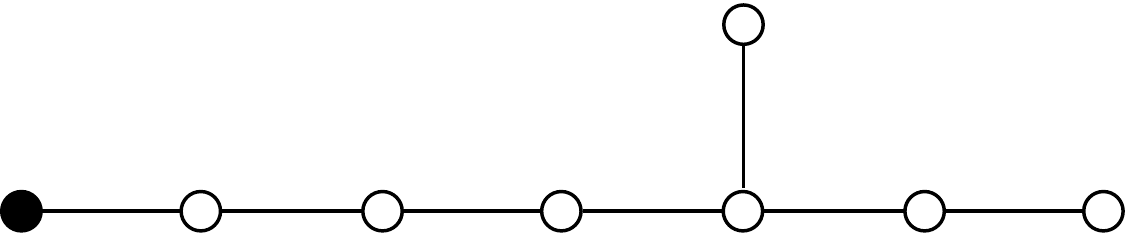}
&
$E_{7}\oplus \mathfrak{su}(2)$\\
\bottomrule
\end{tabular}
\end{table}

Using these diagrams, the matrix $I_{p,q}$ defined by Equation~(\ref{supq}),
and the three matrices
\begin{eqnarray}
  J_{n}&=&\left(
    \begin{array}{@{}c@{\qquad}c@{}}
      0&Id^{n\times n} \\
      -Id^{n\times n}&0
    \end{array}
  \right),
  \\ [1 em]
  K_{p,q}&=&\left(
    \begin{array}{@{}c@{\qquad}c@{\qquad}c@{\qquad}c@{}}
      Id^{p\times p}&0&0&0 \\
      0& -Id^{q\times q}&0&0 \\
      0&0&Id^{p\times p}&0 \\
      0&0&0&-Id^{q\times q}
    \end{array}
  \right),
  \\ [1 em]
  L_{p,q}=K_{p,q}\,J_{p+q}&=& \left(
    \begin{array}{@{}c@{\qquad}c@{\qquad}c@{\qquad}c@{}}
      0&0&Id^{p\times p}&0 \\
      0& 0&0&-Id^{q\times q} \\
      -Id^{p\times p}&0&0&0 \\
      0&Id^{q\times q}&0&0
    \end{array}
  \right),
\end{eqnarray}
we may check that the involutive automorphisms
of the classical Lie algebras are all conjugate to one of the types
listed in Table~\ref{table:AllAutomorphisms}.

\begin{table}
  \caption[List of all involutive automorphisms (up to conjugation) of
  the classical compact real Lie algebras.]{List of all involutive
  automorphisms (up to conjugation) of the classical compact real Lie
  algebras~\cite{Helgason}. The first column gives the
  complexification $\mathfrak{u}_0^{\mathbb{C}}$ of the  compact real
  algebra $\mathfrak{u}_0$, the second $\mathfrak{u}_0$, the third the
  involution $\tau$ that $\mathfrak{u}_0$ defines in
  $\mathfrak{u}^\CC$, and the fourth a non-compact real subalgebra
  $\mathfrak{g}_0$ of $\mathfrak{u}^\CC$ aligned with the compact
  one. In the second table, the second column  displays the involution
  that $\mathfrak{g}_0$ defines on $\mathfrak{u}^\CC$, the third the
  involutive automorphism of $\mathfrak{u}_0$, i.e, the Cartan
  conjugation \index{Cartan conjugation|bb} $\theta=\sigma\tau$, and
  the last column indicates the common compact subalgebra
  $\mathfrak{k}_0$ of $\mathfrak{u}_0=\mathfrak{k}_0\oplus
  i\,\mathfrak{p}_0$ and $\mathfrak{g}_0=\mathfrak{k}_0\oplus
  \mathfrak{p}_0$.}
  \renewcommand{\arraystretch}{1.2}
  \vspace{0.5 em}
  \centering
  \begin{tabular}{c|c|c|c}
    \toprule
    $\mathfrak{u}^{\CC}$ &
    $\mathfrak{u}_0$ &
    $\tau$ &
    $\mathfrak{g}_0$ \\
    \midrule
    $\mathfrak{sl}(n,\CC)$ &
    $ \mathfrak{su}(n) $ &
    $-X^\dagger$ &
    $A\,I$\ $\mathfrak{sl}(n,\RR)$ \\ [1 em]
    $\mathfrak{sl}(2n,\CC)$ &
    $ \mathfrak{su}(2n) $ &
    $-X^\dagger$ &
    $A\,II$\ $\mathfrak{su}^*(2n)$ \\ [1 em]
    $\mathfrak{sl}(p+q,\CC)$ &
    $ \mathfrak{su}(p+q) $ &
    $-X^\dagger$ &
    $A\,III$\ $\mathfrak{su}(p,q)$ \\ [1 em]
    $\mathfrak{so}(p+q,\CC)$ &
    $ \mathfrak{so}(p+q,\RR) $ &
    $\overline{X}$ &
    $B\,I,\, D\,I$\ $\mathfrak{so}(p,q)$ \\ [1 em]
    $\mathfrak{so}(2n,\CC)$ &
    $ \mathfrak{so}(2n,\RR) $ &
    $\overline{X}$ &
    $D\,III\ \mathfrak{so}^*(2n)$ \\ [1 em]
    $\mathfrak{sp}(n,\CC)$ &
    $ \mathfrak{usp}(n) $ &
    $-J_n\overline{X}J_n$ &
    $C\,I$\ $\mathfrak{sp}(n,\RR)$ \\ [1 em]
    $\mathfrak{sp}(p+q,\CC)$ &
    $ \mathfrak{usp}(p+q) $ &
    $-J_{p+q}\overline{X}J_{p+q}$ &
    $C\,III$\ $\mathfrak{sp}(p,q)$ \\
    \bottomrule
  \end{tabular}
  \\ \vspace*{0.5 em}
  \begin{tabular}{c|c|c|c}
    \toprule
    $\mathfrak{u}^{\CC}$ &
    $\sigma$ &
    $\theta$ &
    $\mathfrak{k}_0$ \\ 
    \midrule
    $\mathfrak{sl}(n,\CC)$ &
    ${\overline {X}}^{\textrm{\scriptsize\strut }} $ &
    $-X^t$ &
    $ \mathfrak{so}(n,\RR)$ \\ [1 em]
    $\mathfrak{sl}(2n,\CC)$ &
    $-J_n{\overline {X}}J_n$ &
    $J_n{ {X}}^{t}J_n$ &
    $ \mathfrak{usp}(2n)$ \\ [1 em]
    $\mathfrak{sl}(p+q,\CC)$ &
    $-I_{p,q}{ {X}^{\dagger }}I_{p,q}$ &
    $I_{p,q}{ {X} }I_{p,q}$ &
    $ \mathfrak{so}(n,\RR)$ \\ [1 em]
    $\mathfrak{so}(p+q,\CC)$ &
    $I_{p,q}{\overline {X}} I_{p,q}$ &
    $I_{p,q} XI_{p,q}$ &
    $ \mathfrak{so}(p,\RR)\oplus \mathfrak{so}(q,\RR)$ \\ [1 em]
    $\mathfrak{so}(2n,\CC)$ &
    $-J_n{\overline {X}}J_n$ &
    $-J_n{{X}}J_n$ &
    $ \mathfrak{su}(n)\oplus\mathfrak{u}(1)$ \\ [1 em]
    $\mathfrak{sp}(n,\CC)$&${\overline {X}} $ &
    $-J_n{X}J_n$ &
    $  \mathfrak{su}(n)\oplus\mathfrak{u}(1)$ \\ [1 em]
    $\mathfrak{sp}(p+q,\CC)$ &
    $-K_{p,q}X^\dagger K_{p,q}$ &
    $L_{p,q}X^t L_{p,q}$ &
    $ \mathfrak{sp}(p)\oplus \mathfrak{sp}(q)_{\textrm{\scriptsize\strut}}$ \\
    \bottomrule
  \end{tabular}
  \label{table:AllAutomorphisms}
  \renewcommand{\arraystretch}{1.0}
\end{table}

For completeness we remind the reader of the definitions of matrix
algebras ($\mathfrak{su}(p,\,q)$ has been defined in Equation~(\ref{supq})):
\begin{eqnarray}
  \mathfrak{su}^{*}(2n)&=&
  \left\{ X\left\vert X J_n-J_n\overline{X}=0,\, \Tr X=0,\,X\in
  \CC^{2n\times 2n}\right.\right\}
  \nonumber
  \\
  &=&\left\{\left(
      \begin{array}{@{}r@{\quad}r@{}}
        {A}& C
        \\
        -\overline{ C}&{\overline A}
      \end{array}
    \right) \left\vert
      \begin{array}{l}
        { A}, \, { C}\in\CC^{n\times n}
        \\
        \Real [\Tr {A}]=0
      \end{array}
    \right. \right\},
  \label{sustar}
  \\ [0.5 em]
  \mathfrak{so}(p,\,q)&=&
  \left\{ X\left\vert XI_{p,q}+I_{p,q}X^t=0,\,X=-X^t,\,
  X\in \RR^{(p+q)\times (p+q)}\right.\right\}
  \nonumber
  \\
  &=&\left\{\left(
      \begin{array}{@{}r@{\quad}r@{}}
        { A}& C \\
        { C}^t&{ B}
      \end{array}
    \right)\left\vert
      \begin{array}{l}
        A =-A^t\in\RR^{p\times p}, \, B =-B^t\in\RR^{q\times q}, \\
        { C}\in\RR^{p\times q}
      \end{array}
    \right.\right\},
  \label{sopq}
  \\ [0.5 em]
  \mathfrak{so}^{*}(2n)&=&
  \left\{ X\left\vert X^t J_n+J_n\overline{X}=0,\,
  X=-X^t,\,X\in \CC^{2n\times 2n}\right.\right\}
  \nonumber
  \\
  &=&\left\{\left(
      \begin{array}{@{}r@{\quad}r@{}}
        { A}& B \\
        -\overline{ B} &\overline{ A}
      \end{array}
    \right)\left\vert { A}=-{ A}^t,\,
    {B}= { B}^\dagger\in\CC^{n\times n}\right.\right\},
  \label{sostar}
  \\ [0.5 em]
  \mathfrak{sp}(n,\,\RR )&=&\left\{ X \left\vert X^t J_n+J_nX=0,\,
  \Tr X=0,\,X\in \RR^{2n\times 2n}\right.\right\}\nonumber
  \\
  &=&\left\{\left(
      \begin{array}{@{}r@{\quad}r@{}}
        { A}& B \\
        { C} &-{ A}^t
      \end{array}
    \right)\left\vert { A}, \, { B} ={B}^t,\,
    { C}={ C}^t\in\RR^{n\times n}\right.\right\},
  \label{spnR}
  \\ [0.5 em]
  \mathfrak{sp}(n,\, \CC) )&=&\left\{ X\left\vert X^t J_n+J_nX=0,\,
  \Tr X=0,\,X\in \CC^{2n\times 2n}\right.\right\}
  \nonumber
  \\
  &=&\left\{\left(
      \begin{array}{@{}r@{\quad}r@{}}
        { A}& B \\
        { C}&-{ A}^t
      \end{array}
    \right)\left\vert { A}, \, { B} ={B}^t,\,
    { C}={ C}^t\in\CC^{n\times n}\right.\right\},
  \label{spnC}
  \\ [0.5 em]
  \mathfrak{sp}(p,\,q)&=&\left\{ X\left\vert X^t
  K_{p,q}+K_{p,q}\overline{X}=0,\, \Tr X=0,\,X\in
  \CC^{(p+q)\times (p+q)}\right.\right\}
  \nonumber
  \\
  &=&\left\{\left(
      \begin{array}{@{}r@{\quad}r@{\quad}r@{\quad}r@{}}
        { A}& P& Q& R \\
        { P}^\dagger& B&{ R}^t& S \\
        -\overline{ Q}&\overline{ R}&\overline{ A}&-\overline{ P} \\
        { R}^\dagger&-\overline{ S}&- { P}^t&\overline{ B}
      \end{array}
    \right)\left\vert
      \begin{array}{l}
        { A},\, { Q}\in \CC^{p\times p} \\
        { P},\,  { R} \in \CC^{p\times q},\, { S}\in \CC^{q\times p} \\
        { A}=-{ A}^\dagger,\, { B}=-{ B}^\dagger \\
        { Q}={ Q}^t,\, { S}={ S}^t
      \end{array}
    \right.\right\},
  \label{sppqI}
  \\ [0.5 em]
  \mathfrak{usp}(2p,\,2q)&=&
  \mathfrak{su}(2p,\,2q)\cap \mathfrak{sp}(2p+2q).
\end{eqnarray}
Alternative definitions are:
\begin{equation}
  \begin{array}{rcl}
    \mathfrak{sp}(p,q)&=&
    \{ X\in \mf{gl}(p+q,\HH)\vert \overline{ X } \, I_{p,q}+I_{p,q}\, X=0\},
    \\ [0.25 em]
    \mathfrak{sp}(n,\RR)&=&
    \{ X\in \mf{gl}(2\,n,\RR)\vert { X}^t \, J_{n}+ J_{n}\, X=0\},
    \\ [0.25 em]
    \mathfrak{sl}(n,\HH)&=&
    \{ X\in \mf{gl}(n,\HH)\vert \overline{ X} + X=0\},
    \\ [0.25 em]
    \mf{so}^{*}(2\,n)&=&
    \{ X\in \mathfrak{su}(n,n)\vert { X}^t\,K_{n} + K_{n}\, X=0\}.
  \end{array}
\end{equation}

For small dimensions we have the following isomorphisms:
\begin{equation}
  \begin{array}{rcl}
    \mathfrak{so}(1,2) &\simeq&
    \mathfrak{su}(1,1)\simeq \mathfrak{sp}(1,\RR)\simeq
    \mathfrak{sl}(2,\RR),
    \\
    \mathfrak{sl}(2,\CC)&\simeq&
    \mathfrak{so}(1,3),
    \\
    \mf{so}^{*}(4)&\simeq&
    \mathfrak{su}(2)\oplus\mathfrak{su}(1,1),
    \\
    \mf{so}^{*}(6)&\simeq&
    \mathfrak{su}(3,1),
    \\
    \mathfrak{sp}(1,1)&\simeq&
    \mathfrak{so}(1,4),
    \\
    \mathfrak{sl}(2,\HH)&\simeq&
    \mathfrak{so}(1,5),
    \\
    \mathfrak{su}(2,2)&\simeq&
    \mathfrak{so}(2,4),
    \\
    \mathfrak{sl}(4,\RR)&\simeq&
    \mathfrak{so}(3,3),
    \\ 
    \mf{so}^*(8)&\simeq&
    \mathfrak{so}(2,6).
  \end{array}
\end{equation}


\subsection{Tits--Satake diagrams}
\index{Tits--Satake diagram|bb}

The classification of real forms of a semi-simple Lie algebra,
using Vogan diagrams, rests on the construction of a maximally
compact Cartan subalgebra. On the other hand, the Iwasawa
decomposition \index{Iwasawa decomposition} emphasizes the role of a
maximally noncompact Cartan subalgebra. The consideration of these
Cartan subalgebras leads to another way to classify real forms of
semi-simple Lie algebras, developed mainly by Araki~\cite{Araki}, and
based on so-called Tits--Satake diagrams~\cite{Tits,Satake}.


\subsubsection[Example 1: $\mathfrak{su}(3,2)$]%
              {Example 1: \boldmath $\mathfrak{su}(3,2)$}

\subsubsection*{Diagonal description}

At the end of Section~\ref{section:CTsu32}, we obtained a matrix
representation of a maximally noncompact Cartan subalgebra of
$\mathfrak{su}(3,2)$ in terms of the natural description of this
algebra. To facilitate the forthcoming discussion, we find it
useful to use an equivalent description, in which the matrices
representing this Cartan subalgebra are diagonal, as this
subalgebra will now play a central role. It is obtained by
performing a similarity transformation
 $ X\mapsto S^{T} X\, S$, where 
\begin{equation}
  S= \left(
    \begin{array}{@{}r@{\quad}r@{\quad}r@{\quad}r@{\quad}r@{}}
      1 & 0 & 0 & 0 & 0 \\
      0 & \frac{1}{\sqrt{2}} & 0 & 0 & \frac{1}{\sqrt{2}} \\
      0 & 0 & \frac{1}{\sqrt{2}} & \frac{1}{\sqrt{2}} & 0 \\
      0 & 0 & -\frac{1}{\sqrt{2}} & \frac{1}{\sqrt{2}} & 0 \\
      0 & -\frac{1}{\sqrt{2}} & 0 & 0 & \frac{1}{\sqrt{2}}
    \end{array}
  \right).
\end{equation}
In this new ``diagonal'' description, the conjugation $\sigma$ (see
Equation~(\ref{sn})) becomes 
\begin{equation}
  \sigma ( X)=-\tilde I_{3,2} X^\dagger \tilde I_{3,2},
\end{equation}
where 
\begin{equation}
  \tilde I_{3,2}=S^{T}I_{3,2}S = \left(
    \begin{array}{@{}r@{\quad}r@{\quad}r@{\quad}r@{\quad}r@{}}
      1 & 0 & 0 & 0 & 0 \\
      0 & 0 & 0 & 0 & 1 \\
      0 & 0 & 0 & 1 & 0 \\
      0 & 0 & 1 & 0 & 0 \\
      0 & 1 & 0 & 0 & 0
    \end{array}
  \right).
\end{equation}
The Cartan involution has the following realisation: 
\begin{equation}
  \theta (X)=\tilde I_{3,2} X\, \tilde I_{3,2}.
\end{equation}
In terms of the four matrices introduced in Equation~(\ref{H14}), the
generators defining this Cartan subalgebra $\cH$ reads 
\begin{equation}
  \begin{array}{rcl@{\qquad}rcl}
    h_{1}&=& H_{3}, &
    h_{2}&=& H_{2}+ H_{3}+ H_{4},
    \\
    h_{3}&=& i(2\,H_{1}+2\,H_{2}+ H_{3}), &
    h_{4}&=&i(2\,H_{1}+ H_{2}+ H_{3}+ H_{4}).
    \label{hH}
  \end{array}
\end{equation}
Let us emphasize that we have numbered the basis generators of
$\cH=\cA \oplus \cT$ by first
choosing those in $\cA$, then those in $\cT$.

\subsubsection*{Cartan involution and roots}

The standard matrix representation of $\mathfrak{su}(5)$
constitutes a compact real Lie subalgebra of
$\mathfrak{sl}(5,\CC)$ aligned with the diagonal description of
the real form $\mathfrak{su}(3,2)$. Moreover, its Cartan subalgebra
$\HO$ generated by purely imaginary combinations of the four
diagonal matrices $ H_{k}$ is such that its complexification
$\HC$ contains $\cH$. Accordingly, the roots it defines act both on
$\HO$ and $\cH$. Note that on $\fHr=i\,\HO$, the roots take only
real values.

Our first task is to compute the action of the Cartan involution
$\theta$ on the root lattice. With this aim in view, we introduce
two distinct bases on $\fHr^{*}$. The first one is
$\{F^{1},F^{2},F^{3},F^{4}\}$, which is dual to the basis
$\{ H_{1}, H_{2}, H_{3}, H_{4}\}$ and is adapted to the
relation $\fHr=i\, \HO$. The second one is
$\{f^{1},f^{2},f^{3},f^{4}\}$, dual to the basis
$\{ h_{1},\, h_{2},\,-i h_{3},\,-i h_{4}\}$, which is
adapted to the decomposition $\fHr=\cA \oplus i\,\cT$. The Cartan
involution acts on these root space bases as 
\begin{equation}
  \theta\{f^{1},f^{2},f^{3},f^{4}\}=\{-f^{1},-f^{2},f^{3},f^{4}\}.
\end{equation}
From the relations~(\ref{hH}) it is easy to obtain the
expression of the $\{F^{k}\}\, $ $(k=1,\cdots,4)$ in terms of the
$\{f^{k}\}$ and thus also the expressions for the simple roots
$\alpha_1 = 2F^1 - F^2$, $\alpha_2 = -F^1 + 2 F^2 - F^3 $,
$\alpha_3 = - F^2 + 2 F^3 - F^4$ and $\alpha_4 = - F^3 + 2 F^4$,
defined by $\HO$,
\begin{equation}
  \begin{array}{rcl}
    \alpha_{1}&=&-f^{2}+2\,f^{3}+3\,f^{4},
    \\
    \alpha_{2}&=&- f^{1}+ f^{2}+f^{3}-f^{4},
    \\
    \alpha_{3}&=&2\,f^{1},
    \\
    \alpha_{4}&=&-f^{1}+f^{2}-f^{3}+f^{4}.
  \end{array}
\end{equation}
It is then straightforward to obtain the action of $\theta$ on the roots,
which, when expressed in terms of the $\HO$ simple roots
themselves, is given by
\begin{equation}
  \begin{array}{rcl}
    \theta[\alpha_{1}]&=&\alpha_{1}+\alpha_{2}+\alpha_{3}+\alpha_{4},
    \\
    \theta[\alpha_{2}]&=&-\alpha_{4},
    \\
    \theta[\alpha_{3}]&=&-\alpha_{3},
    \\
    \theta[\alpha_{4}]&=&-\alpha_{2}.
  \end{array}
  \label{thal1}
\end{equation}
We see that the root $\alpha_3$ is real while $\alpha_1$,
$\alpha_2$ and $\alpha_4$ are complex. As a check of these
results, we may, for instance, verify that 
\begin{equation}
  \theta E_{\alpha_{1}}=
  \tilde I_{3,2}\,K^{1}_{2}\,\tilde I_{3,2}=K^{1}_{5}=
  E_{\alpha_{1}+\alpha_{2}+\alpha_{3}+\alpha_{4}}.
\end{equation}
In fact, this kind of computation provides a simpler way to obtain
Equation~(\ref{thal1}).

The basis $\{f^{1},f^{2},f^{3},f^{4}\}$ allows to define a
different ordering on the root lattice, merely by considering the
corresponding lexicographic order. In terms of this new ordering
we obtain for instance $\alpha_{1}<0$ since the first nonzero
component of $\alpha_1$ (in this case $-1$ along $f^2$) is
strictly negative. Similarly, one finds $\alpha_{2}<0$,
$\alpha_{3}>0$, $\alpha_{4}<0$, $\alpha_{1}+\alpha_{2}<0$,
$\alpha_{2}+\alpha_{3}>0$, $\alpha_{3}+\alpha_{4}>0$,
$\alpha_{1}+\alpha_{2}+\alpha_{3}>0$,
$\alpha_{2}+\alpha_{3}+\alpha_{4}>0$,
$\alpha_{1}+\alpha_{2}+\alpha_{3}+\alpha_{4}>0$. A basis of simple
roots, according to this ordering, is given by

\begin{equation}
  \begin{array}{rcl}
    \tilde\alpha_{1}&=&-\alpha_{4}=f^{1}-f^{2}+f^{3}-f^{4},
    \\
    \tilde \alpha_{2} &=&
    \alpha_{1}+\alpha_{2}+\alpha_{3}+\alpha_{4}=f^{2}+2\,f^{3}+3\,f^{4},
    \\
    \tilde\alpha_{3}&=&-\alpha_{1}=f^{2}-2\,f^{3}-3\,f^{4},
    \\
    \tilde\alpha_{4}&=&-\alpha_{2}=f^{1}-f^{2}-f^{3}+f^{4}.
  \end{array}
\end{equation}
(We have put $\tilde\alpha_{4}$ in fourth position, rather than in
second, to follow usual conventions.) The action of $\theta$ on
this basis reads 
\begin{equation}
  \theta[\tilde\alpha_{1}]=-\tilde\alpha_{4},
  \qquad
  \theta[\tilde \alpha_{2}] =-\tilde\alpha_{3},
  \qquad
  \theta[\tilde\alpha_{3}]=-\tilde\alpha_{2},
  \qquad
  \theta[\tilde\alpha_{4}]=-\tilde\alpha_{1}.
\end{equation}
These new simple roots are now all complex.

\subsubsection*{Restricted roots}

The restricted roots are obtained by considering only the action of the
roots on the noncompact Cartan generators $h_1$ and $h_2$.
The two-dimensional vector space spanned by the restricted roots
can be identified with the subspace spanned by $f_1$ and $f_2$;
one simply projects out $f_3$ and $f_4$. In the notations
$\beta_1 = f_1 - f_2$ and $\beta_2 = f_2$, one gets as positive
restricted roots: 
\begin{equation}
  \beta_1,
  \qquad
  \beta_2,
  \qquad
  \beta_1 + \beta_2,
  \qquad
  \beta_1 + 2 \beta_2,
  \qquad
  2 \beta_2,
  \qquad
  2(\beta_1 + \beta_2),
\end{equation}
which are the positive roots of the
$(BC)_2$ (non-reduced) root system. The first four roots are
degenerate twice, while the last two roots are nondegenerate. For
instance, the two simple roots $\tilde\alpha_{1}$ and
$\tilde\alpha_{4}$ project on the same restricted root $\beta_1$,
while the two simple roots $\tilde\alpha_{2}$ and
$\tilde\alpha_{3}$ project on the same restricted root $\beta_2$.

Counting multiplicities, there are ten restricted roots -- the same
number as the number of positive roots of $\mathfrak{sl}(5,\CC)$.
No root of $\mathfrak{sl}(5,\CC)$ projects onto zero. The
centralizer of $\cA$ consists only of $\cA \oplus \cT$.


\subsubsection[Example 2: $\mathfrak{su}(4,1)$]%
              {Example 2: \boldmath $\mathfrak{su}(4,1)$}

\subsubsection*{Diagonal description}

Let us now perform the same analysis within the framework of
$\mathfrak{su}(4,1)$. Starting from the natural
description~(\ref{natsu}) of $\mathfrak{su}(4,1)$, we first make a
similarity transformation using the matrix 
\begin{equation}
  S=\left(
    \begin{array}{@{}r@{\quad}r@{\quad}r@{\quad}r@{\quad}r@{}}
      1 & 0 & 0 & 0 & 0 \\
      0 & 1 & 0 & 0 & 0 \\
      0 & 0 & 1 & 0 & 0 \\
      0 & 0 & 0 & \frac 1{\sqrt{2}} & \frac{1}{\sqrt{2}} \\
      0 & 0 & 0 & -\frac{1}{\sqrt{2}} & \frac{1}{\sqrt{2}}
    \end{array}
  \right),
\end{equation}
so that a maximally noncompact Cartan subalgebra can
be taken to be diagonal and is explicitly given by 
\begin{equation}
  h_{1}= H_{4},
  \qquad
  h_{2}=i\, H_{1},
  \qquad
  h_{3}=i\, H_{2},
  \qquad h_{4}=i(2 \, H_{3}+ H_{4}). 
  \label{hHsu42}
\end{equation}
The corresponding $\mathfrak{su}(4,1)$ in the
$\mathfrak{sl}(5,\CC)$ algebra is still aligned with the natural
matrix representation of $\mathfrak{su}(5)$. The Cartan
involution is given by $ X\mapsto \tilde I_{4,1} X\,\tilde
I_{4,1}$ where $\tilde I_{4,1}=S^{T} \,I_{4,1}\, S$. One has $\cH
= \cA \oplus \cT$ where the noncompact part $\cA$ is
one-dimensional and spanned by $h_1$, while the compact part $\cT$
is three-dimensional and spanned by $h_2$, $h_3$ and $h_4$.

\subsubsection*{Cartan involution and roots}

In terms of the $f^i$'s, the standard simple roots now read
\begin{equation}
  \begin{array}{rcl}
    \alpha_{1}&=&2\,f^{2}-f^{3},
    \\
    \alpha_{2}&=&- f^2 + 2 \, f^{3}-2 \,f^{4},
    \\
    \alpha_{3}&=&-f^{1}- f^2 + 3 \, f^4,
    \\
    \alpha_{4}&=&2 \, f^{1}.
  \end{array}
\end{equation}
The Cartan involution acts as
\begin{equation}
  \begin{array}{rcl}
    \theta[\alpha_{1}]&=&\alpha_{1},
    \\
    \theta[\alpha_{2}]&=&\alpha_{2},
    \\
    \theta[\alpha_{3}]&=&\alpha_{3} + \alpha_4,
    \\
    \theta[\alpha_{4}]&=&-\alpha_{4}, 
  \end{array}
\end{equation}
showing that $\alpha_1$ and $\alpha_2$ are
imaginary, $\alpha_4$ is real, while $\alpha_3$ is complex.

A calculation similar to the one just described above, using as
ordering rules the lexicographic ordering defined by the dual of the
basis in Equation~(\ref{hHsu42}), leads to the new system of simple
roots,
\begin{equation}
  \begin{array}{rcl}
    \tilde \alpha_{1}&=&- \alpha_1 - \alpha_2 - \alpha_3,
    \\
    \tilde \alpha_{2}&=& \alpha_1 + \alpha_2,
    \\
    \tilde \alpha_{3}&=& - \alpha_2,
    \\
    \tilde \alpha_{4}&=& \alpha_2 + \alpha_3 + \alpha_4,
  \end{array}
\end{equation}
which transform as
\begin{equation}
  \begin{array}{rcl}
    \theta[\tilde\alpha_{1}]&=&
    -\tilde\alpha_{4}-\tilde\alpha_{2}-\tilde\alpha_{3},
    \\
    \theta[\tilde \alpha_{2}]& =&\tilde\alpha_{2},
    \\
    \theta[\tilde\alpha_{3}]&=&\tilde\alpha_{3},
    \\
    \theta[\tilde\alpha_{4}]&=&
    -\tilde\alpha_{1}-\tilde\alpha_{2}-\tilde\alpha_{3}
  \end{array}
\end{equation}
under the Cartan involution. Note that in this system, the simple roots
$\tilde\alpha_{2}$ and $\tilde\alpha_{3}$ are imaginary and hence
fixed by the Cartan involution, while the other two simple roots
are complex.

\subsubsection*{Restricted roots}

The restricted roots are obtained by considering the action of the
roots on the single noncompact Cartan generator $h_1$. The
one-dimensional vector space spanned by the restricted roots can
be identified with the subspace spanned by $f_1$; one now simply
projects out $f_2$, $f_3$ and $f_4$. With the notation $\beta_1
= f_1$, we get as positive restricted roots 
\begin{equation}
  \beta_1,
  \qquad
  2 \beta_1,
\end{equation}
which are the positive roots of the $(BC)_1$
(non-reduced) root system. The first root is six
times degenerate, while the second one is nondegenerate. The simple roots
$\tilde\alpha_1$ and $\tilde\alpha_4$ project on the same
restricted root $\beta_1$, while the imaginary root
$\tilde\alpha_2$ and $\tilde\alpha_3$ project on zero (as does
also the non-simple, positive, imaginary root $\tilde\alpha_2+
\tilde\alpha_3$).

Let us finally emphasize that the centralizer of $\cA$ in
$\mathfrak{su}(4,1)$ is now given by $\cA\oplus \cM$, where $\cM$
is the center of $\cA$ in $\cK$ (i.e., the subspace generated by
the compact generators that commute with $ H_{4}$) and contains
more than just the three compact Cartan generators $ h_{2}$,
$ h_{3}$ and $ h_{4}$. In fact, $\cM$ involves also the root
vectors $ E_{\be}$ whose roots restrict to zero. Explicitly, expressed
in the basis of Equation~(\ref{e5}), these roots read
$\beta=\epsilon_p-\epsilon_q$ with $p,q=1,2, \mbox{ or } 3$ and are
orthogonal to $\alpha_{4}$. The algebra $\cM$ constitutes a rank 3,
9-dimensional Lie algebra, which can be identified with
$\mathfrak{su}(3)\oplus \mf{u}(1)$.


\subsubsection{Tits--Satake diagrams: Definition}
\index{Tits--Satake diagram}

We may associate with each of the constructions of these simple root bases a
Tits--Satake diagram as follows. We start with a Dynkin diagram \index{Dynkin diagram} of
the complex Lie algebra and paint in black ($\bullet$) the imaginary
simple roots, i.e., the ones fixed by the Cartan involution. \index{Cartan involution} The
others are represented by a white vertex ($\circ$). Moreover, some
double arrows are introduced in the following way. It can be easily proven
(see Section~\ref{section:formal considerations}) that for real
semi-simple Lie algebras, there always exists a basis of simple roots
$B$ that can be split into two subsets: $B_{0} =\{\alpha_{r+1},\ldots,
\alpha_{l}\}$ whose elements are fixed by $\theta$ (they correspond to
the black vertices) and $B\setminus B_{0}=\{\alpha_{1},\ldots,
\alpha_{r}\}$ (corresponding to white vertices) such that 
\begin{equation}
  \forall \alpha_{k}\in B\setminus B_{0}:
  \theta[\alpha_{k}]=-\alpha_{\bpi(k)}+\!\!
  \sum_{j=r+1}^l a_{k}^j\,\alpha_{j}, 
  \label{thalgen}
\end{equation}
where $\bpi$ is an involutive permutation of the $r$
indices of the elements of $B\setminus B_{0}$. Accordingly, $\bpi$
contains cycles of one or two elements. In the Tits--Satake
diagram, we connect by a double arrow all pairs of distinct simple
roots $\alpha_{k}$ and $\alpha_{\bpi(k)}$ in the same two-cycle
orbit. For instance, for $\mathfrak{su}(3,2)$ and
$\mathfrak{su}(4,1)$, we obtain the diagrams in
Figure~\ref{figure:TSsu3241}.

\epubtkImage{TSsu3241.png}{%
  \begin{figure}[htbp]
    \centerline{\includegraphics[width=100mm]{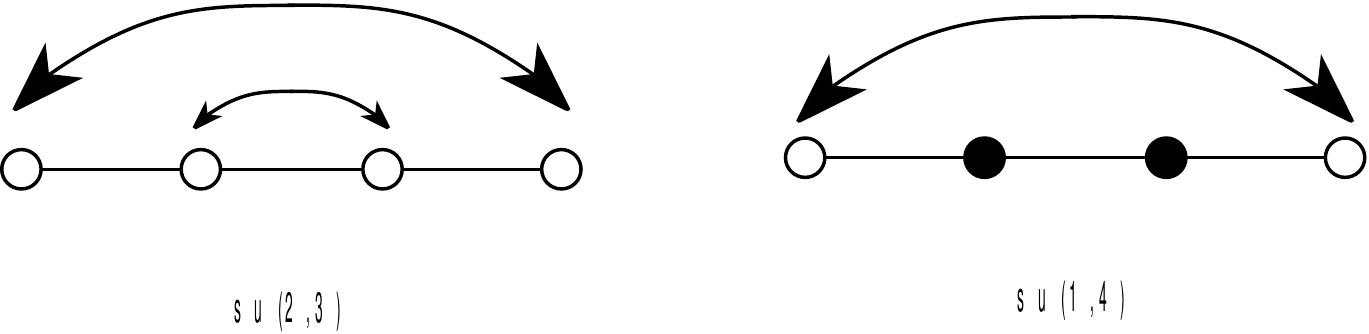}}
    \caption{Tits--Satake diagrams for $\mathfrak{su}(3,2)$ and
      $\mathfrak{su}(4,1)$.}
    \label{figure:TSsu3241}
  \end{figure}}


\subsubsection{Formal considerations}
\label{section:formal considerations}

Tits--Satake diagrams \index{Tits--Satake diagram} provide
a lot of information about real semi-simple Lie algebras. For
instance, we can read from them the full action of the Cartan
involution as we now briefly pass to show. More information may
be found in~\cite{Araki, Helgason}.

The Cartan involution \index{Cartan involution} allows one to define a closed
subsystem\epubtkFootnote{A system $\Delta$ is closed if $\alpha,\,
  \beta\in \Delta$ implies that $-\alpha\in\Delta $ and $ \alpha+\beta
  \in \Delta$.} $\Delta_{0}$ of $\Delta$: 
\begin{equation}
  \label{Delta0}
  \Delta_{0}=\{\alpha\in\Delta\vert\theta[\alpha]= \alpha \},
\end{equation}
which is the system of imaginary roots. These project to zero
when restricted to the maximally noncompact Cartan subalgebra. As
we have seen in the examples, it is useful to use an ordering
adapted to the Cartan involution. This can be obtained by
considering a basis of $\cH$ constituted firstly by elements of
$\cA$ followed by elements of $\cT$. If we use the lexicographic
order defined by the dual of this basis, we obtain a root ordering
such that if $\alpha\not \in \Delta_{0}$ is positive,
$\theta[\alpha]$ is negative since the real part comes first and
changes sign. Let $B$ be a simple root basis built with respect to
this ordering and let $B_{0}=B\cap \Delta_{0}$. Then we have 
\begin{equation}
  \label{ordB}
  B=\{\alpha_{1}, \ldots,\alpha_{l}\}
  \qquad \mbox{and} \qquad
  B_{0}=\{\alpha_{r+1}, \ldots,\alpha_{l}\}.
\end{equation}
The subset $B_{0}$ is a basis for $\Delta_{0}$. To see this,
let $B\setminus B_{0}=\{\alpha_{1},\ldots, \alpha_{r}\}$. If
$\beta=\sum_{k=1}^{l}b^k\,\alpha_{k}$ is, say, a positive root
(i.e., with coefficients $b^k\geq 0 $) belonging to
$\Delta_{0}$, then $\beta -\theta[\beta]=0$ is given by a sum of
positive roots, weighted by non-negative coefficients,
$\sum_{k=1}^{r}b^k\,(\alpha_{k}-\theta[\alpha_{k}])$. As a
consequence, the coefficients $b^k $ are all zero for $k=1,\cdots, r$
and $B_{0}$ constitutes a basis of $\Delta_{0}$, as claimed.

To determine completely $\theta$ we just need to know its action
on a basis of simple roots. For those belonging to $B_{0}$ it is
the identity, while for the other ones we have to compute the
coefficients $a_{k}^j$ in Equation~(\ref{thalgen}). These are obtained
by solving the linear system given by the scalar products of these
equations with the elements of $B_{0}$, 
\begin{equation}
  (\theta[\alpha_{k}]+\alpha_{\pi(k)}\vert \alpha_{q})=\!\!
  \sum_{j=r+1}^l a_{k}^j\,(\alpha_{j}\vert \alpha_{q}).
\end{equation}
Solving these equations for the unknown
coefficients $a_{k}^j$ is always possible because the Killing
metric is nondegenerate on $B_{0}$.

The black roots of a Tits--Satake diagram \index{Tits--Satake diagram}
represent $B_{0}$ and
constitute the Dynkin diagram of the compact part $\cM$ of the
centralizer of $\cA$. Because $\cM$ is compact, it is the direct
sum of a semi-simple compact Lie algebra and one-dimensional,
Abelian ${\mathfrak{u}}(1)$ summands. The rank of $\cM$ (defined as
the dimension of its maximal Abelian subalgebra; diagonalizability
is automatic here because one is in the compact case) is equal to
the sum of the rank of its semi-simple part and of the number of
${\mathfrak{u}}(1)$ terms, while the dimension of $\cM$ is equal to
the dimension of its semi-simple part and of the number of
${\mathfrak{u}}(1)$ terms. The Dynkin diagram of $\cM$ reduces to
the Dynkin diagram of its semi-simple part.

The rank of the compact subalgebra $\cM$ is given by 
\begin{equation}
  \rank \cM=\rank \cG-\rank \cP, 
  \label{rankM}
\end{equation}
where $\rank \cP$, called as we have indicated above the
\emph{real rank} of $\cG$, is given by the number of cycles of the
permutation $\bpi$ (since two simple white roots joined by a
double-arrow project on the same simple restricted root~\cite{Araki,
  Helgason}). These two sets of data allow one to determine the
dimension of $\cM$ (without missing ${\mathfrak{u}}(1)$
generators)~\cite{Araki, Helgason}. Another useful information, which
can be directly read off from the Tits--Satake diagrams
\index{Tits--Satake diagram} is the dimension of the noncompact
subspace $\cP$ appearing in the splitting $\cG=\cK\oplus \cP$. It is
given (see Section~\ref{section:formal}) by 
\begin{equation}
  \dim \cP=\frac 12(\dim \cG-\dim \cM + \rank \cP). 
  \label{dimP}
\end{equation}
This can be illustrated in the two previous
examples. For $\mathfrak{su}(3,2)$, one gets $\dim  \cG =
24$, $\rank \cG = 4$ and $\rank  \cP = 2$. It
follows that $\rank \cM = 2$ and since $\cM$ has no
semi-simple part (no black root), it reduces to $\cM =
{\mathfrak{u}}(1) \oplus {\mathfrak{u}}(1)$ and has dimension 2.
This yields $\dim  \cP = 12$, and, by substraction,
$\dim  \cK = 12$ ($\cK$ is easily verified to be equal to
$\mathfrak{su}(3) \oplus \mathfrak{su}(2) \oplus {\mathfrak{u}}(1)$).
Similarly, for $\mathfrak{su}(4,1)$, one gets $\dim  \cG =
24$, $\rank \cG = 4$ and $\rank  \cP = 1$. It
follows that $\rank \cM = 3$ and since the semi-simple part
of $\cM$ is read from the black roots to be $\mathfrak{su}(3)$,
which has rank two, one deduces $\cM = \mathfrak{su}(3) \oplus
{\mathfrak{u}}(1)$ and $\dim  \cM = 9$. This yields
$\dim  \cP = 8$, and, by substraction, $\dim  \cK =
16$ ($\cK$ is easily verified to be equal to $\mathfrak{su}(4)
\oplus {\mathfrak{u}}(1)$ in this case).

Finally, from the knowledge of $\theta$, we may obtain the
restricted root space by projecting the root space according to
\begin{equation}
  \Delta \rightarrow \overline\Delta : \alpha\mapsto \bar\alpha =
  \frac 12(\alpha-\theta[\alpha])
\end{equation}
and restricting their action on $\cA$ since $\alpha$ and
$-\theta(\alpha)$ project on the same restricted
root~\cite{Araki, Helgason}.


\subsubsection[Illustration: $F_{4}$]%
              {Illustration: \boldmath $F_{4}$}

The Lie algebra $F_{4}$ is a 52-dimensional simple Lie algebra of
rank~4. Its root vectors can be expressed in terms of the elements of
an orthonormal basis $\{e_{k}\vert k=1,\ldots,4\}$ of a
four-dimensional Euclidean space:
\begin{equation}
  \Delta_{ F_{4} } = \left\{\pm e_{i}\pm e_{j} \vert i<j\}\cup
  \{\pm e_{i}\}\cup \{\frac 12 (\pm e_{1}\pm e_{2}\pm e_{3}\pm e_{4}) \right\}.
\end{equation}
A basis of simple roots is 
\begin{equation}
  \alpha_{1} = e_{2}-e_{3},
  \qquad
  \alpha_{2} =e_{3}-e_{4},
  \qquad
  \alpha_{3} =e_{4},
  \qquad
  \alpha_{4} =\frac 12(e_{1}-e_{2}-e_{3}-e_{4}).
\end{equation}
The corresponding Dynkin diagram can be obtained from
Figure~\ref{figure:naf4} by ignoring the painting of the vertices. To
the real Lie algebra, denoted $ {F\, II}$ in~\cite{Cartan}, is
associated the Tits--Satake diagram \index{Tits--Satake diagram} of the left hand side of
Figure~\ref{figure:naf4}. We immediately obtain from this diagram the
following information:
\begin{equation}
  \rank \cP=1,
  \qquad
  \rank \cM=3,
  \qquad
  \cM=\mathfrak{so}(7),
  \qquad
  \dim {\cP}=\frac 12(52-21+1)=16.
\end{equation}
Accordingly, ${ F\, II}$ has signature $-21$ (compact part) + $1$
(rank of $\cP$) $= - 20$ and is denoted $F_{4(-20)}$. Moreover,
solving a system of three equations, we obtain:
$\theta[\alpha_{4}]=-\alpha_{4}-\,\alpha_{1}-2\,\alpha_{2}-3\,\alpha_{3}$,
i.e.,
\begin{equation}
  \theta[e_{1}]=-e_{1}
  \qquad \mbox{and} \qquad
  \theta[e_{k}]=e_{k} \quad \mbox{if } k=2,3,4.
\end{equation}
This shows that the projection defining the reduced root
system $\Sigma$ consists of projecting any given root orthogonally
onto its $e_{1}$ component. Thus we obtain $\Sigma=\{\pm \frac 12
e_{1},\, \pm e_{1}\}$, with multiplicity 8 for $\frac 12 e_{1}$
(resulting from the projection of the eight roots $\{\frac
12(e_{1}\pm e_{2}\pm e_{3}\pm e_{4})\}$) and 7 for $ e_{1}$
(resulting from the projection of the seven roots $\{ e_{1}\pm
e_{k}\vert k=2,3,4\}\cup \{ e_{1}\}$).

\epubtkImage{naf4.png}{%
  \begin{figure}[htbp]
    \centerline{\includegraphics[width=120mm]{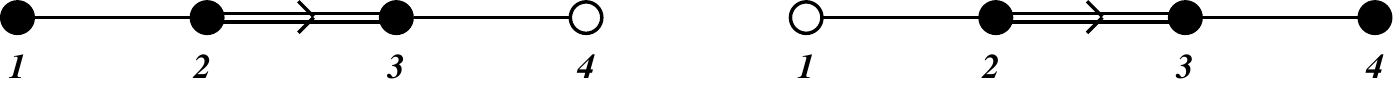}}
    \caption{On the left, the Tits--Satake diagram of the real form
      $F_{4(-20)}$. On the right, a non-admissible Tits--Satake diagram.}
    \label{figure:naf4}
  \end{figure}}

Let us mention that, contrary to the Vogan diagrams, any ``formal
Tits--Satake diagram'' is not admissible. For instance if we
consider the right hand side diagram of Figure~\ref{figure:naf4} we
get 
\begin{equation}
  \theta[e_{1}]=-e_{2},\, \theta[e_2] = - e_1,
  \qquad \mbox{and} \qquad
  \theta[e_{k}]=e_{k}
  \qquad \mbox{if }
  k=3\mbox{ or }4.
\end{equation}
But this means that for the root $\alpha=e_{1}$, $\alpha +
\theta^{*}[\alpha]=e_{1}-e_{2}$ is again a root, which is
impossible as we shall see below.


\subsubsection{Some more formal considerations}
\label{section:formal}

Let us recall some crucial aspects of the discussion so far. Let $\Gs$
be a real form of the complex semi-simple Lie algebra $\GC$ and
$\sigma$ be the conjugation it defines. We have seen that there always
exists a compact real Lie algebra $\Ut$ such that the corresponding
conjugation $\tau$ commutes with $\sigma$. Moreover, we may choose a
Cartan subalgebra $\cH$ of $\Ut$ such that its complexification $\HC$
is invariant under $\sigma$, i.e., $\sigma(\HC)=\HC$. Then the real
form $\Gs$ is said to be \emph{normally related} to $(\Uh,\, \cH)$. As
previously, we denote by the same letter $\theta$ the involution
defined by duality on $(\HC)^*$ (and also on the root lattice with
respect to $\HC$: $\Delta$) by $\theta = \tau \sigma$.

When $\Gs$ and $\Ut$ are normally related, we may decompose the
former into compact and noncompact components $\Gs=\cK\oplus \cP$
such that $\Ut=\cK \oplus i\, \cP$. As mentioned,
the starting point consists of choosing a maximally Abelian
noncompact subalgebra $\cA\subset \cP$ and extending it to a Cartan
subalgebra $\cH=\cT \oplus\cA$, where $\cT\subset \cK$. This Cartan
subalgebra allows one to consider the real Cartan subalgebra 
\begin{equation}
  \HR=i \cT\oplus \cP = \!\! \sum_{\alpha \in \Delta}\RR \, H_{\al}.
\end{equation}
Let us remind the reader that, in this case, the Cartan involution
$\theta=\sigma\,\tau=\tau\,\sigma$ is such that $\theta\vert_{\cK}=+1$
and $\theta\vert_{\cP}=-1$. From Equation~(\ref{thtEa}) we obtain
\begin{equation}
  \theta( E_{\al})=\rho_{\alpha}\, E_{\theta[\alpha]},
\end{equation}
and using $\theta^{2}=1$ we deduce that 
\begin{equation}
  \rho_{\alpha}\,\rho_{\theta[\alpha]}= 1.
\end{equation}
Furthermore, Equation~(\ref{comEE}) and the fact that the structure
constants are rational yield the following relations: 
\begin{equation}
  \begin{array}{rcl}
    \rho_{\alpha}\,\rho_{\beta}N_{\theta[\alpha],\,\theta[\beta]}&=&
    \rho_{\alpha+\beta}N_{\alpha,\,\beta},
    \\ [0.25 em]
    \theta( H_{\al})&=& H_{\theta[\alpha]},
    \\ [0.15 em]
    \rho_{\alpha}\,\rho_{-\alpha}&=& 1.
  \end{array}
  \label{rho1} 
\end{equation}
On the other hand, the commutativity of $\tau$ and $\sigma$ implies
\begin{equation}
  \sigma( H_{\alpha})=- H_{\sigma[\alpha]},
  \qquad
  \sigma( E_{\al})=\kappa_{\alpha}\, E_{\sigma[\alpha]},
\end{equation}
with
\begin{equation}
  \kappa_{\alpha}=-\bar\rho_{\alpha},
  \qquad
  \sigma[\alpha]=-\theta[\alpha].
\end{equation}
In particular, if the root $\alpha$ belongs to $\Delta_{0}$,
defined in Equation~(\ref{Delta0}), then $\theta[\alpha]=\alpha$ and thus
$\rho^{2}_{\alpha}=1$, i.e., 
\begin{equation}
  \rho_{\alpha}=-\kappa_{\alpha}=\pm 1.
\end{equation}

Let us denote by $\Delta_{0,\,-}$ and $\Delta_{0,\,+}$ the
subsets of $\Delta_{0}$ corresponding to the imaginary noncompact
and imaginary compact roots, respectively. We have 
\begin{equation}
  \Delta_{0,\,-}=\{\alpha \in \Delta_{0}\vert\rho_{\alpha}=-1\}
  \qquad \mbox{and} \qquad
  \Delta_{0,\,+}=\{\alpha \in \Delta_{0}\vert \rho_{\alpha}=+1\}.
\end{equation}
Obviously, for
$\alpha\in \Delta_{0,\,-}$, $ E_{\al}$ belongs to $\cP^\CC$, while for
$\alpha\in \Delta_{0,\,+}$, $ E_{\al}$ belongs to $\cK^\CC$. Moreover,
if $\alpha\in \Delta \setminus \Delta_{0}$ we find 
\begin{equation}
  E_{\al}+\theta( E_{\al})\in \cK^\CC
  \qquad \mbox{and} \qquad
  E_{\al}-\theta( E_{\al})\in \cP^\CC.
\end{equation}
These remarks lead to the following explicit constructions of the
complexifications of $\cK$ and $\cP$: 
\begin{equation}
  \begin{array}{rcl}
    \cK^{\CC}&=& \displaystyle
    \cT^\CC \oplus \!\!
    \bigoplus_{\alpha \in \Delta_{0,\,+} } \!\!\CC\,
    E_{\al}\oplus \!\!
    \bigoplus_{\alpha \in \Delta\setminus \Delta_{0}} \!\!
    \CC\,( E_{\al}+\theta( E_{\al})),
    \\ [1.5 em]
    \cP^{\CC}&=& \displaystyle
    \cA^\CC \oplus \!\!
    \bigoplus_{\alpha \in \Delta_{0,\,-} } \!\! \CC\,
    E_{\al} \oplus \!\!
    \bigoplus_{\alpha \in \Delta\setminus \Delta_{0}} \!\!
    \CC\,( E_{\al}-\theta( E_{\al})). 
  \end{array}
\end{equation}
Furthermore, since $\theta$ fixes all the elements of $\Delta_{0}$,
the subspace $\bigoplus_{\alpha \in \Delta_{0,\,-} }\CC\, E_{\al} $
belongs to the centralizer\epubtkFootnote{Geometrically, this results
  from the orthogonality of roots $\alpha$ and $\beta$ such that $
  H_{\al}\in\cK$ and $ H_{\be}\in\cP$, or, equivalently, because
  $\alpha( H_{\be})=\theta[\alpha]( H_{\be})=\alpha(\theta(
  H_{\be}))=-\alpha( H_{\be})$.} of $\cA$ and thus is empty if $\cA$
is maximally Abelian in $\cP$. Taking this remark into account, we
immediately obtain the dimension formulas~(\ref{rankM}, \ref{dimP}).

Using, as before, the basis in Equation~(\ref{ordB}) we obtain for the roots
belonging to $B\setminus B_{0}$, i.e., for an index $i\leq r$: 
\begin{equation}
  -\theta[\alpha_{i}]= \!\!\!\! \sum_{j=1,\ldots,r} \!\!\!\!
  p_{i}^{j}\alpha_{j}+ \!\!\!\! \sum_{j=r+1,\ldots,l} \!\!\!\!
  q_{i}^{j}\alpha_{j}
  \qquad \mbox{with} \qquad
  p_{i}^{j},\, q_{i}^{j}\in \NN.
\end{equation}
Thus
\begin{equation}
  \alpha_{i}=(-\theta)^{2}[\alpha_{i}]= \!\!\!\!\!\!
  \sum_{\scriptsize
    \begin{array}{l}
      j=1,\ldots,r \\
      k=1,\ldots,r
    \end{array}} \!\!\!\!\!\!
  p_{i}^{j}p_{j}^{k}\alpha_{k} + \!\!\!\!\!\!
  \sum_{\scriptsize
    \begin{array}
      {l}j=1,\ldots,r \\
      k=r+1,\ldots,l
    \end{array}} \!\!\!\!\!\!
  p_{i}^{j}q_{j}^{k}\alpha_{k}- \!\!\!\!\!\!
  \sum_{j=r+1,\ldots,l} \!\!\!\!\!\! q_{i}^{j}\alpha_{j}.
\end{equation}
As $\sum_{j=1,\ldots,r}p_{i}^{j}p_{j}^{k}=\delta_{i}^j$, where the
coefficients $ p_{i}^{j}$ are non-negative integers, the matrix
$(p_{i}^{j})$ must be a permutation matrix and it follows that 
\begin{equation}
  \theta[\alpha_{i}]=-\alpha_{\bpi(i)}
  \qquad
  \pmod{\Delta_{0}},
\end{equation}
where $\bpi$ is an involutive permutation of $\{1,\ldots,r\}$.

A fundamental property of $\Delta$ is 
\begin{equation}
  \forall \alpha \in \Delta :
  \theta[\alpha]+\alpha\not \in \Delta . 
  \label{fundD}
\end{equation}
To show this, note that if $\alpha\in\Delta_{0}$, it would
imply that $2\,\alpha$ belongs to $\Delta$, which is impossible
for the root lattice of a semi-simple Lie algebra. If $\alpha \in
\Delta \setminus \Delta_{0}$ and $\theta[\alpha]+\alpha \in
\Delta $, then $\theta[\alpha]+\alpha \in \Delta_{0}$. Thus we
obtain using Equation~(\ref{Nrel}) and taking into account that $\cA$
is maximal Abelian in $\cP$, that $\rho_{\alpha}=+1$, i.e., 
\begin{equation}
  \sigma( E_{\sigma[\alpha]-\alpha})=+ E_{\theta[\alpha]-\alpha}
\end{equation}
and 
\begin{equation}
  \begin{array}{rcl}
    \Lb{ E_{\alpha}}{\sigma( E_{-\alpha})}&=&
    \rho_{-\alpha}\,N_{\alpha,-\sigma[\alpha]}\,
    E_{\alpha-\sigma[\alpha]},
    \\ [0.25 em]
    \Lb{\sigma( E_{\alpha})}{ E_{-\alpha}}&=&
    \overline{\rho_{-\alpha}}\,N_{\alpha,-\sigma[\alpha]}\,
    E_{\sigma[\alpha]-\alpha}
    \\ [0.25 em]
    &=&\rho_{-\alpha}\,N_{\sigma[\alpha],-\alpha}\,
    E_{\sigma[\alpha]-\alpha}.
  \end{array}
\end{equation}
From this result we deduce 
\begin{equation}
  \rho_{\alpha}\,N_{\sigma[\alpha],-\alpha}=
  \overline{\rho_{-\alpha}}\,N_{\alpha,-\sigma[\alpha]}=
  -\rho_{\alpha}\,N_{\alpha,-\sigma[\alpha]},
\end{equation}
i.e., $\rho_{\alpha}=-\overline{\rho_{-\alpha}}$ which is incompatible
with equation~(\ref{rho1}). Thus, the statement~(\ref{fundD}) follows.


\subsection[The real semi-simple algebras $\mathfrak{so}(k,l)$]%
           {The real semi-simple algebras \boldmath $\mathfrak{so}(k,l)$}
\label{section:restrictedrootsystemSO}

The dimensional reduction from 10 to 3 dimensions of $\mc{N}=1$
supergravity coupled to $m$ Maxwell multiplets leads to a
nonlinear sigma model  \index{nonlinear sigma model} ${\cal G}/{\cal K}(\mcg)$ with Lie$(\cal
G)$=$\mathfrak{so}(8,8+m)$ (see Section~\ref{section:KMBilliardsII}). To investigate the geometry of these
cosets, we shall construct their Tits--Satake diagrams. \index{Tits--Satake diagram}

The $\mathfrak{so}(n,\, \CC)$ Lie algebra can be represented by
$n\times n$ antisymmetric complex matrices. The compact real form
is $\mathfrak{so}(k+l,\,\RR)$, naturally represented as the set of
$n\times n$ antisymmetric real matrices. One way to describe the
real subalgebras $\mathfrak{so}(k,l)$, aligned with the compact
form $\mathfrak{so}(k+l,\,\RR)$, is to consider 
$\mathfrak{so}(k,l)$ as the set of infinitesimal rotations
expressed in Pauli coordinates, i.e., to represent the hyperbolic
space on which they act as a Euclidean space whose first $k$
coordinates, $x^a$, are real while the last $l$ coordinates $y^b$
are purely imaginary. Writing the matrices of $\mathfrak{so}(k,l)$ in
block form as 
\begin{equation}
  X=\left(
    \begin{array}{@{}r@{\quad}r@{}}
      { A}&i\, C \\
      -i\, C^t&{ B}
    \end{array}
  \right),
\end{equation}
where
\begin{equation}
  { A} =-{A}^t\in\RR^{k\times k},
  \qquad
  { B} =-{B}^t\in\RR^{l\times l},
  \qquad
  { C} \in\RR^{k\times l},
  \label{natso}
\end{equation}
we may obtain a maximal Abelian subspace $\cA$ by allowing $C$ to have
nonzero elements only on its diagonal, i.e., to be of the form: 
\begin{equation}
  { C}=\left(
    \begin{array}{@{}c@{\quad}c@{\quad}c@{}}
      a_{1}&\cdots&0 \\
      &\ddots& \\
      0&\cdots&a_{l} \\
      \vdots&\cdots&\vdots
    \end{array}
  \right)
  \qquad \mbox{or} \qquad
  {C}=\left(
    \begin{array}{@{}c@{\quad}c@{\quad}c@{\quad}c@{\quad}c@{}}
      a_{1}&\cdots&\ &\cdots&0 \\
      &\ddots& & & \\
      0&\cdots&a_{k}&\cdots&0
    \end{array}
  \right),
\end{equation}
with $k>l$ or $l< k$, respectively.

To proceed, let us denote by $ H_{j}$ the matrices whose entries are
everywhere vanishing except for a $2\times 2$ block,
\begin{displaymath}
  \left(
    \begin{array}{@{}r@{\quad}r@{}}
      0&1 \\
      -1&0
    \end{array}
  \right),
\end{displaymath}
on the diagonal. These matrices have the following realisation in terms of
the ${K^{i}}_j$ (defined in Equation~(\ref{GLgenerators})):
\begin{equation}
  \label{Csason} H_{j}={K^{2j-1}}_{2j}-{K^{2j}}_{2j-1}.
\end{equation}
They constitute a set of $\mathfrak{so}(k+l)$ commuting
generators that provide a Cartan subalgebra; it will be the Cartan
subalgebra fixed by the Cartan involution defined by the real
forms that we shall now discuss.


\subsubsection[Dimensions $l=2\,q+1<k=2\,p$]%
              {Dimensions \boldmath $l=2\,q+1<k=2\,p$}

Motivated by the dimensional reduction of supergravity, we shall
assume $k=2p$, even. We first consider $l=2\,q+1<k$. Then by
reordering the coordinates as follows,
\begin{equation}
  \{x_{1},y_{1};\cdots;\, x_{l},y_{l};\, x_{l+1},x_{l+2};\cdots;\,
  x_{2p-2},x_{2p-1};\,x_{2p}\},
\end{equation}
we obtain a Cartan subalgebra of $\mathfrak{so}(2q+1,2p)$, with
noncompact generators first, and aligned with the one introduced in
Equation~(\ref{Csason}) by considering the basis $\{i\,
H_{1},\cdots,\, i\, H_{l},\, H_{l+1},\cdots,$
$H_{q+p}\}$ \epubtkFootnote{If $p=q+1$, this basis consists only of
  noncompact generators.}. These generators are all orthogonal to
each other. Let us denote the elements of the dual basis by
$\{f_{A}\vert A=1,\cdots,p+q\}$, and split them into two subsets:
$\{f_{a}\vert a=1,\cdots,2q+1\}$ and $\{f_{\alpha}\vert \alpha =2 q
+2,\cdots,p+q\}$. The action of the Cartan involution on these
generators is very simple,
\begin{equation}
  \label{thfBpq} \theta[f_{a}]=-f_{a},
  \qquad \mbox{and} \qquad
  \theta[f_{\alpha}]=+f_{\alpha}.
\end{equation}
The root system of $\mathfrak{so}(2q+1,2p)$ is $B_{(p+q)}$,
represented by $\Delta=\{\pm f_{A}\pm f_{B}\vert
A<B=1,\cdots,p+q\}\cup\{\pm f_{A}\vert A1,\cdots,p+q\}$. A simple root
basis can be taken as: $$\{\alpha_{1}=f_{1}-f_{2}, \cdots,\,
\alpha_{p+q-1}=f_{p+q-1}-f_{p+q},\, \alpha_{p+q}=f_{p+q}\}.$$ It is
then straigthforward to obtain the action of the Cartan involution on
the simple roots:
\begin{equation}
  \begin{array}{rcl@{\qquad}l}
    \theta[\alpha_{A}]&=&-\alpha_{A} &
    \mbox{for } A=1,\cdots,2q,
    \nonumber
    \\
    \theta[\alpha_{2q+1}]&=&
    -\alpha_{2q+1}-2(\alpha_{2q+2}+\cdots+\alpha_{q+p}),
    \\
    \theta[\alpha_{A}]&=&+\alpha_{A} &
    \mbox{for } A=2q+2,\cdots,q+p.
  \end{array}
\end{equation}
The corresponding Tits--Satake diagrams are displayed in
Figure~\ref{figure:TSBpq}.

\epubtkImage{TSBpq.png}{%
  \begin{figure}[htbp]
    \centerline{\includegraphics[width=100mm]{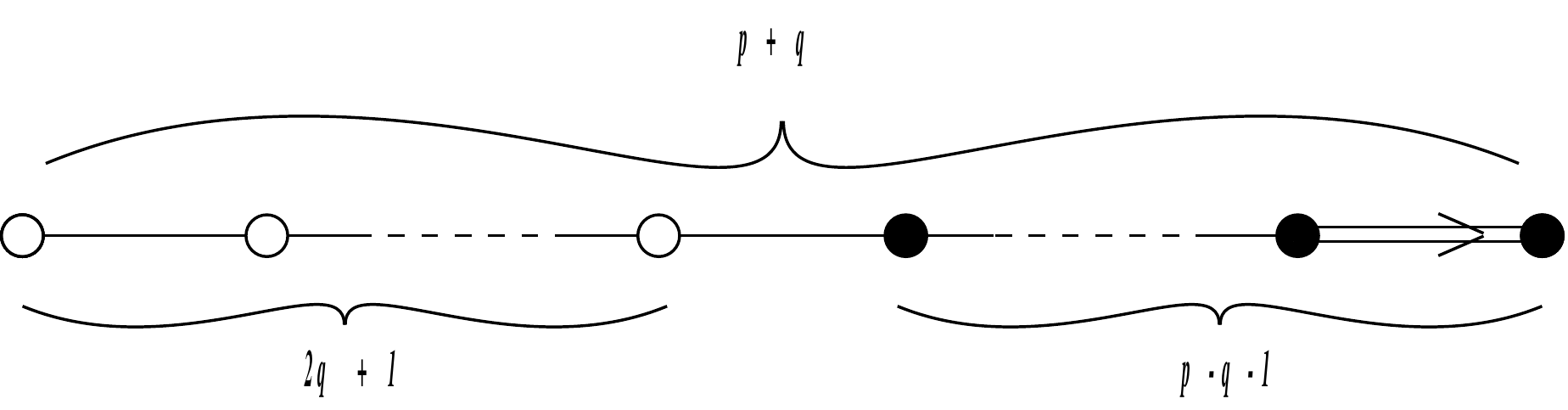}}
    \caption{Tits--Satake diagrams for the $\mathfrak{so}(2p,2q+1)$ Lie
      algebra with $q<p$. If $p=q+1$, all nodes are white.}
    \label{figure:TSBpq}
  \end{figure}}

From Equation~(\ref{thfBpq}) we also obtain without effort that the set of
restricted roots consists of the $4q(2q+1)$ roots $\{\pm f_{a}\pm f_{b}\}$,
each of multiplicity one, and the $4q+2$ roots $\{\pm f_{a}\}$, each
of multiplicity $2(p-q)-1$. These constitute a $B_{2q+1}$ root system.


\subsubsection[Dimensions $l=2\,q+1>k=2\,p$]%
              {Dimensions \boldmath $l=2\,q+1>k=2\,p$}

Following the same procedure as for the previous case, we obtain
a Cartan subalgebra consisting of $2p$ noncompact generators
and $q-p$ compact generators. The corresponding Tits--Satake diagrams are
displayed in Figure~\ref{figure:TSBpq2}.

\epubtkImage{TSBpq2.png}{%
  \begin{figure}[htbp]
    \centerline{\includegraphics[width=100mm]{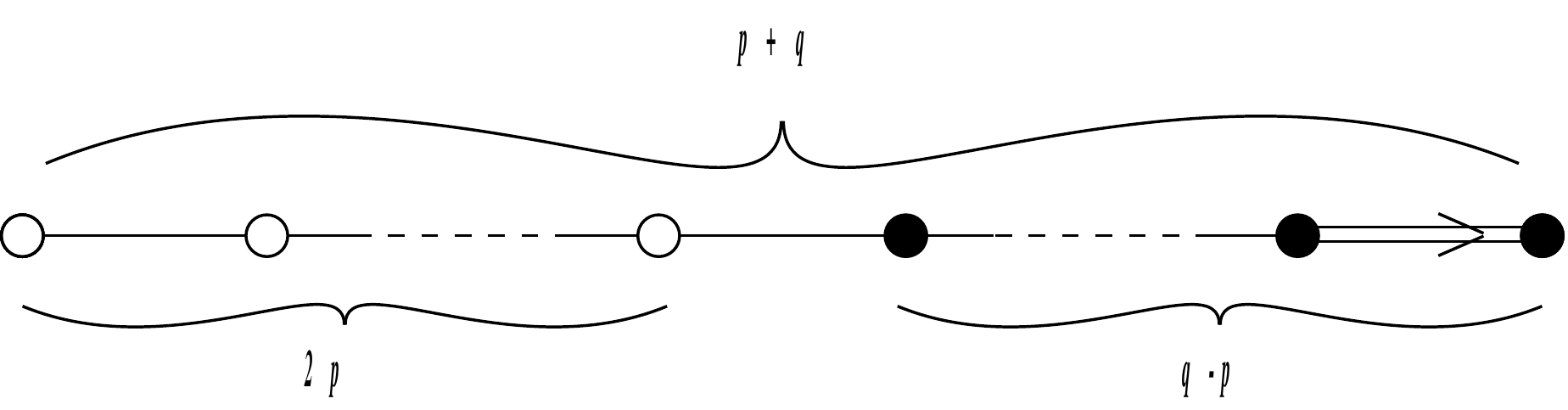}}
    \caption{Tits--Satake diagrams for the $\mathfrak{so}(2p,2q+1)$ Lie
      algebra with $q\geq p$. If $q=p$, all nodes are white.}
    \label{figure:TSBpq2}
  \end{figure}}

The restricted root system \index{restricted root system} is now of
type $B_{2p}$, with $4p(2p-1)$ long roots of multiplicity one and $4p$
short roots of multiplicity $2(q-p)+1$.


\subsubsection[Dimensions $l=2\,q,\,k=2\,p$]%
              {Dimensions \boldmath $l=2\,q,\,k=2\,p$}

Here the root system is of type $D_{p+q}$, represented by
$\Delta=\{\pm f_{A}\pm f_{B}\vert \, A<B=1,\cdots,\, p+q\}$,
where the orthonormal vectors $f_{A}$ again constitute a basis
dual to the natural Cartan subalgebra of $\mathfrak{so}(k+l)$.
Now, $k=2p$ and $l=2 q$ are both assumed even, and we may always
suppose $k\geq l$. The Cartan involution to be considered acts as
previously on the $f_{A}$: 
\begin{equation}
  \theta[f_{a}]=-f_{a},
  \qquad
  a=1,\cdots,\, 2q
\end{equation}
and
\begin{equation}
  \theta[f_{\alpha}]=+f_{\alpha},
  \qquad \alpha=2q+1,\cdots,\, p+q
  \qquad
  \mbox{for } q<p.
\end{equation}
The simple roots can be chosen as
\begin{displaymath}
  \{\alpha_{1}=f_{1}-f_{2},\cdots,\,\alpha_{p+q-1}=f_{p+q-1}-f_{p+q},\,
  \alpha_{p+q}=f_{p+q-1}+f_{p+q}\},
\end{displaymath}
on which the Cartan involution has the following action:

\begin{itemize}
\item For $q=p$
  \begin{equation}
    \theta[\alpha_{A}] = -\alpha_{A}
    \qquad
    \mbox{for } A=1,\cdots,\, q+p.
  \end{equation}
\item For $q=p-1$
  \begin{equation}
    \begin{array}{rcl@{\qquad}l}
      \theta[\alpha_{A}]&=&-\alpha_{A} &
      \mbox{for } A=1,\cdots,\, 2q=q+p-1,
      \\
      \theta[\alpha_{q+p-1}]&=&-\alpha_{q+p},
      \\
      \theta[\alpha_{q+p}]&=&-\alpha_{q+p-1}.
    \end{array}
  \end{equation}
\item For $q<p-1$
  \begin{equation}
    \begin{array}{rcl@{\qquad}l}
      \theta[\alpha_{A}]&=&-\alpha_{A} &
      A=1,\cdots,\, 2q-1,
      \\
      \theta[\alpha_{2q}]&=&
      -\alpha_{2q}-2(\alpha_{2q+1}+\cdots,\alpha_{q+p-2}) \\
      & & - \alpha_{q+p-1}-\alpha_{q+p},
      \\
      \theta[\alpha_{A}]&=&+\alpha_{A} &
      A=2q+1,\cdots,\, q+p,
    \end{array}
  \end{equation}
\end{itemize}

The corresponding Tits--Satake diagrams are obtained in the same way as
before and are displayed in Figure~\ref{figure:TSDpq}.

\epubtkImage{TSDpq.png}{%
  \begin{figure}[htbp]
    \centerline{\includegraphics[width=80mm]{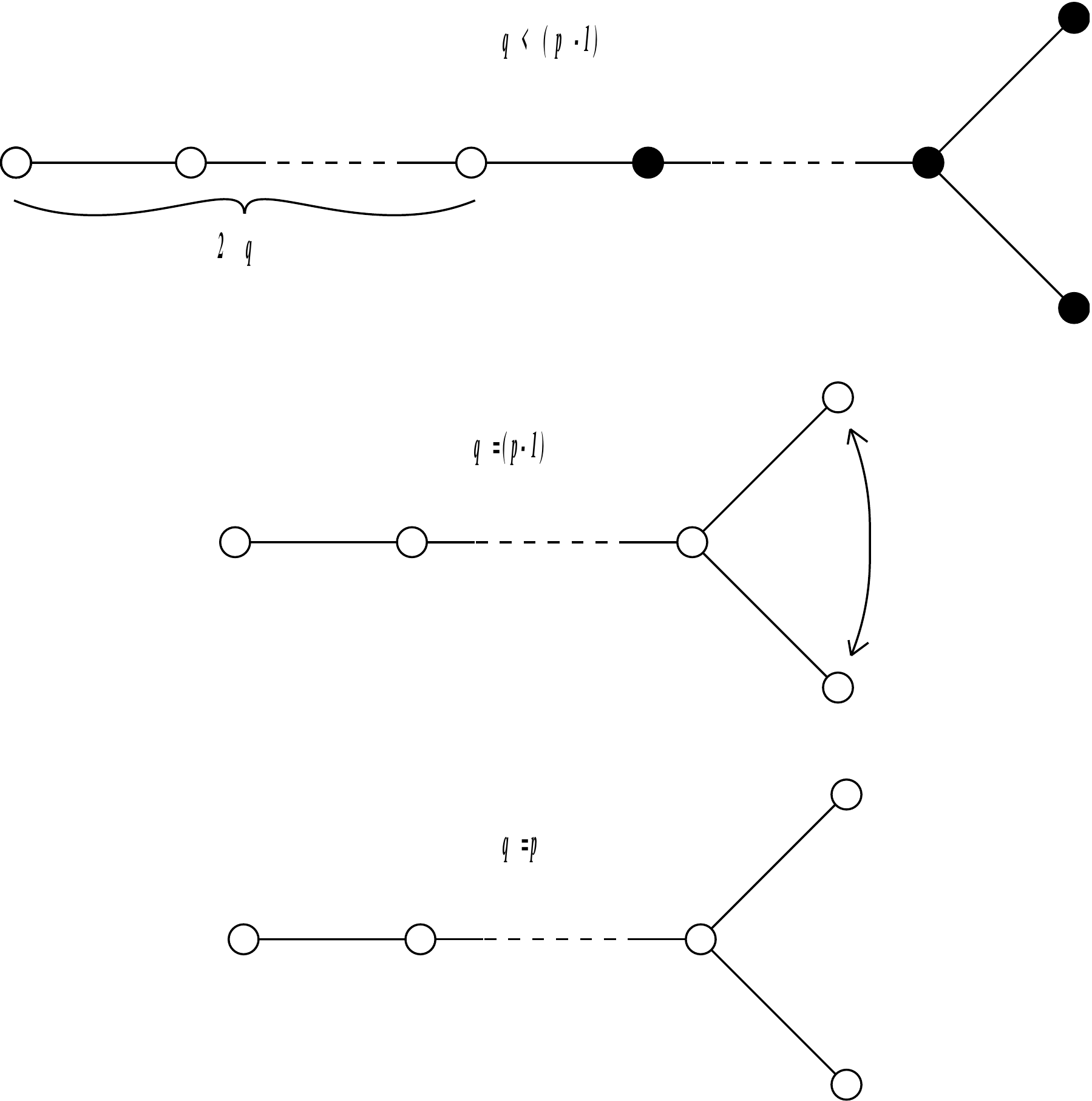}}
    \caption{Tits--Satake diagrams for the $\mathfrak{so}(2p,2q)$ Lie
      algebra with $q< p-1$, $q=p-1$ and $q=p$.}
    \label{figure:TSDpq}
  \end{figure}}

When $q<p$, the restricted root system is again of type $B_{2q}$,
with $4q(2q-1)$ long roots of multiplicity one and $4q$ short roots of
multiplicity $2(p-q)$. For $p=q$, the short roots disappear and the
restricted root system is of $D_{2p}$ type, with all roots having
multiplicity one.


\subsection{Summary -- Tits--Satake diagrams for non-compact real forms}
\label{section:summary-titssatakediagrams}
\index{Tits--Satake diagram}

To summarize the analysis, we provide the Tits--Satake diagrams for
all noncompact real forms of all simple Lie
algebras~\cite{Araki,Helgason}. We do not give explicitly the
Tits--Satake diagrams of the compact real forms as these are simply
obtained by painting in black all the roots of the standard Dynkin
diagrams.

\begin{theorem}
  The simple real Lie algebras are:
  
  \begin{itemize}
  \item The Lie algebras $\GR$ where ${\mathfrak{g}}$ is one of the
    complex simple Lie algebras $A_{n} \, (n\geq 1),\, B_{n} \, (n\geq 2),
    \, C_{n} \, (n\geq 3),\, D_{n} \, (n\geq 4), \, G_{2} ,\, F_{4}, \, E_{6},\
    E_{7}, \mbox{ or } E_{8}$, and the compact real forms of these.
  \item The classical real Lie algebras of types $\mf{su}$, $\mf{so}$,
    $\mf{sp}$ and $\mf{sl}$. These are listed in
    Table~\ref{table:ClassicalRealLie}.
  \item The twelve exceptional real Lie algebras, listed in
    Table~\ref{table:ExceptionalRealLie} (our conventions are due to
    Cartan).
  \end{itemize}
  
\end{theorem}

\begin{table}
  \caption{All classical real Lie algebras of $\mf{su}$, $\mf{so}$,
  $\mf{sp}$ and $\mf{sl}$ type.}
  \label{table:ClassicalRealLie}
  \renewcommand{\arraystretch}{1.2}
  \vspace{0.5 em}
  \centering
  \begin{tabular}{cl|c|c}
    \toprule
    \multicolumn{2}{c|}{Algebra} &
    Real rank &
    Restricted root lattice \\
    \midrule
    $\mathfrak{su}(p,q)$ &
    $p\geq q >0\, p+q\geq 2$ &
    $q$ &
    $(BC)_{q}$ if $p>q$, $C_{q}$ if $p=q$ \\ [0.5 em]
    $\mathfrak{so}(p,q)$ &
    $p> q >0\, p+q= 2n +1\geq 5$ &
    $q$ &
    $B_{q}$ \\
    &
    $p\geq q >0\, p+q=2n\geq 8 $ &
    $q$ &
    $B_{q}$ if $p>q$, $D_{q}$ if $p=q$ \\ [0.5 em]
    $\mathfrak{sp}(p,q)$ &
    $p\geq q >0\, p+q\geq  3$ &
    $q$ &
    $(BC)_{q}$ if $p>q$, $C_{q}$ if $p=q$ \\ [0.5 em]
    $\mathfrak{sp}(n,\RR)$ &
    $n \geq  3$ &
    $n$ &
    $C_{n}$ \\ [0.5 em]
    $\mf{so}^{*}(2\,n)$ &
    $n\geq 5 $ &
    $[n/2] $ &
    $C_{\frac n 2}$ if $n$ even, $(BC)_{\frac{n-1}2}$ if $n$ odd \\ [0.5 em]
    $\mathfrak{sl}(n,\RR)$ &
    $n\geq 3$ &
    $n-1$ &
    $A_{n-1}$ \\ [0.5 em]
    $\mathfrak{sl}(n,\HH)$ &
    $n\geq 2$ &
    $n-1$ &
    $A_{n-1}$ \\
    \bottomrule
  \end{tabular}
  \renewcommand{\arraystretch}{1.0}
\end{table}

\begin{table}
  \caption{All exceptional real Lie algebras.}
  \label{table:ExceptionalRealLie}
  \renewcommand{\arraystretch}{1.2}
  \vspace{0.5 em}
  \centering
  \begin{tabular}{c|c|c}
    \toprule
    Algebra & Real rank & Restricted root lattice \\
    \midrule
    $G$  &  2 & $G_{2}$ \\
    $F~I$  &  4 & $F_{4}$ \\
    $F~II$  &  1 & $(BC)_{1}$ \\
    $E~I$  &  6 & $E_{6}$ \\
    $E~II$  &  4 & $F_{4}$ \\
    $E~III$  &  2 & $(BC)_{2}$ \\
    $E~IV$  &  2 & $A_{2}$ \\
    $E~V$  &  7 & $E_{7}$ \\
    $E~VI$  &  4 & $F_{4}$ \\
    $E~VII$  &  3 & $C_{3}$ \\
    $E~VIII$  &  8 & $E_{8}$ \\
    $E~IX$  &  4 & $F_{4}$ \\
    \bottomrule
  \end{tabular}
  \renewcommand{\arraystretch}{1.0}
\end{table}


\begin{table}[htbp]
  \caption{Tits--Satake diagrams ($A_{n}$ series)}
  \label{table:A(n)series}
  \vspace{0.5 em}
  \centering
\begin{tabular}{l|c|c}
\toprule
$A_{n}$ series $n\geq 1$ & Tits--Satake diagram & Restricted root system\\
\midrule
\raisebox{5mm}{
\parbox[t][15mm][t]{40mm}{
$\mathfrak{sl}(n,\RR),\  n\geq 3$\\
\strut\\
$A~I$}}
&
\centering \includegraphics[width=45mm]{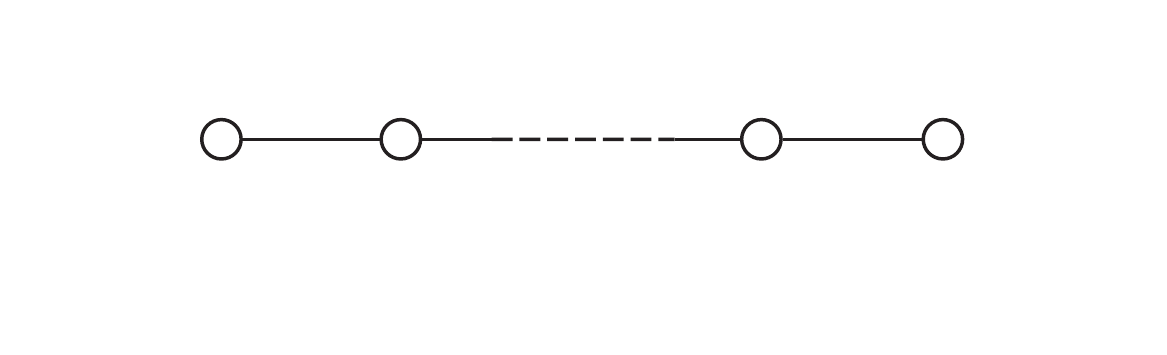}
&
\parbox[t][15mm][t]{45mm}{
\centering \includegraphics[width=45mm]{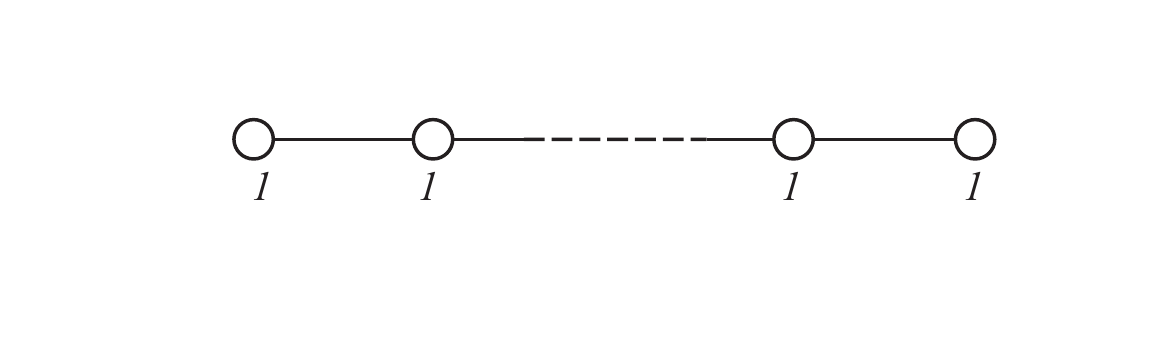}\\
$A_{n}$}\\
\midrule
\raisebox{5mm}{
\parbox[t][15mm][t]{40mm}{
$\mathfrak{su}^{*}(n+1),\ n=2k+1 $\\
\strut\\
$A~II$}}
&
\parbox[t][15mm][t]{45mm}{
\centering \includegraphics[width=45mm]{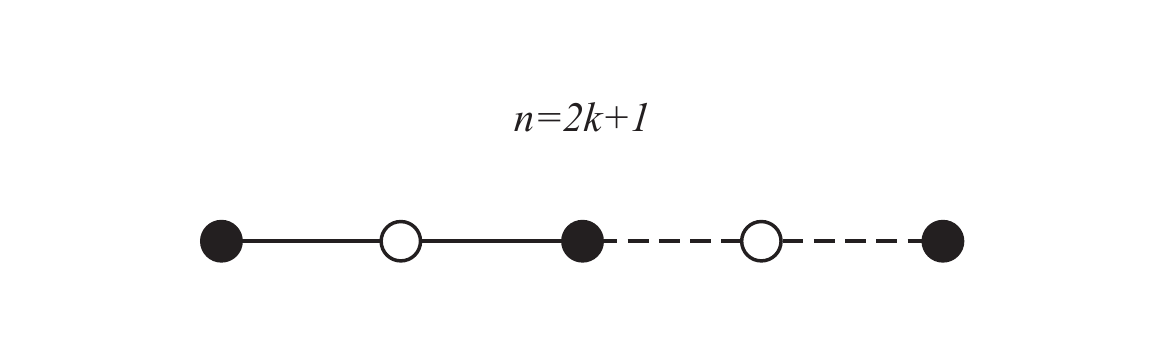}\\
\scriptsize (k+1) black and $k$ white roots alternate.}
&
\parbox[t][15mm][t]{45mm}{
\centering \includegraphics[width=45mm]{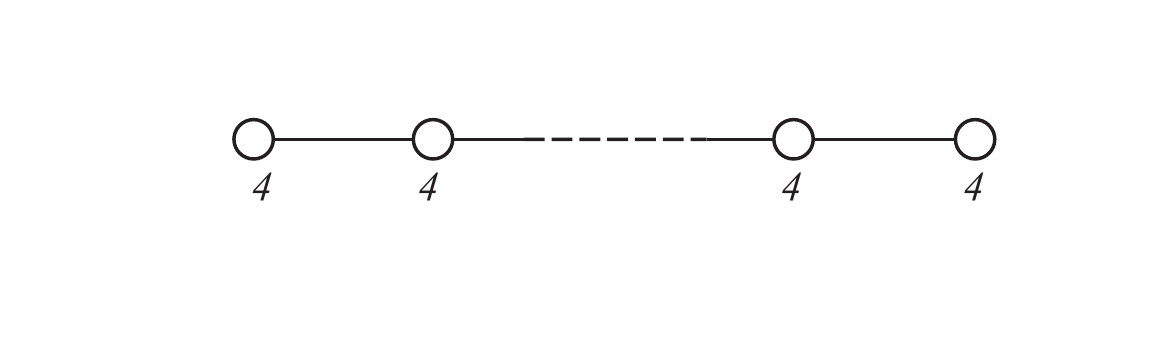}\\
$A_{2k}$}\\
\midrule
\raisebox{5mm}{
\parbox[t][15mm][t]{40mm}{
$\mathfrak{su}(p, n+1-p)$\\
\strut\\
$A~III$}}
&
\parbox[t][15mm][t]{45mm}{
\centering \includegraphics[width=45mm]{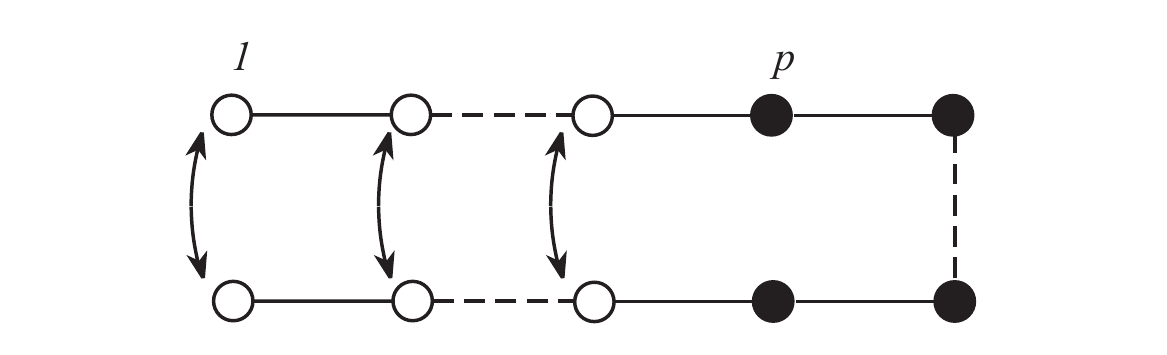}\\
\scriptsize The $p(>0)$ first and $p$ last roots \\are white and connected.}
&
\parbox[t][15mm][t]{45mm}{
\centering \includegraphics[width=45mm]{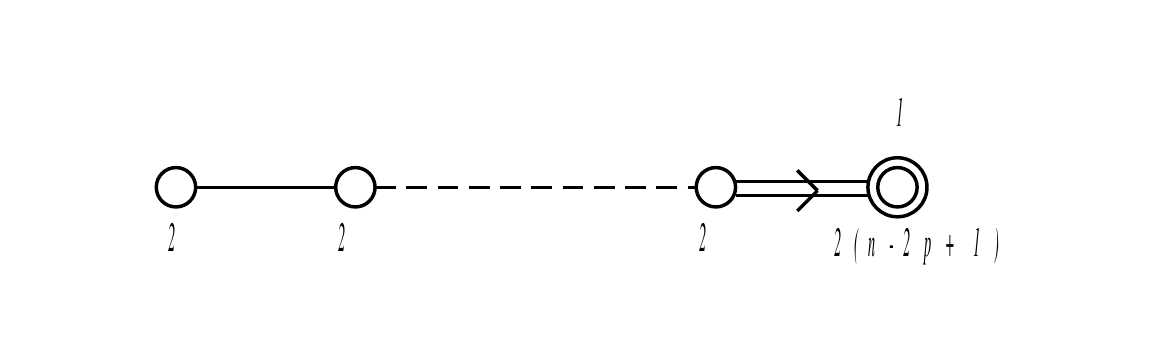}\\
$BC_{p}$}
\\
\midrule
\raisebox{5mm}{
\parbox[t][15mm][t]{40mm}{
$\mathfrak{su}(\frac{n+1}2,\frac{n+1}2),\ n=2k+1$\\
\strut\\
$A~III$}}
&
\centering \includegraphics[width=45mm]{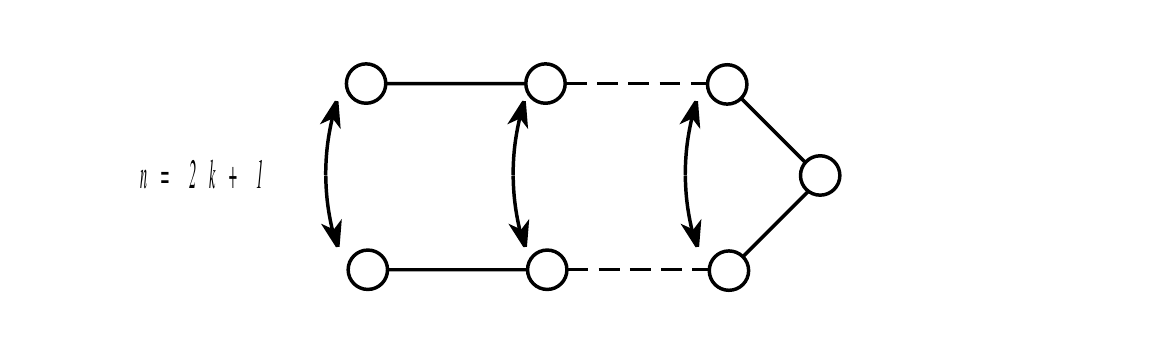}
&
\parbox[t][15mm][t]{45mm}{
\centering \includegraphics[width=45mm]{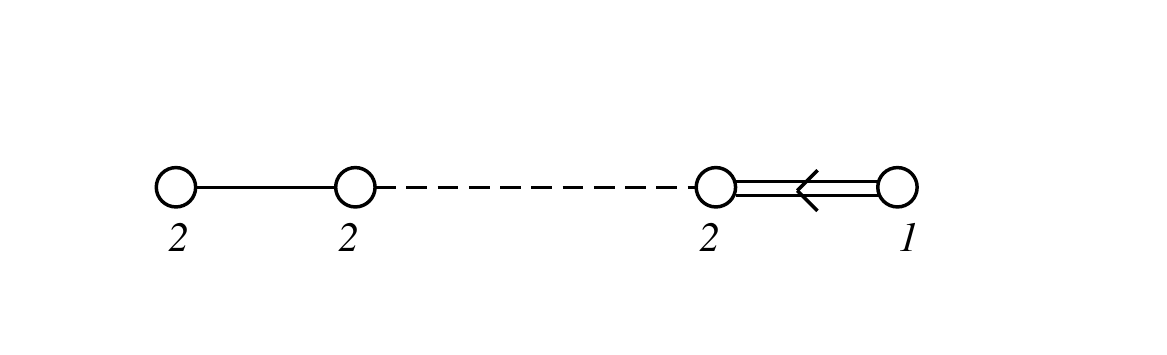}\\
$C_{(k+1)}$}\\
\midrule
\raisebox{5mm}{
\parbox[t][15mm][t]{35mm}{
$\mathfrak{su}(1,n-1),\  n\geq 3 $\\
\strut\\
$A~IV$}}
&
\parbox[t][15mm][t]{45mm}{
\centering \includegraphics[width=45mm]{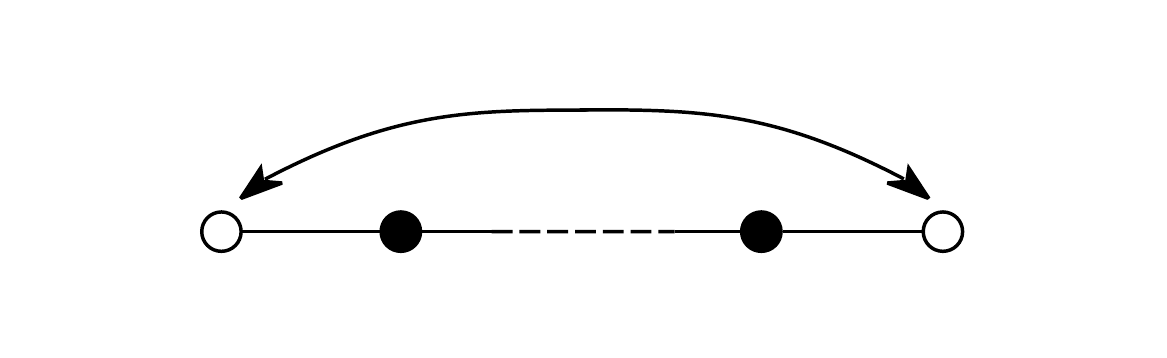}\\
\scriptsize Only the first and last roots \\are white and connected.}
&
\parbox[t][15mm][t]{45mm}{
\centering \includegraphics[width=45mm]{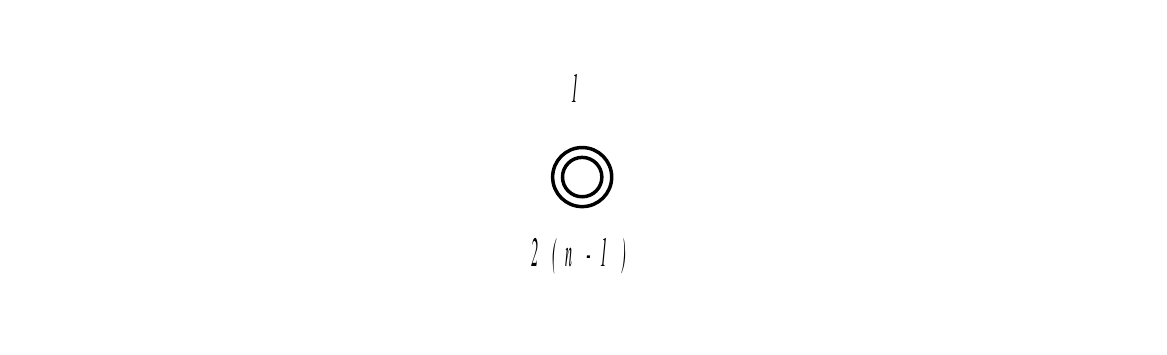}\\
$A_{1}$}\\
\bottomrule
\end{tabular}
\end{table}

\begin{table}[htbp]
  \caption{Tits--Satake diagrams ($B_{n}$ series)}
  \vspace{0.5 em}
  \centering
\begin{tabular}{l|c|c}
\toprule
$B_{n}$ series $n\geq 4$ & Tits--Satake diagram & Restricted root system\\
\midrule
\raisebox{5mm}{
\parbox[t][15mm][t]{40mm}{
$\mathfrak{so}(p,2n-p+1),\ p\geq 1$\\
\strut\\
$B~I$}}
&
\parbox[t][15mm][t]{45mm}{
\centering \includegraphics[width=45mm]{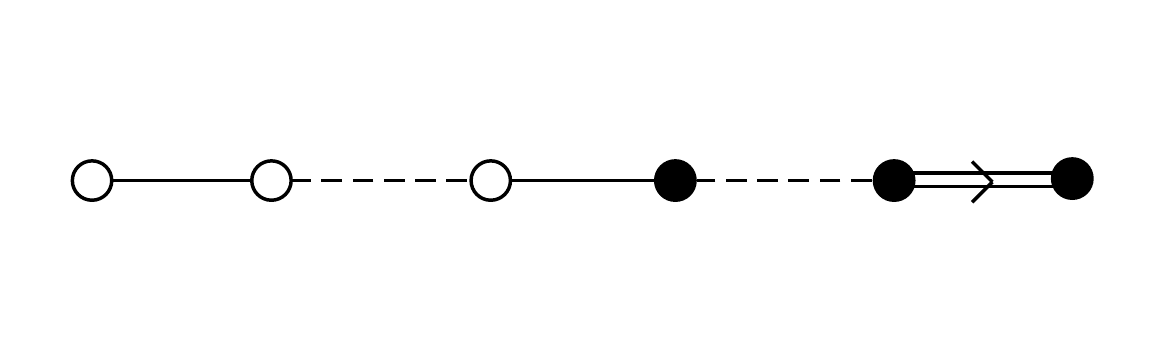}\\
\scriptsize The $p(\geq 2)$ first roots are white.}
&
\parbox[t][15mm][t]{45mm}{
\centering \includegraphics[width=45mm]{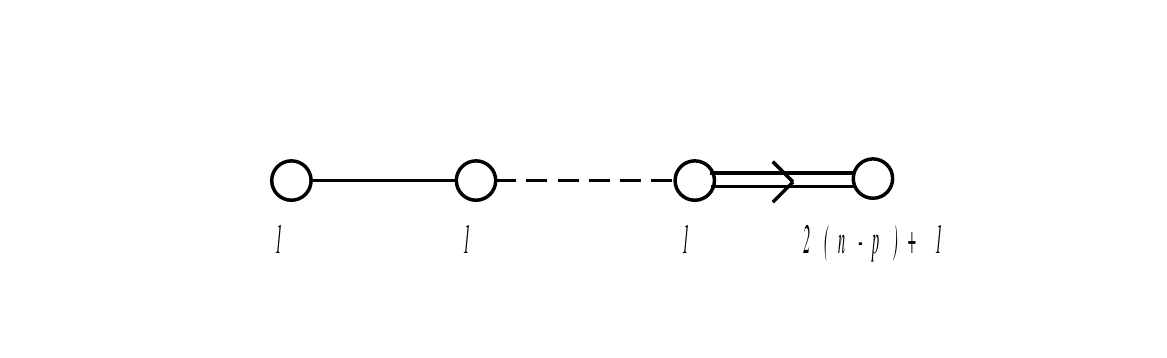}\\
$B_{p}$}\\
\midrule
\raisebox{5mm}{
\parbox[t][15mm][t]{40mm}{
$\mathfrak{so}(1,2n)$\\
\strut\\
$B~II$}}
&
\parbox[t][15mm][t]{45mm}{
\centering \includegraphics[width=45mm]{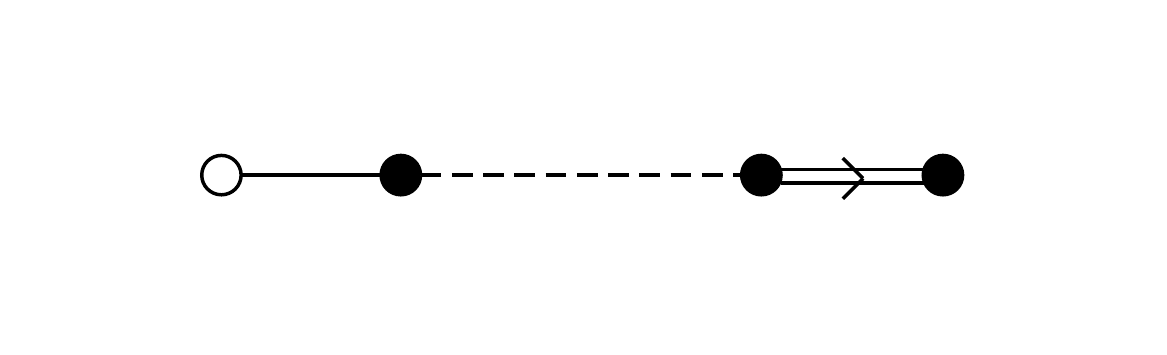}\\
\scriptsize Only the first root is white.}
&
\parbox[t][15mm][t]{45mm}{
\centering \includegraphics[width=45mm]{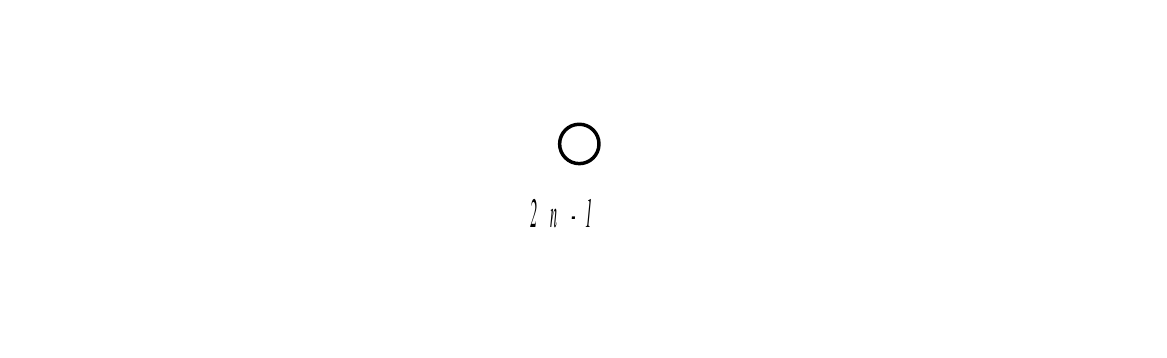}\\
$A_{1}$}\\
\bottomrule
\end{tabular}
\end{table}

\begin{table}[htbp]
  \caption{Tits--Satake diagrams ($C_{n}$ series)}
  \vspace{0.5 em}
  \centering
\begin{tabular}{l|c|c}
\toprule
$C_{n}$ series $n\geq 3$ & Tits--Satake diagram & Restricted root system\\
\midrule
\raisebox{5mm}{
\parbox[t][15mm][t]{40mm}{
$\mathfrak{sp}(n,\RR)$\\
\strut\\
$C~I$}}
&
\centering \includegraphics[width=45mm]{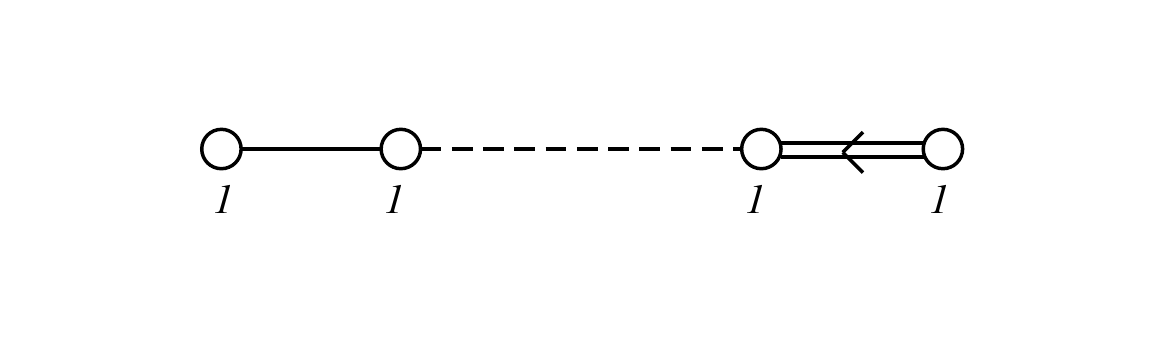}
&
\parbox[t][15mm][t]{45mm}{
\centering \includegraphics[width=45mm]{BSTCI}\\
$C_{n}$}\\
\midrule
\raisebox{5mm}{
\parbox[t][15mm][t]{40mm}{
$\mathfrak{sp}(p,n-p)$\\
\strut\\
$C~II$}}
&
\parbox[t][25mm][t]{45mm}{
\includegraphics[width=45mm]{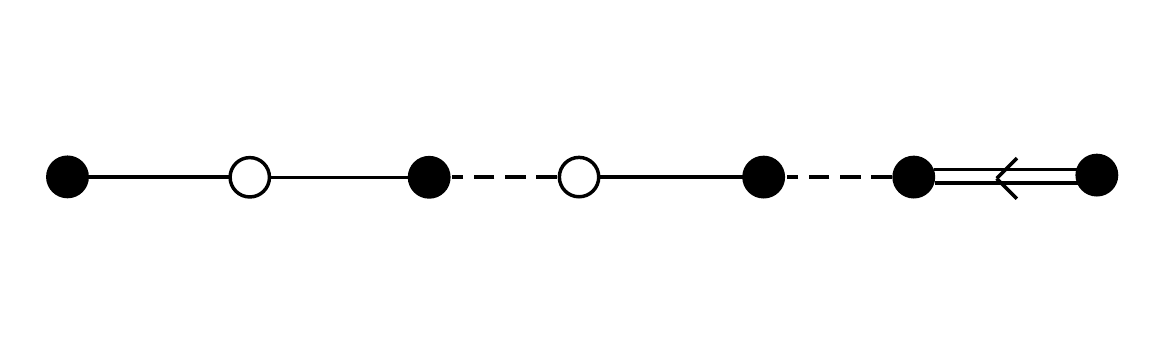}\\
\scriptsize The $2p$ first roots are alternatively \\
white and black, the $n-2p$ remaining are black}
&
\parbox[t][15mm][t]{45mm}{
\centering \includegraphics[width=45mm]{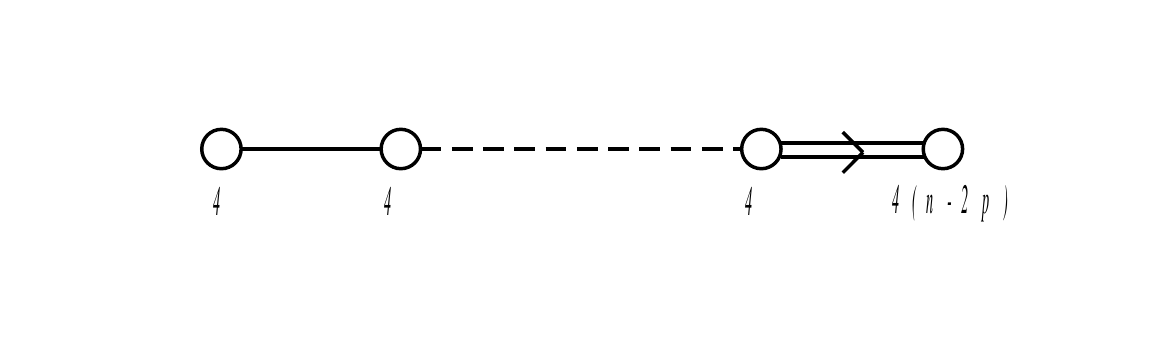}\\
$B_{p}$}\\
\midrule
\raisebox{5mm}{
\parbox[t][15mm][t]{40mm}{
$\mathfrak{sp}(\frac n2,\frac n2),\, n=2k$\\
\strut\\
$C~II$}}
&
\centering \includegraphics[width=45mm]{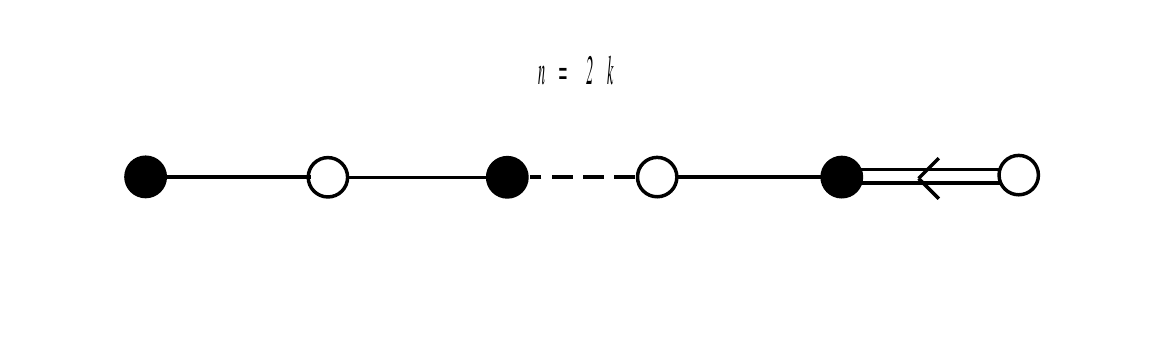}
&
\parbox[t][15mm][t]{45mm}{
\centering \includegraphics[width=45mm]{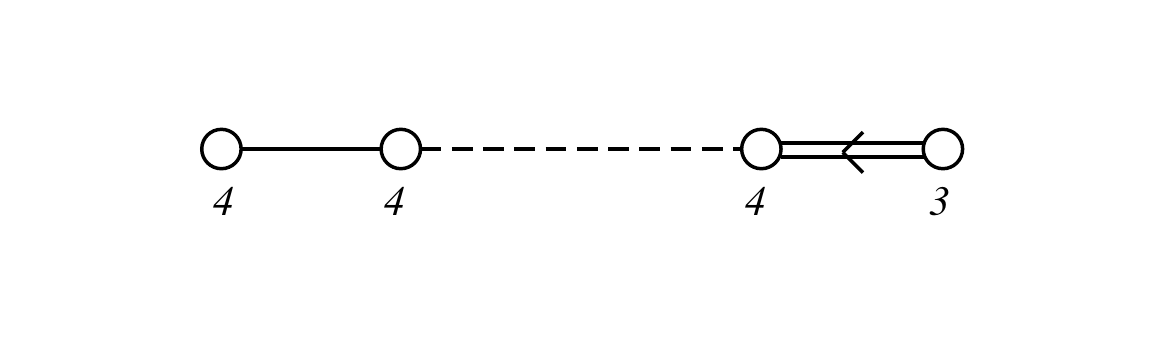}\\
$C_{\frac n2}$}\\
\bottomrule
\end{tabular}
\end{table}

\begin{table}[htbp]
  \caption{Tits--Satake diagrams ($D_{n}$ series)}
  \vspace{0.5 em}
  \centering
\begin{tabular}{l|c|c}
\toprule
$D_{n}$ series $n\geq 4$ & Tits--Satake diagram & Restricted root system\\
\midrule
\raisebox{5mm}{
\parbox[t][15mm][t]{40mm}{
$\mathfrak{so}(p,2n-p),\  p\leq n-2$\\
\strut\\
$D~I$}}
&
\parbox[t][15mm][t]{45mm}{
\centering \includegraphics[width=45mm]{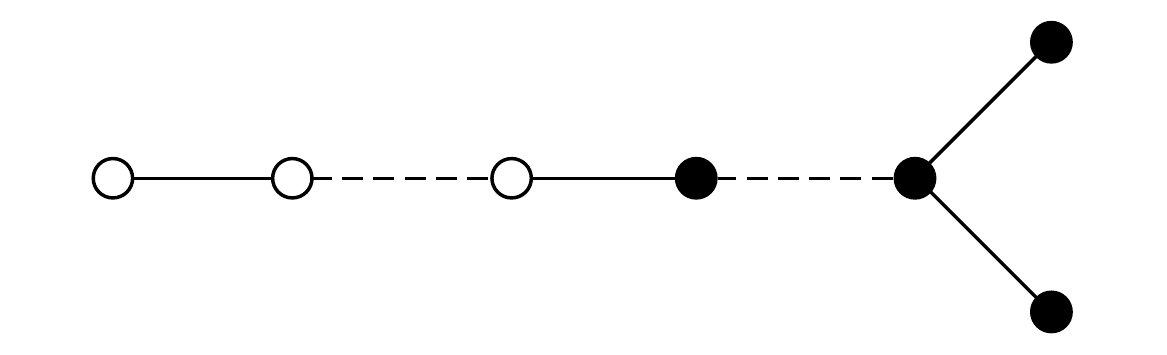}\\
\scriptsize The $p \leq n-2 $ first roots are white.}
&
\parbox[t][15mm][t]{45mm}{
\centering \includegraphics[width=45mm]{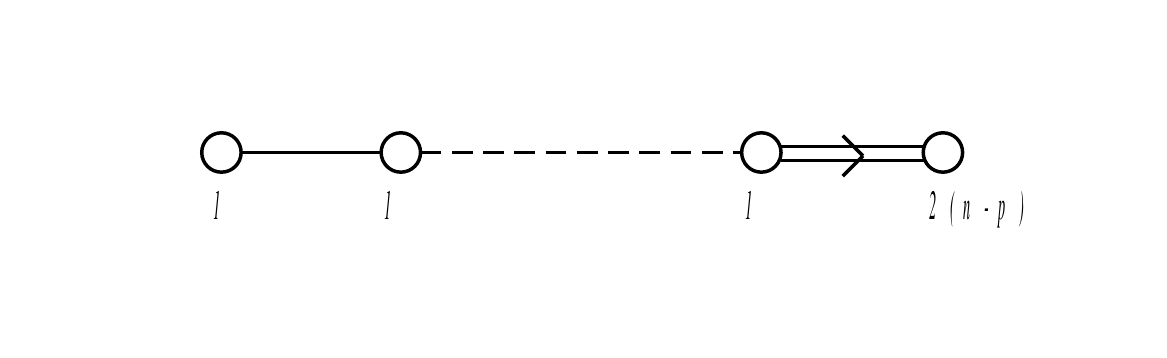}\\
$B_{p}$}\\
\midrule
\raisebox{5mm}{
\parbox[t][15mm][t]{40mm}{
$\mathfrak{so}(n-1,n+1)$\\
\strut\\
$D~I$}}
&
\centering \includegraphics[width=45mm]{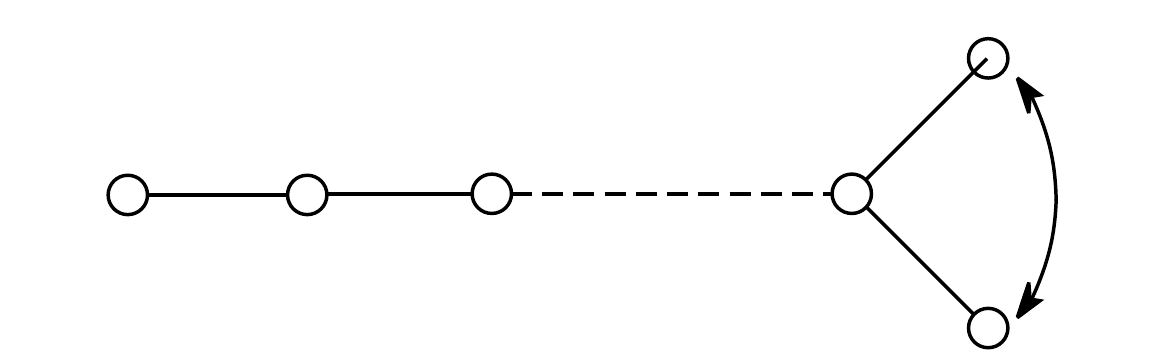}
&
\parbox[t][15mm][t]{45mm}{
\centering \includegraphics[width=45mm]{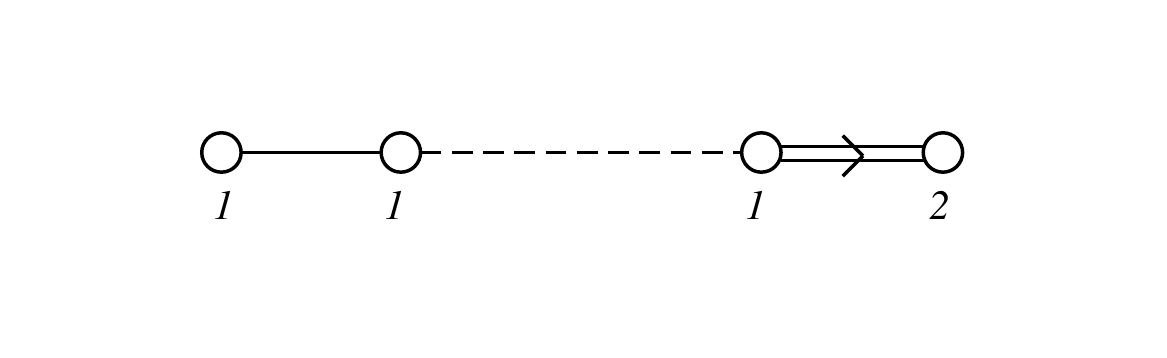}\\
$B_{(n-1)}$}\\
\midrule
\raisebox{5mm}{
\parbox[t][15mm][t]{40mm}{
$\mathfrak{so}(n,n)$\\
\strut\\
$D~I$}}
&
\centering \includegraphics[width=45mm]{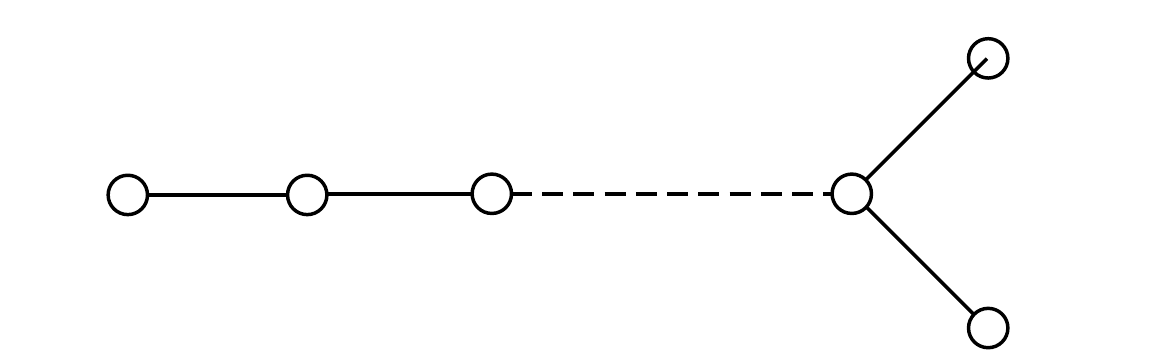}
&
\parbox[t][15mm][t]{45mm}{
\centering \includegraphics[width=45mm]{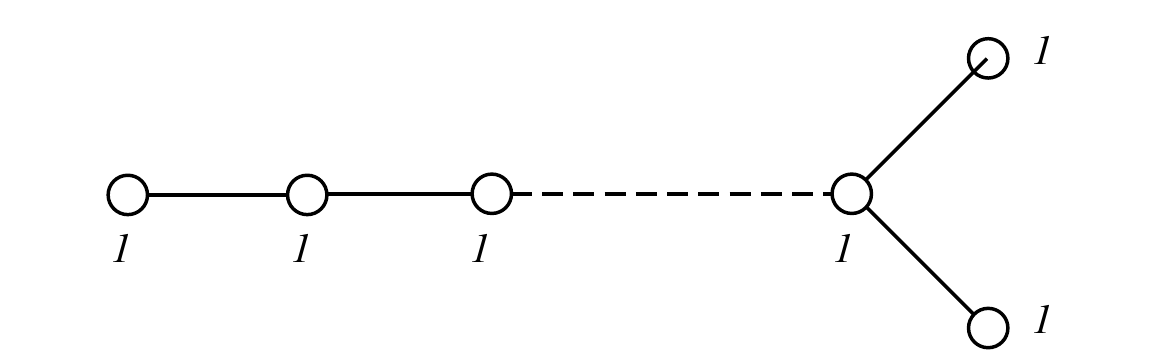}\\
$D_{n}$}\\
\midrule
\raisebox{5mm}{
\parbox[t][15mm][t]{40mm}{
$\mathfrak{so}(1,2n-1)$\\
\strut\\
$D~II$}}
&
\centering \includegraphics[width=45mm]{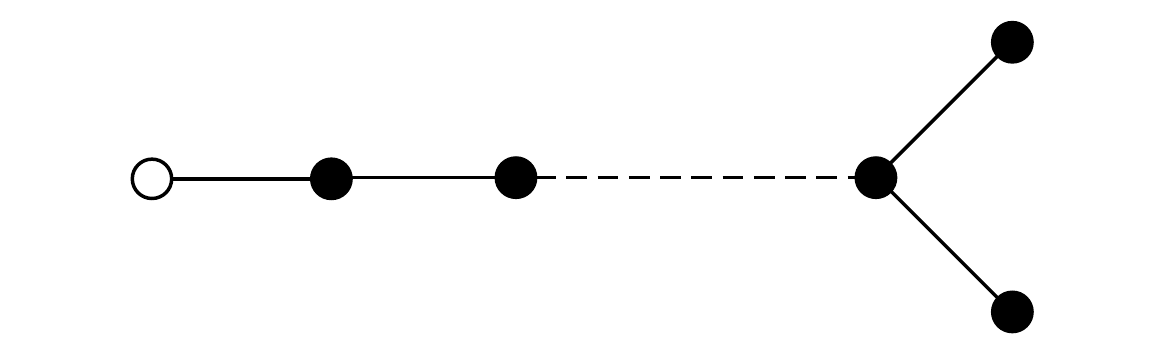}
&
\parbox[t][15mm][t]{45mm}{
\centering \includegraphics[width=45mm]{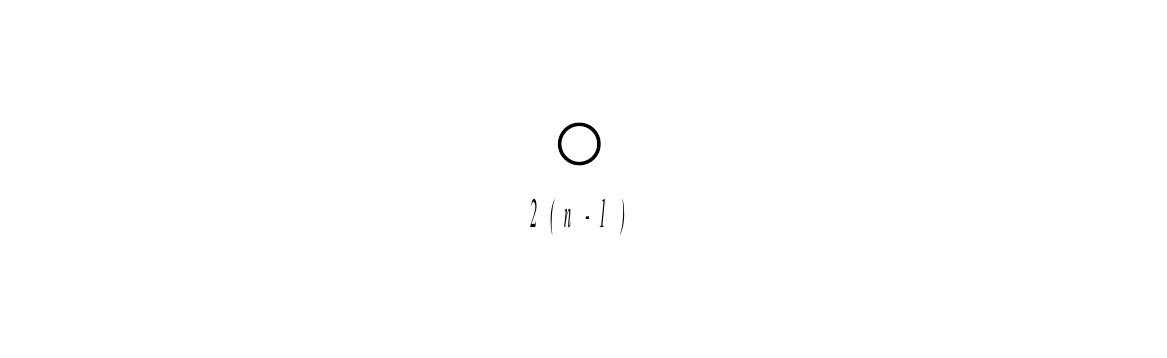}\\
$A_{1}$}\\
\midrule
\raisebox{5mm}{
\parbox[t][15mm][t]{40mm}{
$\mathfrak{so}(^{*}(2\,n)),\ n=2k$\\
\strut\\
$D~III$}}
&
\centering \includegraphics[width=45mm]{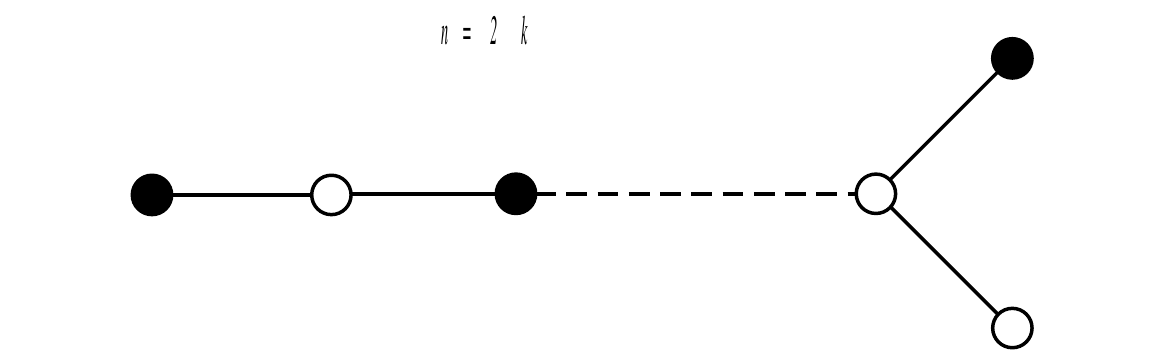}
&
\parbox[t][15mm][t]{45mm}{
\centering \includegraphics[width=45mm]{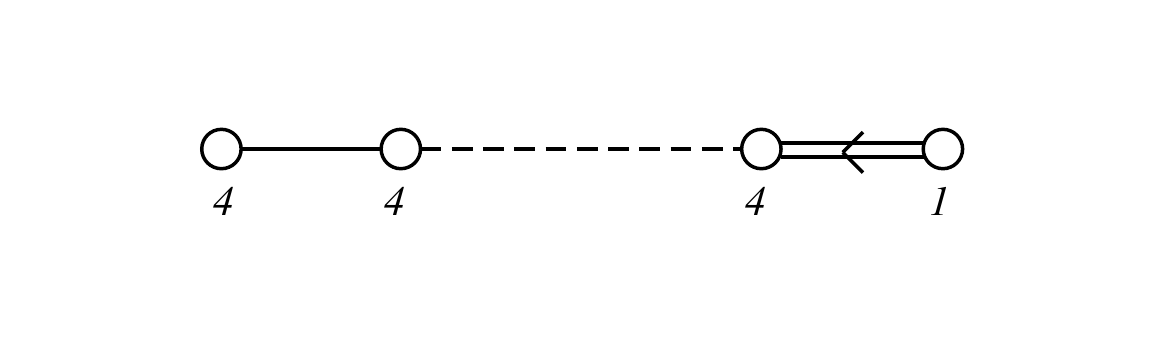}\\
$C_{2k-1}$}\\
\midrule
\raisebox{5mm}{
\parbox[t][15mm][t]{45mm}{
$\mathfrak{so}(^*(2n)),\ n=2k+1$\\
\strut\\
$D~III$}}
&
\centering \includegraphics[width=45mm]{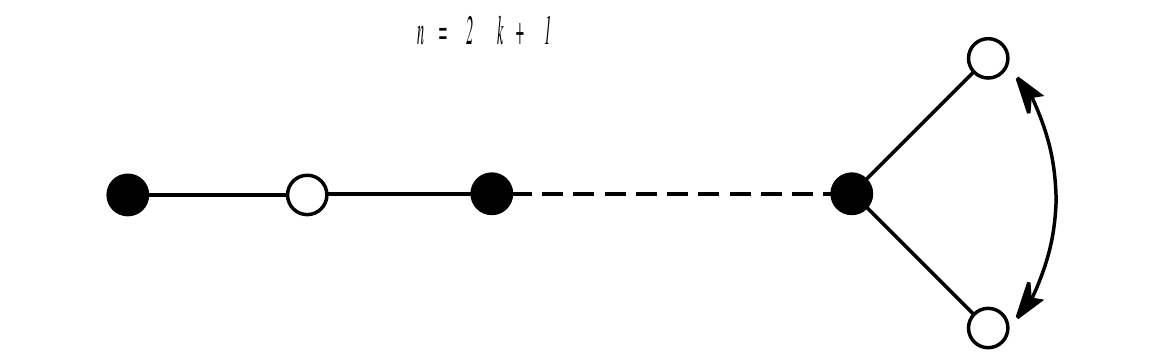}
&
\parbox[t][15mm][t]{45mm}{
\centering \includegraphics[width=45mm]{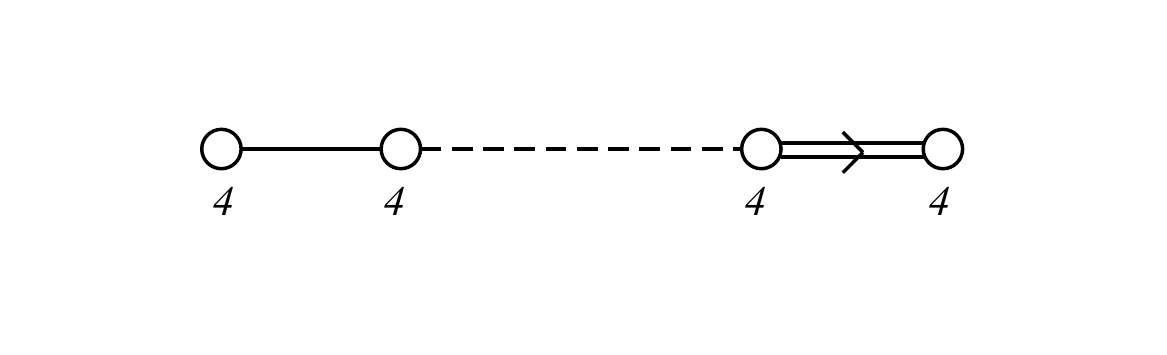}\\
$BC_{2k}$}\\
\bottomrule
\end{tabular}
\end{table}

\begin{table}[htbp]
  \caption{Tits--Satake diagrams ($G_{2}$ series)}
  \vspace{0.5 em}
  \centering
\begin{tabular}{l|c|c}
\toprule
$G_{2}$ series & Tits--Satake diagram & Restricted root system\\
\midrule
\raisebox{12mm}{
\parbox[t][15mm][t]{25mm}{
$G_{2(2)}$\\
\strut\\
$G$}}
&
\raisebox{1mm}{
\parbox[t][10mm][t]{45mm}{
\centering \includegraphics[width=20mm]{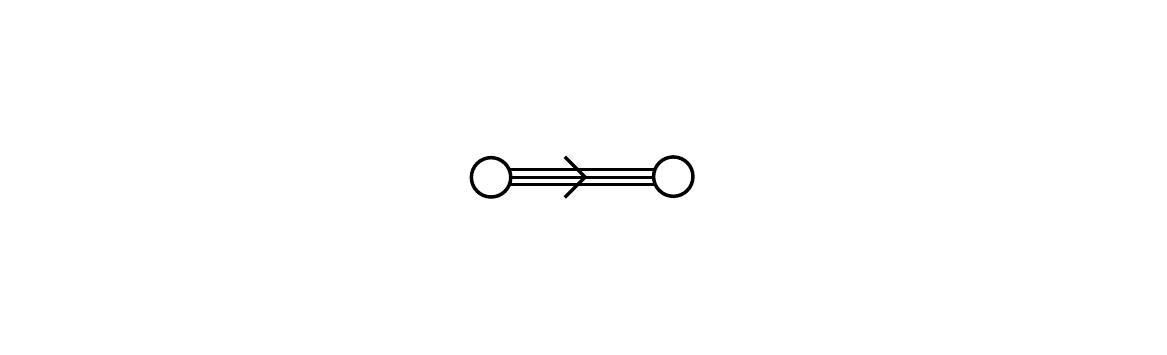}}}
&
\parbox[t][10mm][t]{45mm}{
\centering \includegraphics[width=20mm]{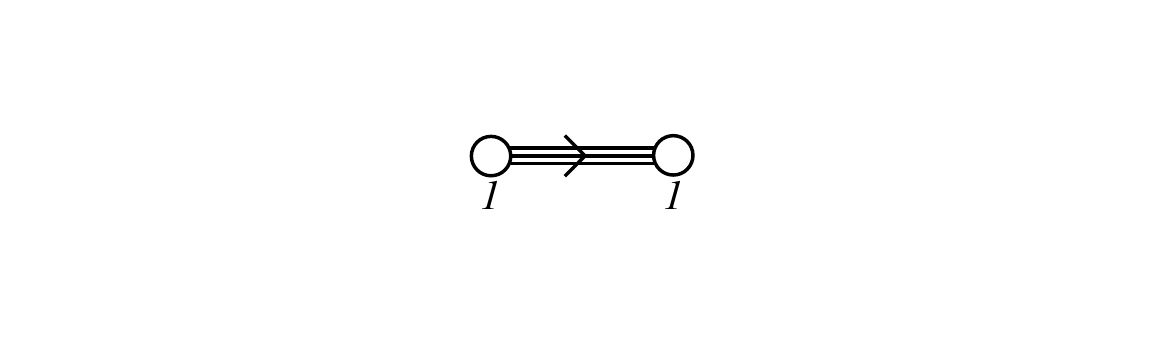}}\\
\bottomrule
\end{tabular}
\end{table}

\begin{table}[htbp]
  \caption{Tits--Satake diagrams ($F_{4}$ series)}
  \vspace{0.5 em}
  \centering
\begin{tabular}{l|c|c}
\toprule
$F_{4}$ series & Tits--Satake diagram & Restricted root system\\
\midrule
\raisebox{10mm}{
\parbox[t][15mm][t]{25mm}{
$F_{4(4)}$\\
\strut\\
$F~I$}}
&
\includegraphics[width=45mm]{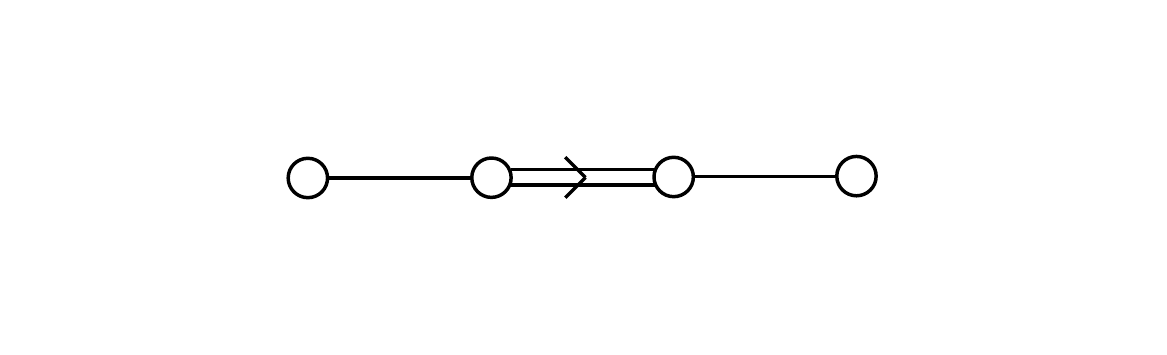}
&
\includegraphics[width=45mm]{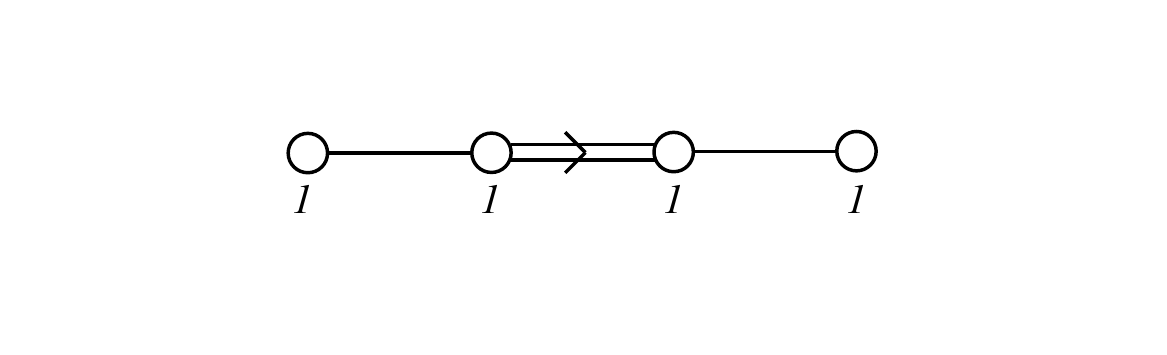}\\
\midrule
\raisebox{10mm}{
\parbox[t][15mm][t]{25mm}{
$F_{4(-20)}$\\
\strut\\
$F~II$}}
&
\includegraphics[width=45mm]{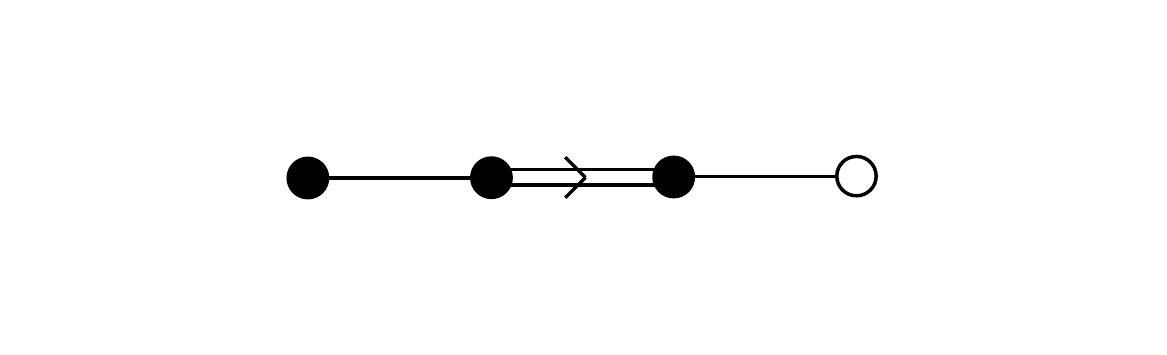}
&
\includegraphics[width=45mm]{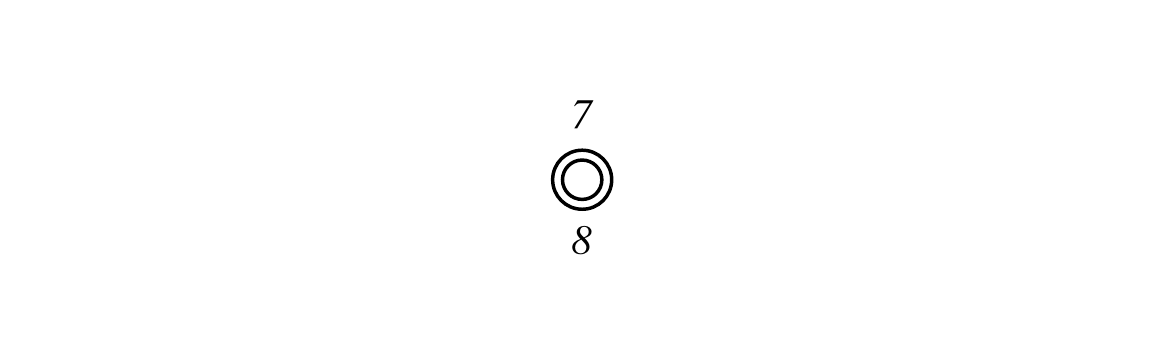}\\
\bottomrule
\end{tabular}
\end{table}

\begin{table}[htbp]
  \caption{Tits--Satake diagrams ($E_{6}$ series)}
  \vspace{0.5 em}
  \centering
\begin{tabular}{l|c|c}
\toprule
$E_{6}$ series & Tits--Satake diagram & Restricted root system\\
\midrule
\raisebox{10mm}{
\parbox[t][15mm][t]{25mm}{
$E_{6(6)}$\\
\strut\\
$E~I$}}
&
\includegraphics[width=45mm]{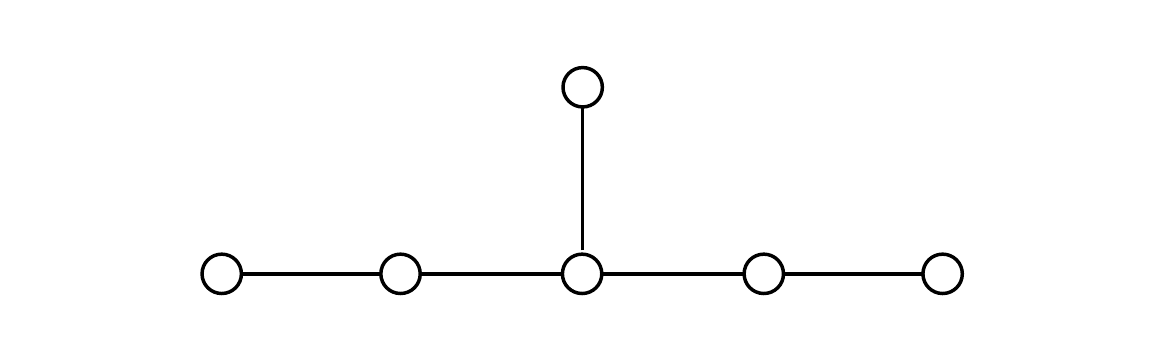}
&
\includegraphics[width=45mm]{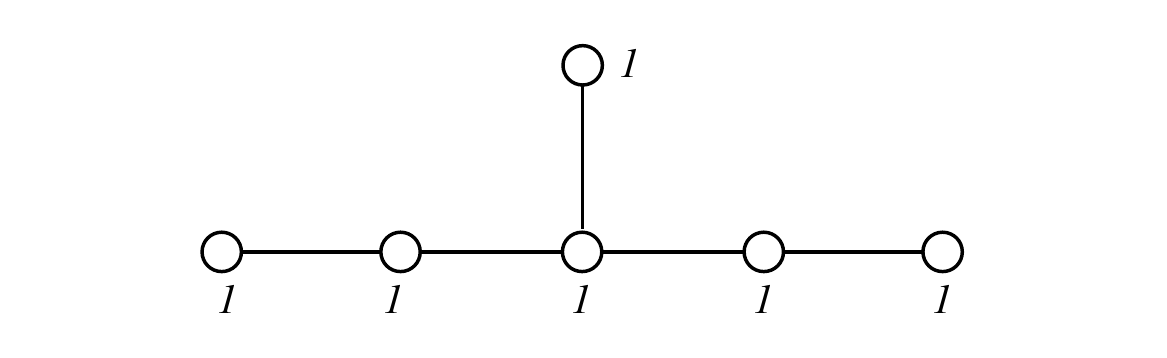}\\
\midrule
\raisebox{10mm}{
\parbox[t][15mm][t]{25mm}{
$E_{6(2)}$\\
\strut\\
$E~II$}}
&
\includegraphics[width=45mm]{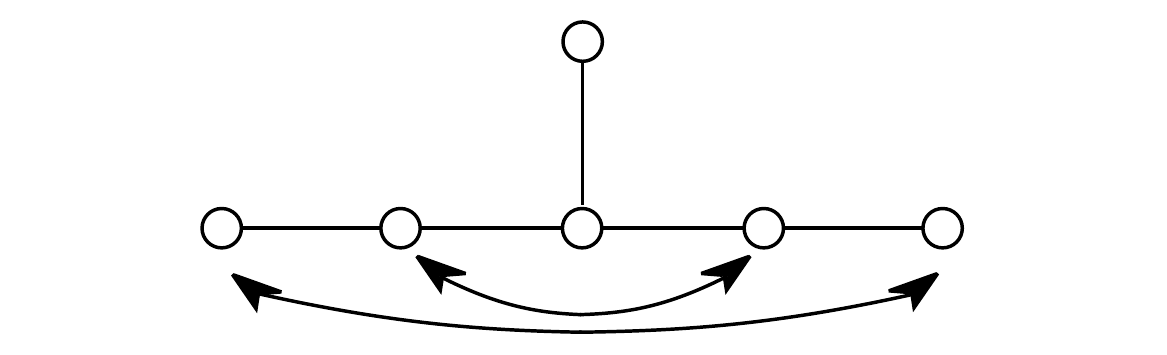}
&
\includegraphics[width=45mm]{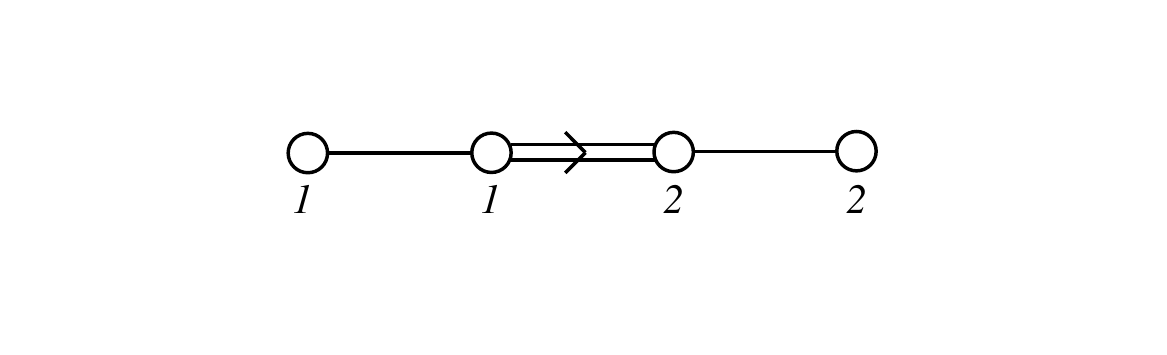}\\
\midrule
\raisebox{10mm}{
\parbox[t][15mm][t]{25mm}{
$E_{6(-14)}$\\
\strut\\
$E~III$}}
&
\includegraphics[width=45mm]{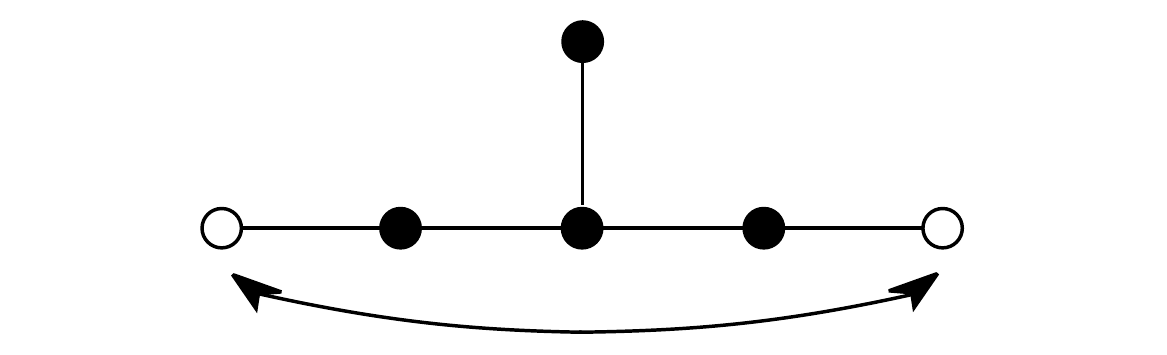}
&
\includegraphics[width=45mm]{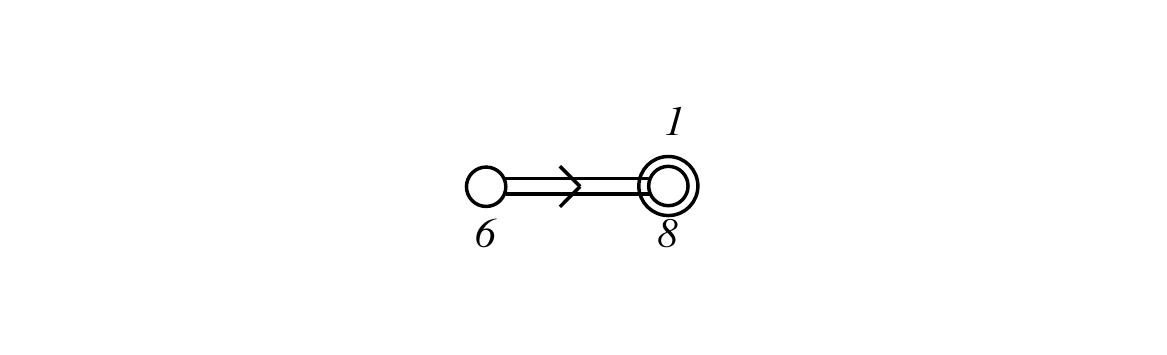}\\
\midrule
\raisebox{10mm}{
\parbox[t][15mm][t]{25mm}{
$E_{6(-26)}$\\
\strut\\
$E~IV$}}
&
\includegraphics[width=45mm]{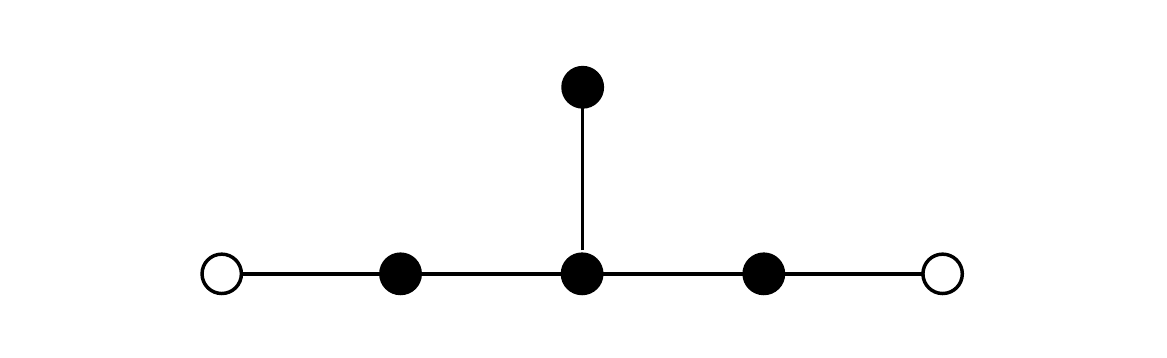}
&
\includegraphics[width=45mm]{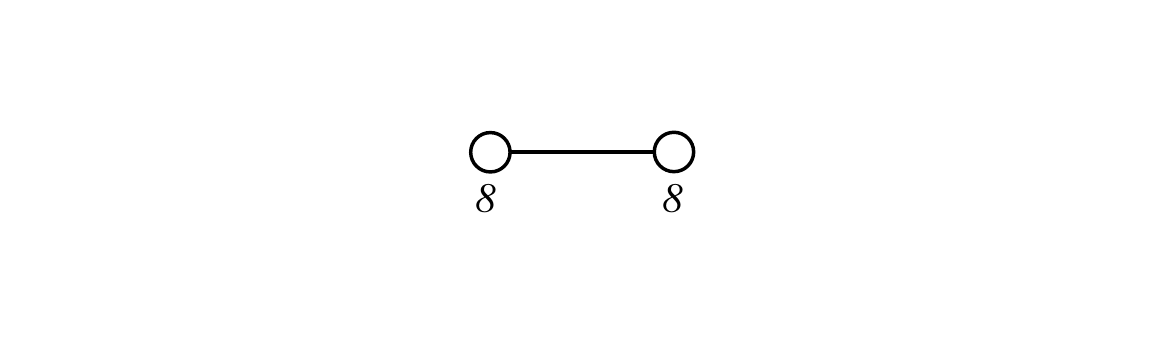}\\
\bottomrule
\end{tabular}
\end{table}

\begin{table}[htbp]
  \caption{Tits--Satake diagrams ($E_{7}$ series)}
  \vspace{0.5 em}
  \centering
\begin{tabular}{l|c|c}
\toprule
$E_{7}$ series & Tits--Satake diagram & Restricted root system\\
\midrule
\raisebox{10mm}{
\parbox[t][15mm][t]{25mm}{
$E_{7(7)}$\\
\strut\\
$E~V$}}
&
\parbox[t][15mm][t]{50mm}{
\includegraphics[width=45mm]{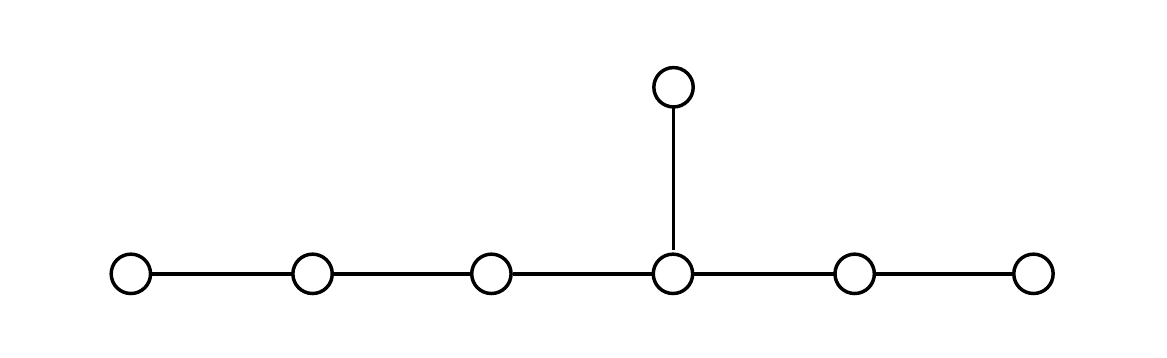}
}
&
\includegraphics[width=45mm]{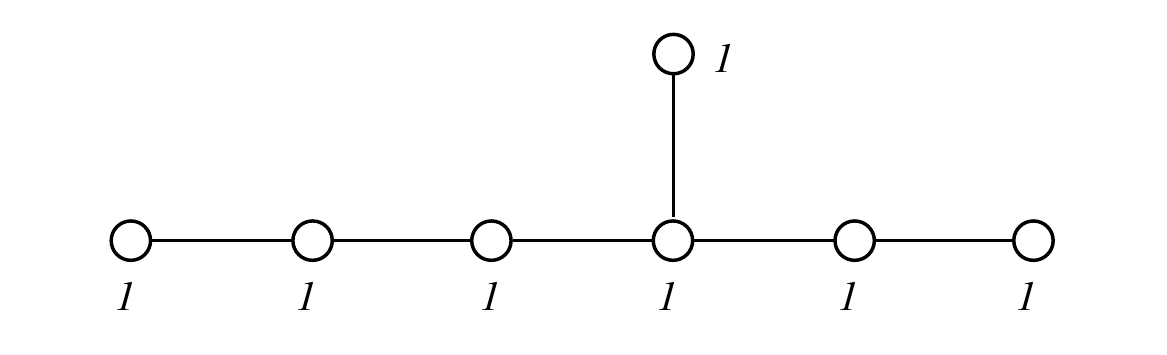}\\
\midrule
\raisebox{10mm}{
\parbox[t][15mm][t]{25mm}{
$E_{7(-5)}$\\
\strut\\
$E~VI$}}
&
\includegraphics[width=45mm]{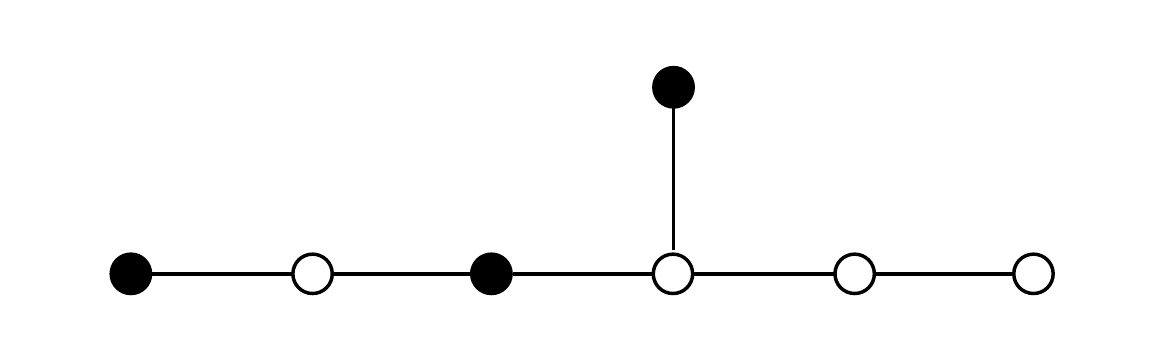}
&
\includegraphics[width=45mm]{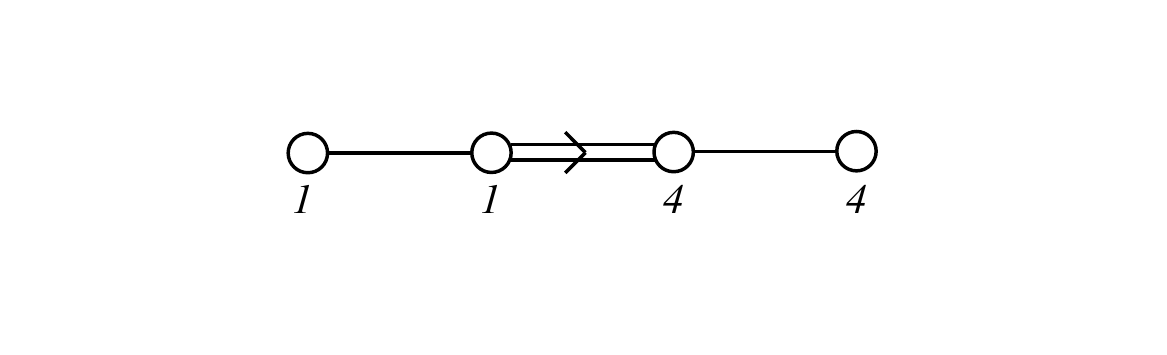}\\
\midrule
\raisebox{10mm}{
\parbox[t][15mm][t]{25mm}{
$E_{7(-25)}$\\
\strut\\
$E~VII$}}
&
\includegraphics[width=45mm]{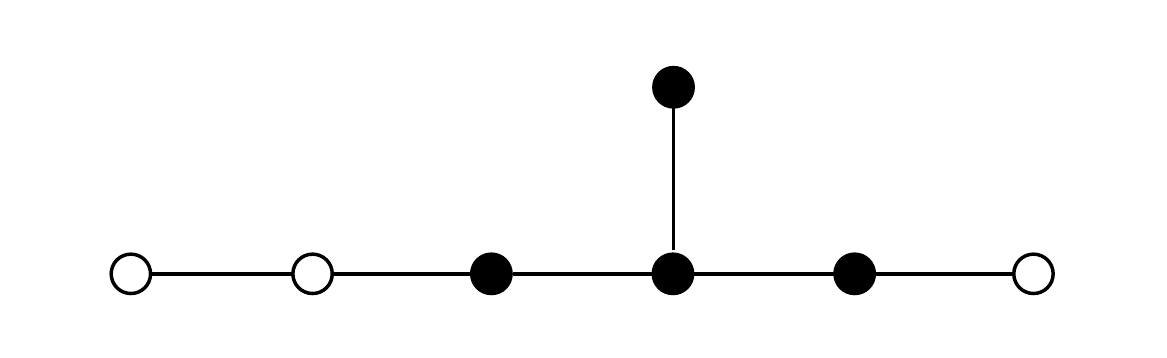}
&
\parbox[t][15mm][t]{60mm}{
\includegraphics[width=45mm]{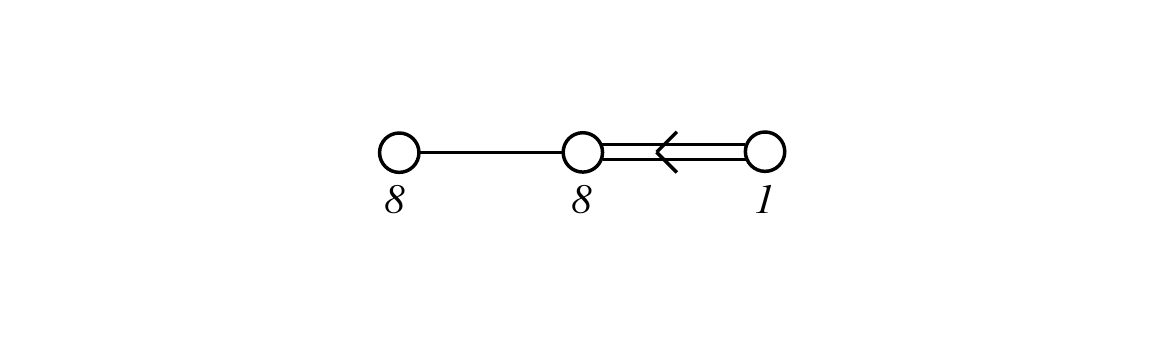}
}\\
\bottomrule
\end{tabular}
\end{table}

\begin{table}[htbp]
  \caption{Tits--Satake diagrams ($E_{8}$ series)}
  \vspace{0.5 em}
  \centering
\begin{tabular}{l|c|c}
\toprule
$E_{8}$ series & Tits--Satake diagram & Restricted root system\\
\midrule
\raisebox{10mm}{
\parbox[t][15mm][t]{25mm}{
$E_{8(8)}$\\
\strut\\
$E~VIII$}}
&
\includegraphics[width=50mm]{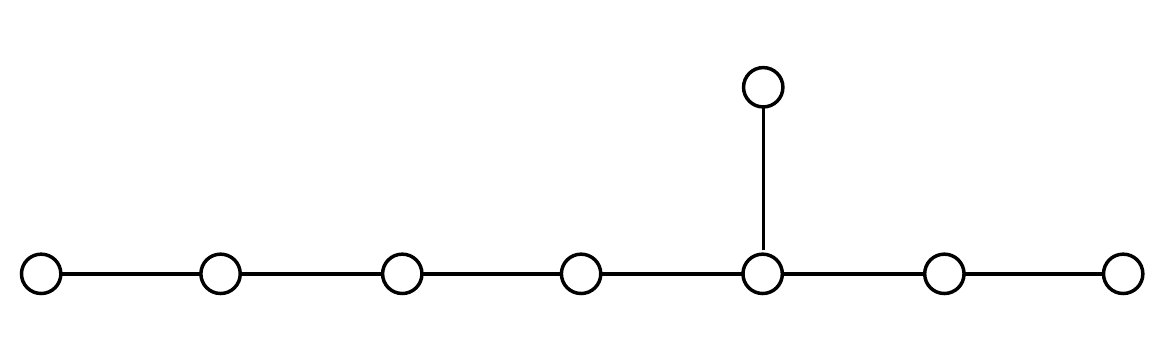}
&
\includegraphics[width=50mm]{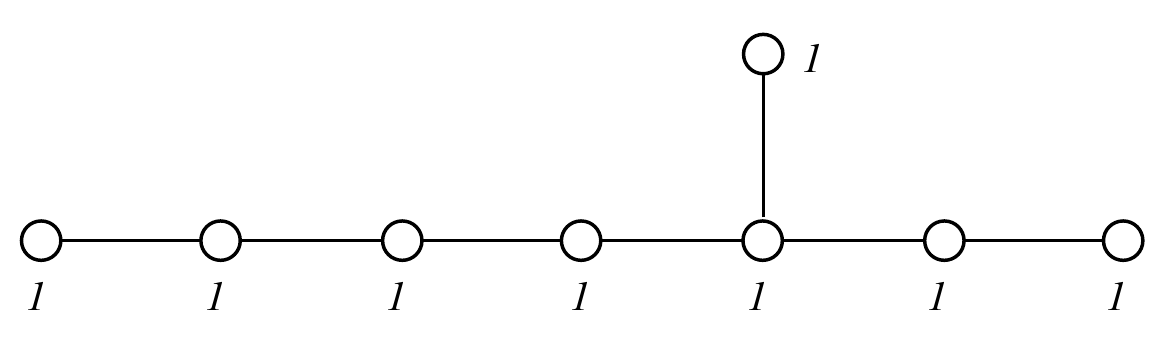}\\
\midrule
\raisebox{10mm}{
\parbox[t][15mm][t]{25mm}{
$E_{8(-24)}$\\
\strut\\
$E~IX$}}
&
\includegraphics[width=50mm]{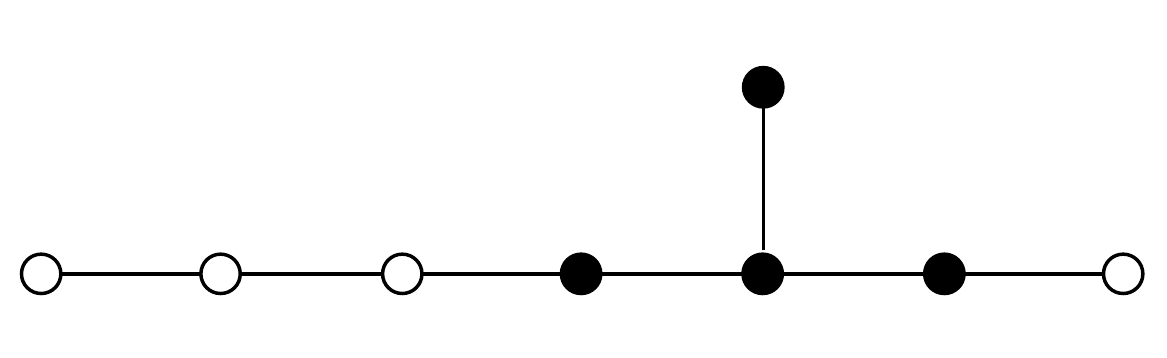}
&
\includegraphics[width=60mm]{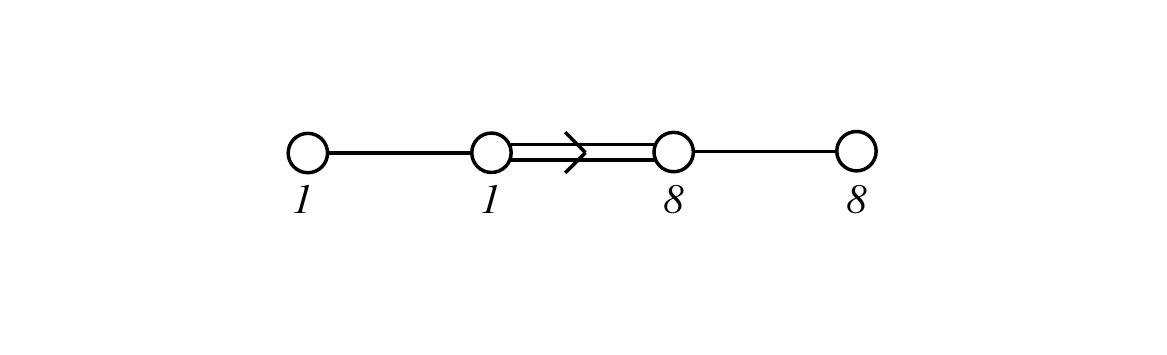}\\
\bottomrule
\end{tabular}
\end{table}

\clearpage


\section{Kac--Moody Billiards II -- The Case of Non-Split Real Forms}
\label{section:KMBilliardsII}
\setcounter{equation}{0}

We will now make use of the results from the previous section to
extend the analysis of Kac--Moody billiards to include also
theories whose U-duality symmetries are described by algebras
$\mf{u}_3$ that are non-split. The key concepts are that of \emph
{restricted root systems}, \emph{restricted Weyl group} -- and the
associated concept of \emph{maximal split subalgebra} -- as well
as the Iwasawa decomposition \index{Iwasawa decomposition} already
encountered above. These play a prominent role in our discussion as
they determine the billiard structure. We mainly
follow~\cite{HenneauxJulia}.


\subsection{The restricted Weyl group and the maximal split ``subalgebra''}
\label{section:SplitVSNonSplit}

Let $\mf{u}_3$ be any real form of the complex Lie algebra
$\mf{u}_3^{\mbb{C}}$, $\theta$ its Cartan involution, \index{Cartan involution} and let 
\begin{equation}
  \mf{u}_3=\mf{k}_3\oplus \mf{p}_3
  \label{CartandecompositionU3}
\end{equation}
be the corresponding Cartan decomposition. Furthermore, let 
\begin{equation}
  \cH_3 = \cT_3 \oplus \cA_3 
  \label{noncompactCartansubalgebra}\end{equation}
be a maximal noncompact Cartan subalgebra, with $\cT_3$
(respectively, $\cA_3$) its compact (respectively, noncompact)
part. The real rank of $\mf{u}_3$ is, as we have seen, the dimension
of $\cA_3$. Let now $\Delta$ denote the root system of
$\mf{u}_3^{\mbb{C}}$, $\Sigma$ the restricted root system
\index{restricted root system} and $m_\lambda$ the multiplicity of the
restricted root $\lambda$.

As explained in Section~\ref{section:RootSystemsEuclidean}, the
restricted root system of the real form $\mf{u}_3$ can be either
reduced or non-reduced. If it is reduced, it  corresponds to one of
the root systems of the finite-dimensional simple Lie algebras. On the
other hand, if the restricted root system is non-reduced, it is
necessarily of $(BC)_n$-type~\cite{Helgason} (see
Figure~\ref{figure:BC23} for a graphical presentation of the $BC_3$
root system).

\subsubsection*{The restricted Weyl group}

By definition, the restricted Weyl group is the Coxeter
group generated by the fundamental reflections,
Equation~(\ref{fundamentalWeyl}), with respect to the simple roots of
the restricted root system. \index{restricted root system} The restricted Weyl group preserves
multiplicities~\cite{Helgason}.

\subsubsection*{The maximal split ``subalgebra'' \boldmath $\mf{f}$}

Although multiplicities are an essential ingredient for
describing the full symmetry $\mf{u}_3$, they turn out to be
irrelevant for the construction of the gravitational billiard. For
this reason, it is useful to consider the \emph{maximal split
``subalgebra''} $\mf{f}$, which is defined as the real,
semi-simple, split Lie algebra with the same root system as the
restricted root system \index{restricted root system} as $\mf{u}_3$ (in the $(BC)_n$-case, we
choose for definiteness the root system of $\mf{f}$ to be of
$B_n$-type). The real rank of $\mf{f}$ coincides with the rank of
its complexification $\mf{f}^\mathbb{C}$, and one can find a
Cartan subalgebra $\cH_{\mf{f}}$ of $\mf{f}$, consisting of all generators of $\mf{h}_3$ 
which are diagonalizable over the reals. This subalgebra
$\cH_{\mf{f}}$ has the same dimension as the maximal noncompact
subalgebra $\cA_3$ of the Cartan subalgebra $\cH_3$ of $\mf{u}_3$.

By construction, the root space decomposition of $\mf{f}$ with respect
to $\cH_{\mf{f}}$ provides the same root system as the restricted
root space decomposition of $\mf{u}_3$ with respect to $\cA_3$, except
for multiplicities, which are all trivial (i.e., equal to one) for
$\mf{f}$. In the $(BC)_n$-case, there is also the possibility that
twice a root of $\mf{f}$ might be a root of $\mf{u}_3$. It is only
when $\mf{u}_3$ is itself split that $\mf{f}$ and $\mf{u}_3$
coincide.

One calls $\mf{f}$ the ``split symmetry algebra''. It contains as
we shall see all the information about the billiard
region~\cite{HenneauxJulia}. How $\mf{f}$ can be embedded as a
subalgebra of $\mf{u}_3$ is not a question that shall be of our
concern here.

\subsubsection*{The Iwasawa decomposition and scalar coset Lagrangians}
\index{Iwasawa decomposition}

The purpose of this section is to use the Iwasawa
decomposition for $\mf{u}_3$ described in
Section~\ref{section:Iwasawa} to derive the scalar Lagrangian based
on the coset space  \index{coset space} $\mc{U}_3/\mc{K}(\mc{U}_3)$. The important point is
to understand the origin of the similarities between the two
Lagrangians in Equation~(\ref{LagrangiansplitU3}) and
Equation~(\ref{LagrangianU3}) below.

The full algebra $\mf{u}_3$ is subject to the root space
decomposition 
\begin{equation}
  \mf{u}_3 = \ag{0}\oplus \bigoplus_{\lambda\in \Sigma} \agl
  \label{rootspacedecomposition}
\end{equation}
with respect to the restricted root system. For each restricted root
$\lambda$, the space $\agl$ has dimension $m_\lambda$. The nilpotent
algebra $\mf{n}_3\subset \mf{u}_3$, consisting of positive root
generators only, is the direct sum 
\begin{equation}
  \mf{n}_3=\bigoplus_{\lambda\in\Sigma^{+}}\agl
  \label{nilpotentpart}
\end{equation}
over positive roots. The Iwasawa decomposition \index{Iwasawa decomposition} of the U-duality
algebra $\mf{u}_3$ reads
\begin{equation}
  \mf{u}_3=\mf{k}_3\oplus \cA_3 \oplus \mf{n}_3
  \label{IwasawadecompositionU3}
\end{equation}
(see Section~\ref{section:Iwasawa}). It is $\cA_3$ that appears in
Equation~(\ref{IwasawadecompositionU3}) and \emph{not} the full
Cartan subalgebra \index{Cartan subalgebra} $\mf{h}_3$ since the compact part of $\mf{h}_3$
belongs to $\mf{k}_3$.

This implies that when constructing a Lagrangian based on the
coset space $\mc{U}_3/\mc{K}(\mc{U}_3)$, the only part of
$\mf{u}_3$ that will show up in the Borel gauge is the Borel
subalgebra 
\begin{equation}
  \mf{b}_3=\cA_3 \oplus \mf{n}_3.
  \label{BorelsubalgebraU3}
\end{equation}
Thus, there will be a number of dilatons equal to the dimension of
$\cA_3$, i.e., equal to the real rank of $\mf{u}_3$, and axion fields
for the restricted roots (with multiplicities).

More specifically, an ($x$-dependent) element of the coset space
$\mc{U}_3/\mc{K}(\mc{U}_3)$ takes the form 
\begin{equation}
  \mc{V}(x)=\Exp \left[ \phi (x) \cdot \cA_3\right]\,
  \Exp \left[\chi (x)\cdot \mf{n}_3\right],
  \label{elementofcosetU3}
\end{equation}
where the dilatons $\phi$ and the axions $\chi$ are coordinates on the
coset space, and where $x$ denotes an arbitrary set of parameters on
which the coset element might depend. The corresponding Lagrangian
becomes 
\begin{equation}
  \mc{L}_{\mc{U}_3/\mc{K}(\mc{U}_3)}= \!\!\!
  \sum_{i=1}^{\dim \cA_3} \!\!\! \pa_x \phi^{(i)}(x) \pa_x \phi^{(i)}(x)+ \!\!\!
  \sum_{\al\in \Sigma^{+}} \! \sum_{s_\al=1}^{\mult \al} \!\!\!
  e^{2\al(\phi)}\left[\pa_x\chi^{(\al)}_{[s_\al]}(x)+\cdots\right]
  \left[\pa_x\chi^{(\al)}_{[s_\al]}(x)+\cdots\right],
  \label{LagrangianU3}
\end{equation}
where the sums over $s_{\al}=1,\cdots, \mult \al$ are sums
over the multiplicities of the positive restricted roots $\al$.

By comparing Equation~(\ref{LagrangianU3}) with the corresponding
expression~(\ref{LagrangiansplitU3}) for the split case, it
is clear why it is the maximal split subalgebra of the U-duality
algebra that is relevant for the gravitational billiard. Were it not
for the additional sum over multiplicities,
Equation~(\ref{LagrangianU3}) would exactly be the Lagrangian for the
coset space $\mc{F}/\mc{K}(\mc{F})$, where $\mf{k}_{\mf{f}}=\mathrm{Lie}\,
\mc{K}(\mc{F})$ is the maximal compact subalgebra 
\index{maximal compact subalgebra} of $\mf{f}$ (note that
$\mf{k}_{\mf{f}}\neq \mf{k}_3$). Recall now that from the point of
view of the billiard, the positive roots correspond to walls that
deflect the particle motion in the Cartan subalgebra. Therefore,
multiplicities of roots are irrelevant since these will only result in
several walls stacked on top of each other without affecting the
dynamics. (In the $(BC)_n$-case, the wall associated with $2 \lambda$
is furthermore subdominant with respect to the wall associated with
$\lambda$ when both $\lambda$ and $2 \lambda$ are restricted roots, so
one can keep only the wall associated with $\lambda$. This follows
from the fact that in the $(BC)_n$-case the root system of $\mf{f}$ is
taken to be of $B_n$-type.)


\subsection{``Split symmetry controls chaos''}

The main point of this section is to illustrate and explain the
statement ``split symmetry controls chaos''~\cite{HenneauxJulia}. To
this end, we will now extend the analysis of
Section~\ref{section:KMBilliardsI} to non-split real forms, using the
Iwasawa decomposition. As we have seen, there are two main cases to be
considered:

\begin{itemize}
\item The restricted root system $\Sigma$ of $\mf{u}_3$ is of
  \emph{reduced} type, in which case it is one of the standard root
  systems for the Lie algebras $A_n, B_n, C_n, D_n, G_2, F_4, E_6,
  E_7$ or $E_8$.
\item The restricted root system, \index{restricted root system}
  $\Sigma$, of $\mf{u}_3$ is of \emph{non-reduced} type, in which case
  it is of $(BC)_n$-type.
\end{itemize}

In the first case, the billiard \index{cosmological billiard} is
governed by the overextended algebra $\mf{f}^{++}$, where $\mf{f}$ is
the ``maximal split subalgebra'' of $\mf{u}_3$. Indeed, the coupling
to gravity of the coset Lagrangian of Equation~(\ref{LagrangianU3})
will introduce, besides the simple roots of $\mf{f}$ (electric walls)
the affine root of $\mf{f}$ (dominant magnetic wall) and the
overextended root (symmetry wall), just as in the split case (but for
$\mf{f}$ instead of $\mf{u}_3$). This is therefore a straightforward
generalization of the analysis in Section~\ref{section:KMBilliardsI}.

The second case, however, introduces a new phenomenon, the
\emph{twisted overextensions} \index{overextension} of
Section~\ref{section:KacMoody}. This is because the highest root of
the $(BC)_n$ system differs from the highest root of the $B_n$
system. Hence, the dominant magnetic wall will provide a twisted
affine root, to which the symmetry wall will attach itself as
usual~\cite{HenneauxJulia}.

We illustrate the two possible cases in terms of explicit
examples. The first one is the simplest case for which a twisted
overextension appears, namely the case of pure four-dimensional
gravity coupled to a Maxwell field. This is the bosonic sector of
$\mc{N}=2$ supergravity in four dimensions, which has the
non-split real form $\mf{su}(2,1)$ as its U-duality symmetry. The
restricted root system of $\mf{su}(2,1)$ is the non-reduced
$(BC)_1$-system, and, consequently, as we shall see explicitly,
the billiard is governed by the twisted overextension
$A_{2}^{(2)+}$.

The second example is that of heterotic supergravity, which
exhibits an $SO(8,24)/(SO(8)$ $\times SO(24))$ coset
symmetry in three dimensions. The U-duality algebra is thus
$\mf{so}(8, 24)$, which is non-split. In this example,
however, the restricted root system is $B_8$, which is reduced,
and so the billiard is governed by a standard overextension of the
maximal split subalgebra $\mf{so}(8,9)\subset
\mf{so}(8, 24)$.


\subsubsection[$(BC)_1$ and $\mc{N}=2$, $D=4$ pure supergravity]%
              {\boldmath $(BC)_1$ and $\mc{N}=2$, $D=4$ pure supergravity}

We consider $\mc{N}=2$ supergravity in four dimensions where the
bosonic sector consists of gravity coupled to a Maxwell field. It
is illuminating to compare the construction of the billiard in the
two limiting dimensions, $D=4$ and $D=3$.

In maximal dimension the metric contains three scale factors,
$\be^1,\be^2$ and $\be^3$, which give rise to three symmetry wall
forms, 
\begin{equation}
  s_{21}(\be)=\be^2-\be^1,
  \qquad
  s_{32}(\be)=\be^{3}-\be^2,
  \qquad
  s_{31}(\be)=\be^3-\be^1,
  \label{N2sugraSymmetrywalls}
\end{equation}
where only $s_{21}$ and
$s_{32}$ are dominant. In four dimensions the curvature walls
read
\begin{equation}
  c_{123}(\be)\equiv c_{1}(\be)=2\be^1,
  \qquad
  c_{231}(\be)\equiv c_{2}(\be)=2\be^2,
  \qquad
  c_{312}(\be)\equiv c_{3}(\be)=2\be^3. 
  \label{CurvaturewallsD4}
\end{equation}
Finally we
have the electric and magnetic wall forms of the Maxwell field.
These are equal because there is no dilaton. Hence, the wall forms
are 
\begin{equation}
  e_{1}(\be)=m_{1}(\be)=\be^1,
  \qquad
  e_{2}(\be)=m_{2}(\be)=\be^2,
  \qquad
  e_{3}(\be)=m_{3}(\be)=\be^3.
  \label{ElectricMagneticWallsD4}
\end{equation}
The billiard region
$\mc{B}_{\mc{M}_\be}$ is defined by the set of dominant wall
forms, 
\begin{equation}
  \mc{B}_{\mc{M}_\be}=\{\be \in\mc{M}_\be \, |\, e_{1}(\be),
  s_{21}(\be), s_{32}(\be)\, > \, 0\}.
  \label{dominantwallsD4}
\end{equation}
The first dominant wall form,
$e_{1}(\be)$, is twice degenerate because it occurs once as an
electric wall form and once as a magnetic wall form. Because of
the existence of the curvature wall, $c_{1}(\be)=2\be^1$, we see
that $2\al_{1}$ is also a root.

The same billiard emerges after reduction to three spacetime
dimensions, where the algebraic structure is easier to exhibit. As
before, we perform the reduction along the first spatial
direction. The associated scale factor is then replaced by the
Kaluza--Klein dilaton $\hvp$ such that 
\begin{equation}
  \be^1=\f{1}{\sqrt{2}}\hvp. 
  \label{dilatonD3}
\end{equation}
 The remaining
scale factors change accordingly, 
\begin{equation}
  \be^2=\hb^2-\f{1}{\sqrt{2}}\hvp,
  \qquad
  \be^3=\hb^3-\f{1}{\sqrt{2}}\hvp, 
  \label{scalefactorsD3}
\end{equation}
and the two symmetry walls become 
\begin{equation}
  s_{21}(\hb,\hvp)=\hb^2-\sqrt{2}\hvp,
  \qquad
  \hat{s}_{32}(\hb)=\hb^3-\hb^2. 
  \label{symmetryelectricwallsD3}
\end{equation}
In addition to the dilaton $\hvp$, there are three axions: one
($\hat{\chi}$) arising from the dualization of the Kaluza--Klein
vector, one ($\hat{\chi}^{E}$) coming from the component $A_1$ of the
Maxwell vector potential and one ($\hat{\chi}^{C}$) coming from
dualization of the Maxwell vector potential in 3 dimensions (see,
e.g.,~\cite{Cremmer:1997ct} for a review). There are then a total of
four scalars. These parametrize the coset space
$SU(2,1)/S(U(2)\times U(1))$~\cite{Julia:1980gr}.

The Einstein--Maxwell Lagrangian in four dimensions yields indeed
in three dimensions the Einstein--scalar Lagrangian, where the
Lagrangian for the scalar fields is given by 
\begin{equation}
  \mc{L}_{SU(2,1)/S(U(2)\times U(1))}=
  \pa_{\mu}\hvp\pa^{\mu}\hvp+e^{2 e_{1}(\hvp)}
  \left(\pa_{\mu}\hat{\chi}^{E}\pa^{\mu}\hat{\chi}^{E}+
  \pa_{\mu}\hat{\chi}^{C}\pa^{\mu}\hat{\chi}^{C}\right) +
  e^{4e_{1}(\hvp)}\left(\pa_{\mu}\hat{\chi}\pa^{\mu}\hat{\chi}\right) +\cdots
  \label{LagrangianSU(3,2)} 
\end{equation}
with
\begin{displaymath}
  e_1(\hvp) = \f{1}{\sqrt{2}}\hvp.
\end{displaymath}
Here, the ellipses denotes terms that are not relevant for
understanding the billiard \index{cosmological billiard}
structure. The U-duality algebra of $\mc{N}=2$ supergravity
compactified to three dimensions is therefore
\begin{equation}
  \mf{u}_3=\mf{su}(2,1),
  \label{UdualityN2}
\end{equation}
which is a non-split real form of the
complex Lie algebra $\mf{sl}(3,\mbb{C})$. This is in agreement
with Table~1 of~\cite{Julia:1980gr}. The restricted
root system \index{restricted root system} of $\mf{su}(2,1)$ is of
$(BC)_1$-type (see Table~\ref{table:A(n)series} in
Section~\ref{section:summary-titssatakediagrams}) and has four roots:
$\alpha_1$, $2\alpha_1$, $-\alpha_1$ and $-2\alpha_1$. One may take
$\alpha_1$ to be the simple root, in which case $\Sigma_+ = \{
\alpha_1, 2 \alpha_1 \}$ and $2 \alpha_1$ is the highest root. The
short root $\alpha_1$ is degenerate twice while the long root $2
\alpha_1$ is nondegenerate. The Lagrangian~(\ref{LagrangianSU(3,2)})
coincides with the Lagrangian~(\ref{LagrangianU3}) for
$\mf{su}(2,1)$ with the identification 
\begin{equation}
  \hat{\al}_{1}\equiv e_{1}. 
  \label{firstsimplerootD4}
\end{equation}
We clearly see from the Lagrangian that the simple root
$\hat{\al}_{1}$ has multiplicity $2$ in the restricted root
system, since the corresponding wall appears twice. The maximal
split subalgebra may be taken to be $A_1 \equiv \mf{su}(1,1)$ with
root system $\{\hat{\al}_{1}, -\hat{\al}_{1} \}$.

Let us now see how one goes from $\mf{su}(2,1)$ described by the
scalar Lagrangian to the full algebra, by including the
gravitational scale factors. Let us examine in particular how the
twist arises. For the standard root system of $A_1$ the highest
root is just $\hat{\al}_{1}$. However, as we have seen, for the
$(BC)_1$ root system the highest root is
$\theta_{(BC)_1}=2\hat{\al}_{1}$, with 
\begin{equation}
  (\theta_{(BC)_1}|\theta_{(BC)_1})=4(\hat{\al}_{1}|\hat{\al}_{1})=2.
  \label{normhigestrootBC1}
\end{equation}
So we see that because of
$(\hat{\al}_{1}|\hat{\al}_{1})=\f{1}{2}$, the highest root
$\theta_{(BC)_1}$ already comes with the desired normalization.
The affine root is therefore 
\begin{equation}
  \hat{\al}_{2}(\hvp,\hb)=
  \hat{m}^{\hat{\chi}}_{2}(\hb,\hvp)=
  \hb^2-\theta_{(BC)_1}=\hb^2-\sqrt{2}\hvp,
  \label{affinerootBC1}
\end{equation}
whose norm is 
\begin{equation}
  (\hat{\al}_{2}|\hat{\al}_{2})=2. 
  \label{normaffinerootBC1}
\end{equation}
The scalar product between $\hat{\al}_{1}$ and $\hat{\al}_{2}$ is
$(\hat{\al}_{1}|\hat{\al}_{2})=-1$ and the Cartan matrix at this
stage becomes $(i,j=1,2)$ 
\begin{equation}
  A_{ij}[A_2^{(2)}]=
  2\f{(\hat{\al}_{i}|\hat{\al}_{j})}{(\hat{\al}_{i}|\hat{\al}_{i})}=\left(
    \begin{array}{@{}r@{\quad}r@{}}
      2 & -4 \\
      -1 & 2
    \end{array}
  \right),
  \label{twistedaffineA1}
\end{equation}
which may be identified not with the affine extension of $A_1$ but
with the Cartan matrix \index{Cartan matrix} of the \emph{twisted} affine Kac--Moody algebra
$A_2^{(2)}$. It is the underlying $(BC)_1$ root system that is solely
responsible for the appearance of the twist. Because of the fact that
$\theta_{(BC)_1}=2\hat{\al}_{1}$ the two simple roots of the affine
extension come with different length and hence the asymmetric Cartan
matrix in Equation~(\ref{twistedaffineA1}). It remains to include the
overextended root 
\begin{equation}
  \hat{\al}_{3}(\hb)=\hat{s}_{32}(\hb)=\hb^{3}-\hb^{2},
  \label{overextendedrootBC1}
\end{equation}
which has non-vanishing scalar product only with $\hat{\al}_{2}$,
$(\hat{\al}_{2}|\hat{\al}_{3})=-1$, and so its node in the Dynkin
diagram is attached to the second node by a single link. The
complete Cartan matrix \index{Cartan matrix} is 
\begin{equation}
  A[A_{2}^{(2)+}]=\left(
    \begin{array}{@{}r@{\quad}r@{\quad}r@{}}
      2 & -4 & 0 \\
      -1 & 2 & -1 \\
      0 & -1 & 2
    \end{array}
  \right),
  \label{twistedoverextendedA1}
\end{equation}
which is the Cartan matrix of the Lorentzian extension
$A_{2}^{(2)+}$ of $A_{2}^{(2)}$ henceforth referred to as the
\emph{twisted overextension} of $A_{1}$. Its Dynkin diagram \index{Dynkin diagram} is
displayed in Figure~\ref{figure:A1pTwist}.

The algebra $A_2^{(2)+}$ was already analyzed in
Section~\ref{section:KacMoody}, where it was shown that its Weyl
group coincides with the Weyl group of the algebra $A_1^{++}$. Thus,
in the BKL-limit the dynamics of the coupled Einstein--Maxwell system
in four-dimensions is equivalent to that of pure four-dimensional
gravity, although the set of dominant walls are different. Both
theories are chaotic.

\epubtkImage{A1pTwist.png}{%
\begin{figure}[htbp]
\centerline{\includegraphics[width=40mm]{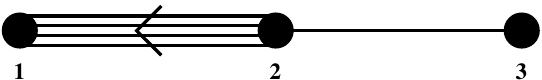}}
\caption{The Dynkin diagram of $A_2^{(2)+}$. Label $1$ denotes
the simple root $\hat{\al}_{(1)}$ of the restricted root system of
$\mf{u}_3=\mf{su}(2,1)$. Labels $2$ and $3$ correspond to
the affine and overextended roots, respectively. The arrow points
towards the short root which is normalized such that
$(\hat{\al}_{1}|\hat{\al}_{1})=\f{1}{2}$.}
\label{figure:A1pTwist}
\end{figure}}


\subsubsection[Heterotic supergravity and $\mf{so}(8,24)$]%
              {Heterotic supergravity and \boldmath $\mf{so}(8,24)$}

Pure $\mc{N} = 1$ supergravity in $D=10$ dimensions has a billiard \index{cosmological billiard}
description in terms of the hyperbolic Kac--Moody algebra
$DE_{10}=D_8^{++}$ \index{$DE_{10}$} \cite{ArithmeticalChaos}. This algebra is the
overextension \index{overextension} of the U-duality algebra, $\mf{u}_3=D_8\equiv
\mf{so}(8,8)$, appearing upon compactification to three
dimensions. In this case, $\mf{so}(8,8)$ is the split form
of the complex Lie algebra $D_8$, so we have
$\mf{f}=\mf{u}_3$.

By adding one Maxwell field to the theory we modify the billiard
to the hyperbolic Kac--Moody algebra $BE_{10}=B_8^{++}$, \index{$BE_{10}$|bb} which is
the overextension of the split form $\mf{so}(8,9)$ of
$B_8$~\cite{ArithmeticalChaos}. This is the case relevant
for (the bosonic sector of) Type I supergravity in ten dimensions.
In both these cases the relevant Kac--Moody algebra is the
overextension of a split real form and so falls under the
classification given in Section~\ref{section:KMBilliardsI}.

Let us now consider an interesting example for which the relevant
U-duality algebra is non-split. For the heterotic string, the
bosonic field content of the corresponding supergravity is given
by pure gravity coupled to a dilaton, a 2-form and an $E_8 \times
E_8$ Yang--Mills gauge field. Assuming the gauge field to be in
the Cartan subalgebra, this amounts to adding 16 $\mc{N}=1$ vector
multiplets in the bosonic sector, i.e, to adding 16 Maxwell
fields to the ten-dimensional theory discussed above. Geometrically,
these 16 Maxwell fields correspond to the Kaluza--Klein vectors arising
from the compactification on $T^{16}$ of the 26-dimensional bosonic
left-moving sector of the heterotic string~\cite{Gross:1984dd}.
 
Consequently, the relevant U-duality algebra is
$\mf{so}(8,8+16)=\mf{so}(8,24)$ which is a non-split real form. But we
know that the billiard for the heterotic string is governed by the
same Kac--Moody algebra as for the Type I case mentioned above, namely
$BE_{10}\equiv\mf{so}(8,9)^{++}$, and not $\mf{so}(8,24)^{++}$ as one
might have expected~\cite{ArithmeticalChaos}. The only difference is
that the walls associated with the one-forms are degenerate 16
times. We now want to understand this apparent discrepancy using the
machinery of non-split real forms exhibited in previous sections. The
same discussion applies to the $SO(32)$-superstring.

In three dimensions the heterotic supergravity Lagrangian is given by
a pure three-dimensional Einstein--Hilbert term coupled to a nonlinear
sigma model for the coset $SO(8,24)/(SO(8)\times SO(24))$. This
Lagrangian can be understood by analyzing the Iwasawa decomposition of
$\mf{so}(8,24)=\mathrm{Lie} \left[SO(8,24)\right]$. The maximal compact
subalgebra is 
\begin{equation}
  \mf{k}_3=\mf{so}(8)\oplus \mf{so}(24).
  \label{maximalcompactsubalgebraHeteroticCase}
\end{equation}
This subalgebra does not appear in the sigma model  \index{nonlinear sigma model} since it is divided
out in the coset construction (see Equation~(\ref{elementofcosetU3}))
and hence we only need to investigate the Borel subalgebra \index{Borel subalgebra}
$\cA_3\oplus \mf{n}_3$ of $\mf{so}(8,24)$ in more detail.

As was emphasized in Section~\ref{section:SplitVSNonSplit}, an
important feature of the Iwasawa decomposition is that the full Cartan
subalgebra $\mf{h}_3$ does not appear explicitly but only the maximal
noncompact Cartan subalgebra $\cA_3$, associated with the restricted
root system. This is the maximal Abelian subalgebra of
$\mf{u}_3=\mf{so}(8,24)$, whose adjoint action can be diagonalized
over the reals. The remaining Cartan generators of $\mf{h}_3$ are
compact and so their adjoint actions have imaginary eigenvalues. The
general case of $\mf{so}(2q,2p)$ was analyzed in detail in
Section~\ref{section:restrictedrootsystemSO} where it was found that
if $q<p$, the restricted root system is of type $B_{2q}$. For the case
at hand we have $q=4$ and $p=12$ which implies that the restricted
root system \index{restricted root system} of $\mf{so}(8,24)$ is (modulo multiplicities)
$\Sigma_{\mf{so}(8,24)}=B_8$.

The root system of $B_8$ is eight-dimensional and hence there are eight Cartan
generators that may be simultaneously diagonalized over the real
numbers. Therefore the real rank of $\mf{so}(8,24)$ is
eight, i.e., 
\begin{equation}
  \rank_{\mbb{R}}\, \mf{u}_3 = \dim \cA_3=8. 
  \label{RealrankHeterotic}
\end{equation}
Moreover, it was shown in Section~\ref{section:restrictedrootsystemSO}
that the restricted root system of $\mf{so}(2q,2p)$ has $4q(2q-1)$
long roots which are nondegenerate, i.e., with multiplicity one, and
$4q$ long roots with multiplicities $2(p-q)$. In the example under
consideration this corresponds to seven nondegenerate simple roots
$\al_{1},\cdots,\al_{7}$ and one short simple root $\al_{8}$ with
multiplicity $16$. The Dynkin diagram \index{Dynkin diagram} for the restricted root system
$\Sigma_{\mf{so}(8,24)}$ is displayed in
Figure~\ref{figure:B8restricted} with the multiplicity indicated in
brackets over the short root. It is important to note that the
restricted root system $\Sigma_{\mf{so}(8,24)}$ differs from the
standard root system of $\mf{so}(8,9)$ precisely because of the
multiplicity 16 of the simple root $\al_{8}$.

\epubtkImage{B8restricted.png}{%
  \begin{figure}[htbp]
    \centerline{\includegraphics[width=100mm]{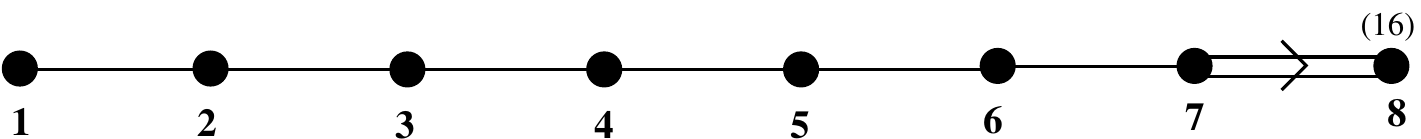}}
    \caption{The Dynkin diagram representing the restricted root
      system $\Sigma_{\mf{so}(8,24)}$ of $\mf{so}(8,24)$. Labels $1, \cdots,
      7$ denote the long simple roots that are nondegenerate while the
      eighth simple root is short and has multiplicity 16.}
    \label{figure:B8restricted}
  \end{figure}}

Because of these properties of $\mf{so}(8,24)$ the Lagrangian for the coset 
\begin{equation}
  \f{SO(8,24)}{SO(8)\times SO(24)} 
  \label{3DcosetHeterotic}
\end{equation}
takes a form very similar to the Lagrangian for the coset 
\begin{equation}
  \f{SO(8,9)}{SO(8)\times SO(9)}.
  \label{3DcosetSplitSubalgebraHeterotic}
\end{equation}
The algebra constructed from the restricted root system $B_8$ is the
maximal split subalgebra 
\begin{equation}
  \mf{f}=\mf{so}(8,9).
  \label{MaximallysplitsubalgebraHeterotic}
\end{equation}
Let us now take a closer look at the Lagrangian in three spacetime
dimensions. We parametrize an element of the coset by 
\begin{equation}
  \mc{V}(x^{\mu}) =
  \Exp \left[ \sum_{i=1}^{8} \phi^{(i)}(x^{\mu})\al_{i}^{\vee}\right]
  \Exp \left[\sum_{\ga\in \Delta_+} \chi^{(\ga)}(x^{\mu}) E_{\ga}\right]
  \in \f{SO(8,24)}{SO(8)\times SO(24)},
  \label{CosetElementHeterotic}
\end{equation}
where $x^{\mu}\, (\mu=0,1,2)$ are the coordinates of the external
three-dimensional spacetime, $\al_{i}^{\vee}$ are the noncompact
Cartan generators and $\Delta_+$ denotes the full set of positive
roots of $\mf{so}(8, 24)$.

The Lagrangian constructed from the coset representative in
Equation~(\ref{CosetElementHeterotic}) becomes (again, neglecting
corrections to the single derivative terms of the form
``$\pa_x\chi$'')
\begin{eqnarray}
  \mc{L}_{\mc{U}_3/\mc{K}(\mc{U}_3)} &=&
  \sum_{i=1}^{8}\pa_\mu \phi^{(i)}(x) \, \pa^{\mu} \phi^{(i)}(x) +
  \sum_{j=1}^{7} e^{\al_{j}(\phi)} \, \pa_{\mu}\chi^{(j)}(x) \, \pa^{\mu}\chi^{(j)}(x)
  \nonumber \\
  & &
  + e^{\al_{8}(\phi)}\left(\sum_{k=1}^{16}\pa_{\mu}\chi^{(8)}_{[k]}(x)
  \, \pa^{\mu}\chi^{(8)}_{[k]}(x)\right) + \!\!\!
  \sum_{\al \in \tilde{\Sigma}^{+}} \! \sum_{s_{\al} = 1}^{\mult(\al)} 
  \!\!\! e^{\al(\phi)} \, \pa_\mu \chi^{(\al)}_{[s_{\al}]}(x) \,
  \pa^{\mu}\chi^{(\al)}_{[s_{\al}]}(x),\qquad
  \label{3DLagrangianHeterotic} 
\end{eqnarray}
where $\tilde{\Sigma}^{+}$ denotes all non-simple positive roots of
$\Sigma$, i.e., 
\begin{equation}
  \tilde{\Sigma}^{+} = \Sigma^{+}/ \bar{B}
  \label{PositiverootsB8modded}
\end{equation}
with 
\begin{equation}
  \bar{B}=\{\al_{1},\cdots, \al_{8}\}. 
  \label{PositiverootsB8}
\end{equation}
This Lagrangian is equivalent to the Lagrangian for
$SO(8,9)/(SO(8)\times SO(9))$ except for the existence of
the non-trivial root multiplicities.

The billiard for this theory can now be computed with the same
methods that were treated in detail in
Section~\ref{section:3Dbilliard}. In the BKL-limit, the simple roots
$\al_{1}, \cdots, \al_{8}$ become the non-gravitational dominant wall
forms. In addition to this we get one magnetic and one gravitational
dominant wall form:
\begin{equation}
  \begin{array}{rcl}
    \al_{0}&=& \be^{1}-\theta(\phi),
    \\ [0.25 em]
    \al_{-1} &=& \be^2-\be^1,
  \end{array}
  \label{overextensionrootsHeterotic}
\end{equation}
where $\theta(\phi)$ is the highest root of $\mf{so}(8,9)$:
\begin{equation}
  \theta=\al_{1}+2\al_{2}+\cdots + 2\al_{7}+\al_{8}. 
  \label{highestrootB8}
\end{equation}
The affine root $\al_{0}$ attaches with a single link to the second
simple root $\al_{2}$ in the Dynkin diagram of $B_8$. Similarly the
overextended root $\al_{-1}$ attaches to $\al_{0}$ with a single
link so that the resulting Dynkin diagram \index{Dynkin diagram}corresponds to $BE_{10}$ \index{$BE_{10}$} (see Figure~\ref{figure:BE10restricted}). It is important to note
that the underlying root system is still an overextension of the
restricted root system and hence the multiplicity of the simple
short root $\al_{8}$ remains equal to 16. Of course, this does not
affect the dynamics in the BKL-limit because the multiplicity of
$\al_{8}$ simply translates to having multiple electric walls on
top of each other and this does not alter the billiard motion.

This analysis again showed explicitly how it is always the split
symmetry that controls the chaotic behavior in the BKL-limit. It
is important to point out that when going beyond the strict BKL-limit, 
one expects more and more roots of the algebra to play a
role. Then it is no longer sufficient to study only the maximal
split subalgebra $\mf{so}(8, 9)^{++}$ but instead the
symmetry of the theory is believed to contain the full algebra
$\mf{so}(8, 24)^{++}$. In the spirit of~\cite{DHN2} one may then
conjecture that the dynamics of the heterotic supergravity should be
equivalent to a null geodesic on the coset space $SO(8,
24)^{++}/\mc{K}(SO(8, 24)^{++})$~\cite{CompatibilityConditions}. 

\epubtkImage{BE10restricted.png}{%
\begin{figure}[htbp]
  \centerline{\includegraphics[width=110mm]{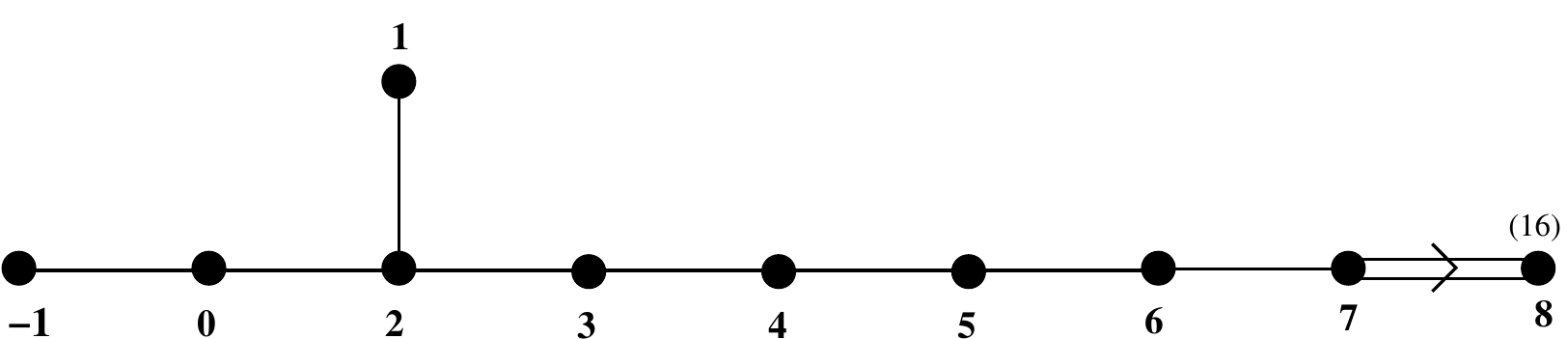}}
  \caption{The Dynkin diagram representing the overextension
    $B_8^{++}$ of the restricted root system $\Sigma=B_8$ of
    $\mf{so}(8,24)$. Labels $-1,0,1, \cdots, 7$ denote the long
    simple roots that are nondegenerate while the eighth simple root
    is short and has multiplicity 16.}
  \label{figure:BE10restricted}
\end{figure}}


\subsection{Models associated with non-split real forms}
\label{section:classificationnonsplit}

In this section we provide a list of all theories coupled to
gravity which, upon compactification to three dimensions, display
U-duality algebras that are \emph{not} maximal
split~\cite{HenneauxJulia}. This therefore completes the
classification of Section~\ref{section:KMBilliardsI}.

One can classify the various theories through the number $\mc{N}$ of
supersymmetries that they possess in $D=4$ spacetime
dimensions. All $p$-forms can be dualized to scalars or to
1-forms in four dimensions so the theories all take the form of
pure supergravities coupled to collections of Maxwell multiplets.
The analysis performed for the split forms in
Section~\ref{section:understanding} were all concerned with the cases
of $\mc{N}=0$ or $\mc{N}=8$ supergravity in $D=4$. We consider all
pure four-dimensional supergravities ($\mc{N}=1,\cdots, 8$) as well as
pure $\mc{N}=4$ supergravity coupled to $k$ Maxwell multiplets.

As we have pointed out, the main new feature in the non-split
cases is the possible appearance of so-called \emph{twisted
overextensions}. These arise when the restricted root system 
\index{restricted root system} of
$\mc{U}_3$ is of non-reduced type hence yielding a twisted affine
Kac--Moody algebra in the affine extension of $\mf{f}\subset
\mf{u}_3$. It turns out that the only cases for which the
restricted root system is of non-reduced ($(BC)$-type) is for the
pure $\mc{N}=2, 3$ and $\mc{N}=5$ supergravities. The example of
$\mc{N}=2$ was already discussed in detail before, where it was
found that the U-duality algebra is $\mf{u}_3=\mf{su}(2,1)$
whose restricted root system is $(BC)_1$, thus giving rise to the
twisted overextension $A_2^{(2)+}$. It turns out that for the
$\mc{N}=3 $ case the same twisted overextension appears. This is
due to the fact that the U-duality algebra is
$\mf{u}_3=\mf{su}(4,1)$ which has the same restricted root system
as $\mf{su}(2,1)$, namely $(BC)_1$. Hence, $A_1^{(2)+}$ controls the
BKL-limit also for this theory.

The case of $\mc{N}=5$ follows along similar lines. In $D=3$ the
non-split form $E_{6(-14)}$ of $E_{6}$ appears, whose maximal
split subalgebra is $\mf{f}=C_2$. However, the relevant Kac--Moody
algebra is not $C_2^{++}$ but rather $A_4^{(2)+}$ because the
restricted root system of $E_{6(-14)}$ is $(BC)_2$.

In Table~\ref{table:nonsplit} we display the algebraic structure
for all pure supergravities in four dimensions as well as for
$\mc{N}=4$ supergravity with $k$ Maxwell multiplets. We give the
relevant U-duality algebras $\mf{u}_3$, the restricted root
systems $\Sigma$, the maximal split subalgebras $\mf{f}$ and,
finally, the resulting overextended Kac--Moody algebras $\mf{g}$.

Let us end this section by noting that the study of real forms of
hyperbolic Kac--Moody algebras has been pursued
in~\cite{BenMessaoud:2006ug}.

\begin{table}
  \caption[Classification of theories whose U-duality symmetry in
    three dimensions is described by a non-split real form
    $\mf{u}_3$.]{Classification of theories whose U-duality symmetry in
    three dimensions is described by a non-split real form
    $\mf{u}_3$. The leftmost column indicates the number $\mc{N}$ of
    supersymmetries that the theories possess when compactified to
    four dimensions, and the associated number $k$ of Maxwell
    multiplets. The middle column gives the restricted root system
    $\Sigma$ of $\mf{u}_3$ and to the right of this we give the
    maximal split subalgebras $\mf{f}\subset\mf{u}_3$, constructed
    from a basis of $\Sigma$. Finally, the rightmost column displays
    the overextended Kac--Moody algebras that governs the billiard
    dynamics.}
  \renewcommand{\arraystretch}{1.2}
  \vspace{0.5 em}
  \centering
  \begin{tabular}{l|cccc}
    \toprule
    $(\mc{N},k) $ & $\mf{u}_3$ & $\Sigma$ & $\mf{f}$ & $\mf{g}$ \\
    \midrule
    (1,0) & $\mf{sl}(2,\mbb{R})$ & $A_1$ & $A_1$ & $A_1^{++}$ \\
    (2,0) & $\mf{su}(2,1)$ & $(BC)_1$ & $A_1$ & $A_2^{(2)+}$ \\
    (3,0) & $\mf{su}(4,1)$ & $(BC)_1$ & $A_1$ & $A_1^{(2)+}$ \\
    (4,0) & $\mf{so}(8,2)$ & $C_2$ & $C_2$ & $C_2^{++}$ \\
    $(4,k<6)$ & $\mf{so}(8,k+2)$ & $B_{k+2}$ & $B_{k+2}$ & $B_{k+2}^{++}$ \\
    $(4,6)$ & $\mf{so}(8,8)$ & $D_8$ & $D_8$ & $DE_{10}=D_8^{++}$ \\
    $(4,k>6)$ & $\mf{so}(8,k+2)$ & $B_8$ & $B_8$ & $BE_{10}=B_8^{++}$ \\
    (5,0) & $E_{6(-14)}$ & $(BC)_2$ & $C_2$ & $A_4^{(2)+}$ \\
    (6,0) & $E_{7(-5)}$ & $F_4$ & $F_4$ & $F_4^{++}$ \\
    (8,0) & $E_{8(+8)}$ & $E_8$ & $E_8$ & $E_{10}=E_8^{++}$ \\
    \bottomrule
  \end{tabular}
  \label{table:nonsplit}
  \renewcommand{\arraystretch}{1.0}
\end{table}

\newpage


\section{Level Decomposition in Terms of Finite Regular Subalgebras}
\label{section:LevelDecomposition}
\setcounter{equation}{0}

We have shown in the previous sections that Weyl groups of
Lorentzian Kac--Moody algebras naturally emerge when analysing
gravity in the extreme BKL regime. \index{level decomposition|bb} This has led to the conjecture
that the corresponding Kac--Moody algebra is in fact a symmetry of
the theory (most probably enlarged with new
fields)~\cite{HyperbolicKaluzaKlein}. The idea is that the BKL
analysis is only the ``revelator'' of that huge symmetry, which would
exist independently of that limit, without making the BKL
truncations. Thus, if this conjecture is true, there should be a way
to rewrite the gravity Lagrangians in such a way that the Kac--Moody
symmetry is manifest. This conjecture itself was made previously (in
this form or in similar ones) by other authors on the basis of
different considerations~\cite{Julia:1980gr,Nicolai:1986jk,E11andMtheory}. To
explore this conjecture, it is desirable to have a concrete method of
dealing with the infinite-dimensional structure of a Lorentzian
Kac--Moody algebra $\mf{g}$. In this section we present such a method.

The method by which we shall deal with the infinite-dimensional
structure of a Lorentzian Kac--Moody algebra $\mf{g}$ is based on a
certain gradation of $\mf{g}$ into finite-dimensional subspaces
$\mf{g}_{\ell}$. More precisely, we will define a so-called
\emph{level decomposition} of the adjoint representation of
$\mf{g}$ such that each level $\ell$ corresponds to a finite
number of representations of a finite regular subalgebra $\mf{r}$
of $\mf{g}$. Generically the decomposition will take the form of
the adjoint representation of $\mf{r}$ plus a (possibly infinite)
number of additional representations of $\mf{r}$. This type of
expansion of $\mf{g}$ will prove to be very useful when
considering sigma models invariant under $\mf{g}$ for which we may
use the level expansion to consistently truncate the theory to any
finite level $\ell$ (see Section~\ref{section:sigmamodels}).

We begin by illustrating these ideas for the finite-dimensional
Lie algebra $\mf{sl}(3,\mbb{R})$ after which we generalize the
procedure to the indefinite case in Sections~\ref{section:FormalC},
\ref{section:DecompAE3AE3} and~\ref{section:DecompE10E10}.


\subsection[A finite-dimensional example: $\mf{sl}(3,\mbb{R})$]
           {A finite-dimensional example: \boldmath $\mf{sl}(3,\mbb{R})$}
\label{section:finitedimensionexample}

The rank~2 Lie algebra $\mf{g}=\mf{sl}(3,\mbb{R})$ is
characterized by the Cartan matrix
\begin{equation}
  A[\mf{sl}(3,\mbb{R})]=
  \left(
    \begin{array}{@{}r@{\quad}r@{}}
      2 & -1 \\
      -1 & 2
    \end{array}
  \right),
  \label{CartanMatrixsl(3)}
\end{equation}
whose Dynkin diagram \index{Dynkin diagram} is displayed in Figure~\ref{figure:A2x}.

\epubtkImage{A2.png}{%
  \begin{figure}[htbp]
    \centerline{\includegraphics[width=35mm]{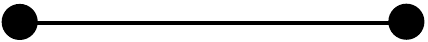}}
    \caption{The Dynkin diagram of $\mf{sl}(3,\mbb{R})$.}
    \label{figure:A2x}
  \end{figure}}

Recall from Section~\ref{section:FiniteRealLieAlgebras} that
$\mf{sl}(3, \mbb{R})$ is the split real form of $\mf{sl}(3,\mbb{C})
\equiv A_2$, and is thus defined through the same Chevalley--Serre
presentation as for $\mf{sl}(3,\mbb{C})$, but with all coefficients
restricted to the real numbers.

The Cartan generators $\{ h_1, h_2\} $ will indifferently be denoted
by $ \{ \al_{1}^{\vee}, \al_{2}^{\vee}\}$. As we have seen, they form
a basis of the Cartan subalgebra $\mf{h}$, while the simple roots
$\{\al_{1},\al_{2}\}$, associated with the raising operators $e_1$ and
$e_2$, form a basis of the dual root space $\mf{h}^{\star}$. Any root
$\ga\in \mf{h}^{\star}$ can thus be decomposed in terms of the simple
roots as follows,
\begin{equation}
  \ga=m\al_{1}+\ell\al_{2},
  \label{rootdecompositionsl(3)}
\end{equation}
and the only values of $(m,n)$ are $(1,0)$, $(0,1)$, $(1,1)$ for
the positive roots and minus these values for the negative ones.

The algebra $\mf{sl}(3,\mbb{R})$ defines through the adjoint
action a representation of $\mf{sl}(3,\mbb{R})$ itself, called the
adjoint representation, which is eight-dimensional and denoted
$\mathbf{8}$. The weights of the adjoint representation are the
roots, plus the weight $(0,0)$ which is doubly degenerate. The
lowest weight of the adjoint representation is 
\begin{equation}
  \Lambda_{\mf{g}}=-\al_{1}-\al_{2},
  \label{Lowestweightsl(3)}
\end{equation}
corresponding to the generator $[f_1, f_2]$. We display the weights of
the adjoint representation in Figure~\ref{figure:AdjointSL3}.

\epubtkImage{AdjointSL3.png}{%
  \begin{figure}[htbp]
    \centerline{\includegraphics[width=90mm]{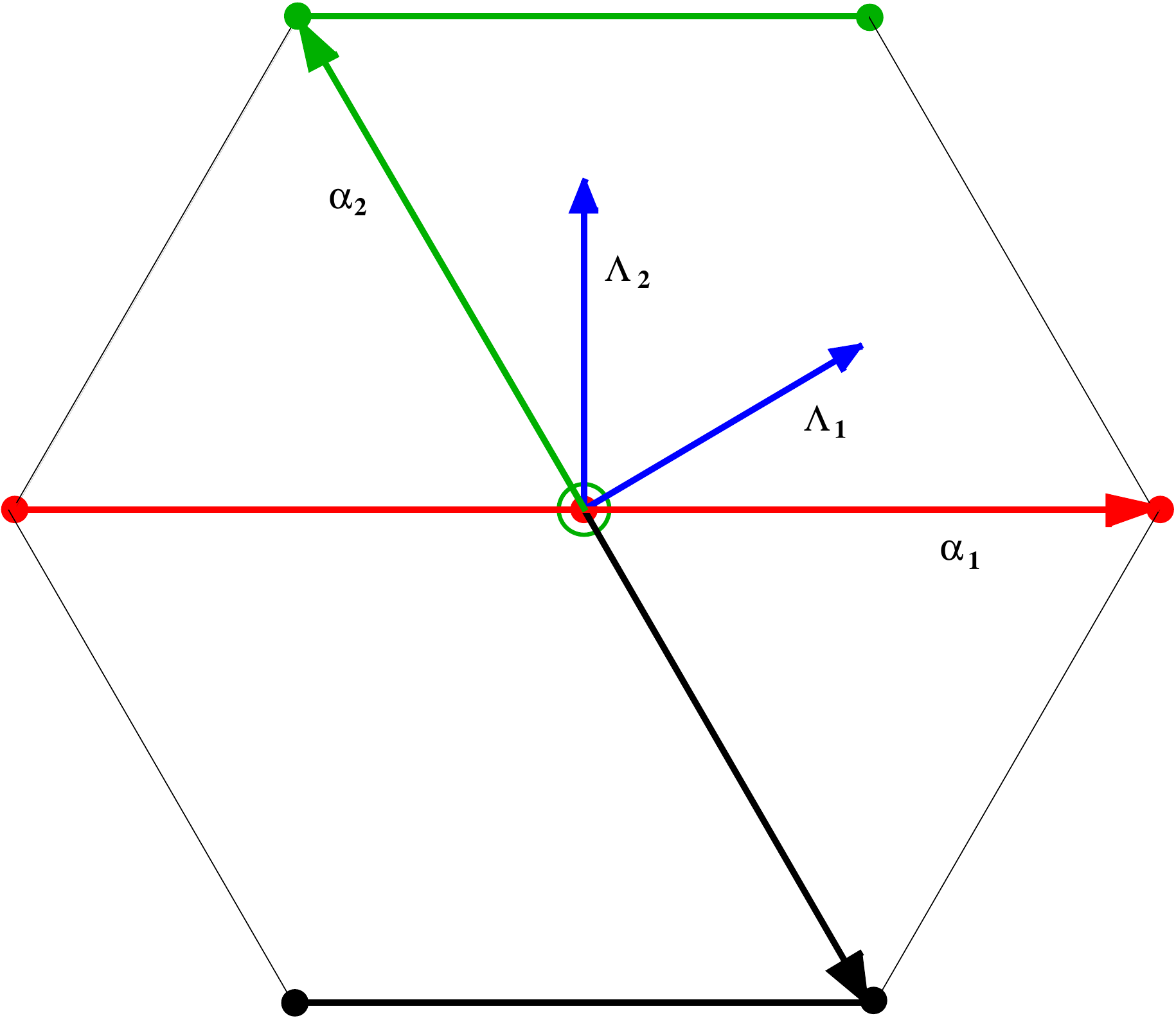}}
    \caption{Level decomposition of the adjoint representation
      $\mc{R}_\mathrm{ad}=\mathbf{8}$ of $\mf{sl}(3,\mbb{R})$ into representations
      of the subalgebra $\mf{sl}(2,\mbb{R})$. The labels $1$ and $2$
      indicate the simple roots $\al_{1}$ and $\al_{2}$. Level zero
      corresponds to the horizontal axis where we find the adjoint
      representation $\mc{R}^{(0)}_\mathrm{ad}=\mathbf{3}_0$ of
      $\mf{sl}(2,\mbb{R})$ (red nodes) and the singlet representation
      $\mc{R}^{(0)}_s=\mathbf{1}_0$ (green circle about the origin). At
      level one we find the two-dimensional representation
      $\mc{R}^{(1)}=\mathbf{2}_1$ (green nodes). The black arrow denotes
      the negative level root $-\al_{2}$ and so gives rise to the
      level $\ell=-1$ representation $\mc{R}^{(-1)}=\mathbf{2}_{(-1)}$. The
      blue arrows represent the fundamental weights $\Lambda_{1}$ and
      $\Lambda_{2}$.}
    \label{figure:AdjointSL3}
  \end{figure}}

The idea of the level decomposition \index{level decomposition} is to decompose the adjoint
representation into representations of one of the regular
$\mf{sl}(2,\mbb{R})$-subalgebras associated with one of the two
simple roots $\al_{1}$ or $\al_{2}$, i.e., either
$\{e_1,\al_{1}^{\vee},f_1\}$ or $\{e_2,\al_{2}^{\vee},f_2\}$.
For definiteness we choose the level to count the number $\ell$ of
times the root $\al_{2}$ occurs, as was anticipated by the
notation in Equation~(\ref{rootdecompositionsl(3)}). Consider the subspace
of the adjoint representation spanned by the vectors with a fixed
value of $\ell$. This subspace is invariant under the action of
the subalgebra $\{e_1,\al_{1}^{\vee},f_1\}$, which only changes
the value of $m$. Vectors at a definite level transform
accordingly in a representation of the regular
$\mf{sl}(2,\mbb{R})$-subalgebra 
\begin{equation}
  \mf{r} \equiv
  \mbb{R}e_1\oplus \mbb{R}\al_{1}^{\vee}\oplus \mbb{R}f_1.
\end{equation}

Let us begin by analyzing states at level $\ell=0$, i.e., with
weights of the form $\ga=m\al_{1}$. We see from
Figure~\ref{figure:AdjointSL3} that we are restricted to move along
the horizontal axis in the root diagram. By the defining Lie algebra
relations we know that $\ad_{f_1}(f_1)=0$, implying that
$\Lambda_\mathrm{ad}^{(0)}=-\al_{1}$ is a lowest weight of the
$\mf{sl}(2,\mbb{R})$-representation. Here, the superscript $0$
indicates that this is a level $\ell=0$ representation. The
corresponding complete irreducible module is found by acting on
$f_1$ with $e_1$, yielding 
\begin{equation}
  [e_1,f_1]=\al_{1}^{\vee},
  \qquad
  [e_1,\al_{1}^{\vee}]=-2e_1,
  \qquad [e_1,e_1]=0.
  \label{Representationsl(2)}
\end{equation}
We can then conclude that $\Lambda_\mathrm{ad}^{(0)}=-\al_{1}$ is the lowest
weight of the three-dimensional adjoint representation $\mathbf{3}_0$ of
$\mf{sl}(2,\mbb{R})$ with weights $\{\Lambda_\mathrm{ad}^{(0)}, 0,
-\Lambda_\mathrm{ad}^{(0)}\}$, where the subscript on $\mathbf{3}_0$ again
indicates that this representation is located at level $\ell=0$ in the
decomposition. The module for this representation is
$\mc{L}(\Lambda_\mathrm{ad}^{(0)})=\spn \{f_1,\al_{1}^{\vee},e_1\}$.

This is, however, not the complete content at level zero since we
must also take into account the Cartan generator
$\al_{2}^{\vee}$ which remains at the origin of the root
diagram. We can combine $\al_{2}^{\vee}$ with $\al_{1}^{\vee}$
into the vector 
\begin{equation}
  h=\al_{1}^{\vee}+2\al_{2}^{\vee},
  \label{singletSL2}
\end{equation}
which constitutes the one-dimensional singlet representation
$\mathbf{1}_0$ of $\mf{r}$ since it is left invariant under all
generators of $\mf{r}$, 
\begin{equation}
  [e_1,h]=[f_1,h]=[\al_{1}^{\vee},h]=0,
  \label{singletSL2commutators}
\end{equation}
as follows trivially from the Chevalley relations. Thus level zero
contains the representations $\mathbf{3}_0$ and $\mathbf{1}_0$.

Note that the vectors at level~0 not only transform in a
(reducible) representation of $\mf{sl}(2,\mbb{R})$, but also form
a subalgebra since the level is additive under taking commutators.
The algebra in question is $\mf{gl}(2,\mbb{R}) = \mf{sl}(2,\mbb{R})
\oplus \mathbb{R}$. Accordingly, if the generator $\al_{2}^{\vee}$ is
added to the subalgebra $\mf{r}$, through the combination in
Equation~(\ref{singletSL2}), so as to take the entire $\ell = 0$
subspace, $\mf{r}$ is enlarged from $\mf{sl}(2,\mbb{R})$ to
$\mf{gl}(2,\mbb{R})$, the generator $h$ being somehow the ``trace''
part of $\mf{gl}(2,\mbb{R})$. This fact will prove to be important in
subsequent sections.

Let us now ascend to the next level, $\ell=1$. The weights of
$\mf{r}$ at level 1 take the general form
$\ga=m\al_{1}+\al_{2}$ and the lowest weight is
$\Lambda^{(1)}=\al_{2}$, which follows from the vanishing of the
commutator 
\begin{equation}
  [f_1,e_2]=0.
  \label{vanishingcommutatorlevel1}
\end{equation}
Note that $m \geq 0$ whenever $\ell >0$ since $
m\al_{1}+\ell\al_{2}$ is then a positive root. The complete
representation is found by acting on the lowest weight
$\Lambda^{(1)}$ with $e_1$ and we get that the commutator
$[e_1,e_2]$ is allowed by the Serre relations, \index{Serre relations} while
$[e_1,[e_1,e_2]]$ is killed, i.e.,
\begin{equation}
  \begin{array}{rcl}
    [e_1,e_2] &\neq & 0, \\ [0 em]
    [e_1,[e_1,e_2]]] &=& 0.
  \end{array}
  \label{Serrerelationslevel1} 
\end{equation}
The non-vanishing commutator $e_{\theta}\equiv [e_1,e_2]$ is the
vector associated with the highest root $\theta$ of
$\mf{sl}(3,\mbb{R})$ given by 
\begin{equation}
  \theta=\al_{1}+\al_{2}.
  \label{highestrootsl(3)}
\end{equation}
This is just the negative of the
lowest weight $\Lambda_{\mf{g}}$. The only representation at
level one is thus the two-dimensional representation $\mathbf{2}_1$
of $\mf{r}$ with weights $\{\Lambda^{(1)},\theta\}$. The
decomposition stops at level one for $\mf{sl}(3,\mbb{R})$ because
any commutator with two $e_2$'s vanishes by the Serre relations. \index{Serre relations}
The negative level representations may be found simply by applying
the Chevalley involution and the result is the same as for level
one.

Hence, the total level decomposition \index{level decomposition} of $\mf{sl}(3,\mbb{R})$ in
terms of the subalgebra $\mf{sl}(2,\mbb{R})$ is given by 
\begin{equation}
  \mathbf{8}=\mathbf{3}_0\oplus \mathbf{1}_0 \oplus \mathbf{2}_1 \oplus \mathbf{2}_{(-1)}. 
  \label{leveldecompositionsl(3)}
\end{equation}
Although extremely simple (and familiar), this example illustrates
well the situation encountered with more involved cases below. In the
following analysis we will not mention the negative levels any longer
because these can always be obtained simply through a reflection with
respect to the $\ell =0$ ``hyperplane'', using the Chevalley
involution.


\subsection{Some formal considerations}
\label{section:FormalC}

Before we proceed with a more involved example, let us formalize
the procedure outlined above. We mainly follow the excellent
treatment given in~\cite{AxelThesis}, although we restrict ourselves to the
cases where $\mf{r}$ is a \emph{finite} regular subalgebra \index{regular subalgebra} of
$\mf{g}$.

In the previous example, we performed the decomposition of the
roots (and the ensuing decomposition of the algebra) with respect
to one of the simple roots which then defined the level. In
general, one may consider a similar decomposition of the roots of
a rank $r$ Kac--Moody algebra  \index{Kac--Moody algebra} with respect to an arbitrary number
$s<r$ of the simple roots and then the level $\ell$ is generalized
to the ``multilevel'' $\mathbf{\ell}=(\ell_1,\cdots, \ell_s)$.


\subsubsection{Gradation}

We consider a Kac--Moody algebra $\mf{g}$ of rank $r$ and
we let $\mf{r}\subset \mf{g}$ be a finite regular rank $m< r$
subalgebra of $\mf{g}$ whose Dynkin diagram is obtained by
deleting a set of nodes $\mc{N}=\{n_1, \cdots, n_s\} \, (s=r-m)$
from the Dynkin diagram of $\mf{g}$.

Let $\ga$ be a root of $\mf{g}$, 
\begin{equation}
  \ga=\sum_{i \notin \mc{N}}
  m_i \al_{i} + \sum_{a \in \mc{N}}\ell_a \al_{a}.
\end{equation}
To this decomposition of the roots corresponds a decomposition of the
algebra, which is called a \emph{gradation} of $\mf{g}$ and which can
be written formally as 
\begin{equation}
  \mf{g}=\bigoplus_{\mathbf{\ell}\in \mbb{Z}^{s}} \mf{g}_{\mathbf{\ell}},
  \label{gradation}
\end{equation}
where for a given $\ell$, $\mf{g}_{\mathbf{\ell}}$ is the subspace
spanned by all the vectors $e_\gamma$ with that definite value $\ell$
of the multilevel, 
\begin{equation}
  [h,e_\gamma] = \gamma(h) e_\gamma,
  \qquad
  l_a(\gamma) = \ell_a.
\end{equation}
Of course, if $\mf{g}$ is finite-dimensional this sum terminates for
some finite level, as in Equation~(\ref{leveldecompositionsl(3)}) for
$\mf{sl}(3,\mbb{R})$. However, in the following we shall mainly be
interested in cases where Equation~(\ref{gradation}) is an infinite
sum.

We note for further reference that the following structure is
inherited from the gradation:
\begin{equation}
  [\mf{g}_{\ell},\mf{g}_{\ell^{\prime}}]\subseteq \mf{g}_{\ell+\ell^{\prime}}. 
  \label{gradation2}
\end{equation}
This implies that for $\ell=0$ we have 
\begin{equation}
  [\mf{g}_0,\mf{g}_{\ell^{\prime}}]\subseteq \mf{g}_{\ell^{\prime}},
\end{equation}
which means that $\mf{g}_{\ell^{\prime}}$ is a representation of
$\mf{g}_0$ under the adjoint action. Furthermore, $\mf{g}_0$ is a
subalgebra. Now, the algebra $\mf{r}$ is a subalgebra of $\mf{g}_0$
and hence we also have 
\begin{equation}
  [\mf{r},\mf{g}_{\ell^{\prime}}]\subseteq \mf{g}_{\ell^{\prime}},
\end{equation}
so that \emph{the subspaces $\mf{g}_{\mathbf{\ell}}$ at definite values of
  the multilevel are invariant subspaces under the adjoint action of
  $\mf{r}$}. In other words, the action of $\mf{r}$ on
$\mf{g}_{\ell}$ does not change the coefficients $\ell_a$.

At level zero, $\mathbf{\ell}=(0,\cdots,0)$, the representation of the
subalgebra $\mf{r}$ in the subspace $\mf{g}_0$ contains the
adjoint representation of $\mf{r}$, just as in the case of
$\mf{sl}(3,\mbb{R})$ discussed in
Section~\ref{section:finitedimensionexample}. All positive and
negative roots of $\mf{r}$ are relevant. Level zero contains in
addition $s$ singlets for each of the Cartan generator associated to
the set $\mc{N}$.

Whenever one of the $\ell_a$'s is positive, all the other ones
must be non-negative for the subspace $\mf{g}_{\mathbf{\ell}}$ to be
nontrivial and only positive roots appear at that value of the
multilevel.


\subsubsection[Weights of $\mf{g}$ and weights of $\mf{r}$]%
              {Weights of \boldmath $\mf{g}$ and weights of $\mf{r}$}

Let $V$ be the module of a representation $\mc{R}(\mf{g})$ of $\mf{g}$
and $\Lambda \in \mf{h}^{\star}_{\mf{g}}$ be one of the weights
occurring in the representation. We define the action of $h\in
\mf{h}_{\mf{g}}$ in the representation $\mc{R}(\mf{g})$ on $x\in V$ as 
\begin{equation}
  h \cdot x = \Lambda(h) x 
\end{equation}
(we consider representations of $\mf{g}$
for which one can speak of ``weights''~\cite{Kac}). Any
representation of $\mf{g}$ is also a representation of $\mf{r}$.
When restricted to the Cartan subalgebra $\mf{h}_{\mf{r}}$ of
$\mf{r}$, $\Lambda$ defines a weight $\bar{\Lambda} \in
\mf{h}^{\star}_{\mf{r}}$, which one can realize geometrically as
follows.

The dual space $\mf{h}^{\star}_{\mf{r}}$ may be viewed as the
$m$-dimensional subspace $\Pi$ of $\mf{h}^{\star}_{\mf{g}}$
spanned by the simple roots $\al_{i}$, $i \notin \mc{N}$. The
metric induced on that subspace is positive definite since
$\mf{r}$ is finite-dimensional. This implies, since we assume
that the metric on $\mf{h}^{\star}_{\mf{g}}$ is nondegenerate,
that $\mf{h}^{\star}_{\mf{g}}$ can be decomposed as the direct sum
\begin{equation}
  \mf{h}^{\star}_{\mf{g}} = \mf{h}^{\star}_{\mf{r}} \oplus \Pi^\perp.
\end{equation}
To that decomposition corresponds the decomposition 
\begin{equation}
  \Lambda = \Lambda^\| + \Lambda^\perp
\end{equation}
of any weight, where $\Lambda^\| \in \mf{h}^{\star}_{\mf{r}} \equiv
\Pi$ and $\Lambda^\perp \in \Pi^\perp$. Now, let $h = \sum k_i
\alpha_{i}^\vee \in \mf{h}_{\mf{r}}$ ($i \notin \mc{N}$). One
has $\Lambda(h) = \Lambda^\| (h) + \Lambda^\perp (h) = \Lambda^\|
(h)$ because $\Lambda^\perp (h) = 0$: The component perpendicular
to $\mf{h}^{\star}_{\mf{r}}$ drops out. Indeed, $\Lambda^\perp
(\alpha_{i}^\vee) = \frac{2 (\Lambda^\perp \vert
\alpha_{i})}{(\alpha_{i} \vert \alpha_{i})} = 0$ for $i
\notin \mc{N}$.

It follows that one can identify the weight $\bar{\Lambda} \in
\mf{h}^{\star}_{\mf{r}}$ with the orthogonal projection
$\Lambda^\| \in \mf{h}^{\star}_{\mf{r}}$ of $\Lambda \in
\mf{h}^{\star}_{\mf{g}}$ on $\mf{h}^{\star}_{\mf{r}}$. This is
true, in particular, for the fundamental weights $\Lambda_{i}$.
The fundamental weights $\Lambda_{i}$ project on $0$ for $i \in
\mc{N}$ and project on the fundamental weights
$\bar{\Lambda}_{i}$ of the subalgebra ${\mf{r}}$ for $i \notin
\mc{N}$. These are also denoted $\lambda_{i}$. For a general
weight, one has 
\begin{equation}
  \Lambda=\sum_{i \notin \mc{N}} p_i \Lambda_{i} + \!\!
  \sum_{a \in \mc{N}} k_a \Lambda_{a}
  \label{generalweight}
\end{equation}
and 
\begin{equation}
  \bar{\Lambda} = \Lambda^\| = \sum_{i \notin \mc{N}} p_i \lambda_{i}.
\end{equation}
The coefficients $p_i$ can easily be extracted by taking the scalar
product with the simple roots, 
\begin{equation}
  p_i = \frac{2}{(\alpha_i \vert \alpha_i)}(\alpha_i \vert \Lambda),
\end{equation}
a formula that reduces to 
\begin{equation}
  p_i = (\alpha_i \vert \Lambda)
\end{equation}
in the simply-laced case. Note that $(\Lambda^\| \vert \Lambda^\|)> 0$
even when $\Lambda$ is non-spacelike.


\subsubsection{Outer multiplicity}
\index{outer multiplicity}

There is an interesting relationship between root
multiplicities in the Kac--Moody algebra $\mf{g}$ and weight
multiplicites of the corresponding $\mf{r}$-weights, which we
will explore here.

For finite Lie algebras, the roots always come with multiplicity
one. This is in fact true also for the real roots of indefinite
Kac--Moody algebras. However, as pointed out in
Section~\ref{section:KacMoody}, the imaginary roots can have
arbitrarily large multiplicity. This must therefore be taken into
account in the sum~(\ref{gradation}).

Let $\ga\in\mf{h}^{\star}_{\mf{g}}$ be a root of $\mf{g}$. There
are two important ingredients:
\begin{itemize}
\item The multiplicity $\mult(\ga)$ of each
  $\ga\in\mf{h}^{\star}_{\mf{g}}$ at level $\mathbf{\ell}$ as a
  \emph{root} of $\mf{g}$.
\item The multiplicity $\mult_{\mc{R}^{(\ell)}_{\ga}}(\ga)$ of the
  corresponding weight $\bar{\ga} \in \mf{h}^{\star}_{\mf{r}}$ at
  level $\mathbf{\ell}$ as a \emph{weight} in the representation
  $\mc{R}^{(\ell)}_\ga$ of $\mf{r}$. (Note that two distinct roots at
  the same level project on two distinct ${\mf{r}}$-weights, so that
  given the ${\mf{r}}$-weight and the level, one can reconstruct the
  root.)
\end{itemize}
It follows that the root multiplicity of $\ga$ is given as a sum
over its multiplicities as a weight in the various representations
$\{\mc{R}^{(\ell)}_q \, | \, q=1,\cdots, N_{\ell}\}$ at level $\ell$.
Some representations can appear more than once at each level, and
it is therefore convenient to introduce a new measure of
multiplicity, called the \emph{outer multiplicity} \index{outer multiplicity|bb} 
$\mu(\mc{R}^{(\ell)}_{q})$, which counts the number of times each
representation $\mc{R}^{(\ell)}_q$ appears at level $\ell$. So,
for each representation at level $\mathbf{\ell}$ we must count the
individual weight multiplicities in that representation and also
the number of times this representation occurs. The total
multiplicity of $\ga$ can then be written as 
\begin{equation}
  \mult(\ga)=\sum_{q=1}^{N_{\ell}}\mu(\mc{R}^{(\ell)}_q)
  \mult_{\mc{R}^{(\ell)}_q}(\ga). 
\label{totalmultiplicity}
\end{equation}
This simple formula might provide useful information on which
representations of $\mf{r}$ are allowed within $\mf{g}$ at a given
level. For example, if $\ga$ is a real root of $\mf{g}$, then it
has multiplicity one. This means that in the
formula~(\ref{totalmultiplicity}), only the representations of
$\mf{r}$ for which $\ga$ has weight multiplicity equal to one are
permitted. The others have $\mu(\mc{R}^{(\ell)}_q)=0$.
Furthermore, only one of the permitted representations does
actually occur and it has necessarily outer multiplicity equal to
one, $\mu(\mc{R}^{(\ell)}_q)=1$.

The subspaces $\mf{g}_{\mathbf{\ell}}$ can now be written explicitly
as 
\begin{equation}
  \mf{g}_{\mathbf{\ell}}=\bigoplus_{q=1}^{N_{\mathbf{\ell}}}
  \left[ \bigoplus_{k=1}^{\mu(\mc{R}^{(\ell)}_{q})}
  \mc{L}^{[k]}(\Lambda_q^{(\mathbf{\ell})})\right],
  \label{decompositionsubspaces}
\end{equation}
where $\mc{L}(\Lambda^{(\ell)}_q)$ denotes the module of the
representation $\mc{R}^{(\ell)}_q$ and $N_{\mathbf{\ell}}$ is the
number of inequivalent representations at level $\ell$. It is
understood that if $\mu(\mc{R}^{(\ell)}_{q})=0$ for some $\ell$
and $q$, then $\mc{L}(\Lambda^{(\mathbf{\ell})}_{q})$ is absent from
the sum. Note that the superscript $[k]$ labels multiple modules
associated to the same representation, e.g., if
$\mu(\mc{R}^{(\ell)}_q)=3$ this contributes to the sum with a
term
\begin{equation}
  \mc{L}^{[1]}(\Lambda^{(\ell)}_q)\oplus
  \mc{L}^{[2]}(\Lambda^{(\ell)}_q)\oplus
  \mc{L}^{[3]}(\Lambda^{(\ell)}_q).
\end{equation}

Finally, we mention that the multiplicity $\mult(\al)$ of
a root $\al\in\mf{h}^{\star}$ can be computed recursively using
the \emph{Peterson recursion relation}, defined as~\cite{Kac} 
\begin{equation}
  (\al\vert\al-2\,\rho)c_{\al}= \!\!\!
  \sum_{\scriptsize
    \begin{array}{l}
      \gamma+\gamma'=\al \\
      \gamma,\gamma'\in Q_+
    \end{array}} \!\!\!
  (\gamma\vert\gamma')c_{\gamma}c_{\gamma'}, 
  \label{PetersonRelation} 
\end{equation}
where $Q_+$ denotes the set of all positive integer linear
combinations of the simple roots, i.e., the positive part of the root
lattice, and $\rho$ is the Weyl vector (defined in
Section~\ref{section:KacMoody}). The coefficients $c_{\gamma}$ are
defined as
\begin{equation}
  c_{\ga}=\sum_{k\geq1}\frac 1k\mult\left(\f{\ga}{k}\right),
\end{equation}
and, following~\cite{Bergshoeff:2007qi}, we call this the
\emph{co-multiplicity}. Note that if $\ga/k$ is not a root, this gives
no contribution to the co-multiplicity. Another feature of the
co-multiplicity is that even if the multiplicity of some root $\ga$ is
zero, the associated co-multiplicity $c_{\ga}$ does not necessarily
vanish. Taking advantage of the fact that all real roots have
multiplicity one it is possible, in principle, to compute recursively
the multiplicity of any imaginary root. Since no closed formula exists
for the outer multiplicity $\mu$, one must take a detour via the
Peterson relation and Equation~(\ref{totalmultiplicity}) in order to
find the outer multiplicity of each representation at a given
level. We give in Table~\ref{table:multiplicities} a list of root
multiplicities and co-multiplicities of roots of $AE_3$ up to height
10.

\begin{table}
  \caption{Multiplicities $m_{\al}=\mult(\al)$ and
    co-multiplicities $c_{\al}$ of all roots $\al$ of $AE_3$ up to
    height 10.}
  \renewcommand{\arraystretch}{1.2}
  \vspace{0.5 em}
  \centering
  \begin{minipage}[t]{8 cm}
    \begin{tabular}{l|l|l|l|l|l}
      \toprule
      $\ell$&$m_1$&$m_2$&$c_{\al}$&$m_{\al}$& $\al^{2}$ \\
      \midrule
      0& 0& 1& 1&1&2 \\
      0& 0& $k>1$& $1/k$&0&$2\,k^2$ \\
      \midrule
      0& 1& 0& 1&1&2 \\
      0& $k>1$& 1& $1/k$&0&$2\,k^2$ \\
      \midrule
      1&0& 0& 1&1&2 \\
      $k>0$&0& 0& $1/k$&0&$2\,k^2$ \\
      \midrule
      0& 1& 1& 1&1&2 \\
      0& $k>1$& $k>1$& $1/k$&0&$2\,k^2$ \\
      \midrule
      1& 1& 0& 1&1&0 \\
      2& 2& 0& 3/2&1&0 \\
      3& 3& 0& 4/3&1&0 \\
      4& 4& 0& 7/4&1&0 \\
      5& 5& 0& 6/5&1&0 \\
      \midrule
      1& 1& 1& 1&1&0 \\
      2& 2& 2& 3/2&1&0 \\
      3& 3& 3& 4/3&1&0 \\
      \midrule
      1& 2& 0& 1&1&2 \\
      2& 4& 0& 1/2&0&8 \\
      3& 6& 0& 1/3&0&2 \\
      \midrule
      2& 1& 0& 1&1&2 \\
      4& 2& 0& 1/2&0&8 \\
      6& 3& 0& 1/3&0&18 \\
      \midrule
      1& 2& 1& 1&1&0 \\
      2& 4& 2& 3/2&1&0 \\
      \midrule
      2& 1& 1& 1&1&2 \\
      4& 2& 2& 1/2&0&8 \\
      \midrule
      1& 2& 2&1&1&2 \\
      2& 4& 4&1/2&0&8 \\
      \midrule
      2& 2& 1& 2&2&-2 \\
      4& 4& 2& 8&7&-8 \\
      \midrule
      2& 3& 0& 1&1&2 \\
      4& 6& 0& 1/2&0&8 \\
      \midrule
      3& 2& 0& 1&1&2 \\
      6& 4& 0& 1/2&0&8 \\
      \bottomrule
    \end{tabular}
  \end{minipage}
  ~
  \begin{minipage}[t]{6 cm}
    \begin{tabular}{l|l|l|l|l|l}
      \toprule
      $\ell$&$m_1$&$m_2$&$c_{\al}$&$m_{\al}$& $\al^{2}$ \\
      \midrule
      2& 3& 1& 2&2&-2 \\
      \midrule
      3& 2& 1& 1&1&0 \\
      \midrule
      2& 4& 1&1&1&2 \\
      \midrule
      2& 3&2& 2&2&-2 \\
      \midrule
      3& 2& 2& 1&1&2 \\
      \midrule
      3& 3& 1& 3&3&-4 \\
      \midrule
      3& 4& 0& 1&1&2 \\
      \midrule
      4& 3& 0& 1&1&2 \\
      \midrule
      2& 3& 3& 1&1&2 \\
      \midrule
      3& 4& 1&3&3&-4 \\
      \midrule
      2& 4& 3& 1&1&2 \\
      \midrule
      3& 3& 2& 3&3&-4 \\
      \midrule
      4& 3& 1& 2&2&-2 \\
      \midrule
      3& 4& 2& 5&5&-6 \\
      \midrule
      3& 5& 1& 1&1&0 \\
      \midrule
      4& 3& 2& 2&2&-2 \\
      \midrule
      4& 4& 1& 5&5&-6 \\
      \midrule
      4& 5& 0& 1&1&2 \\
      \midrule
      5& 4& 0& 1&1&2 \\
      \midrule
      3& 4& 3& 3&3&-4 \\
      \midrule
      3& 5& 2& 3&3&-4 \\
      \midrule
      4& 3& 3& 1&1&2 \\
      \midrule
      4& 5& 1& 5&5&-6 \\
      \midrule
      5& 4& 1& 3&3&-4 \\
      \bottomrule
    \end{tabular}
  \end{minipage}
  \label{table:multiplicities}
  \renewcommand{\arraystretch}{1.0}
\end{table}


\subsection[Level decomposition of $AE_3$]%
           {Level decomposition of \boldmath $AE_3$}
\label{section:DecompAE3AE3}

The Kac--Moody algebra  \index{Kac--Moody algebra}$AE_3=A_1^{++}$ \index{$A_1^{++}$}is one of the simplest
hyperbolic algebras and so provides a nice testing ground for
investigating general properties of hyperbolic Kac--Moody algebras.
From a physical point of view, it is the Weyl group of $AE_3$
which governs the chaotic behavior of pure four-dimensional
gravity close to a spacelike singularity~\cite{HyperbolicKaluzaKlein},
as we have explained. Moreover, as we saw in
Section~\ref{section:Coxeter}, the Weyl group of $AE_3$ is isomorphic
with the well-known arithmetic group $PGL(2,\mbb{Z})$ which has
interesting properties~\cite{FeingoldFrenkel}.

The level decomposition \index{level decomposition}of $\mf{g}=AE_3$ follows a similar route
as for $\mf{sl}(3,\mbb{R})$ above, but the result is much more
complicated due to the fact that $AE_3$ is infinite-dimensional.
This decomposition has been treated before in~\cite{DHNReview}.
Recall that the Cartan matrix \index{Cartan matrix}for $AE_3$ is given by 
\begin{equation}
  \left(
    \begin{array}{@{}r@{\quad}r@{\quad}r@{}}
      2 & -2 & 0 \\
      -2 & 2 & -1 \\
      0 & -1 & 2
    \end{array}
  \right),
  \label{CartanMatrixAE3}
\end{equation}
and the associated Dynkin diagram \index{Dynkin diagram}is given in
Figure~\ref{figure:A1pp}.

\epubtkImage{A1pp.png}{%
  \begin{figure}[htbp]
    \centerline{\includegraphics[width=50mm]{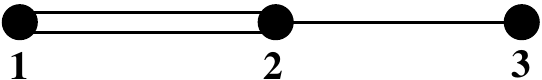}}
    \caption{The Dynkin diagram of the hyperbolic Kac--Moody algebra
      $AE_3\equiv A_1^{++}$. The labels indicate the simple roots
      $\al_{1}, \al_{2}$ and $\al_{3}$. The nodes ``2'' and ``3''
      correspond to the subalgebra $\mf{r}=\mf{sl}(3,\mbb{R})$ with
      respect to which we perform the level decomposition.}
    \label{figure:A1pp}
  \end{figure}}

We see that there exist three rank 2 regular subalgebras \index{regular subalgebra}that we
can use for the decomposition: $A_2, A_1\oplus A_1$ or $A_1^{+}$.
We will here focus on the decomposition into representations of
$\mf{r}=A_2=\mf{sl}(3,\mbb{R})$ because this is the one relevant
for pure gravity in four
dimensions~\cite{HyperbolicKaluzaKlein}\epubtkFootnote{The
  decomposition of $AE_3$ into representations of $A_1^{+}$ was done
  in~\cite{FeingoldFrenkel}.}. The level $\ell$ is then the coefficient
in front of the simple root $\al_{1}$ in an expansion of an arbitrary
root $\ga\in \mf{h}_{\mf{g}}^{\star}$, i.e., 
\begin{equation}
  \ga =\ell \al_{1}+m_2\al_{2}+m_3\al_{3}. 
  \label{rootAE3}
\end{equation}

We restrict henceforth our analysis to positive levels only,
$\ell\geq 0$. Before we begin, let us develop an intuitive idea of
what to expect. We know that at each level we will have a set of
finite-dimensional representations of the subalgebra $\mf{r}$. The
corresponding weight diagrams will then be represented in a
Euclidean two-dimensional lattice in exactly the same way as in
Figure~\ref{figure:AdjointSL3} above. The level $\ell$ can be
understood as parametrizing a third direction that takes us into
the full three-dimensional root space of $AE_3$. We display the
level decomposition \index{level decomposition}up to positive level two in
Figure~\ref{figure:AE3Dec}\epubtkFootnote{D.P.\ would like to thank
  Bengt E.W.\ Nilsson and Jakob Palmkvist for helpful discussions during the
creation of Figure~\ref{figure:AE3Dec}.}.

\epubtkImage{AE3Dec.png}{%
  \begin{figure}
    \centerline{\includegraphics[width=150mm]{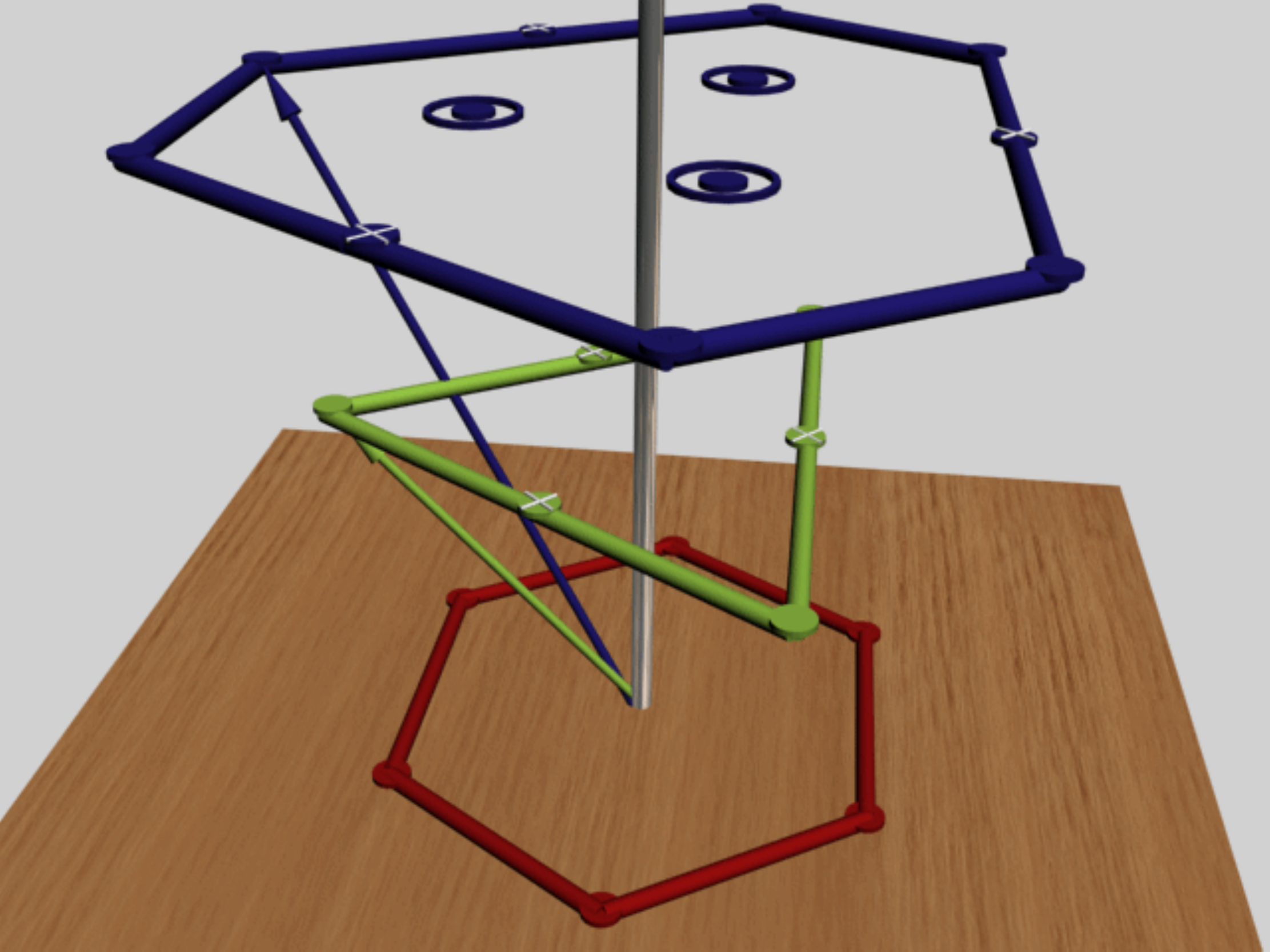}}
    \caption{Level decomposition of the adjoint representation of
      $AE_3$. We have displayed the decomposition up to positive level
      $\ell=2$. At level zero we have the adjoint representation
      $\mc{R}^{(0)}_1=\mathbf{8}_0$ of $\mf{sl}(3,\mbb{R})$ and the singlet
      representation $\mc{R}^{(0)}_2=\mathbf{1}_0$ defined by the simple
      Cartan generator $\al_{1}^{\vee}$. Ascending to level one with
      the root $\al_{1}$ (green vector) gives the lowest weight
      $\Lambda^{(1)}$ of the representation $\mc{R}^{(1)}=\mathbf{6}_1$.
      The weights of $\mc{R}^{(1)}$ labelled by white crosses are on the
      lightcone and so their norm squared is zero. At level two we find
      the lowest weight $\Lambda^{(2)}$ (blue vector) of the
      15-dimensional representation $\mc{R}^{(2)}=\mathbf{15}_2$. Again,
      the white crosses label weights that are on the lightcone. The
      three innermost weights are inside of the lightcone and the rings
      indicate that these all have multiplicity 2 as weights of
      $\mc{R}^{(2)}$. Since these also have multiplicity 2 as
      \emph{roots} of $\mf{h}^{\star}_{\mf{g}}$ we find that the outer
      multiplicity of this representation is one, $\mu(\mc{R}^{(2)})=1$.}
    \label{figure:AE3Dec}
  \end{figure}}

From previous sections we recall that $AE_3$ is hyperbolic so its
root space is of Lorentzian signature. This implies that there is
a lightcone in $\mf{h}^{\star}_{\mf{g}}$ whose origin lies at the
origin of the root diagram for the adjoint representation of
$\mf{r}$ at level $\ell=0$. The lightcone separates real roots
from imaginary roots and so it is clear that if a representation at
some level $\ell$ intersects the walls of the lightcone, this
means that some weights in the representation will correspond to
imaginary roots of $\mf{h}_{\mf{g}}^{\star}$ but will be real as
weights of $\mf{h}_{\mf{r}}^{\star}$. On the other hand if a
weight lies outside of the lightcone it will be real both as a
root of $\mf{h}_{\mf{g}}^{\star}$ and as a weight of
$\mf{h}_{\mf{r}}^{\star}$.

\subsubsection[Level $\ell=0$]{Level \boldmath $\ell=0$}

Consider first the representation content at level zero.
Given our previous analysis we expect to find the adjoint
representation of $\mf{r}$ with the additional singlet
representation from the Cartan generator $\al_{1}^{\vee}$. The
Chevalley generators of $\mf{r}$ are
$\{e_2,f_2,e_3,f_3,\al_{2}^{\vee},\al_{3}^{\vee}\}$ and the
generators associated to the root defining the level are
$\{e_1,f_1,\al_{1}^{\vee}\}$. As discussed previously, the
additional Cartan generator $\al_{1}^{\vee}$ that sits at the
origin of the root space enlarges the subalgebra from
$\mf{sl}(3,\mbb{R})$ to $\mf{gl}(3,\mbb{R})$. A canonical
realisation of $\mf{gl}(3,\mbb{R})$ is obtained by defining the
Chevalley generators in terms of the matrices ${K^{i}}_j \,
(i,j=1,2,3)$ whose commutation relations are 
\begin{equation}
  [{K^{i}}_j,{K^{k}}_l]=\delta^{k}_j {K^{i}}_l-\delta^{i}_l {K^{k}}_j. 
  \label{gl(3)}
\end{equation}
All the defining Lie algebra
relations of $\mf{gl}(3,\mbb{R})$ are then satisfied if we make
the identifications 
\begin{equation}
  \begin{array}{rcl@{\qquad}rcl@{\qquad}rcl}
    && & && & \al_{1}^{\vee}&=& {K^{1}}_1-{K^{2}}_2-{K^{3}}_3, \\
    e_2&=&{K^{2}}_1, &
    f_2&=&{K^{1}}_2, &
    \al_{2}^{\vee}&=&{K^{2}}_2-{K^{1}}_1, \\
    e_3&=&{K^{3}}_2, &
    f_3&=&{K^{2}}_3, &
    \al_{3}^{\vee}&=&{K^{3}}_3-{K^{2}}_2.
  \end{array}
  \label{realisationgl(3)} 
\end{equation}
Note that the trace
${K^{1}}_1+{K^{2}}_2+{K^{3}}_3$ is equal to $-4 \al_{2}^{\vee} - 2
\al_{3}^{\vee} - 3 \al_{1}^{\vee}$. The generators $e_1$ and $f_1$
can of course not be realized in terms of the matrices ${K^{i}}_j$
since they do not belong to level zero. The invariant bilinear
form $(\ |\ )$ at level zero reads 
\begin{equation}
  ({K^{i}}_j|{K^{k}}_l)=
  \delta^{i}_l\delta^{k}_j-\delta^{i}_j\delta^{k}_l,
  \label{Killingformlevelzero}
\end{equation}
where the coefficient in front of the second term on the right hand
side is fixed to $-1$ through the embedding of $\mf{gl}(3, \mbb{R})$
in $AE_3$.

The commutation relations in Equation~(\ref{gl(3)}) characterize the
adjoint representation of $\mf{gl}(3,\mbb{R})$ as was expected at
level zero, which decomposes as the representation
$\mc{R}^{(0)}_\mathrm{ad} \oplus \mc{R}^{(0)}_{s}$ of
$\mf{sl}(3,\mbb{R})$ with $\mc{R}^{(0)}_\mathrm{ad}=\mathbf{8}_0$ and
$\mc{R}^{(0)}_{s}=\mathbf{1}_0$.

\subsubsection{Dynkin labels}

It turns out that at each positive level $\ell$, the
weight that is easiest to identify is the lowest weight. For
example, at level one, the lowest weight is simply $\alpha_1$ from
which one builds all the other weights by adding appropriate
positive combinations of the roots $\alpha_2$ and $\alpha_3$. It
will therefore turn out to be convenient to characterize the
representations at each level by their (conjugate) Dynkin labels \index{Dynkin labels|bb}
$p_2$ and $p_3$ defined as the coefficients of minus the
(projected) lowest weight $-\bar{\Lambda}^{(\ell)}_\mathrm{lw}$ expanded
in terms of the fundamental weights $\lambda_{2}$ and
$\lambda_{3}$ of $\mf{sl}(3,\mbb{R})$ (blue arrows in
Figure~\ref{figure:15ofSL3}),
\begin{equation}
  -\bar{\Lambda}^{(\ell)}_\mathrm{lw} = p_2 \lambda_{2} + p_3\lambda_{3}. 
  \label{DynkinLabels}
\end{equation}
Note that for any weight $\Lambda$ we have the inequality 
\begin{equation}
  (\Lambda|\Lambda)\leq (\bar{\Lambda}|\bar{\Lambda})
  \label{projectionInequality}
\end{equation}
since $(\Lambda|\Lambda)= (\bar{\Lambda}|\bar{\Lambda}) - \vert
(\Lambda^\perp|\Lambda^\perp) \vert$.

The Dynkin labels \index{Dynkin labels}can be computed using the
scalar product $(\ |\ )$ in $\mf{h}_{\mf{g}}^{\star}$ in the following way:
\begin{equation}
  p_2=-(\al_{2}|\Lambda^{(\ell)}_\mathrm{lw}),
  \qquad
  p_3=-(\al_{3}|\Lambda^{(\ell)}_\mathrm{lw}). 
  \label{DynkinLabels2}
\end{equation}
For the level zero sector we therefore have 
\begin{equation}
  \begin{array}{rcl}
    \mathbf{8}_0 & : & [p_2,p_3]=[1,1],
    \\
    \mathbf{1}_0 & : & [p_2,p_3]=[0,0].
  \end{array}
  \label{DynkinLabelslevelzero}
\end{equation}

The module for the representation $\mathbf{8}_0$ is realized by the
eight traceless generators ${K^{i}}_j$ of $\mf{sl}(3,\mbb{R})$
and the module for the representation $\mathbf{1}_0$ corresponds to
the ``trace'' $\al_{1}^{\vee}$.

Note that the highest weight $\Lambda_\mathrm{hw}$ of a given
representation of $\mf{r}$ is not in general equal to minus the
lowest weight $\Lambda$ of the same representation. In fact,
$-\Lambda_\mathrm{hw}$ is equal to the lowest weight of the
\emph{conjugate} representation. This is the reason our Dynkin
labels are really the conjugate Dynkin labels \index{Dynkin labels}in standard
conventions. It is only if the representation is self-conjugate
that we have $\Lambda_\mathrm{hw}=-\Lambda$. This is the case for example
in the adjoint representation $\mathbf{8}_0$.

It is interesting to note that since the weights of a
representation at level $\ell$ are related by Weyl reflections to
weights of a representation at level $-\ell$, it follows that the
negative of a lowest weight $\Lambda^{(\ell)}$ at level $\ell$ is
actually equal to the \emph{highest} weight
$\Lambda_\mathrm{hw}^{(-\ell)}$ of the conjugate representation at level
$-\ell$. Therefore, the Dynkin labels at level $\ell$ as defined
here are the standard Dynkin labels of the representations at
level $- \ell$.

\subsubsection[Level $\ell=1$]{Level \boldmath $\ell=1$}

We now want to exhibit the representation content at the
next level $\ell=1$. A generic level one commutator is of the form
$[e_1,[\cdots [\cdots ]]]$, where the ellipses denote (positive)
level zero generators. Hence, including the generator $e_1$
implies that we step upwards in root space, i.e., in the direction
of the forward lightcone. The root vector $e_1$ corresponds to a
lowest weight of $\mf{r}$ since it is annihilated by $f_2$ and
$f_3$, 
\begin{equation}
  \begin{array}{rcl}
    \ad_{f_2}(e_1) & = & [f_2,e_1] = 0,
    \\
    \ad_{f_3}(e_1) & = & [f_3,e_1]  = 0,
  \end{array}
  \label{leveloneadjointaction} 
\end{equation}
which follows from the defining relations of $AE_3$.

Explicitly, the root associated to $e_1$ is simply the root
$\al_{1}$ that defines the level expansion. Therefore the lowest
weight of this level one representation is 
\begin{equation}
  \Lambda^{(1)}_\mathrm{lw}=\bar{\al}_{1}, 
  \label{lowestweightlevelone}
\end{equation}
Although $\al_{1}$ is a real \emph{positive} root of
$\mf{h}_{\mf{g}}^{\star}$, its projection $\bar{\al}_{(1)}$ is a
\emph{negative} weight of $\mf{h}_{\mf{r}}^{\star}$. Note that
since the lowest weight $\Lambda^{(1)}_1$ is real, the
representation $\mc{R}^{(1)}$ has outer multiplicity \index{outer multiplicity} one,
$\mu(\mc{R}^{(1)})=1$.

Acting on the lowest weight state with the raising operators of
$\mf{r}$ yields the six-dimensional representation
$\mc{R}^{(1)}=\mathbf{6}_1$ of $\mf{sl}(3,\mbb{R})$. The root
$\al_{1}$ is displayed as the green vector in
Figure~\ref{figure:AE3Dec}, taking us from the origin at level zero to
the lowest weight of $\mc{R}^{(1)}$. The Dynkin labels of this
representation are 
\begin{equation}
  \begin{array}{rcl}
    p_2(\mc{R}^{(1)})&=&-(\al_{2}|\al_{1}) = 2,
    \\
    p_3(\mc{R}^{(1)})&=&-(\al_{3}|\al_{1}) = 0,
  \end{array}
  \label{Dynkinlabelslevel1} 
\end{equation}
which follows directly from the
Cartan matrix of $AE_3$. Three of the weights in $\mc{R}^{(1)}$
correspond to roots that are located on the lightcone in root
space and so are null roots of $\mf{h}^{\star}_{\mf{g}}$. These
are $\alpha_{1}+ \alpha_{2}$, $\alpha_{1}+ \alpha_{2} +
\alpha_{3}$ and $\alpha_{1}+ 2\alpha_{2} + \alpha_{3}$ and
are labelled with white crosses in Figure~\ref{figure:AE3Dec}. The
other roots present in the representation, in addition to 
$\alpha_{1}$, are $\alpha_{1}+ 2\alpha_{2}$ and $\alpha_{1}+
2\alpha_{2} +2 \alpha_{3}$, which are real. This representation
therefore contains no weights inside the lightcone.

The $\mf{gl}(3,\mbb{R})$-generator encoding this representation is
realized as a symmetric 2-index tensor $E^{ij}$ which indeed
carries six independent components. In general we can easily
compute the dimensionality of a representation given its Dynkin
labels using the \emph{Weyl dimension formula} which for
$\mf{sl}(3,\mbb{R})$ takes the form~\cite{Fuchs} 
\begin{equation}
  d_{\Lambda_\mathrm{hw}}\left(\mf{sl}(3,\mbb{R})\right)=
  (p_2+1)(p_3+1)\left(\f{1}{2}(p_2+p_3)+1\right).
  \label{WeylDimensionFomulaA2}
\end{equation}
In particuar, for
$(p_2,p_3)=(2,0)$ this gives indeed $d_{\Lambda^{(1)}_\mathrm{hw,1}}\!\!=6$.

It is convenient to encode the Dynkin labels, \index{Dynkin labels}and, consequently,
the index structure of a given representation module, in a Young
tableau. We follow conventions where the first Dynkin label gives
the number of columns with 1 box and the second Dynkin label gives
the number of columns with 2 boxes\epubtkFootnote{Since we are, in fact, 
using conjugate Dynkin labels, these conventions are equivalent to
the standard ones if one replaces covariant indices by
contravariant ones, and vice-versa.}. For the representation $\mathbf{6}_1$ the first Dynkin
label is 2 and the second is 0, hence the associated Young
tableau is
\begin{equation}
  \mathbf{6}_1
  \quad \Longleftrightarrow \quad
  {\footnotesize
    \setlength{\tabcolsep}{0.55 em}
    \begin{tabular}{cc}
      \cline{1-2}
      \multicolumn{1}{|c|}{} &
      \multicolumn{1}{c|}{} \\
      \cline{1-2}
    \end{tabular}}\,.
  \label{YoungtableauLevel1}
\end{equation}

At level $\ell=-1$ there is a corresponding negative generator
$F_{ij}$. The generators $E^{ij}$ and $F_{ij}$ transform
contravariantly and covariantly, respectively, under the level
zero generators, i.e., 
\begin{equation}
  \begin{array}{rcl}
    [{K^{i}}_j,E^{kl}]&=&\delta^{k}_{j}E^{il}+\delta^{l}_{j}E^{ki},
    \\ [0 em]
    [{K^{i}}_j,F_{kl}]&=& -\delta^{i}_k F_{jl}-\delta^{i}_l F_{kj}.
  \end{array}
\end{equation}
The internal commutator on level one can be obtained by first
identifying 
\begin{equation}
  e_1\equiv E^{11},
  \qquad
  f_1\equiv F_{11}, 
  \label{levelonegenerators}
\end{equation}
and then by demanding $[e_1,f_1]=\al_{1}^{\vee}$ we find 
\begin{equation}
  [E^{ij},F_{kl}]=2\delta^{(i}_{(k}{K^{j)}}_{l)}-\delta^{(i}_{k}\delta^{k)}_{l}
  ({K^{1}}_1+{K^{2}}_2+{K^{3}}_3),
\end{equation}
which is indeed compatible with the realisation of $\al_{1}^{\vee}$
given in Equation~(\ref{realisationgl(3)}). The Killing form at
level~1 takes the form 
\begin{equation}
  \left( F_{ij}|E^{kl}\right)=\delta_i^{(k}\delta_j^{l)}.
  \label{KillingformLevel1}
\end{equation}

\subsubsection{Constraints on Dynkin labels}
\index{Dynkin labels}

As we go to higher and higher levels it is useful to
employ a systematic method to investigate the representation
content. It turns out that it is possible to derive a set of
equations whose solutions give the Dynkin labels for the
representations at each level~\cite{DHN2}.

We begin by relating the Dynkin labels to the expansion
coefficients $\ell,m_2$ and $m_3$ of a root $\ga\in
\mf{h}_{\mf{g}}^{\star}$, whose projection $\bar{\ga}$ onto
$\mf{h}_{\mf{r}}^{\star}$ is a lowest weight vector for some
representation of $\mf{r}$ at level $\ell$. We let $a=2,3$ denote
indices in the root space of the subalgebra $\mf{sl}(3,\mbb{R})$
and we let $i=1,2,3$ denote indices in the full root space of
$AE_3$. The formula for the Dynkin labels then gives 
\begin{equation}
  p_{a}=-(\al_{a}|\ga)=-\ell A_{a1}-m_2A_{a2}-m_3A_{a3},
  \label{DynkinlabelsArbitraryLevel}
\end{equation}
where $A_{ij}$ is the
Cartan matrix for $AE_3$, given in Equation~(\ref{CartanMatrixAE3}).
Explicitly, we find the following relations between the
coefficients $m_2, m_3$ and the Dynkin labels: 
\begin{equation}
  \begin{array}{rcl}
    p_2 &=& 2\ell-2m_2+m_3,
    \\
    p_3 &=& m_2-2m_3.
  \end{array}
  \label{Dynkinlabelrelation} 
\end{equation}
These formulae restrict the possible Dynkin labels for each $\ell$ since
the coefficients $m_2$ and $m_3$ must necessarily be non-negative
integers. Therefore, by inverting Equation~(\ref{Dynkinlabelrelation}) we
obtain two Diophantine equations that restrict the possible Dynkin
labels, 
\begin{equation}
  \begin{array}{rcl}
    m_2&=& \displaystyle \f{4}{3}\ell-\f{2}{3}p_2-\f{1}{3}p_3 \geq 0,
    \\
    m_3&=& \displaystyle \f{2}{3}\ell-\f{1}{3}p_2-\f{2}{3}p_3 \geq 0.
  \end{array}
  \label{DiophantineAE3}
\end{equation}
In addition to these constraints we
can also make use of the fact that we are decomposing the adjoint
representation of $AE_3$. Since the weights of the adjoint
representation are the roots of the algebra we know that the
lowest weight vector $\Lambda$ must satisfy 
\begin{equation}
  (\Lambda | \Lambda)\leq 2. 
  \label{RootConstraint}
\end{equation}
Taking $\Lambda=\ell
\al_{1}+m_2 \al_{2}+m_3 \al_{3}$ then gives the following
constraint on the coefficients $\ell, m_2$ and $m_3$:
\begin{equation}
  (\Lambda|\Lambda)=2\ell^2+2m_2^2+2m_3^2-4\ell m_2-2m_2 m_3 \leq 2. 
  \label{RootConstraint2}
\end{equation}
We are interested in
finding an equation for the Dynkin labels, so we insert
Equation~(\ref{DiophantineAE3}) into Equation~(\ref{RootConstraint2}) to obtain
the constraint
\begin{equation}
  p_2^2+p_3^2+p_2 p_3-\ell^2\leq 3.
  \label{ConstraintDynkinLabels}
\end{equation}
The inequalities in Equation~(\ref{DiophantineAE3}) and
Equation~(\ref{ConstraintDynkinLabels}) are sufficient to determine
the representation content at each level $\ell$. However, this
analysis does not take into account the outer multiplicities, which
must be analyzed separately by comparing with the known root
multiplicities of $AE_3$ as given in
Table~\ref{table:multiplicities}. We shall return to this issue later.

\subsubsection[Level $\ell=2$]{Level \boldmath $\ell=2$}

Let us now use these results to analyze the case for
which $\ell=2$. The following equations must then be satisfied:
\begin{equation}
  \begin{array}{rcl}
    a 8-2p_2-p_3&\geq & 0,
    \\
    4-p_2-2p_3&\geq & 0,
    \\
    p_2^2+p_3^2+p_2 p_3& \leq & 7.
  \end{array}
  \label{Level2Constraints} 
\end{equation}
The only admissible solution is $p_2=2$ and $p_3=1$. This
corresponds to a 15-dimensional representation $\mathbf{15}_2$ with
the following Young tableau
\begin{equation}
  \mathbf{6}_1
  \quad \Longleftrightarrow \quad
  {\footnotesize
    \setlength{\tabcolsep}{0.55 em}
    \begin{tabular}{ccc}
      \cline{1-3}
      \multicolumn{1}{|c|}{} &
      \multicolumn{1}{c|}{} &
      \multicolumn{1}{c|}{} \\
      \cline{1-3}
      \multicolumn{1}{|c|}{} \\
      \cline{1-1}
    \end{tabular}}\,.
  \label{YoungtableauLevel2}
\end{equation}
Note that $p_2=p_3=0$ is also a solution to
Equation~(\ref{Level2Constraints}) but this violates the constraint
that $m_2$ and $m_3$ be integers and so is not allowed.

Moreover, the representation $[p_2,p_3]=[0,2]$ is also a solution
to Equation~(\ref{Level2Constraints}) but has not been taken into account
because it has vanishing outer multiplicity. This can be
understood by examining Figure~\ref{figure:15ofSL3} a little
closer. The representation $[0,2]$ is six-dimensional and has
highest weight $2\lambda_{3}$, corresponding to the middle node
of the top horizontal line in Figure~\ref{figure:15ofSL3}. This
weight lies outside of the lightcone and so is a real root of
$AE_3$. Therefore we know that it has root multiplicity one and
may therefore only occur once in the level decomposition. Since
the weight $2\lambda_{3}$ already appears in the larger
representation $\mathbf{15}_2$ it cannot be a highest weight in
another representation at this level. Hence, the representation
$[0,2]$ is not allowed within $AE_3$. A similar analysis reveals that
also the representation $[p_2, p_3]=[1, 0]$, although allowed by
Equation~(\ref{Level2Constraints}), has vanishing outer multiplicity.

The level two module is realized by the tensor ${E_i}^{jk}$ whose
index structure matches the Young tableau above. Here we have used
the $\mf{sl}(3,\mbb{R})$-invariant antisymmetric tensor
$\epsilon^{abc}$ to lower the two upper antisymmetric indices
leading to a tensor ${E_{i}}^{jk}$ with the properties
\begin{equation}
  {E_i}^{jk} = {E_i}^{(jk)},
  \qquad
  {E_i}^{ik}=0.
\end{equation}
This corresponds to a positive root generator and by the Chevalley
involution we have an associated negative root generator
${F^i}_{jk}$ at level $\ell=-2$. Because the level decomposition
gives a gradation of $AE_3$ we know that all higher level
generators can be obtained through commutators of the level one
generators. More specifically, the level two tensor ${E_i}^{jk}$
corresponds to the commutator
\begin{equation}
  [E^{ij},E^{kl}]=\epsilon^{mk(i}{E_m}^{j)l}+\epsilon^{ml(i}{E_m}^{j)k},
\end{equation}
where $\epsilon^{ijk}$ is the
totally antisymmetric tensor in three dimensions. Inserting the
result $p_2=2$ and $p_3=1$ into Equation~(\ref{DiophantineAE3}) gives
$m_2=1$ and $m_3=0$, thus providing the explicit form of the root
taking us from the origin of the root diagram in
Figure~\ref{figure:AE3Dec} to the lowest weight of $\mathbf{15}_2$ at
level two:
\begin{equation}
  \Lambda^{(2)}=2\al_{1}+\al_{2}.
  \label{LowestweightLevel2}
\end{equation}
This is a real root of $AE_3$,
$(\ga |\ga)=2$, and hence the representation $\mathbf{15}_2$ has
outer multiplicity one. We display the representation $\mathbf{15}_2$
of $\mf{sl}(3,\mbb{R})$ in Figure~\ref{figure:15ofSL3}. The lower
leftmost weight is the lowest weight $\Lambda^{(2)}$. The
expansion of the lowest weight $\Lambda_\mathrm{lw}^{(2)}$ in terms of
the fundamental weights $\lambda_{2}$ and $\lambda_{3}$ is
given by the (conjugate) Dynkin labels
\begin{equation}
  -\Lambda_\mathrm{hw}^{(2)}=
  p_2\lambda_{2} + p_3\lambda_{3}=2\lambda_{2}+\lambda_{3}.
  \label{HighestWeightLevel2}
\end{equation}
The three innermost weights all
have multiplicity 2 as weights of $\mf{sl}(3,\mbb{R})$, as
indicated by the black circles. These lie inside the lightcone of
$\mf{h}^{\star}_{\mf{g}}$ and so are timelike roots of $AE_3$.

\epubtkImage{15ofSL3.png}{%
  \begin{figure}
    \centerline{\includegraphics[width=90mm]{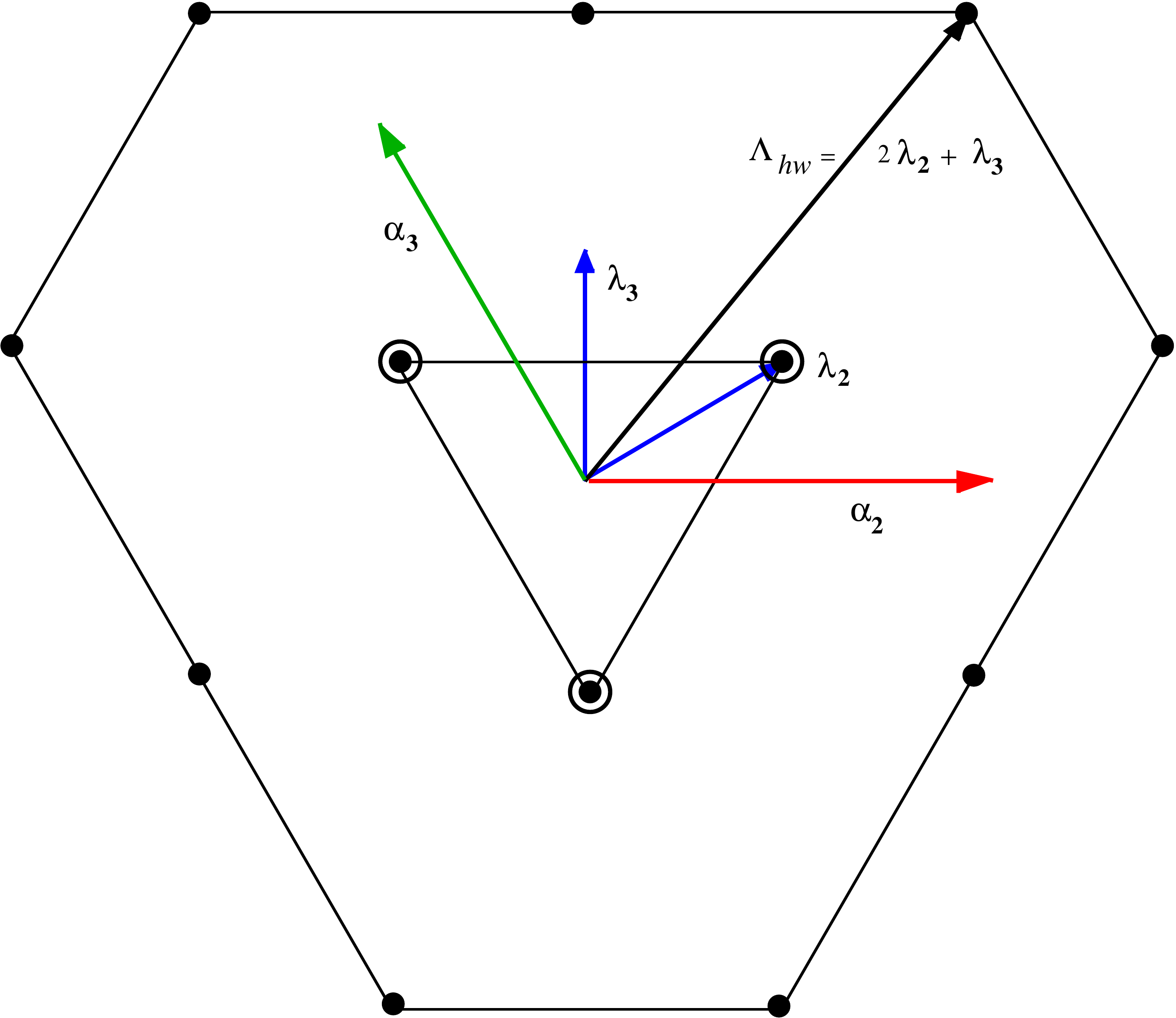}}
    \caption{The representation $\mathbf{15}_2$ of $\mf{sl}(3,\mbb{R})$
      appearing at level two in the decomposition of the adjoint
      representation of $AE_3$ into representations of
      $\mf{sl}(3,\mbb{R})$. The lowest leftmost node is the lowest
      weight of the representation, corresponding to the real root
      $\Lambda^{(2)}=2\al_{1}+\al_{2}$ of $AE_3$. This
      representation has outer multiplicity one. }
    \label{figure:15ofSL3}
  \end{figure}}

\subsubsection[Level $\ell=3$]{Level \boldmath $\ell=3$}

We proceed quickly past level three since the analysis
does not involve any new ingredients. Solving
Equation~(\ref{DiophantineAE3}) and
Equation~(\ref{ConstraintDynkinLabels}) for $\ell=3$ yields two
admissible $\mf{sl}(3,\mbb{R})$ representations, $\mathbf{27}_3$ and
$\mathbf{8}_3$, represented by the following Dynkin labels and Young
tableaux:
\begin{equation}
  \begin{array}{rcl}
    \mathbf{27}_3 &:& [p_2,p_3]=[2,2]
    \quad \Longleftrightarrow \quad
    {\footnotesize
      \setlength{\tabcolsep}{0.55 em}
      \begin{tabular}{cccc}
        \cline{1-4}
        \multicolumn{1}{|c|}{} &
        \multicolumn{1}{c|}{} &
        \multicolumn{1}{c|}{} &
        \multicolumn{1}{c|}{} \\
        \cline{1-4}
        \multicolumn{1}{|c|}{} &
        \multicolumn{1}{c|}{} \\
        \cline{1-2}
      \end{tabular}}\,,
    \\ [1.0 em]
    \mathbf{8}_3 &:& [p_2,p_3]=[1,1]
    \quad \Longleftrightarrow \quad
    {\footnotesize
      \setlength{\tabcolsep}{0.55 em}
      \begin{tabular}{cc}
        \cline{1-2}
        \multicolumn{1}{|c|}{} &
        \multicolumn{1}{c|}{} \\
        \cline{1-2}
        \multicolumn{1}{|c|}{} \\
        \cline{1-1}
      \end{tabular}}\,.
  \end{array}
  \label{RepresentationsLevel3} 
\end{equation}
The lowest weight vectors for
these representations are 
\begin{equation}
  \begin{array}{rcl}
    \Lambda^{(3)}_{\mathbf{15}}& =& 3\al_{1}+2\al_{2},
    \\
    \Lambda^{(3)}_{\mathbf{8}} &=& 3\al_{1}+3\al_{2}+\al_{3}.
  \end{array}    
  \label{LowestWeightsLevel3} 
\end{equation}
The lowest weight vector for
$\mathbf{27}_3$ is a real root of $AE_3$, $(\Lambda^{(3)}_{\mathbf{27}}
|\Lambda^{(3)}_{\mathbf{27}} )=2$, while the lowest weight vectors
for $\mathbf{8}_3$ is timelike,
$(\Lambda^{(3)}_{\mathbf{8}}|\Lambda^{(3)}_{\mathbf{8}})=-4$. This
implies that the entire representation $\mathbf{8}_3$ lies inside the
lightcone of $\mf{h}^{\star}_{\mf{g}}$. Both representations have
outer multiplicity one.

Note that $[0,3]$ and $[3,0]$ are also admissible solutions but
have vanishing outer multiplicities by the same arguments as for
the representation $[0,2]$ at level 2.

\subsubsection[Level $\ell=4$]{Level \boldmath $\ell=4$}

At this level we encounter for the first time a
representation with non-trivial outer multiplicity. It is a
15-dimensional representation with the following Young tableau
structure:
\begin{equation}
  \mathbf{\bar{15}}_4 : [p_2,p_3]=[1,2]
  \quad \Longleftrightarrow \quad
  {\footnotesize
    \setlength{\tabcolsep}{0.55 em}
    \begin{tabular}{ccc}
      \cline{1-3}
      \multicolumn{1}{|c|}{} &
      \multicolumn{1}{c|}{} &
      \multicolumn{1}{c|}{} \\
      \cline{1-3}
      \multicolumn{1}{|c|}{} &
      \multicolumn{1}{c|}{} \\
      \cline{1-2}
    \end{tabular}}\,.
  \label{RepresentationLevel4}
\end{equation}
The lowest weight vector is
\begin{equation}
  \Lambda^{(4)}_{\mathbf{\bar{15}}}=4\al_{1}+4\al_{2}+\al_{3},
  \label{LowestWeightLevel4}
\end{equation}
which is an imaginary root of $AE_3$,
\begin{equation}
  (\Lambda^{(4)}_{\mathbf{\bar{15}}}|\Lambda^{(4)}_{\mathbf{\bar{15}}})=-6.
  \label{NormLowestWeightLevel4}
\end{equation}
From Table~\ref{table:multiplicities} we find that this root has
multiplicity 5 as a root of $AE_3$,
\begin{equation}
  \mult(\Lambda^{(4)}_{\mathbf{\bar{15}}})=5.
  \label{MultiplicityLevel4}
\end{equation}
In order for Equation~(\ref{decompositionsubspaces}) to make sense,
this multiplicity must be matched by the total multiplicity of
$\Lambda^{(4)}_{\mathbf{\bar{15}}}$ as a weight of $\mf{sl}(3,\mbb{R})$
representations at level four. The remaining representations at this
level are 
\begin{equation}
  \begin{array}{rcl}
    \mathbf{24}_4 &:& [3,1]
    \quad \Longleftrightarrow \quad
    {\footnotesize
      \setlength{\tabcolsep}{0.55 em}
      \begin{tabular}{cccc}
        \cline{1-4}
        \multicolumn{1}{|c|}{} &
        \multicolumn{1}{c|}{} &
        \multicolumn{1}{c|}{} &
        \multicolumn{1}{c|}{} \\
        \cline{1-4}
        \multicolumn{1}{|c|}{} \\
        \cline{1-1}
      \end{tabular}}\,,
    \\ [1 em]
    \mathbf{\bar{3}}_4 &:& [0,1]
    \quad \Longleftrightarrow \quad
    {\footnotesize
      \setlength{\tabcolsep}{0.55 em}
      \begin{tabular}{c}
        \cline{1-1}
        \multicolumn{1}{|c|}{} \\
        \cline{1-1}
        \multicolumn{1}{|c|}{} \\
        \cline{1-1}
      \end{tabular}}\,,
    \\ [1 em]
    \mathbf{6}_4 &:& [2,0]
    \quad \Longleftrightarrow \quad
    {\footnotesize
      \setlength{\tabcolsep}{0.55 em}
      \begin{tabular}{cc}
        \cline{1-2}
        \multicolumn{1}{|c|}{} &
        \multicolumn{1}{c|}{} \\
        \cline{1-2}
      \end{tabular}}\,,
    \\ [1 em]
    \mathbf{42}_4 &:& [2,3]
    \quad \Longleftrightarrow \quad
    {\footnotesize
      \setlength{\tabcolsep}{0.55 em}
      \begin{tabular}{cccccc}
        \cline{1-6}
        \multicolumn{1}{|c|}{} &
        \multicolumn{1}{c|}{} &
        \multicolumn{1}{c|}{} &
        \multicolumn{1}{c|}{} &
        \multicolumn{1}{c|}{} &
        \multicolumn{1}{c|}{} \\
        \cline{1-6}
        \multicolumn{1}{|c|}{} &
        \multicolumn{1}{c|}{} &
        \multicolumn{1}{c|}{} \\
        \cline{1-3}
      \end{tabular}}\,.
  \end{array}
  \label{MoreRepresentationsLevel4} 
\end{equation}
By drawing these representations explicitly, one sees that the root
$4\al_{1}+4\al_{2}+\al_{3}$, representing the weight
$\Lambda^{(4)}_{\mathbf{\bar{15}}}$, also appears as a weight (but not as
a lowest weight) in the representations $\mathbf{42}_4$ and
$\mathbf{24}_4$. It occurs with weight multiplicity 1 in the $\mathbf{24}_4$
but with weight multiplicity 2 in the $\mathbf{42}_4$. Taking also into
account the representation $\mathbf{\bar{15}}_4$ in which it is the
lowest weight we find a total weight multiplicity of 4. This implies
that, since in $AE_3$
\begin{equation}
  \mult(4\al_{1}+4\al_{2}+\al_{3})=5,
  \label{RootmultiplicityLevel4}
\end{equation}
the outer multiplicity of $\mathbf{\bar{15}}_4$ must be 2, i.e., 
\begin{equation}
  \mu\left(\Lambda^{(2)}_{\mathbf{\bar{15}}}\right)=2.
  \label{OutermultiplicityLevel4}
\end{equation}
When we go to higher and higher levels, the outer multiplicities of
the representations located entirely inside the lightcone in
$\mf{h}_{\mf{g}}$ increase exponentially.


\subsection[Level decomposition of $E_{10}$]%
           {Level decomposition of \boldmath $E_{10}$}
\label{section:DecompE10E10}

As we have seen, the Kac--Moody algebra \index{Kac--Moody algebra} 
$E_{10}$ is one of the four hyperbolic algebras of maximal rank, the
others being $BE_{10}, DE_{10}$ and $CE_{10}$. It can be constructed
as an overextension \index{overextension} of $E_{8}$ and is therefore
often denoted by $E_{8}^{++}$. Similarly to $E_8$ in the rank 8 case,
$E_{10}$ is the unique indefinite rank 10 algebra with an even
self-dual root lattice, namely the Lorentzian lattice $\Pi_{1,9}$.

Our first encounter with $E_{10}$ in a physical application was in
Section~\ref{section:KMBilliardsI} where we have showed that the
Weyl group of $E_{10}$ describes the chaos that
emerges when studying eleven-dimensional supergravity close to a
spacelike singularity~\cite{ArithmeticalChaos}.

In Section~\ref{section:E10SigmaModel}, we will discuss how to
construct a Lagrangian manifestly invariant under global
$E_{10}$-transformations and compare its dynamics to that of
eleven-dimensional supergravity. The level decomposition 
\index{level decomposition}associated with the removal of the
``exceptional node'' labelled ``10'' in Figure~\ref{figure:E10a} will
be central to the analysis. It turns out that the low-level structure
in this decomposition precisely reproduces the bosonic field content
of eleven-dimensional supergravity~\cite{DHN2}.

Moreover, decomposing $E_{10}$ with respect to different regular
subalgebras reproduces also the bosonic field contents of the Type
IIA and Type IIB supergravities. The fields of the IIA
theory are obtained by decomposition in terms of representations
of the $D_9=\mf{so}(9,9,\mbb{R})$ subalgebra obtained by removing
the first simple root $\al_{1}$~\cite{E10andIIA}. Similarly the
IIB-fields appear at low levels for a decomposition with respect
to the $A_9\oplus
A_{1}=\mf{sl}(9,\mbb{R})\oplus\mf{sl}(2,\mbb{R})$ subalgebra found
upon removal of the second simple root $\al_{2}$~\cite{E10andIIB}. The
extra $A_1$-factor in this decomposition ensures that the
$SL(2,\mbb{R})$-symmetry of IIB supergravity is recovered.

For these reasons, we investigate now these various level
decompositions.


\subsubsection[Decomposition with respect to $\mf{sl}(10,\mbb{R})$]%
              {Decomposition with respect to \boldmath $\mf{sl}(10,\mbb{R})$}

Let $\al_{1},\cdots,\al_{10}$ denote the simple roots of
$E_{10}$ \index{$E_{10}$}and $\al_{1}^{\vee}, \cdots, \al_{10}^{\vee}$ the
Cartan generators. These span the root space $\mf{h}^{\star}$ and
the Cartan subalgebra $\mf{h}$, respectively. Since $E_{10}$ is
simply laced the Cartan matrix \index{Cartan matrix}is given by the scalar products
between the simple roots:
\begin{equation}
  A_{ij}[E_{10}]=(\al_{i}|\al_{j})= \left(
    \begin{array}{@{}r@{\quad}r@{\quad}r@{\quad}r@{\quad}r@{\quad}r@{\quad}r@{\quad}r@{\quad}r@{\quad}r@{}}
      2 & -1 & 0 & 0 & 0 & 0 & 0 & 0 & 0 & 0 \\
      -1 & 2 & -1 & 0 & 0 & 0 & 0 & 0 & 0 & 0 \\
      0 & -1 & 2 & -1 & 0 & 0 & 0 & 0 & 0 & -1 \\
      0 & 0 & -1 & 2 & -1 & 0 & 0 & 0 & 0 & 0 \\
      0 & 0 & 0 & -1 & 2 & -1 & 0 & 0 & 0 & 0 \\
      0 & 0 & 0 & 0 & -1 & 2 & -1 & 0 & 0 & 0 \\
      0 & 0 & 0 & 0 & 0 & -1 & 2 & -1 & 0 & 0 \\
      0 & 0 & 0 & 0 & 0 & 0 & -1 & 2 & -1 & 0 \\
      0 & 0 & 0 & 0 & 0 & 0 & 0 & -1 & 2 & 0 \\
      0 & 0 & -1 & 0 & 0 & 0 & 0 & 0 & 0 & 2 \\
    \end{array}
  \right).
\end{equation}
The associated Dynkin diagram \index{Dynkin diagram}is displayed in
Figure~\ref{figure:E10a}. We will perform the decomposition with
respect to the $\mf{sl}(10,\mbb{R})$ subalgebra represented by the
horizontal line in the Dynkin diagram so the level $\ell$ of an
arbitrary root $\al\in\mf{h}^{\star}$ is given by the coefficient in
front of the exceptional simple root, i.e.,
\begin{equation}
  \ga=\sum_{i=1}^{9}m^{i}\al_{i}+\ell\al_{10}.
  \label{rootE10}
\end{equation}

As before, the weight that is easiest to identify for each
representation $\mc{R}(\Lambda^{(\ell)})$ at positive level $\ell$
is the lowest weight $\Lambda^{(\ell)}_\mathrm{lw}$. We denote by
$\bar{\Lambda}_\mathrm{lw}^{(\ell)}$ the projection onto the spacelike
slice of the root lattice defined by the level $\ell$. The
(conjugate) Dynkin labels $p_1,\cdots, p_9$ characterizing the
representation $\mc{R}(\Lambda^{(\ell)})$ are defined as before as
minus the coefficients in the expansion of
$\bar{\Lambda}_\mathrm{lw}^{(\ell)}$ in terms of the fundamental weights
$\lambda^{i}$ of $\mf{sl}(10,\mbb{R})$:
\begin{equation}
  -\bar{\Lambda}_\mathrm{lw}^{(\ell)}=\sum_{i=1}^{9}p_{i}\lambda^{i}.
  \label{Dynkinlabels}
\end{equation}

\epubtkImage{E10.png}{%
  \begin{figure}[t]
    \centerline{\includegraphics[width=110mm]{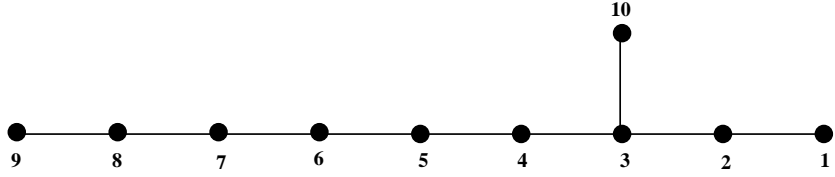}}
    \caption{The Dynkin diagram of $E_{10}$. Labels $i=1,\cdots, 9$
      enumerate the nodes corresponding to simple roots $\al_{i}$ of the
      $\mf{sl}(10,\mathbb{R})$ subalgebra and ``$10$'' labels the
      exceptional node.}
    \label{figure:E10a}
  \end{figure}}

The Killing form on each such slice is positive definite so the
projected weight $\bar{\Lambda}_\mathrm{hw}^{(\ell)}$ is of course real.
The fundamental weights of $\mf{sl}(10,\mbb{R})$ can be computed
explicitly from their definition as the duals of the simple roots:
\begin{equation}
  \lambda^{i}=\sum_{j=1}^{9}B^{ij}\al_{j},
  \label{fundamentalweightsA9}
\end{equation}
where $B^{ij}$ is the inverse of the Cartan matrix of $A_9$,
\begin{equation} \left(B_{ij}[A_{9}]\right)^{-1}= \f{1}{10}\left(
    \begin{array}{@{}r@{\quad}r@{\quad}r@{\quad}r@{\quad}r@{\quad}r@{\quad}r@{\quad}r@{\quad}r@{}}
      9 & 8 & 7 & 6 & 5 & 4 & 3 & 2 & 1 \\
      8 & 16 & 14 & 12 & 10 & 8 & 6 & 4 & 2 \\
      7 & 14 & 21 & 18 & 15 & 12 & 9 & 6 & 3 \\
      6 & 12 & 18 & 24 & 20 & 16 & 12 & 8 & 4 \\
      5 & 10 & 15 & 20 & 25 & 20 & 15 & 10 & 5 \\
      4 & 8 & 12 & 16 & 20 & 24 & 18 & 12 & 6 \\
      3 & 6 & 9 & 12 & 15 & 18 & 21 & 14 & 7 \\
      2 & 4 & 6 & 8 & 10 & 12 & 14 & 16 & 8 \\
      1 & 2 & 3 & 4 & 5 & 6 & 7 & 8 & 9 \\
    \end{array}
  \right).
  \label{CartanMatrixA9}
\end{equation}
Note that all the entries of $B^{ij}$ are positive which will prove to
be important later on. As we saw for the $AE_3$ case we want to find
the possible allowed values for $(m_1,\cdots, m_9)$, or, equivalently,
the possible Dynkin labels $[p_1,\cdots,p_9]$ for each level $\ell$.

The corresponding diophantine equation, Equation~(\ref{DiophantineAE3}),
for $E_{10}$ was found in~\cite{DHN2} and reads
\begin{equation}
  m^{i}=B^{i3}\ell-\sum_{j=1}^{9}B^{ij}p_j \geq 0.
  \label{DiophantineE10}
\end{equation}
Since the two sets $\{p_i\}$ and $\{m^{i}\}$ both consist of
non-negative integers and all entries of $B^{ij}$ are positive, these
equations put strong constraints on the possible representations that
can occur at each level. Moreover, each lowest weight vector
$\Lambda^{(\ell)}=\ga$ must be a root of $E_{10}$, so we have the
additional requirement
\begin{equation}
  (\Lambda^{(\ell)}|\Lambda^{(\ell)})=
  \sum_{i,j=1}^{9}B^{ij}p_i p_j-\f{1}{10}\ell^2 \leq 2. 
  \label{rootconstraintE10}
\end{equation}

The representation content at each level is represented by
$\mf{sl}(10,\mbb{R})$-tensors whose index structure are encoded in
the Dynkin labels \index{Dynkin labels}$[p_1,\cdots,p_9]$. At level $\ell=0$ we have the
adjoint representation of $\mf{sl}(10,\mbb{R})$ represented by the
generators ${K^{a}}_b$ obeying the same commutation relations as
in Equation~(\ref{gl(3)}) but now with $\mf{sl}(10,\mbb{R})$-indices.

All higher (lower) level representations will then be tensors
transforming contravariantly (covariantly) under the level
$\ell=0$ generators. The resulting representations are displayed
up to level~3 in Table~\ref{table:A9reps}. We see that the level~1
and 2 representations have the index structures of a 3-form and a
6-form respectively. In the $E_{10}$-invariant sigma model, to be
constructed in Section~\ref{section:sigmamodels}, these
generators will become associated with the time-dependent physical
``fields'' $A_{abc}(t)$ and $A_{a_1\cdots a_6}(t)$ which are
related to the electric and magnetic component of the 3-form in
eleven-dimensional supergravity. Similarly, the level 3 generator
$E^{a|b_1\cdots b_9}$ with mixed Young symmetry will be associated
to the dual of the spatial part of the eleven-dimensional
vielbein. This field is therefore sometimes referred to as the
``dual graviton''.

\begin{table}
  \caption{The low-level representations in a decomposition of the
    adjoint representation of $E_{10}$ into representations of its
    $A_9$ subalgebra obtained by removing the exceptional node in the
    Dynkin diagram in Figure~\ref{figure:E10a}.}
  \renewcommand{\arraystretch}{1.2}
  \vspace{0.5 em}
  \centering
  \begin{tabular}{l|cccc}
    \toprule
    $\ell$ & $\Lambda^{(\ell)}=[p_1,\cdots, p_9]$ &
    $\Lambda^{(\ell)}=(m_1,\cdots,m_{10})$ &
    $A_9$-representation &
    $E_{10}$-generator \\
    \midrule
    1 &
    $[0,0,1,0,0,0,0,0,0]$ &
    $(0,0,0,0,0,0,0,0,0,1)$ &
    $\mathbf{120}_1$ &
    $E^{abc}$ \\
    2 &
    $[0,0,0,0,0,1,0,0,0]$ &
    $(1,2,3,2,1,0,0,0,0,2)$ &
    $\mathbf{210}_2$ &
    $E^{a_1\cdots a_6}$ \\
    3 &
    $[1,0,0,0,0,0,0,1,0]$ &
    $(1,3,5,4,3,2,1,0,0,3)$ &
    $\mathbf{440}_3$ &
    $E^{a|b_1\cdots b_8}$ \\
    \bottomrule
  \end{tabular}
  \label{table:A9reps}
  \renewcommand{\arraystretch}{1.0}
\end{table}

\subsubsection*{Algebraic structure at low levels}

Let us now describe in a little more detail the
commutation relations between the low-level generators in the level
decomposition of $E_{10}$ (see Table \ref{table:A9reps}). We recover
the Chevalley generators of $A_9$ through the following realisation:
\begin{equation}
  e_{i}={K^{i+1}}_{i},
  \qquad
  f_{i}={K^{i}}_{i+1},
  \qquad
  h_{i}={K^{i+1}}_{i+1}-{K^{i}}_{i}
  \qquad
  (i=1,\cdots , 9),
  \label{A9generators}
\end{equation}
where, as before, the ${K^{i}}_j$'s obey the commutation relations
\begin{equation}
  [{K^{i}}_j,{K^{k}}_l]=\delta^{k}_j {K^{i}}_l-\delta^{i}_l{K^{k}}_j. 
  \label{gl(10)}
\end{equation}
At levels $\pm 1$ we have the positive root generators $E^{abc}$ and
their negative counterparts $F_{abc}=-\tau(E^{abc})$, where $\tau$
denotes the Chevalley involution as defined in
Section~\ref{section:KacMoody}. Their transformation properties under
the $\mf{sl}(10,\mbb{R})$-generators ${K^{a}}_b$ follow from the
index structure and reads explicitly 
\begin{equation}
  \begin{array}{rcl}
    [{K^{a}}_{b},E^{cde}] & = & \displaystyle
    3\delta^{[c}_{b}E^{de]a},
    \\ [0.5 em]
    [{K^{a}}_{b},F_{cde}] & = & \displaystyle
    -3{\delta^{a}}_{[c}F_{de]b},
    \\ [0.25 em]
    [E^{abc},F_{def}] & = & \displaystyle
    18\delta^{[ab}_{[de}{K^{c]}}_{f]}-
    2\delta^{abc}_{def}\sum_{a=1}^{10}{K^{a}}_{a},
  \end{array}
  \label{level1commutationrelations}
\end{equation}
where we defined
\begin{equation}
  \begin{array}{rcl}
    \delta^{ab}_{cd}&=& \displaystyle
    \f{1}{2}(\de^{a}_{c}\de^{b}_{d}-\de^{b}_{c}\de^{a}_{d})
    \\ [0.75 em]
    \delta^{abc}_{def}&=& \displaystyle
    \f{1}{3!}(\delta^{a}_{d}\de^{b}_{e}\de^{c}_{f}\pm 5
    \mathrm{\ permutations}).
  \end{array}
  \label{deltadefinitions} 
\end{equation}
The ``exceptional'' generators $e_{10}$ and $f_{10}$ are fixed by
Equation~(\ref{A9generators}) to have the following realisation:
\begin{equation}
  e_{10}=E^{123},
  \qquad
  f_{10}=F_{123}.
  \label{realisationexceptionalgenerator}
\end{equation}
The corresponding Cartan generator is obtained by requiring
$[e_{10},f_{10}]=h_{10}$ and upon inspection of the last equation in
Equation~(\ref{level1commutationrelations}) we find
\begin{equation}
  h_{10}=-\f{1}{3} \!\! \sum_{i\neq 1,2,3} \!\!
  {K^{a}}_{a}+\f{2}{3}({K^{1}}_{1}+{K^{2}}_{2}+{K^{3}}_{3}),
  \label{ExceptionalCartanGenerator}
\end{equation}
enlarging $\mf{sl}(10,\mbb{R})$ to $\mf{gl}(10,\mbb{R})$.

The bilinear form at level zero is
\begin{equation}
  ({K^{i}}_j|{K^{k}}_l)=\delta^{i}_l\delta^{k}_j-\delta^{i}_j\delta^{k}_l
  \label{KillingformlevelzeroForE10}
\end{equation}
and can be extended level by level to the full algebra by using its
invariance, $([x,y]|z)=(x|[y,z])\,$ for $ x,y,z\in E_{10}$ (see
Section~\ref{section:KacMoody}). For level~1 this yields
\begin{equation}
  \left(E^{abc}|F_{def}\right) = 3! \de^{abc}_{def}, 
  \label{BilinearformLevel1}
\end{equation}
where the normalization was chosen such that
\begin{equation}
  \left(e_{10}|f_{10}\right)=\left(E^{123}|F_{123}\right)=1.
  \label{BilinearformExceptionalGenerators}
\end{equation}

Now, by using the graded structure of the level decomposition we
can infer that the level~2 generators can be obtained by commuting
the level~1 generators
\begin{equation}
  [\mf{g}_1,\mf{g}_1]\subseteq \mf{g}_2.
  \label{GradingLevel2}
\end{equation}
Concretely, this means that the level~2 content should be found from
the commutator
\begin{equation}
  [E^{a_1a_2a_3},E^{a_4a_5a_6}].
\end{equation}
We already know that the only representation at this level is $\mathbf{210}_2$, realized by an
antisymmetric 6-form. Since the normalization of this generator is
arbitrary we can choose it to have weight one and hence we find
\begin{equation}
  E^{a_1\cdots a_6}=[E^{a_1a_2a_3},E^{a_4a_5a_6}]. 
\end{equation}
The bilinear form is lifted to level 2 in a similar way as before
with the result
\begin{equation}
  \left(E^{a_1\cdots a_6}|F_{b_1\cdots b_6}\right)=
  6! \delta^{a_1\cdots a_6}_{b_1\cdots b_6}.
  \label{BilinearFormLevel2}
\end{equation}
Continuing these arguments, the level~3-generators can be obtained
from
\begin{equation}
  [[\mf{g}_1,\mf{g}_1],\mf{g}_1]\subseteq \mf{g}_3.
  \label{GradingLevel3}
\end{equation}
From the index structure one would expect to find a 9-form generator
$E^{a_1\cdots a_9}$ corresponding to the Dynkin labels
$[0,0,0,0,0,0,0,0,1]$. However, we see from Table~\ref{table:A9reps}
that only the representation $[1,0,0,0,0,0,0,1,0]$ appears at
level~3. The reason for the disappearance of the representation
$[0,0,0,0,0,0,0,0,1]$ is because the generator $E^{a_1\cdots a_9}$ is
not allowed by the Jacobi identity. A detailed explanation for this
can be found in~\cite{Fischbacher:2005fy}. The right hand side of
Equation~(\ref{GradingLevel3}) therefore only contains the index
structure compatible with the generators $E^{a|b_1\cdots b_8}$,
\begin{equation}
  [[E^{ab_1b_2},E^{b_3b_4b_5}],E^{b_6b_7b_8}]=-E^{[a|b_1b_2]b_3\cdots b_8}, 
  \label{Level3Generator}
\end{equation}
where the minus sign is purely conventional.

For later reference, we list here some additional commutators that
are useful~\cite{DamourNicolaiLowLevels}: 
\begin{equation}
  \begin{array}{rcl}
    [E^{a_1\cdots a_6}, F_{b_1b_2b_3}] &=& \displaystyle
    -5!\delta^{[a_1a_2a_3}_{b_1b_2b_3}E^{a_1a_2a_3]},
    \\ [0.5 em]
    [E^{a_1\cdots a_6}, F_{b_1\cdots b_6}]&=& \displaystyle
    6\cdot 6!\delta^{[a_1\cdots a_5}_{[b_1\cdots b_5}
    {K^{a_6]}}_{b_6]}-\f{2}{3}\cdot 6!\delta^{a_1\cdots
    a_6}_{b_1\cdots b_6} \sum_{a=1}^{10}{K^{a}}_a,
    \\ [1 em]
    [E^{a_1|a_2\cdots a_9}, F_{b_1b_2b_3}]&=& \displaystyle
    -7\cdot 48 \left(\delta^{a_1[a_2a_3}_{b_1b_2b_3}E^{a_4\cdots
    a_9]}-\delta^{[a_2a_3a_4}_{b_1b_2b_3}E^{a_5\cdots a_9]a_1}\right),
    \\ [1 em]
    [E^{a_1|a_2\cdots a_9}, F_{b_1\cdots b_6}] &=& \displaystyle
    -8! \left(\delta^{a_1[a_2\cdots a_6}_{b_1\cdots b_6}E^{a_7a_8a_9]}-
    \delta^{[a_2\cdots a_7}_{b_1\cdots b_6}E^{a_8a_9]a_1}\right).
  \end{array}
\end{equation}


\subsubsection{``Gradient representations''}

So far, we have only discussed the representations occurring at the
first four levels in the $E_{10}$ \index{$E_{10}$}decomposition. This is due to
the fact that a physical interpretation of higher level fields is
yet to be found. There are, however, among the infinite number of
representations, a subset of three (infinite) towers of
representations with certain appealing properties. These are the
``gradient representations'', so named due to their conjectured
relation to the emergence of space, through a Taylor-like
expansion in spatial gradients~\cite{DHN2}. We explain here how
these representations arise and we emphasize some of their
important properties, leaving a discussion of the physical
interpretation to Section~\ref{section:sigmamodels}.

The gradient representations are obtained by searching for
``affine representations'', for which the coefficient $m^{9}$ in
front of the overextended simple root of $E_{10}$ vanishes, i.e.,
the lowest weights of the representations correspond to the
following subset of $E_{10}$ roots,
\begin{equation}
  \ga=\sum_{i=1}^{8}m^{i}\al_i+\ell \al_{10}.
\end{equation}

The Dynkin labels allowed by this restricting are parametrized by
an integer $k$ which is related to the level at which a specific
representation occurs in the following way:
\begin{eqnarray}
  \ell &=&3k+1
  \qquad
  [0,0,1,0,0,0,0,0,k],
  \\
  \ell &=&3k+2
  \qquad
  [0,0,0,0,0,1,0,0,k],
  \\
  \ell &=&3k+3
  \qquad
  [1,0,0,0,0,0,0,1,k].
\end{eqnarray} 
One easily verifies that these representations
fulfill the diophantine constraints~(\ref{DiophantineE10}) and
the lowest weight has length squared,
Equation~(\ref{rootconstraintE10}), equal to 2 and is thus indeed a real
root of $E_{10}$. For $k=0$ these representations reduce to the
ones for $\ell=1, 2$ and $3$, and hence the gradient
representations correspond to generalizations of these standard
low-level structures. The corresponding generators have,
respectively at levels $\ell=3k+1, 3k+2$ or $3k+3$, additional
sets of $k$ ``9-tuples'' of antisymmetric indices, 
\begin{equation}
  \begin{array}{l}
    [0,0,1,0,0,0,0,0,k]
    \quad \Longrightarrow \quad
    E^{a_1\cdots a_9, b_1\cdots b_9, \cdots, c_1c_2c_3},
    \\ [0.25 em]
    [0,0,0,0,0,1,0,0,k] 
    \quad \Longrightarrow \quad
    E^{a_1\cdots a_9, b_1\cdots b_9, \cdots, c_1\cdots c_6},
    \\ [0.25 em]
    [1,0,0,0,0,0,0,1,k] 
    \quad \Longrightarrow \quad
    E^{a_1\cdots a_9, b_1\cdots b_9, \cdots, |c| d_1\cdots d_8}
  \end{array}
\end{equation}
(with the irreducibility conditions expressing that
antisymmetrizations involving one more index over the explicit
antisymmetry are zero). Since these are $\mf{sl}(10,
\mbb{R})$-representations we can use the rank~10 antisymmetric epsilon
tensor $\epsilon_{a_1\cdots a_{10}}$ to dualize these representations,
for instance for the $\ell=3k+1$ tower we get
\begin{equation}
  {E^{a_1a_2a_3}}_{b_1\cdots b_k}=
  \epsilon_{b_1c_1\cdots c_9}
  \epsilon_{b_2 d_1\cdots d_9} \cdots
  \epsilon_{b_k e_1\cdots e_9}
  E^{c_1\cdots c_9, d_1\cdots d_9, \cdots, e_1\cdots e_9, a_1a_2a_3},
\end{equation}
where the lower indices $b_1\cdots b_k$ are now completely
symmetric and furthermore, obey appropriate tracelessness
conditions when contracted with an upper index.

Thus, in this way we obtain the three infinite towers of $E_{10}$
generators
\begin{equation}
  {E^{a_1a_2a_3}}_{b_1\cdots b_k},
  \qquad
  {E^{a_1\cdots a_6}}_{b_1\cdots b_k},
  \qquad
  {E^{a_1|a_2\cdots a_9}}_{b_1\cdots b_k}.
\end{equation}
The lowest weight vectors of these representations are all spacelike
and so these representations always come with outer multiplicity one.

The existence of these towers of representations is not special
for $E_{10}$ among the exceptional algebras, although the
symmetric Young structure of the lower indices is actually a very
special and important feature of $E_{10}$. In
Section~\ref{section:sigmamodels} we will discuss the tantalizing
possibility that these representations encode an infinite set of
spatial gradients that describe the emergence, or ``unfolding'', of
space.

To illustrate the difference from other exceptional algebras, we
consider, for instance, a similar search for affine
representations within $E_{11}$ (see,
e.g.~\cite{NicolaiFischbacher}). The same sets of 9-tuples appear,
but now these should be dualized with the rank~11 epsilon tensor of
$\mf{sl}(11, \mbb{R})$, leaving us with three towers of generators
that have $k$ \emph{pairs of antisymmetric indices}, i.e.,
\begin{equation}
  {E^{\mu_1\mu_2\mu_3}}_{[\nu_1\rho_1]\cdots [\nu_k\rho_k]},
  \qquad
  {E^{\mu_1\cdots \mu_6}}_{[\nu_1\rho_1]\cdots [\nu_k\rho_k]},
  \qquad
  {E^{\mu_1|\si_2\cdots \si_9}}_{[\nu_1\rho_1]\cdots [\nu_k\rho_k]},
\end{equation}
where all indices are $\mf{sl}(11, \mbb{R})$-indices and so
run from 1 to 11. No interpretation in terms of spatial gradients
exist for these generators. Note, however, that these representations
have recently been interpreted as dual to
scalars~\cite{Riccioni:2006az}.

Finally, we note that because all these representations were found
by setting $m^{9}=0$, we are really dealing with representations
that also exist within $E_9$, in the sense that when restricting
all indices to $\mf{sl}(9, \mbb{R})$-indices, these generators can
be found in a level decomposition of $E_9$ with respect to its
$\mf{sl}(9, \mbb{R})$-subalgebra. However, it is important to note
that in $E_{10}$ and $E_{11}$ the affine representations
constitute merely a small subset of all representations occurring
in the level decomposition, while in $E_9$ they are actually the
only ones and so they provide (together with their transposed
partners) the full structure of the algebra. Moreover, in $E_9$
the epsilon tensor is of rank~9 so all the 9-tuples of
antisymmetric indices are ``swallowed'' by the epsilon tensor.
This reflects the fact that for affine algebras the level
decomposition corresponds to an infinite repetition of the
low-level representations.


\subsubsection[Decomposition with respect to $\mf{so}(9,9)$ and
  $\mf{sl}(2,\mbb{R})\oplus\mf{sl}(9,\mbb{R})$]%
              {Decomposition with respect to \boldmath $\mf{so}(9,9)$ and
  $\mf{sl}(2,\mbb{R})\oplus\mf{sl}(9,\mbb{R})$}

A level decomposition can be performed with respect to any of the
regular subalgebras encoded in the Dynkin diagram. We mention here
two additional cases which are specifically interesting for our
purposes, since they give rise to low-level field contents that
coincide with the bosonic spectrum of Type IIA and IIB
supergravity. The relevant decompositions are the following: 
\begin{equation}
  \begin{array}{l}
    \mbox{IIA}
    \quad \Longleftrightarrow \quad
    \mf{so}(9,9) \subset E_{10},
    \\ [0.25 em]
    \mbox{IIB}
    \quad \Longleftrightarrow \quad
    \mf{sl}(2,\mbb{R})\oplus\mf{sl}(9,\mbb{R}) \subset E_{10}.
  \end{array}
  \label{IIAandIIBdecompositions}
\end{equation}
The corresponding levels are defined as
\begin{equation}
  \begin{array}{rcl}
    \mf{so}(9,9) & : & \displaystyle
    \ga=\ell\al_{1}+\sum_{i=2}^{10}m^{i}\al_{i} \in \mf{h}^{\star},
    \\ [0.25 em]
    \mf{sl}(2,\mbb{R})\oplus\mf{sl}(9,\mbb{R}) & : & \displaystyle
    \ga=m^{1}\al_{1}+\ell\al_{2}+\sum_{j=3}^{10}m^{j}\al_{j} \in \mf{h}^{\star}. 
  \end{array}
  \label{IIAandIIBlevels} 
\end{equation}
It turns out that in the $\mf{so}(9,9)$ decomposition the even levels
correspond to vectorial representations of $\mf{so}(9,9)$ while the
odd levels give spinorial representations. This implies that the
fields in the NS-NS sector arise at even levels and the R-R fields
correspond to odd level representations of $\mf{so}(9,9)$.

On the contrary, in the $\mf{sl}(2,\mbb{R})\oplus\mf{sl}(9,\mbb{R})$
decomposition the additional factor of $\mf{sl}(2,\mbb{R})$ causes
mixing between the R-R and NS-NS fields at each level. This is to be
expected since we know that for example the fundamental string (F1)
and the D1-brane couples to the NS-NS 2-form $B_{2}$ and the R-R
2-form $C_2$, respectively, which transform as a doublet under the
$SL(2,\mbb{R})$-symmetry of Type~IIB supergravity.

In the $\mf{sl}(2,\mbb{R})\oplus\mf{sl}(9,\mbb{R})$ the level
$\ell=0$ content is of course just the adjoint representation in
the same way as in the $\mf{sl}(10,\mbb{R})$ decomposition
considered above. In the other case instead we find the adjoint
representation $M^{ab}=-M^{ba}$ of $\mf{so}(9,9)$ with commutation
relations
\begin{equation}
  [M^{ab},M^{cd}]=\eta^{ca}M^{bd}-\eta^{cb}M^{ad}-\eta^{da}M^{bc}+\eta^{db}M^{ac},
  \label{adjointSO(9,9)}
\end{equation}
where $\eta^{ab}$ is the split diagonal metric with $(9,9)$-signature.

The procedure follows a similar structure as for the previous
cases so we will not give the details here. We refer the
interested reader to~\cite{E10andIIA, E10andIIB} for a
detailed account. The result of the decompositions up to level~3
for the two cases discussed here is displayed in
Tables~\ref{table:IIAdec} and~\ref{table:IIBdec}.

\begin{table}
  \caption{The low-level representations in a decomposition of the
    adjoint representation of $E_{10}$ into representations of its
    $\mf{so}(9,9)$ subalgebra obtained by removing the first node in the
    Dynkin diagram in Figure~\ref{figure:E10a}. Note that the lower
    indices at levels~1 and 3 are spinor indices of $\mf{so}(9,9)$.}
  \renewcommand{\arraystretch}{1.2}
  \vspace{0.5 em}
  \centering
  \begin{tabular}{l|cc}
    \toprule
    $\ell$ & $\Lambda^{(\ell)}=[p_1,\cdots, p_9] $ & $E_{10}$-generator \\
    \midrule
    1 & $[0,0,0,0,0,0,0,1,0]$ &$E_{\al}$ \\
    2 & $[0,0,1,0,0,0,0,0,0]$ & $E^{a_1a_2a_3}$ \\
    3 & $[1,0,0,0,0,0,0,1,0]$ & ${E^{a}}_\be$ \\
    \bottomrule
  \end{tabular}
  \label{table:IIAdec}
  \renewcommand{\arraystretch}{1.0}
\end{table}

\begin{table}
  \caption{The low-level representations in a decomposition of the
    adjoint representation of $E_{10}$ into representations of its
    $\mf{sl}(2,\mbb{R})\oplus\mf{sl}(9,\mbb{R})$ subalgebra obtained by
    removing the second node in the Dynkin diagram in
    Figure~\ref{figure:E10a}. The index $\al$ at levels~1 and 3
    corresponds to the fundamental representation of
    $\mf{sl}(2,\mbb{R})$.}
  \renewcommand{\arraystretch}{1.2}
  \vspace{0.5 em}
  \centering
  \begin{tabular}{l|cc}
    \toprule
    $\ell$ & $\Lambda^{(\ell)}=[p_1,\cdots, p_9]\otimes \mc{R}[{A_1}]$  & $E_{10}$-generator \\
    \midrule
    1 & $[0,0,0,0,0,0,0,1,0]\otimes\mathbf{2}$ &${E^{ab}}_{\al}$ \\
    2 & $[0,0,0,0,0,1,0,0,0]\otimes\mathbf{1}$ & $E^{a_1\cdots a_4}$ \\
    3 & $[1,0,0,0,0,0,0,1,0]\otimes \mathbf{2}$ & ${E^{a_1\cdots a_6}}_{\al}$ \\
    \bottomrule
  \end{tabular}
  \label{table:IIBdec}
  \renewcommand{\arraystretch}{1.0}
\end{table}

\clearpage


\section{Hidden Symmetries Made Manifest -- Infinite-Dimensional Coset Spaces}
\label{section:sigmamodels}
\setcounter{equation}{0}

As we have indicated above, the emergence of hyperbolic Coxeter groups
in the BKL-limit \index{BKL-limit}has been argued to be the tip of an iceberg signaling
the existence of a huge number of symmetries underlying gravitational
theories. However, while the appearance of hyperbolic Coxeter groups
is a solid fact that will in our opinion survive future developments,
the exact way in which the conjectured infinite-dimensional symmetry
acts is still a matter of debate and research.

The aim of this section is to describe one line of investigation for
making the infinite-dimensional symmetry manifest. This approach is
directly inspired by the results obtained through toroidal dimensional
reduction of gravitational theories, where the scalar fields form
coset manifolds exhibiting explicitly larger and larger symmetries as
one goes down in dimensions. In the case of eleven-dimensional
supergravity, reduction on an $n$-torus $T^{n}$ reveals a chain of
exceptional U-duality symmetries
$\mc{E}_{n(n)}$~\cite{Cremmer:1978ds, Cremmer:1979up}, culminating
with $\mc{E}_{8(8)}$ in three dimensions~\cite{Marcus:1983hb}. This
has lead to the conjecture~\cite{Julia:1980gr} that the chain of
enhanced symmetries should in fact remain unbroken and give rise to
the infinite-dimensional duality groups $\mc{E}_{9(9)},
\mc{E}_{10(10)}$ and $\mc{E}_{11(11)}$, as one reduces the theory to
two, one and zero dimensions, respectively.

The connection between the symmetry groups controlling the billiards
in the BKL-limit, and the symmetry groups appearing in toroidal
dimensional reduction to three dimensions, where coset spaces play a
central role, has led to the attempt to reformulate eleven-dimensional
supergravity as a $(1+0)$-dimensional nonlinear sigma model based on
the infinite-dimensional coset manifold
$\mc{E}_{10(10)}/\mc{K}(\mc{E}_{10(10)})$~\cite{DHN2}. This sigma
model describes the geodesic flow of a particle moving on
$\mc{E}_{10(10)}/\mc{K}(\mc{E}_{10(10)})$, whose dynamics can be seen
to match the dynamics of the associated (suitably truncated)
supergravity theory. Another, related, source of inspiration for the
idea pushed forward in~\cite{DHN2} has been the earlier proposal to
reformulate eleven-dimensional supergravity as a nonlinear realisation
of the even bigger symmetry $\mc{E}_{11(11)}$~\cite{E11andMtheory},
containing $\mc{E}_{10(10)}$ as a subgroup.

A central tool in the analysis of~\cite{DHN2} is the level
decomposition studied in
Section~\ref{section:LevelDecomposition}. Although proposed some time
ago and crowned with partial successes at low levels, the attempt to
reformulate eleven-dimensional supergravity as an infinite-dimensional
nonlinear sigma model, faces obstacles that have not yet been
overcome at higher levels. This indicates that novel ideas are needed
in order to make further progress towards a complete understanding of
the role of infinite-dimensional symmetry groups in gravitational
theories.

We begin by describing some general aspects of nonlinear sigma models
for finite-dimensional coset spaces. We then explain how to generalize
the construction to the infinite-dimensional case. We finally apply
the construction in detail to the case of eleven-dimensional
supergravity where the conjectured symmetry group is
$\mc{E}_{10(10)}$. This is one of the most extensively investigated
models in the literature in view of its connection with M-theory. The
techniques presented, however, can be applied to all gravitational
models exhibiting the $\mc{U}_3$-duality symmetries discussed in
Sections~\ref{section:KMBilliardsI} and~\ref{section:KMBilliardsII}.


\subsection{Nonlinear sigma models on finite-dimensional coset spaces}
\label{section:nonlinearsigmamodels}
\index{coset space}

A nonlinear sigma model  \index{nonlinear sigma model|bb} describes
maps $\xi$ from one Riemannian
space $X$, equipped with a metric $\ga$, to another
Riemannian space, the ``target space'' $M$, with metric $g$.
Let $x^{m}\, (m=1, \cdots, p=\dim X)$ be coordinates on $X$ and
$\xi^{\al} \, (\al=1, \cdots, q=\dim M)$ be coordinates on $M$.
Then the standard action for this sigma model is
\begin{equation}
  S=\int_{X} d^{p}x \, \sqrt{\ga}\, \ga^{mn}(x)\,
  \pa_m \xi^{\al}(x)\, \pa_n \xi^{\be}(x)\, g_{\al\be}\left(\xi(x)\right). 
  \label{sigmamodel}
\end{equation}
Solutions to the equations of motion resulting from this action
will describe the maps $\xi^{\al}$ as functions of $x^{m}$.

A familiar example, of direct interest to the analysis below, is
the case where $X$ is one-dimensional, parametrized by the
coordinate $t$. Then the action for the sigma model reduces to
\begin{equation}
  S_{\mathrm{geodesic}}=\int dt \, A \,
  \f{d\xi^{\al}(t)}{dt}\f{d\xi^{\be}(t)}{dt}
  g_{\al\be}\left(\xi(t)\right),
\end{equation}
where $ A$ is $\gamma^{11}\sqrt{\ga}$ and ensures reparametrization
invariance in the variable $t$. Extremization with respect to $A$
enforces the constraint
\begin{equation} 
  \f{d\xi^{\al}(t)}{dt}\f{d\xi^{\be}(t)}{dt} g_{\al\be}\left(\xi(t)\right)= 0,
\end{equation}
ensuring that solutions to this model are null geodesics
on $M$. We have already encountered such a sigma model before,
namely as describing the free lightlike motion of the billiard
ball in the $(\dim M-1)$-dimensional scale-factor space. \index{scale-factor space} In that
case $A$ corresponds to the inverse ``lapse-function'' $N^{-1}$
and the metric $ g_{\al\be}$ is a constant Lorentzian metric.


\subsubsection{The Cartan involution and symmetric spaces}
\label{section:Involutive}

In what follows, we shall be concerned with sigma models on
symmetric spaces $\mc{G}/\mc{K}(\mc{G})$ where $\mc{G}$ is a Lie
group with semi-simple real Lie algebra $\cG$ and $\mc{K}(\mc{G})$
its maximal compact subgroup \index{maximal compact subgroup|bb} with
real Lie algebra $\cK$, corresponding to the maximal compact
subalgebra \index{maximal compact subalgebra|bb} of $\cG$. Since
elements of the coset are obtained by factoring out
$\mc{K}(\mc{G})$, this subgroup is referred to as the ``local
gauge symmetry group'' (see below). Our aim is to provide an
algebraic construction of the metric on the coset and of the
Lagrangian.

We have investigated real forms in
Section~\ref{section:FiniteRealLieAlgebras} and have found that the
Cartan involution \index{Cartan involution}$\theta$ induces a Cartan decomposition of $\cG$
into even and odd eigenspaces:
\begin{equation}
  \mf{g}=\mf{k}\oplus\mf{p}
  \label{CartanDecomposition}
\end{equation}
(direct sum of vector spaces), where 
\begin{equation}
  \begin{array}{rcl}
    \mf{k}&=& \{ x\in\mf{g}\, |\, \theta(x)=x \},
    \\ [0.25 em]
    \mf{p}&=& \{ y\in \mf{g}\, |\, \theta(y)=-y\}
  \end{array}
\end{equation}
play central roles. The decomposition~(\ref{CartanDecomposition}) is
orthogonal, in the sense that
$\mf{p}$ is the orthogonal complement of $\mf{k}$ with respect to the
invariant bilinear form $(\cdot |\cdot ) \equiv B(\cdot ,\cdot )$,
\begin{equation}
  \mf{p}=\{y\in \mf{g}\, |\,\forall x\in \mf{k} : (y|x)=0 \}.
\end{equation}
The commutator relations split in a way characteristic for symmetric
spaces,
\begin{equation}
  [\mf{k}, \mf{k}]\subset\mf{k},
  \qquad
  [\mf{k}, \mf{p}]\subset\mf{p},
  \qquad
  [\mf{p}, \mf{p}]\subset \mf{k}.
\end{equation}
The subspace $\mf{p}$ is not a subalgebra. Elements of $\mf{p}$
transform in some representation of $\mf{k}$, which depends on the Lie
algebra $\mf{g}$. We stress that if the commutator $[\mf{p}, \mf{p}]$
had also contained elements in $\mf{p}$ itself, this would not have
been a symmetric space.

The coset space $\mc{G}/\mc{K}(\mc{G})$  \index{coset space|bb} is defined as the set of
equivalence classes $[g]$ of $\mc{G}$ defined by the equivalence
relation 
\begin{equation}
  g\sim g^{\p}
  \qquad
  \mbox{iff } g {g^{\p}}^{-1}\in \mck \mbox{ and } g, g^{\p}\in \mcg, 
\end{equation}
i.e.,
\begin{equation}
  [g]=\{k g \, | \, \forall k\in \mck\}.
\end{equation}

\subsubsection*{Example: The coset space \boldmath $SL(n,\mbb{R})/SO(n)$}

As an example to illustrate the Cartan involution we
consider the coset space  \index{coset space} $SL(n,\mbb{R})/SO(n)$. The group $SL(n,
\mbb{R})$ contains all $n\times n$ real matrices with determinant
equal to one. The associated Lie algebra $\mf{sl}(n,\mbb{R})$ thus
consists of real $n\times n$ traceless matrices. In this case the
Cartan involution is simply (minus) the ordinary matrix transpose
$(\, )^{T}$ on the Lie algebra elements:
\begin{equation}
  \tau : a
  \quad \longmapsto \quad
  -a^{T}\qquad a\in\mf{sl}(n,\mbb{R}).
\end{equation}
This implies that all antisymmetric traceless $n\times n$ matrices
belong to $\mf{k}=\mf{so}(n)$. The Cartan involution $\theta$ is the
differential at the identity of an involution $\Theta$ defined on the
group itself, such that for real Lie groups (real or complex matrix
groups), $\theta$ is just the inverse conjugate transpose. Defining 
\begin{equation}
  \mc{K}(\mc{G})=\{ g \in\mc{G} \, |\, \Theta g=g \}
\end{equation}
then gives in this example the group $\mc{K}(\mc{G})=SO(n)$. The
Cartan decomposition of $\mf{sl}(n, \mbb{R})$ thus splits all elements
into symmetric and antisymmetric matrices, i.e., for $a\in \mf{sl}(n,
\mbb{R})$ we have 
\begin{equation}
  \begin{array}{rcl}
    a-a^{T} &\in& \mf{so}(n),
    \\ [0.25 em]
    a+a^{T} &\in& \mf{p}.
  \end{array}
\end{equation}


\subsubsection{Nonlinear realisations}
\index{nonlinear realisation|bb}

The group $\mc{G}$ naturally acts through (here, right)
multiplication on the quotient space $\mc{G}/\mc{K}(\mc{G})$\epubtkFootnote{Strictly speaking, the coset space  \index{coset space} defined in this way should be written as $\mc{K}(\mc{G})\backslash \mc{G}$. However, we follow what has become common practice in the literature and denote it by $\mc{G}/\mc{K}(\mc{G})$.} as
\begin{equation}
  [h] \mapsto [h g].
  \label{rigidsymmetry}
\end{equation}
This definition 
makes sense because if $h \sim h'$, i.e., $h' = k h$ for some $k \in
\mc{K}(\mc{G})$, then $h'g \sim hg$ since $h'g = (kh)g = k(hg)$
(left and right multiplications commute).

In order to describe a dynamical theory on the quotient space
$\mc{G}/\mc{K}(\mc{G})$, it is convenient to introduce as
dynamical variable the group element $\mathrm{V}(x)\in \mc{G}$ and
to construct the action for $\mathrm{V}(x)$ in such a way that the equivalence relation
\begin{equation}
  \forall k(x)\in \mck : \mathrm{V}(x) \sim k(x) \mathrm{V}(x)
  \label{gaugesymmetryK}
\end{equation}
corresponds to a gauge symmetry. The physical (gauge invariant)
degrees of freedom are then parametrized indeed by points of the coset
space. We also want to impose Equation~(\ref{rigidsymmetry}) as a rigid
symmetry. Thus, the action should be invariant under
\begin{equation}
  \mathrm{V} (x)
  \quad \longmapsto \quad
  k(x)\mathrm{V}(x) g,
  \qquad k(x)\in \mc{K}(\mc{G}), \, g\in\mc{G}. 
\label{LinearRealization}
\end{equation}

One may develop the formalism without fixing the
$\mc{K}(\mc{G})$-gauge symmetry, or one may instead fix the gauge
symmetry by choosing a specific coset representative $\mathrm{V} (x)\in
\mcgk$. When $\mck$ is a maximal compact subgroup \index{maximal compact subgroup} of $\mcg$ there are
no topological obstructions, and a standard choice, which is always
available, is to take $\mathrm{V} (x)$ to be of upper triangular form as
allowed by the Iwasawa decomposition. \index{Iwasawa decomposition} This is usually called the
\emph{Borel gauge} and will be discussed in more detail later. In this
case an arbitrary global transformation,
\begin{equation}
  \mathrm{V} (x)
  \quad \longmapsto \quad
  \mathrm{V} (x)^{\p}=\mathrm{V} (x) g, \qquad g\in\mcg,
\end{equation}
will destroy the gauge choice because $\mathrm{V} ^{\p}(x)$
will generically not be of upper triangular form. Then, a
compensating local $\mck$-transformation is needed that restores
the original gauge choice. The total transformation is thus
\begin{equation}
  \mathrm{V}(x)
  \quad \longmapsto \quad
  \mathrm{V}  (x)^{\p\p}=
  k\left(\mathrm{V}  (x), g \right) \mathrm{V} (x) g,
  \qquad
  k\left(\mathrm{V}(x), g \right)\in \mck,\, g\in\mcg,
\end{equation}
where $\mathrm{V}^{\pp}(x)$ is again in the upper triangular
gauge. Because now $k\left(\mathrm{V} (x), g \right)$ depends nonlinearly
on $\mathrm{V} (x)$, this is called a \emph{nonlinear realisation}
\index{nonlinear realisation} of $\mc{G}$.


\subsubsection[Three ways of writing the quadratic
  $\mck_{\mathrm{local}}\times \mcg_{\mathrm{rigid}}$-invariant action]%
              {Three ways of writing the quadratic \boldmath
  $\mck_{\mathrm{local}}\times \mcg_{\mathrm{rigid}}$-invariant action}

Given the field $\mathrm{V}(x)$, we can form the Lie algebra valued
one-form (Maurer--Cartan form) \index{Maurer--Cartan form|bb} 
\begin{equation}
  d\mathrm{V} (x) \, \mathrm{V} (x)^{-1} =
  dx^{\mu} \, \pa_{\mu}\mathrm{V} (x)\mathrm{V} (x)^{-1}. 
\end{equation}
Under the Cartan decomposition, this element
splits according to Equation~(\ref{CartanDecomposition}),
\begin{equation}
  \pa_{\mu}\mathrm{V} (x) \, \mathrm{V} (x)^{-1} =
  \mathrm{Q}_{\mu}(x)+\mathrm{P}_{\mu} (x),
\end{equation}
where $\mathrm{Q}_{\mu}(x)\in \mf{k}$ and
$\mathrm{P}_{\mu}(x)\in\mf{p}$. We can use the Cartan involution $\theta$ to
write these explicitly as projections onto the odd and even
eigenspaces as follows: 
\begin{equation}
  \begin{array}{rcl}
    \mathrm{Q}_{\mu}(x)&=& \displaystyle
    \f{1}{2}\left[\pa_{\mu}\mathrm{V} (x)\mathrm{V} (x)^{-1}+
    \theta\left(\pa_{\mu}\mathrm{V} (x)\mathrm{V} (x)^{-1}\right)\right] \in \mf{k},
    \\ [0.75 em]
    \mathrm{P}_{\mu}(x)&=& \displaystyle
    \f{1}{2}\left[\pa_{\mu}\mathrm{V} (x)\mathrm{V} (x)^{-1}-
    \theta\left(\pa_{\mu}\mathrm{V} (x)\mathrm{V} (x)^{-1}\right)\right] \in \mf{p}. 
  \end{array}
  \label{CartanProjections}
\end{equation}
If we define a \emph{generalized transpose} $\mc{T}$ by
\begin{equation}
  (\, )^{\mc{T}}\equiv -\theta (\, ),
\end{equation}
then $\mathrm{P}_{\mu}(x)$ and $\mathrm{Q}_{\mu}(x)$ correspond to
symmetric and antisymmetric elements, respectively,
\begin{equation}
  \mathrm{P}_{\mu}(x)^{\mc{T}}=\mathrm{P} _{\mu}(x),
  \qquad
  \mathrm{Q}_{\mu}(x)^{\mc{T}}=-\mathrm{Q}_{\mu}(x).
\end{equation}
Of course, in the special case when $\mf{g}=\mf{sl}(n, \mbb{R})$ and
$\mf{k}=\mf{so}(n)$, the generalized transpose $(\, )^{\mc{T}}$
coincides with the ordinary matrix transpose $(\, )^{T}$. The Lie
algebra valued  one-forms with components $\pa_{\mu}\mathrm{V}
(x)\mathrm{V} (x)^{-1}$, $\mathrm{Q}_{\mu}(x)$ and $\mathrm{P} _{\mu}(x)$
are invariant under rigid right multiplication, $\mathrm{V} (x) \mapsto
\mathrm{V} (x) g$.

Being an element of the Lie algebra of the gauge group,
$\mathrm{Q}_{\mu}(x)$ can be interpreted as a gauge connection for
the local symmetry $\mck$. Under a local transformation
$k(x)\in\mck$, $\mathrm{Q}_{\mu}(x)$ transforms as
\begin{equation}
  \mck : \mathrm{Q}_{\mu}(x)
  \quad \longmapsto \quad
  k(x)\mathrm{Q}_{\mu}(x)k(x)^{-1}+\pa_{\mu}k(x) k(x)^{-1},
\end{equation}
which indeed is the characteristic transformation property of a gauge
connection. On the other hand, $\mathrm{P}_{\mu}(x)$ transforms
covariantly,
\begin{equation}
  \mck : \mathrm{P}_{\mu}(x)
  \quad \longmapsto \quad
  k(x)\mathrm{P} _{\mu}(x)k(x)^{-1},
\end{equation}
because the element $\pa_{\mu}k(x) k(x)^{-1}$ is projected out due to
the negative sign in the construction of $\mathrm{P}_{\mu}(x)$ in
Equation~(\ref{CartanProjections}).

Using the bilinear form $(\cdot |\cdot)$ we can now form a manifestly
$\mck_{\mathrm{local}}\times\mcg_{\mathrm{rigid}}$-invariant expression by
simply ``squaring'' $\mathrm{P} _{\mu}(x)$, i.e., the $p$-dimensional
action takes the form (cf.~Equation~(\ref{sigmamodel}))
\begin{equation}
  S_{\mcgk}=\int_{X} d^{p}x \, \sqrt{\ga}\ga^{\mu\nu}
  \left(\mathrm{P}_{\mu}(x) \big|\mathrm{P}_{\nu}(x)\right).
  \label{Firstaction}
\end{equation}

We can rewrite this action in a number of ways. First, we note
that since $\mathrm{Q}_{\mu}(x)$ can be interpreted as a gauge
connection we can form a ``covariant derivative'' $\mathrm{D}_{\mu}$
in a standard way as
\begin{equation}
  \mathrm{D}_{\mu}\mathrm{V} (x)\equiv
  \pa_{\mu}\mathrm{V} (x)-\mathrm{Q}_{\mu}(x)\mathrm{V} (x),
\label{covariantderivate}
\end{equation}
which, by virtue of Equation~(\ref{CartanProjections}), can
alternatively be written as
\begin{equation}
  \mathrm{D}_{\mu}\mathrm{V} (x)=\mathrm{P}_{\mu}(x)\mathrm{V}(x).
\end{equation}
We see now that the action can indeed be interpreted as a gauged
nonlinear sigma model, in the sense that the local invariance is
obtained by minimally coupling the globally $\mc{G}$-invariant
expression $(\pa_{\mu}\mathrm{V} (x)\mathrm{V} (x)^{-1}| \pa^{\mu}\mathrm{V}
(x)\mathrm{V} (x)^{-1})$ to the gauge field $\mathrm{Q}_{\mu}(x)$ through
the ``covariantization'' $\pa_{\mu}\rightarrow \mathrm{D}_{\mu}$,
\begin{equation}
  \left(\pa_{\mu}\mathrm{V} (x)\mathrm{V} (x)^{-1}
  \big| \pa^{\mu}\mathrm{V} (x)\mathrm{V} (x)^{-1}\right)
  \quad \longrightarrow \quad
  \left(\mathrm{D}_{\mu}\mathrm{V} (x)\mathrm{V} (x)^{-1}
  \big| \mathrm{D}^{\mu}\mathrm{V} (x)\mathrm{V} (x)^{-1}\right) =
  \left(\mathrm{P}_{\mu}(x) \big|\mathrm{P}_{\nu}(x)\right).
\end{equation}
Thus, the action then takes the form
\begin{equation}
  S_{\mcgk}=\int_{X} d^{p}x \, \sqrt{\ga}\ga^{\mu\nu}
  \left(\mathrm{D}_{\mu}\mathrm{V} (x)\mathrm{V} (x)^{-1}\big|
  \mathrm{D}_{\nu}\mathrm{V} (x)\mathrm{V} (x)^{-1}\right).
\end{equation}

We can also form a generalized ``metric'' $\mathrm{M}(x)$
\index{generalized metric|bb} that does not
transform at all under the local symmetry, but only transforms under
rigid $\mc{G}$-transformations. This is done, using the generalized
transpose (extended from the algebra to the group through the
exponential map~\cite{Helgason}), in the following way,
\begin{equation}
  \mathrm{M} (x)\equiv \mathrm{V} (x)^{\mc{T}}\mathrm{V} (x),
  \label{generalizedmetric}
\end{equation}
which is clearly invariant under local transformations
\begin{equation}
  \mck : \mathrm{M} (x)
  \quad \longmapsto \quad
  \left(k(x)\mathrm{V} (x)\right)^{\mc{T}}
  \left(k(x)\mathrm{V} (x)\right)=
  \mathrm{V} (x)^{\mc{T}}\left(k(x)^{\mc{T}}k(x)\right)\mathrm{V} (x)=
  \mathrm{M} (x)
\end{equation}
for $k(x)\in\mck$, and transforms as follows under global
transformations on $\mathrm{V} (x)$ from the right,
\begin{equation}
  \mcg : \mathrm{M} (x)
  \quad \longmapsto \quad
  g^{\mc{T}}\mathrm{M} (x) g,
  \qquad
  g\in \mcg.
\end{equation}
A short calculation shows that the relation
between $\mathrm{M} (x)\in \mc{G}$ and $\mathrm{P} (x)\in \mf{p}$ is
given by 
\begin{eqnarray}
  \f{1}{2}\mathrm{M} (x)^{-1}\pa_{\mu}\mathrm{M} (x) &=&
  \f{1}{2}\left(\mathrm{V} (x)^{\mc{T}}\mathrm{V} (x)\right)^{-1}
  \pa_{\mu}\mathrm{V} (x)^{\mc{T}}\mathrm{V} (x)+
  \left(\mathrm{V} (x)^{\mc{T}}\mathrm{V} (x)\right)^{-1}
  \mathrm{V}(x)^{\mc{T}}\pa_{\mu}\mathrm{V} (x)
  \nonumber
  \\
  &=& \f{1}{2}\mathrm{V} (x)^{-1}\left[\left(\pa_{\mu}\mathrm{V}(x)
  \mathrm{V} (x)^{-1}\right)^{\mc{T}}+\pa_{\mu}\mathrm{V} (x)
  \mathrm{V}(x)^{-1}\right]\mathrm{V} (x)
  \nonumber
  \\
  &=& \mathrm{V} (x)^{-1}\mathrm{P}_{\mu}(x)\mathrm{V} (x).
  \label{MPrelation} 
\end{eqnarray}
Since the factors of $\mathrm{V} (x)$ drop out in the squared
expression,
\begin{equation}
  \left( \mathrm{V} (x)^{-1}\mathrm{P} _{\mu}(x)\mathrm{V} (x)\big|
  \mathrm{V} (x)^{-1}\mathrm{P}^{\mu}(x)\mathrm{V} (x)\right)=
  \left(\mathrm{P}_{\mu}(x)\big|\mathrm{P} ^{\mu}(x)\right),
\end{equation}
Equation~(\ref{MPrelation}) provides a third way to write the
$\mck_{\mathrm{local}}\times\mcg_{\mathrm{rigid}}$-invariant action,
completely in terms of the generalized metric $\mathrm{M} (x)$,
\begin{equation}
  S_{\mcgk}=\f{1}{4}\int_{X} d^{p}x \,
  \sqrt{\ga}\ga^{\mu\nu}\left(\mathrm{M} (x)^{-1}
  \pa_{\mu}\mathrm{M} (x) \big|\mathrm{M} (x)^{-1}
  \pa_{\nu}\mathrm{M} (x)\right). 
  \label{Maction}
\end{equation}
(We call $\mathrm{M}$ a ``generalized metric'' because in the
$GL(n, \mbb{R})/SO(n)$-case, it does correspond to the metric, the field
$\mathrm{V}$ being the ``vielbein''; see Section~\ref{GL10/SL10Sigma}.)

All three forms of the action are manifestly gauge invariant under
$\mck_{\mathrm{local}}$. If desired, one can fix the gauge, and 
thereby eliminating the redundant degrees of freedom.


\subsubsection{Equations of motion and conserved currents}

Let us now take a closer look at the equations of motion resulting
from an arbitrary variation $\delta \mathrm{V} (x)$ of the action
in Equation~(\ref{Firstaction}). The Lie algebra element $\delta \mathrm{V}
(x)\mathrm{V} (x)^{-1}\in \mf{g}$ can be decomposed according to the
Cartan decomposition,
\begin{equation}
  \delta \mathrm{V} (x)\mathrm{V} (x)^{-1}=\Sigma(x) +\Lambda(x),
  \qquad
  \Sigma(x)\in \mf{k}, \, \Lambda(x)\in \mf{p}.
\end{equation}
The variation $\Sigma(x)$ along the gauge orbit will be automatically
projected out by gauge invariance of the action. Thus we can set
$\Sigma(x)=0$ for simplicity. Let us then compute $\de \mathrm{P}
_{\mu}(x)$. One easily gets
\begin{equation}
  \de \mathrm{P} _{\mu}(x)= \pa_{\mu}\Lambda(x)-[\mathrm{Q}_{\mu}(x), \Lambda(x)].
\end{equation}
Since $\Lambda(x)$ is a Lie algebra valued scalar we can freely set
$\pa_{\mu}\Lambda(x)\rightarrow \nabla_{\mu}\Lambda(x)$ in the
variation of the action below, where $\nabla^{\mu}$ is a covariant
derivative on $X$ compatible with the Levi--Civita connection. Using
the symmetry and the invariance of the bilinear form one then finds 
\begin{equation}
  \delta S_{\mcgk}=
  \int_{X}d^{p}x \, \sqrt{\ga}\ga^{\mu\nu}2
  \left[\left(-\nabla_{\nu}\mathrm{P}_{\mu}(x)+
  [\mathrm{Q}_{\nu}(x), \mathrm{P}_{\mu}(x)]\big|\Lambda(x)\right)\right].
\end{equation}
The equations of motion are therefore equivalent to
\begin{equation}
  \mathrm{D}^{\mu}\mathrm{P} {\mu}(x)=0,
\end{equation}
with
\begin{equation}
 \mathrm{D}_{\mu}\mathrm{P}_{\nu}(x)=
 \nabla_{\mu}\mathrm{P}_{\nu}(x)-[\mathrm{Q}_{\mu}(x), \mathrm{P}_{\nu}(x)],
\end{equation}
and simply express the covariant conservation of $\mathrm{P} _{\mu}(x)$.

It is also interesting to examine the dynamics in terms of the
generalized metric $\mathrm{M}(x)$. The equations of motion for
$\mathrm{M}(x)$ are
\begin{equation}
  \f{1}{2}\nabla^{\mu}\left(\mathrm{M}(x)^{-1}\pa_{\mu}\mathrm{M} (x)\right)=0.
\end{equation}
These equations ensure the conservation of the current
\begin{equation}
  \mc{J}_{\mu}\equiv \f{1}{2} \mathrm{M} (x)^{-1}
  \pa_{\mu}\mathrm{M} (x)=
  \mathrm{V}(x)^{-1}\mathrm{P}_{\mu}(x)\mathrm{V}(x),
  \label{ConservedCurrent}
\end{equation}
i.e.,
\begin{equation}
  \nabla^{\mu}\mc{J}_{\mu}=0.
\end{equation}
This is the conserved Noether current associated with the rigid
$\mcg$-invariance of the action.


\subsubsection[Example: $SL(2,\mbb{R})/SO(2)$ (hyperbolic space)]%
              {Example: \boldmath $SL(2,\mbb{R})/SO(2)$ (hyperbolic space)}

Let us consider the example of the coset space $SL(2,\mbb{R}/SO(2)$,
which, although very simple, is nevertheless quite
illustrative. Recall from Section~\ref{section:SL2Example} that the
Lie algebra $\mf{sl}(2,\mbb{R})$ is constructed from the Chevalley
triple $(e, h, f)$,
\begin{equation}
  \mf{sl}(2, \mbb{R})=\mbb{R}f\oplus\mbb{R}h\oplus \mbb{R}e, 
  \label{SL2Rdecomposition}
\end{equation}
with the following standard commutation relations
\begin{equation}
  [h, e]=2e,
  \qquad
  [h, f]=-2f,
  \qquad
  [e, f]=h
\end{equation}
and matrix realisation
\begin{equation}
  e=\left(
    \begin{array}{@{}r@{\quad}r@{}}
      0 & 1 \\
      0 & 0
    \end{array}
  \right),
  \qquad
  h=\left(
    \begin{array}{@{}r@{\quad}r@{}}
      1 & 0 \\
      0 & -1
    \end{array}
  \right),
  \qquad
  f=\left(
    \begin{array}{@{}r@{\quad}r@{}}
      0 & 0 \\
      1 & 0
    \end{array}
  \right).
  \label{adjointSL2R}
\end{equation}
In the Borel gauge, $\mathrm{V} (x)$ reads
\begin{equation}
  \mathrm{V} (x)=
  \Exp \left[\f{\phi(x)}{2} h\right]\,\Exp \left[\chi(x) e\right] =
  \left(
    \begin{array}{@{}c@{\quad}c@{}}
      e^{\phi(x)/2} & \chi(x)e^{\phi(x)/2} \\
      0 & e^{-\phi(x)/2}
    \end{array}
  \right),
\end{equation}
where $\phi(x)$ and $\chi(x)$ represent coordinates on the
coset space, i.e., they describe the sigma model map
\begin{equation}
  X \ni x
  \quad \longmapsto \quad
  \left(\phi(x), \chi(x)\right) \in SL(2, \mbb{R})/SO(2).
\end{equation}

An arbitrary transformation on $\mathrm{V} (x)$ reads
\begin{equation}
  \mathrm{V} (x)
  \quad \longmapsto \quad
  k(x)\mathrm{V} (x) g,
  \qquad
  k(x)\in SO(2), \, g\in SL(2, \mbb{R}),
\end{equation}
which in infinitesimal form becomes
\begin{equation}
  \delta_{\delta k(x), \delta g} \mathrm{V} (x)=
  \delta k(x)\mathrm{V} (x)+\mathrm{V} (x)\delta g,
  \qquad
  \delta k(x)\in \mf{so}(2),\, \delta g\in \mf{sl}(2, \mbb{R}).
\end{equation}
Let us then check how $\mathrm{V} (x)$ transforms under the generators
$\delta g=e, f, h$. As expected, the Borel generators $h$ and $e$
preserve the upper triangular structure 
\begin{equation}
  \begin{array}{rcl}
    \delta_{e}\mathrm{V} (x)&=&
    \mathrm{V} (x) e = \left(
      \begin{array}{@{}c@{\quad}c@{}}
        0 & e^{\phi(x)/2} \\
        0 & 0
      \end{array}
    \right),
    \\ [1 em]
    \delta_{h}\mathrm{V} (x)&=&
    \mathrm{V} (x) h = \left(
      \begin{array}{@{}c@{\quad}c@{}}
        e^{\phi(x)/2} & -\chi(x)e^{\phi(x)/2} \\
        0 & -e^{\phi(x)/2}
      \end{array}
    \right),
  \end{array}
\end{equation}
while the negative root generator $f$ does not respect the form of
$\mathrm{V} (x)$,
\begin{equation}
  \delta_{f}\mathrm{V} (x)= \mathrm{V} (x) f = \left(
    \begin{array}{@{}c@{\quad}c@{}}
      \chi(x)e^{\phi(x)/2} & 0 \\
      e^{-\phi(x)/2} & 0
    \end{array}
  \right).
\end{equation}
Thus, in this case we need a compensating transformation to restore
the upper triangular form. This transformation needs to cancel the
factor $e^{-\phi(x)/2}$ in the lower left corner of the matrix
$\delta_f \mathrm{V} (x)$ and so it must necessarily depend on
$\phi(x)$. The transformation that does the job is
\begin{equation}
  \delta k(x)=\left(
    \begin{array}{@{}c@{\quad}c@{}}
      0 & e^{-\phi(x)} \\
      -e^{-\phi(x)} & 0
    \end{array}
  \right) \in \mf{so}(2),
\end{equation}
and we find
\begin{eqnarray}
  \delta_{\delta k(x), f}\mathrm{V} (x)&=&
  \delta k(x)\mathrm{V} (x)+\mathrm{V} (x) f
  \nonumber
  \\
  &=& \left(
    \begin{array}{@{}c@{\quad}c@{}}
      \chi(x)e^{\phi(x)/2} & e^{-3\phi(x)/2} \\
      0 & -\chi(x)e^{-\phi(x)/2}
    \end{array}
  \right) \in SL(2, \mbb{R})/SO(2).
\end{eqnarray}

Finally, since the generalized transpose $(\,)^{\mc{T}}$ in this case
reduces to the ordinary matrix transpose, the ``generalized'' metric
becomes
\begin{equation}
  \mathrm{M} (x)=\mathrm{V} (x)^{T}\mathrm{V} (x)=\left(
    \begin{array}{@{}c@{\quad}c@{}}
      e^{\phi(x)} & \chi(x)e^{\phi(x)} \\
      \chi(x)e^{\phi(x)} & \chi(x)^{2}e^{\phi(x)}+e^{-\phi(x)}
    \end{array}
  \right).
\end{equation}
The Killing form $(\cdot |\cdot)$ corresponds to taking the
trace in the adjoint representation of Equation~(\ref{adjointSL2R})
and the action~(\ref{Maction}) therefore takes the form
\begin{equation}
  S_{SL(2, \mbb{R})/SO(2)}= \f{1}{2}\int_{X}d^{p}x \,
  \sqrt{\ga}\ga^{\mu\nu}
  \left[\pa_{\mu}\phi(x)\,\pa_{\nu}\phi(x)+e^{2\phi(x)}
  \pa_{\mu}\chi(x\,)\pa_{\nu}\chi(x)\right].
  \label{actionSL2R}
\end{equation}


\subsubsection[Parametrization of $\mcgk$]%
              {Parametrization of \boldmath $\mcgk$}
\label{subsection:parametrization}

The Borel gauge choice is always accessible when the group $\mck$ is
the maximal compact subgroup of $\mcg$. In the noncompact case this is
no longer true since one cannot invoke the Iwasawa decomposition (see,
e.g.~\cite{Keurentjes:2005jw} for a discussion of the subtleties
involved when $\mck$ is noncompact). This point will, however, not be
of concern to us in this paper. We shall now proceed to write down the
sigma model action in the Borel gauge for the coset space $\mcgk$,
with $\mck$ being the maximal compact subgroup. Let
$\Pi=\{\al_1^{\vee}, \cdots, \al_n^{\vee}\}$ be a basis of the Cartan
subalgebra $\mf{h}\subset\mf{g}$, and let $\Delta_+\subset
\mf{h}^{\star}$ denote the set of positive roots. The Borel subalgebra 
\index{Borel subalgebra} of $\mf{g}$ can then be written as
\begin{equation}
  \mf{b}=\sum_{i=1}^{n}\mbb{R}\al_i^{\vee}+\!\!
  \sum_{\al\in\Delta_+} \!\! \mbb{R}E_{\al},
\end{equation}
where $E_{\al}$ is the positive root generator spanning the
one-dimensional root space $\mf{g}_{\al}$ associated to the root
$\al$. The coset representative is then chosen to be
\begin{equation}
  \mathrm{V} (x)=\mathrm{V}_1(x)\mathrm{V}_2(x)=
  \Exp \left[\sum_{i=1}^{n}\phi_{i}(x)\al_{i}^{\vee}\right]\,
  \Exp \left[\sum_{\al\in\Delta_+}\chi_{\al}(x)E_{\al}\right] \in \mcgk. 
  \label{GeneralCosetRepresentative}
\end{equation}
Because $\mf{g}$ is a finite Lie algebra, the sum over positive roots
is finite and so this is a well-defined construction.

From Equation~(\ref{GeneralCosetRepresentative}) we may compute the Lie
algebra valued one-form $\mfVV$ explicitly. Let us do this in some
detail. First, we write the general expression in terms of
$\tV_1(x)$ and $\tV_2(x)$,
\begin{equation}
  \mfVV=\pa_{\mu}\tV_1(x) \tV_1(x)^{-1}+
  \tV_1(x)\left(\pa_{\mu}\tV_2(x) \tV_2(x)^{-1}\right)\tV_1(x)^{-1}. 
  \label{Oneformsplit}
\end{equation}
To compute the individual terms in this expression we need to make use
of the Baker--Hausdorff formulas: 
\begin{equation}
  \begin{array}{rcl}
    \pa_{\mu}e^{A}e^{-A}&=& \displaystyle
    \pa_{\mu}A+\f{1}{2!}[A, \pa_{\mu}A]+\f{1}{3!}[A, [A, \pa_{\mu}A]]+\cdots,
    \\ [0.5 em]
    e^{A}B e^{-A}&=& \displaystyle
    B+[A, B]+\f{1}{2!}[A, [A, B]]+\cdots.
  \end{array}
  \label{BakerHausdorff} 
\end{equation}
The first term in Equation~(\ref{Oneformsplit}) is easy to compute
since all generators in the exponential commute. We find
\begin{equation}
  \pa_{\mu}\tV_1(x) \tV_1(x)^{-1}=
  \sum_{i=1}^{n}\pa_{\mu}\phi_i(x)\al_{i}^{\vee} \in \mf{h}.
\end{equation}
Secondly, we compute the corresponding expression for $\tV_2(x)$. Here
we must take into account all commutators between the positive root
generators $E_{\al}\in\mf{n}_+$. Using the first of the
Baker--Hausdorff formulas above, the first terms in the series become
\begin{eqnarray}
  \pa_{\mu}\tV_2(x) \tV_2(x)^{-1}&=&
  \pa_{\mu}\Exp \left[\sum_{\al\in\Delta_+}\chi_{\al}(x)E_{\al}\right]\,
  \Exp \left[-\!\!\!\sum_{\al^{\p}\in\Delta_+}\!\!\!
  \chi_{\al^{\p}}(x)E_{\al^{\p}}\right]
  \nonumber
  \\
  &=& \!\! \sum_{\al\in\Delta_+} \!\! \pa_{\mu}\chi_{\al}(x)E_{\al}+
  \f{1}{2!} \!\! \sum_{\al, \al^{\p}\in\Delta_+} \!\!\!\!
  \chi_{\al}(x)\pa_{\mu}\chi_{\al^{\p}}(x)[E_{\al}, E_{\al^{\p}}]
  \nonumber
  \\
  & & + \f{1}{3!} \!\! \sum_{\al, \al^{\p}, \al^{\pp}\in \Delta_+}
  \!\!\!\! \chi_{\al}(x) \chi_{\al^{\p}}(x)\pa_{\mu}
  \chi_{\al^{\p\p}}(x)[E_{\al}, [E_{\al^{\p}}, E_{\al^{\p\p}}]]+\cdots.
  \label{longsum}
\end{eqnarray}
Each multi-commutator $[E_{\al}, [E_{\al^{\p}}, \cdots]\cdots,
  E_{\al^{\p\p\p}}]$ corresponds to some new positive root generator,
say $E_{\ga}\in \mf{n}_+$. However, since each term in the
expansion~(\ref{longsum}) is a sum over all positive roots, the
specific generator $E_{\ga}$ will get a contribution from all
terms. We can therefore write the sum in ``closed form'' with the
coefficient in front of an arbitrary generator $E_{\ga}$ taking the
form
\begin{equation}
  \mc{R}_{\ga, \mu}(x)\equiv
  \pa_{\mu}\chi_{\ga}(x)+\f{1}{2!}
  \underbrace{\chi_{\zeta}(x)\pa_{\mu}\chi_{\zeta^{\p}}(x)}_{\zeta+\zeta^{\p}=\ga}+
  \cdots+\f{1}{k_{\ga}!}
  \underbrace{\chi_{\eta}(x)\chi_{\eta^{\p}}(x)\cdots
  \chi_{\eta^{\p\p}}(x)\pa_{\mu}\chi_{\eta^{\p\p\p}}(x)}_{\eta+\eta{\p}+
  \cdots+\eta^{\pp}+\eta^{\pp\p}=\ga},
\end{equation}
where $k_{\ga}$ denotes the number corresponding to the last term in
the Baker--Hausdorff expansion in which the generator $E_{\ga}$
appears. The explicit form of $\mc{R}_{\ga, \mu}(x)$ must be computed
individually for each root $\ga\in\Delta_+$.

The sum in Equation~(\ref{longsum}) can now be written as
\begin{equation}
  \pa_{\mu}\tV_2(x) \tV_2(x)^{-1}=
  \!\! \sum_{\al\in\Delta_+}\!\! \mc{R}_{\al,\mu}(x)E_{\al}. 
  \label{sumclosedform}
\end{equation}
To proceed, we must conjugate this expression with $\tV_1(x)$ in order
to compute the full form of Equation~(\ref{Oneformsplit}). This
involves the use of the second Baker--Hausdorff formula in
Equation~(\ref{BakerHausdorff}) for each term in the sum,
Equation~(\ref{sumclosedform}). Let $h$ denote an arbitrary element of
the Cartan subalgebra,
\begin{equation}
  h=\sum_{i=1}^{n}\phi_{i}(x)\al_{i}^{\vee} \in \mf{h}.
\end{equation}
Then the commutators we need are of the form
\begin{equation}
  [h, E_{\al}]=\al(h)E_{\al},
\end{equation}
where $\al(h)$ denotes the value of the root $\al\in\mf{h}^{\star}$
acting on the Cartan element $h\in\mf{h}$,
\begin{equation}
  \al(h)=\sum_{i=1}^{n}\phi_i(x)\al(\al_i^{\vee})=
  \sum_{i=1}^{n}\phi_i(x)\left<\al,\al_i^{\vee}\right>\equiv
  \sum_{i=1}^{n}\phi_i(x)\al_i.
\end{equation}
So, for each term in the sum in Equation~(\ref{sumclosedform}) we
obtain 
\begin{eqnarray}
  \tV_1(x)E_{\al}\tV_1(x)^{-1}&=&
  E_{\al}+\sum_{i}\phi_i(x)\al_i E_{\al} +
  \f{1}{2}\sum_{i,j} \phi_i(x)\phi_j(x)\al_i\al_j E_{\al}+\cdots
  \nonumber
  \\
  &=& \Exp \left[\sum_i \phi_i(x)\al_i\right] E_{\al}
  \nonumber
  \\
  &=& {e}^{\al(h)} E_{\al}. 
\end{eqnarray}
We can now write down the complete expression for the element $\mfVV$,
\begin{equation}
  \mfVV=\sum_{i=1}^{n}\pa_{\mu}\phi_i(x)\al_{i}^{\vee}+
  \!\! \sum_{\al\in\Delta_+} \!\! {e}^{\al(h)} \mc{R}_{\al, \mu}(x)E_{\al}.
\end{equation}
Projection onto the coset $\mf{p}$ gives (see
Equation~(\ref{CartanProjections}) and
Section~\ref{section:CompactAndRealForms})
\begin{equation}
  \mathrm{P}_{\mu}(x)=
  \sum_{i=1}^{n}\pa_{\mu}\phi_i(x)\al_i^{\vee}+
  \f{1}{2}\!\sum_{\al\in\Delta_+} \!\! e^{\al(h)}\mc{R}_{\al, \mu}(x)
  \left(E_{\al}+E_{-\al}\right),
\end{equation}
where we have used that $E_{\al}^{\mc{T}}=E_{-\al}$ and
$(\al_i^{\vee})^{\mc{T}}=\al_i^{\vee}$.

Next we want to compute the explicit form of the action in
Equation~(\ref{Firstaction}). Choosing the following normalization for the
root generators,
\begin{equation}
  (E_{\al}|E_{\al^{\p}})=
  \delta_{\al,-\al^{\p}},
  \qquad
  (\al_i^{\vee}|\al_j^{\vee})= \delta_{ij},
  \label{NormalizationRootGenerators}
\end{equation}
which implies
\begin{equation}
  (E_{\al}| E_{\al^{\p}}^{\mc{T}})=
  (E_{\al}|E_{-\al^{\p}})=\delta_{\al,\al^{\p}}
\end{equation}
one finds the form of the $\mck_{\mathrm{local}} \times
\mcg_{\mathrm{rigid}}$-invariant action in the parametrization of
Equation~(\ref{GeneralCosetRepresentative}),
\begin{equation}
  S_{\mcgk}=\int_{X} d^{p}x \, \sqrt{\ga}\ga^{\mu\nu}
  \left[\sum_{i=1}^{n}\pa_{\mu}\phi_i(x)\pa_{\nu}\phi_i(x)+
  \f{1}{2} \!\sum_{\al\in\Delta_{+}} \!\!
  e^{2\al(h)}\mc{R}_{\al, \mu}(x)\mc{R}_{\al, \nu}(x)\right]. 
  \label{FiniteSigmaModel}
\end{equation}


\subsection{Geodesic sigma models on infinite-dimensional coset spaces}
\label{section:InfiniteSigmaModels}

In the following we shall both ``generalize and specialize'' the
construction from Section~\ref{section:nonlinearsigmamodels}. The
generalization amounts to relaxing the restriction that the algebra
$\mf{g}$ be finite-dimensional. Although in principle we could
consider $\mf{g}$ to be any indefinite Kac--Moody algebra, we shall be
focusing on the case where it is of Lorentzian type. The analysis
will also be a specialization, in the sense that we consider only
\emph{geodesic} sigma models, meaning that the Riemannian space
$X$ is the one-dimensional worldline of a particle, parametrized
by one variable $t$. This restriction is of course motivated by
the billiard description of gravity close to a spacelike
singularity, where the dynamics at each spatial point is
effectively described by a particle geodesic in the fundamental
Weyl chamber of a Lorentzian Kac--Moody algebra.

The motivation is that the construction of a geodesic sigma model
that exhibits this Kac--Moody symmetry in a manifest way, would
provide a link to understanding the role of the full algebra
$\mf{g}$ beyond the BKL-limit.


\subsubsection{Formal construction}

For definiteness, we consider only the case when the Lorentzian
algebra $\mf{g}$ is a split real form, although this is not really
necessary as the Iwasawa decomposition \index{Iwasawa decomposition} 
holds also in the non-split case.

A very important difference from the finite-dimensional case is
that we now have nontrivial \emph{multiplicities} of the
imaginary roots (see Section~\ref{section:KacMoody}).
Recall that if a root $\al\in\Delta$ has multiplicity $m_{\al}$,
then the associated root space $\mf{g}_{\al}$ is
$m_{\al}$-dimensional. Thus, it is spanned by $m_{\al}$ generators
$E_{\al}^{[s]}\, (s=1, \cdots, m_{\al})$,
\begin{equation}
  \mf{g}_{\al}=\mbb{R}E_{\al}^{[1]}+ \cdots + \mbb{R} E_{\al}^{[m_{\al}]}.
\end{equation}
The root multiplicities are not known in
closed form for any indefinite Kac--Moody algebra, but must be
computed recursively as described in
Section~\ref{section:LevelDecomposition}.

Our main object of study is the coset representative $\mc{V}(t)\in
\mcgk$, which must now be seen as ``formal'' exponentiation of the
infinite number of generators in $\mf{p}$. We can then proceed as
before and choose $\mc{V}(t)$ to be in the Borel gauge, i.e., of
the form
\begin{equation}
  \mc{V}(t)=
  \Exp \left[\sum_{\mu=1}^{\dim \mf{h}} \be^{\mu}(t)\al_{\mu}^{\vee}\right]\,
  \Exp \left[\sum_{\al\in\Delta_+}\sum_{s=1}^{m_{\al}}
  \xi^{[s]}_{\al}(t)E_{\al}^{[s]}\right] \in \mcgk. 
\label{InfiniteCosetRepresentative}
\end{equation}
Here, the index $\mu$ does not correspond to ``spacetime'' but instead
is an index in the Cartan subalgebra \index{Cartan subalgebra}$\mf{h}$, or, equivalently,
in ``scale-factor space'' \index{scale-factor space} (see Section~\ref{section:BKL}).
In the following we shall dispose of writing the sum over $\mu$
explicitly. The second exponent in
Equation~(\ref{InfiniteCosetRepresentative}) contains a formal infinite
sum over all positive roots $\Delta_+$. We will describe in detail
in subsequent sections how it can be suitably truncated. The coset
representative $\mc{V}(t)$ corresponds to a nonlinear realisation
\index{nonlinear realisation} of $\mcg$ and transforms as
\begin{equation}
  \mcg : \mc{V}(t)
  \quad \longmapsto \quad
  k\left(\mc{V}(t), g\right)\mc{V}(t) g,
  \qquad
  k\left(\mc{V}(t), g\right)\in\mck,\, g\in\mcg.
\end{equation}

A $\mf{g}$-valued ``one-form'' can be constructed analogously to
the finite-dimensional case,
\begin{equation}
  \pa \mc{V}(t)\mc{V}(t)^{-1}=\mc{Q}(t)+\mc{P}(t),
\end{equation}
where $\pa\equiv \pa_t$. The first term on the right hand side
represents a $\mf{k}$-connection that is fixed under the Chevalley
involution,
\begin{equation}
  \tau(\mc{Q})=\mc{Q},
\end{equation}
while $\mc{P}(t)$ lies in the orthogonal complement $\mf{p}$ and so is
anti-invariant,
\begin{equation}
  \tau(\mc{P})=-\mc{P}
\end{equation}
(for the split form, the Cartan involution coincides with the
Chevalley involution). \index{Chevalley involution} Using the
projections onto the coset $\mf{p}$
and the compact subalgebra $\mf{k}$, as defined in
Equation~(\ref{CartanProjections}), we can compute the forms of
$\mc{P}(t)$ and $\mc{Q}(t)$ in the Borel gauge, and we find 
\begin{equation}
  \begin{array}{rcl}
    \mc{P}(t)&=& \displaystyle
    \pa \be^{\mu}(t)\al_{\mu}^{\vee}+\f{1}{2}\sum_{\al\in\Delta_+}
    \sum_{s=1}^{m_{\al}}e^{\al(\be)}\mf{R}^{[s]}_{\al}(t)
    \left(E^{[s]}_{\al}+E^{[s]}_{-\al}\right),
    \\ [0.75 em]
    \mc{Q}(t)&=& \displaystyle
    \f{1}{2}\sum_{\al\in\Delta_+}
    \sum_{s=1}^{m_{\al}}e^{\al(\be)}\mf{R}^{[s]}_{\al}(t)
    \left(E^{[s]}_{\al}-E^{[s]}_{-\al}\right),
  \end{array}
  \label{InfiniteQandP} 
\end{equation}
where $\mf{R}^{[s]}_{\al}(t)$ is the analogue of $\mc{R}_{\al}(x)$ in
the finite-dimensional case, i.e., it takes the form
\begin{equation}
  \mf{R}^{[s]}_{\al}(t)=\pa \xi^{[s]}_{\al}(t)+
  \f{1}{2}\underbrace{\xi_{\zeta}^{[s]}(t)
  \pa \xi^{[s]}_{\zeta^{\p}}(t)}_{\zeta+\zeta^{\p}=\al}+\cdots,
  \label{InfiniteCosetCoefficients}
\end{equation}
which still contains a finite number of terms for each positive root
$\al$. The value of the root $\al\in\mf{h}^{\star}$ acting on
$\be=\be^{\mu}(t)\al_{\mu}^{\vee}\in\mf{h}$ is
\begin{equation}
  \al(\be)=\al_{\mu}\be^{\mu}.
\end{equation}
Note that here the notation $\al_{\mu}$ does not correspond to a
simple root, but denotes the components of an arbitrary root vector
$\al\in\mf{h}^{\star}$.

The action for a particle moving on the infinite-dimensional coset
space $\mcgk$ can now be constructed using the invariant bilinear
form $(\cdot |\cdot )$ on $\mf{g}$,
\begin{equation}
  S_{\mcgk}=\int dt\,n(t)^{-1}\left(\mc{P}(t)|\mc{P}(t)\right),
\end{equation}
where $n(t)$ ensures invariance under reparametrizations of $t$. The
variation of the action with respect to $n(t)$ constrains the motion
to be a \emph{null geodesic} on $\mcgk$,
\begin{equation}
  \left(\mc{P}(t)|\mc{P}(t)\right)=0.
\end{equation}

Defining, as before, a covariant derivative $\mf{D}$ with respect to
the local symmetry $\mck$ as
\begin{equation}
  \mf{D}\mc{P}(t)\equiv \pa \mc{P}(t)-\left[\mc{Q}(t), \mc{P}(t)\right],
\end{equation}
the equations of motion read simply
\begin{equation}
  \mf{D}\left(n(t)^{-1} \mc{P}(t)\right)=0.
  \label{InfiniteEOM}
\end{equation}
The explicit form of the action in the parametrization of
Equation~(\ref{InfiniteCosetRepresentative}) becomes
\begin{equation}
  S_{\mcgk}=\int dt \, n(t)^{-1}
  \left[G_{\mu\nu}\,\pa\be^{\mu}(t)\,\pa\be^{\nu}(t)+
  \f{1}{2}\sum_{\al\in\Delta_+}\sum_{s=1}^{m_{\al}}e^{2\al(\be)}\,
  \mf{R}^{[s]}_{\al}(t)\,\mf{R}^{[s]}_{\al}(t)\right],
  \label{InfiniteSigmaModel}
\end{equation}
where $G_{\mu\nu}$ is the flat Lorentzian metric, defined by the
restriction of the bilinear form $(\cdot |\cdot )$ to the Cartan
subalgebra $\mf{h}\subset\mf{g}$. The metric $G_{\mu\nu}$ is exactly
the same as the metric in scale-factor space \index{scale-factor space} (see
Section~\ref{section:BKL}), and the kinetic term for the Cartan
parameters $\be^{\mu}(t)$ reads explicitly
\begin{equation}
  G_{\mu\nu}\,\pa\be^{\mu}(t)\,\pa\be^{\nu}(t)=
  \sum_{i=1}^{\dim \mf{h}-1}\pa \be^{i}(t)\,\pa\be^{i}(t)-
  \left(\sum_{i=1}^{\dim \mf{h}-1} \!\! \pa\be^{i}(t)\right)
  \left(\sum_{j=1}^{\dim \mf{h}-1} \!\! \pa\be^{j}(t)\right)+
  \pa\phi(t)\,\pa\phi(t).
  \label{MetricCartanSubalgebra}
\end{equation}

Although $\mf{g}$ is infinite-dimensional we still have the notion
of ``formal integrability'', owing to the existence of an infinite
number of conserved charges, defined by the equations of motion in
Equation~(\ref{InfiniteEOM}). We can define the generalized metric
\index{generalized metric} for any $\mf{g}$ as
\begin{equation}
  \mc{M}(t)\equiv \mc{V}(t)^{\mc{T}}\mc{V}(t),
\end{equation}
where the transpose $(\, )^{\mc{T}}$ is defined as before in
terms of the Chevalley involution,
\begin{equation}
  (\,)^{\mc{T}}=-\tau(\,).
\end{equation}
Then the equations of motion $\mf{D}\mc{P}(t)=0$ are equivalent to the
conservation $\pa \mf{J}=0$ of the current
\begin{equation}
  \mf{J}\equiv \f{1}{2}\mc{M}(t)^{-1}\pa \mc{M}(t).
\end{equation}
This can be formally solved in closed form
\begin{equation}
  \mc{M}(t)=e^{t\mf{J}^{\mc{T}}}\mc{M}(0)e^{t\mf{J}},
\end{equation}
and so an arbitrary group element $g\in\mcg$ evolves according to
\begin{equation}
  g(t)=k(t)g(0)e^{t\mf{J}},
  \qquad
  k(t)\in\mck.
  \label{GeneralSolution}
\end{equation}

Although the explicit form of $\mc{P}(t)$ contains infinitely many
terms, we have seen that each coefficient $\mf{R}_{\al}^{[s]}(t)$
can, in principle, be computed exactly for each root $\al$. This,
however, is not the case for the current $\mf{J}$. To find the
form of $\mf{J}$ one must conjugate $\mc{P}(t)$ with the coset
representative $\mc{V}(t)$ and this requires an infinite number of
commutators to get the correct coefficient in front of any
generator in $\mf{J}$.


\subsubsection{Consistent truncations}
\label{section:ConsistentTruncations}

One method for dealing with infinite expressions like
Equation~(\ref{InfiniteQandP}) consists in considering successive
finite expansions allowing more and more terms, while still respecting
the dynamics of the sigma model. 

This leads us to the concept of a \emph{consistent truncation} of the
sigma model for $\mcgk$. We take as definition of such a truncation
any sub-model $S^{\p}$ of $S_{\mcgk}$ whose solutions are also
solutions of the original model.

There are two main approaches to finding suitable truncations that
fulfill this latter criterion. These are the so-called
\emph{subgroup truncations} and the \emph{level truncations},
which will both prove to be useful for our purposes, and we
consider them in turn below.

\subsubsection*{Subgroup truncation}

The first consistent truncation we shall treat is the
case when the dynamics of a sigma model for some global group
$\mcg$ is restricted to that of an appropriately chosen subgroup
$\bar{\mcg}\subset\mcg$. We consider here only subgroups
$\bar{\mcg}$ which are obtained by exponentiation of regular
subalgebras $\bar{\mf{g}}$ of $\mf{g}$. The concept of regular
embeddings of Lorentzian Kac--Moody algebras 
\index{Kac--Moody algebra} was discussed in detail in
Section~\ref{section:KacMoody}.

To restrict the dynamics to that of a sigma model based on the
coset space $\bar{\mcg}/\mc{K}(\bar{\mcg})$, we first assume that
the initial conditions $g(t)\big|_{t=0}=g(0)$ and $\pa g(t)\big|_{t=0}$
are such that the following two conditions are satisfied:

\begin{enumerate}
\item The group element $g(0)$ belongs to $\bar{\mcg}$.
\item The conserved current $\mf{J}$ belongs to $\bar{\mf{g}}$.
\end{enumerate}

When these conditions hold, then $g(0)e^{t\mf{J}}$ belongs to $\bar{\mcg}$
for all $t$. Moreover, there always exists $\bar{k}(t)\in
\mc{K}(\bar{\mcg})$ such that
\begin{equation}
  \bar{g}(t)\equiv \bar{k}(t) g(0) e^{t\mf{J}} \in \bar{\mcg}/\mc{K}(\bar{\mcg}),
\end{equation}
i.e, $\bar{g}(t)$ belongs to the Borel subgroup of $\bar{\mcg}$.
Because the embedding is regular, $\bar{k}(t)$ belongs to $\mck$
and we thus have that $\bar{g}(t)$ also belongs to the Borel subgroup
of the full group $\mcg$. 

Now recall that from Equation~(\ref{GeneralSolution}), we know that
$\bar{g}(t)=\bar{k}(t)g(0)e^{t\mf{J}}$ is a solution to the
equations of motion for the sigma model on
$\bar{\mcg}/\mc{K}(\bar{\mcg})$. But since we have found that
$\bar{g}(t)$ preserves the Borel gauge for $\mcgk$, it follows
that $\bar{k}(t)g(0)e^{t\mf{J}}$ is a solution to the equations of
motion for the full sigma model. Thus, the dynamical evolution of the
subsystem $S^{\p}=S_{\bar{\mcg}/\mc{K}(\bar{\mcg})}$ preserves the
Borel gauge of $\mcg$. These arguments show that initial conditions in $\bar{\mcg}$
remain in $\bar{\mcg}$, and hence the dynamics of a sigma model on
$\mcgk$ can be consistently truncated to a sigma model on
$\bar{\mcg}/\mc{K}(\bar{\mcg})$.

Finally, we recall that because the embedding
$\bar{\mf{g}}\subset\mf{g}$ is regular, the restriction of the
bilinear form on $\mf{g}$ coincides with the bilinear form on
$\bar{\mf{g}}$. This implies that the Hamiltonian constraints
\index{Hamiltonian constraint} for
the two models, arising from time reparametrization invariance of
the action, also coincide. 

We shall make use of subgroup truncations in
Section~\ref{section:cosmologicalsolutions}.

\subsubsection*{Level truncation and height truncation}

Alternative ways of consistently truncating the
infinite-dimensional sigma model rest on the use of
\emph{gradations} of $\mf{g}$,
\begin{equation}
  \mf{g}=\cdots + \mf{g}_{-2}+\mf{g}_{-1}+\mf{g}_0+\mf{g}_1+\mf{g}_2+\cdots,
\end{equation}
where the sum is infinite but each subspace is finite-dimensional.
One also has
\begin{equation}
  [\mf{g}_{\ell^{\p}}, \mf{g}_{\ell^{\p\p}}]\subseteq \mf{g}_{\ell^{\p}+\ell^{\p\p}}.
\end{equation}
Such a gradation was for instance constructed in
Section~\ref{section:LevelDecomposition} and was based on a so-called
\emph{level decomposition} of the adjoint representation of $\mf{g}$
into representations of a finite regular subalgebra \index{regular subalgebra}
$\mf{r}\subset \mf{g}$. We will now use this construction to truncate
the sigma model based of $\mcgk$, by ``terminating'' the gradation of
$\mf{g}$ at some finite level $\bar{\ell}$. More specifically, the
truncation will involve setting to zero all coefficients
$\mf{R}_{\al}^{[s]}(t)$, in the expansion of $\mc{P}(t)$,
corresponding to roots $\al$ whose generators $E_{\al}^{[s]}$ belong
to subspaces $\mf{g}_{\ell}$ with $\ell>\bar{\ell}$. Part of this
section draws inspiration from the treatment in~\cite{DHN2, DHNReview,
  AxelThesis}.

The level $\ell$ might be the height, or it might count the number
of times a specified single simple root appears. In that latter
case, the actual form of the level decomposition must of course be
worked out separately for each choice of algebra $\mf{g}$ and each
choice of decomposition. We will do this in detail in
Section~\ref{section:E10SigmaModel} for a specific level decomposition
of the hyperbolic algebra $E_{10}$. Here, we shall display the general
construction in the case of the \emph{height truncation}, which exists
for any algebra.

Let $\al$ be a positive root, $\al\in\Delta_+$. It has the
following expansion in terms of the simple roots
\begin{equation}
  \al=\sum_{i}m_i\al_i
  \qquad
  (m_i\geq 0).
\end{equation}
Then the \emph{height} of $\al$ is defined as (see
Section~\ref{section:KacMoody})
\begin{equation}
  \htx (\al)=\sum_{i}m_i.
\end{equation}
The height can thus be seen as a linear integral map $\htx : \Delta
\rightarrow \mbb{Z}$, and we shall sometimes use the notation
$\htx (\al)=h_{\al}$ to denote the value of the map
$\htx $ acting on a root $\al\in \Delta$.

To achieve the height truncation, we assume that the sum over all
roots in the expansion of $\mc{P}(t)$, Equation~(\ref{InfiniteQandP}), is
ordered in terms of increasing height. Then we can consistently
set to zero all coefficients $\mf{R}_{\al}^{[s]}(t)$ corresponding
to roots with greater height than some, suitably chosen, finite
height $\bar{h}$. We thus find that the finitely truncated coset
element $\mc{P}_0(t)$ is
\begin{equation}
  \mc{P}_0(t)\equiv \mc{P}(t)\big|_{\htx \leq \bar{h}}=
  \pa \be^{\mu}(t)\al_{\mu}^{\vee}+
  \f{1}{2}\sum_{\substack{\al \in \Delta_{+}\\ \htx (\al) \leq \bar{h}}}
  \sum_{s=1}^{m_{\al}}e^{\al(\be)}\mf{R}^{[s]}_{\al}(t)
  \left(E^{[s]}_{\al}+E^{[s]}_{-\al}\right),
\end{equation}
which is equivalent to the statement
\begin{equation}
  \mf{R}_{\ga}^{[s]}(t)=0
  \qquad
  \forall \ga\in\Delta_+ ,\, \htx (\ga)>\bar{h}.
\end{equation}

For further use, we note here some properties of the coefficients
$\mf{R}_{\al}^{[s]}(t)$. By examining the structure of
Equation~(\ref{InfiniteCosetCoefficients}), we see that
$\mf{R}_{\al}^{[s]}(t)$ takes the form of a temporal derivative
acting on $\xi_{\al}^{[s]}(t)$, followed by a sequence of terms
whose individual components, for example $\xi_{\zeta}^{[s]}(t)$,
are all associated with roots of \emph{lower} height than $\al$,
$\htx (\zeta)< \htx (\al)$. It will prove useful to think
of $\mf{R}_{\al}^{[s]}(t)$ as representing a kind of
``generalized'' derivative operator acting on the field
$\xi_{\al}^{[s]}$. Thus we define the operator $\mc{D}$ by
\begin{equation}
  \mc{D}\xi_{\al}^{[s]}(t)\equiv
  \pa \xi^{[s]}_{\al}(t)+\mc{F}_{\al}^{[s]}
  \left(\xi\pa\xi, \xi^2\pa\xi,\cdots \right),
\end{equation}
where $\mc{F}_{\al}^{[s]}(t)$ is a polynomial function of the
coordinates $\xi(t)$, whose explicit structure follows from
Equation~(\ref{InfiniteCosetCoefficients}). It is common in the
literature to refer to the level truncation as ``setting all higher
level covariant derivatives to zero'', by which one simply means that
all $\mc{D}\xi_{\ga}^{[s]}(t)$ corresponding to roots $\ga$ above a
given finite level $\bar{\ell}$ should vanish. Following~\cite{DHN2}
we shall call the operators $\mc{D}$ ``covariant derivatives''.

It is clear from the equations of motion $\mf{D}\mc{P}(t)=0$, that if
all covariant derivatives $\mc{D}\xi_{\ga}^{[s]}(t)$ above a given
height are set to zero, this choice is preserved by the dynamical
evolution. Hence, the height (and any level) truncation is indeed a
consistent truncation. Let us here emphasize that it is \emph{not}
consistent by itself to merely put all fields $\xi_{\ga}^{[s]}(t)$
above a certain level to zero, but one must take into account the fact
that combinations of lower level fields may parametrize a higher level
generator in the expansion of $\mc{P}(t)$, and therefore it is crucial
to define the truncation using the derivative operator
$\mc{D}\xi_{\ga}^{[s]}(t)$.


\subsection[Eleven-dimensional supergravity and $\mc{E}_{10}/\mc{K}(\mc{E}_{10})$]%
           {Eleven-dimensional supergravity and \boldmath $\mc{E}_{10}/\mc{K}(\mc{E}_{10})$}
\label{section:E10SigmaModel}

We shall now illustrate the results of the previous sections by
explicitly constructing an action for the coset space
$\mc{E}_{10}/\mc{K}(\mc{E}_{10})$. \index{$E_{10}$} We employ the level
decomposition of $E_{10}=\mathrm{Lie}\, \mc{E}_{10}$ in terms of its
regular $\mf{sl}(10, \mbb{R})$-subalgebra (see
Section~\ref{section:LevelDecomposition}), to write the coordinates
on the coset space  \index{coset space} as (time-dependent) $\mf{sl}(10, \mbb{R})$-tensors. It is
then shown that for a truncation of the sigma model  \index{nonlinear sigma model} at level
$\ell=3$, these fields can be interpreted as the physical fields
of eleven-dimensional supergravity. This ``dictionary'' ensures
that the equations of motion arising from the sigma model on
$\mc{E}_{10}/\mc{K}(\mc{E}_{10})$ are equivalent to the (suitably
truncated) equations of motion of eleven-dimensional
supergravity~\cite{DHN2}.


\subsubsection{Low level fields}
\label{section:lowlevelfields}

We perform the level decomposition \index{level decomposition}of $E_{10}$ with respect to the
$\mf{sl}(10, \mbb{R})$-subalgebra obtained by removing the
exceptional node in the Dynkin diagram in
Figure~\ref{figure:E10a}. This procedure was described in
Section~\ref{section:LevelDecomposition}. When using this
decomposition, a sum over (positive) roots becomes a sum over all $\mf{sl}(10,
\mbb{R})$-indices in each (positive) representation appearing in
the decomposition.

We recall that up to level three the following representations
appear 
\begin{equation}
  \begin{array}{l}
    \ell=0:
    \qquad
    {K^{a}}_b,
    \\
    \ell=1:
    \qquad
    E^{abc}=E^{[abc]},
    \\
    \ell=2:
    \qquad
    E^{a_1\cdots a_6}=E^{[a_1\cdots a_6]},
    \\
    \ell=3:
    \qquad
    E^{a|b_1\cdots b_8}=E^{a|[b_1\cdots b_8]},
  \end{array}
  \label{LowLevelRepsE10} 
\end{equation}
where all indices are $\mf{sl}(10,
\mbb{R})$-indices and so run from 1 to 10. The level zero
generators ${K^{a}}_b$ correspond to the adjoint representation of
$\mf{sl}(10, \mbb{R})$ and the higher level generators correspond
to an infinite tower of raising operators of $E_{10}$. As
indicated by the square brackets, the level one and two
representations are completely antisymmetric in all indices, while
the level three representation has a mixed Young tableau symmetry:
It is antisymmetric in the eight indices $b_1 \cdots b_8$ and is
also subject to the constraint
\begin{equation}
  E^{[a|b_1\cdots b_8]}=0.
  \label{LevelThreeConstraint}
\end{equation}
In the scale factor space ($\be$-basis), the roots of $E_{10}$
corresponding to the above generators act as follows on $\be\in
\mf{h}$: 
\begin{equation}
  \begin{array}{rcl}
    {K^{a}}_{b} & \quad \Longleftrightarrow & \quad
    \al_{ab}(\be)=\be^{a}-\be^{b}
    \qquad (a>b),
    \\
    E^{abc} & \quad \Longleftrightarrow & \quad
    \al_{abc}(\be) =\be^{a}+\be^{b}+\be^c,
    \\
    E^{a_1\cdots a_6} & \quad \Longleftrightarrow & \quad
    \al_{a_1\cdots a_6}(\be)=\be^{a_1}+\cdots +\be^{a_6},
    \\
    E^{a|a b_1\cdots b_7 } & \quad \Longleftrightarrow & \quad
    \al_{ab_1\cdots b_7}(\be) =2\be^{a}+\be^{b_1}+\cdots +\be^{b_7},
    \\
    E^{a_1|a_2\cdots a_9} & \quad \Longleftrightarrow & \quad
    \al_{a_1\cdots a_9}(\be) =\be^{a_1}+\cdots + \be^{a_9}.
  \end{array}
\end{equation}
We can use the scalar product in root space,
$\mf{h}^{\star}_{E_{10}}$, to compute the norms of these roots. Recall
from Section~\ref{section:KMBilliardsI} that the metric on
$\mf{h}^{\star}_{E_{10}}$ is the inverse of the metric in
Equation~(\ref{MetricCartanSubalgebra}), and for $E_{10}$ it takes
the form 
\begin{equation}
  (\om|\om)=G^{ij}\om_{i}\om_{j}=
  \sum_{i=1}^{10}\om_i \om_i -
  \f{1}{9}\left(\sum_{i=1}^{10}\om_i\right)\left(\sum_{j=1}^{10}\om_j\right),
  \qquad
  \om\in\mf{h}^{\star}_{E_{10}}.
\end{equation}
The level zero, one and two generators correspond to real roots of
$E_{10}$,
\begin{equation}
  (\al_{ab}|\al_{cd})=2,
  \qquad
  (\al_{abc}|\al_{def})=2,
  \qquad
  (\al_{a_1\cdots a_6}|\al_{b_1\cdots b_6})=2.
\end{equation}
We have split the roots corresponding to the level three generators
into two parts, depending on whether or not the special index $a$
takes the same value as one of the other indices. The resulting two
types of roots correspond to real and null roots, respectively,
\begin{equation}
  (\al_{ab_1\cdots b_7}|\al_{cd_1\cdots d_7})=2,
  \quad
  (\al_{a_1\cdots a_9}|\al_{b_1\cdots b_9})=0.
\end{equation}
Thus, the first time that generators corresponding to imaginary roots
appear in the level decomposition is at level three. This will prove
to be important later on in our analysis.


\subsubsection[The $GL(10, \mbb{R})/SO(10)$-sigma model]%
              {The \boldmath $GL(10, \mbb{R})/SO(10)$-sigma model}
\label{GL10/SL10Sigma}
\index{nonlinear sigma model}

Because of the importance and geometric significance of level zero, we
shall first develop the formalism for the $GL(10,
\mbb{R})/SO(10)$-sigma model. A general group element $H$ in the
subgroup $GL(10, \mbb{R})$ reads 
\begin{equation}
  H = \Exp \left[{h_{a}}^b {K^{a}}_b \right]
\end{equation}
where ${h_{a}}^b$ is a $10\times 10$ matrix (with
$a$ being the row index and $b$ the column index). Although the
${K^{a}}_b $'s are generators of $E_{10}$ and can, within this framework, at best be
viewed as infinite matrices, it will prove convenient -- for streamlining
the calculations -- to view them in the present section also as
$10\times 10$ matrices, since we confine our attention to the
finite-dimensional subgroup $GL(10, \mbb{R})$. Namely, ${K^{a}}_b $ is
treated as a 
$10\times 10$ matrix with 0's everywhere except 1 in position
$(a,b)$ (see Equation~(\ref{GLgenerators})). The final formulation in terms of the variables
${h_{a}}^b(t)$ -- which are $10\times 10$ matrices irrespectively as to 
whether one deals with $GL(10, \mbb{R})$ \emph{per se} or as a subgroup of
$E_{10}$ -- does not depend on this interpretation.

It is also useful to describe $GL(10, \mbb{R})$ as the set of linear
combinations ${m_i}^j K{^i}_{j}$ where the $10\times 10$ matrix
${m_i}^j$ is invertible. The product of the $K{^i}_{j}$'s is
given by
\begin{equation}
K{^i}_{j}\, K{^k}_{m} = {\delta^k}_j \, {K^i}_m.
\end{equation}
One easily verifies that if $M = {m_i}^j K{^i}_{j}$ and $N
= {n_i}^j K{^i}_{j}$ belong to $GL(10, \mbb{R})$, then $MN =
{(mn)_i}^j K{^i}_{j}$ 
where $mn$ is the standard product of the $10\times 10$ matrices $m$
and $n$. Furthermore, $\Exp \left({h_i}^j {K^i}_j \right) =
{\left(e^h\right)_i}^j {K^i}_j$ where $e^h$ is the standard matrix
exponential.

Under a general transformation, the representative $H(t)$ is
multiplied from the left by a time-dependent $SO(10)$ group
element $R$ and from the right by a constant linear $GL(10, \mbb{R})$-group element $L$.
Explicitly, the transformation takes the form (suppressing the
time-dependence for notational convenience)
\begin{equation} H
  \rightarrow H' = R H L. 
  \label{transformationvielbein}
\end{equation}
In terms of components, with $H = {e_a}^b {K^a}_b$, ${e_a}^b =
{(e^h)_a}^b$, $R = {R_a}^b {K^a}_b$ and $L = {L_a}^b {K^a}_b$, one
finds
\begin{equation}
  {e'_a}^b = {R_a}^c {e_c}^d {L_d}^b,
\end{equation}
where we have
set $H' = {e'_a}^b {K^a}_b$. The indices on the coset
representative have different covariance properties. To emphasize
this fact, we shall write a bar over the first index, ${e_a}^b
\rightarrow {e_{\bar{a}}}^b$. Thus, barred indices transform under
the local $SO(10)$ gauge group and are called ``local'', or also
``flat'', indices, while unbarred indices transform under the
global $GL(10, \mbb{R})$ and are called ``world'', or also ``curved'',
indices. The gauge invariant matrix product $M= H^T H$ is equal to
\begin{equation}
  M = g^{ab} K_{ab},
\end{equation}
with $K_{ab} \equiv {K^c}_b \delta_{ac}$ and
\begin{equation}
  g^{ab} = \sum_{\bar{c}} {e_{\bar{c}}}^a {e_{\bar{c}}}^b. 
  \label{sigmamodelMetric}
\end{equation}
The $g^{ab}$ do not transform under local $SO(10)$-transformations and
transform as a (symmetric) contravariant tensor under rigid $GL(10,
\mbb{R})$-transformations,
\begin{equation}
  {g'}^{ab}= g^{cd} {L_c}^a {L_d}^b.
\end{equation}
They are components of a nondegenerate symmetric matrix that can be
identified with an inverse Euclidean metric.

Indeed, the coset space $GL(10, \mbb{R})/SO(10)$ can be
identified with the space of symmetric tensors of Euclidean
signature, i.e., the space of metrics. This is because two
symmetric tensors of Euclidean signature are equivalent under a
change of frame, and the isotropy subgroup, say at the identity,
is evidently $SO(10)$. In that view, the coset representative
${e_a}^b$ is the spatial vielbein.

The action for the coset space $GL(10, \mbb{R})/SO(10)$ with the
metric of Equation~(\ref{KillingformlevelzeroForE10}) is easily found to
be
\begin{equation}
  \mc{L}_0=\f{1}{4}\left(g^{ac}(t)g^{bd}(t)-g^{ab}(t)g^{cd}(t)\right)
  \pa {g}_{ab}(t) \, \pa {g}_{cd}(t).
 \label{LagrangianforGL10SO10}
\end{equation}
Note that the quadratic form multiplying the time derivatives is just
the ``De Witt supermetric''~\cite{DeWitt:1967yk}. Note also for future
reference that the invariant form $\pa H \, H^{-1}$ reads explicitly
\begin{equation}
  \pa {H}\, H^{-1} = {\pa {e}_{\bar{a}}}^{\ph{a} b} \,
  {e_b}^{\bar{c}} \, {K^a}_c,
\end{equation}
where ${e_b}^{\bar{n}}$ is the inverse vielbein.


\subsubsection[Sigma model fields and
  $SO(10)_{\mathrm{local}}\times GL(10,\mbb{R})_{\mathrm{rigid}}$-covariance]%
              {Sigma model fields and \boldmath
  $SO(10)_{\mathrm{local}}\times GL(10,\mbb{R})_{\mathrm{rigid}}$-covariance}

We now turn to the full nonlinear sigma model  \index{nonlinear sigma model} for 
$\mc{E}_{10}/\mc{K}(\mc{E}_{10})$. Rather than exponentiating the
Cartan subalgebra separately as in
Equation~(\ref{InfiniteCosetRepresentative}), it will here prove
convenient to instead single out the level zero subspace
$\mf{g}_{0}=\mf{gl}(10, \mbb{R})$. This permits one to control
easily $SO(10)_{\mathrm{local}}\times GL(10, \mbb{R})_{\mathrm{rigid}}$-covariance. 
To make this level zero 
covariance manifest, we shall furthermore assume that the Borel
gauge has been fixed only for the non-zero levels, and we keep all
level zero fields present. The residual gauge freedom is then just
multiplication by an $SO(10)$ rotation from the left.

Thus, we take a coset representative of the form
\begin{equation}
  \mc{V}(t)=\Exp \left[{h_{a}}^b(t) {K^{a}}_b \right]\,
  \Exp \left[\f{1}{3!}\mc{A}_{abc}(t)E^{abc}+
  \f{1}{6!}\mc{A}_{a_1\cdots a_6}(t)E^{a_1\cdots a_6}+
  \f{1}{9!}\mc{A}_{a|b_1\cdots b_8}(t)E^{a|b_1\cdots b_8}+\cdots \right],
  \label{E10cosetrepresentative}
\end{equation}
where the sum in the first exponent would be restricted to all $a\geq
b$ if we had taken a full Borel gauge also at level zero. The
parameters $\mc{A}_{abc}(t), \mc{A}_{a_1\cdots a_6}(t)$ and
$\mc{A}_{a|b_1\cdots b_8}(t)$ are coordinates on the coset space
$\mc{E}_{10}/\mc{K}(\mc{E}_{10})$ and will eventually be interpreted
as physical time-dependent fields of eleven-dimensional supergravity.

How do the fields transform under $SO(10)_{\mathrm{local}}\times
GL(10,\mbb{R})_{\mathrm{rigid}}$? Let $R \in SO(10)$, $L \in GL(10,
\mbb{R})$ and decompose $\mc{V}$ according to
Equation~(\ref{E10cosetrepresentative}) as the product 
\begin{equation}
  \mc{V} = H T, 
\end{equation}
with 
\begin{equation}
  \begin{array}{rcl}
    H &= & \displaystyle
    \Exp \left[{h_{a}}^b(t) {K^{a}}_b \right] \in GL(10, \mbb{R}),
    \\ [1 em]
    T & = & \displaystyle
    \Exp \left[\f{1}{3!}\mc{A}_{abc}(t)E^{abc}+
    \f{1}{6!} \mc{A}_{a_1\cdots a_6}(t)E^{a_1\cdots a_6}+
    \f{1}{9!}\mc{A}_{a|b_1\cdots b_8}(t)E^{a|b_1\cdots b_8}+\cdots \right].
  \end{array}
\end{equation}
One has
\begin{equation}
  \mc{V} \rightarrow \mc{V}' = R (H T) L = (R H L) (L^{-1} T L).
\end{equation}
Now, the first matrix $H' = RHL$ clearly belongs to $GL(10, \mbb{R})$,
since it is the product of a rotation matrix by two $GL(10,
\mbb{R})$-matrices. It has exactly the same transformation as in
Equation~(\ref{transformationvielbein}) above in the context of the
nonlinear sigma model for $GL(10, \mbb{R})/SO(10)$. Hence, the
geometric interpretation of ${e_{\bar{a}}}^b = {(e^h)_{\bar{a}}}^b$ as
the vielbein remains.

Similarly, the matrix $T' \equiv L^{-1} T L$ has exactly the same
form as $T$, 
\begin{eqnarray}
  T' &=&\Exp \left( L^{-1} \left[\f{1}{3!}\mc{A}_{abc}(t)E^{abc}+
  \f{1}{6!}\mc{A}_{a_1\cdots a_6}(t)E^{a_1\cdots a_6} +
  \f{1}{9!}\mc{A}_{a|b_1\cdots b_8}(t)E^{a|b_1\cdots b_8}+\cdots \right] L \right)
  \nonumber
  \\
  &=&\Exp \left[\f{1}{3!}\mc{A}'_{abc}(t)E^{abc}+
  \f{1}{6!}\mc{A}'_{a_1\cdots a_6}(t)E^{a_1\cdots a_6} +
  \f{1}{9!}\mc{A}'_{a|b_1\cdots b_8}(t)E^{a|b_1\cdots b_8}+\cdots \right], 
\end{eqnarray}
where the variables
$\mc{A}'_{abc}$, $\mc{A}'_{a_1\cdots a_6}$, ..., are obtained from
the variables $\mc{A}_{abc}$, $\mc{A}_{a_1\cdots a_6}$, ..., by
computing $L^{-1} E^{abc} L$, $L^{-1} E^{a_1 \cdots a_6} L$, ...,
using the commutation relations with ${K^a}_b$. Explicitly, one
gets
\begin{equation}
  \mc{A}'_{abc} = {(L^{-1})_a}^e {(L^{-1})_b}^f
  {(L^{-1})_c}^g \mc{A}_{efg},
  \qquad
  \mc{A}'_{a_1\cdots a_6} =
  {(L^{-1})_{a_1}}^{b_1} \cdots {(L^{-1})_{a_6}}^{b_6}
  \mc{A}_{b_1\cdots b_6},
  \qquad
  \mbox{ etc.}
\end{equation}
Hence, the fields
$\mc{A}_{abc}$, $\mc{A}_{a_1\cdots a_6}$, ... do not transform
under local $SO(10)$ transformations. However, they do transform under
rigid $GL(10, \mbb{R})$-transformations as tensors of the type
indicated by their indices. Their indices are world indices and
not flat indices.


\subsubsection{``Covariant derivatives''}

The invariant form $\pa {\mc{V}} \, \mc{V}^{-1}$ reads
\begin{equation}
  \pa {\mc{V}} \, \mc{V}^{-1} =
  \pa {H} \, H^{-1} + H (\pa {T} \, T^{-1}) H^{-1}.
  \label{invariantform10.15}
\end{equation}
The first term is the
invariant form encountered above in the discussion of the level
zero nonlinear sigma model for $GL(10, \mbb{R})/SL(10)$. So let us focus on
the second term. It is clear that $\pa {T} \, T^{-1}$ will contain
only positive generators at level $\geq 1$. So we set, in a
manner similar to Equation~(\ref{sumclosedform}),
\begin{equation}
  \pa {T} \, T^{-1} =
  \sum_{\alpha \in \Delta_+} \sum_s\mc{D} \mc{A}_{\alpha, s} \, E_{\alpha,s},
\end{equation}
 where the sum is over positive roots at levels
one and higher and takes into account multiplicities (through the
extra index $s$). The expressions $\mc{D} \mc{A}_{\alpha, s}$ are
linear in the time derivatives $\pa {\mc{A}}$. As before, we call them
``covariant derivatives''. They are computed by making use of the
Baker--Hausdorff formula, as in
Section~\ref{subsection:parametrization}. Explicitly, up to level 3,
one finds
\begin{equation}
  \dot{T} \, T^{-1} =
  \f{1}{3!}\mc{D}\mc{A}_{abc}(t)E^{abc}+
  \f{1}{6!}\mc{D}\mc{A}_{a_1\cdots a_6}(t)E^{a_1\cdots a_6}+
  \f{1}{9!}\mc{D}\mc{A}_{a|b_1\cdots b_8}(t)E^{a|b_1\cdots b_8}+\cdots,
\end{equation}
with 
\begin{equation}
  \begin{array}{rcl}
    \mc{D}\mc{A}_{abc}(t) &=& \displaystyle
    \pa \mc{A}_{abc}(t),
    \\ [0.25 em]
    \mc{D}\mc{A}_{a_1\cdots a_6}(t)&=& \displaystyle
    \pa \mc{A}_{a_1\cdots a_6}(t) +10\mc{A}_{[a_1a_2a_3}(t)
    \pa\mc{A}_{a_4a_5a_6]}(t),
    \\ [0.25 em]
    \mc{D}\mc{A}_{a|b_1\cdots b_8}(t) &=& \displaystyle
    \pa \mc{A}_{a|b_1\cdots b_8}(t)+42\mc{A}_{\left< ab_1b_2\right.}(t)
    \pa \mc{A}_{\left. b_3\cdots b_8\right>}(t)-42
    \pa\mc{A}_{\left< ab_1b_2\right.}(t)\mc{A}_{\left. b_3\cdots b_8\right>}(t),
    \\ [0.25 em]
    & & \displaystyle
    + 280 \mc{A}_{\left< ab_1b_2\right.}(t)
    \mc{A}_{b_3b_4b_5}(t)\pa\mc{A}_{\left. b_6b_7b_8\right> }(t),
  \end{array}
  \label{CovariantDerivatives}
\end{equation}
as computed in~\cite{DHN2}. The notation $\left< a_1 \cdots
a_k\right>$ denotes projection onto the Young tableaux symmetry
carried by the field upon which the covariant derivative
acts\epubtkFootnote{As an example, consider the projection
$P_{\al\be\ga}\equiv T_{\left<\al\be\ga\right>}$ of a three index
tensor $T_{\al\be\ga}$ onto the Young tableaux
\begin{displaymath}
  {\footnotesize
    \setlength{\tabcolsep}{0.55 em}
    \begin{tabular}{cc}
      \cline{1-2}
      \multicolumn{1}{|c|}{} &
      \multicolumn{1}{c|}{} \\
      \cline{1-2}
      \multicolumn{1}{|c|}{} \\
      \cline{1-1}
    \end{tabular}}\,.
\end{displaymath}
This projection is given by
\begin{displaymath}
  P_{\al\be\ga}=\f{1}{3}(T_{\al\be\ga}+T_{\be \al\ga}-
  T_{\ga\be\al}-T_{\be\ga\al}),
\end{displaymath}
which clearly satisfies
\begin{displaymath}
  P_{\al\be\ga}=-P_{\ga\be \al},
  \qquad
  P_{[\al\be\ga]}=0.
\end{displaymath}
Note also that $P_{\al\be\ga}\neq P_{\be\al\ga}$.}. It should be
stressed that the covariant derivatives $\mc{D}\mc{A}$ have the same
transformation properties under $SO(10)$ (under which they are inert)
and $GL(10, \mbb{R})$ as $\mc{A}$ since the $GL(10, \mbb{R})$
transformations do not depend on time.


\subsubsection[The $\mc{K}(\mc{E}_{10})\times \mc{E}_{10}$-invariant
  action at low levels]%
              {The \boldmath $\mc{K}(\mc{E}_{10})\times \mc{E}_{10}$-invariant
  action at low levels}

The action can now be computed using the bilinear form $(\cdot
|\cdot )$ on $E_{10}$,
\begin{equation}
  S_{\mc{E}_{10}/\mc{K}(\mc{E}_{10})}= \int dt\, n(t)^{-1}
  \left(\mc{P}(t)\big|\mc{P}(t)\right),
\end{equation}
where $\mc{P}$ is
obtained by projecting orthogonally onto the subalgebra
$\mf{k}_{E_{10}}$ by using the generalized transpose,
\begin{equation}
  ({K^{a}}_b)^{\mc{T}}={K^{b}}_a,
  \qquad
  (E^{abc})^{\mc{T}}=F_{abc}, \cdots \mbox{etc.},
\end{equation}
where
as above $(\ )^{\mc{T}}=-\om(\ )$ (with $\om$ being the Chevalley
involution). \index{Chevalley involution}
We shall compute the action up to, and including, level~3,
\begin{equation}
  S_{\mc{E}_{10}/\mc{K}(\mc{E}_{10})}= \int dt\, n(t)^{-1}
  \left(\mc{L}_0+\mc{L}_1+\mc{L}_2+\mc{L}_3+ \cdots\right).
\end{equation}

From Equation~(\ref{invariantform10.15}) and the fact that generators at
level zero are orthogonal to generators at levels $\not=0$, we see
that $\mc{L}_0$ will be constructed from the level zero part
$\dot{H} \, H^{-1}$ and will coincide with the Lagrangian
(\ref{LagrangianforGL10SO10}) for the nonlinear sigma model
$GL(10, \mbb{R})/SO(10)$,
\begin{equation}
  \mc{L}_0=\f{1}{4}\left(g^{ac}(t)g^{bd}(t)-g^{ab}(t)g^{cd}(t)\right)
  \pa g_{ab}(t) \, \pa g_{cd}(t).
\end{equation}

To compute the other terms, we use the following trick. The
Lagrangian must be a $GL(10, \mbb{R})$ scalar. One can easily compute it
in the frame where $H= 1$, i.e., where the metric $g_{ab}$ is
equal to $\delta_{ab}$. One can then covariantize the resulting
expression by replacing everywhere $\delta_{ab}$ by $g_{ab}$. To
illustrate the procedure consider the level $1$ term. One has, for
$H = 1$ and at level $1$, $\pa {\mc{V}} \, \mc{V}^{-1} =
\f{1}{3!}\mc{D}\mc{A}_{abc}(t)E^{abc}$ and thus, with the same gauge 
conditions, $\mc{P}(t) = \f{1}{2 \cdot
3!}\mc{D}\mc{A}_{abc}(t)\left(E^{abc} + F^{abc} \right)$ (where we
have raised the indices of $F_{abc}$ with $\delta^{ab}$, $F_{123}
\equiv F^{123}$ etc). Using $(E^{a_1 a_2 a_3}|F^{b_1 b_2 b_3})=
\delta^{a_1 b_1} \delta^{a_2 b_2} \delta^{a_3 b_3} \pm \mbox
{permutations that make the expression antisymmetric}$ (3!
terms; see Section~\ref{section:DecompE10E10}), one then gets $\mc{L}_1 = \f{1}{2 \cdot
3!}\mc{D}\mc{A}_{abc}(t) \, \mc{D}\mc{A}_{def}(t) \, \delta^{ad}
\delta^{be} \delta^{cf}$ in the frame where $g_{ab} =
\delta_{ab}$. This yields the level 1 Lagrangian in a general
frame, 
\begin{eqnarray}
  \mc{L}_1 &=& \f{1}{2\cdot 3!} g^{a_1c_1}g^{a_2c_2}g^{a_3c_3}
  \mc{D}\mc{A}_{a_1a_2a_3}(t)\,\mc{D}\mc{A}_{c_1c_2c_3}(t)
  \nonumber
  \\
  &=& \f{1}{2\cdot 3!}\mc{D}\mc{A}_{a_1a_2a_3}(t)\,
  \mc{D}\mc{A}^{a_1a_2a_3}(t). 
\end{eqnarray}
By a similar analysis, the level~2 and 3 contributions are 
\begin{equation}
  \begin{array}{rcl}
    \mc{L}_2 &=& \displaystyle
    \f{1}{2\cdot 6!}\,\mc{D}\mc{A}_{a_1\cdots a_6}(t)\,
    \mc{D}\mc{A}^{a_1\cdots a_6}(t),
    \\ [0.75 em]
    \mc{L}_3 &=& \displaystyle
    \f{1}{2\cdot 9!}\,\mc{D}\mc{A}_{a|b_1\cdots b_8}(t)\,
    \mc{D}\mc{A}^{a|b_1\cdots b_8}(t).
  \end{array}
\end{equation}
Collecting all terms, the final form of the action for
$\mc{E}_{10}/\mc{K}(\mc{E}_{10})$ up to and including level
$\ell=3$ is 
\begin{eqnarray}
  S_{\mc{E}_{10}/\mc{K}(\mc{E}_{10})}&=&
  \int dt\, n(t)^{-1}\left[ \f{1}{4}
  \left(g^{ac}(t)g^{bd}(t)-g^{ab}(t)g^{cd}(t)\right)
  \pa g_{ab}(t)\,\pa g_{cd}(t) \right.
  \nonumber
  \\
  & & \qquad \qquad \quad~~
  + \f{1}{2\cdot 3!}\mc{D}\mc{A}_{a_1a_2a_3}(t)\,
  \mc{D}\mc{A}^{a_1a_2a_3}(t)+ \f{1}{2\cdot 6!}
  \mc{D}\mc{A}_{a_1\cdots a_6}(t)\,\mc{D}\mc{A}^{a_1\cdots a_6}(t)
  \nonumber
  \\
  & & \qquad \qquad \quad~~
  \left. +\f{1}{2\cdot 9!}\mc{D}\mc{A}_{a|b_1\cdots b_8}(t)\,
  \mc{D}\mc{A}^{a|b_1\cdots b_8}(t)+\cdots \right],
  \label{E10action}
\end{eqnarray}
which agrees with the action found in~\cite{DHN2}.


\subsubsection{The correspondence}
\label{correspondence}

We shall now relate the equations of motion for the
$\mc{E}_{10}/\mc{K}(\mc{E}_{10})$ sigma model to the equations of
motion of eleven-dimensional supergravity. As the precise
correspondence is not yet known, we shall here only
sketch the main ideas. These work remarkably well at low levels
but need unknown ingredients at higher levels.

We have seen that the sigma model for 
$\mc{E}_{10}/\mc{K}(\mc{E}_{10})$ can be consistently truncated
level by level. More precisely, one can consistently set equal to zero all
covariant derivatives of the fields above a given level and get a
reduced system whose solutions are solutions of the full system.
We shall show here that the consistent truncations of
$\mc{E}_{10}/\mc{K}(\mc{E}_{10})$ at levels 0, 1 and 2
yields equations of motion that coincide with the equations of
motion of appropriate consistent truncations of eleven-dimensional
supergravity, using a prescribed dictionary presented below. We will
also show that the correspondence extends up to parts of level~3.

We recall that in the gauge $N^i=0$ (vanishing shift) and
$A_{0bc}=0$ (temporal gauge), the bosonic fields of eleven-dimensional
supergravity are the spatial metric $\g_{ab}(x^{0},x^{i})$, the
lapse $N(x^{0},x^{i})$ and the spatial components
$A_{abc}(x^0, x^{i})$ of the vector potential 3-form. The
physical field is $F = dA$ and its electric and magnetic
components are, respectively, denoted $F_{0abc}$ and $F_{abcd}$.
The electric field involves only time derivatives of
$A_{abc}(x^0, x^{i})$, while the magnetic field involves spatial
gradients.

\subsubsection*{Levels 0 and 1}

If one keeps only levels zero and one, the sigma model
action~(\ref{E10action}) reduces to
\begin{eqnarray}
  S[g_{ab}(t), \mc{A}_{abc}(t), n(t)] &=& \int dt\,n(t)^{-1}
  \left[\f{1}{4}\left(g^{ac}(t)\,g^{bd}(t)-g^{ab}(t)\,g^{cd}(t)\right)
  \pa {g}_{ab}(t)\,\pa {g}_{cd}(t) \right.
  \nonumber
  \\
  && \qquad \qquad \quad~~
  \left. + \f{1}{2\cdot 3!}\pa {\mc{A}}_{a_1a_2a_3}(t)\,
  \pa {\mc{A}}^{a_1a_2a_3}(t)\right].
  \label{levels0and1}
\end{eqnarray}

Consider now the consistent homogeneous truncation of
eleven-dimensional supergravity in which the spatial metric, the
lapse and the vector potential depend only on time (no spatial
gradient). Then the reduced action for this truncation is
precisely Equation~(\ref{levels0and1}) provided one makes the
identification $t=x^{0}$ and 
\begin{eqnarray}
  g_{ab}(t)&=& \g_{ab}(t),
  \\
  \mc{A}_{abc}(t) &=& A_{abc}(t),
  \\
  n(t) &=& \frac{N(t)}{\sqrt{\g(t)}}
\end{eqnarray}
(see, for instance,~\cite{Demaret}). Also the Hamiltonian
constraints (the only one left) coincide. Thus, there is a perfect
match between the sigma model truncated at level one and
supergravity ``reduced on a 10-torus''. If one were to drop level
one, one would find perfect agreement with pure gravity. In the
following, we shall make the gauge choice $N = \sqrt{\g}$, equivalent
to $n=1$.

\subsubsection*{Level 2}

At levels 0 and 1, the supergravity fields $\g_{ab}$ and $A_{abc}$
depend only on time. When going beyond this truncation, one needs
to introduce some spatial gradients. Level~2 introduces spatial
gradients of a very special type, namely allows for a homogeneous
magnetic field. This means that $A_{abc}$ acquires a space
dependence, more precisely, a linear one (so that its gradient
does not depend on $x$). However, because there is no room for
$x$-dependence on the sigma model side, where the only independent
variable is $t$, we shall use the trick to describe the magnetic
field in terms of a dual potential $A_{a_1 \cdots a_6}$. Thus,
there is a close interplay between duality, the sigma model
formulation, and the introduction of spatial gradients.

There is no tractable, fully satisfactory variational formulation
of eleven-dimensional supergravity where both the 3-form
potential and its dual appear as independent variables in the
action, with a quadratic dependence on the time derivatives (this
would be double-counting, unless an appropriate self-duality
condition is imposed~\cite{Cremmer:1997ct,
 Cremmer:1998px}). This means that from now on, we shall not compare
the actions of the sigma model and of supergravity but, rather, only
their respective equations of motion. As these involve the
electromagnetic field and not the potential, we rewrite the
correspondence found above at levels 0 and 1 in terms of the metric
and the electromagnetic field as 
\begin{equation}
  \begin{array}{rcl}
    g_{ab}(t)&=& \g_{ab}(t),
    \\ [0.25 em]
    \mc{D}\mc{A}_{abc}(t) &=& F_{0abc}(t).
  \end{array}
\end{equation}
The equations of motion for the nonlinear sigma model, obtained from
the variation of the Lagrangian Equation~(\ref{E10action}), truncated
at level two, read explicitly 
\begin{equation}
  \begin{array}{rcl}
    \displaystyle
    \f{1}{2}\pa \left(n(t)^{-1}g^{ac}(t)\,\pa g_{cb}(t)\right) &=&
    \displaystyle
    \f{n(t)^{-1}}{4}\left(\mc{D}\mc{A}^{ac_1c_2}(t)\,
    \mc{D}\mc{A}_{bc_1c_2}(t)-\f{1}{9}
    {\delta^{a}}_b \mc{D}\mc{A}^{c_1c_2c_3}(t)\,
    \mc{D}\mc{A}_{c_1c_2c_3}(t) \right)
    \\ [1 em]
    & & \displaystyle
    +\f{n(t)^{-1}}{2\cdot 5!}
    \left(\mc{D}\mc{A}^{a c_1\cdots c_5}(t)\,
    \mc{D}\mc{A}_{b c_1\cdots c_5}(t)-\f{1}{9} {\delta^{a}}_b
    \mc{D}\mc{A}^{c_1\cdots c_6}(t)\,
    \mc{D}\mc{A}_{c_1\cdots c_6}(t)\right),
    \\ [1 em]
    \pa\left(n(t)^{-1}\mc{D}\mc{A}^{a_1a_2a_3}(t)\right)&=&
    \displaystyle
    - \f{1}{3!}\, n(t)^{-1} \mc{D}\mc{A}^{a_1\cdots a_6}(t)\,
    \mc{D}\mc{A}_{a_4a_5a_6}(t),
    \\ [1 em]
    \pa\left(n(t)^{-1}\mc{D}\mc{A}^{a_1\cdots a_6}(t)\right)&=& 0.
  \end{array}
  \label{EOMLevel2}
\end{equation}

In addition, we have the constraint obtained by varying $n$, 
\begin{eqnarray}
  \left(\mc{P}(t)|\mc{P}(t)\right)& =& 
  \f{1}{4}\left(g^{ac}(t)\,g^{bd}(t)-g^{ab}(t)\,g^{cd}(t)\right)
  \pa g_{ab}(t)\,\pa g_{cd}(t)
  \nonumber
  \\
  & & + \f{1}{2\cdot 3!}\mc{D}\mc{A}^{a_1a_2a_3}(t)\,
  \mc{D}\mc{A}_{a_1a_2a_3}(t)+ \f{1}{2\cdot 6!}
  \mc{D}\mc{A}^{a_1\cdots a_6}(t)\,\mc{D}\mc{A}_{a_1\cdots a_6}(t) 
  \nonumber
  \\
  &=&0.
  \label{HamiltonianConstraintLevel2}
\end{eqnarray}

On the supergravity side, we truncate the
equations to metrics $\g_{ab}(t)$ and electromagnetic fields
$F_{0abc}(t)$, $F_{abcd}(t)$ that depend only on time. We take, as in
Section~\ref{section:BKL}, the spacetime metric to be of the form
\begin{equation}
  ds^2=-N^2(t)\,dt^2+\g_{ab}(t)\,dx^{a}\,dx^{b},
\end{equation}
but now with $x^0=t$. In the following we use Greek letters $\lambda,
\sigma, \rho, \cdots $ to denote eleven-dimensional spacetime indices,
and Latin letters $a, b, c, \cdots $ to denote ten-dimensional spatial
indices.

The equations of motion and the Hamiltonian constraint for
eleven-dimensional supergravity have been explicitly written
in~\cite{Demaret}, so they can be expediently compared with the
equations of motion of the sigma model. The dynamical equations for
the metric read 
\begin{eqnarray}
  \f{1}{2}\pa\left(\sqrt{\g}N^{-1}\g^{ac}\pa \g_{cb}\right) &=&
  \frac{1}{12} N \sqrt{\g} F^{a \rho \sigma \tau}F_{b \rho \sigma \tau} -
  \frac{1}{144}N \sqrt{\g} \, {\delta^a}_b\, F^{\lambda \rho \sigma \tau}
  F_{\lambda \rho \sigma \tau}
  \nonumber
  \\
  &=& \f{1}{4}N^{-1}\sqrt{\g}F^{0ac_1c_2}F_{0bc_1c_2}-
  \f{1}{36}N^{-1}\sqrt{\g}\, {\delta^{a}}_b\, F^{0c_1c_2c_3}F_{0c_1c_2c_3}
  \nonumber
  \\
  & & +\f{1}{12}N\sqrt{\g}F^{ac_1c_2c_3}F_{bc_1c_2c_3}-
  \f{1}{144}N\sqrt{\g}\, {\delta^{a}}_b\, F^{c_1c_2c_3c_4}F_{c_1c_2c_3c_4},
  \qquad
  \label{Einstein00}
\end{eqnarray}
and for the electric and magnetic fields we have, respectively, the
equations of motion and the Bianchi identity,
\begin{equation}
  \begin{array}{rcl}
    \displaystyle
    \pa\left(F^{0abc} N \sqrt{\g} \right) &=&
    \displaystyle
    \frac{1}{144} \varepsilon^{0 a b c d_1 d_2 d_3 e_1 e_2 e_3 e_4}
    F_{0 d_1 d_2 d_3} F_{e_1 e_2 e_3 e_4},
    \\ [0.75 em]
    \displaystyle
    \pa F_{a_1 a_2 a_3 a_4} &=& 0.
  \end{array}
  \label{ChernSimons} 
\end{equation}
Furthermore we have the Hamiltonian constraint \index{Hamiltonian constraint} 
\begin{equation}
  \f{1}{4}\left(\g^{ac}\g^{bd}-\g^{ab}\g^{cd}\right)
  \pa \g_{ab}\,\pa \g_{cd} + \frac{1}{12} F^{0abc}F_{0abc} +
  \frac{1}{48}N^2 F^{abcd}F_{abcd} = 0.
  \label{HamiltonianC0}
\end{equation}
(We shall not consider the other constraints here; see remarks as at
the end of this section.)

One finds again perfect agreement between the sigma model equations,
Equation~(\ref{EOMLevel2}) and~(\ref{HamiltonianConstraintLevel2}),
and the equations of eleven-dimensional supergravity,
Equation~(\ref{Einstein00}) and~(\ref{HamiltonianC0}), provided one
extends the above dictionary through~\cite{DHN2}
\begin{equation}
  \mc{D}\mc{A}^{a_1 \cdots a_6}(t) = - \frac{1}{4!}
  \varepsilon^{a_1 \cdots a_6 b_1 b_2 b_3 b_4}F_{b_1 b_2 b_3 b_4}(t).
\end{equation}
This result appears to be quite remarkable, because the Chern--Simons
term in Equation~(\ref{ChernSimons}) is in particular reproduced with
the correct coefficient, which in eleven-dimensional supergravity is
fixed by invoking supersymmetry.

\subsubsection*{Level 3}

Level~3 should correspond to the introduction of further
controlled spatial gradients, this time for the metric. Because
there is no room for spatial derivatives as such on the sigma
model side, the trick is again to introduce a dual graviton field.
When this dual graviton field is non-zero, the metric does depend
on the spatial coordinates.

Satisfactory dual formulations of non-linearized gravity do not exist. At
the linearized level, however, the problem is well understood
since the pioneering work by Curtright~\cite{Curtright:1980yk} (see
also~\cite{E11andMtheory, Bekaert:2002uh, Boulanger:2003vs}). In
eleven spacetime dimensions, the dual graviton field is {\em
described precisely by a tensor $\mc{A}_{a \vert b_1 \cdots b_8}$
with the mixed symmetry of the Young tableau $[1, 0, 0, 0, 0, 0,
0, 1,0]$ appearing at level~3 in the sigma model description}.
Exciting this field, i.e., assuming $\mc{D}\mc{A}_{a \vert b_1
\cdots b_8} \not=0$ amounts to introducing spatial gradients for the
metric -- and, for that matter, for the other fields as well -- as
follows. Instead of considering fields that are homogeneous on a
torus, one considers fields that are homogeneous on non-Abelian
group manifolds. This introduces spatial gradients (in coordinate
frames) in a well controlled manner.

Let $\theta^{a}$ be the group invariant one-forms, with structure
equations
\begin{equation}
  d \theta^a = \frac{1}{2} {C^a}_{bc} d \theta^b \wedge d \theta^c.
\end{equation}
We shall assume that ${C^a}_{ac} = 0$ (``Bianchi class
A''). Truncation at level~3 assumes that the metric and the electric
and magnetic fields depend only on time in this frame and that the
${C^a}_{bc}$ are constant (corresponding to a group). The
supergravity equations have been written in that case
in~\cite{Demaret} and can be compared with the sigma model
equations. There is almost a complete match between both sets of
equations provided one extends the dictionary at level~3 through
\begin{equation}
  \mc{D}\mc{A}^{a| b_1\cdots b_8}(t)=
  \f{3}{2}\varepsilon^{b_1\cdots b_8 cd}{{C}^{a}}_{cd}
\end{equation}
(with the equations of motion of the sigma model implying indeed that
$\mc{D}\mc{A}^{a| b_1\cdots b_8}$ does not depend on time). Note that
to define an invertible mapping between the level three fields and the
${C^{a}}_{bc}$, it is important that ${C^{a}}_{bc}$ be traceless;
there is no ``room'' on level three on the sigma model side to
incorporate the trace of ${C^{a}}_{bc}$.

With this correspondence, the match works perfectly for real roots up
to, and including, level three. However, it fails for fields
associated with imaginary roots (level~3 is the first time imaginary
roots appear, at height 30)~\cite{DHN2}. In fact, the terms that match
correspond to ``$SL(10, \mbb{R})$-covariantized $E_8$'', i.e., to
fields associated with roots of $E_8$ and their images under the Weyl
group of $SL(10, \mbb{R})$.

Since the match between the sigma model equations and supergravity
fails at level~3 under the present line of investigation, we shall
not provide the details but refer instead to~\cite{DHN2} for more
information. The correspondence up to level~3 was also checked
in~\cite{DamourNicolaiLowLevels} through a slightly different
approach, making use of a formulation with local frames, i.e., using
local flat indices rather than global indices as in the present
treatment.

Let us note here that higher level fields of $E_{10}$, corresponding
to imaginary roots, have been considered from a different point of
view in~\cite{Brown:2004jb}, where they were associated with certain
brane configurations (see also~\cite{Brown:2004ar, Bagnoud:2006mz}).

\subsubsection*{The dictionary}

One may view the above failure at level~3 as a serious flaw
to the sigma model approach to exhibiting the $E_{10}$
symmetry\epubtkFootnote{This does not exclude that other approaches
  would be successful. That $E_{10}$, or perhaps $E_{11}$, does encode
  a lot of information about M-theory is a fact, but that this should
  be translated into a sigma model reformulation of the theory appears
  to be questionable.}. Let us, however, be optimistic for a moment
and assume that these problems will somehow get resolved, perhaps by
changing the dictionary or by including higher order terms. So, let us
proceed.

What would be the meaning of the higher level fields? As discussed in
Section~\ref{section:higherlevels}, there are indications that fields
at higher levels contain higher order spatial gradients and therefore
enable us to reconstruct completely, through something similar to a
Taylor expansion, the most general field configuration from the fields
at a given spatial point.

From this point of view, the relation between the supergravity degrees
of freedom $\g_{ij}(t,x)$ and $F_{(4)}(t,x)=dA_{(3)}(t, x)$ would be
given, at a specific spatial point $x=\mathbf{x}_0$ and in a suitable
spatial frame $\theta^{a}(x)$ (that would also depend on $x$), by the
following ``dictionary'':
\begin{equation}
  \begin{array}{rcl}
    g_{ab}(t) &=& \g_{ab}(t, \mathbf{x}_0),
    \\ [0.75 em]
    \mc{D}\mc{A}_{abc}(t)& =& F_{tabc}(t, \mathbf{x}_0),
    \\ [0.75 em]
    \mc{D}\mc{A}^{a_1\cdots a_6}(t) &=& \displaystyle
    -\f{1}{4!}\varepsilon^{a_1\cdots a_6 bcde}F_{bcde}(t, \mathbf{x}_0),
    \\ [1 em]
    \mc{D}\mc{A}^{a| b_1\cdots b_8}(t)&=& \displaystyle
    \f{3}{2}\varepsilon^{b_1\cdots b_8 cd}{{C}^{a}}_{cd}(\mathbf{x}_0),
  \end{array}
\end{equation}
which reproduces in the homogeneous case what we have seen up to
level~3.

This correspondence goes far beyond that of the algebraic
description of the BKL-limit \index{BKL-limit}in terms of Weyl reflections in the
simple roots of a Kac--Moody algebra. \index{Kac--Moody algebra}
Indeed, the dynamics of the
billiard is controlled entirely by the walls associated with {\em
simple} roots and thus does not transcend height one. Here, we go
to a much higher height and successfully extend (unfortunately
incompletely) the intriguing connection between eleven-dimensional
supergravity and $E_{10}$.


\subsubsection{Higher levels and spatial gradients}
\label{section:higherlevels}

We have seen that the correspondence between the
$\mc{E}_{10}$-invariant sigma model and eleven-dimensio\-nal
supergravity fails when we include spatial gradients beyond first
order. It is nevertheless believed that the information about
spatial gradients is somehow encoded within the algebraic
description: One idea is that space is ``smeared out'' among
the infinite number of fields contained in $\mc{E}_{10}$ and it is
for this reason that a direct dictionary for the inclusion of
spatial gradients is difficult to find. If true, this would imply
that we can view the level expansion on the algebraic side as
reflecting a kind of ``Taylor expansion'' in spatial gradients on
the supergravity side. Below we discuss some speculative ideas
about how such a correspondence could be realized in practice.

\subsubsection*{The ``gradient conjecture''}

One intriguing suggestion put forward in~\cite{DHN2}
was that fields associated to certain ``affine representations''
of $E_{10}$ \index{$E_{10}$}could be interpreted as spatial derivatives acting on
the level one, two and three fields, thus providing a direct
conjecture for how space ``emerges'' through the level
decomposition of $E_{10}$. The representations in question
are those for which the Dynkin label associated with the
overextended root of $E_{10}$ vanishes, and hence these
representations are realized also in a level decomposition \index{level decomposition}of the
regular $E_9$-subalgebra obtained by removing the overextended
node in the Dynkin diagram of $E_{10}$.

The affine representations were discussed in
Section~\ref{section:LevelDecomposition} and we recall that they are
given in terms of three infinite towers of generators, with the
following $\mf{sl}(10, \mbb{R})$-tensor structures,
\begin{equation}
  {E^{a_1a_2a_3}}_{b_1\cdots b_k},
  \qquad
  {E^{a_1\cdots a_6}}_{b_1\cdots b_k},
  \qquad
  {E^{a_1|a_2\cdots a_9}}_{b_1\cdots b_k},
\end{equation}
where the upper indices have the same Young tableau symmetries as the
$\ell=1, 2$ and $3$ representations, while the lower indices are all
completely symmetric. In the sigma model these generators of
$\mc{E}_{10}$ are parametrized by fields exhibiting the same index
structure, i.e., ${\mc{A}_{a_1a_2a_2}}^{b_1\cdots b_k}(t)$,
${\mc{A}_{a_1\cdots a_6}}^{b_1\cdots b_k}(t)$ and
${\mc{A}_{a_1|a_2\cdots a_9}}^{b_1\cdots b_k}(t)$.

The idea is now that the three towers of fields have precisely the
right index structure to be interpreted as spatial gradients of
the low level fields 
\begin{equation}
  \begin{array}{rcl}
    {\mc{A}_{a_1a_2a_2}}^{b_1\cdots b_k}(t)&=&
    \pa^{b_1}\cdots \pa^{b_k}\mc{A}_{a_1a_2a_3}(t),
    \\ [0.25 em]
    {\mc{A}_{a_1\cdots a_6}}^{b_1\cdots b_k}(t)&=&
    \pa^{b_1}\cdots \pa^{b_{k}}\mc{A}_{a_1\cdots a_6}(t),
    \\ [0.25 em]
    {\mc{A}_{a_1|a_2\cdots a_9}}^{b_1\cdots b_k}(t)&=&
    \pa^{b_1}\cdots \pa^{b_{k}}\mc{A}_{a_1|a_2\cdots a_9}(t).
  \end{array}
\end{equation}

Although appealing and intuitive as it is, this conjecture is
difficult to prove or to check explicitly, and not much progress
in this direction has been made since the original proposal.
However, recently~\cite{E9multiplets} this problem was attacked
from a rather different point of view with some very interesting
results, indicating that the gradient conjecture may need to be
substantially modified. For completeness, we briefly review here
some of the main features of~\cite{E9multiplets}.

\subsubsection*{U-duality and the Weyl Group of \boldmath $E_9$}

Recall from Section~\ref{section:KacMoody} that
the infinite-dimensional Kac--Moody algebras $E_9$ and $E_{10}$ can
be obtained from $E_8$ through prescribed extensions of the $E_8$
Dynkin diagram: $E_9=E_8^{+}$ is obtained by extending with one
extra node, and $E_{10}=E_8^{++}$ by extending with two extra
nodes. This procedure can be continued and after extending $E_8$
three times, one finds the Lorentzian Kac--Moody algebra
$E_{11}=E_8^{+++}$, which is also believed to be relevant as a
possible underlying symmetry of M-theory~\cite{E11andMtheory,
SymmetryOfMtheories}.

These algebras are part of the chain of exceptional regular
embeddings,
\begin{equation}
  \cdots E_{8} \subset E_9 \subset E_{10} \subset E_{11} \cdots,
\end{equation}
which was used in~\cite{FromE11toE10} to show that a sigma model for
the coset space $\mc{E}_{11}/\mc{K}(\mc{E}_{11})$ can be consistently
truncated to a sigma model for the coset space
$\mc{E}_{10}/\mc{K}(\mc{E}_{10})$, which coincides with
Equation~(\ref{E10action}). This result builds upon previous work
devoted to general constructions of sigma models invariant under
Lorentzian Kac--Moody algebras of
$\mf{g}^{+++}$-type~\cite{SymmetryOfMtheories, BPSEnglert,
  BraneDynamicsEnglert, IntersectingEnglert}.

It was furthermore shown in~\cite{FromE11toE10} that by performing a
suitable Weyl reflection before truncation, yet another sigma
model based on $\mc{E}_{10}$ could be obtained. It differs from
Equation~(\ref{E10action}) because the parameter along the geodesic is a
\emph{spacelike}, not timelike, variable in spacetime. This
follows from the fact that the sigma model is constructed from the
coset space $\mc{E}_{10}/\mc{K}^{-}(\mc{E}_{10})$, where
$\mc{K}^{-}(\mc{E}_{10})$ coincides with the noncompact group $SO(1,
9)$ at level zero, and not $SO(10)$ as is the case for
Equation~(\ref{E10action}). The two sigma model actions were referred
to in~\cite{FromE11toE10} as $S_{\mathrm{cosmological}}$ and
$S_{\mathrm{brane}}$, since solutions to the first model translate to
time-dependent (cosmological) solutions of eleven-dimensional
supergravity, while the second model gives rise to stationary (brane)
solutions, which are smeared in all but one spacelike direction. In
particular, the $\ell=1$ and $\ell=2$ fields correspond to potentials
for the $M2$- and $M5$-branes, respectively.

In~\cite{E9multiplets}, solutions associated to the infinite tower
of affine representations for the brane sigma model based on
$\mc{E}_{10}/\mc{K}^{-}(\mc{E}_{10})$ were investigated. The idea
was that by restricting the indices to be $\mf{sl}(9,
\mbb{R})$-indices, any such representation coincides with a
generator of $E_9$, and so different fields in these affine towers
must be related by Weyl reflections in $E_{9}$.

The Weyl group $\mf{W}[E_9]$ is a subgroup of the U-duality group
$\mc{E}_9(\mbb{Z})$ of M-theory compactified on $T^{9}$ to two
spacetime dimensions. Moreover, the continuous group
$\mc{E}_9=\mc{E}_{9}(\mbb{R})$ is the M-theory analogue of the Geroch
group, i.e., it is a symmetry of the space of solutions of $N=16$
supergravity in two dimensions~\cite{E9Geroch}. Under these
considerations it is natural to expect that the fields associated with
the affine representations should somehow be related to the infinite
number of ``dual potentials'' appearing in connection with the Geroch
group in two dimensions. Indeed, the authors of~\cite{E9multiplets}
were able to show, using the embedding $\mf{W}[E_9]\subset
\mf{W}[E_{10}]$, that given, e.g., a representation in the $\ell=1$
affine tower, there exists a $\mf{W}[E_9]\subset
\mc{E}_{9}(\mbb{Z})$-transformation that relates the associated field
to the lowest $\ell=1$ generator $E^{a_1a_2a_3}$. The resulting
solution, however, is different from the standard brane solution
obtained from the $\ell=1$-field because the new solution is smeared
in all directions \emph{except two spacelike directions}, i.e., the
solution is an $M2$-brane solution which depends on two spacelike
variables.

Thus, by taking advantage of the embedding $E_9\subset E_{10}$, it
was shown that the three towers of ``gradient representations''
encode a kind of ``de-compactification'' of one spacelike
variable. In a way this therefore indicates that part of the
gradient conjecture must be correct, in the sense that the towers
of affine representations indeed contain information about the
emergence of spacelike directions. On the other hand, it also
seems that the correspondence is more complicated than was
initially believed, perhaps deeply connected to U-duality in some,
as of yet, unknown way.


\subsection{Further comments}


\subsubsection{Massive type~IIA supergravity}

We have just seen that some of the
higher level fields might have an interpretation in terms of
spatial gradients. This would account for a subclass of
representations at higher levels. The existence of other
representations at each level besides the ``gradient
representations'' shows that the sigma model contains further
degrees of freedom besides the supergravity fields, conjectured
in~\cite{DHN2} to correspond to M-theoretic degrees of freedom and
(quantum) corrections.

The gradient representations have the interesting properties that
their highest $\mf{sl}(10, \mbb{R})$-weight is a real root. There are
other representations with the same properties. An interesting
interpretation of some of those has been put forward recently using
dimensional reduction, as corresponding to the $(D-1)$-forms that
generate the cosmological constant for maximal gauged supergravities
in $D$ spacetime dimensions~\cite{Bergshoeff:2007qi, Riccioni:2007au,
  E9multiplets}. (A cosmological constant that appears as a constant
  of integration can be described by a
  $(D-1)$-form~\cite{Aurilia:1980xj, Henneaux:1984ji}.) For
definiteness, we shall consider here only the representations at
level~4, related to the mass term of type~IIA theory.

There are two representations at level~4, both of them with a highest
weight which is a real root of $E_{10}$, namely $[0,0,1,0,0,0,0,0,1]$
and $[2,0,0,0,0,0,0,0,0]$~\cite{NicolaiFischbacher}. The lowest weight
of the first one is, in terms of the scale factors, $2(\be^1 + \be^2 +
\be^3) + \be^4 + \be^5 + \be^6 + \be^7 + \be^8 + \be^9$. The lowest
weight of the second one is $3 \be^1 + \be^2 + \be^3 + \be^4 + \be^5 +
\be^6 + \be^7 + \be^8 + \be^9 + \be^{10}$. Both weights are easily
verified to have squared length equal to 2 and, since they are on the
root lattice, they are indeed roots by the criterion for roots of
hyperbolic algebras. The first representation is described by a tensor
with mixed symmetry $\mc{A}_{a_1 a_2 a_3 \vert b_1 b_2 \cdots b_9}$
corresponding, as we have seen, to the conjectured gradient
representation (with one derivative) of the level~1 field $\mc{A}_{a_1
  a_2 a_3}$. We shall thus focus on the second representation,
described by a tensor $\mc{A}_{a_1 \vert b_1 \vert c_1 c_2 \cdots
  c_{10}}$.

By dimensional reduction along the first direction, the
representation $[2,0,0,0,0,0,0,0,0]$ splits into various $\mf{sl}(9, \mbb{R})$
representations, one of which is described by the completely
antisymmetric field $\mc{A}_{c_2 \cdots c_{10}}$, i.e., a 9-form (in ten
spacetime dimensions). It is obtained by taking $a_1= b_1 = c_1 =
1$ in $\mc{A}_{a_1 \vert b_1 \vert c_1 c_2 \cdots c_{10}}$ and
corresponds precisely to the lowest weight $\gamma = 3 \be^1 +
\be^2 + \be^3 + \be^4 + \be^5 + \be^6 + \be^7 + \be^8 + \be^9 +
\be^{10}$ given above. If one rewrites the corresponding term
$\sim \mc{D}\mc{A}_{1 \vert 1 \vert 1 c_2 \cdots c_{10}}^2 e^{2
\gamma} $ in the Lagrangian in terms of ten-dimensional scale
factors and dilatons, one reproduces, using the field equations
for $\mc{A}_{1 \vert 1 \vert 1 c_2 \cdots c_{10}}$, the mass term of
massive Type~IIA supergravity.

The fact that $E_{10}$ \index{$E_{10}$} contains information about the massive Type~IIA
theory is in our opinion quite profound because, contrary to the
low level successes which are essentially a covariantization of known
$E_8$ results, this is a true $E_{10}$ test. The understanding of the
massive Type~IIA theory in the light of infinite Kac--Moody algebras
was studied first in~\cite{Schnakenburg:2002xx}, where the embedding
of the mass term in a nonlinear realisation of $E_{11}$ was
constructed. The precise connection between the mass term and an
$E_{10}$ positive real root was first explicitly made
in Section $6.5$ of~\cite{InvarianceUnderCompactification}. It is interesting to note
that even though the corresponding representation does not appear in
$E_9$, it is present in $E_{10}$ without having to go to $E_{11}$. The mass term of Type~IIA was also studied from the point of view of the $E_{10}$ coset model in~\cite{AxelThesis}.

This analysis suggests an interesting possibility for evading the
no-go theorem of~\cite{Bautier:1997yp} on the impossibility to
generate a cosmological constant in eleven-dimensional supergravity.
This should be tried by introducing new degrees of freedom
described by a mixed symmetry tensor $\mc{A}_{a_1 \vert b_1 \vert c_1
c_2 \cdots c_{10}}$. If this tensor can be consistently coupled
to gravity (a challenge in the context of field theory with a
finite number of fields!), it would provide the eleven-dimensional
origin of the cosmological constant in massive Type~IIA. There
would be no contradiction with~\cite{Bautier:1997yp} since in
eleven dimensions, the new term would not be a standard
cosmological constant, but would involve dynamical degrees of
freedom. This is, of course, quite speculative.

Finally, there are extra fields at higher levels besides spatial
gradients and the massive Type~IIA term. These might correspond to
higher spin degrees of freedom~\cite{DHN2, Boulanger:2003vs,
  Bunster:2006rt, West:2007mh}.


\subsubsection{Including fermions}

Another attractive aspect of the $E_{10}$-sigma model formulation is
that it can easily account for the fermions of supergravity up to the
levels that work in the bosonic sector. The fermions transform in
representations of the compact subalgebra $\mf{k}_{E_{10}}\subset
E_{10}$. An interesting feature of the analysis is that
$E_{10}$-covariance leads to $\mf{k}_{E_{10}}$-covariant derivatives
that coincides with the covariant derivatives dictated by
supersymmetry. This has been investigated in detail
in~\cite{deBuyl:2005zy, Damour:2005zs, deBuyl:2005mt, Damour:2006xu,
  Kleinschmidt:2006dy}, to which we refer the interested reader.


\subsubsection{Quantum corrections}

If the gradient conjecture is
correct (perhaps with a more sophisticated dictionary), then one
sees that the sigma model action would contain spatial derivatives of
higher order. It has been conjectured that these could perhaps
correspond to higher quantum corrections~\cite{DHN2}. This is
supported by the fact that the known quantum corrections of
M-theory do correspond to roots of $E_{10}$~\cite{Damour:2005zb}.

The idea is that with each correction curvature term of the form $R^N
\sqrt{-{}^{(11)}\g}$, where $R^N$ is a generic monomial of order
$N$ in the Riemann tensor, one can associate a linear form in the
scale factors $\be^\mu$'s in the BKL-limit. \index{BKL-limit} This linear form will be a
root of $E_{10}$ only for certain values of $N$. Hence compatibility
of the corresponding quantum correction with the $E_{10}$ structure
constrains the power $N$.

The evaluation of the curvature components in the BKL-limit goes
back to the paper by BKL themselves in four dimensions~\cite{BKL}
and was extended to higher dimensions in~\cite{BK1, Demaret1}. It
was rederived in~\cite{Damour:2005zb} for the purpose of
evaluating quantum corrections. It is shown in these references
that the leading terms in the curvature expressed in an
orthonormal frame adapted to the slicing are, in the BKL-limit, $
R_{\perp a \perp b}$ and $R_{abab}$ ($a \not=b$) which behave as
\begin{equation}
  R_{\perp a \perp b} \sim e^{2 \sigma},
  \qquad
  R_{abab} \sim e^{2 \sigma},
\end{equation}
where $\sigma$ is the sum of all the scale factors
\begin{equation}
  \sigma = \be^1 + \be^2 + \cdots + \be^{10},
\end{equation}
and where we have set $R_{\perp a \perp b}=N^{-2}R_{0a0b}$. This
implies that
\begin{equation}
  R \sim e^{2 \sigma},
  \qquad
  R^N \sim e^{2N \sigma},
  \qquad
  R^N \sqrt{- {}^{(11)}\g} \sim e^{2(N - 1)\sigma}.
\end{equation}

Now, $\sigma$ is not on the root lattice. It is not an integer
combination of the simple roots and it has length squared equal to
$- 10/9$. Integer combinations of the simple roots contains
$3\ell$ $\beta^\mu$'s, where $\ell$ is the level. Since 10 and
3 are relatively primes, the only multiples of $\sigma$ that are
on the root lattice are of the form $3k \sigma$, $k = 1, 2, 3 ,
\cdots $. These are negative, imaginary roots. The smallest value
is $k=1$, corresponding to the imaginary root
\begin{equation}
 \om(\be)=3 \sigma
\label{fundamentalweight}
\end{equation}
at level $-10$, with squared length $-10$. It follows that the only
quantum corrections compatible with the $E_{10}$ structure must have
$N-1 = 3k$, i.e., $N = 3k+1$~\cite{Damour:2005zb}, since it it only in
this case that $R^N \sqrt{- {}^{(11)}\g} \sim e^{-2 \ga}$ has $\ga
= -(N-1) \sigma$ on the root lattice. The first corrections are thus
of the form $R^4$, $R^7$, $R^{10}$ etc. This in in remarkable
agreement with the quantum computations of~\cite{Green:2005ba} (see
also~\cite{Russo:1997mk}).

The analysis of~\cite{Damour:2005zb} was completed
in~\cite{CompatibilityConditions} where it was observed that the
imaginary root (\ref{fundamentalweight}) was actually one of the
fundamental weights of $E_{10}$, namely, the fundamental weight
conjugate to the exceptional root that defines the level. In the case
of $E_{10}$, the root lattice and the weight lattice coincides, but
this observation was useful in the analysis of the quantum corrections
for other theories where the weight lattice is strictly larger than
the root lattice. The compatibility conditions seem in those cases to
be that quantum corrections should be associated with vectors on the
weight lattice. (See also~\cite{Lambert:2006he, Lambert:2006ny,
    Bao:2007er, Michel:2007vh}.)

Finally, we note that recent work devoted to investigations of
U-duality symmetries of compactified higher curvature corrections
indicates that the results reported here in the context of $E_{10}$
might require reconsideration~\cite{SwedishGuys}.


\subsubsection{Understanding duality}

The previous analysis has revealed that the hyperbolic Kac--Moody
algebra $E_{10}$ contains a large amount of information about the
structure and the properties of M-theory. How this should ultimately
be incorporated in the final formulation of the theory is, however,
not clear. 

The sigma model approach exhibits some important drawbacks and 
therefore it does not appear
to be the ultimate formulation of the theory. In addition to the
absence of a complete dictionary enabling one to go satisfactorily
beyond level~3 (the level where the first imaginary root appears),
more basic difficulties already appear at low levels. These are:

\begin{description}
\item[The Hamiltonian constraint] ~\\
  There is an obvious discrepancy between the Hamiltonian constraint 
  \index{Hamiltonian constraint} of the sigma model
  and the Hamiltonan constraint of supergravity. In the sigma model
  case, all terms are positive, except for the kinetic term of the
  scale factors, which contain a negative sign related to the
  conformal factor. On the supergravity side, the kinetic term of the
  scale factors matches correctly, but there are extra negative
  contributions coming from level~3 (something perhaps not too
  surprising if level~3 is to be thought as a dual formulation of
  gravity and hence contains in particular dual scale factors). How
  this problem can be cured by tractable redefinitions is far from
  obvious.
\item[Gauge invariance] ~\\
  The sigma model formulation corresponds
  to a partially gauge-fixed formulation since there are no arbitrary
  functions of time in the solutions of the equations of motion
  (except for the lapse function $n(t)$). The only gauge freedom left
  corresponds to time-independent gauge transformation (this is the
  equivalent of the ``temporal gauge'' of electromagnetism). The
  constraints associated with the spatial diffeomorphisms and with the
  3-form gauge symmetry have not been eliminated. How they are
  expressed in terms of the sigma model variables and how they fit
  with the $E_{10}$-symmetry is a question that should be
  answered. Progress along these lines may be found in recent
  work~\cite{Constraints}.
\item[Electric-magnetic duality] ~\\
  The sigma model approach
  contains both the graviton and its dual, as well as both the 3-form
  and its dual 6-form. Since these obey second-order equations of
  motion, there is a double-counting of degrees of freedom. For
  instance, the magnetic field of the 3-form would also appear as a
  spatial gradient of the 3-form at level~4, but nothing in the
  formalism tells that this is the same magnetic field as the time
  derivative of the 6-form at level~2. A generalized self-duality
  condition should be imposed~\cite{Cremmer:1997ct, Cremmer:1998px},
  not just in the 3-form sector but also for the graviton. Better yet,
  one might search for a duality-invariant action without
  double-counting. Such actions have been studied both for
  p-forms~\cite{Deser:1976iy, Deser:1997mz} and for
  gravity~\cite{Boulanger:2003vs, Henneaux:2004jw, Julia:2005ze} and
  are not manifestly spacetime covariant (this is not an issue here
  since manifest spacetime covariance has been given up anyway in the
  $(1+0)$-dimensional $\mc{E}_{10}$-sigma model). One must pick a
  spacetime coordinate, which might be time, or one spatial
  direction~\cite{Henneaux:1988gg, Schwarz:1993vs}. We feel that a
  better understanding of duality might yield an important
  clue~\cite{Boulanger:2003vs, Bunster:2006rt, Bunster:2007sn,
  Riccioni:2007hm}.
\end{description}

\newpage


\section[Cosmological Solutions from $E_{10}$]%
        {Cosmological Solutions from \boldmath $E_{10}$}
\label{section:cosmologicalsolutions}
\setcounter{equation}{0}

In this last main section we shall show that the low level
equivalence between the $\mc{E}_{10}/\mc{K}(\mc{E}_{10})$ sigma
model and eleven-dimensional supergravity can be put to practical
use for finding exact solutions of eleven-dimensional
supergravity. This is a satisfactory result because even in the
cosmological context of homogeneous fields $G_{\alpha \beta}(t)$,
$F_{\alpha \beta \gamma \delta}(t)$ that depend only on time
(``Bianchi~I cosmological models''~\cite{Demaret}), the equations of
motion of eleven-dimensional supergravity remain notoriously
complicated, while the corresponding sigma model is, at least
formally, integrable.

We will remain in the strictly cosmological sector where it is
assumed that all spatial gradients can be neglected so that all
fields depend only on time. Moreover, we impose diagonality of the
spatial metric. These conditions must of course be compatible with
the equations of motion; if the conditions are imposed initially,
they should be preserved by the time evolution.

A large class of solutions to eleven-dimensional supergravity
preserving these conditions were found in~\cite{Demaret}. These
solutions have zero magnetic field but have a restricted number of
electric field components turned on. Surprisingly, it was found
that such solutions have an elegant interpretation in terms of so
called \emph{geometric configurations}, 
\index{geometric configuration} denoted $(n_m, g_3)$, of
$n$ points and $g$ lines (with $n \leq 10$) drawn on a plane with
certain pre-determined rules. That is, for each geometric
configuration (whose definition is recalled below) one can
associate a diagonal solution with some non-zero electric field
components $F_{tijk}$, determined by the configuration. In this
section we re-examine this result from the point of view of the
sigma model based on $\mc{E}_{10}/\mc{K}(\mc{E}_{10})$.

We show, following~\cite{Henneaux:2006gp}, that each configuration
$(n_m, g_3)$ encodes information about a (regular) subalgebra
$\mgb$ of $E_{10}$, \index{$E_{10}$} and the supergravity solution
associated to the configuration $(n_m g_3)$ can be obtained by
restricting the $\mc{E}_{10}$-sigma model to the subgroup $\bar{\mcg}$
whose Lie algebra is $\mgb$. Therefore, we will here make use of both
the level truncation and the subgroup truncation simultaneously; first
by truncating to a certain level and then by restricting to the
relevant $\mgb$-algebra generated by a subset of the representations
at this level. Large parts of this section are based
on~\cite{Henneaux:2006gp}.


\subsection{Bianchi~I models and eleven-dimensional supergravity}

On the supergravity side, we will restrict the metric and the
electromagnetic field to depend on time only, 
\begin{equation}
  \begin{array}{rcl}
    ds^2 & = & - N^2(t) \, dt^2 + \g_{ab}(t)\,dx^a\,dx^b,
    \\ [0.25 em]
    F_{\lambda \rho \sigma \tau} & =&
    F_{\lambda \rho \sigma \tau}(t).
  \end{array}
\end{equation}
Recall from Section~\ref{section:E10SigmaModel} that with these
ans\"atze the dynamical equations of motion of eleven-dimensional
supergravity reduce to~\cite{Demaret}
\begin{eqnarray}
  \f{1}{2}\pa\left(\sqrt{\g}N^{-1}\g^{ac}\pa \g_{cb}\right) &=&
  \frac{1}{12} N \sqrt{\g} F^{a \rho \sigma \tau} F_{b \rho \sigma \tau} -
  \frac{1}{144}N \sqrt{\g} \, {\delta^a}_b\, F^{\lambda \rho \sigma \tau}
  F_{\lambda \rho \sigma \tau},
  \label{Einstein0}
  \\
  \pa\left(F^{tabc} N \sqrt{\g} \right) &=&
  \frac{1}{144} \varepsilon^{t a b c d_1 d_2 d_3 e_1 e_2 e_3 e_4}
  F_{t d_1 d_2 d_3} F_{e_1 e_2 e_3 e_4},
  \\
  \pa F_{a_1 a_2 a_3 a_4} &=& 0. 
\end{eqnarray}
This corresponds to the truncation of the sigma model at
level~2 which, as we have seen, completely matches the
supergravity side. We also defined $\pa\equiv \pa_t$ as in
Section~\ref{section:E10SigmaModel}. Furthermore we have the following
constraints,
\begin{eqnarray}
  \f{1}{4}\left(\g^{ac}\g^{bd}-\g^{ab}\g^{cd}\right)
  \pa\g_{ab}\,\pa\g_{cd} + \frac{1}{12} F^{tabc}F_{tabc} +
  \frac{1}{48}N^2 F^{abcd}F_{abcd} &=& 0,
  \label{HamiltonianC}
  \\
  \frac{1}{6} N F^{tbcd}F_{abcd} &=& 0,
  \label{momentum}
  \\
  \varepsilon^{tabc_1 c_2 c_3 c_4 d_1 d_2 d_3 d_4}
  F_{c_1 c_2 c_3 c_4} F_{d_1 d_2 d_3 d_4} &=& 0,
\end{eqnarray}
which are, respectively, the Hamiltonian constraint, momentum
constraint and Gauss' law. Note that Greek indices $\al, \be, \ga,
\cdots$ correspond to the full eleven-dimensional spacetime, while
Latin indices $a, b, c, \cdots$ correspond to the ten-dimensional
spatial part.

We will further take the metric to be purely time-dependent and
diagonal,
\begin{equation}
  ds^2 = - N^{2}(t) \, dt^2 + \sum_{i = 1}^{10} a_i^2(t) (dx^i)^2.
  \label{diagonalmetric}
\end{equation}
This form of the metric has manifest invariance under the ten distinct
spatial reflections 
\begin{equation}
  \begin{array}{rcl}
    x^j &\rightarrow & -x^j,
    \\ [0.25 em]
    x^{i\neq j} & \rightarrow & x^{i\neq j},
  \end{array}
\end{equation}
and in order to ensure compatibility with the Einstein equations, the
energy-momentum tensor of the 4-form field strength must also be
diagonal.


\subsubsection{Diagonal metrics and geometric configurations}

Assuming zero magnetic field (this restriction will be lifted below),
one way to achieve diagonality of the energy-momentum tensor is to
assume that the non-vanishing components of the electric field
$F^{\perp a b c}=N^{-1}F_{tabc}$ are determined by \emph{geometric
  configurations} $(n_m,g_3)$ with $n \leq 10$~\cite{Demaret}.

A geometric configuration \index{geometric configuration|bb} $(n_m,g_3)$ is a set of $n$ points and $g$
lines with the following incidence rules~\cite{Kantor, Hilbert, Page}:

\begin{enumerate}
\item Each line contains three points. \label{rule_1}
\item Each point is on $m$ lines. \label{rule_2}
\item Two points determine at most one line. \label{rule_3}
\end{enumerate}

It follows that two lines have at most one point
in common. It is an easy exercise to verify that $m n = 3 g $. An
interesting question is whether the lines can actually be realized
as straight lines in the (real) plane, but, for our purposes, it
is not necessary that it should be so; the lines can be bent.

Let $(n_m,g_3)$ be a geometric configuration with $n \leq 10$
points. We number the points of the configuration $1, \cdots, n$.
We associate to this geometric configuration a pattern of
electric field components $F^{\perp a b c}$ with the following
property: $F^{\perp a b c}$ can be non-zero only if the triple
$(a,b,c)$ is a line of the geometric configuration. If it is
not, we take $F^{\perp a b c} = 0$. It is clear that this
property is preserved in time by the equations of motion (in the
absence of magnetic field). Furthermore, because of Rule~\ref{rule_3}
above, the products $F^{\perp a b c} F^{\perp a' b' c'} g_{bb'}
g_{cc'}$ vanish when $a \not= a'$ so that the energy-momentum tensor
is diagonal.


\subsection[Geometric configurations and regular subalgebras of $E_{10}$]%
           {Geometric configurations and regular subalgebras of \boldmath $E_{10}$}
\label{geometric}
\index{regular subalgebra}
\index{regular subalgebra}

We prove here that the conditions on the electric field embodied
in the geometric configurations $(n_m, g_3)$ have a direct
Kac--Moody algebraic \index{Kac--Moody algebra}interpretation. They simply correspond to
a consistent truncation of the $E_{10}$ nonlinear sigma model to a
$\bar{\mf{g}}$ nonlinear sigma model, where $\bar{\mf{g}}$ is a
rank $g$ Kac--Moody subalgebra of $E_{10}$ (or a quotient of such a
Kac--Moody subalgebra by an appropriate ideal when the relevant
Cartan matrix has vanishing determinant), with three crucial
properties: (i) It is regularly embedded in $E_{10}$ (see
Section~\ref{section:KacMoody} for the definition of regular
subalgebras), (ii) it is generated by electric roots only, and (iii)
every node $P$ in its Dynkin diagram \index{Dynkin diagram}$\mathbb{D}_{\mgb}$ is linked to
a number $k$ of nodes that is independent of $P$ (but depend on the
algebra). We find that the Dynkin diagram $\mathbb{D}_{\mgb}$ of
$\mgb$ is the \emph{line incidence diagram} \index{line incidence diagram|bb} of the geometric
configuration $(n_m, g_3)$, in the sense that (i) each line of $(n_m,
g_3)$ defines a node of $\mathbb{D}_{\mgb}$, and (ii) two nodes of
$\mathbb{D}_{\mgb}$ are connected by a single bond iff the
corresponding lines of $(n_m, g_3)$ have no point in common. This
defines a geometric duality between a configuration $(n_m, g_3)$ and
its associated Dynkin diagram $\mathbb{D}_{\mgb}$. In the following we
shall therefore refer to configurations and Dynkin diagrams
\index{Dynkin diagram}related in this way as \emph{dual}.

None of the algebras $\mgb$ relevant to the truncated models turn out
to be hyperbolic: They can be finite, affine, or Lorentzian with
infinite-volume Weyl chamber. \index{Weyl chamber} Because of this,
the solutions are non-chaotic. After a finite number of collisions,
they settle asymptotically into a definite Kasner regime (both in the
future and in the past).


\subsubsection{General considerations}

In order to match diagonal Bianchi~I cosmologies with the sigma
model, one must truncate the
$\mathcal{E}_{10}/\mathcal{K}(\mathcal{E}_{10})$ action in such a
way that the sigma model metric $g_{ab}$ is diagonal. This will be
the case if the subalgebra $\mgb$ to which one truncates has no
generator $K^i{}_{j}$ with $i \not=j$. Indeed, recall from
Section~\ref{section:sigmamodels} that the off-diagonal
components of the metric are precisely the exponentials of the
associated sigma model fields. The set of simple roots of $\mgb$
should therefore not contain any root at level zero.

Consider ``electric'' regular subalgebras of $E_{10}$, for which
the simple roots are all at level one, where the 3-form electric
field variables live. These roots can be parametrized by three
indices corresponding to the indices of the electric field, with
$i_1<i_2<i_3$. We denote them $\alpha_{i_1i_2i_3}$. For
instance, $\alpha_{123} \equiv \alpha_{10}$. In terms of the
$\beta$-parametrization of~\cite{ArithmeticalChaos, DHNReview}, one
has $\alpha_{i_1i_2i_3} = \beta^{i_1} + \beta^{i_2} +
\beta^{i_3}$.

Now, for $\mgb$ to be a regular subalgebra, \index{regular subalgebra}it must fulfill, as we
have seen, the condition that the difference between any two of
its simple roots is not a root of $E_{10}$: $\alpha_{i_1i_2i_3} -
\alpha_{i'_1i'_2i'_3} \notin \Phi_{E_{10}}$ for any pair
$\alpha_{i_1i_2i_3}$ and $\alpha_{i'_1i'_2i'_3}$ of simple roots
of $\mgb$. But one sees by inspection of the commutator of
$E^{i_1i_2i_3}$ with $F_{i'_1i'_2i'_3}$ in
Equation~(\ref{level1commutationrelations}) that $\alpha_{i_1i_2i_3} -
\alpha_{i'_1i'_2i'_3}$ is a root of $E_{10}$ if and only if the sets
$\{i_1, i_2, i_3\}$ and $\{i'_1, i'_2, i'_3\}$ have exactly two points
in common. For instance, if $i_1= i'_1$, $i_2 = i'_2$ and $i_3 \not=
i'_3$, the commutator of $E^{i_1i_2i_3}$ with $F_{i'_1i'_2i'_3}$
produces the off-diagonal generator $K^{i_3}{}_{i'_3}$
corresponding to a level zero root of $E_{10}$. In order to fulfill
the required condition, one must avoid this case, i.e., one must
choose the set of simple roots of the electric regular subalgebra
$\mgb$ in such a way that given a pair of indices $(i_1, i_2)$, there
is at most one $i_3$ such that the root $\alpha_{i j k}$ is a simple
root of $\mgb$, with $(i,j,k)$ being the re-ordering of $(i_1, i_2, i_3 )$
such that $i<j<k$.

To each of the simple roots $\alpha_{i_1i_2i_3}$ of $\mgb$, one can
associate the line $(i_1,i_2,i_3)$ connecting the three points
$i_1$, $i_2$ and $i_3$. If one does this, one sees that the above
condition is equivalent to the following statement: \emph{The set of
points and lines associated with the simple roots of $\mgb$ must
fulfill the third rule defining a geometric configuration,
namely, that two points determine at most one line}. Thus, this
geometric condition has a nice algebraic interpretation in terms
of regular subalgebras of $E_{10}$.

The first rule, which states that each line contains 3 points, is
a consequence of the fact that the $E_{10}$-generators at level
one are the components of a 3-index antisymmetric tensor. The
second rule, that each point is on $m$ lines, is less fundamental
from the algebraic point of view since it is not required to hold
for $\mgb$ to be a regular subalgebra. It was imposed
in~\cite{Demaret} in order to allow for solutions isotropic in the
directions that support the electric field. We keep it here as it
yields interesting structure.


\subsubsection{Incidence diagrams and Dynkin diagrams}
\label{section:incidenceanddynkin}

We have just shown that each geometric configuration \index{geometric configuration} $(n_m,g_3)$
with $n \leq 10$ defines a regular subalgebra $\mgb$ of $E_{10}$. In
order to determine what this subalgebra $\mgb$ is, one needs,
according to the theorem recalled in Section~\ref{section:KacMoody}, to
compute the Cartan matrix \index{Cartan matrix}
\begin{equation}
  C = [C_{i_1 i_2 i_3, i'_1 i'_2 i'_3}] =
  \left[\left(\alpha_{i_1 i_2 i_3} \vert \alpha_{i'_1 i'_2 i'_3}\right)\right]
\end{equation}
(the real roots of $E_{10}$ have length squared equal to 2).
According to that same theorem, the algebra $\mgb$ is then just the
rank $g$ Kac--Moody algebra with Cartan matrix $C$, unless $C$ has zero
determinant, in which case $\mgb$ might be the quotient of that
algebra by a nontrivial ideal.

Using for instance the root parametrization
of~\cite{ArithmeticalChaos, DHNReview} and the expression of the
scalar product in terms of this parametrization, one
easily verifies that the scalar product $\left(\alpha_{i_1 i_2
i_3} \vert \alpha_{i'_1 i'_2 i'_3}\right)$ is equal to
\begin{equation}
  \left(\alpha_{i_1 i_2 i_3} \vert \alpha_{i'_1 i'_2 i'_3}\right) =
  \left\{
    \begin{array}{l@{\qquad}l}
      2 & \mbox{if all three indices coincide},
      \\
      1 & \mbox{if two and only two indices coincide},
      \\
      0 & \mbox{if one and only one index coincides},
      \\
      -1 & \mbox{if no indices coincide}.
    \end{array}
  \right.
\end{equation}
The second possibility does not arise in our case since we deal with
geometric configurations. For completeness, we also list the scalar
products of the electric roots $\alpha_{ijk}$ ($i<j<k$) with the
symmetry roots $\alpha_{\ell m}$ ($\ell < m$) associated with the
raising operators $K^m {}_{\ell}$: 
\begin{equation}
  \left(\alpha_{ijk} \vert \alpha_{\ell m}\right) =
  \left\{
    \begin{array}{l@{\qquad}l}
      -1 & \mbox{if }\ell\in \{i,j,k\}\mbox{ and }m \notin \{i,j,k\},
      \\
      0 & \mbox{if }\{\ell, m\} \subset \{i,j,k\}\mbox{ or }\{\ell, m\}
      \cap \{i,j, k\} = \emptyset,
      \\
      1 & \mbox{if }\ell\notin \{i,j,k\}\mbox{ and }m \in \{i,j,k\},
    \end{array}
  \right.
\end{equation}
as well as the scalar products of the symmetry
roots among themselves, 
\begin{equation}
  \left(\alpha_{ij} \vert \alpha_{\ell m}\right) =
 \left\{
    \begin{array}{l@{\qquad}l}
      - 1 & \mbox{if }j = \ell\mbox{ or }i = m,
      \\
      0 & \mbox{if }\{\ell, m\} \cap \{i,j\} = \emptyset,
      \\
      1 & \mbox{if }i = \ell\mbox{ or }j \not= m,
      \\
      2 & \mbox{if }\{\ell, m\} = \{i,j\}.
    \end{array}
  \right.
\end{equation} 
Given a geometric configuration \index{geometric configuration} $(n_m,g_3)$, one can
associate with it a ``line incidence diagram'' \index{line incidence diagram} that encodes the
incidence relations between its lines. To each line of
$(n_m,g_3)$ corresponds a node in the incidence diagram. Two nodes
are connected by a single bond if and only if they correspond to
lines with no common point (``parallel lines''). Otherwise, they
are not connected\epubtkFootnote{One may also consider a point incidence
diagram defined as follows: The nodes of the point incidence
diagram are the points of the geometric configuration. Two nodes
are joined by a single bond if and only if there is no straight
line connecting the corresponding points. The point incidence
diagrams of the configurations $(9_3,9_3)$ are given
in~\cite{Hilbert}. For these configurations, projective duality
between lines and points lead to identical line and point incidence
diagrams. Unless otherwise stated, the expression ``incidence
diagram'' will mean ``line incidence diagram''. \index{line incidence diagram}}. By inspection of the
above scalar products, we come to the important conclusion that {\em
  the Dynkin diagram of the regular, rank $g$, Kac--Moody subalgebra
  $\mgb$ associated with the geometric configuration $(n_m,g_3)$ is
  just its line incidence diagram.} We shall call the Kac--Moody
algebra $\mgb$ the algebra ``dual'' to the geometric configuration
$(n_m,g_3)$.

Because the geometric configurations have the property that the
number of lines through any point is equal to a constant $m$, the
number of lines parallel to any given line is equal to a number
$k$ that depends only on the configuration and not on the line.
This is in fact true in general and not only for $n\leq 10$ as can
be seen from the following argument. For a configuration with $n$
points, $g$ lines and $m$ lines through each point, any given line
$\Delta$ admits $3(m-1)$ true secants, namely, $(m-1)$ through
each of its points\epubtkFootnote{A true secant is here defined as a
line, say $\Delta^{\prime}$, distinct from $\Delta$ and with a
non-empty intersection with $\Delta$.}. By definition, these
secants are all distinct since none of the lines that $\Delta$
intersects at one of its points, say $P$, can coincide with a
line that it intersects at another of its points, say
$P^{\prime}$, since the only line joining $P$ to $P^{\prime}$ is
$\Delta$ itself. It follows that the total number of lines that
$\Delta$ intersects is the number of true secants plus $\Delta$
itself, i.e., $3(m-1)+1$. As a consequence, each line in the
configuration admits $k=g-[3(m-1)+1]$ parallel lines, which is
then reflected by the fact that each node in the associated Dynkin
diagram has the same number $k$ of adjacent nodes.


\subsection{Cosmological solutions with electric flux}

Let us now make use of these considerations to construct some
explicit supergravity solutions. We begin by analyzing the
simplest configuration $(3_{1},1_{3})$, of three points and one
line. It is displayed in Figure~\ref{figure:G31}. This case
is the only possible configuration for $n=3$.

\epubtkImage{G31.png}{%
  \begin{figure}[htbp]
    \centerline{\includegraphics[width=50mm]{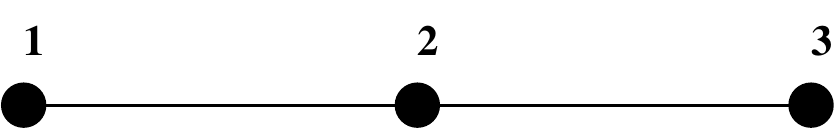}}
    \caption{$(3_{1},1_{3})$: The only allowed configuration for $n=3$.}
    \label{figure:G31}
  \end{figure}}

This example also exhibits some subtleties associated with the
Hamiltonian constraint and the ensuing need to extend $\mgb$ when the
algebra dual to the geometric configuration is
finite-dimensional. We will come back to this issue below.


\subsubsection{General discussion}

In light of our discussion, considering the geometric
configuration $(3_{1},1_{3})$ is equivalent to turning on only the
component $\mc{A}_{123}(t)$ of the 3-form that parametrizes the
generator $E^{123}$ in the coset representative $\mc{V}(t)\in
\mc{E}_{10}/\mc{K}(\mc{E}_{10})$. Moreover, in order to have the
full coset description, we must also turn on the diagonal metric
components corresponding to the Cartan generator $h = [E^{123},
F_{123}]$. The algebra has thus basis $\{e,f,h\}$ with
\begin{equation}
  e\equiv E^{123},
  \qquad
  f\equiv F_{123},
  \qquad
  h=[e,f] = - \! \f{1}{3}\sum_{a\neq 1,2,3} \!\!\!
  {K^{a}}_{a}+\f{2}{3}({K^{1}}_{1}+{K^{2}}_{2}+{K^{3}}_{3}),
  \label{A1generators}
\end{equation}
where the form of $h$ followed directly from the general commutator
between $E^{abc}$ and $F_{def}$ in
Section~\ref{section:LevelDecomposition}. The Cartan matrix is just
$(2)$ and is nondegenerate. It defines an $A_1=\mf{sl}(2, \mbb{R})$
regular subalgebra. The Chevalley--Serre relations, which are
guaranteed to hold according to the general argument, are easily
verified. The configuration $(3_{1},1_{3})$ is thus dual to $A_{1}$,
\begin{equation}
  \mf{g}_{(3_1,3_1)}=A_1.
\end{equation}
This $A_1$ algebra is simply the $\mf{sl}(2, \mbb{R})$-algebra
associated with the simple root $\alpha_{10}$. Because the
Killing form of $A_1$ restricted to the Cartan subalgebra
$\mf{h}_{A_1}=\mbb{R}h$ is positive definite, one cannot find a
solution of the Hamiltonian constraint if one turns on only the
fields corresponding to $A_1$. One needs to enlarge $A_1$ (at
least) by a one-dimensional subalgebra $\mathbb{R} l$ of
$\mh_{E_{10}}$ that is timelike. As will be discussed further
below, we take for $l$ the Cartan element ${K^4}_4 + {K^5}_5 +
{K^6}_6 + {K^7}_7 + {K^8}_8 + {K^9}_9 + {K^{10}}_{10}$, which
ensures isotropy in the directions not supporting the electric
field. Thus, the appropriate regular subalgebra of $E_{10}$ in
this case is $A_1 \oplus \mathbb{R}l$.

The need to enlarge the algebra $A_1$ was discussed in the
paper~\cite{CosmologyNicolai} where a group theoretical interpretation
of some cosmological solutions of eleven-dimensional supergravity was
given. In that paper, it was also observed that $\mathbb{R}l$ can be
viewed as the Cartan subalgebra \index{Cartan subalgebra}of the (non-regularly embedded)
subalgebra $A_1$ associated with an imaginary root at level~21, but
since the corresponding field is not excited, the relevant subalgebra
is really $\mathbb{R}l$.


\subsubsection{The solution}
\label{section:solution}

In order to make the above discussion a little less abstract we
now show how to obtain the relevant supergravity solution by
solving the $\mc{E}_{10}$-sigma model equations of motion and then
translating these, using the dictionary from
Section~\ref{section:sigmamodels}, to supergravity solutions. For this
particular example the analysis was done in~\cite{CosmologyNicolai}.

In order to better understand the role of the timelike generator
$l\in \mf{h}_{E_{10}}$ we begin the analysis by omitting it. The
truncation then amounts to considering the coset representative
\begin{equation}
  \mc{V}(t)=e^{\phi(t)h}\, e^{\mc{A}_{123}(t)E^{123}} \in
  \mc{E}_{10}/\mc{K}(\mc{E}_{10}).
\end{equation}
The projection $\mc{P}(t)$ onto the coset becomes
\begin{eqnarray}
  \mc{P}(t)&=& \f{1}{2}\left[\pa \mc{V}(t) \, \mc{V}(t)^{-1}+
  \left(\pa\mc{V}(t) \, \mc{V}(t)^{-1}\right)^{\mc{T}}\right]
  \nonumber
  \\
  &=& \pa\phi(t) h +\f{1}{2}e^{2\phi(t)}\pa\mc{A}_{123}(t)
  \left(E^{123}+F_{123}\right), 
\end{eqnarray}
where the exponent is the linear form $\al(\phi)=2\phi$ representing
the exceptional simple root $\al_{123}$ of $E_{10}$. More precisely,
it is the linear form $\al$ acting on the Cartan generator $\phi(t)
h$, as follows:
\begin{equation}
  \al(\phi h)=\phi \left< \al, h\right>=
  \phi \left<\al, \al^{\vee}\right>=\al^2 \phi=2\phi.
\end{equation}
The Lagrangian becomes 
\begin{eqnarray}
  \mc{L}&=& \f{1}{2}\left(\mc{P}(t)|\mc{P}(t)\right)
  \nonumber
  \\
  &=& \pa\phi(t)\pa\phi(t)+\f{1}{4}e^{4\phi(t)}\,
  \pa \mc{A}_{123}(t)\,\pa\mc{A}_{123}(t). 
  \label{A1truncation} 
\end{eqnarray}
For convenience we have chosen the gauge $n=1$ of the free parameter
in the $\mc{E}_{10}/\mc{K}(\mc{E}_{10})$-Lagrangian (see
Section~\ref{section:sigmamodels}). Recall that for the level one
fields we have $\mc{D}\mc{A}_{abc}(t)=\pa \mc{A}_{abc}(t)$, which is
why only the partial derivative of $\mc{A}_{123}(t)$ appears in the
Lagrangian.

The reason why this simple looking model contains information
about eleven-dimensional supergravity is that the $A_{1}$
subalgebra represented by $(e, f, h)$ is embedded in $E_{10}$
through the level~1-generator $E^{123}$, and hence this
Lagrangian corresponds to a consistent subgroup truncation of the
$\mc{E}_{10}$- sigma model.

Let us now study the dynamics of the Lagrangian in
Equation~(\ref{A1truncation}). The equations of motion for
$\mc{A}_{123}(t)$ are
\begin{equation}
  \pa \left(\f{1}{2} e^{4\phi(t)}\pa \mc{A}_{123}(t)\right)=0
  \quad \Longrightarrow \quad
  \f{1}{2} e^{4\phi(t)}\pa \mc{A}_{123}(t)=a,
  \label{A123Equation}
\end{equation}
where $a$ is a constant. The equations for the $\ell=0$ field
$\phi$ may then be written as
\begin{equation}
  \pa^{2}\phi(t)=2a^2 e^{-4\phi(t)}.
\end{equation}
Integrating once yields
\begin{equation}
  \pa\phi(t) \, \pa\phi(t) + a^2 e^{-4\phi(t)}=E,
\end{equation}
where $E$ plays the role of the energy for the dynamics of
$\phi(t)$. This equation can be solved exactly with the
result~\cite{CosmologyNicolai}
\begin{equation}
  \phi(t)=\f{1}{2}\ln \left[\f{2a}{\sqrt{E}}\cosh \sqrt{E} t\right] \equiv
  \f{1}{2}\ln H(t).
  \label{phisolution}
\end{equation}
We must also take into account the Hamiltonian constraint
\begin{equation}
  \mc{H}=\left(\mc{P}|\mc{P} \right) = 0,
\end{equation}
arising from the variation of $n(t)$ in the $\mc{E}_{10}$-sigma
model. The Hamiltonian becomes 
\begin{eqnarray}
  \mc{H} & =& 2\pa \phi(t) \, \pa \phi(t) +
  \f{1}{2}e^{4\phi(t)}\pa \mc{A}_{123}(t) \, \pa \mc{A}_{123}(t)
  \nonumber
  \\
  &=& 2\left(\pa \phi(t)\pa \phi(t)+a^2 e^{-4\phi(t)}\right)
  \nonumber
  \\
  &=& 2E. 
\end{eqnarray}
It is therefore impossible to satisfy the Hamiltonian constraint
unless $E=0$. This is the problem which was discussed above, and the
reason why we need to enlarge the choice of coset representative to
include the timelike generator $l\in \mf{h}_{E_{10}}$. We choose $l$
such that it commutes with $h$ and $E^{123}$,
\begin{equation}
  [l, h]=[l, E^{123}]=0,
\end{equation}
and such that isotropy in the directions not supported by the electric
field is ensured. Most importantly, in order to solve the problem of
the Hamiltonian constraint, $l$ must be timelike, 
\begin{equation}
  l^2= \left( l | l \right) < 0, 
  \label{timelikegenerator}
\end{equation}
where $(\cdot |\cdot )$ is the scalar product in the Cartan subalgebra
of $E_{10}$. The subalgebra to which we truncate the sigma model is
thus given by 
\begin{equation}
  \mgb=A_{1}\oplus \mathbb{R} l \subset E_{10},
  \label{A1plustimelike}
\end{equation}
and the corresponding coset representative is
\begin{equation}
  \tilde{\mc{V}}(t) =
  e^{\phi(t)h + \tilde{\phi}(t)l}e^{\mc{A}_{123}(t)E^{123}}.
  \label{timelikecoset}
\end{equation}
The Lagrangian now splits into two disconnected parts, corresponding
to the direct product $SL(2, \mbb{R})/SO(2)\times \mbb{R}$,
\begin{equation}
  \tilde{\mc{L}}=
  \left(\pa\phi(t)\,\pa\phi(t)+\f{1}{4}e^{4\phi(t)}
  \pa\mc{A}_{123}(t)\,\pa \mc{A}_{123}(t)\right) +
  \f{l^2}{2}\pa\tilde{\phi}(t)\,\pa\tilde{\phi}(t).
  \label{newLagrangian}
\end{equation}
The solution for $\tilde{\phi}$ is therefore simply linear in time,
\begin{equation}
  \tilde{\phi}=|l^2|\sqrt{\tilde{E}}\, t. 
  \label{phitildesolution}
\end{equation}
The new Hamiltonian now gets a contribution also from the Cartan
generator $l$,
\begin{equation}
  \tilde{\mc{H}}=2E-|l^2|\tilde{E}.
  \label{newHamiltonian}
\end{equation}
This contribution depends on the norm of $l$ and since $l^2<0$, it is
possible to satisfy the Hamiltonian constraint, provided that we set 
\begin{equation}
  \tilde{E}=\f{2}{|l^2|}E.
\end{equation}

We have now found a consistent truncation of the
$\mc{K}(\mc{E}_{10})\times \mc{E}_{10}$-invariant sigma model which
exhibits $SL(2, \mbb{R})\times SO(2)\times \mbb{R}$-invariance. We
want to translate the solution to this model,
Equation~(\ref{phisolution}), to a solution of eleven-dimensional
supergravity. The embedding of $\mf{sl}(2, \mbb{R})\subset E_{10}$
in Equation~(\ref{A1generators}) induces a natural ``Freund--Rubin'' type
($1+3+7$) split of the coordinates in the physical metric, where
the 3-form is supported in the three spatial directions
$x^{1},x^{2},x^{3}$. We must also choose an embedding of the
timelike generator $l$. In order to ensure isotropy in the
directions $x^{4}, \cdots, x^{10}$, where the electric field has
no support, it is natural to let $l$ be extended only in the
``transverse'' directions and we take~\cite{CosmologyNicolai}
\begin{equation}
  l ={K^{4}}_{4}+\cdots +{K^{10}}_{10},
  \label{timelikegeneratorembedding}
\end{equation}
which has norm
\begin{equation}
  (l | l )=\left( {K^{4}}_{4}+\cdots +{K^{10}}_{10} |
  {K^{4}}_{4}+\cdots +{K^{10}}_{10}\right ) =-42. 
  \label{timelikenorm}
\end{equation}
To find the metric solution corresponding to our sigma model, we first
analyze the coset representative at $\ell=0$,
\begin{equation}
  \tilde{\mc{V}}(t)\big|_{\ell=0}=
  \Exp \left[\phi(t) h +\tilde{\phi}(t)l\right].
\end{equation}
In order to make use of the dictionary from
Section~\ref{correspondence} it is necessary to rewrite this in a way
more suitable for comparison, i.e., to express the Cartan generators
$h$ and $l$ in terms of the $\mf{gl}(10, \mbb{R})$-generators
${K^{a}}_b$. We thus introduce parameters ${\xi^{a}}_b(t)$ and
$\tilde{\xi}^{a}{}_{b}(t)$ representing, respectively, $\phi$ and
$\tilde{\phi}$ in the $\mf{gl}(10, \mbb{R})$-basis. The level zero
coset representative may then be written as 
\begin{eqnarray}
  \tilde{\mc{V}}(t)\big|_{\ell=0} &=&\Exp \left[\sum_{a=1}^{10}
  \left({\xi^{a}}_{a}(t)+\tilde{\xi}^{a}{}_{a}(t)\right){K^{a}}_{a}\right]
  \nonumber
  \\
  &=&\Exp \left[\sum_{a=4}^{10}\left({\xi^{a}}_{a}(t)+
  \tilde{\xi}^{a}{}_{a}(t)\right){K^{a}}_{a}+
  \left({\xi^{1}}_{1}(t){K^{1}}_{1}+{\xi^{2}}_{2}(t){K^{2}}_{2}+
  {\xi^{3}}_{3}(t){K^{3}}_{3}\right)\right],
  \label{levelzero} 
\end{eqnarray}
where in the second line we have split the sum in order to highlight
the underlying spacetime structure, i.e., to emphasize that
$\ti{\xi}^{a}{}_{b}$ has no non-vanishing components in the
directions $x^{1}, x^{2}, x^{3}$. Comparing this to
Equation~(\ref{A1generators}) and
Equation~(\ref{timelikegeneratorembedding}) gives the diagonal
components of ${\xi^{a}}_{b}$ and $\tilde{\xi}^{a}{}_{b}$,
\begin{equation}
  {\xi^{1}}_{1}={\xi^{2}}_{2}={\xi^{3}}_{3}=2\phi/3,
  \qquad
  {\xi^{4}}_{4}= \cdots={\xi^{10}}_{10}=-\phi/3,
  \qquad
  {\tilde{\xi}^{4}}{}_{4}=\cdots = {\tilde{\xi}^{10}}{}_{10}=\ti{\phi}.
  \label{chiandxicomponents}
\end{equation}
Now, the dictionary from Section~\ref{section:sigmamodels} identifies
the physical spatial metric as follows:
\begin{equation}
  \g_{ab}(t)=
  {e_{a}}^{\bar{a}}(t){e_{b}}^{\bar{b}}(t) \delta_{\bar{a}\bar{b}}=
  {(e^{\xi(t)+\ti{\xi}(t)})_{a}}^{\bar{a}}
  {(e^{\xi(t)+\ti{\xi}(t)})_{b}}^{\bar{b}}\delta_{\bar{a}\bar{b}}
\end{equation}
By observation of Equation~(\ref{chiandxicomponents}) we find the
components of the metric to be 
\begin{equation}
  \begin{array}{rcl}
    \g_{11}&=&\g_{22}=\g_{33}=e^{4\phi/3},
    \\ [0.25 em]
    \g_{44}&=&\cdots=\g_{(10)(10)}=e^{-2\phi/3+2\ti{\phi}}.
  \end{array}
  \label{metriccomponents} 
\end{equation}
This result shows clearly how the embedding of $h$ and $l$ into
$E_{10}$ is reflected in the coordinate split of the metric. The gauge
fixing $N=\sqrt{\g}$ (or $n=1$) gives the $\g_{tt}$-component of the
metric,
\begin{equation}
  \g_{tt}=N^2=e^{14\ti{\phi}-2\phi/3}. 
  \label{ttcomponent}
\end{equation}
Next we consider the generator $E^{123}$. The dictionary tells us that
the field strength of the 3-form in eleven-dimensional supergravity
at some fixed spatial point $\mathbf{x}_0$ should be identified as
\begin{equation}
  F_{t123}(t,\mathbf{x}_{0})=\mc{D}\mc{A}_{123}(t)=\pa \mc{A}_{123}(t). 
\end{equation}
It is possible to eliminate the $\mc{A}_{123}(t)$ in favor of the
Cartan field $\phi(t)$ using the first integral of its equations
of motion, Equation~(\ref{A123Equation}),
\begin{equation}
  \f{1}{2}e^{-4\phi(t)}\pa\mc{A}_{123}(t)= a. 
  \label{eomA}
\end{equation}
In this way we may write the field strength in terms of $a$ and the
solution for $\phi$,
\begin{equation}
  F_{t123}(t,\mathbf{x}_{0})=2 a e^{4\phi(t)}=2 a H^{-2}(t).
  \label{Fsolution}
\end{equation}
Finally, we write down the solution for the spacetime metric
explicitly:
\begin{eqnarray}
  ds^{2}&=&-e^{14\ti{\phi}+2\phi/3}\,dt^{2}+
  e^{4\phi/3}\left[(dx^1)^{2}+(dx^2)^2+(dx^3)^2\right]+
  e^{2\ti{\phi}-2\phi/3}\sum_{\bar{a}=4}^{10}(dx^{\bar{a}})^{2}
  \nonumber
  \\
  &=& - H^{1/3}(t)e^{\f{1}{3}\sqrt{\ti{E}t}}\,dt^{2}+
  H^{-2/3}(t)\left[(dx^1)^{2}+(dx^2)^2+(dx^3)^2\right]+
  H^{1/3}(t)e^{\f{\sqrt{\ti{E}}}{21}t}\sum_{\bar{a}=4}^{10}(dx^{\bar{a}})^{2},
  \nonumber
  \\
  \label{SM2Brane} 
\end{eqnarray}
where
\begin{equation}
  H(t)=\f{2a}{\sqrt{E}}\cosh \sqrt{E} t.
\end{equation}
This solution coincides with the cosmological solution first found
in~\cite{Demaret} for the geometric configuration $(3_1, 3_1)$, and it
is intriguing that it can be exactly reproduced from a manifestly
$\mc{E}_{10}\times \mc{K}(\mc{E}_{10})$-invariant action, a priori
unrelated to any physical model.

Note that in modern terminology, this solution is an $SM2$-brane
solution (see, e.g.,~\cite{AcceleratingOhta} for a review) since
it can be interpreted as a spacelike (i.e., time-dependent)
version of the $M2$-brane solution. From this point of view the
world volume of the $SM2$-brane is extended in the directions
$x^1, x^2$ and $x^3$, and so is Euclidean.

In the BKL-limit \index{BKL-limit}this solution describes two asymptotic Kasner
regimes, at $t\rightarrow \infty$ and at $t\rightarrow -\infty$.
These are separated by a collision against an electric wall,
corresponding to the blow-up of the electric field
$F_{t123}(t)\sim H^{-2}(t)$ at $t=0$. In the billiard
\index{cosmological billiard}picture the
dynamics in the BKL-limit is thus given by free-flight motion
interrupted by one geometric reflection against the electric wall,
\begin{equation}
  e_{123}(\be)=\be^1+\be^2+\be^3,
\end{equation}
which is the exceptional simple root of $E_{10}$. This indicates that
in the strict BKL-limit, electric walls and $SM2$-branes are actually
equivalent.


\subsubsection{Intersecting spacelike branes from geometric configurations}

Let us now examine a slightly more complicated example. We consider
the configuration $(6_{2},4_{3})$, shown in
Figure~\ref{figure:G62}. This configuration has four lines and six
points. As such the associated supergravity model describes a
cosmological solution with four components of the electric field
turned on, or, equivalently, it describes a set of four intersecting
$SM2$-branes~\cite{Henneaux:2006gp}.

\epubtkImage{G62.png}{%
  \begin{figure}[htbp]
    \centerline{\includegraphics[width=50mm]{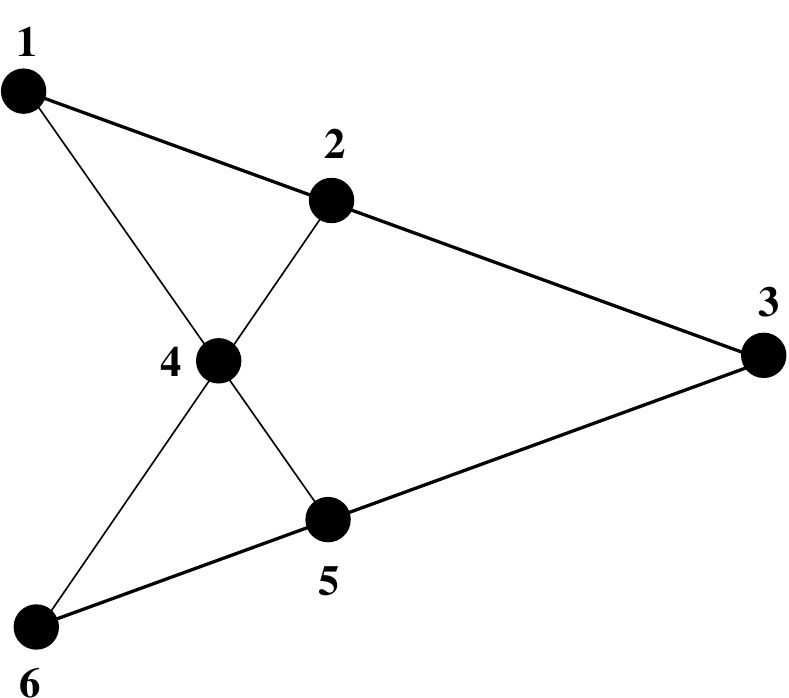}}
    \caption{The configuration $(6_{2},4_{3})$, dual to the Lie
      algebra $A_1\oplus A_1\oplus A_1\oplus A_1$.}
    \label{figure:G62}
  \end{figure}}

From the configuration we read off the Chevalley--Serre
generators associated to the simple roots of the dual algebra:
\begin{equation}
  e_{1}=E^{123},
  \qquad
  e_{2}=E^{145},
  \qquad
  e_{3}=E^{246},
  \qquad e_{4}=E^{356}.
  \label{G62generators}
\end{equation}
The first thing to note is that all generators have one index in
common since in the graph any two lines share one node. This implies
that the four lines in $(6_{2},4_{3})$ define four \emph{commuting}
$A_1$ subalgebras,
\begin{equation}
  (6_{2},4_{3})
  \quad \Longleftrightarrow \quad
  \mf{g}_{(6_{2},4_{3})}=A_{1}\oplus A_{1} \oplus A_{1}\oplus A_{1}.
  \label{A1^4}
\end{equation}
One can make sure that the Chevalley--Serre relations are indeed
fulfilled for this embedding. For instance, the Cartan element $h =
[E^{b_1 b_2 b_3}, F_{b_1 b_2 b_3}]$ (no summation on the fixed,
distinct indices $b_{1},b_{2},b_{3}$) reads
\begin{equation}
  h=-\f{1}{3}\!\sum_{a\neq b_{1},b_{2},b_{3}} \!\!\!
  {K^{a}}_{a}+\f{2}{3}({K^{b_{1}}}_{b_{1}}+
  {K^{b_{2}}}_{b_{2}}+{K^{b_{3}}}_{b_{3}}).
  \label{GeneralCartanElement}
\end{equation}
Hence, the commutator $[h,E^{b_{i} cd}]$ vanishes whenever $E^{b_{i}
  cd}$ has only one $b$-index, 
\begin{eqnarray}
  [h,E^{b_{i} cd}]&=&-\f{1}{3}[({K^{c}}_{c}+{K^{d}}_{d}),E^{b_{i}cd}]+
  \f{2}{3}[({K^{b_{1}}}_{b_{1}}+{K^{b_{2}}}_{b_{2}}+{K^{b_{3}}}_{b_{3}}),E^{b_{i}cd}]
  \nonumber
  \\
  &=& \left(-\f{1}{3}-\f{1}{3}+\f{2}{3}\right)E^{b_{i}cd}=0
  \qquad
  (i=1,2,3). 
  \label{vanishingcommutator} 
\end{eqnarray}
Furthermore, multiple commutators of the step operators are
immediately killed at level $2$ whenever they have one index or more
in common, e.g.,
\begin{equation}
  [E^{123},E^{145}]=E^{123145}=0.
  \label{vanishingcommutator2}
\end{equation}
To fulfill the Hamiltonian constraint, one must extend the algebra
by taking a direct sum with $\mathbb{R}l$, $l={K^7}_7 + {K^8}_8 +
{K^9}_9 + {K^{10}}_{10}$. Accordingly, the final algebra is
$A_{1}\oplus A_{1} \oplus A_{1}\oplus A_{1} \oplus \mathbb{R}l$.
Because there is no magnetic field, the momentum constraint and
Gauss' law are identically satisfied.

By investigating the sigma model solution corresponding to the
algebra $\mf{g}_{(6_2, 4_3)}$, augmented with the timelike
generator $l$,
\begin{equation}
  \mgb=A_1\oplus A_1\oplus A_1 \oplus A_1\oplus \mbb{R}l,
\end{equation}
we find a supergravity solution which generalizes the one found
in~\cite{Demaret}. The solution describes a set of four intersecting
$SM2$-branes, with a five-dimensional transverse spacetime in the
directions $t, x^{7},x^{8},x^{9},x^{10}$.

Let us write down also this solution explicitly. The full set of
generators for $\mf{g}_{(6_2, 4_3)}$ is 
\begin{equation}
  \begin{array}{rcl@{\qquad}rcl@{\qquad}rcl@{\qquad}rcl}
    e_{1}&=&E^{123}, &
    e_{2}&=&E^{145}, &
    e_{3}&=&E^{246}, &
    e_{4}=E^{356}
    \\ [0.25 em]
    f_{1}&=&F_{123}, &
    f_{2}&=&F_{145}, &
    f_{3}&=&F_{246}, &
    f_{4}=F_{356}
    \\ [0.5 em]
    h_{1}&=&\multicolumn{10}{l}{\displaystyle
    -\f{1}{3}\!\sum_{a\neq 1,2,3}\!\!\!
    {K^{a}}_{a}+\f{2}{3}({K^{1}}_{1}+{K^{2}}_{2}+{K^{3}}_{3}),}
    \\ [0.5 em]
    h_{2}&=&\multicolumn{10}{l}{-\displaystyle
    \f{1}{3}\!\sum_{a\neq 1,4,5}\!\!\!
    {K^{a}}_{a}+\f{2}{3}({K^{1}}_{1}+{K^{4}}_{4}+{K^{5}}_{5}),}
    \\ [0.5 em]
    h_{3}&=&\multicolumn{10}{l}{\displaystyle
    -\f{1}{3}\!\sum_{a\neq 2,4,6}\!\!\!
    {K^{a}}_{a}+\f{2}{3}({K^{2}}_{2}+{K^{4}}_{4}+{K^{6}}_{6}),}
    \\ [0.5 em]
    h_{4}&=&\multicolumn{10}{l}{\displaystyle
    -\f{1}{3}\!\sum_{a\neq3,5,6}\!\!\!
    {K^{a}}_{a}+\f{2}{3}({K^{3}}_{3}+{K^{5}}_{5}+{K^{6}}_{6}).}
  \end{array}
  \label{(6_2,4_3)generators} 
\end{equation}
The coset element for this configuration then becomes
\begin{equation}
  \mc{V}(t)=e^{\phi_{1}(t)h_{1}+\phi_{2}(t)h_{2}+\phi_{3}(t)h_{3}+
  \phi_{4}(t)h_{4}+\ti{\phi}(t)l}\,
  e^{\mc{A}_{123}(t)E^{123}+\mc{A}_{145}(t)E^{145}+\mc{A}_{246}(t)E^{246}+
  \mc{A}_{356}(t)E^{356}}.
  \label{fourA1coset}
\end{equation}
We must further choose the timelike Cartan generator, $l\in
\mf{h}_{E_{10}}$, appropriately. Examination of
Equation~(\ref{(6_2,4_3)generators}) reveals that the four electric
fields are supported only in the spatial directions
$x^{1},\cdots,x^{6}$ so, again, in order to ensure isotropy in the
directions transverse to the $S$-branes, we choose the timelike Cartan
generator as follows:
\begin{equation}
  l={K^{7}}_{7}+{K^{8}}_{8}+{K^{9}}_{9}+{K^{10}}_{10},
  \label{timelikeCartan}
\end{equation}
which implies
\begin{equation}
  l^2=\left( l | l \right)=
  \left({K^{7}}_{7}+{K^{8}}_{8}+{K^{9}}_{9}+{K^{10}}_{10}|
  {K^{7}}_{7}+{K^{8}}_{8}+{K^{9}}_{9}+{K^{10}}_{10}\right) = -12. 
  \label{qminus}
\end{equation}
The Lagrangian for this system becomes
\begin{equation}
  \mc{L}_{(6_2,4_3)}=\mc{L}_{1}+\mc{L}_{2}+\mc{L}_{3}+\mc{L}_{4}+
  \f{l^2}{2}\pa \ti{\phi}(t)\,\pa \ti{\phi}(t), 
  \label{Lagrangian4A1}
\end{equation}
where $\mc{L}_1, \mc{L}_2, \mc{L}_3$ and $\mc{L}_4$ represent the
$SL(2, \mbb{R})\times SO(2)$-invariant Lagrangians corresponding to
the four $A_1$-algebras. The solutions for $\phi_1(t), \cdots,
\phi_4(t)$ and $\ti{\phi}(t)$ are separately identical to the ones for
$\phi(t)$ and $\ti{\phi}(t)$, respectively, in
Section~\ref{section:solution}. From the embedding into $E_{10}$,
provided in Equation~(\ref{(6_2,4_3)generators}), we may read off the
solution for the spacetime metric,
\begin{eqnarray}
  ds^{2}_{(6_2,4_3)}&=&
  -(H_{1}H_{2}H_{3}H_{4})^{1/3}e^{\f{2}{3}\sqrt{E_{-}}t}\,dt^{2}+
  (H_{1}H_{4})^{-2/3}(H_{2}H_{3})^{1/3}(dx^{1})^{2}
  \nonumber
  \\
  & &+(H_{1}H_{3})^{-2/3}(H_{2}H_{4})^{1/3}(dx^{2})^{2}+
  (H_{1}H_{2})^{-2/3}(H_{3}H_{4})^{1/3}(dx^{3})^{2}
  \nonumber
  \\
  & &+(H_{3}H_{4})^{-2/3}(H_{1}H_{2})^{1/3}(dx^{4})^{2}+
  (H_{2}H_{4})^{-2/3}(H_{1}H_{3})^{1/3}(dx^{5})^{2}
  \nonumber
  \\
  & &+(H_{2}H_{3})^{-2/3}(H_{1}H_{4})^{1/3}(dx^{6})^{2}+
  (H_{1}H_{2}H_{3}H_{4})^{1/3}e^{\f{1}{6}\sqrt{E_{-}}t}
  \sum_{\bar{a}=7}^{10}(dx^{\bar{a}})^2.
  \label{fourSM2branes} 
\end{eqnarray}
As announced, this describes four intersecting $SM2$-branes with a
$1+4$-dimensional transverse spacetime. For example the brane that
couples to the field associated with the first Cartan generator is extended
in the directions $x^{1},x^{2},x^{3}$. By restricting to the case
$\phi_{1}=\phi_2=\phi_3=\phi_4\equiv \phi$ the metric simplifies to 
\begin{eqnarray}
  ds^{2}_{(6_2,4_3)}&=&
  -\left(\f{2a}{\sqrt{E}}\right)^{4/3}\cosh^{4/3}\sqrt{E}t\,
  e^{\f{2}{3}\sqrt{\ti{E}}t}dt^{2}+
  \left(\f{2a}{\sqrt{E}}\right)^{-2/3}\cosh^{-2/3}\sqrt{E}t
  \sum_{a^{\prime}=1}^{6}(dx^{a^{\prime}})^{2}
  \nonumber
  \\
  & &+\left(\f{2a}{\sqrt{E}}\right)^{4/3} \cosh^{4/3}\sqrt{E} t \,
  e^{\f{1}{6}\sqrt{\ti{E}}t}\sum_{\bar{a}=7}^{10}(dx^{\bar{a}})^{2},
  \label{restrictedsolution} 
\end{eqnarray}
which coincides with the cosmological solution found in~\cite{Demaret}
for the configuration $(6_{2},4_3)$. We can therefore conclude that
the algebraic interpretation of the geometric configurations found in
this paper generalizes the solutions given in the aforementioned
reference.

In a more general setting where we excite more roots of $E_{10}$,
the solutions of course become more complex. However, as long as
we consider \emph{commuting} subalgebras there will naturally be
no coupling in the Lagrangian between fields parametrizing
different subalgebras. This implies that if we excite a direct sum
of $m$ $A_{1}$-algebras the total Lagrangian will split according
to
\begin{equation}
  \mc{L}=\sum_{k=1}^{m} \mc{L}_{k} + \ti{\mc{L}},
  \label{LagrangianSplitting}
\end{equation}
where $\mc{L}_{k}$ is of the same form as
Equation~(\ref{A1truncation}), and $\ti{\mc{L}}$ is the Lagrangian for
the timelike Cartan element, needed in order to satisfy the
Hamiltonian constraint. It follows that the associated solutions are 
\begin{equation}
  \begin{array}{rcl}
    \phi_{k}(t)&=&\displaystyle
    \f{1}{2}\ln\left[\f{a_{k}}{E_{k}}\cosh \sqrt{E_{k}} t\right]
    \qquad
    (k=1,\cdots,m),
    \\ [1.0 em]
    \ti{\phi}(t)&=& |l^2|\sqrt{\ti{E}}t.
  \end{array}
  \label{intersectingsolutions} 
\end{equation}
Furthermore, the resulting structure of the metric depends on the
embedding of the $A_{1}$-algebras into $E_{10}$, i.e., which
level~1-generators we choose to realize the step-operators and hence
which Cartan elements that are associated to the $\phi_{k}$'s. Each
excited $A_{1}$-subalgebra will turn on an electric 3-form that
couples to an $SM2$-brane and hence the solution for the metric will
describe a set of $m$ intersecting $SM2$-branes.

As an additional nice example, we mention here the configuration
$(7_3, 7_3)$, also known as the \emph{Fano plane}, which consists of
7 lines and 7 points (see Figure~\ref{figure:Fano}). This
configuration is well known for its relation to the octonionic
multiplication table~\cite{Baez}. For our purposes, it is interesting
because none of the lines in the configuration are parallell. Thus,
the algebra dual to the Fano plane is a direct sum of seven
$A_1$-algebras and the supergravity solution derived from the sigma
model describes a set of seven intersecting $SM2$-branes.

\epubtkImage{G7.png}{%
  \begin{figure}[htbp]
    \centerline{\includegraphics[width=60mm]{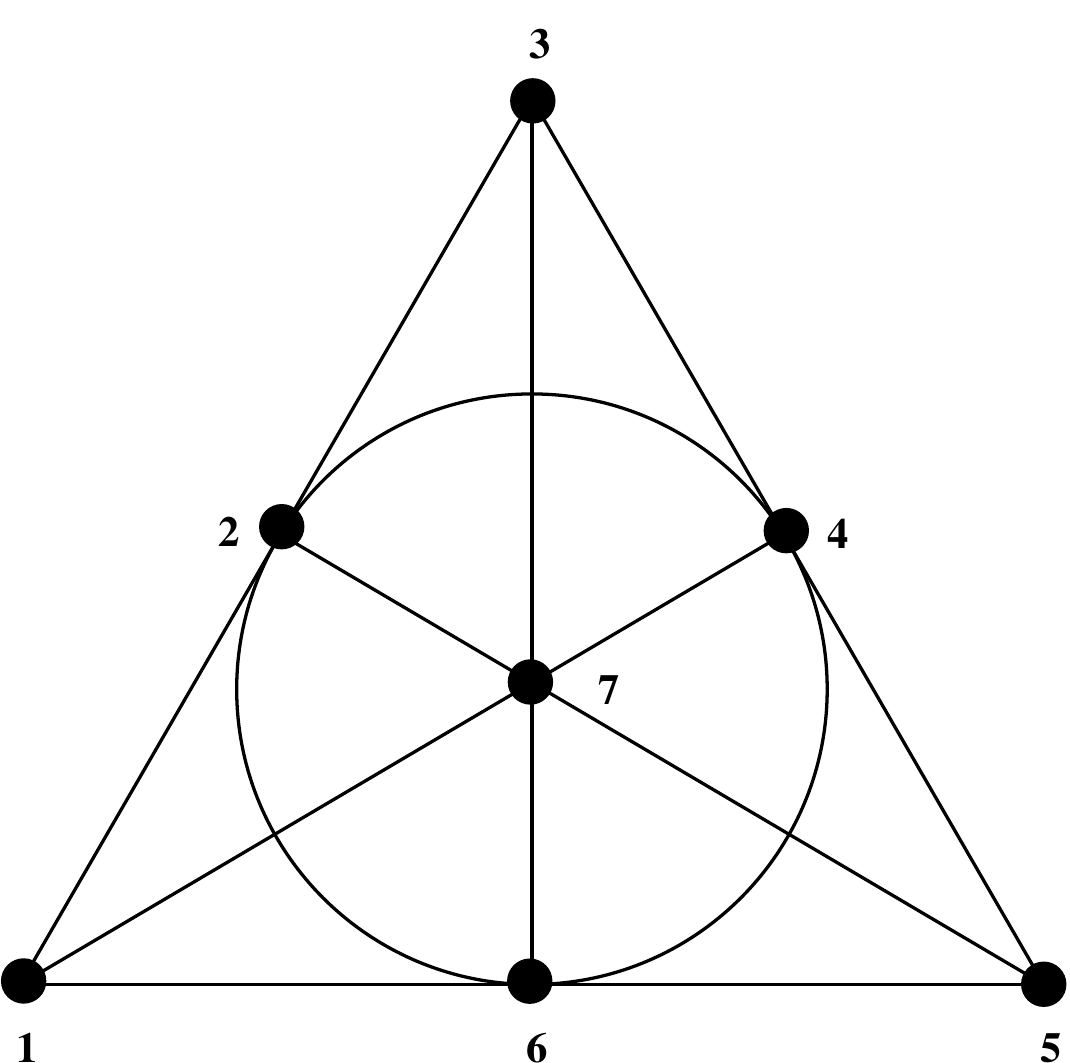}}
    \caption{The Fano Plane, $(7_3, 7_3)$, dual
      to the Lie algebra $A_1\oplus A_1\oplus A_1\oplus A_1\oplus A_1\oplus A_1\oplus A_1$.}
    \label{figure:Fano}
  \end{figure}}


\subsubsection{Intersection rules for spacelike branes}

For multiple brane solutions, there are rules for how these branes may
intersect in order to describe allowed
BPS-solutions~\cite{IntersectingArgurio}. These intersection rules \index{intersection rules|bb} also
apply to spacelike branes~\cite{IntersectingOhta} and hence they apply
to the solutions considered here. In this section we will show that the
intersection rules for multiple $S$-brane solutions are encoded in the
associated geometric configurations~\cite{Henneaux:2006gp}.

For two spacelike $q$-branes, $A$ and $B$, in $M$-theory the rules
are
\begin{equation}
  SMq_{A}\cap SMq_{B}=\f{(q_{A}+1)(q_{B}+1)}{9}-1.
  \label{intersectionrules}
\end{equation}
So, for example, if we have two $SM2$-branes the result is
\begin{equation}
  SM2\cap SM2=0,
  \label{intersectingSM2}
\end{equation}
which means that they are allowed to intersect on a 0-brane. Note
that since we are dealing with spacelike branes, a zero-brane is
extended in one spatial direction, so the two $SM2$-branes may
therefore intersect in one spatial direction only. We see from
Equation~(\ref{fourSM2branes}) that these rules are indeed fulfilled
for the configuration $(6_2,4_3)$.

In~\cite{IntersectingEnglert} it was found in the context of
$\mf{g}^{+++}$-algebras that the intersection rules \index{intersection rules} for extremal
branes are encoded in orthogonality conditions between the various
roots from which the branes arise. This is equivalent to saying
that the subalgebras that we excite are commuting, and hence the
same result applies to $\mf{g}^{++}$-algebras in the cosmological
context\epubtkFootnote{This was also pointed out
  in~\cite{CosmologyNicolai}.}. From this point of view, the
intersection rules can also be read off from the geometric
configurations in the sense that the configurations encode information
about whether or not the algebras commute.

The next case of interest is the Fano plane, $(7_3,7_3)$. As
mentioned above, this configuration corresponds to the direct sum
of 7 commuting $A_{1}$ algebras and so the gravitational
solution describes a set of 7 intersecting $SM2$-branes. The
intersection rules \index{intersection rules} are guaranteed to be
satisfied for the same reason as before.


\subsection{Cosmological solutions with magnetic flux}

We will now briefly sketch how one can also obtain the $SM5$-brane
solutions from geometric configurations and regular subalgebras of
$E_{10}$. In order to do this we consider ``magnetic'' subalgebras
of $E_{10}$, constructed only from simple root generators at level
two in the level decomposition of $E_{10}$. To the best of our
knowldege, there is no theory of geometric configurations
developed for the case of having 6 points on each line, which
would be needed here. However, we may nevertheless continue to
investigate the simplest example of such a configuration, namely
$(6_1, 1_6)$, displayed in Figure~\ref{figure:MagneticConfiguration}.

\epubtkImage{MagneticConfiguration.png}{%
  \begin{figure}[htbp]
    \centerline{\includegraphics[width=100mm]{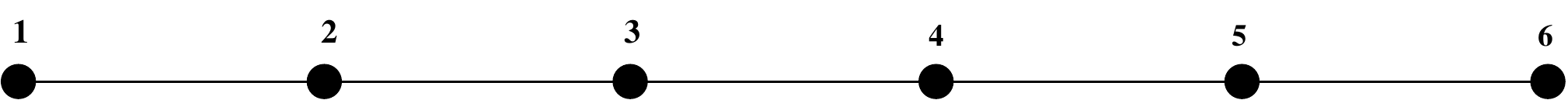}}
    \caption{The simplest ``magnetic configuration'' $(6_1, 1_6)$,
      dual to the algebra $A_1$.The associated supergravity solution
      describes an $SM5$-brane, whose world volume is extended in the
      directions $x^{1}, \cdots, x^{6}$.}
    \label{figure:MagneticConfiguration}
  \end{figure}}

The algebra dual to this configuration is an
$A_1$-subalgebra of $E_{10}$ with the following generators:
\begin{equation}
  \begin{array}{rcl}
    e&=&E^{123456}=F_{123456},
    \\ [0.25 em]
    h&\equiv&\displaystyle
    [E^{123456},F_{123456}]=-\f{1}{6}\sum_{a\neq 1,\cdots,6}
    {K^{a}}_{a}+ \f{1}{3}({K^{1}}_{1}+\cdots+{K^{6}}_{6}).
  \end{array}
  \label{SM5generators} 
\end{equation}
Although the embedding of this algebra
is different from the electric cases considered previously, the
sigma model solution is still associated to an $SL(2,
\mbb{R})/SO(2)$ coset space and therefore the solutions for
$\phi(t)$ and $\ti{\phi}(t)$ are the same as before. Because of
the embedding, however, the sigma model translates to a different
type of supergravity solution, namely a spacelike five-brane whose
world volume is extended in the directions $x^{1}, \cdots, x^{6}$.
The metric is given by
\begin{equation}
  ds^{2}=- H^{-4/3}(t)\,e^{\f{2}{3}\sqrt{E_{-}} t}\,
  dt^{2}+H^{-1/3}(t)\sum_{a^{\prime}=1}^{6}(dx^{a^{\prime}})^2+
  H^{1/6}(t)e^{\f{1}{6}\sqrt{E_{-}}t}\sum_{\bar{a}=7}^{10}(dx^{\bar{a}})^2.
  \label{SM5brane}
\end{equation}
This solution coincides with the $SM5$-brane found by Strominger and
Gutperle in~\cite{Strominger}\epubtkFootnote{In~\cite{Strominger}
  they were dealing with a hyperbolic internal space so there was an
  additional $\sinh$-function in the transverse spacetime.}. Note that
the correct power of $H(t)$ for the five-brane arises here entirely
due to the embedding of $h$ into $E_{10}$ through
Equation~(\ref{SM5generators}).

Because of the existence of electric-magnetic duality on the
supergravity side, it is suggestive to expect the existence of a
duality between the two types of configurations $(n_m, g_3)$ and
$(n_m, g_6)$, of which we have here seen the simplest realisation
for the configurations $(3_1, 1_3)$ and $(6_1, 1_6)$.


\subsection{The Petersen algebra and the Desargues configuration}

We want to end this section by considering an example which is
more complicated, but very interesting from the algebraic point of
view. There exist ten geometric configurations of the form
$(10_3, 10_3)$, i.e., with exactly ten points and ten
lines. In~\cite{Demaret}, these were associated to supergravity
solutions with ten components of the electric field turned on. This
result was re-analyzed by some of the present authors
in~\cite{Henneaux:2006gp} where it was found that many of these
configurations have a dual description in terms of Dynkin diagrams of
rank~10 Lorentzian Kac--Moody subalgebras of $E_{10}$. One would
therefore expect that solutions of the sigma models for these algebras
should correspond to new solutions of eleven-dimensional
supergravity. However, since these algebras are infinite-dimensional,
the corresponding sigma models are difficult to solve without further
truncation. Nevertheless, one may argue that explicit solutions should
exist, since the algebras in question are all non-hyperbolic, so we know
that the supergravity dynamics is non-chaotic.

We shall here consider one of the $(10_3, 10_3)$-configurations in
some detail, referring the reader to~\cite{Henneaux:2006gp} for a
discussion of the other cases. The configuration we will treat is
the well known \emph{Desargues configuration}, displayed in
Figure~\ref{figure:GDesargues}. The Desargues configuration is
associated with the 17th century French mathematician
\emph{G\'{e}rard Desargues} to illustrate the following ``Desargues
theorem'' (adapted from~\cite{Page}):

\begin{quote}
  \emph{Let the three lines defined by $\{4,1\},\{5,2\}$ and
    $\{6,3\}$ be concurrent, i.e., be intersecting at one point, say
    $\{7\}$. Then the three intersection points $8\equiv
    \{1,2\}\cap\{4,5\}, 9\equiv \{2,3\}\cap\{5,6\}$ and $10\equiv
    \{1,3\}\cap\{4,6\}$ are colinear. }
\end{quote}

\epubtkImage{C3.png}{%
\begin{figure}[htbp]
\centerline{\includegraphics[width=70mm]{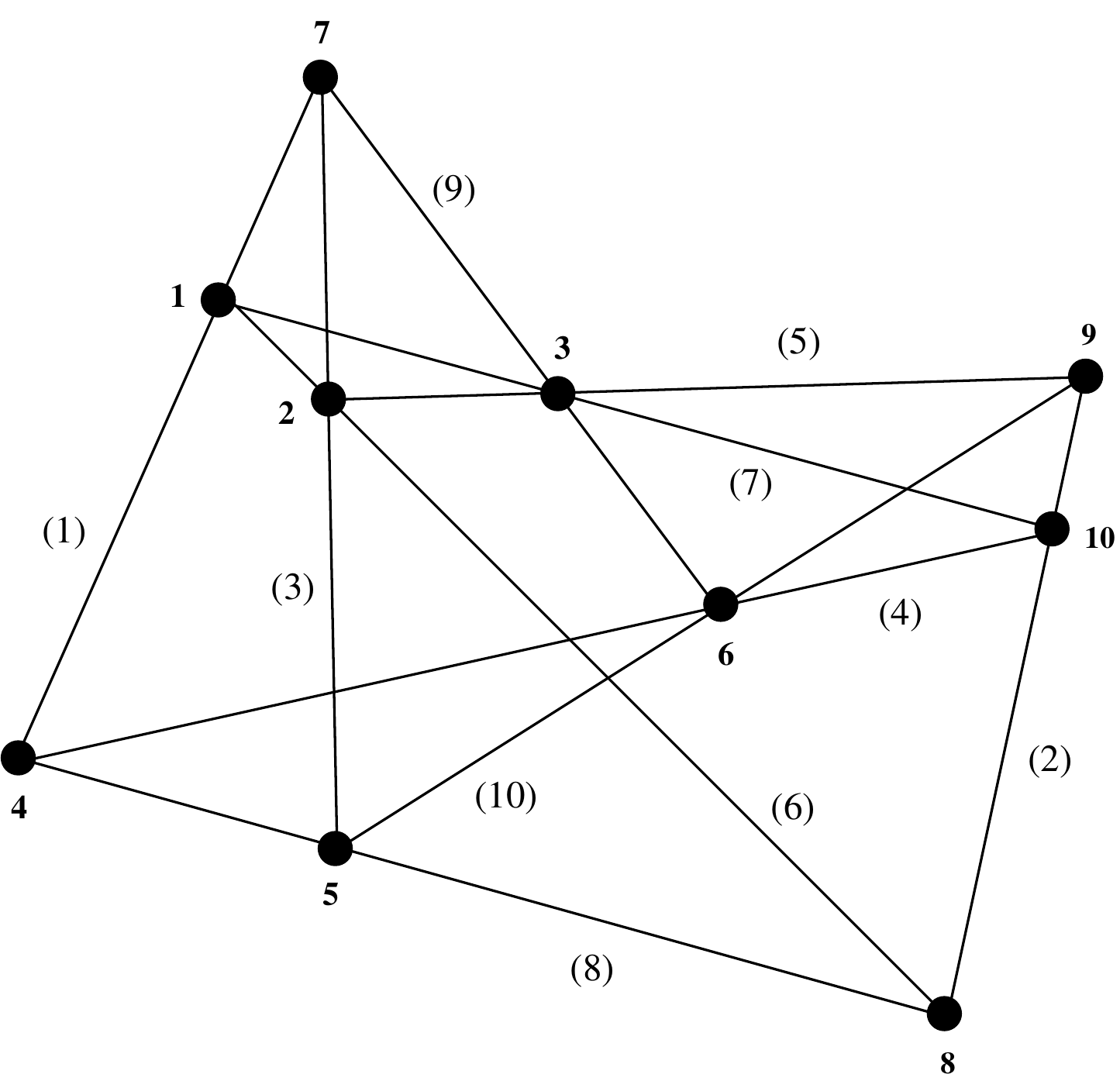}}
\caption{$(10_{3},10_{3})_{3}$: The Desargues configuration, dual
to the Petersen graph.} \label{figure:GDesargues}
\end{figure}}

Another way to say this is that the two triangles
$\{1,2,3\}$ and $\{4,5,6\}$ in Figure~\ref{figure:GDesargues} are
in perspective from the point $\{7\}$ and in perspective from the
line $\{8,10,9\}$.

As we will see, a new fascinating feature emerges for this case, namely
that the Dynkin diagram dual to this configuration \emph{also}
corresponds in itself to a geometric configuration. In fact, the
Dynkin diagram dual to the Desargues configuration turns out to be
the famous \emph{Petersen graph}, denoted $(10_{3},15_{2})$, which
is displayed in Figure~\ref{figure:D2(Petersen)}.

To construct the Dynkin diagram we first observe that each line in
the configuration is disconnected from three other lines, e.g., 
$\{4,1,7\}$ have no nodes in common with the lines $\{2,3,9\}$,
$\{5,6,9\}$, $\{8,10,9\}$. This implies that all nodes in the
Dynkin diagram will be connected to three other nodes. Proceeding
as in Section~\ref{section:incidenceanddynkin} leads to the Dynkin
diagram in Figure~\ref{figure:D2(Petersen)}, which we identify as the
Petersen graph. The corresponding Cartan matrix \index{Cartan matrix} is
\begin{equation}
  A(\mf{g}_\mathrm{Petersen})=\left(
    \begin{array}{@{}r@{\quad}r@{\quad}r@{\quad}r@{\quad}r@{\quad}r@{\quad}r@{\quad}r@{\quad}r@{\quad}r@{}}
      2 & -1 & 0 & 0 & 0 & 0 & 0 & 0 & -1 & -1 \\
      -1 & 2 & -1 & 0 & 0 & -1 & 0 & 0 & 0 & 0 \\
      0 & -1 & 2 & -1 & 0 & 0 & 0 & -1 & 0 & 0 \\
      0 & 0 & -1 & 2 & -1 & 0 & 0 & 0 & 0 & -1 \\
      0 & 0 & 0 & -1 & 2 & -1 & 0 & 0 & -1 & 0 \\
      0 & -1 & 0 & 0 & -1 & 2 & -1 & 0 & 0 & 0 \\
      0 & 0 & 0 & 0 & 0 & -1 & 2 & -1 & 0 & -1 \\
      0 & 0 & -1 & 0 & 0 & 0 & -1 & 2 & -1 & 0 \\
      -1 & 0 & 0 & 0 & -1 & 0 & 0 & -1 & 2 & 0 \\
      -1 & 0 & 0 & -1 & 0 & 0 & -1 & 0 & 0 & 2 \\
    \end{array}
  \right),
  \label{PetersenCartanMatrix}
\end{equation}
which is of Lorentzian signature with
\begin{equation}
  \det A(\mf{g}_\mathrm{Petersen})=-256.
  \label{PetersenDeterminant}
\end{equation}
The Petersen graph was invented
by the Danish mathematician \emph{Julius Petersen} in the end of
the 19th century. It has several embeddings on the
plane, but perhaps the most famous one is as a star inside a
pentagon as depicted in Figure~\ref{figure:D2(Petersen)}. One of its
distinguishing features from the point of view of graph theory is
that it contains a \emph{Hamiltonian path} but no
\emph{Hamiltonian cycle}\epubtkFootnote{We recall that a
Hamiltonian path is defined as a path in an undirected graph which
intersects each node once and only once. A Hamiltonian cycle is
then a Hamiltonian path which also returns to its initial node.}.

\epubtkImage{D2PetersenStar.png}{%
  \begin{figure}[htbp]
    \centerline{\includegraphics[width=60mm]{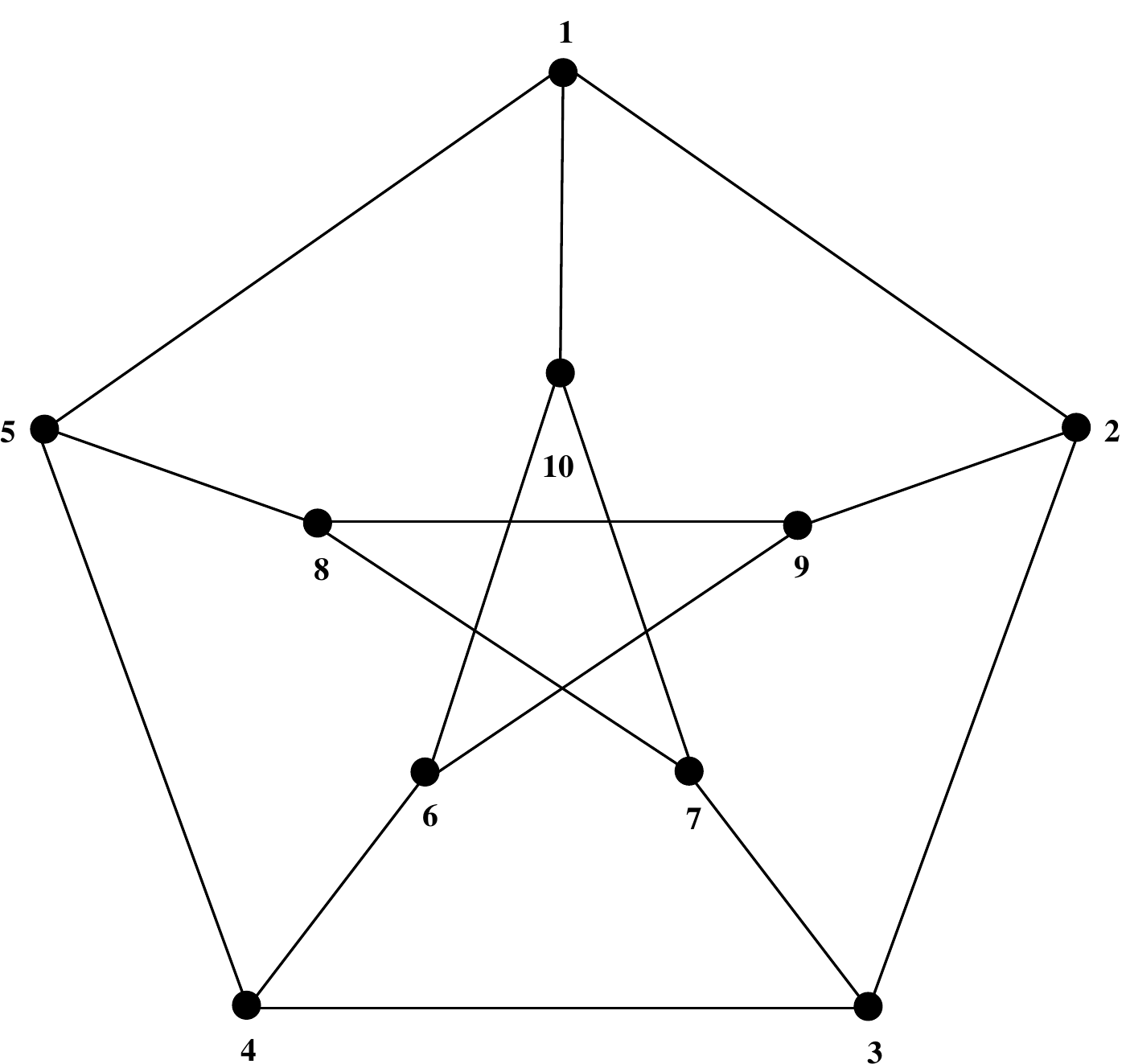}}
    \caption{This is the so-called Petersen graph. It is the
      Dynkin diagram dual to the Desargues configuration, and is in fact a
      geometric configuration itself, denoted $(10_{3},15_{2})$.}
    \label{figure:D2(Petersen)}
  \end{figure}}

\epubtkImage{D2Petersen.png}{%
  \begin{figure}[htbp]
    \centerline{\includegraphics[width=45mm]{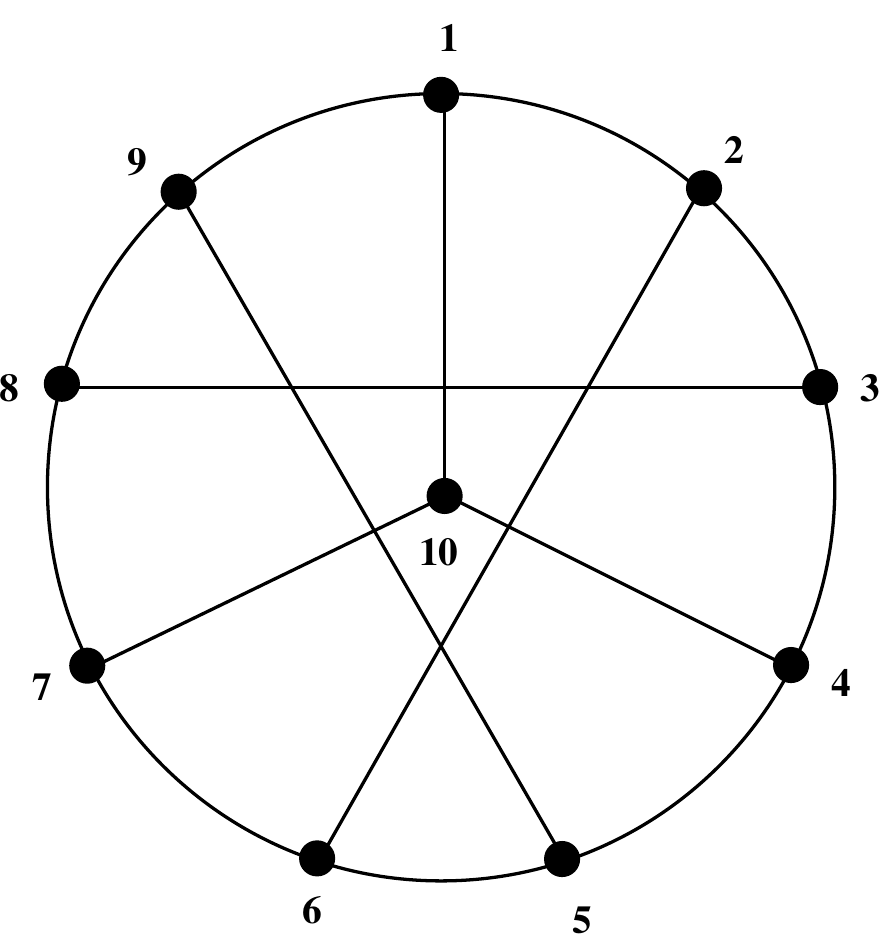}}
    \caption{An alternative drawing of the Petersen graph in the
      plane. This embedding reveals an $S_3$ permutation symmetry about
      the central point.}
    \label{figure:D2(Petersen)2}
  \end{figure}}

Because the algebra is Lorentzian (with a metric that coincides
with the metric induced from the embedding in $E_{10}$), it does
not need to be enlarged by any further generator to be compatible
with the Hamiltonian constraint.\\
\indent It is interesting to examine the symmetries of the various
embeddings of the Petersen graph in the plane and the connection to
the Desargues configurations. The embedding in
Figure~\ref{figure:D2(Petersen)} clearly exhibits a
$\mathbb{Z}_{5}\times \mathbb{Z}_2$-symmetry, while the Desargues
configuration in Figure~\ref{figure:GDesargues} has only a
$\mathbb{Z}_2$-symmetry. Moreover, the embedding of the Petersen graph
shown in Figure~\ref{figure:D2(Petersen)2} reveals yet another
symmetry, namely an $S_3$ permutation symmetry about the central
point, labeled ``10''. In fact, the external automorphism group of
the Petersen graph is $S_5$, so what we see in the various embeddings
are simply subgroups of $S_5$ made manifest. It is not clear how these
symmetries are realized in the Desargues configuration that seems to
exhibit much less symmetry.


\subsection{Further comments}

\begin{itemize}
\item The analysis of the present section exhibits subgroups of the
  Coxeter group \index{Coxeter group}$E_{10}$, with the property that their Coxeter
  exponents $m_{ij}$ (see Section~\ref{section:Coxeter}) are either 2
  or 3, but never infinity\epubtkFootnote{When no Coxeter exponent
    $m_{ij}$ is equal to infinity, the Coxeter group is called
    2-\emph{spherical}. 2-spherical Coxeter subgroups of $E_{10}$
    are rare~\cite{Caprace}.}. Furthermore, the associated Coxeter
  graphs all have incidence index $\mc{I}=3$, meaning that each node
  in the graph is connected to three and only three other nodes. A
  classification of all rank~10 and 11 Coxeter groups with these
  properties has been given in~\cite{Henneaux:2006bw}.
\item Integrability of sigma models for ``cosmological billiards'' in
  relation to dimensional reduction to three dimensions has been
  extensively investigated in~\cite{Fre:2003ep, Fre:2003tg,
  Fre:2005sr, Fre:2005si, Fre:2007hd}.
\end{itemize}

\newpage


\section{Conclusions}
\label{section:conclusions}
\setcounter{equation}{0}

In this review, we have investigated the remarkable structures that
emerge when studying gravitational theories in the BKL-limit,
i.e., close to a spacelike singularity. Although it has been known for
a long time that in this limit the dynamics can be described in
terms of billiard motion in hyperbolic space, it is only recently that
the connection between the billiards and Coxeter groups have been
uncovered. Furthermore, the relevant Coxeter groups turn out to
be the Weyl groups of the Lorentzian Kac--Moody algebra obtained by
double extension (sometimes twisted) of the U-duality algebra
appearing upon dimensional reduction to three dimensions.

These results, which in our opinion are solid and here to stay,
necessitate some mathematical background which is not part of the average physicist's working knowledge. For this reason, we have also
devoted a few sections to the development of the necessary
mathematical concepts.

We have then embarked on the exploration of more speculative
territory. A natural question that arises is whether or not the
emergence of Weyl groups of Kac--Moody algebras in the BKL-limit has a
profound meaning independently of the BKL-limit (which would serve
only as a ``revelator'') and could indicate that the gravitational
theories under investigation -- possibly supplemented by additional
degrees of freedom -- possess these infinite Kac--Moody algebras as
``hidden symmetries'' (in any regime). The existence of these
infinite-dimensional symmetries was also advocated in the pioneering
work~\cite{Julia:1980gr} and more
recently~\cite{E11andMtheory, Schnakenburg:2001ya,
  Schnakenburg:2002xx, Henry-Labordere:2002dk, Henry-Labordere:2002xh,
  SymmetryOfMtheories, Henry-Labordere:2003rd, Schnakenburg:2004vd,
  West:2004st} from a somewhat different point of view. It is also
argued in those references that even bigger symmetries ($E_{11}$ that
contains $E_{10}$, or Borcherds subalgebras) might actually be relevant. In
order to make the conjectured $E_{10}$-symmetry manifest (which is
perhaps itself part of a bigger symmetry), we have investigated a
nonlinear sigma model for the coset space
$\mc{E}_{10}/\mc{K}(\mc{E}_{10})$ using the level decomposition
techniques introduced in~\cite{DHN2}. Although very suggestive and partially successful, this approach exhibits limitations which, in spite
of many efforts, have not yet been overcome. It is likely that new
ideas are needed, or that the implementation of the symmetry must be
made in a more subtle fashion, where duality will perhaps play a more
central role.

Independently of the way they are actually implemented, it appears
that infinite-dimensional Kac--Moody algebras (e.g, $E_{10}$ or,
perhaps, $E_{11}$) do encode important features of gravitational
theories, and the idea that they constitute essential elements of the final
formulation will surely play an important role in future developments.

\newpage


\section{Acknowledgements}
\label{section:acknowledgements}

We are grateful to Mauricio Leston who took an active part in the
early stages of this work. We also acknowledge the precious insight
into the subject gained through discussions and collaborations with
many persons, whose (non-exhaustive) list includes Pierre Bieliavsky,
Martin Cederwall, Thibault Damour, Sophie de Buyl, Francois Englert,
Jarah Evslin, Luca Forte, Laurent Houart, Bernard Julia, Axel
Kleinschmidt, Arjan Keurentjes, Stanislav Kuperstein, Hermann Nicolai,
Bengt E.W.\ Nilsson, Jakob Palmkvist, Louis Paulot, Christoffer
Petersson, Christiane Schomblond, Daniel H.\ Wesley and Peter West. In
addition, we thank Jakob Palmkvist for carefully reading the
manuscript and giving many useful comments. This work was supported in part by
IISN-Belgium (convention 4.4511.06 (M.H.\ and P.S) and convention
4.4505.86 (M.H.\ and D.P)), by the Belgian National Lottery, by the
European Commission FP6 RTN programme MRTN-CT-2004-005104 for which
M.H.\ and D.P.\ are associated with VUB, and by the Belgian Federal
Science Policy Office through the Interuniversity Attraction Pole
P5/27.

\newpage


\appendix


\section{Proof of Some Important Properties of the Bilinear Form}
\label{appendix_1}

We demonstrate in this appendix the properties of the bilinear
form $B$ associated with the geometric realisation of a Coxeter
group $\mf{C}$. Recall that the matrix $$(B_{ij}) = \left( -\cos
\left( \frac{\pi}{m_{ij}} \right) \right)$$ has only 1's on the
diagonal and non-positive numbers off the diagonal. Recall also
that a vector $v$ is said to be positive if and only if \emph{all}
its components $v_i$ are strictly positive, $v_i >0$; this is
denoted $v>0$. Similarly, a vector $v$ is non-negative, $v \geq 0$,
if and only if \emph{all} its components $v_i$ are non-negative,
$v_i \geq 0$. Finally, a vector is non-zero if and only if at
least one of its component is non-zero, which is denoted $v
\not=0$. Our analysis is based on reference~\cite{Kac}. We shall
assume throughout that $B$ is indecomposable.

\begin{maintheorem}
  
  \begin{enumerate}
  \item The Coxeter group $\mf{C}$ is of finite type if and only if
    there exists a positive vector $v_i >0$ such that $\sum_j B_{ij} v_j >0$.
    \label{ml_case_1}
  \item The Coxeter group $\mf{C}$ is of affine type if and only if
    there exists a positive vector $v_i >0$ such that $\sum_j B_{ij} v_j =0$.
    \label{ml_case_2}
  \item The Coxeter group $\mf{C}$ is of indefinite type if and only
    if there exists a positive vector $v_i >0$ such that $\sum_j B_{ij} v_j < 0$.
    \label{ml_case_3}
  \end{enumerate}
  
  \noindent
  These cases are mutually exclusive and exhaust all possibilities.
  
  \begin{newproof}
    The proof follows from a series of lemmata. The
    inequalities $v \geq 0$ define a convex cone, namely the first
    quadrant $Q$. Similarly, the inequalities $Bv \geq 0$ define also a
    convex cone $K_B$. One has indeed:
    \begin{displaymath}
      u, v \in K_B
      \quad \Rightarrow \quad
      \lambda u + (1 - \lambda ) v
      \in K_B
      \qquad
      \forall \lambda \in [0,1].
    \end{displaymath}
    Note that one has also
    \begin{displaymath}
      v \in K_B \quad \Rightarrow \quad \lambda v \in K_B
      \qquad
      \forall \lambda \geq 0
    \end{displaymath}
    and $\ker B = \{ v \vert B v = 0 \} \subset K_B$. There are three
    distinct cases for the intersection $K_B \cap Q$:
    \begin{enumerate}
    \item Case 1: $K_B \cap Q \not= \{0\}, \; K_B \subset Q $.
    \item Case 2: $K_B \cap Q \not= \{0\}, \; K_B \not\subset Q $.
    \item Case 3: $K_B \cap Q = \{0\}$.
    \end{enumerate}
    
    These three distinct cases correspond, as we shall now show, to
    the three distinct cases of the theorem. To investigate these
    distinct cases, we need the following lemmata:

    \begin{lemma}
      The conditions $B v \geq 0$ and $v \geq 0$ imply either $v >0$
      or $v =0$. In other words
      \begin{displaymath}
        K_B \cap Q \equiv
        \{ v \vert B v \geq 0 \} \cap \{ v \vert v \geq 0 \}
        \subset \{v \vert v >0\} \cup \{v = 0\}.
      \end{displaymath}
      
      \begin{newproof}
        Assume that $v \geq 0$ fulfills $Bv \geq 0$ and has at least
        one component equal to zero. We shall show that all its
        components are then zero. Assume $v_i = 0$ for $i = 1, \cdots,
        s$ and $v_i >0$ for $i >s$. One has $1 \leq s \leq n$ (with no
        non-vanishing component $v_i$ if $s = n$). From $Bv \geq 0$
        one gets $(Bv)_i = \sum_{j=1}^n B_{ij} v_j = \sum_{j= s+1}^n
        B_{ij} v_j \geq 0$. Take $i \leq s$. As $j >s$ in the previous
        sum, one has $B_{ij} < 0 $ and thus $\sum_{j= s+1}^n B_{ij}
        v_j \leq 0$, which implies $\sum_{j= s+1}^n B_{ij} v_j =
        0$. As $v_j >0$ ($j >s$), this leads to $B_{ij} = 0$ for $i
        \leq s$ and $j >s$. The matrix $B$ would be decomposable,
        unless $s = n$, i.e. when all components $v_i$ vanish.
      \end{newproof}
      \label{lemma_a_1}
    \end{lemma}
    
    \begin{lemma}
      Consider the system of linear homogeneous inequalities
      \begin{displaymath}
        \lambda_\alpha \equiv \sum_i a_{\alpha i} v_i> 0
      \end{displaymath}
      on the vector $v$. This system possesses a solution if and only
      if there is no set of numbers $\mu_\alpha \geq 0$ that are not
      all zero such that $\sum_\alpha \mu_\alpha a_{\alpha i} = 0$.
      
      \begin{newproof}
        This is a classical result in the theory of linear
        inequalities (see~\cite{Kac}, page~47). 
      \end{newproof}
      \label{lemma_a_2}
    \end{lemma}
    
    We can now study more thoroughly the three cases listed above.

    \subsubsection*{Case~\protect\ref{ml_case_1}: \boldmath $ K_B \cap Q \not=
      \{0\} $, $ K_B \subset Q $}

    In that case, one has
    \begin{displaymath}
      Bv \geq 0 \quad \Rightarrow \quad v >0 \mbox{ or } v = 0
    \end{displaymath}
    by Lemma~\ref{lemma_a_1}. Furthermore, $K_B$ cannot contain a
    nontrivial subspace $W$ since $w \in W$ implies $-w \in W$, but
    only one of the two can be in $Q$ when $w \not=0$. Hence $\ker B
    =0$, i.e., $\det B \not=0$ and
    \begin{displaymath}
      Bv = 0 \quad \Rightarrow \quad v = 0.
    \end{displaymath}
    This excludes in particular the existence of a vector $u> 0$ such
    that $Bu <0$ or $Bu = 0$.

    Finally, the interior of $K_B$ is non-empty since $B$ is
    nondegenerate. Taking a non-zero vector $v$ such that $Bv>0$, one
    concludes that there exists a vector $v>0$ such that $Bv>0$. This
    shows that Case~\ref{ml_case_1} corresponds to the first case in
    the theorem. We shall verify below that $B_{ij}$ is indeed
    positive definite.
    
    \subsubsection*{Case~\protect\ref{ml_case_2}: \boldmath $ K_B \cap Q \not=
      \{0\} $, $ K_B \not\subset Q $}

    $K_B$ reduces in that case to a straight line. Indeed, let $v
    \not=0$ be an element of $K_B \cap Q$ and let $w \not=0$ be in
    $K_B$ but not in $Q$. Let $\ell$ be the straight line joining $w$
    and $v$. Consider the line segment from $w$ to $v$. This line
    segment is contained in $K_B$ and crosses the boundary $\partial
    Q$ of $Q$ at some point $r$. But by Lemma~\ref{lemma_a_1}, this point
    $r$ must be the origin. Thus, $w = \mu v$, for some real number
    $\mu <0$. This implies that the entire line $\ell$ is in $K_B$
    since $v \in K_B \Rightarrow \lambda v \in K_B$ for all $\lambda
    >0$, and also for all $\lambda <0$ since $w \in K_B$.

    Let $q$ be any other point in $K_B$. If $q \notin Q$, the segment
    joining $q$ to $v$ intersects $\partial Q$ and this can only be at
    the origin by Lemma~\ref{lemma_a_1}. Hence $q \in \ell$. If $q \in Q$,
    the segment joining $q$ to $w$ intersects $\partial Q$ and this
    can only be at the origin by Lemma~\ref{lemma_a_1}. Hence, we find
    again that $q \in \ell$. This shows that $K_B$ reduces to the
    straight line $\ell$.
    
    Since $v \in K_B \Rightarrow -v \in K_B$, one has $Bv = 0 \quad
    \forall v \in K_B$. Hence, $Bv \geq 0 \Rightarrow Bv = 0$, which
    excludes the existence of a vector $v >0$ such that $Bv < 0$ (one
    would have $B(-v) >0$ and hence $Bv = 0$). Furthermore, there
    exists $v >0$ such that $Bv = 0$. This shows that Case~\ref{ml_case_2}
    corresponds to the second case in the theorem. We shall verify
    below that $B_{ij}$ is indeed positive semi-definite. Note that
    $\det B = 0$ and that the corank of $B$ is one.
    
    \subsubsection*{Case~\protect\ref{ml_case_3}: \boldmath $ K_B \cap Q = \{0\}$}

    In that case, there is a vector $v >0$ such that $Bv <0$, which
    corresponds to the third case in the theorem. Indeed, consider
    the system of homogeneous linear inequalities
    \begin{displaymath}
      - \sum_j B_{uj} v_j >0,
      \qquad
      v_j >0.
    \end{displaymath}
    By Lemma~\ref{lemma_a_2}, this system possesses a solution if and
    only if there is no non-trivial $\mu_{\alpha} \equiv (\mu_i,
    \bar{\mu}_i) \geq 0 $ such that $\sum_i \mu_i (-B_{ij}) +
    \bar{\mu}_j = 0 $.

    Consider thus the equations $\sum_i \mu_i (-B_{ij}) + \bar{\mu}_j
    = 0 $ for $\mu_\alpha \geq 0$, or, as $B_{ij}$ is symmetric,
    $\sum_j B_{ij} \mu_j = \bar{\mu}_i$. Since $\bar{\mu}_i \geq 0$,
    these conditions are equivalent to $\sum_j B_{ij} \mu_{j} \geq 0$
    (if $\sum_j B_{ij} \mu_{j} \geq 0$, one defines $\bar{\mu}_i$
    through $\sum_j B_{ij} \mu_j = \bar{\mu}_i$), i.e., $\mu \in K_B$.
    But $\mu_i \geq 0$, i.e., $\mu \in Q$, which implies $\mu_i = 0$
    and hence also $\bar{\mu}_i = 0$. The $\mu_{\alpha}$ all vanish
    and the general solution $\mu\geq 0$ to the equations $\sum_i
    \mu_i (-B_{ij}) + \bar{\mu}_j = 0 $ is accordingly trivial.
    
    To conclude the proof of the main theorem, we prove the 
    following proposition:
    
    \begin{proposition}
      The Coxeter group $\mf{C}$ belongs to Case~\ref{ml_case_1} if
      and only if $B$ is positive definite; it belongs to
      Case~\ref{ml_case_2} if and only if $B$ is positive
      semi-definite with $\det B = 0$.

      \begin{newproof}
        If $B$ is positive semi-definite, then it belongs to
        Case~\ref{ml_case_1} or Case~\ref{ml_case_2} since otherwise
        there would be a vector $w >0$ such that $Bw < 0$ and thus
        $B_{ij} w_i w_j < 0$, leading to a contradiction. In the
        finite case, $B$ is positive definite and hence, $\det B
        \not=0$: This corresponds to Case~\ref{ml_case_1}. In the
        affine case, there are zero eigenvectors and $\det B = 0$:
        This corresponds to Case~\ref{ml_case_2}.

        Conversely, assume that the Coxeter group $\mf{C}$ belongs to
        Case~\ref{ml_case_1} or Case~\ref{ml_case_2}. Then there
        exists a vector $w$ such that $Bw \geq 0$. This yields $(B -
        \lambda I) w >0$ for $\lambda <0$ and therefore $B - \lambda
        I$ belongs to Case~\ref{ml_case_1} $\forall \lambda <0$. In
        particular, $\det (B - \lambda I) \not=0 \quad \forall \lambda
        <0$, which shows that the eigenvalues of $B$ are all
        non-negative: $B$ is positive semi-definite. We have seen
        furthermore that it has the eigenvalue zero only in
        Case~\ref{ml_case_2}.
      \end{newproof}
    \end{proposition}
    
    \noindent
    This completes the proof of the main theorem.
  \end{newproof}
\end{maintheorem}

\newpage


\section{Existence and ``Uniqueness'' of the Aligned Compact Real Form}
\label{appendix_2}

We prove in this appendix the crucial result that for any real
form of a complex semi-simple Lie algebra, one can always find a
compact real form aligned with it~\cite{Helgason,Knapp}.

Let $\GO$ be a specific real form of the semi-simple, complex Lie
algebra $\GC$. Let $\mf{c}_{0}$ be a compact real form of $\GC$. We may
introduce on $\GC$ two conjugations. A first one (denoted by
$\sigma$) with respect to $\GO$ and another one (denoted by
$\tau$) with respect to the compact real form $\mf{c}_{0}$. The product
of these two conjugations constitutes an automorphism
$\lambda=\sigma\tau$ of $\GC$. For any automorphism $\varphi$ we
have the identity
\begin{equation}
  \ad (\varphi Z)=\varphi\,\ad Z\,\varphi^{-1},
\end{equation}
and,
as a consequence, the invariance of  the Killing form with respect
to the automorphisms of the Lie algebra: 
\begin{equation}
  B(\varphi Z,\,\varphi Z')=
  \Tr(\ad{(\varphi\, Z)}\ad{(\varphi\, Z')})=
  \Tr(\varphi\,\ad{ Z}\,\varphi^{-1}\,
  \varphi\,\ad{ Z'}\,\varphi^{-1})=B( Z,\, Z').
\end{equation}
The automorphism $\lambda=\sigma\tau$ is symmetric with respect to the
Hermitian product $B^\tau$ defined by $B^\tau( X,  Y) = - B( X,
\tau(  Y))$. Indeed $(\sigma\tau)^{-1}\tau =\tau(\sigma\tau)$
implies that $B^{\tau}(\sigma \tau [ Z
],\, Z')=B^{\tau}( Z,\,\sigma \tau [ Z'  ])$. Thus its square
$\rho =(\sigma\tau)^{2}$ is positive definite. It can be proved
that $\rho^t \, (t\in \RR)$ is a one-parameter group of internal
automorphisms of $\GO$ such that\epubtkFootnote{To convince oneself of
the validity of this commutation relation, it suffices to check it
in a basis where the (finite-dimensional) matrix $\rho$ is
diagonal, using the symmetry of the matrix $\sigma\tau$.}
$\rho^t \,\tau=\tau\,\rho^{-t}$. It follows that
\begin{equation}
  \rho^{\f{1}{4}}\tau\rho^{-\f{1}{4}}\sigma=
  \rho^{\f{1}{2}}\tau\,\sigma=
  \rho^{-\f{1}{2}}\rho\,\tau\,\sigma=
  \rho^{-\f{1}{2}}\sigma\,\tau=
  \sigma\,\tau\,\rho^{-\f{1}{2}}=
  \sigma\,\rho^{\f{1}{4}}\tau\rho^{-\f{1}{4}}.
\end{equation}
In other words, the conjugation $\sigma$ always commutes with the
conjugation $\tilde \tau=\rho^{\frac 14}\tau\rho^{-\frac 14}$, which
is the conjugation with respect to the compact real algebra
$\rho^{\frac 14}[\mf{c}_{0}]$. This shows that the compact real form
$\rho^{\frac 14}[\mf{c}_{0}]$ is aligned with the given real form
$\GO$.

Note also that if there are two Cartan involutions, $\theta$ and
$\theta'$, defined on a real semi-simple Lie algebra, they are
conjugated by an internal automorphism. Indeed, as we just mentioned,
then an automorphism $\phi=((\theta\theta^{\p})^2)^{\f{1}{4}}$ exists,
such that $\theta$ and $\psi=\phi\theta^{\p}\phi^{-1}$ commute. If
$\psi\neq \theta$, the eigensubspaces of eigenvalues $+1$ and $-1$ of
these two involutions are disitnct but, because they commute, a vector
$X$ exists, such that $\theta[X]=X$ and $\psi[X]=-X$. For this vector
we obtain
\begin{equation}
  \begin{array}{rcl}
    0 & < & B^{\theta}(X, X)=-B(X, \theta[X])=-B(X, X),
    \\
    0 & < & B^{\psi}(X, X)= -B(X, \psi[X])=+B(X, X),
  \end{array}
\end{equation}
which constitutes a contradiction, and thus implies $\theta=\psi$. An
important consequence of this is that any real semi-simple Lie algebra
possesses a ``unique'' Cartan involution\epubtkFootnote{The
  uniqueness derives from the fact that the internal automorphism groups of
  $\mf{g}^{\mbb{R}}$ and $\mf{g}$ are identical.}. \index{Cartan involution}
In the same way, if $\mf{g}$ is a complex semi-simple Lie algebra, the only Cartan
involutions of $\mf{g}^{\mbb{R}}$ are obtained from the conjugation
with respect to a compact real form of $\mf{g}$; all compact real
forms being conjugated to each other by internal automorphisms.

\newpage

\bibliography{refsUpdated}

\newpage

\ifpdf
\phantomsection
\addcontentsline{toc}{section}{Index}
\printindex

\else
\printindex
\fi

\end{document}